\documentclass[traditabstract]{aa} 
\usepackage{graphicx,natbib,longtable,txfonts,lscape,lineno}
\newcommand{\FR}{FRI{\sl{CAT}}}
\newcommand{\sFR}{sFRI{\sl{CAT}}}
\newcommand{\ergs}{\>{\rm erg}\,{\rm s}^{-1}}
\newcommand{\kms}{$\rm{\,km \,s}^{-1}$}

\begin{document}

   \title{FRI{\sl{CAT}}: A FIRST catalog of FR~I radio galaxies.}

   \subtitle{}

   \author{A. Capetti\inst{1}
          \and
          F. Massaro\inst{2}
          \and
          R.D. Baldi\inst{3}
          }

   \institute{INAF-Osservatorio Astrofisico di Torino, via Osservatorio 20,
     10025 Pino Torinese, Italy, 
\and
Dipartimento di Fisica, Universit\`a degli
     Studi di Torino, via Pietro Giuria 1, 10125 Torino, Italy, 
\and
Department of
     Physics and Astronomy, University of Southampton, Highfield, SO17 1BJ, UK}
   \date{}

   \abstract {We built a catalog of 219 FR~I radio galaxies (FR~Is),
     called FRI{\sl{CAT}}, selected from a published sample and
     obtained by combining observations from the NVSS, FIRST, and SDSS
     surveys. We included in the catalog the sources with an
     edge-darkened radio morphology, redshift $\leq 0.15$, and
     extending (at the sensitivity of the FIRST images) to a radius
     $r$ larger than 30 kpc from the center of the host. We also
     selected an additional sample (\sFR) of 14 smaller (10 $<r<$ 30
     kpc) FR~Is, limiting to $z<0.05$. The hosts of the FRI{\sl{CAT}}
     sources are all luminous ($-21 \gtrsim M_r \gtrsim -24$), red
     early-type galaxies with black hole masses in the range $10^8
     \lesssim M_{\rm BH} \lesssim 3\times10^9 M_\odot$; the
     spectroscopic classification based on the optical emission line
     ratios indicates that they are all low excitation
     galaxies. Sources in the FRI{\sl{CAT}} are then indistinguishable
     from the FR~Is belonging to the Third Cambridge Catalogue of
     Radio Sources (3C) on the basis of their optical properties.
     Conversely, while the 3C-FR~Is show a strong positive trend
     between radio and [O~III] emission line luminosity, these two
     quantities are unrelated in the FRI{\sl{CAT}} sources; at a given
     line luminosity, they show radio luminosities spanning about two
     orders of magnitude and extending to much lower ratios between
     radio and line power than 3C-FR~Is. Our main conclusion is that
     the 3C-FR~Is just represent the tip of the iceberg of a much
     larger and diverse population of FR~Is.}   \maketitle

\section{Introduction}

\begin{figure*}
\includegraphics[width=6.3cm,height=6.3cm]{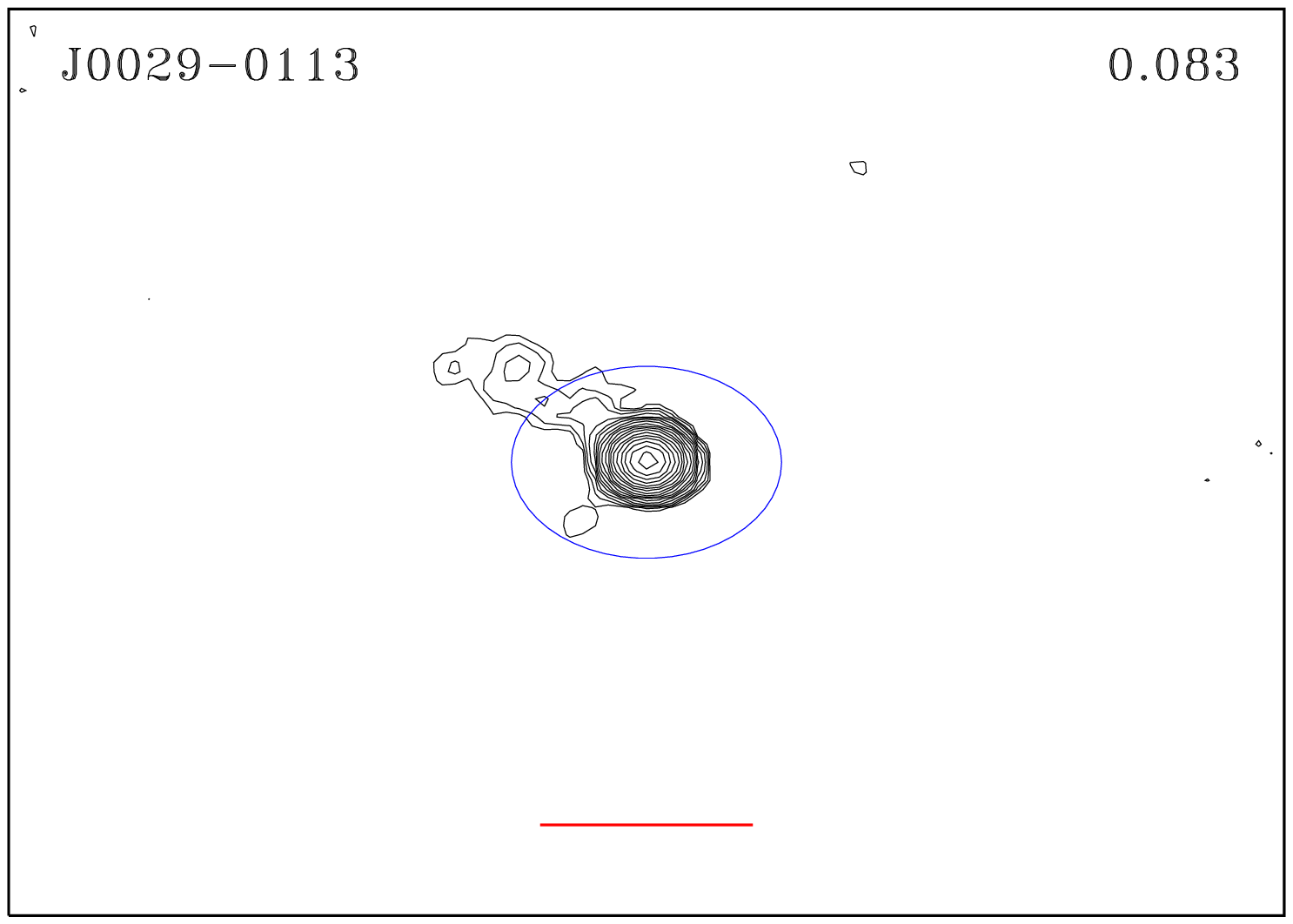} 
\includegraphics[width=6.3cm,height=6.3cm]{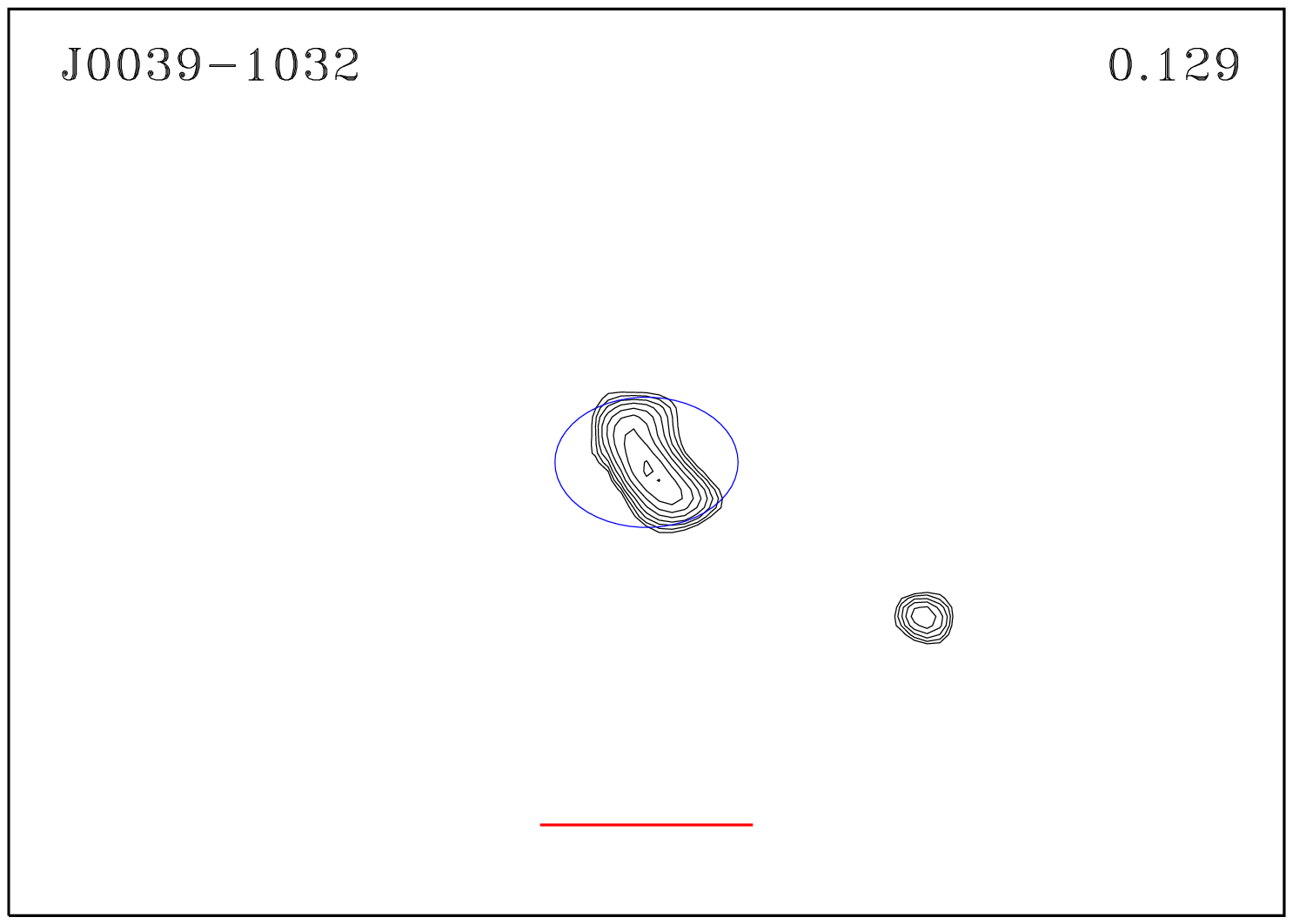} 
\includegraphics[width=6.3cm,height=6.3cm]{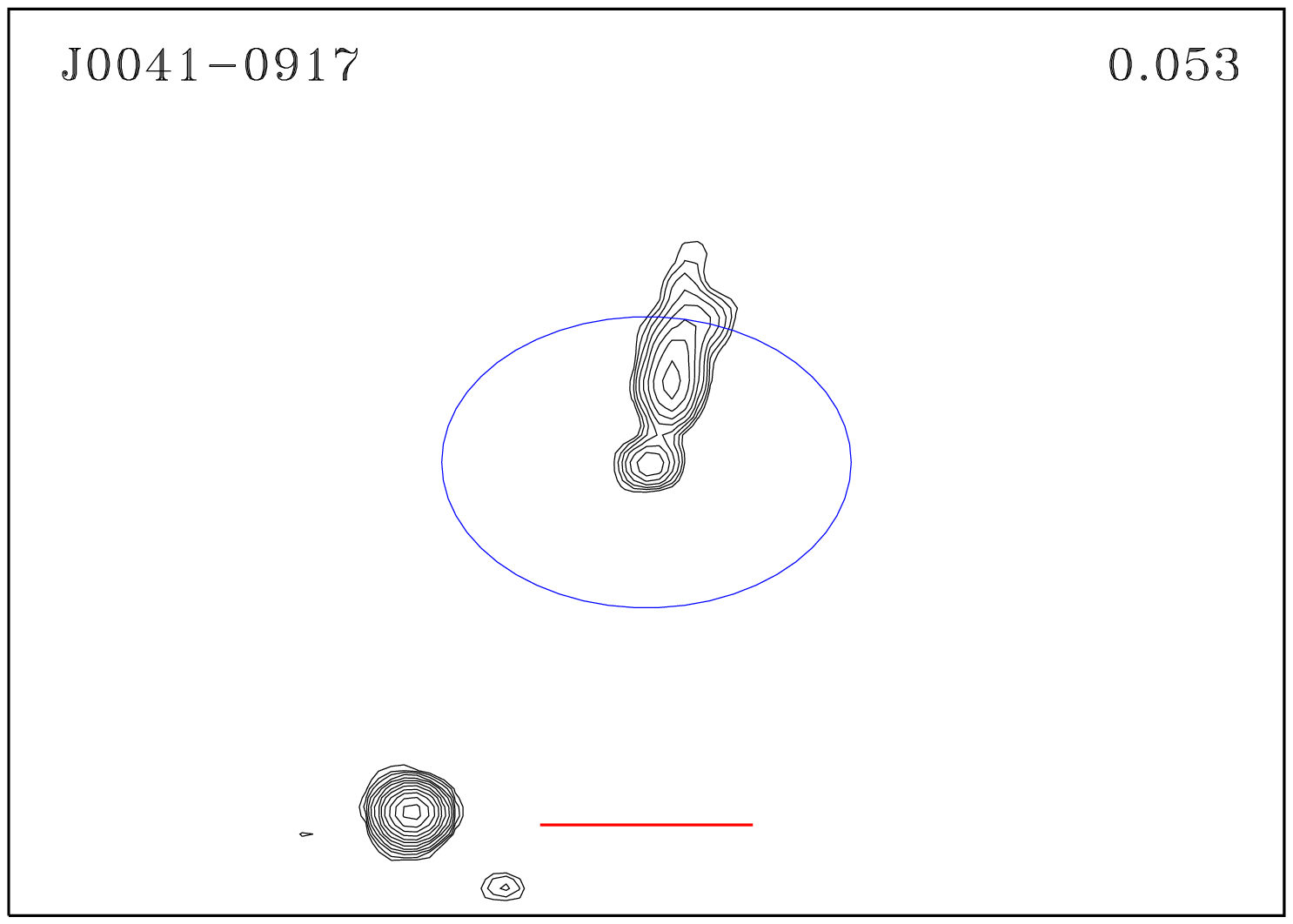} 

\includegraphics[width=6.3cm,height=6.3cm]{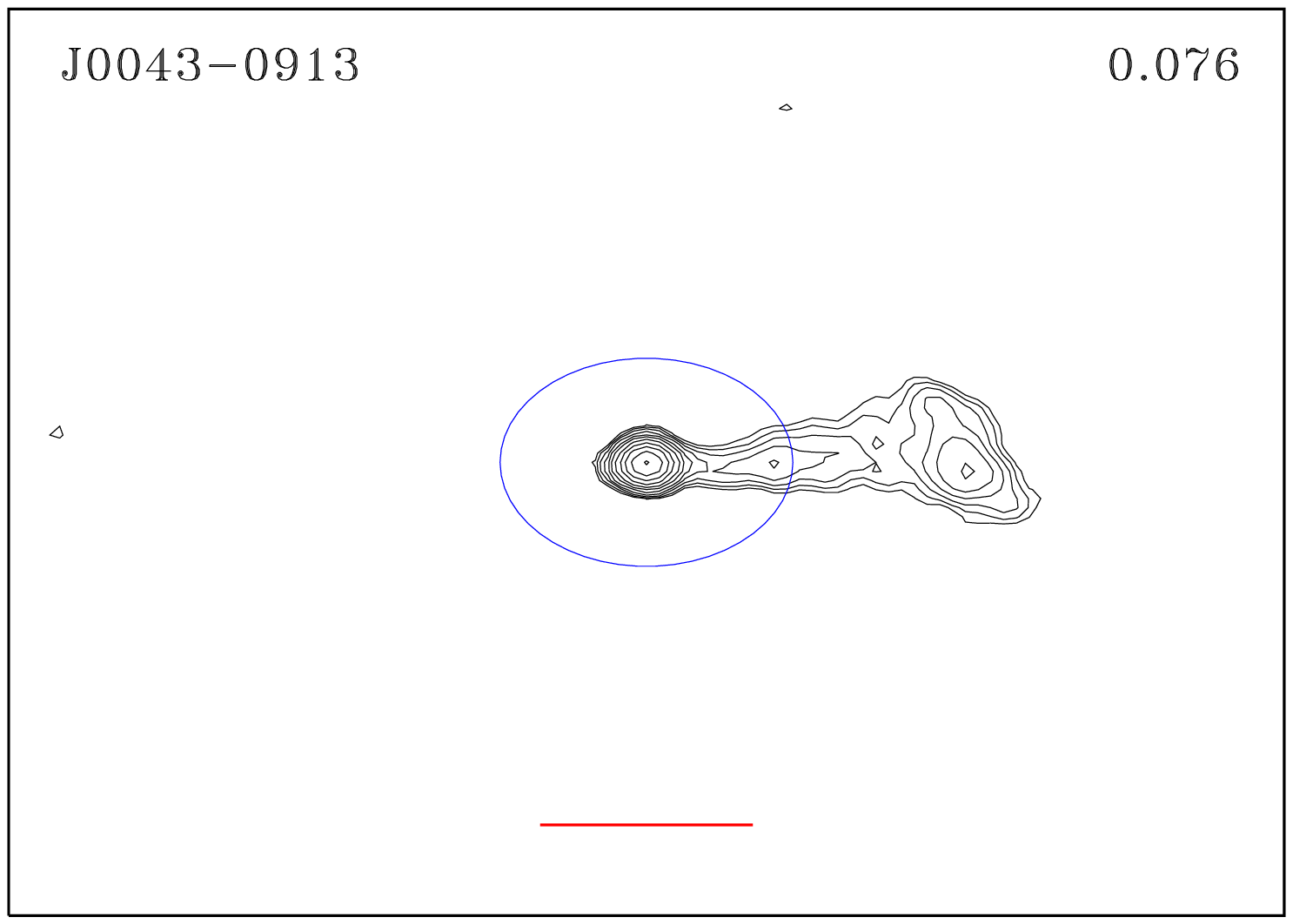} 
\includegraphics[width=6.3cm,height=6.3cm]{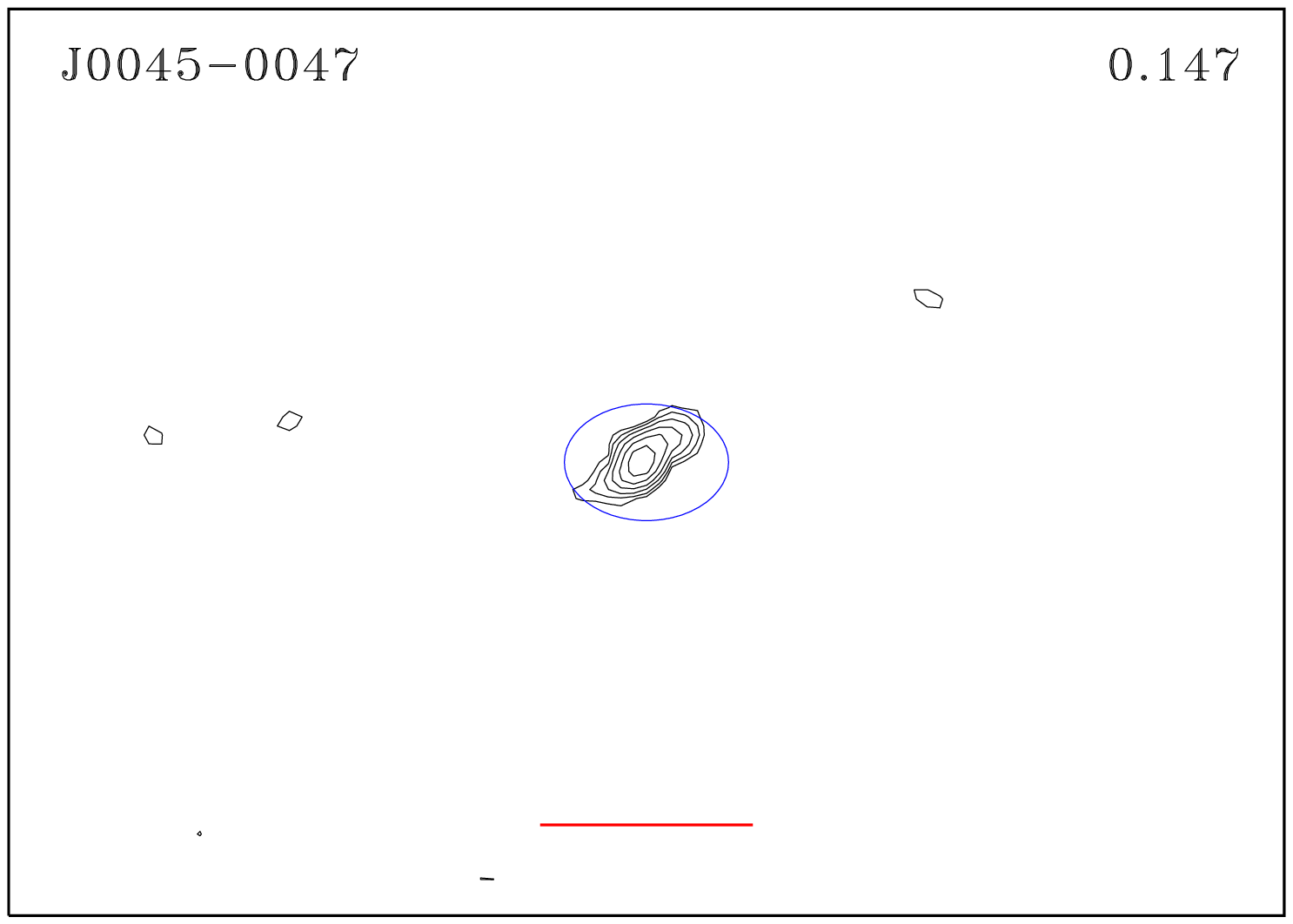} 
\includegraphics[width=6.3cm,height=6.3cm]{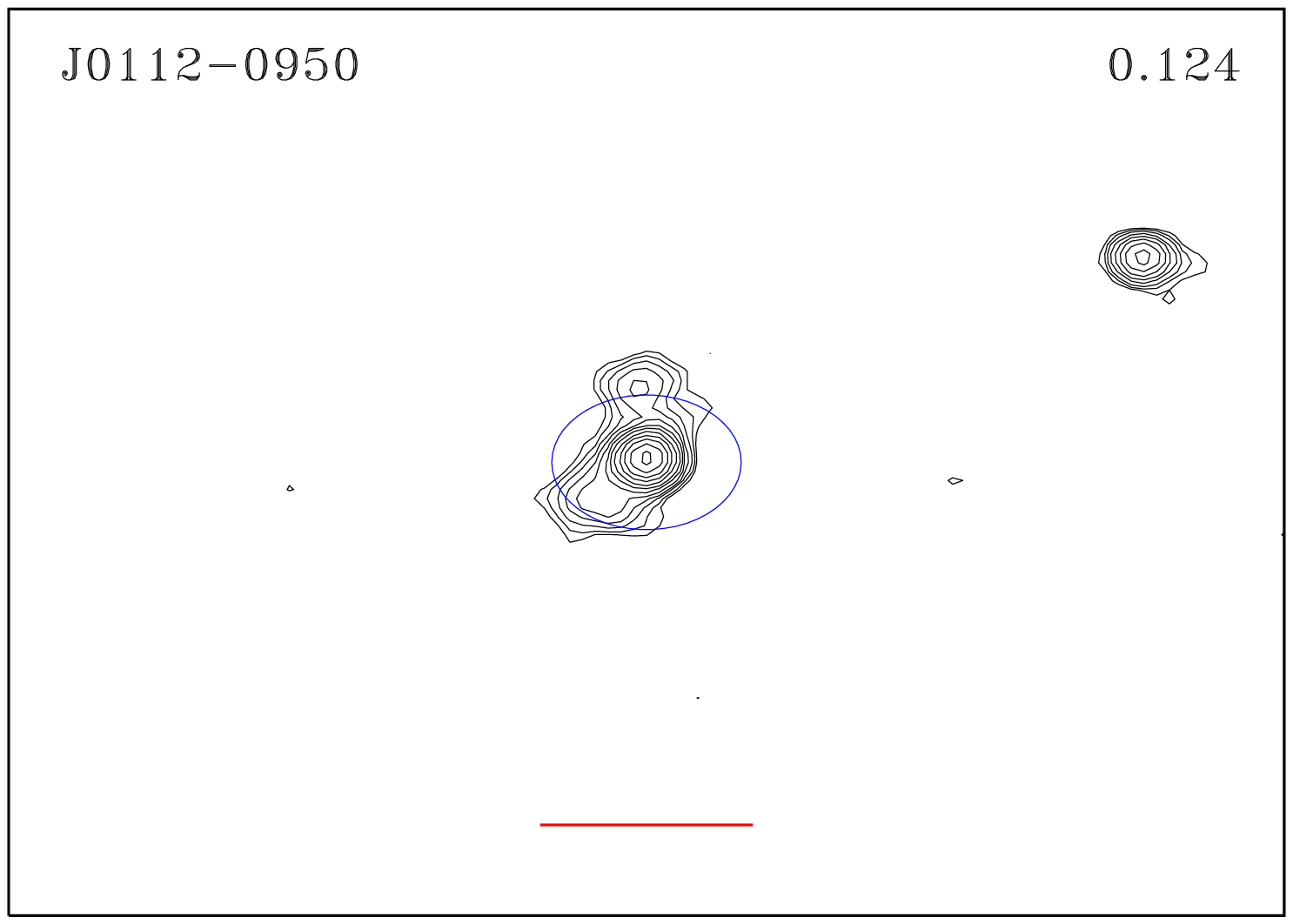} 

\includegraphics[width=6.3cm,height=6.3cm]{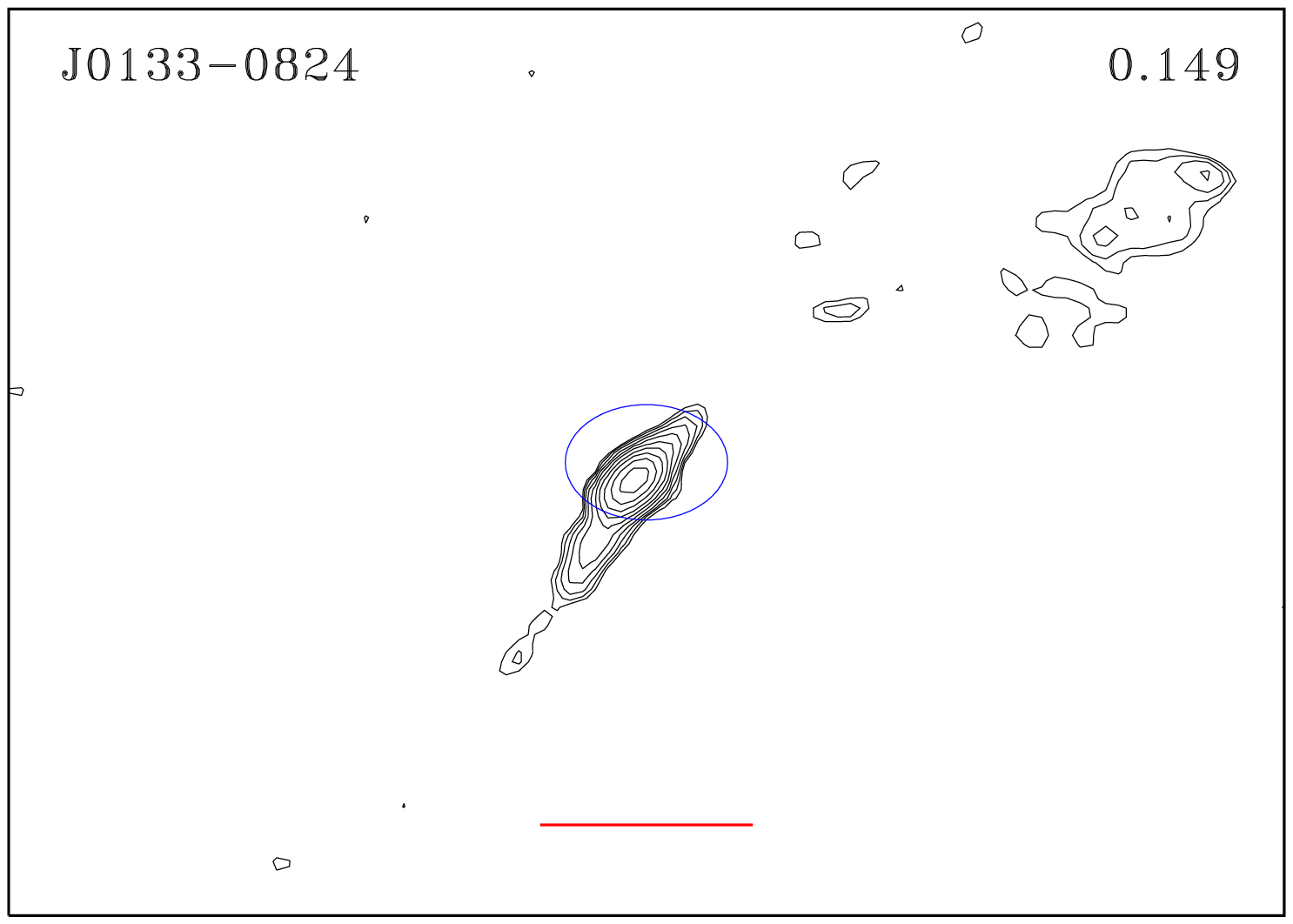} 
\includegraphics[width=6.3cm,height=6.3cm]{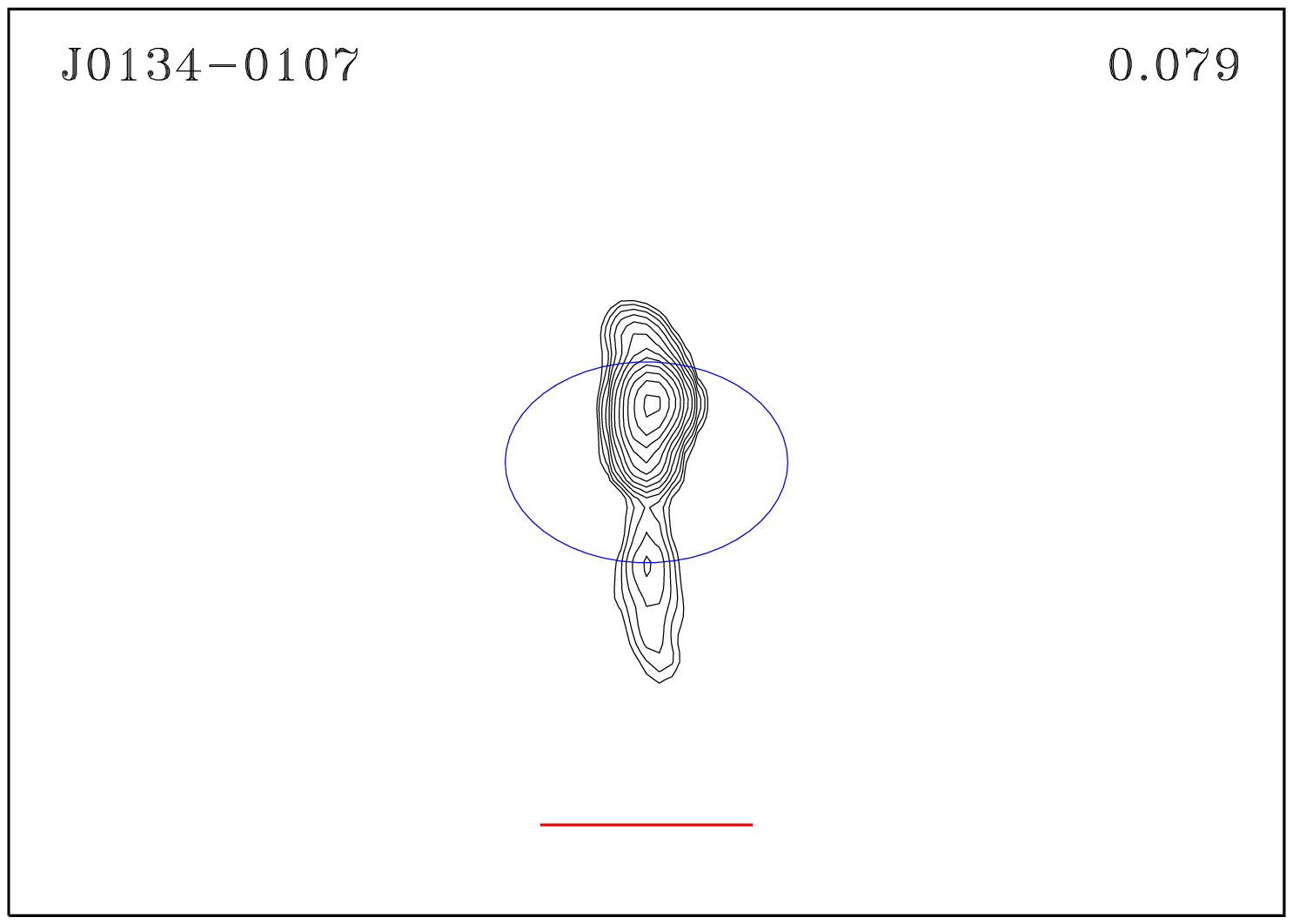} 
\includegraphics[width=6.3cm,height=6.3cm]{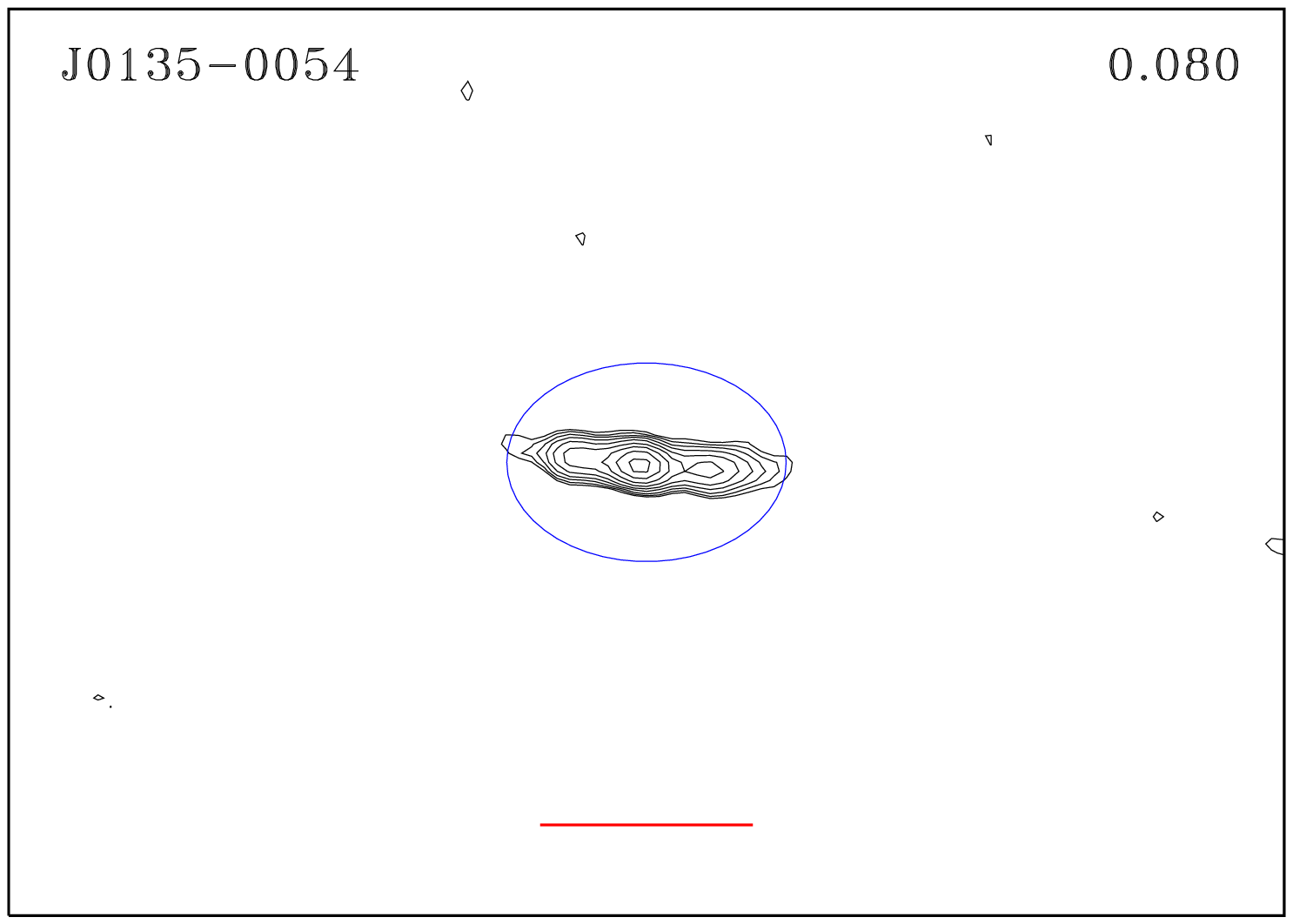} 

\includegraphics[width=6.3cm,height=6.3cm]{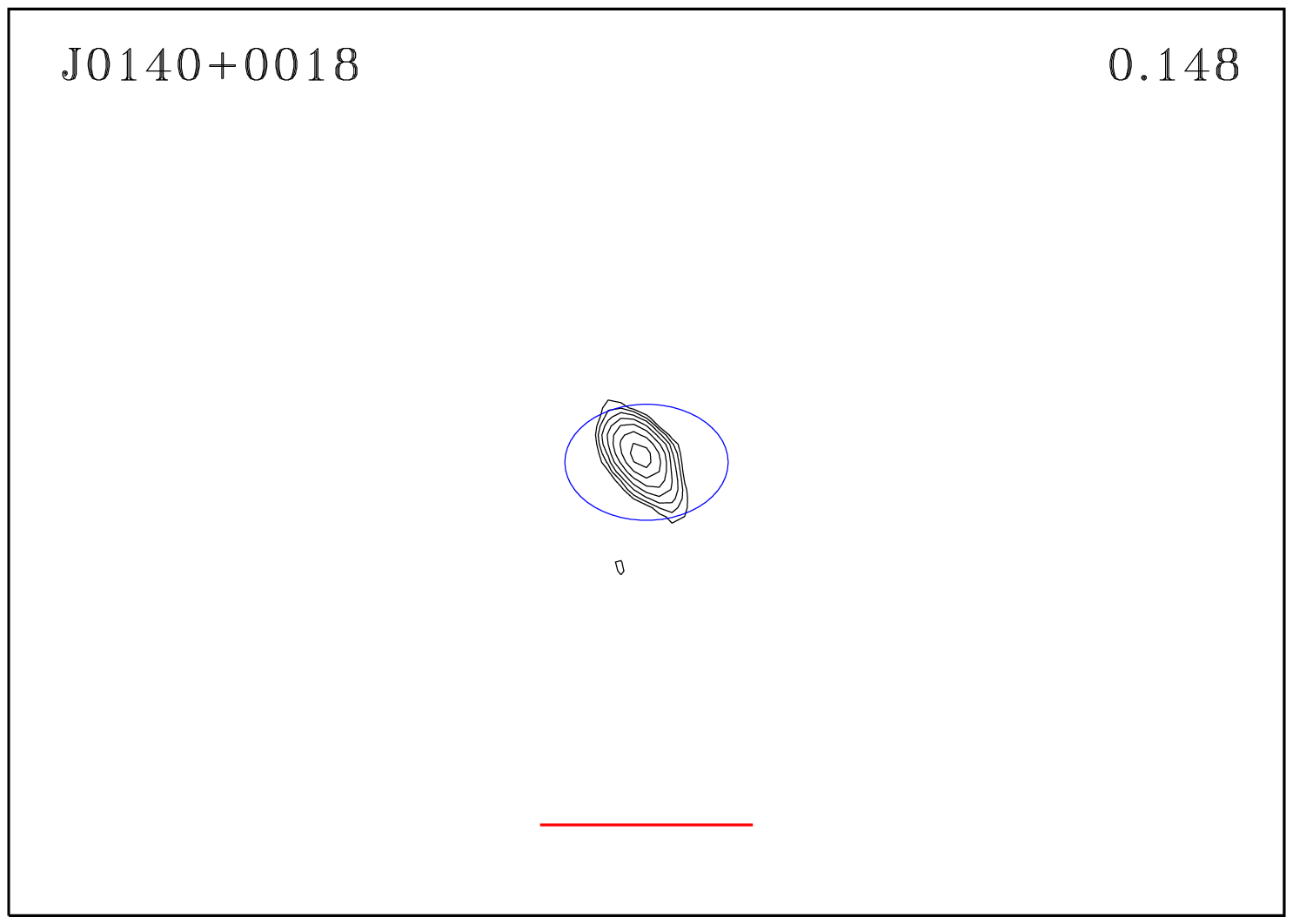} 
\includegraphics[width=6.3cm,height=6.3cm]{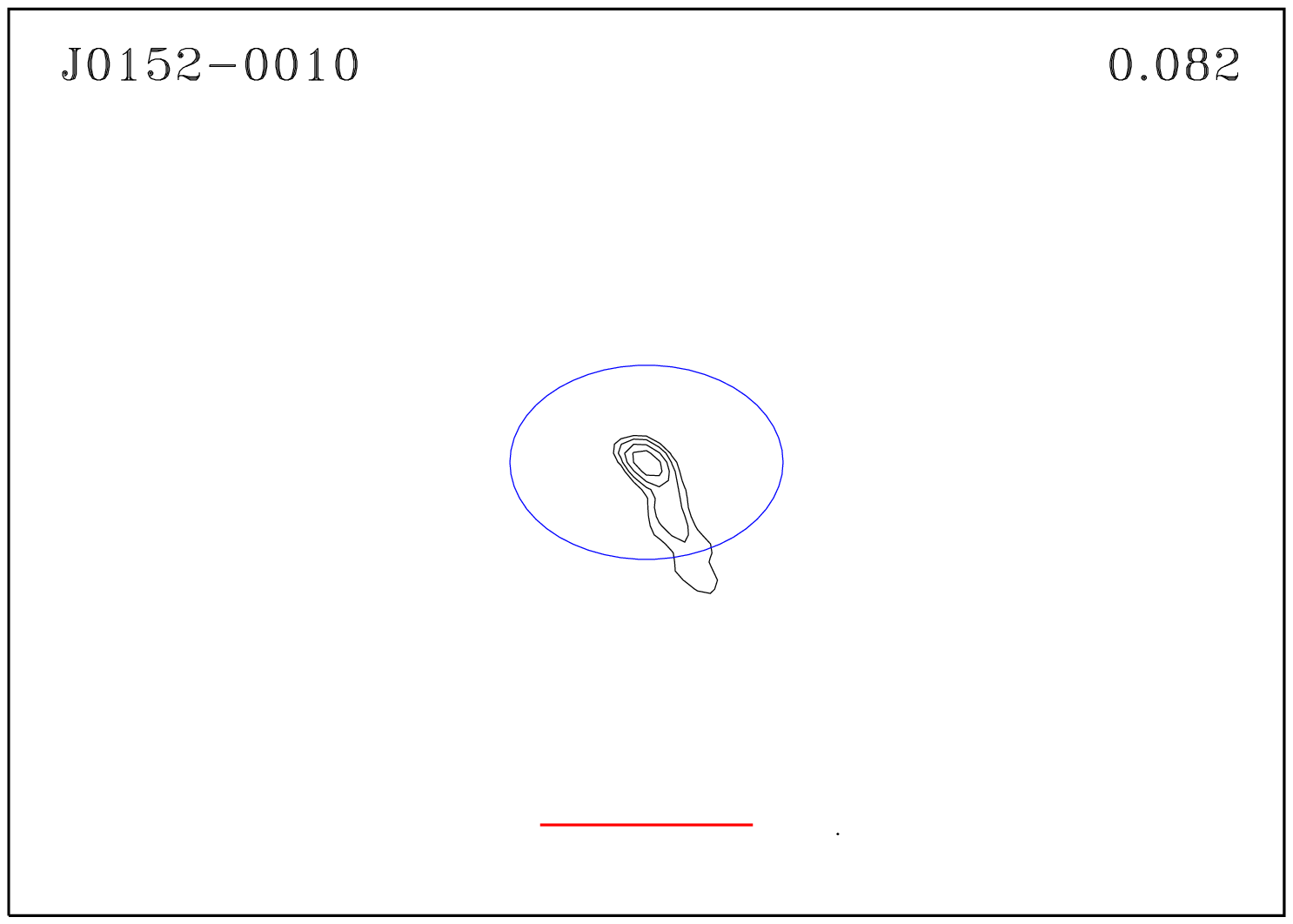} 
\includegraphics[width=6.3cm,height=6.3cm]{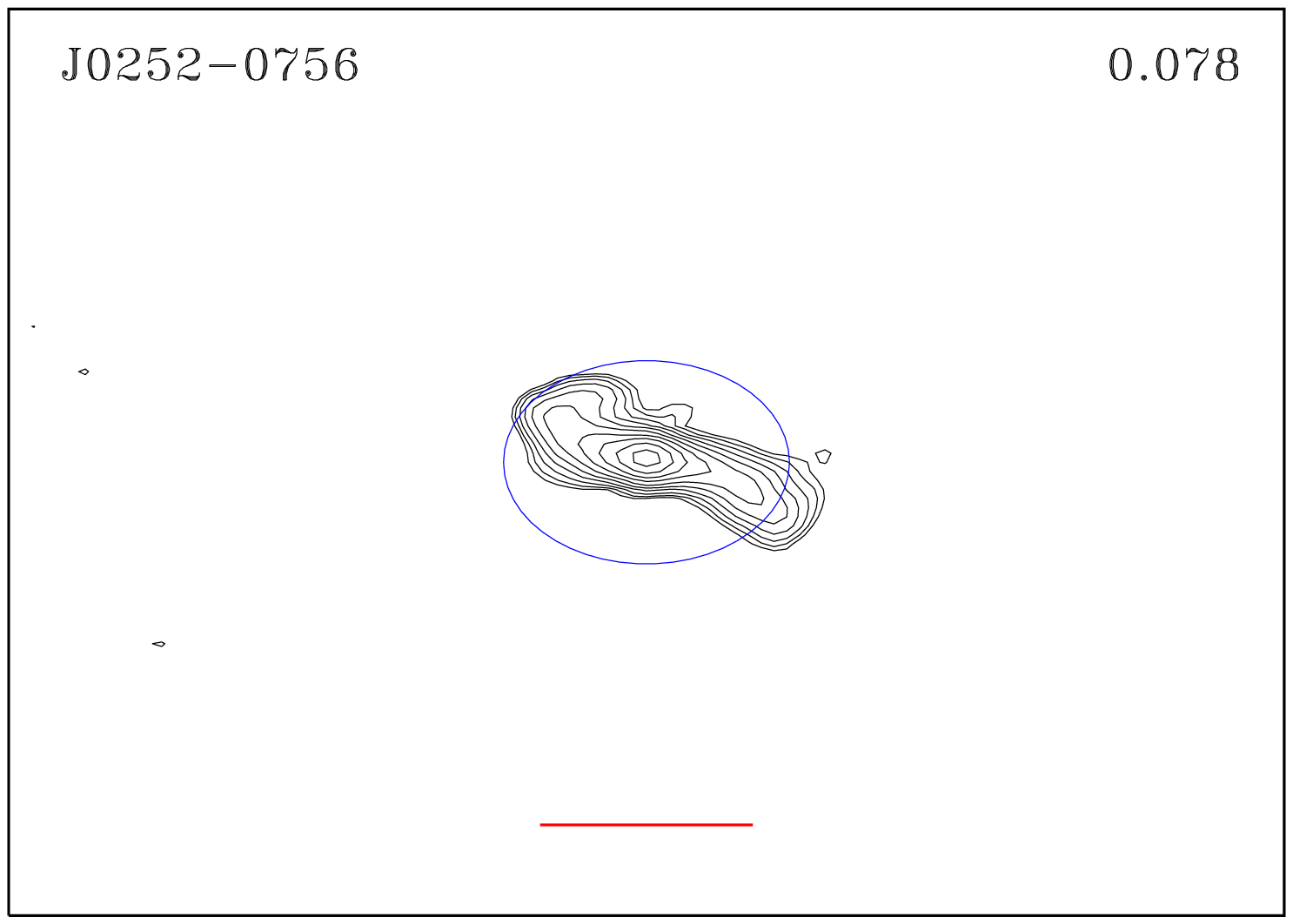} 
\caption{FIRST images of the first 12 \FR\ sources. Contours are drawn
  starting from 0.45 mJy/beam and increase with a geometric progression with a
  common ratio of $\sqrt2$. The field of view is 3'$\times$3'; the red tick at
  the bottom is 30$\arcsec$ long. The blue circle is centered on the host
  galaxy and has a radius of 30 kpc. The sources \FR\ name and redshift are
  reported in the upper corners.}
\label{images}
\end{figure*}

\begin{figure*}
\includegraphics[width=4.5cm,height=4.5cm]{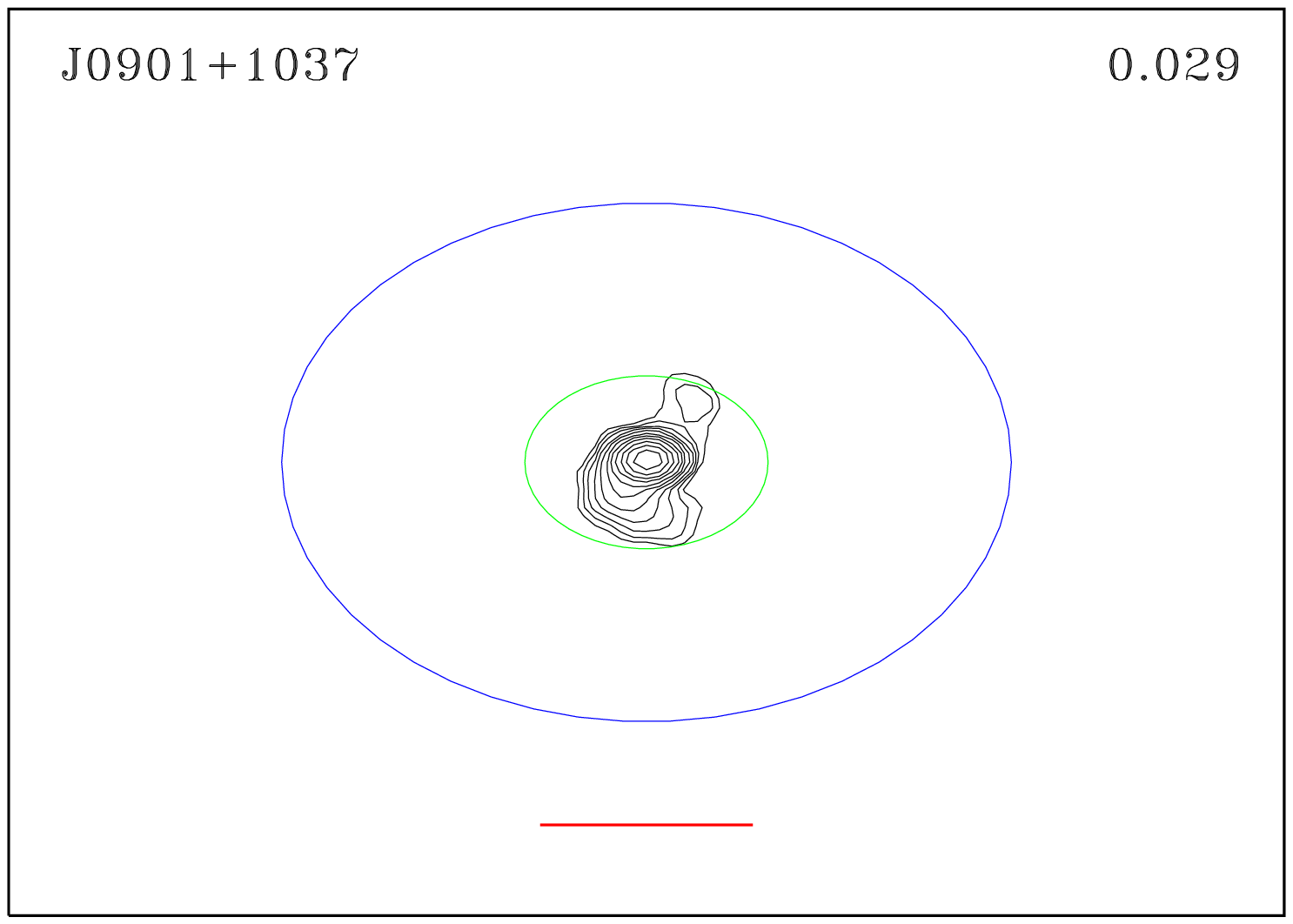}
\includegraphics[width=4.5cm,height=4.5cm]{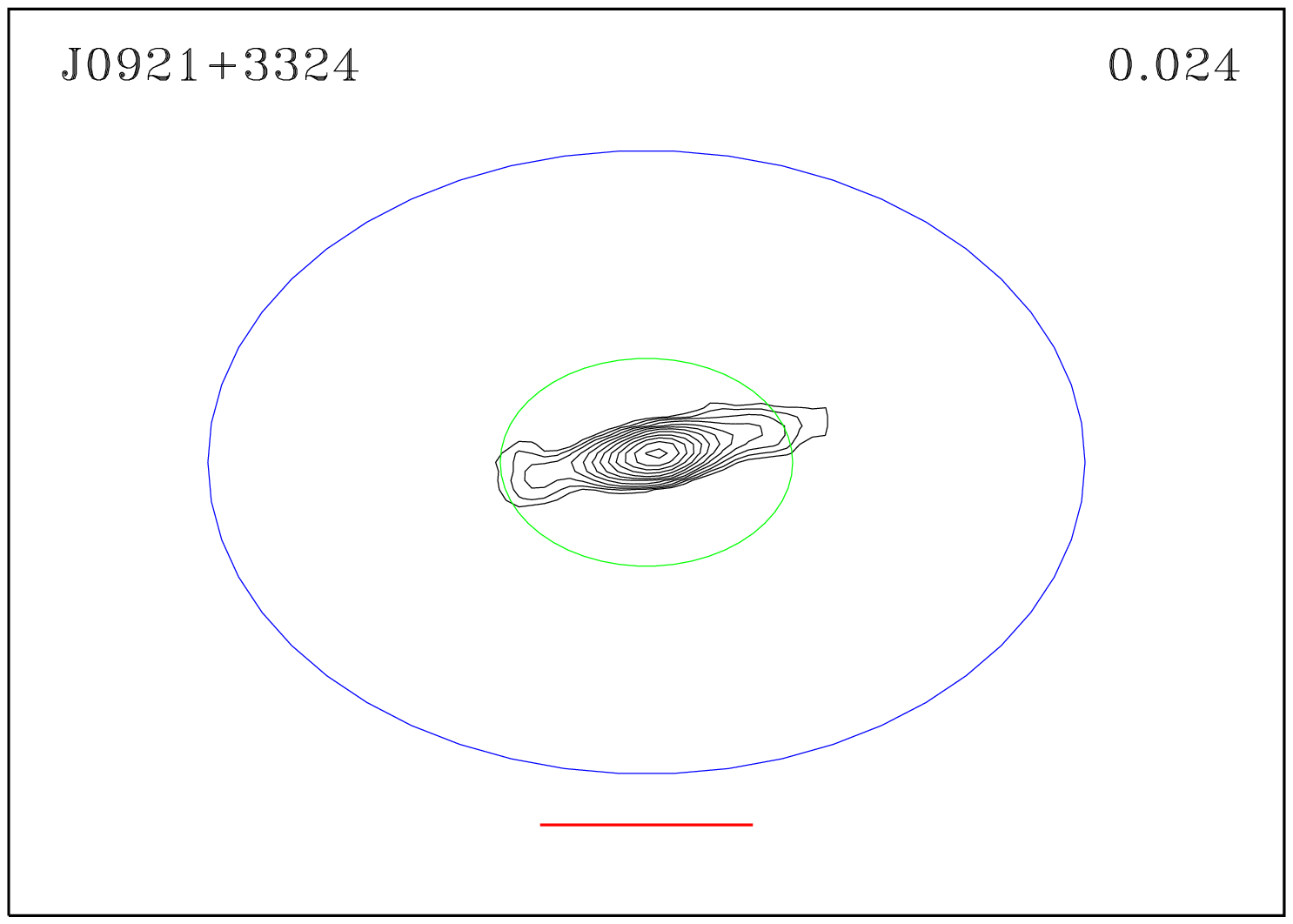}
\includegraphics[width=4.5cm,height=4.5cm]{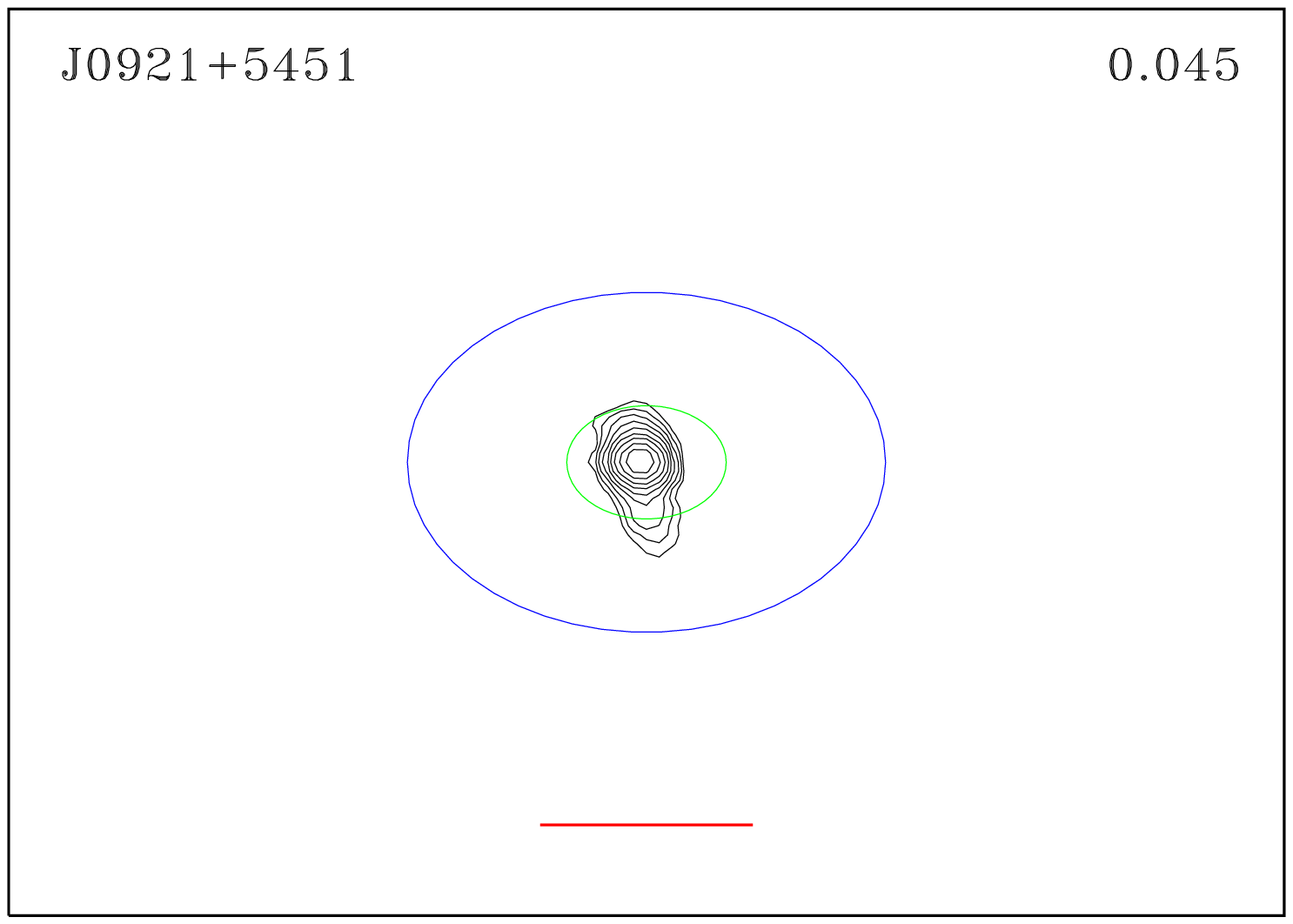}
\includegraphics[width=4.5cm,height=4.5cm]{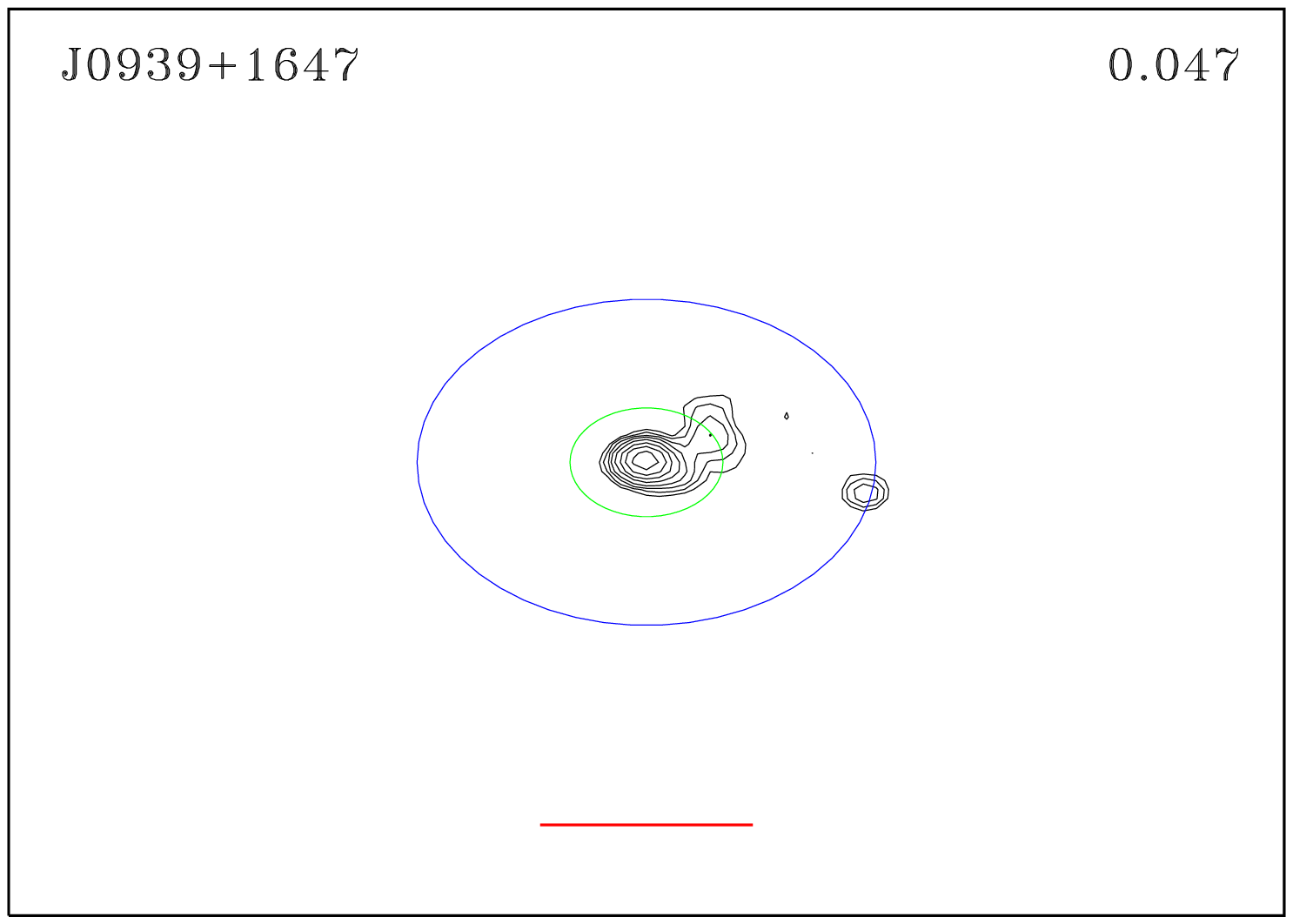}
                                                        
\includegraphics[width=4.5cm,height=4.5cm]{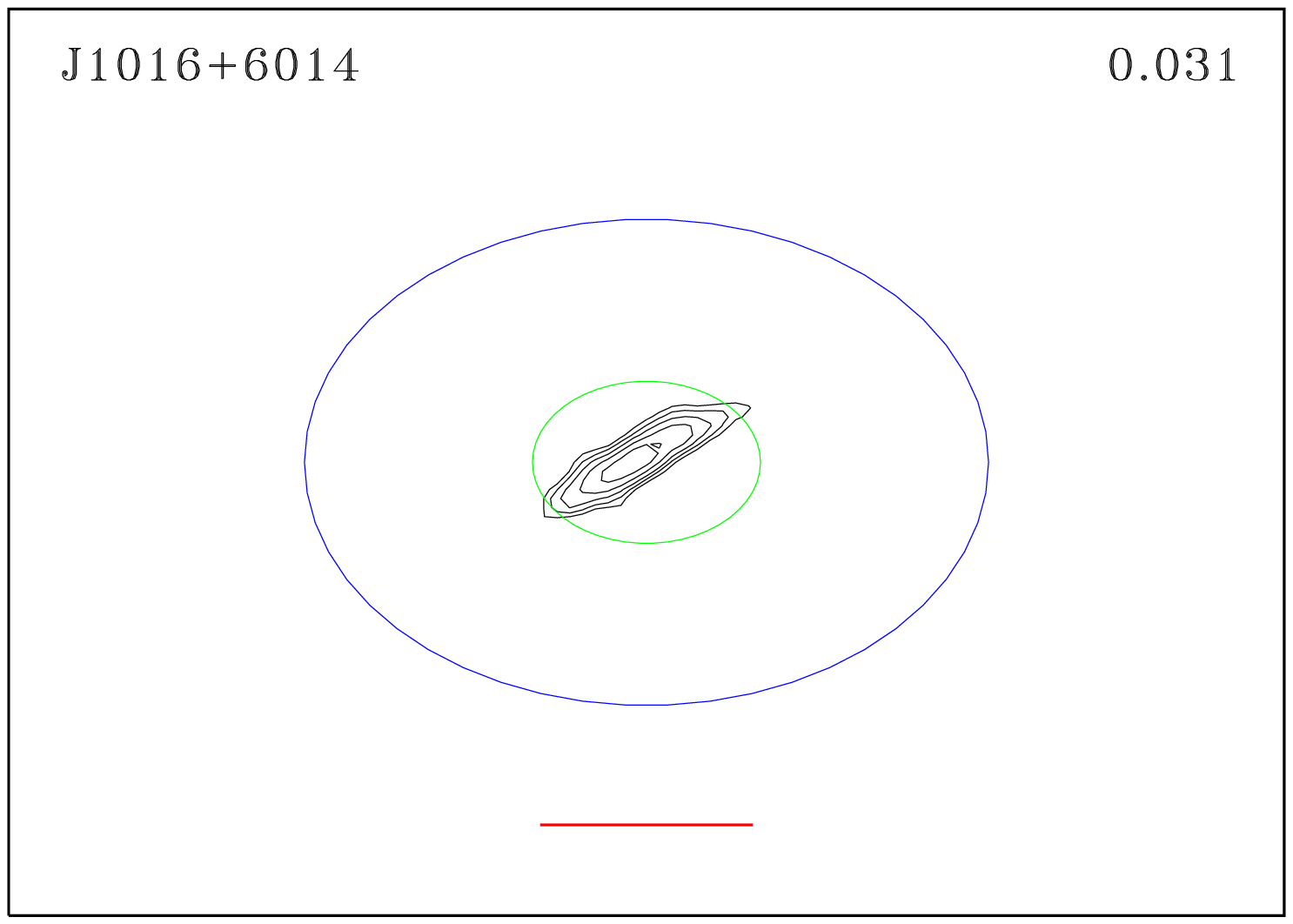}
\includegraphics[width=4.5cm,height=4.5cm]{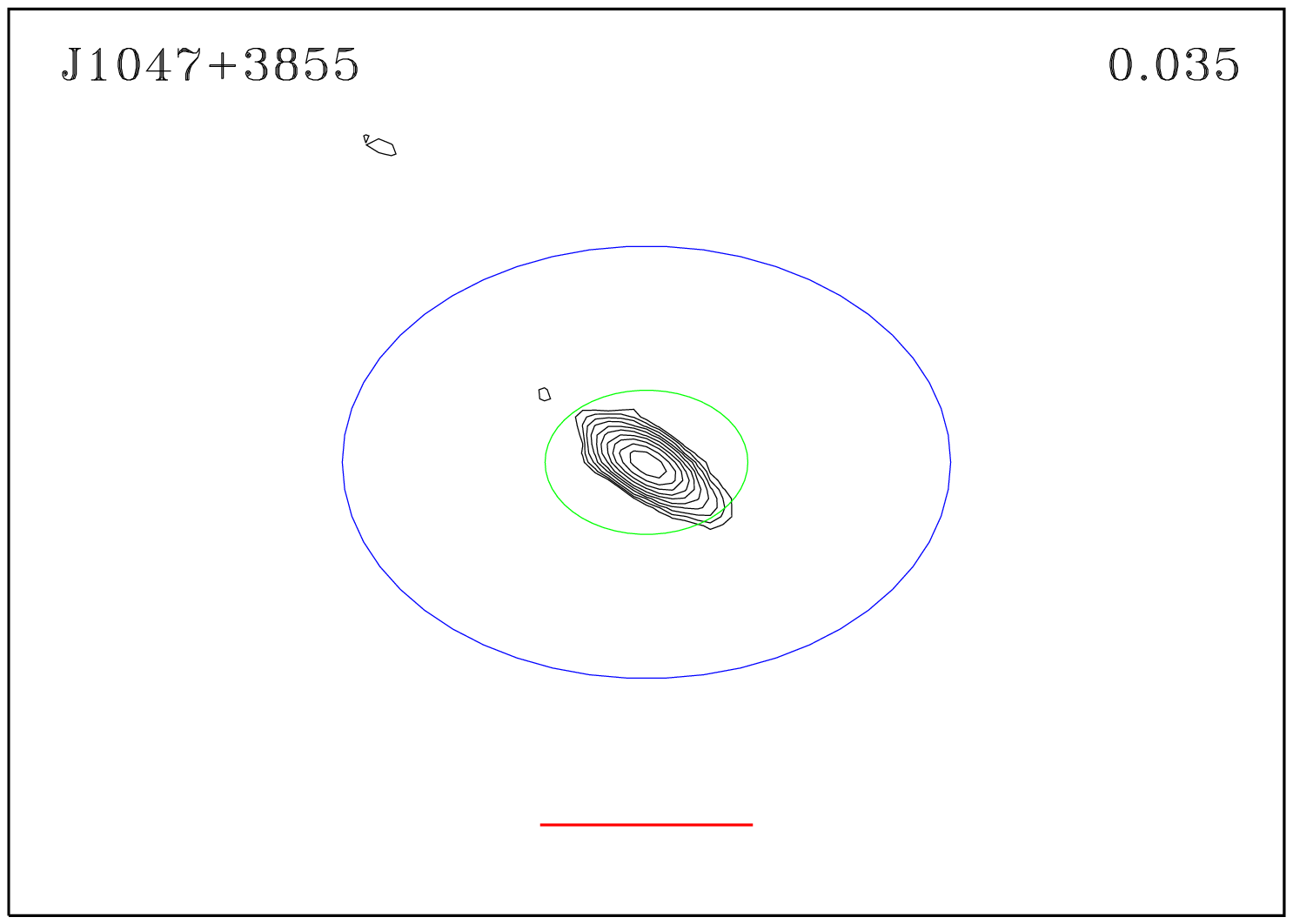}
\includegraphics[width=4.5cm,height=4.5cm]{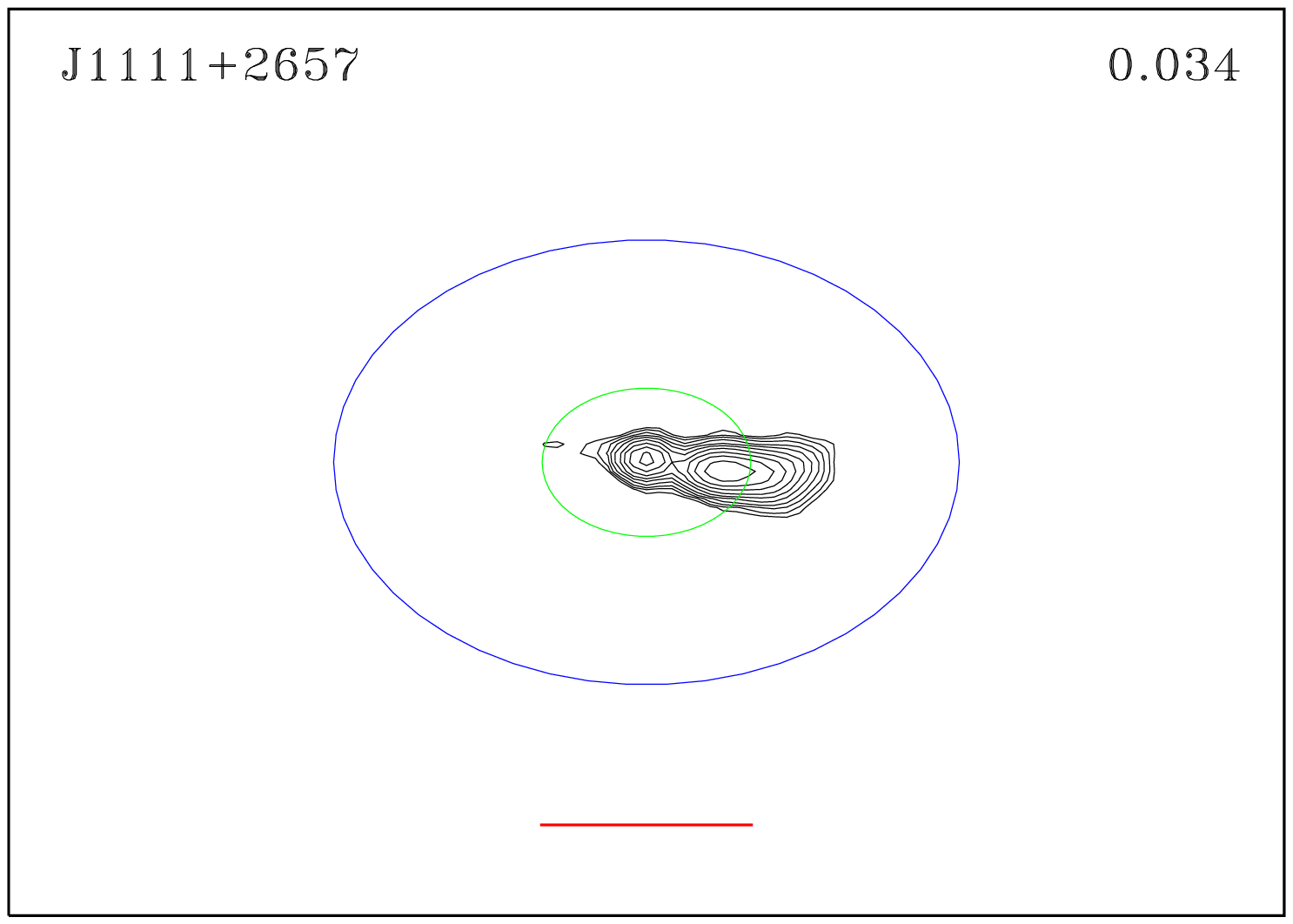}
\includegraphics[width=4.5cm,height=4.5cm]{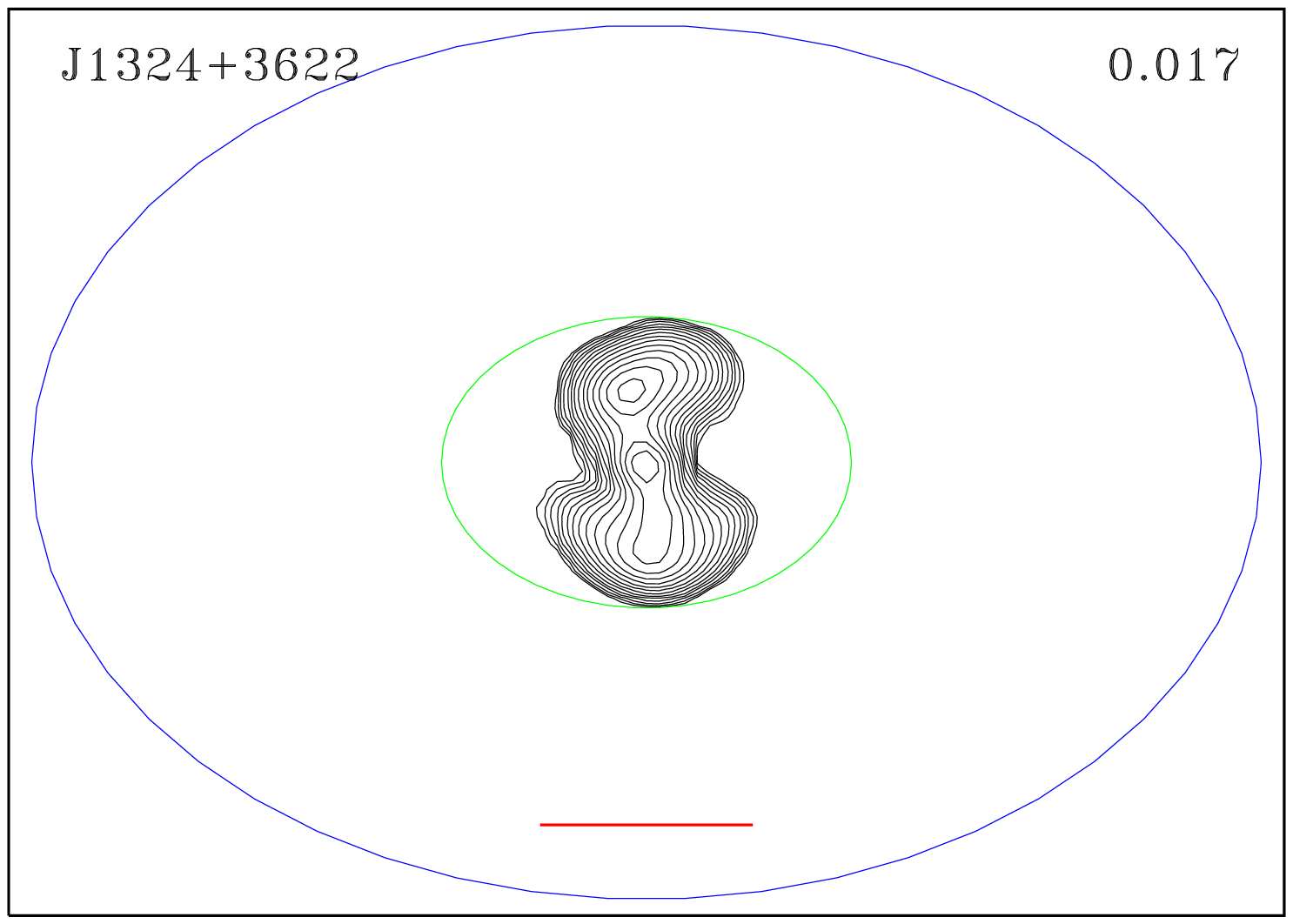}
                                                        
\includegraphics[width=4.5cm,height=4.5cm]{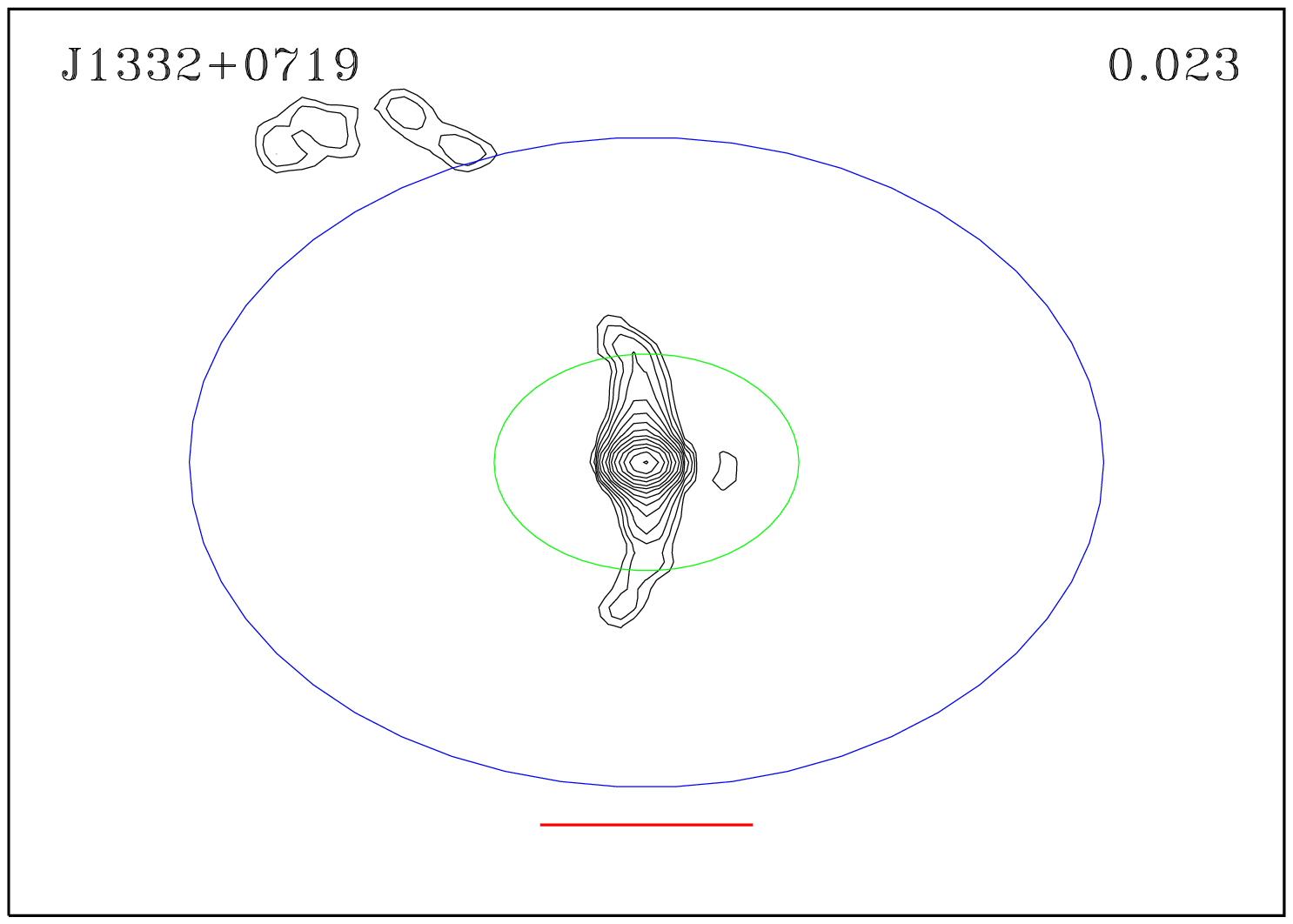}
\includegraphics[width=4.5cm,height=4.5cm]{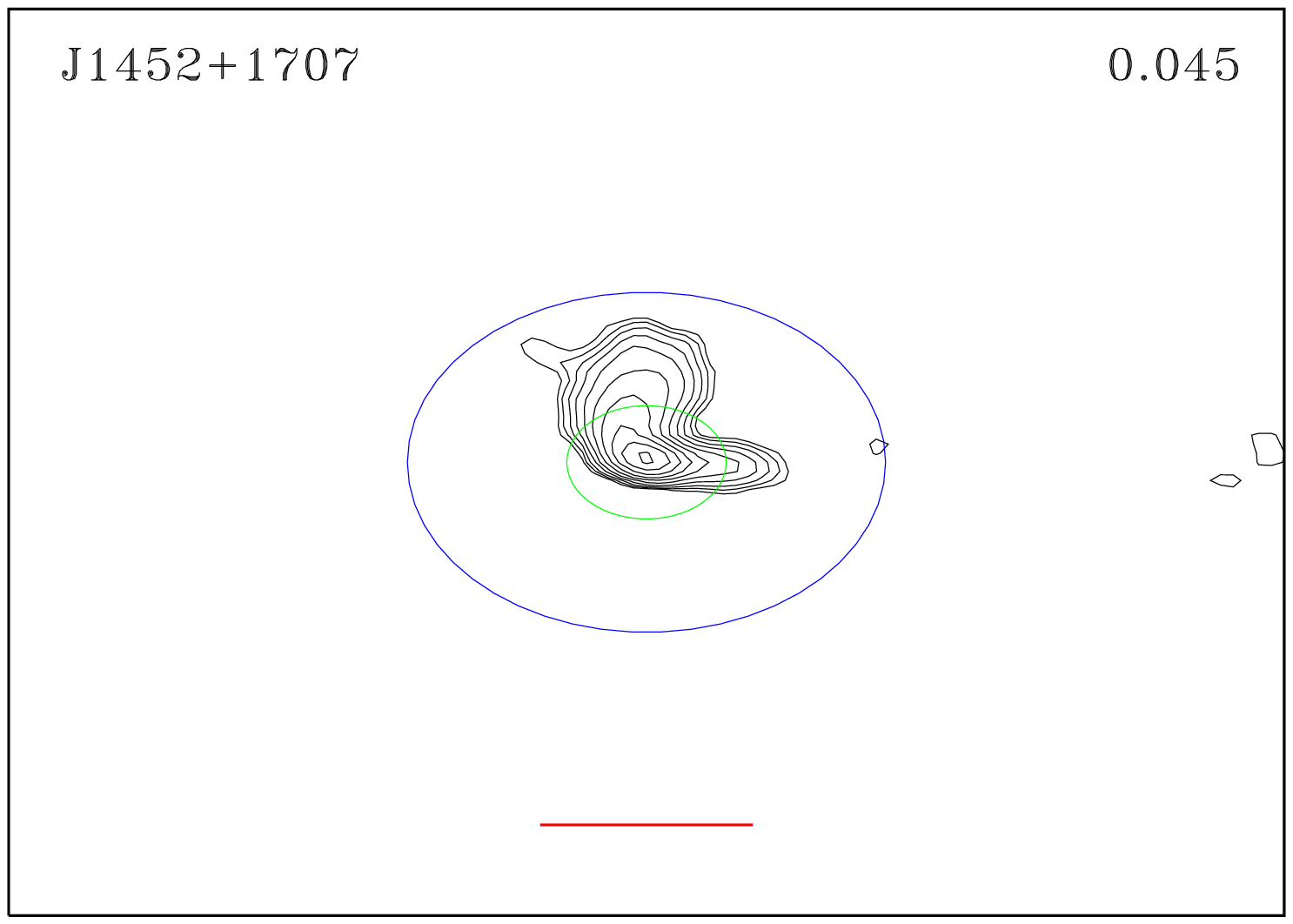}
\includegraphics[width=4.5cm,height=4.5cm]{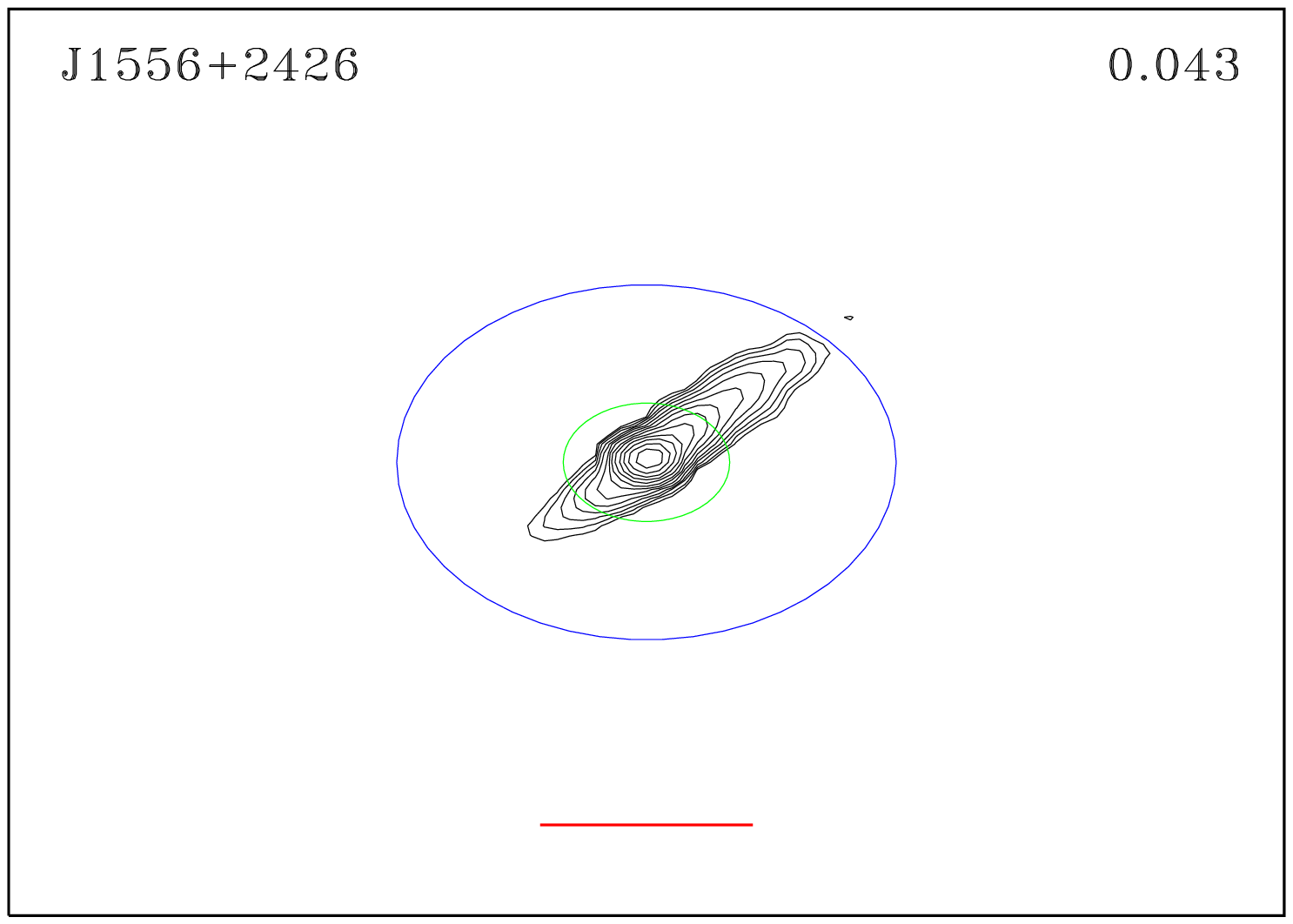}
\includegraphics[width=4.5cm,height=4.5cm]{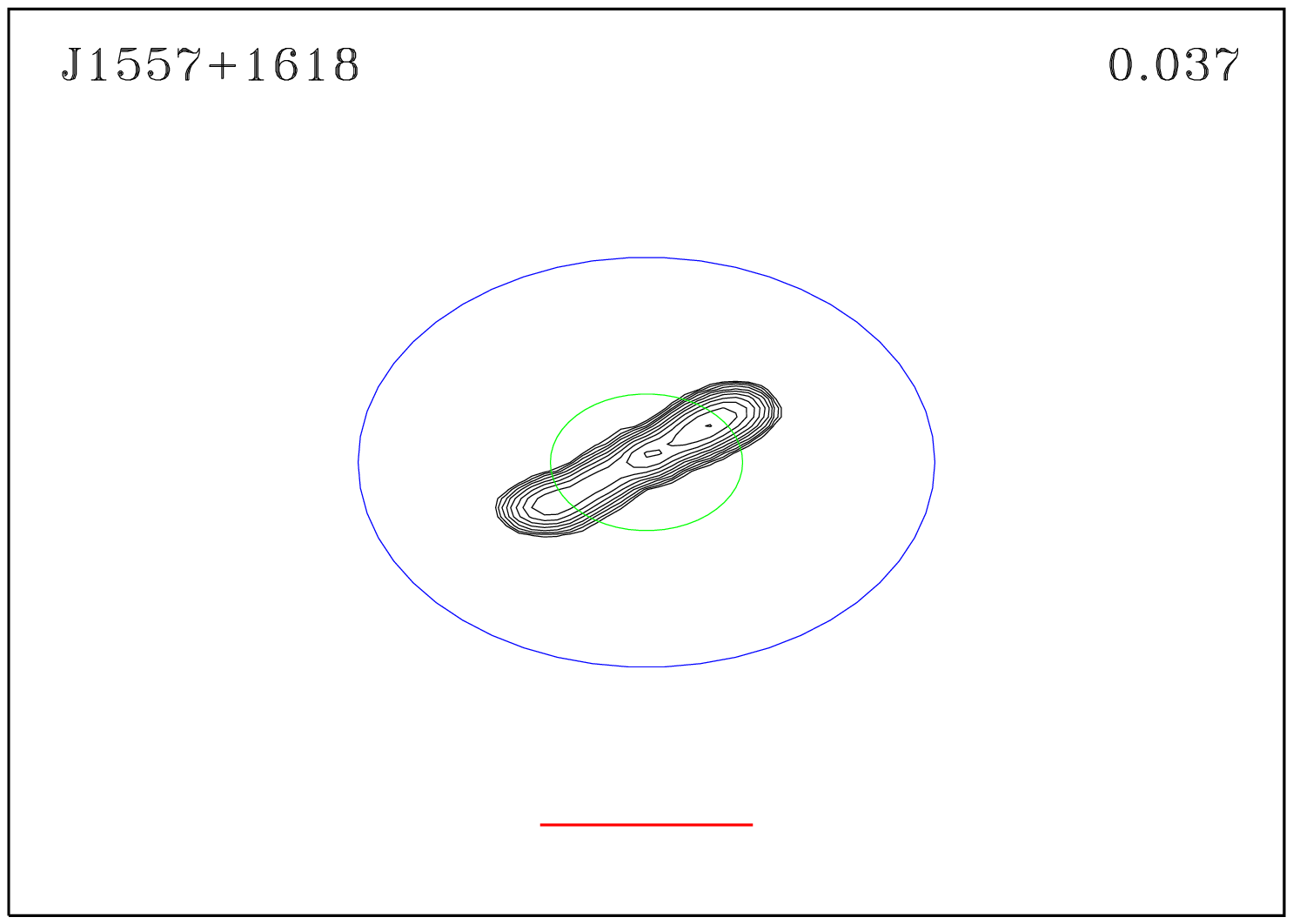}
                                                                         
\includegraphics[width=4.5cm,height=4.5cm]{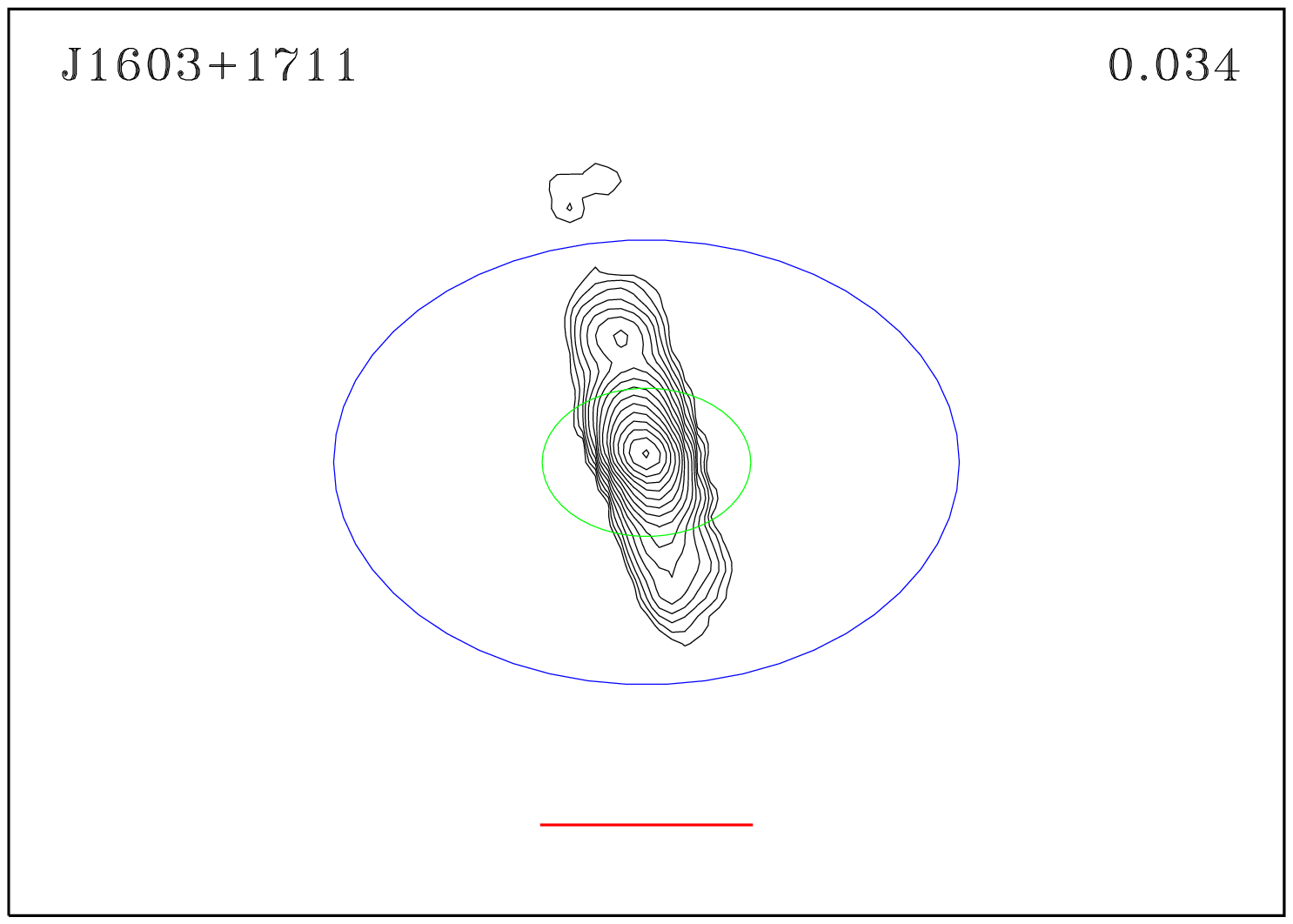}
\includegraphics[width=4.5cm,height=4.5cm]{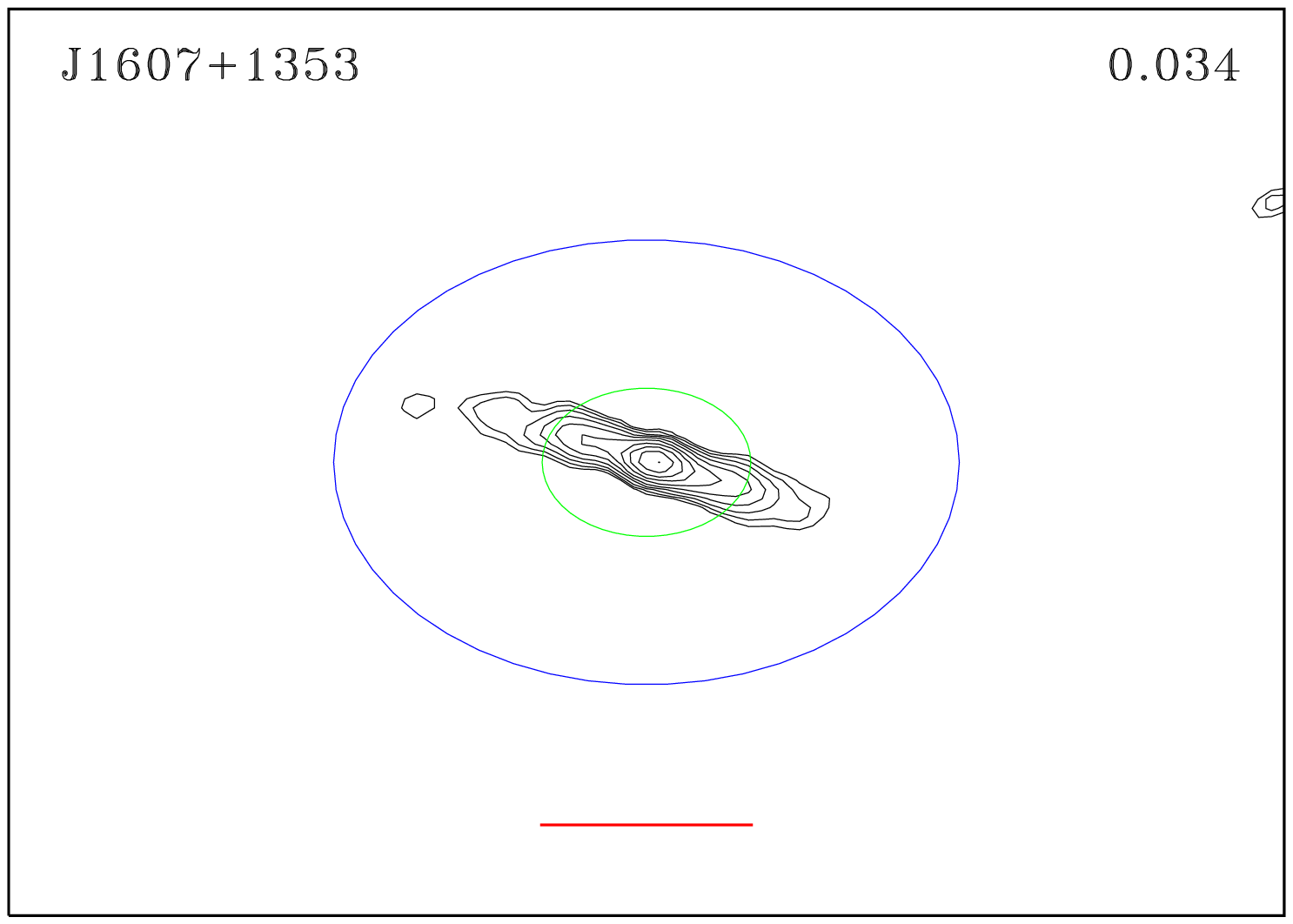}
\caption{FIRST images of the 14 \sFR\ sources selected at $z<0.05$ and
  extended between 10 and 30 kpc. The green (blue) circle is centered on the
  host galaxy and has a radius of 10 (30) kpc.  The sources
  \sFR\ name and redshift are reported in the upper corners.}
\label{imagesn}
\end{figure*}

\citet{fanaroff74} introduced the first classification scheme for
extragalactic radio sources with large-scale structures (i.e., greater
than $\sim$15-20 kpc in size). They proposed to distinguish radio
sources into two main classes on the basis of the relation between
relative positions of regions of high and low surface brightness in
their extended components. This scheme was based on the ratio $R_{FR}$
of the distance between the regions of highest surface brightness on
opposite sides of the central host galaxy to the total extent of the
source up to the lowest brightness contour in the radio images. Radio
sources with $R_{FR}<$0.5 were placed in Class I (i.e., the
edge-darkened FR~Is) and sources with $R_{FR}>$0.5 in Class II (i.e.,
the edge-brightened FR~IIs).

This morphology-based classification scheme was also linked to their
intrinsic power, when Fanaroff and Riley found that all sources in
their sample with luminosity at 178 MHz smaller than
2$\times$10$^{25}$ W Hz$^{-1}$ sr$^{-1}$ (for a Hubble constant of 50
\kms\ Mpc$^{-1}$) were classified as FR~I while the brighter sources
all were FR~II. The luminosity distinction between FR classes is
fairly sharp at 178 MHz but their separation is cleaner in the
optical-radio luminosity plane, implying that the FR~I/FR~II dichotomy
depends on optical and radio luminosity \citep{ledlow96}.

The selection of large and well-defined samples of radio galaxies is
of great importance to properly address several issues, such as
building their luminosity functions, exploring the properties of their
hosts, studying their environment and their cosmic evolution, and
comparing the results obtained for the different classes of
radio-galaxies for radio-quiet active nuclei and for the population of
quiescent galaxies.

In particular, the number of known FR~I radio galaxies is rather
small. For example the Third Cambridge Catalogue of Radio Sources (3C;
\citealt{bennett62a}) includes less than $\sim$30 FR~Is. The second
Bologna sample (B2; \citealt{colla75,fanti78}) is formed by $\sim$100
radio galaxies of lower luminosity than those of the 3C; most of these
have a luminosity below the FR~I/FR~II transition and about half of
them of are FR~I. These samples are not sufficiently large to address
the issues listed above properly. Furthermore, as these samples are
selected with a rather high flux threshold, they present a limited
(and possibly statistically biased) view of the FR~I population.

The advent of large area surveys opens the opportunity to set the
results on several key issues on strong statistical foundations. In
particular, the radio, infrared, and optical observations available
thanks to recent large-area surveys are a unique tool in the analysis
of the radio galaxies and quasars, since they allow us to identify
large numbers of radio sources, obtain spectroscopic redshifts, and
determine the properties of their hosts.  \citet{best05b},
\citet{baldi10b}, and \citet{best12} already used the extensive
multifrequency information available to analyze the properties of the
population of low redshift radio emitting AGN. We here also consider
the radio morphological information and explore the possibility to
create the first catalog of FR~I radio galaxies selected on the basis
of radio and optical data, which we call the \FR.

This paper is organized as follows. In Sect.\ 2 we present the
selection criteria of the sample of FR~Is, whose completeness is
discussed in Sect.\ 3.  The radio and optical properties of the
selected sources are presented in Sect.\ 4 and discussed in Sect.\
5. Sect.\ 6 is devoted to our summary and conclusions.

Throughout the paper we adopt a cosmology with $H_0=67.8 \, \rm km \, s^{-1} \,
Mpc^{-1}$, $\Omega_{\rm M}=0.308$, and $\Omega_\Lambda=0.692$ \citep{ade16}.  

For our numerical results, we use c.g.s. units unless stated
otherwise. Spectral indices $\alpha$ are defined by the usual convention on
the flux density, $S_{\nu}\propto\,\nu^{-\alpha}$. The SDSS magnitudes are in the AB
system and are corrected for the Galactic extinction; {\em Wise} magnitudes
are instead in the Vega system and are not corrected for extinction since, as
shown by, for example, \citet{dabrusco14}, such correction affects only the
magnitude at 3.4 $\mu$ of sources lying at low Galactic latitudes (and by less
than $\sim$3\%).

\section{Sample selection}
\label{sample}

We searched for FR~I radio galaxies in the sample of 18,286 radio
sources built by \citet{best12} (hereafter the BH12 sample) by
limiting our search to the subsample of objects in which, according to
these authors, the radio emission is produced by an active nucleus.
They cross-matched the optical spectroscopic catalogs produced by the
group from the Max Planck Institute for Astrophysics and The Johns
Hopkins University \citep{bri04,tre04} based on data from the data
release 7 of the Sloan Digital Sky Survey (DR7/SDSS;
\citealt{abazajian09}),\footnote{Available at {\tt
    http://www.mpa-garching.mpg.de/SDSS/}.} with the National Radio
Astronomy Observatory Very Large Array Sky Survey (NVSS;
\citealt{condon98}) and the Faint Images of the Radio Sky at Twenty
centimeters survey (FIRST; \citealt{becker95}) adopting a radio flux
density limit of 5 mJy in the NVSS.  We focused on the 3,357 sources
with redshift $z < 0.15$.

We visually inspected all the FIRST images of each individual source
and preserved only those whose radio emission reaches a distance of at
least 30 kpc from the center of the optical host at the sensitivity of
the FIRST images. Such a radius corresponds to 11$\farcs$4 for the
farthest objects; this ensures that all the 741 selected sources are
well resolved with the 5$\arcsec$ resolution of the FIRST images. This
permitted us to properly explore their morphology. The reference
surface brightness level adopted is 0.45 mJy/beam (approximatively
three times the typical rms of the FIRST images) for the objects at
z=0.15. The brightness level is increased by a factor
$\big[(1+0.15)/(1+z)\big]^4$ for closer objects to compensate for the
cosmological surface brightness dimming; this level corresponds to a
correction factor of $\sim$1.75 for $z=0$. We also applied a k
correction by assuming a spectral index of 0.7, which is typical of
the extended radio emission; in this case the correction is rather
small, amounting to at most $\sim$10\%.

We adopted a purely morphological classification based on the radio
structure shown by the FIRST images. The original FR~I definition
corresponds to ``a great diversity of structure'' \citep{fanaroff74},
and it is not always of easy application. We adopted rather strict
criteria for a positive classification for the selection of the FIRST
sample of FR~Is. We limited our selection to the sources showing
one-sided or two-sided jets in which the surface brightness is
generally decreasing along its whole length, lacking of any brightness
enhancement at the jet end. We allowed for bent jets and we thus
included narrow angle tail (NAT; \citealt{rudnick77}) sources;
conversely, we excluded the sources in which a substantial brightening
occurs along the jet, thus excluding, for example, wide angle tail
(WAT; \citealt{owen76}) objects.

The three authors performed this analysis independently and we included only
the sources for which a FR~I classification is proposed by at least two of us.

The resulting sample, to which we refer as \FR, is formed by 219
FR~Is. In Fig. \ref{images} we present the FIRST images of the first
12 \FR\ sources selected to illustrate the outcome of our
selection. Images of all \FR\ objects are available in the
Appendix. Their main properties are presented in Table \ref{tab},
where we report the SDSS name, redshift, and NVSS 1.4 GHz flux density
(from BH12).  The [O~III] line flux, the r-band SDSS AB magnitude,
$m_r$, the Dn(4000) index (see Section 4 for the definition of the
Dn(4000)), and the stellar velocity dispersion $\sigma_*$ are instead
from the MPA-JHU DR7 release of spectrum measurements. The
concentration index $C_r$ was obtained for each source directly from
the SDSS database. For sake of clarity, errors are not shown in the
table; we estimated a median error of 0.08 on $C_r$, of 0.03 on
Dn(4000), of 0.004 magnitudes on $m_r$, and of 9 \kms\ on
$\sigma_*$. Finally we list the resulting radio and line luminosity,
and the black hole masses estimated from the stellar velocity
dispersion and the relation $\sigma_* - M_{\rm BH}$ of
\citet{tremaine02}. The error in the $M_{\rm BH}$ is dominated by the
spread of the relation used (rather than by the errors in the
measurements of $\sigma_*$) resulting in an uncertainty of a factor
$\sim$ 2.

The limited resolution of FIRST imposes a minimum size of 30 kpc to
the FR~Is. We selected (with the same criteria discussed above) a
second sample of FR~Is extending to smaller radii, $10 < r< 30 $ kpc,
to consider also less extended radio sources. We limited this sample
to nearby objects ($z<0.05$) to preserve a sufficient spatial
resolution. The images of these 14 sources, forming the `small' FR~Is
sample (hereafter \sFR), are presented in Fig. \ref{imagesn} and their
properties are listed in Table \ref{tabn}.

\begin{figure*}
\includegraphics[width=9.cm]{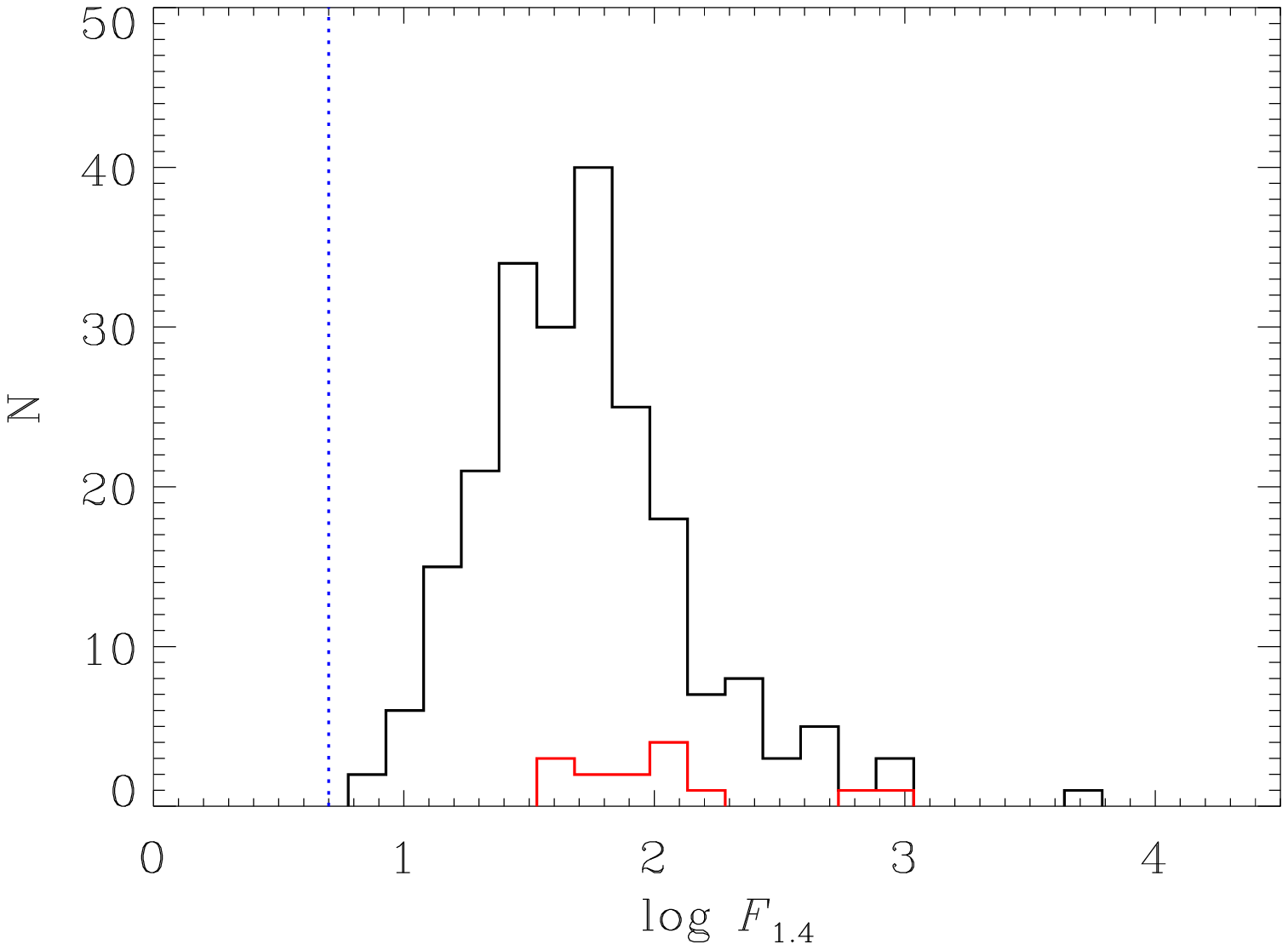}
\includegraphics[width=9.cm]{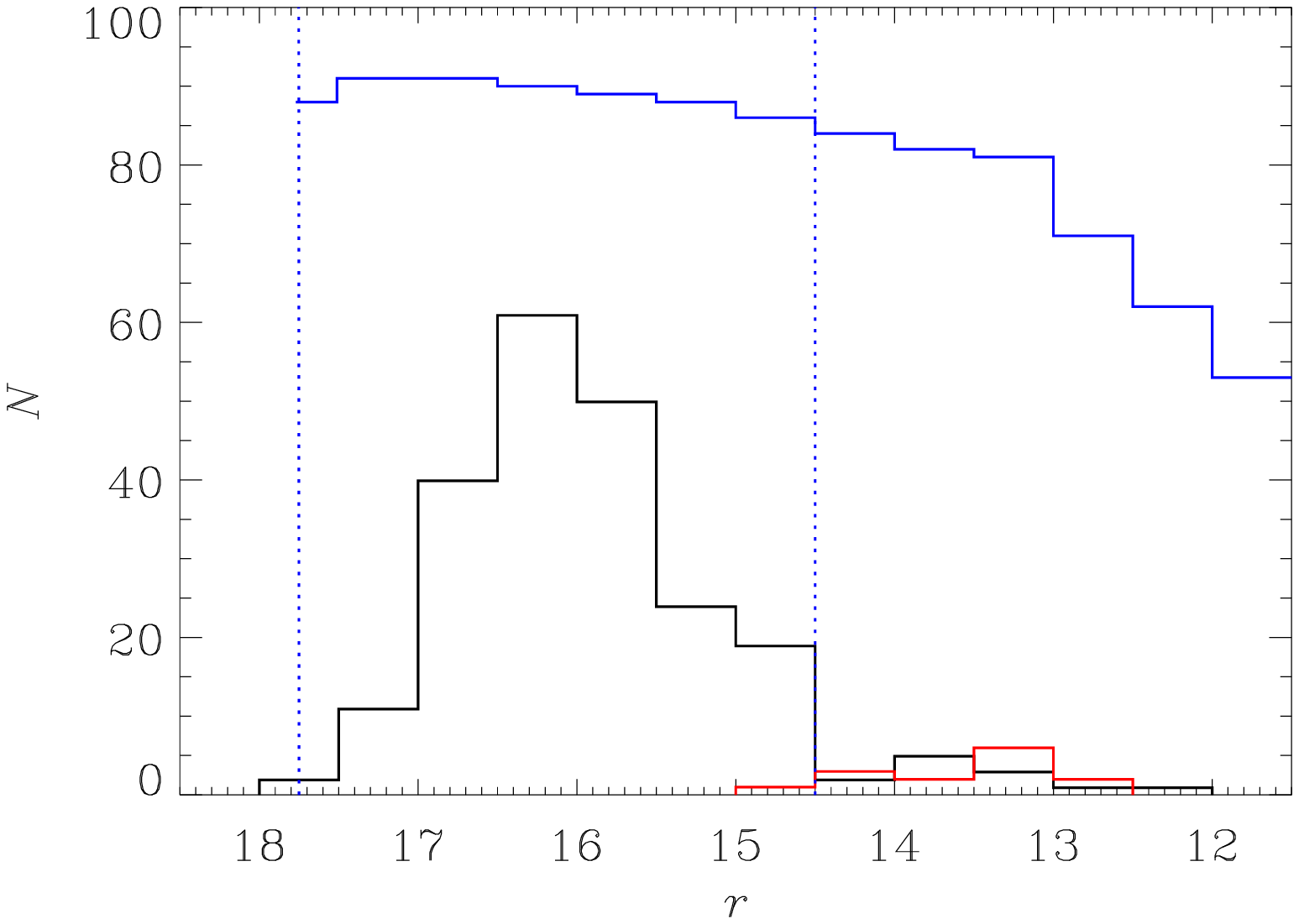}
\caption{Left: distribution of the NVSS fluxes of the 219 \FR\ sources; the
  red histogram is for the 14 \sFR\ sources. The vertical blue dotted line
  indicates the 5 mJy limits of the BH12 sample. Right: the black curve shows the
  $r$-band magnitude distribution of the \FR\ hosts (the red histogram is for
  the \sFR\ sources); the vertical dotted lines indicate the limits defining the
  SDSS main galaxies sample. The blue histogram report the SDSS completeness
  in percentage from \citet{montero09}}
\label{hist}
\end{figure*}

\section{The completeness of the \FR\ samples.}
\label{complete}

We now discuss the completeness of the sample related to the radio and optical
selection.

Concerning the radio selection, the BH12 includes sources with a NVSS
flux density larger than 5 mJy. However, our selection also depends on
the brightness distribution in the FIRST images. Therefore we might be
missing objects characterized by, for example, diffuse emission not
reaching the 3$\sigma$ limit in these higher resolution images;
furthermore, some extended emission might be resolved out, and missed,
by the FIRST maps.

In the left panel of Fig. \ref{hist}, we show the distribution of the
flux density at 1.4 GHz (i.e., $F_{1.4}$) for all the sources
belonging to the \FR; this flux density peaks at $\sim$50 mJy and
extends up to $\sim$5 Jy. The brightest source, \FR~1416+1048, is the
only objects belonging to the 3C sample, 3C~296. Below the peak the
sources density decreases and there are only two objects between 5 and
10 mJy. This flux distribution indicates that indeed the completeness
limit of \FR\ is higher than the original 5 mJy and can be set at
$\sim$50 mJy.

As for the optical selection of the sample, according to
\citet{montero09}, the redshift completeness of the SDSS decreases
with decreasing apparent magnitude, starting from $\sim$90\% at the
SDSS spectroscopic limit of $r=17.77$ and reaching $\sim$50\% at
$r=11.75$. Most of the incompleteness is due to the SDSS fiber
cladding, which prevents fibers on any given plate from being placed
closer than 55\arcsec\ apart. For brighter (and more extended) objects
other effects become important, such as the superposition of bright
saturated stars on the target.

In the right panel of Fig. \ref{hist}, we show the distribution of the
$r$ magnitude of the \FR\ hosts. The vast majority of them fall in the
magnitude range of the SDSS main galaxies sample (\citealt{strauss02};
$17.77 < r < 14.5$); a bright tail of objects (also including most of
the \sFR\ hosts) is present but it drops to zero well before the
redshift completeness is significantly reduced.

Thus both FR~Is catalogs (\FR\ and \sFR) are statistically complete at
level of $\sim$90\% in the optical energy range. However, it is worth
mentioning that this extremely low level of incompleteness is only due
to a random loss of $\sim$10\% of the potential spectroscopic targets
\citep[see, e.g.,][]{zehavi02}.

\section{\FR\ hosts and radio properties}
\label{hosts}

\subsection{Hosts properties}

All selected FR~Is are classified as low excitation galaxies (LEG) by
\citet{best12} based on the ratios of the optical emission lines in
their SDSS spectra. There are only four exceptions and these are
sources that cannot be classified spectroscopically because some of
the diagnostic emission lines cannot be measured in their spectra (see
Tab. \ref{tab} and \ref{tabn}); based on the criteria used by
\citet{best12} their radio emission is powered by an AGN.
Furthermore, \citet{baldi10b} show that the spectroscopically
unclassified objects likely belong to the class of LEG, but with an
even lower contrast of the AGN against the host galaxy emission.

\begin{figure*}
\includegraphics[width=9.5cm]{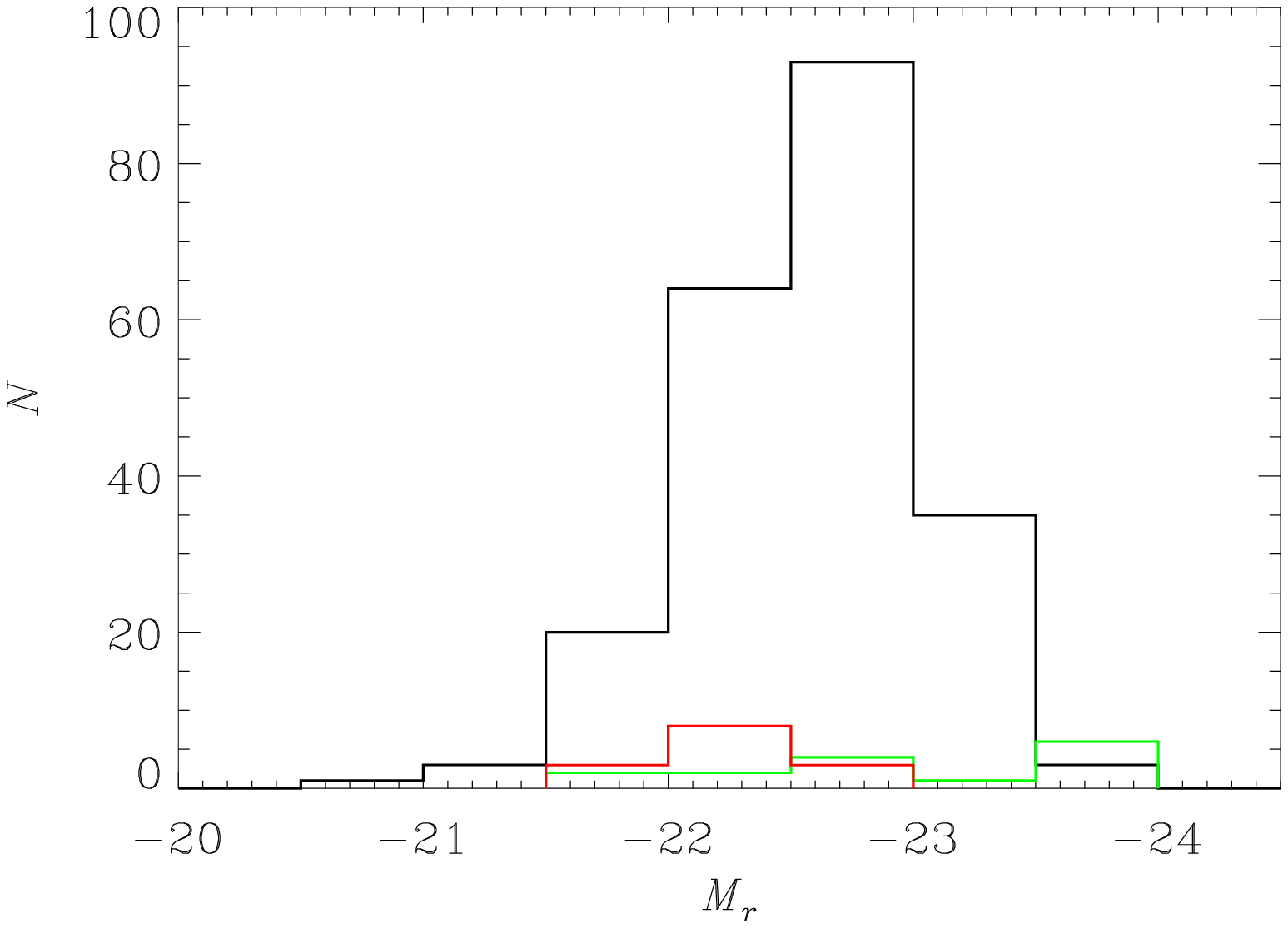}
\includegraphics[width=9.5cm]{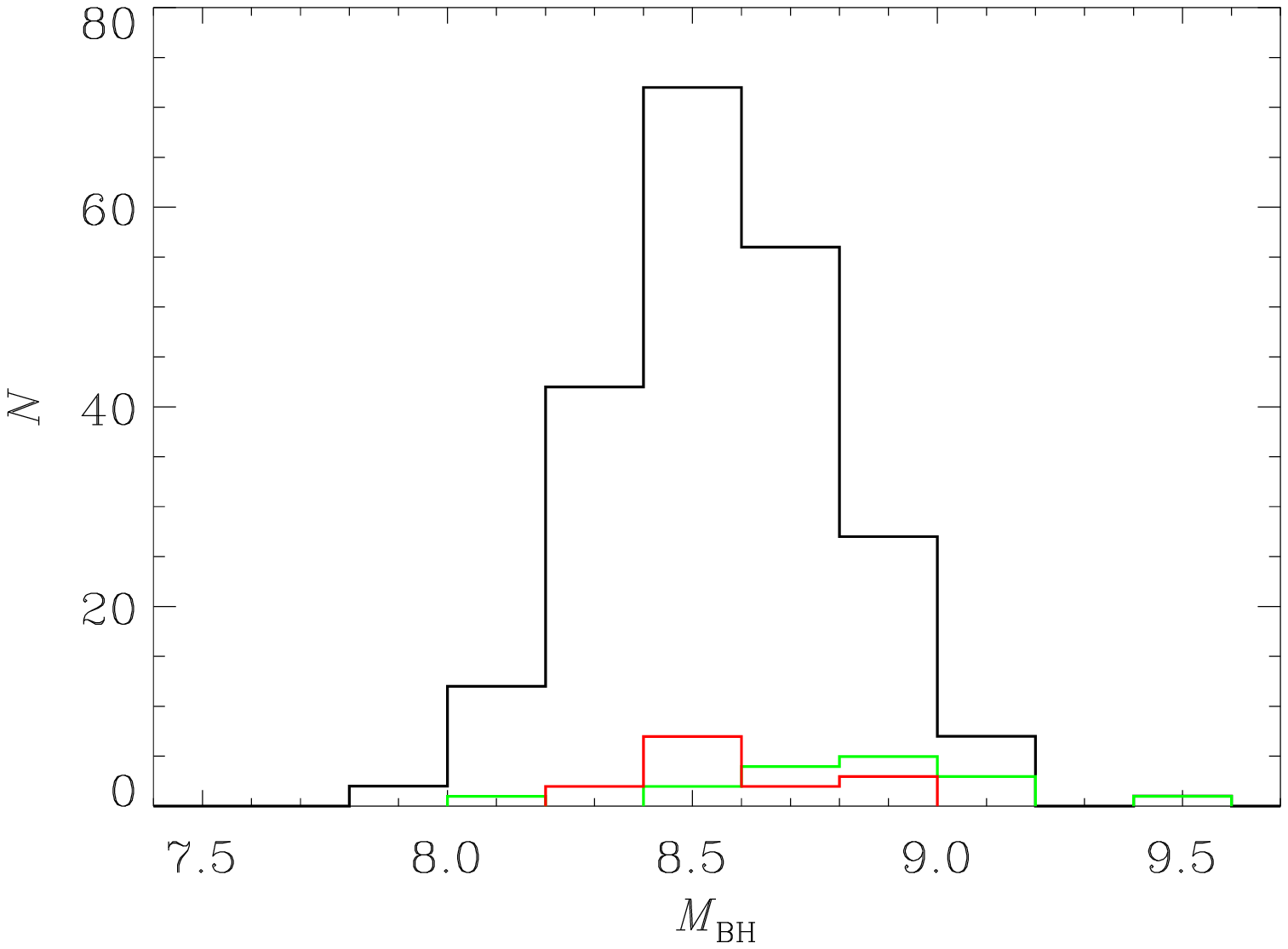}
\caption{Distributions of the $r$ band absolute magnitude (left) and black
  hole masses (right), black for \FR, red for the \sFR, and green for the
  3C-FRIs.}
\label{mhist}
\end{figure*}

The distribution of absolute magnitude of the \FR\ hosts covers the
range $-21 \gtrsim M_r \gtrsim -24$ with a maximum at $M_r \sim -22.5$
(see Fig. \ref{mhist}, left panel). The distribution of black hole
masses (Fig. \ref{mhist}, right panel) covers the range $8.0 \lesssim
\log M_{\rm BH} \lesssim 9.5 M_\odot$, peaking at $\sim10^{8.5}
M_\odot$.

Various diagnostics can be used for a morphological and spectroscopic
classification of the hosts. 

The concentration index $C_r$ is defined as the ratio of the radii
including 90\% and 50\% of the light in the $r$ band,
respectively. Early-type galaxies (ETGs) have higher values of $C_r$
than late-type galaxies. Two thresholds have been suggested to define
ETGs: a more conservative value at $C_r \gtrsim$ 2.86
\citep{nakamura03,shen03} and a more relaxed selection at $C_r
\gtrsim$ 2.6
\citep{strateva01,kauffmann03b,bell03}. \citet{bernardi10} found that
the second threshold of the concentration index corresponds to a mix
of E+S0+Sa types, while the first mainly selects ellipticals galaxies,
removing the majority of Sas, but also some Es and S0s.

The Dn(4000) spectroscopic index is defined according to
\citet{balogh99} as the ratio between the flux density measured on the
``red" side of the Ca~II break (4000--4100 \AA) and that on the
``blue" side (3850--3950 \AA). Low redshift ($z < 0.1$) red galaxies
show Dn(4000)$= 1.98 \pm 0.05$, which is a value that decreases to $=
1.95 \pm 0.05$ for $0.1 < z < 0.15$ galaxies \citep{capetti15}. The
presence of young stars or of nonstellar emission reduces the Dn(4000)
index.

\begin{figure*}
\includegraphics[width=9.5cm]{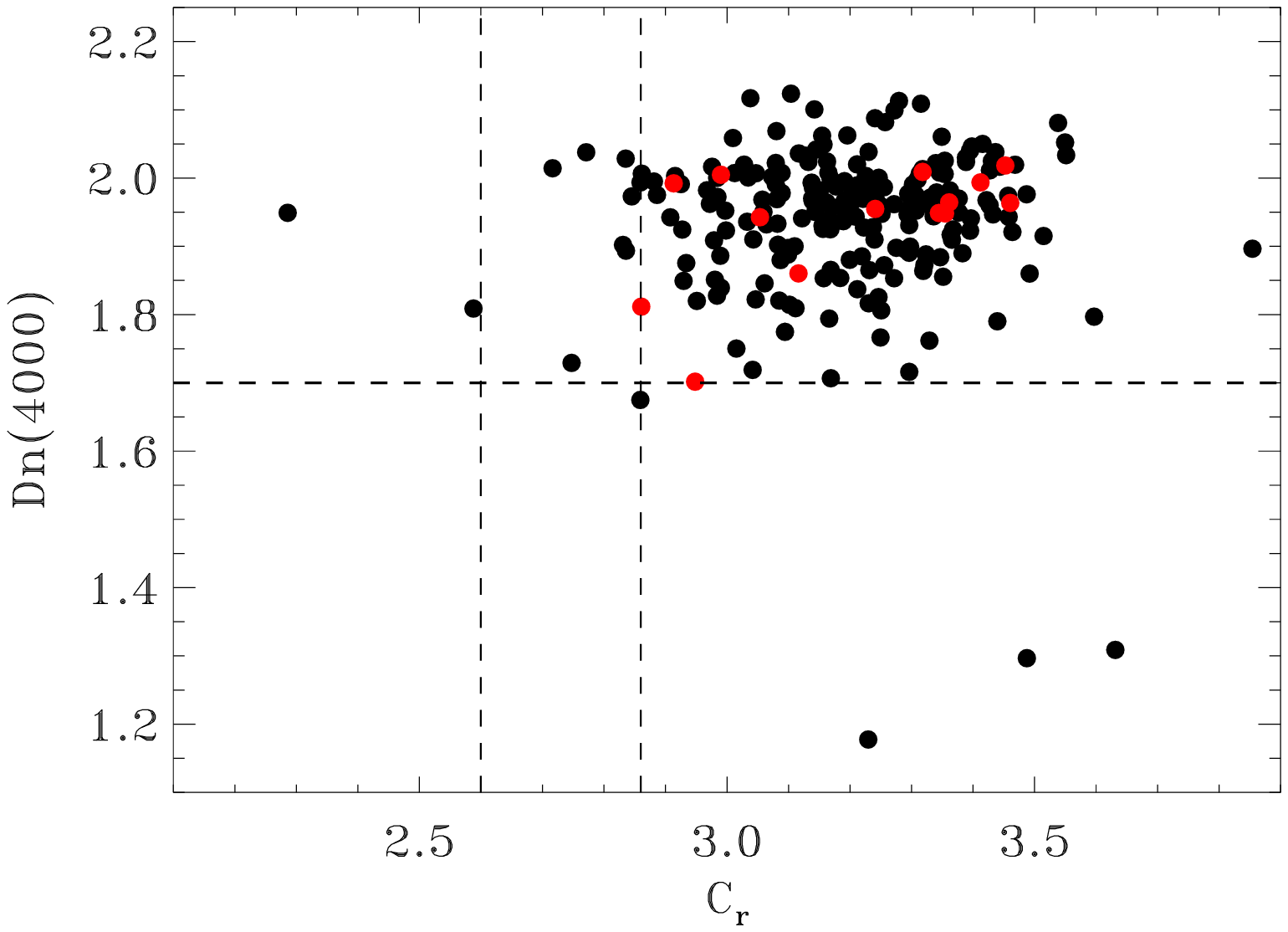}
\includegraphics[width=9.5cm]{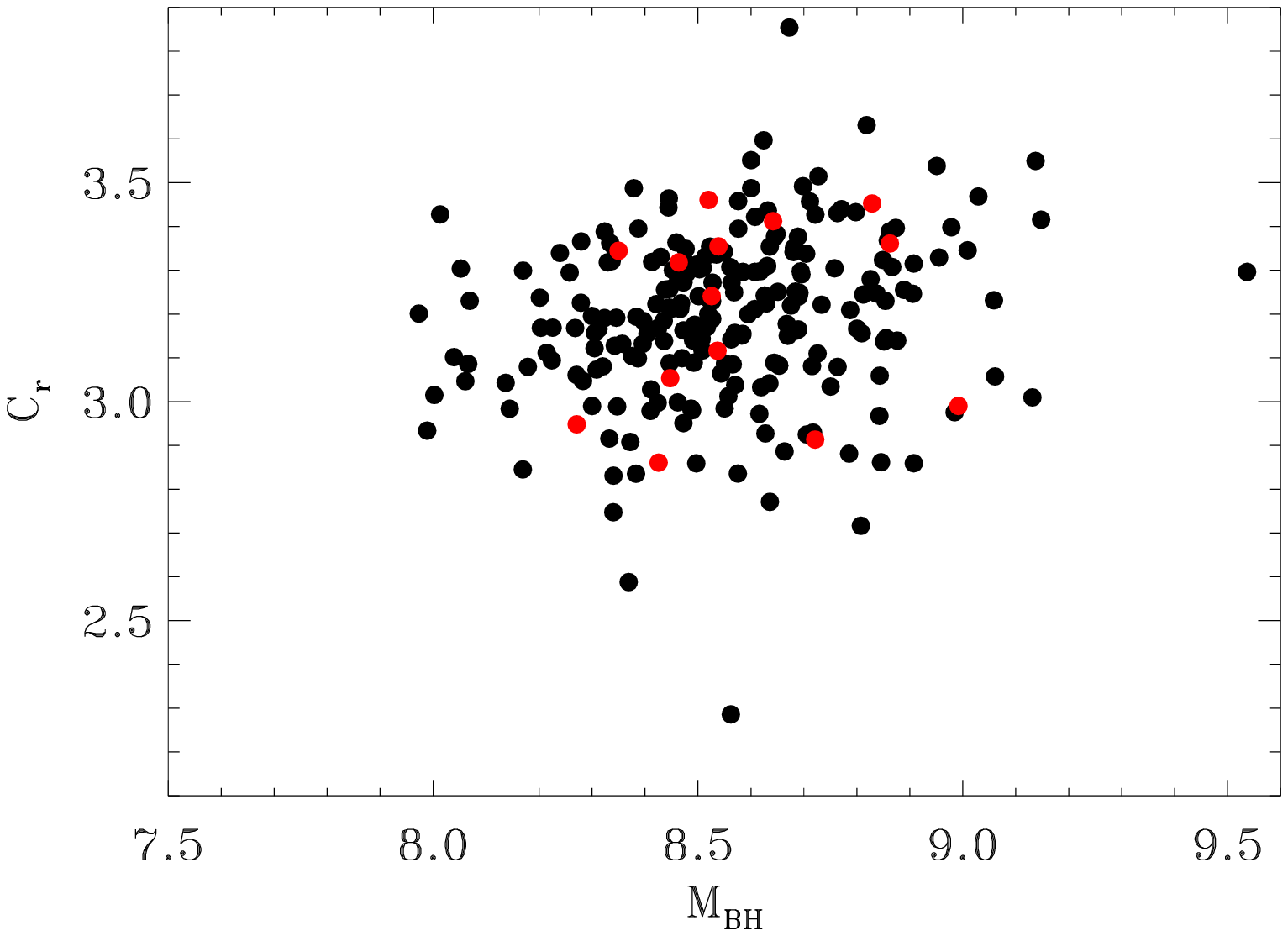}
\caption{Left: concentration index $C_r$ vs. Dn(4000) index for the \FR\
  and the \sFR\ samples (red dots). Right: logarithm of the black hole
  mass (in solar units) vs. concentration index $C_r$.}
\label{crdn}
\end{figure*}

In Fig. \ref{crdn} we show the concentration index $C_r$ versus the
Dn(4000) index (left panel) and versus $M_{\rm BH}$ (right panel) for
the \FR\ sources. The vast majority of the hosts lie in the region of
high $C_r$ and Dn(4000) values, indicating that they are red
ETGs. There are only a few exceptions: \FR~0735+4158 has a low
concentration index ($C_r = 2.29$), but this is due to the presence of
two compact sources close to the host center. \FR~1053+4929,
\FR~1428+4240, and \FR~1518+0613 instead have a low Dn(4000) index,
$\sim$1.3; their spectra are rich in absorption lines, suggesting a
dilution from nonstellar continuum rather than young stars. Indeed all
three sources (that we keep in \FR) are included in the list of low
luminosity BL~Lacs compiled by \citet{capetti15}.

\begin{figure*}
\centerline{ \includegraphics[width=9.5cm]{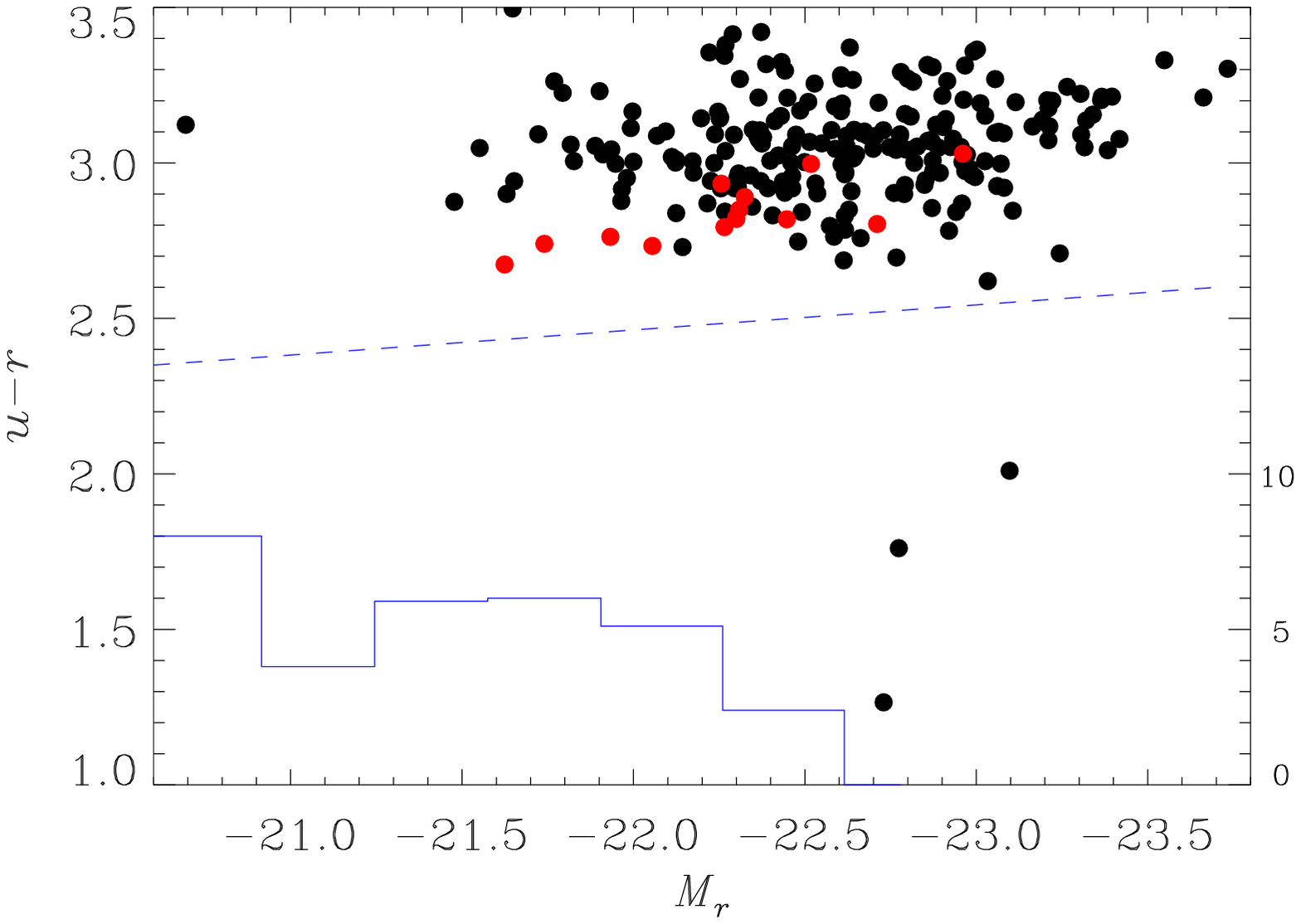} 
\includegraphics[width=9.5cm]{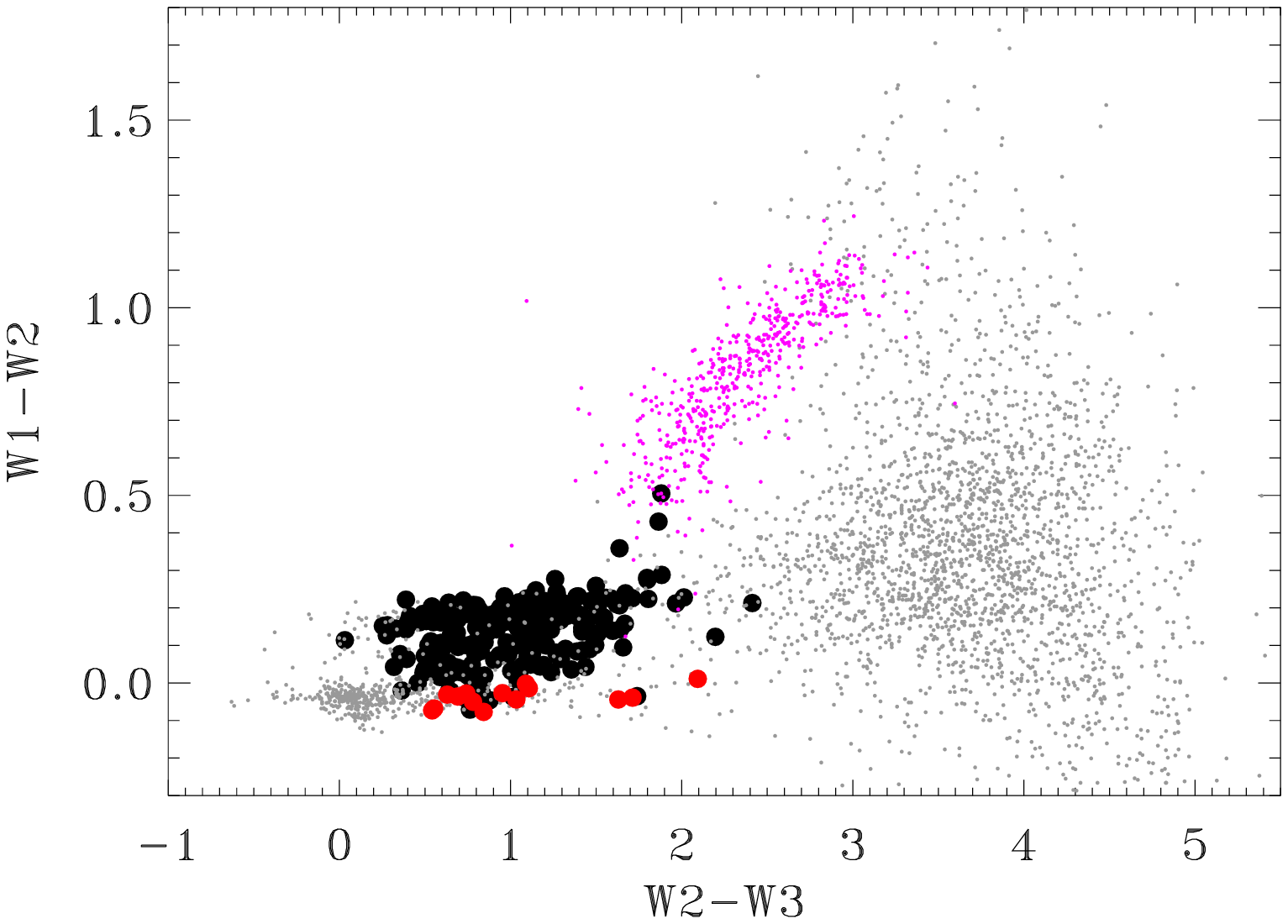} }
\caption{Left: absolute $r$ band magnitude, $M_r$, vs. $u-r$ color for the
  \FR\ hosts (the red dots represent the \sFR\ sample). The blue histogram on
  the bottom shows the percentage of blue ETGs (scale on the right axis) from
  \citet{schawinski09}. The dashed line separates the ``blue'' ETG from the
  red sequence, following their definition. Right: {\em{WISE}} mid-IR colors
  of the \FR\ hosts compared to those of $\sim$ 3000 randomly selected IR
  sources (gray dots) selected at high Galactic latitudes. We also show the
  region occupied by the {\em{Fermi}} blazars (purple dots).}
\label{mrur}
\end{figure*}

The Dn(4000) index refers only to the region covered by the SDSS
spectroscopic aperture, 3$\arcsec$ in diameter. In order to explore
the global properties of the \FR\ hosts, we also consider the $u-r$
color of the galaxies as a whole.  In Fig.~\ref{mrur} we show the
$u-r$ color versus the absolute r-band magnitude $M_r$ of the
hosts. With the exception of the three BL~Lacs, they are all located
above the line separating red and blue ETGs. The fraction of ``blue''
ETGs (represented as the histogram at the bottom of the figure)
decreases with increasing luminosity and these ETGs disappear for $M_r
\lesssim -22.5$ \citep{schawinski09}. The lack of blue ETGs among the
\FR\ hosts is relevant; however, their expected number, based on their
$M_r$ distribution and the ``blue'' fraction of the general ETGs
population, is only 4.3.

The {\em{WISE}} infrared colors further support the passive nature of
the \FR\ hosts.  In Fig. \ref{mrur} we show the comparison between the
mid-IR colors of \FR\ sources and those of $\sim$3000 randomly
selected sources (gray circles) at high Galactic latitudes (i.e.,
$\mid$b$\mid > $40\degr). The associations between the \FR\ and the
{\em{WISE}} catalog were computed adopting a 3\farcs3 angular
separation, which corresponds to the combination of the typical
positional uncertainty of the {\em{WISE}} all sky survey
\citep{wright10} and that of the FIRST \citep{dabrusco14}. In the same
figure we also report the mid-IR colors of the {\em{Fermi}} blazars
for reference of {\em{WISE}} sources whose IR emission is dominated by
nonthermal radiation \citep{massaro11,dabrusco12}. \FR\ sources appear
to have mid-IR colors mostly dominated by their host galaxies (they
fall in the same region as elliptical galaxies; \citealt{wright10})
and not contaminated by the emission of their jets. Only the three
BL~Lacs have $W2-W3 > 0.3$ and they are located at the onset of the
sequence defined by the more luminous objects of this class
\citep{massaro12}

\subsection{Radio properties}

\begin{figure*}
\includegraphics[width=9.5cm]{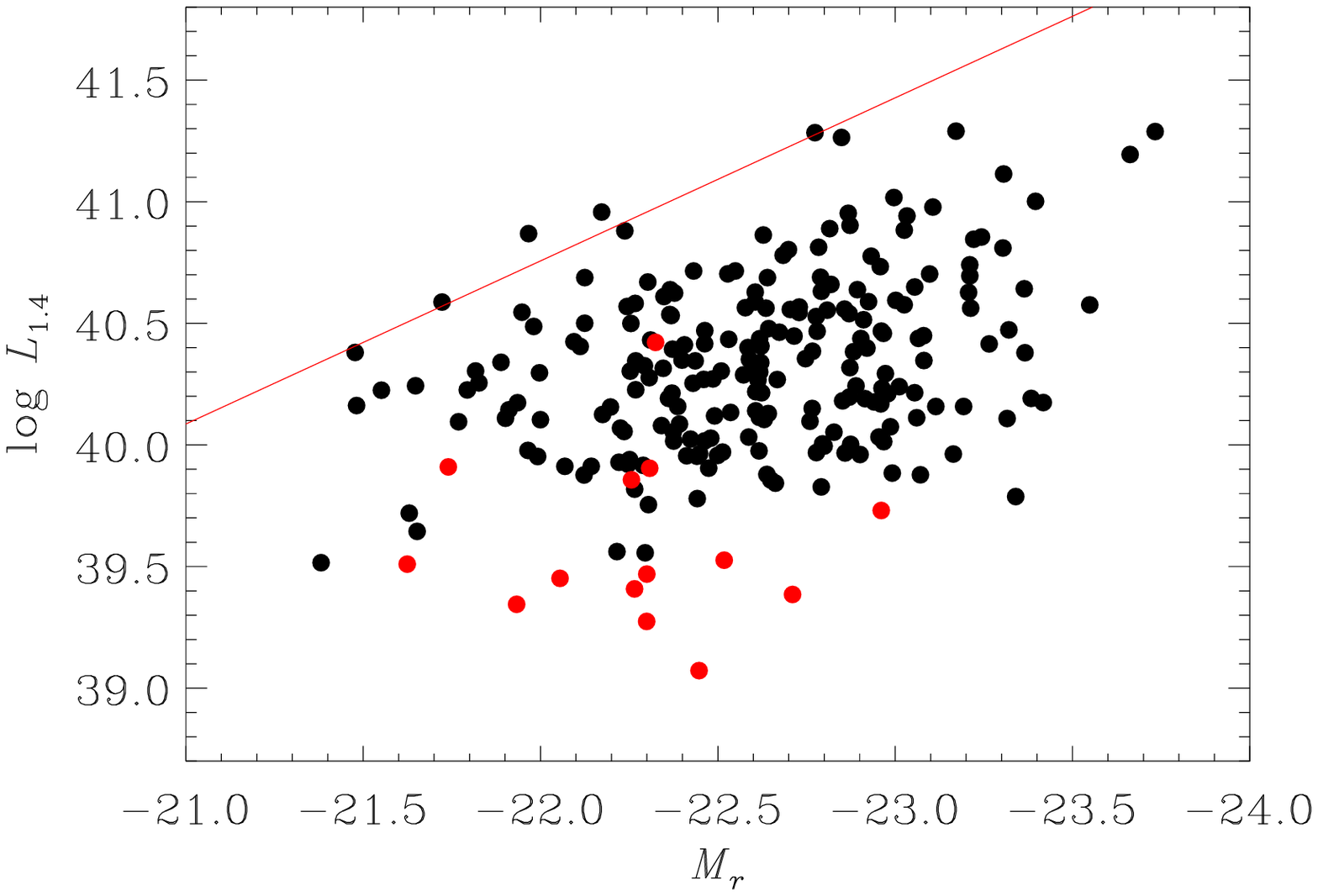}
\includegraphics[width=9.5cm]{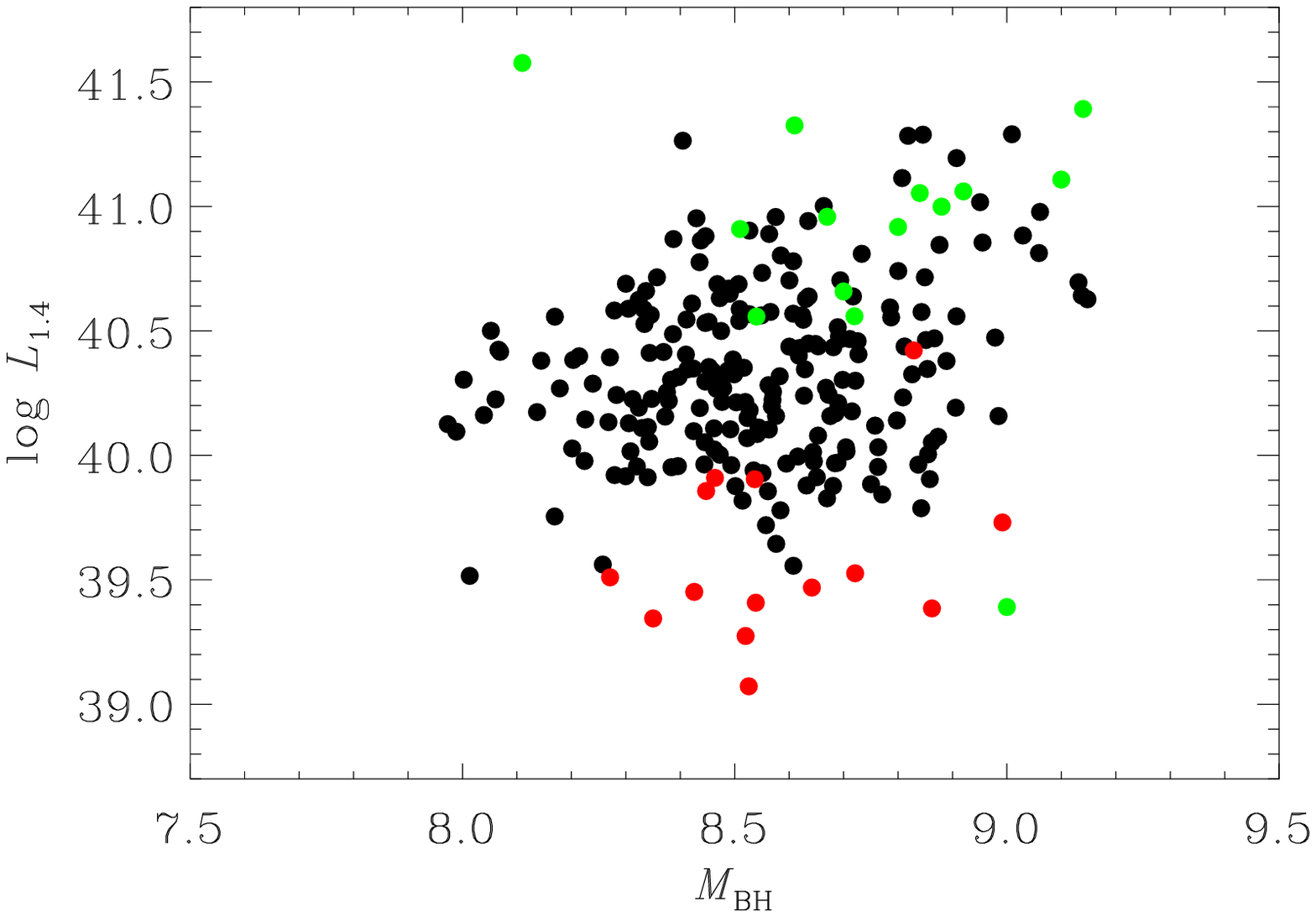}
\caption{Left panel: radio luminosity (NVSS) vs. host absolute
  magnitude , $M_r$, for \FR and \sFR\ (black and red,
  respectively). The solid line shows the separation between FR~I and
  FR~II reported by \citet{ledlow96} to which we applied a correction
  of 0.34 mag to account for the different magnitude definition and
  the color transformation between the SDSS and Cousin systems. Right
  panel: radio luminosity vs. black hole mass. The green points are
  the 3C-FR~Is.}
\label{lr}
\end{figure*}

The distribution of radio luminosity at 1.4 GHz of the \FR\ covers the
range $L_{1.4}$ = $\nu_{\rm r} l_{\rm r}$ = $\sim 10^{39.5} -
10^{41.3}$ $\ergs$ those of the \sFR\ sample instead have $10^{39}
\lesssim L_{1.4} \lesssim 10^{40.4}$ $\ergs$. The \citet{fanaroff74}
separation between FR~Is and FR~IIs translates, with our adopted
cosmology and by assuming a spectral index of 0.7 between 178 MHz and
1.4 GHz, into $L_{1.4} \sim 10^{41.6}$ $\ergs$. All objects included
in our sample fall below this threshold, although it must be kept in
mind that the power separation between FR~Is and FR~IIs is sharper at
178 MHz than at higher frequencies \citep{zirbel95}.

The separation between FR classes is cleaner in the optical-radio
luminosity plane \citep{ledlow96}. Indeed, the bulk of the \FR\
sources lie below the boundary between FR~I and FR~II reported by
\citeauthor{ledlow96}; see left panel Fig. \ref{lr} in the region
populated by FR~I sources. Be aware that we shifted the dividing line
to the right of the diagram to include a correction of 0.12 mag to
scale our total host magnitude to the M$_{24.5}$ used by these
authors, and an additional 0.22 mag to convert the Cousin system into
the SDSS system \citep{fukugita96}. This confirms the indication that
more powerful FR~Is can be associated with more massive galaxies,
while in less luminous hosts the FR~I/FR~II transition occurs at lower
$L_{1.4}$; as a result, a positive trend links $L_{1.4}$ and $M_{\rm
  r}$. A similar trend is seen also between $L_{1.4}$ and $M_{\rm BH}$
(Fig. \ref{lr}, right panel). This is likely to be driven by the
connection between $M_{\rm BH}$ and the host luminosity combined
\citep{marconi03} with the \citeauthor{ledlow96} effect.

\begin{figure}
\includegraphics[width=9.5cm]{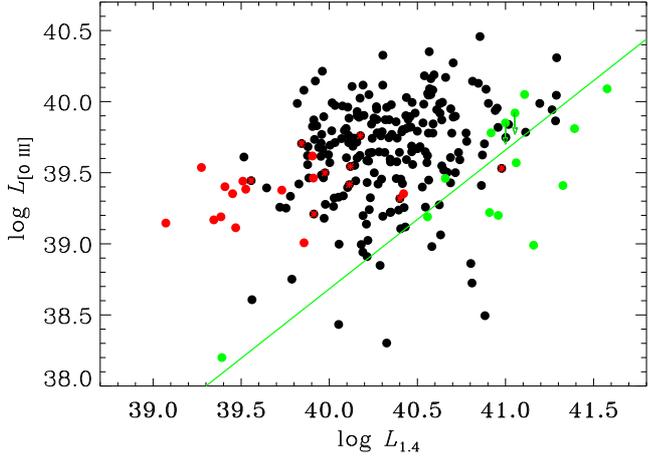}
\caption{Radio (NVSS) vs. [O~III] line luminosity of the \FR, \sFR, and
  3C-FR~I samples (black, red, and green points, respectively). The
  \FR\ sources with $z<0.05$ are represented as black dots with  a
  red asterisk superimposed. The green line shows the linear correlation between these two
  quantities derived from the FR~Is of the 3C sample from
  \citep{buttiglione10}.}
\label{lrlo3}
\end{figure}

FR~Is show a large spread in both radio and [O~III] line luminosities
(see Fig. \ref{lrlo3}), both quantities spanning over two orders of
magnitude. The FR~Is of the \sFR\ sample fall generally in the low end
of the radio luminosity distribution. Within the same volume
($z<0.05$), the sources extending to $10 < r < 30$ kpc have a median
luminosity that are four times smaller than those with $r> 30$ kpc.

\section{Discussion}

The population of the \FR\ hosts is remarkably uniform. They are all
luminous red ETGs, with large black hole masses ($M_{\rm BH} \gtrsim
10^8 M_\odot$), spectroscopically classified as LEGs. All these
properties are shared with the hosts of the `small' FR~Is and the more
powerful 3C-FR~Is. We included in the 3C-FRIs sample the 16 radio
galaxies with $z<0.3$ and a FR~I morphology, according to
\citet{buttiglione10}, and with either a direct $M_{\rm BH}$
measurement or a published stellar velocity dispersion in the
HyperLeda database\footnote{\tt http://leda.univ-lyon1.fr/}. More
quantitatively, the distributions of $M_{\rm BH}$ and $M_{r}$ of the
\FR\ and \sFR\ samples are not statistically distinguishable,
according to the Kolmogoroff-Smirnov test. A small difference might
instead emerge when considering the 3C-FRIs hosts. This latter sample
has a median $M_{\rm BH}$ that is a factor of 1.9 higher with respect
to the \FR\ (and they are 0.2 magnitudes brighter). As discussed in
the previous section, this might be the manifestation of the
\citeauthor{ledlow96} effect. Nonetheless, the null hypothesis that
they are drawn from different populations cannot be rejected at a
3$\sigma$ confidence level.

Even though the hosts of three FR~Is samples are very similar, the \FR\ sources
show a very different properties with respect to what is seen in the 3C sample
for the connection between emission lines and radio luminosities.

The 3C-FRIs show a positive trend between the line and radio
luminosities with a slope consistent with unity (e.g.,
\citealt{buttiglione10}). This indicates that a constant fraction of
the AGN power, as measured from the emission lines, is converted into
radio emission. \citeauthor{buttiglione10} show that the same result,
although with a different normalization, is found when considering the
3C-FR~II radio galaxies.

Conversely, no correlation between $L_{1.4}$ and $L_{\rm [O~III]}$ can
be seen for the \FR\ where, at a given line luminosity, the radio
luminosities span over two orders of magnitude. Similar to what is
seen in the 3C sample, it appears that no source has a $L_{1.4}/L_{\rm
  [O~III]}$ ratio exceeding $\sim$100, producing the scarcely
populated region in the bottom right portion of this diagram; but
objects are found with much lower ratios down to $L_{1.4}/L_{\rm
  [O~III]} \sim$0.5. Furthermore, the radio luminosity grows, not
surprisingly, when the size of the radio source increases. Less
obviously, the FR~Is in the ``small'' sample have a lower (by a factor
of $\sim$ 3) median ratio between line and radio luminosity.

Apparently, the high flux threshold used for the selection of the 3C sources
favored the inclusion of radio galaxies with high ratios between radio and
optical luminosity. A much larger population of FR~Is emerges when lowering
the radio flux limit by three orders of magnitude. The connection between
radio and line luminosity disappears. The spectroscopic and host properties of
the FR~I hosts rule out the possibility of a substantial contribution to the
[O~III] line by star formation that might compromise our ability to reveal
this trend, if indeed this was present. The broad distribution of
$L_{1.4}/L_{\rm[O~III]}$ ratio indicates a corresponding broad range of
conversion of AGN bolometric power into radio emission.

\section{Summary and conclusions}

We built a catalog of 219 FR~I radio galaxies, called FRI{\sl{CAT}}, selected from
the Best \& Heckman (2012) sample, and obtained by combining the SDSS, NVSS, and
FIRST surveys. 

The FR~I classification is purely morphological and based on the visual
inspection of the FIRST radio images. We included the sources in which the
radio emission reaches a distance of at least 30 kpc from the host
(restricting the analysis to those with redshift $z < 0.15$). We adopted
rather strict criteria for a positive FR~I classification; we selected only
sources showing one-sided or two-sided jets in which the surface brightness is
generally decreasing along its whole length, lacking of any brightness
enhancement at the jet end, i.e., with an edge-darkened structure.  The
resulting \FR\ catalog comprises 219 objects.  A second sample of 14 objects,
\sFR, extends the selection to smaller FR~Is by including sources with $10 < r
< 30$ kpc and $z < 0.05$.

These samples have a high level of completeness ($\sim$90\%) in both
their radio and optical selection.  As such, they can be used to
study, for example, their radio and bivariate radio/optical luminosity
functions. One should nonetheless bear in mind the morphological
selection criteria adopted and that the completeness limit in the
radio band is $\sim$50 mJy. These are well suited for our purposes but
it will be certainly interesting to explore the connection of the \FR\
with the remaining $\sim$500 extended radio galaxies we did not
include in our analysis.

The \FR\ hosts are remarkably homogeneous, as they are all luminous
red ETGs with large black hole masses that are spectroscopically
classified as LEGs.  All these properties are shared by the hosts of
more powerful FR~Is in the 3C sample.  They do not show significant
differences from the point of view of their colors with respect to the
general population of massive ETGs. The presence of an active nucleus
(and its level of activity) does not appear to affect the hosts of
FR~Is.

The \FR\ sources differ from the 3C-FRIs for the connection between
emission lines and radio luminosities. While in 3C-FRIs the line and
radio luminosities are correlated (suggesting that a constant fraction
of the AGN power is converted into radio emission) these two
quantities are unrelated in the \FR. We argue that the line/radio
correlation is the result of a selection bias because of the high flux
threshold of the 3C sources that favors the inclusion of radio
galaxies with high ratios between radio and optical
luminosity. \citet{baldi09} reached a similar conclusion from the
comparison of the 3C objects with the radio galaxies associated with
nearby ($z \lesssim 0.01$) optically luminous ETGs.

The 3C-FRIs represent the tip of the iceberg of a much larger and
diverse population of FR~Is. This result highlights the importance of
exploring a broader (and larger) population of FR~Is.  Several other
issues, such as those listed in the Introduction, can now be addressed
by using \FR. In particular we will explore in two forthcoming papers
the environment of \FR\ and how they are related to the class of the
compact FR~0 radio sources \citep{baldi15}.

\begin{acknowledgements}

F.M. gratefully acknowledges the financial support of the Programma Giovani
Ricercatori -- Rita Levi Montalcini -- Rientro dei Cervelli (2012) awarded by
the Italian Ministry of Education, Universities and Research (MIUR).

Part of this work is based on the NVSS (NRAO VLA Sky Survey): The National
Radio Astronomy Observatory is operated by Associated Universities, Inc.,
under contract with the National Science Foundation.

This publication makes use of data products from the Wide-field Infrared
Survey Explorer, which is a joint project of the University of California, Los
Angeles, and the Jet Propulsion Laboratory/California Institute of Technology,
funded by the National Aeronautics and Space Administration.

This research made use of the NASA/ IPAC Infrared Science Archive and
Extragalactic Database (NED), which are operated by the Jet Propulsion
Laboratory, California Institute of Technology, under contract with the
National Aeronautics and Space Administration.

We acknowledge the usage of the HyperLeda database (http://leda.univ-lyon1.fr).

Funding for SDSS-III has been provided by the Alfred P. Sloan Foundation, the
Participating Institutions, the National Science Foundation, and the
U.S. Department of Energy Office of Science. The SDSS-III web site is
http://www.sdss3.org/.  SDSS-III is managed by the Astrophysical Research
Consortium for the Participating Institutions of the SDSS-III Collaboration,
including the University of Arizona, the Brazilian Participation Group,
Brookhaven National Laboratory, University of Cambridge, Carnegie Mellon
University, University of Florida, the French Participation Group, the German
Participation Group, Harvard University, the Instituto de Astrofisica de
Canarias, the Michigan State/Notre Dame/JINA Participation Group, Johns
Hopkins University, Lawrence Berkeley National Laboratory, Max Planck
Institute for Astrophysics, Max Planck Institute for Extraterrestrial Physics,
New Mexico State University, New York University, Ohio State University,
Pennsylvania State University, University of Portsmouth, Princeton University,
the Spanish Participation Group, University of Tokyo, University of Utah,
Vanderbilt University, University of Virginia, University of Washington, and
Yale University.

\end{acknowledgements}

\newpage
\clearpage
\onecolumn
\begin{center}
\begin{longtable}{l l r r l l l l l c l c}

\caption[Properties of the \FR\ sources.]{Properties of the \FR\ sources.} 
\label{tab} \\

\hline \hline 
&  \,\,\,\,\,z & NVSS & [O~III] & \,\,\,\,\,m$_{\rm r}$ & \,\,Dn & \,\,\,$\sigma_*$ & \,\,\,C$_{\rm r}$ & \,\,\,\, $\nu L_r$ &  $L_{\rm[O~III]}$ & $M_{\rm BH}$ \\
\hline	
\endfirsthead

\multicolumn{3}{c}{{\tablename} \thetable{} -- Continued} \\[0.5ex]
\hline \hline 
& \,\,\,\,\,z & NVSS & [O~III] & \,\,\,\,\,m$_{\rm r}$ & \,\,Dn & \,\,\,$\sigma_*$ & \,\,\,C$_{\rm r}$ & \,\,\,\, $\nu L_r$ &  $L_{\rm[O~III]}$ & $M_{\rm BH}$ \\
\hline
\endhead

\hline
  \multicolumn{10}{c}{{Continued on Next Page}} \\
\endfoot

  \\[-1.8ex] 
\endlastfoot


SDSS~J002900.98$-$011341.7 & 0.083 &  282.8 &  158.5 & 14.637 &   1.76 & 321 &   3.33 &  40.86 &  40.46 &   9.0 \\
SDSS~J003930.52$-$103218.6 & 0.129 &   23.7 &    1.9 & 15.996 &   1.97 & 238 &   3.14 &  40.19 &  38.94 &   8.4 \\
SDSS~J004148.22$-$091703.1 & 0.053 &   53.0 &   25.6 & 15.230 &   2.01 & 255 &   3.01 &  39.72 &  39.26 &   8.6 \\
SDSS~J004300.63$-$091346.3 & 0.076 &  148.4 &   40.7 & 15.437 &   1.96 & 244 &   3.32 &  40.50 &  39.79 &   8.5 \\
SDSS~J004530.46$-$004746.9 & 0.147 &   12.4 &   15.1 & 16.266 &   2.02 & 287 &   3.08 &  40.03 &  39.97 &   8.8 \\
SDSS~J011255.11$-$095040.6 & 0.124 &   81.7 &   20.8 & 16.026 &   1.84 & 220 &   2.99 &  40.69 &  39.95 &   8.3 \\
SDSS~J013327.25$-$082416.4 & 0.149 &  218.8 &   31.9 & 16.076 &   2.01 & 331 &   3.35 &  41.29 &  40.31 &   9.0 \\
SDSS~J013412.78$-$010729.4 & 0.079 &  119.6 &   39.6 & 15.152 &   2.03 & 262 &   3.55 &  40.44 &  39.81 &   8.6 \\
SDSS~J013503.43$-$005427.6 & 0.080 &   39.3 &   27.8 & 14.628 &   2.00 & 300 &   3.25 &  39.96 &  39.67 &   8.8 \\
SDSS~J014029.59+001825.8 & 0.148 &   12.3 &    8.2 & 16.642 &   1.95 & 278 &   3.34 &  40.03 &  39.71 &   8.7 \\
SDSS~J015253.79$-$001005.5 & 0.082 &   13.1 &   22.8 & 16.487 &   2.01 & 187 &   3.43 &  39.52 &  39.61 &   8.0 \\
SDSS~J025227.52$-$075605.4 & 0.078 &   99.7 &   10.4 & 15.348 &   1.95 & 237 &   3.00 &  40.35 &  39.22 &   8.4 \\
SDSS~J025437.99+005621.9 & 0.067 &  116.0 &   52.4 & 14.903 &   1.99 & 272 &   3.18 &  40.27 &  39.78 &   8.7 \\
SDSS~J073014.37+393200.4 & 0.142 &   11.3 &    5.3 & 16.716 &   2.07 & 223 &   3.08 &  39.96 &  39.48 &   8.3 \\
SDSS~J073505.25+415827.5 & 0.087 &   67.6 &   21.9 & 17.308 &   1.95 & 256 &   2.29 &  40.28 &  39.65 &   8.6 \\
SDSS~J073719.18+292932.0 & 0.111 &   43.0 &    4.7 & 16.038 &   2.03 & 231 &   2.84 &  40.30 &  39.19 &   8.4 \\
SDSS~J074125.85+480914.3 & 0.120 &   25.7 &   13.0 & 16.540 &   1.94 & 230 &   2.91 &  40.16 &  39.72 &   8.4 \\
SDSS~J074351.25+282128.0 & 0.106 &   20.2 &   15.5 & 16.206 &   1.94 & 252 &   3.34 &  39.94 &  39.68 &   8.5 \\
SDSS~J075309.91+355557.1 & 0.113 &   94.6 &    9.0 & 16.304 &   1.85 & 246 &   2.98 &  40.67 &  39.50 &   8.5 \\
SDSS~J075506.67+262115.9 & 0.123 &   46.0 &    3.1 & 16.483 &   1.94 & 265 &   3.03 &  40.43 &  39.12 &   8.6 \\
SDSS~J080113.28+344030.8 & 0.083 &   45.0 &    5.5 & 15.641 &   2.03 & 226 &   3.13 &  40.06 &  39.00 &   8.3 \\
SDSS~J080326.62+303725.0 & 0.144 &    7.4 &    1.0 & 15.829 &   1.98 & 301 &   2.97 &  39.79 &  38.75 &   8.8 \\
SDSS~J080923.10+211546.2 & 0.142 &   10.6 &   11.8 & 16.910 &   1.88 & 255 &   3.09 &  39.93 &  39.83 &   8.6 \\
SDSS~J081523.21+115715.1 & 0.100 &   40.0 &    3.6 & 15.468 &   1.85 & 251 &   3.27 &  40.18 &  39.00 &   8.5 \\
SDSS~J081604.40+112449.4 & 0.122 &   34.6 &   14.9 & 16.156 &   1.96 & 281 &   3.43 &  40.30 &  39.79 &   8.7 \\
SDSS~J081614.27+425657.6 & 0.127 &  121.0 &   11.3 & 16.628 &   1.98 & 240 &   3.09 &  40.88 &  39.70 &   8.4 \\
SDSS~J081849.74+040631.5 & 0.095 &   61.5 &   18.3 & 15.841 &   1.85 & 233 &   3.18 &  40.32 &  39.64 &   8.4 \\
SDSS~J081932.66+200748.8 & 0.117 &   31.8 &   11.5 & 16.888 &   1.79 & 222 &   3.17 &  40.23 &  39.64 &   8.3 \\
SDSS~J082025.14+160123.2 & 0.148 &   44.2 &   19.3 & 17.505 &   2.00 & 225 &   2.92 &  40.59 &  40.08 &   8.3 \\
SDSS~J082028.09+485347.3 & 0.132 &  188.1 &   12.3 & 15.666 &   2.01 & 295 &   2.72 &  41.11 &  39.78 &   8.8 \\
SDSS~J082603.81+471910.3 & 0.128 &   25.7 &    2.3 & 16.288 &   1.98 & 231 &   3.49 &  40.22 &  39.02 &   8.4 \\
SDSS~J082729.73+530733.4 & 0.119 &   22.9 &   16.7 & 16.942 &   1.88 & 184 &   2.93 &  40.10 &  39.81 &   8.0 \\
SDSS~J082733.62+160053.7 & 0.091 &   53.0 &   12.2 & 15.474 &   1.96 & 250 &   3.20 &  40.21 &  39.43 &   8.5 \\
SDSS~J082926.46+224436.3 & 0.087 &  148.8 &    9.3 & 15.623 &   2.00 & 223 &   3.19 &  40.62 &  39.27 &   8.3 \\
SDSS~J083138.83+223422.9 & 0.087 &  185.3 &   39.4 & 15.441 &   1.89 & 302 &   3.32 &  40.72 &  39.90 &   8.8 \\
SDSS~J083159.69+303930.7 & 0.107 &   20.9 &   52.4 & 15.572 &   1.99 & 246 &   3.18 &  39.96 &  40.21 &   8.5 \\
SDSS~J083224.13+164949.1 & 0.102 &   20.2 &   23.6 & 15.896 &   1.92 & 304 &   3.37 &  39.90 &  39.83 &   8.9 \\
SDSS~J084140.57+254827.9 & 0.140 &   29.1 &   10.6 & 16.353 &   1.97 & 241 &   3.30 &  40.35 &  39.77 &   8.5 \\
SDSS~J084159.65+500551.7 & 0.141 &    7.6 &    3.8 & 16.677 &   2.06 & 259 &   3.15 &  39.78 &  39.33 &   8.6 \\
SDSS~J085321.54+331629.9 & 0.126 &   28.1 &   15.7 & 15.838 &   1.92 & 266 &   2.93 &  40.24 &  39.84 &   8.6 \\
SDSS~J085719.46+241142.6 & 0.133 &   11.8 &   10.9 & 16.694 &   2.06 & 220 &   3.20 &  39.92 &  39.73 &   8.3 \\
SDSS~J090018.16+074535.5 & 0.061 &   70.6 &   31.0 & 14.308 &   1.88 & 261 &   3.20 &  39.97 &  39.46 &   8.6 \\
SDSS~J090245.43+164710.4 & 0.130 &   15.2 &    7.9 & 16.049 &   2.02 & 243 &   3.16 &  40.00 &  39.57 &   8.5 \\
SDSS~J090543.54+401704.8 & 0.128 &   46.0 &    9.7 & 15.924 &   1.89 & 277 &   3.29 &  40.47 &  39.65 &   8.7 \\
SDSS~J091442.02+152155.7 & 0.140 &   28.3 &    7.0 & 16.839 &   1.86 & 235 &   3.32 &  40.35 &  39.60 &   8.4 \\
SDSS~J092049.04+403952.8 & 0.074 &   21.9 &   17.2 & 15.980 &   2.04 & 258 &   3.40 &  39.64 &  39.39 &   8.6 \\
SDSS~J092935.02+625659.3 & 0.121 &   50.9 &   19.4 & 15.785 &   1.90 & 281 &   3.11 &  40.46 &  39.89 &   8.7 \\
SDSS~J093058.74+034827.7 & 0.089 &  111.0 &   22.9 & 15.135 &   2.01 & 276 &   3.16 &  40.51 &  39.68 &   8.7 \\
SDSS~J093305.27+291015.1 & 0.132 &   53.6 &   12.3 & 15.743 &   1.97 & 254 &   3.34 &  40.56 &  39.78 &   8.5 \\
SDSS~J094202.04+105818.3 & 0.136 &   14.4 &    3.5 & 16.613 &   1.90 & 242 &   3.28 &  40.02 &  39.26 &   8.5 \\
SDSS~J094332.99+334158.3 & 0.131 &   23.9 &    2.8 & 15.974 &   1.91 & 276 &   3.24 &  40.21 &  39.14 &   8.7 \\
SDSS~J094614.50+581937.6 & 0.147 &   80.9 &   21.7 & 15.995 &   1.97 & 307 &   3.14 &  40.85 &  40.13 &   8.9 \\
SDSS~J095527.76+034516.8 & 0.091 &   47.0 &     -- & 16.615 &   1.81 & 190 &   3.10 &  40.16 &     -- &   8.0 \\
SDSS~J100451.83+543404.3 & 0.047 &  121.8 &   56.9 & 13.980 &   1.95 & 269 &   3.38 &  39.98 &  39.50 &   8.6 \\
SDSS~J100757.06+280147.9 & 0.148 &   27.3 &   11.5 & 15.864 &   1.99 & 309 &   3.26 &  40.38 &  39.86 &   8.9 \\
SDSS~J100804.13+502642.8 & 0.134 &   53.7 &    1.9 & 16.736 &   2.00 & 218 &   3.23 &  40.58 &  38.98 &   8.3 \\
SDSS~J101114.38+191425.7 & 0.149 &   31.5 &   11.8 & 16.531 &   1.89 & 269 &   3.38 &  40.45 &  39.87 &   8.6 \\
SDSS~J101545.46+311500.2 & 0.125 &   15.6 &    4.4 & 16.869 &   1.77 & 211 &   3.09 &  39.98 &  39.28 &   8.2 \\
SDSS~J101937.94+001955.7 & 0.096 &   43.0 &   31.7 & 15.251 &   1.96 & 275 &   3.16 &  40.17 &  39.89 &   8.7 \\
SDSS~J102008.61+174817.4 & 0.122 &   19.5 &    0.7 & 16.405 &   1.92 & 240 &   3.46 &  40.05 &  38.43 &   8.4 \\
SDSS~J102314.24+483122.0 & 0.148 &   31.1 &   17.0 & 16.331 &   1.98 & 296 &   3.24 &  40.44 &  40.03 &   8.8 \\
SDSS~J102603.83+390524.0 & 0.145 &  108.1 &   11.0 & 17.009 &   1.89 & 258 &   2.84 &  40.96 &  39.82 &   8.6 \\
SDSS~J102703.83+382013.0 & 0.123 &   41.0 &    9.0 & 15.916 &   1.87 & 209 &   3.17 &  40.38 &  39.58 &   8.2 \\
SDSS~J103036.15+355459.8 & 0.124 &   50.0 &    8.5 & 15.487 &   2.05 & 325 &   3.40 &  40.47 &  39.56 &   9.0 \\
SDSS~J103126.60+115250.5 & 0.144 &   25.2 &   12.7 & 16.291 &   1.98 & 259 &   3.15 &  40.32 &  39.87 &   8.6 \\
SDSS~J103258.88+564453.2 & 0.045 &  213.4 &  114.6 & 13.549 &   1.97 & 280 &   3.08 &  40.18 &  39.76 &   8.7 \\
SDSS~J103827.01+414852.9 & 0.125 &   43.0 &   14.7 & 16.366 &   1.82 & 193 &   3.23 &  40.42 &  39.80 &   8.1 \\
SDSS~J103930.43+394718.9 & 0.093 &   23.3 &   15.6 & 15.075 &   1.96 & 274 &   3.34 &  39.88 &  39.56 &   8.7 \\
SDSS~J104045.34+395448.5 & 0.134 &   35.8 &    2.5 & 16.888 &   1.91 & 235 &   2.98 &  40.40 &  39.11 &   8.4 \\
SDSS~J104049.99+561508.1 & 0.134 &   22.3 &   15.9 & 16.123 &   1.77 & 257 &   3.25 &  40.20 &  39.90 &   8.6 \\
SDSS~J104233.38+363946.5 & 0.142 &   53.1 &    2.0 & 16.347 &   1.82 & 243 &   2.95 &  40.63 &  39.06 &   8.5 \\
SDSS~J104855.28+311945.2 & 0.117 &   52.2 &   10.4 & 15.607 &   1.93 & 270 &   3.08 &  40.44 &  39.59 &   8.7 \\
SDSS~J104907.26+551314.9 & 0.126 &   24.0 &   23.8 & 15.440 &   1.89 & 448 &   3.30 &  40.17 &  40.02 &   9.5 \\
SDSS~J104921.13$-$004005.0 & 0.039 &  250.0 &   70.2 & 13.544 &   1.90 & 226 &   2.83 &  40.11 &  39.42 &   8.3 \\
SDSS~J105147.39+552308.3 & 0.074 &  522.0 &   48.5 & 14.624 &   2.08 & 320 &   3.54 &  41.02 &  39.84 &   9.0 \\
SDSS~J105259.97+430255.0 & 0.148 &   52.0 &   23.6 & 16.414 &   1.87 & 225 &   3.32 &  40.66 &  40.17 &   8.3 \\
SDSS~J105344.12+492955.9 & 0.140 &   64.3 &   33.4 & 16.012 &   1.30 & 262 &   3.49 &  40.70 &  40.27 &   8.6 \\
SDSS~J105348.93+402345.9 & 0.128 &  100.1 &    1.2 & 15.592 &   1.97 & 283 &   3.22 &  40.81 &  38.72 &   8.7 \\
SDSS~J105544.98+452401.4 & 0.064 &  133.5 &    6.8 & 14.706 &   2.02 & 213 &   3.34 &  40.29 &  38.85 &   8.2 \\
SDSS~J105702.79+564503.1 & 0.136 &   15.4 &    4.1 & 16.207 &   2.02 & 304 &   3.39 &  40.05 &  39.33 &   8.9 \\
SDSS~J105847.67+164526.0 & 0.116 &   49.4 &   16.4 & 16.261 &   1.88 & 226 &   3.35 &  40.41 &  39.79 &   8.3 \\
SDSS~J110535.78+091956.3 & 0.127 &   22.3 &    6.1 & 16.110 &   2.00 & 251 &   3.23 &  40.15 &  39.44 &   8.5 \\
SDSS~J111020.07+204657.5 & 0.135 &   12.9 &    2.9 & 16.509 &   1.97 & 275 &   3.38 &  39.97 &  39.18 &   8.7 \\
SDSS~J111037.33+541135.7 & 0.141 &   23.5 &   15.7 & 16.449 &   1.99 & 206 &   3.08 &  40.27 &  39.95 &   8.2 \\
SDSS~J111211.37+304352.3 & 0.106 &   43.5 &   24.5 & 15.990 &   1.95 & 244 &   3.29 &  40.27 &  39.87 &   8.5 \\
SDSS~J111337.13+234846.5 & 0.140 &    8.4 &   17.4 & 16.842 &   1.96 & 249 &   3.33 &  39.82 &  39.99 &   8.5 \\
SDSS~J111911.13+081539.8 & 0.076 &   80.0 &   25.6 & 15.054 &   2.05 & 257 &   3.16 &  40.22 &  39.58 &   8.6 \\
SDSS~J112055.83+173854.0 & 0.085 &   21.5 &    9.4 & 15.623 &   1.97 & 205 &   2.85 &  39.75 &  39.25 &   8.2 \\
SDSS~J112352.34+443735.6 & 0.139 &   22.9 &   10.1 & 17.434 &   1.90 & 273 &   3.85 &  40.24 &  39.74 &   8.7 \\
SDSS~J112403.19+475814.9 & 0.140 &   28.2 &    3.0 & 16.675 &   1.97 & 266 &   3.22 &  40.35 &  39.22 &   8.6 \\
SDSS~J112457.40+171744.7 & 0.142 &    9.5 &    5.0 & 16.147 &   2.00 & 285 &   3.03 &  39.88 &  39.46 &   8.8 \\
SDSS~J112603.59+545329.1 & 0.149 &   16.0 &   12.4 & 16.871 &   1.94 & 258 &   3.46 &  40.16 &  39.90 &   8.6 \\
SDSS~J113012.79+235822.1 & 0.140 &   46.5 &   19.8 & 16.474 &   1.80 & 265 &   3.60 &  40.56 &  40.05 &   8.6 \\
SDSS~J113359.23+490343.4 & 0.032 &  732.0 &   84.7 & 13.126 &   1.90 & 264 &   3.30 &  40.40 &  39.32 &   8.6 \\
SDSS~J114210.72+552729.6 & 0.133 &   24.9 &    8.8 & 16.099 &   2.01 & 218 &   3.05 &  40.24 &  39.64 &   8.3 \\
SDSS~J114212.11+101159.0 & 0.103 &   46.6 &    9.1 & 16.086 &   1.89 & 243 &   3.10 &  40.28 &  39.42 &   8.5 \\
SDSS~J114345.53+192333.4 & 0.094 &   37.8 &   17.0 & 15.411 &   2.00 & 237 &   3.17 &  40.10 &  39.61 &   8.4 \\
SDSS~J115109.39+435918.6 & 0.071 &   75.0 &   37.5 & 14.982 &   1.92 & 216 &   3.17 &  40.13 &  39.69 &   8.3 \\
SDSS~J115323.89+305904.8 & 0.136 &   43.2 &    8.9 & 16.911 &   1.97 & 191 &   3.30 &  40.50 &  39.67 &   8.1 \\
SDSS~J115508.97+232623.4 & 0.144 &   26.4 &   12.6 & 17.279 &   2.00 & 245 &   2.98 &  40.34 &  39.87 &   8.5 \\
SDSS~J115729.60+292308.1 & 0.140 &   16.6 &    9.3 & 17.195 &   2.01 & 224 &   3.32 &  40.11 &  39.71 &   8.3 \\
SDSS~J115816.37+340605.9 & 0.131 &   19.8 &   26.3 & 16.311 &   1.85 & 221 &   3.16 &  40.13 &  40.11 &   8.3 \\
SDSS~J115936.05+233947.5 & 0.142 &   31.8 &    5.0 & 16.510 &   1.92 & 282 &   3.51 &  40.41 &  39.45 &   8.7 \\
SDSS~J120021.93$-$020152.7 & 0.146 &   40.4 &   24.8 & 16.336 &   2.04 & 248 &   3.12 &  40.54 &  40.18 &   8.5 \\
SDSS~J120401.47+201356.3 & 0.024 &  402.1 &  111.6 & 12.997 &   1.81 & 270 &   3.25 &  39.91 &  39.21 &   8.7 \\
SDSS~J120425.29+034510.6 & 0.149 &   54.7 &    4.2 & 16.608 &   1.95 & 248 &   3.14 &  40.69 &  39.43 &   8.5 \\
SDSS~J120522.29+050941.4 & 0.136 &  105.0 &    0.6 & 16.002 &   2.02 & 335 &   3.47 &  40.88 &  38.49 &   9.0 \\
SDSS~J120943.62$-$020459.6 & 0.100 &   32.3 &   23.3 & 15.922 &   1.98 & 253 &   3.34 &  40.09 &  39.80 &   8.5 \\
SDSS~J121110.99+060744.1 & 0.139 &   50.9 &   12.8 & 16.154 &   1.94 & 221 &   3.12 &  40.59 &  39.84 &   8.3 \\
SDSS~J121114.07+060833.9 & 0.138 &   78.5 &   11.1 & 16.143 &   1.96 & 238 &   3.18 &  40.78 &  39.78 &   8.4 \\
SDSS~J121121.12+141439.2 & 0.064 &   62.0 &   56.4 & 14.783 &   2.03 & 233 &   3.13 &  39.96 &  39.77 &   8.4 \\
SDSS~J121332.93+072516.9 & 0.137 &   17.2 &   10.9 & 15.739 &   1.92 & 242 &   3.00 &  40.11 &  39.76 &   8.5 \\
SDSS~J121519.19+472142.4 & 0.146 &   31.1 &    2.9 & 17.108 &   0.00 & 193 &   3.09 &  40.42 &  39.25 &   8.1 \\
SDSS~J121534.18+135635.0 & 0.093 &   30.9 &   21.1 & 15.362 &   2.04 & 303 &   3.15 &  40.00 &  39.69 &   8.9 \\
SDSS~J121543.82+170917.6 & 0.095 &  460.0 &   40.0 & 14.538 &   1.99 & 312 &   2.86 &  41.19 &  39.99 &   8.9 \\
SDSS~J121619.95+155417.7 & 0.093 &   77.2 &   19.0 & 15.228 &   1.81 & 210 &   3.11 &  40.40 &  39.64 &   8.2 \\
SDSS~J121640.12+034231.5 & 0.080 &  207.0 &   24.0 & 15.677 &   1.99 & 243 &   3.22 &  40.69 &  39.61 &   8.5 \\
SDSS~J122156.16+020450.8 & 0.126 &  119.5 &    9.5 & 16.885 &   1.92 & 232 &   3.40 &  40.87 &  39.62 &   8.4 \\
SDSS~J122532.09+192615.2 & 0.129 &   25.6 &   23.9 & 17.363 &   1.82 & 192 &   3.05 &  40.22 &  40.05 &   8.1 \\
SDSS~J122622.49+640622.0 & 0.110 &   81.0 &   43.6 & 15.516 &   1.82 & 257 &   3.09 &  40.58 &  40.16 &   8.6 \\
SDSS~J122640.83+430509.2 & 0.074 &   33.5 &   18.4 & 14.836 &   1.98 & 272 &   3.15 &  39.83 &  39.42 &   8.7 \\
SDSS~J123128.93+491537.0 & 0.111 &   52.1 &    5.1 & 16.192 &   1.85 & 217 &   3.06 &  40.39 &  39.24 &   8.3 \\
SDSS~J124135.94+162033.6 & 0.070 &  165.0 &   77.2 & 15.040 &   1.95 & 305 &   3.31 &  40.47 &  39.99 &   8.9 \\
SDSS~J124207.38+502146.6 & 0.148 &  102.0 &   14.3 & 16.365 &   1.97 & 237 &   3.33 &  40.95 &  39.95 &   8.4 \\
SDSS~J124622.48+075327.9 & 0.111 &   19.8 &   23.4 & 15.774 &   1.95 & 275 &   3.25 &  39.97 &  39.89 &   8.7 \\
SDSS~J124647.52+545315.0 & 0.085 &   38.5 &   32.6 & 14.974 &   2.01 & 268 &   3.09 &  40.01 &  39.79 &   8.6 \\
SDSS~J125434.93$-$023412.4 & 0.116 &  122.6 &    2.0 & 15.959 &   1.93 & 260 &   3.30 &  40.80 &  38.86 &   8.6 \\
SDSS~J125953.32+575149.7 & 0.149 &   15.5 &    5.5 & 17.347 &   1.71 & 211 &   3.17 &  40.14 &  39.55 &   8.2 \\
SDSS~J130203.58$-$005012.3 & 0.085 &   73.7 &   30.2 & 14.964 &   2.10 & 243 &   3.27 &  40.29 &  39.76 &   8.5 \\
SDSS~J130248.70+475510.6 & 0.141 &   47.9 &   20.0 & 15.563 &   1.95 & 301 &   3.06 &  40.58 &  40.05 &   8.8 \\
SDSS~J130619.24+111339.7 & 0.086 &  321.0 &   44.5 & 14.926 &   1.72 & 267 &   3.04 &  40.94 &  39.94 &   8.6 \\
SDSS~J131053.44$-$022841.5 & 0.143 &   53.0 &   12.7 & 16.792 &   2.04 & 267 &   2.77 &  40.64 &  39.87 &   8.6 \\
SDSS~J131531.07+525437.3 & 0.121 &   37.1 &    0.5 & 16.470 &   2.01 & 247 &   3.31 &  40.33 &  38.30 &   8.5 \\
SDSS~J131613.54+093236.7 & 0.094 &   42.0 &   23.6 & 15.558 &   1.95 & 293 &   3.43 &  40.14 &  39.74 &   8.8 \\
SDSS~J132017.54+043037.4 & 0.146 &   30.5 &    4.4 & 15.935 &   1.81 & 229 &   2.59 &  40.42 &  39.42 &   8.4 \\
SDSS~J132302.49+172832.9 & 0.120 &   38.7 &   13.9 & 16.128 &   1.92 & 242 &   3.36 &  40.34 &  39.75 &   8.5 \\
SDSS~J132736.13+270816.8 & 0.143 &   34.2 &   11.2 & 16.081 &   2.03 & 267 &   3.35 &  40.45 &  39.82 &   8.6 \\
SDSS~J133038.01+390815.4 & 0.146 &   47.5 &   14.3 & 16.858 &   1.93 & 236 &   3.22 &  40.61 &  39.94 &   8.4 \\
SDSS~J134529.50+054952.9 & 0.127 &   54.2 &   15.8 & 16.498 &   1.97 & 240 &   3.21 &  40.53 &  39.85 &   8.4 \\
SDSS~J134745.19+503203.5 & 0.150 &   18.2 &    1.3 & 16.204 &   2.06 & 244 &   3.35 &  40.21 &  38.91 &   8.5 \\
SDSS~J135214.56+123401.7 & 0.145 &   11.0 &     -- & 16.729 &   2.02 & 239 &   3.44 &  39.96 &     -- &   8.4 \\
SDSS~J135302.04+330528.5 & 0.061 &   80.0 &   22.1 & 14.703 &   1.93 & 208 &   3.24 &  40.03 &  39.32 &   8.2 \\
SDSS~J135511.34+242415.6 & 0.137 &   16.1 &    4.1 & 16.713 &   1.90 & 270 &   3.08 &  40.08 &  39.34 &   8.7 \\
SDSS~J135553.63+262217.9 & 0.141 &   69.8 &    6.9 & 15.904 &   1.94 & 294 &   3.17 &  40.74 &  39.59 &   8.8 \\
SDSS~J135655.28+271120.2 & 0.141 &   16.3 &   13.5 & 16.061 &   1.93 & 253 &   3.06 &  40.11 &  39.88 &   8.5 \\
SDSS~J140313.28+061008.2 & 0.083 &  256.6 &   76.9 & 15.097 &   1.87 & 340 &   3.23 &  40.81 &  40.14 &   9.1 \\
SDSS~J140916.74+060139.4 & 0.139 &   10.8 &   25.3 & 16.844 &   1.91 & 218 &   3.37 &  39.92 &  40.14 &   8.3 \\
SDSS~J141138.22+495304.0 & 0.129 &   27.5 &   16.5 & 16.478 &   2.12 & 230 &   3.10 &  40.25 &  39.88 &   8.4 \\
SDSS~J141243.83+495206.5 & 0.077 &  101.4 &   24.0 & 14.643 &   2.04 & 303 &   3.23 &  40.35 &  39.58 &   8.9 \\
SDSS~J141427.10+282830.5 & 0.140 &   77.1 &    9.4 & 16.420 &   1.72 & 263 &   3.30 &  40.78 &  39.72 &   8.6 \\
SDSS~J141652.94+104826.7 & 0.025 & 4581.1 &  228.7 & 12.057 &   1.97 & 341 &   3.06 &  40.98 &  39.53 &   9.1 \\
SDSS~J142206.79+361434.8 & 0.124 &   13.7 &   17.5 & 16.742 &   1.73 & 226 &   2.75 &  39.91 &  39.87 &   8.3 \\
SDSS~J142521.22+630921.3 & 0.136 &   19.8 &    7.7 & 15.833 &   1.89 & 273 &   3.22 &  40.16 &  39.60 &   8.7 \\
SDSS~J142616.34+005015.3 & 0.125 &   88.7 &   10.1 & 15.878 &   1.97 & 254 &   2.98 &  40.73 &  39.64 &   8.6 \\
SDSS~J142623.76+551804.9 & 0.132 &   52.8 &    8.5 & 16.254 &   1.96 & 205 &   3.30 &  40.56 &  39.62 &   8.2 \\
SDSS~J142649.23+621005.9 & 0.109 &   25.9 &   28.5 & 15.539 &   1.94 & 306 &   3.40 &  40.07 &  39.97 &   8.9 \\
SDSS~J142832.60+424021.0 & 0.129 &   56.4 &   48.0 & 16.182 &   1.18 & 251 &   3.23 &  40.57 &  40.35 &   8.5 \\
SDSS~J143147.54+605109.4 & 0.113 &  163.0 &   27.7 & 15.726 &   1.94 & 251 &   3.19 &  40.90 &  39.99 &   8.5 \\
SDSS~J143257.81+043715.1 & 0.106 &   68.6 &   21.8 & 15.767 &   1.99 & 302 &   3.14 &  40.46 &  39.82 &   8.9 \\
SDSS~J143638.56+011058.8 & 0.137 &   19.2 &   15.9 & 15.944 &   2.02 & 326 &   2.98 &  40.16 &  39.93 &   9.0 \\
SDSS~J143928.78+110613.8 & 0.125 &   39.8 &   12.7 & 16.068 &   1.67 & 247 &   2.86 &  40.39 &  39.74 &   8.5 \\
SDSS~J145215.46+502225.1 & 0.094 &  133.8 &   13.9 & 15.124 &   1.99 & 246 &   3.14 &  40.65 &  39.52 &   8.5 \\
SDSS~J145555.27+115141.4 & 0.032 &  382.0 &  142.0 & 13.224 &   1.98 & 287 &   3.30 &  40.12 &  39.54 &   8.8 \\
SDSS~J150111.50+093547.9 & 0.145 &   15.2 &   15.3 & 16.549 &   1.98 & 246 &   3.09 &  40.10 &  39.96 &   8.5 \\
SDSS~J150148.14+163345.6 & 0.150 &   37.2 &    3.9 & 16.488 &   1.98 & 225 &   3.36 &  40.53 &  39.40 &   8.3 \\
SDSS~J150408.01+565545.4 & 0.148 &   57.3 &   12.4 & 16.708 &   1.98 & 276 &   3.30 &  40.70 &  39.89 &   8.7 \\
SDSS~J150450.51+044054.8 & 0.092 &   53.6 &    7.7 & 15.848 &   1.89 & 227 &   2.99 &  40.23 &  39.24 &   8.3 \\
SDSS~J150957.37+332715.0 & 0.117 &   66.0 &   11.7 & 15.956 &   1.97 & 266 &   3.24 &  40.54 &  39.65 &   8.6 \\
SDSS~J150959.74+332746.1 & 0.110 &   66.0 &   25.0 & 16.560 &   1.90 & 232 &   3.10 &  40.49 &  39.92 &   8.4 \\
SDSS~J151744.96+310015.8 & 0.136 &   49.0 &    7.1 & 16.224 &   1.94 & 291 &   3.21 &  40.55 &  39.57 &   8.8 \\
SDSS~J151845.72+061356.1 & 0.102 &  487.0 &   26.0 & 15.589 &   1.31 & 297 &   3.63 &  41.28 &  39.86 &   8.8 \\
SDSS~J152045.04+483922.9 & 0.078 &   80.6 &   19.8 & 15.917 &   2.12 & 257 &   3.04 &  40.26 &  39.50 &   8.6 \\
SDSS~J152122.54+042030.1 & 0.052 &  452.0 &   51.4 & 13.936 &   1.85 & 280 &   2.93 &  40.64 &  39.55 &   8.7 \\
SDSS~J152126.99+483943.2 & 0.074 &   63.3 &   17.4 & 15.628 &   2.10 & 256 &   3.14 &  40.10 &  39.40 &   8.6 \\
SDSS~J152235.19+155707.6 & 0.145 &   25.2 &   15.2 & 16.575 &   2.11 & 298 &   3.28 &  40.33 &  39.96 &   8.8 \\
SDSS~J152326.91+283732.5 & 0.082 &  733.0 &   48.9 & 15.021 &   1.93 & 234 &   3.16 &  41.26 &  39.94 &   8.4 \\
SDSS~J152500.83+332359.8 & 0.082 &   75.5 &   52.6 & 15.232 &   1.84 & 243 &   3.21 &  40.27 &  39.96 &   8.5 \\
SDSS~J152522.33+314037.1 & 0.079 &   51.4 &   33.7 & 15.541 &   2.01 & 250 &   3.35 &  40.07 &  39.74 &   8.5 \\
SDSS~J152715.31+133650.9 & 0.144 &   47.2 &   25.9 & 16.172 &   2.00 & 291 &   2.88 &  40.59 &  40.19 &   8.8 \\
SDSS~J152737.36+412947.1 & 0.135 &   14.3 &     -- & 16.562 &   2.00 & 222 &   3.07 &  40.02 &     -- &   8.3 \\
SDSS~J152945.60+304235.6 & 0.114 &   88.0 &   27.8 & 15.252 &   2.05 & 356 &   3.55 &  40.64 &  40.00 &   9.1 \\
SDSS~J153138.76+064045.5 & 0.101 &   40.1 &   47.2 & 14.959 &   1.83 & 312 &   3.25 &  40.19 &  40.12 &   8.9 \\
SDSS~J153215.31+433844.5 & 0.145 &   26.8 &   15.9 & 16.593 &   1.96 & 250 &   3.17 &  40.35 &  39.98 &   8.5 \\
SDSS~J153621.11+084112.1 & 0.126 &   68.1 &   18.6 & 16.255 &   2.00 & 266 &   3.31 &  40.63 &  39.92 &   8.6 \\
SDSS~J153932.09+013710.5 & 0.116 &   20.1 &   26.6 & 16.279 &   1.99 & 278 &   2.93 &  40.02 &  39.99 &   8.7 \\
SDSS~J154155.16+012517.4 & 0.085 &  113.0 &   24.5 & 15.291 &   1.96 & 276 &   3.25 &  40.48 &  39.67 &   8.7 \\
SDSS~J154709.22+353846.1 & 0.079 &  213.9 &   36.2 & 14.574 &   2.06 & 355 &   3.01 &  40.70 &  39.78 &   9.1 \\
SDSS~J155222.36+223311.9 & 0.068 &   44.8 &   40.0 & 14.802 &   2.04 & 267 &   3.44 &  39.88 &  39.68 &   8.6 \\
SDSS~J155311.93+273320.6 & 0.147 &  115.0 &    9.0 & 15.829 &   1.98 & 271 &   2.89 &  41.00 &  39.75 &   8.7 \\
SDSS~J155401.99+150946.8 & 0.132 &   50.5 &   19.7 & 16.590 &   1.99 & 240 &   3.21 &  40.54 &  39.98 &   8.5 \\
SDSS~J155721.38+544015.9 & 0.047 &   90.0 &   91.8 & 13.930 &   1.79 & 289 &   3.44 &  39.84 &  39.71 &   8.8 \\
SDSS~J160816.32+373743.1 & 0.103 &   40.4 &   13.3 & 16.018 &   1.99 & 248 &   3.30 &  40.21 &  39.59 &   8.5 \\
SDSS~J161037.77+532421.0 & 0.064 &   61.2 &   48.6 & 14.847 &   2.02 & 287 &   3.43 &  39.95 &  39.71 &   8.8 \\
SDSS~J161114.11+265524.2 & 0.032 &  102.8 &  111.1 & 13.437 &   1.97 & 263 &   3.42 &  39.56 &  39.44 &   8.6 \\
SDSS~J161242.69+295404.7 & 0.053 &   36.1 &    5.6 & 14.666 &   1.98 & 215 &   3.29 &  39.56 &  38.61 &   8.3 \\
SDSS~J162700.42+275547.7 & 0.132 &   14.5 &    8.7 & 16.157 &   1.96 & 264 &   2.97 &  40.00 &  39.63 &   8.6 \\
SDSS~J162806.20+084538.0 & 0.143 &   35.8 &    8.3 & 16.376 &   1.97 & 279 &   3.46 &  40.47 &  39.69 &   8.7 \\
SDSS~J162918.66+133824.0 & 0.118 &   72.0 &   15.6 & 16.093 &   1.96 & 249 &   3.31 &  40.59 &  39.78 &   8.5 \\
SDSS~J163043.14+163910.8 & 0.090 &   52.0 &    7.4 & 15.700 &   2.03 & 224 &   3.39 &  40.19 &  39.20 &   8.3 \\
SDSS~J164053.90+324728.4 & 0.136 &   50.0 &    5.4 & 16.458 &   1.97 & 226 &   3.19 &  40.56 &  39.45 &   8.3 \\
SDSS~J164548.45+393227.4 & 0.141 &   25.4 &   37.4 & 17.304 &   1.75 & 186 &   3.02 &  40.30 &  40.33 &   8.0 \\
SDSS~J164845.08+254119.5 & 0.115 &  102.6 &    8.8 & 16.201 &   2.02 & 228 &   3.13 &  40.72 &  39.50 &   8.4 \\
SDSS~J165304.98+400702.5 & 0.148 &   87.7 &   19.4 & 16.423 &   1.96 & 256 &   3.27 &  40.89 &  40.09 &   8.6 \\
SDSS~J165425.53+414121.2 & 0.147 &   17.2 &    6.8 & 17.283 &   1.91 & 201 &   3.04 &  40.17 &  39.62 &   8.1 \\
SDSS~J165448.44+261841.3 & 0.100 &   18.8 &   44.1 & 15.677 &   1.97 & 256 &   3.31 &  39.86 &  40.08 &   8.6 \\
SDSS~J165500.19+390847.9 & 0.139 &   54.8 &   17.5 & 15.887 &   2.05 & 358 &   3.42 &  40.63 &  39.99 &   9.1 \\
SDSS~J165744.77+215611.1 & 0.141 &   25.5 &   23.2 & 16.862 &   1.86 & 277 &   3.49 &  40.30 &  40.12 &   8.7 \\
SDSS~J170011.22+323514.7 & 0.102 &  185.0 &    9.1 & 15.734 &   1.87 & 239 &   3.26 &  40.86 &  39.41 &   8.4 \\
SDSS~J170115.59+240608.4 & 0.138 &   26.0 &    5.2 & 17.079 &   2.08 & 240 &   3.26 &  40.30 &  39.46 &   8.4 \\
SDSS~J170543.99+583001.2 & 0.114 &   26.9 &     -- & 16.436 &   1.97 & 183 &   3.20 &  40.13 &     -- &   8.0 \\
SDSS~J170602.20+201757.8 & 0.122 &   29.3 &    5.1 & 15.824 &   1.93 & 295 &   3.16 &  40.23 &  39.32 &   8.8 \\
SDSS~J171137.98+580330.2 & 0.147 &   42.0 &   18.6 & 16.354 &   2.11 & 312 &   3.32 &  40.56 &  40.06 &   8.9 \\
SDSS~J171223.15+640157.1 & 0.080 &  150.0 &   10.8 & 15.848 &   2.02 & 235 &   3.03 &  40.55 &  39.26 &   8.4 \\
SDSS~J171243.95+620245.0 & 0.122 &   47.2 &   19.4 & 16.243 &   1.86 & 274 &   3.35 &  40.43 &  39.90 &   8.7 \\
SDSS~J173223.73+552452.8 & 0.062 &   54.8 &   42.6 & 15.093 &   2.09 & 247 &   3.24 &  39.88 &  39.62 &   8.5 \\
SDSS~J212005.00$-$075350.1 & 0.139 &   48.2 &   22.4 & 16.844 &   2.02 & 263 &   3.21 &  40.57 &  40.09 &   8.6 \\
SDSS~J214239.29$-$080423.8 & 0.128 &   13.9 &    8.9 & 16.904 &   1.95 & 231 &   3.19 &  39.95 &  39.61 &   8.4 \\
SDSS~J223143.19$-$082431.7 & 0.083 &  766.0 &   61.1 & 14.149 &   2.01 & 301 &   2.86 &  41.29 &  40.04 &   8.8 \\
SDSS~J234702.42$-$010300.9 & 0.133 &   34.3 &    5.8 & 17.507 &   1.83 & 202 &   2.98 &  40.38 &  39.46 &   8.1 \\

\hline
\hline
\end{longtable}
\end{center}
Column description: (1) source name; (2)
redshift; (3) NVSS 1.4 GHz flux density [mJy]; (4) [O~III] flux [in
10$^{-17}$ erg cm$^{-2}$ s$^{-1}$ units]; (5) SDSS DR7 r band AB magnitude; (6)
concentration index $C_r$; (7) Dn(4000) index; (8) stellar velocity dispersion
[\kms]; (9) logarithm of the radio luminosity [erg s$^{-1}$]; (10) logarithm
of the [O~III] line luminosity [erg s$^{-1}$]; (11) logarithm of the black
hole mass [in solar units].
\twocolumn

\newpage
\clearpage
\onecolumn
\begin{center}
\begin{longtable}{l l r r l l l l l l l l l}

\caption[Properties of the \sFR\ sources.]{Properties of the \sFR\ sources.} 
\label{tabn} \\

\hline \hline 
& \,\,\,\,\,z & NVSS & [O~III] & \,\,\,\,\,m$_{\rm r}$ & \,\,Dn & \,\,\,$\sigma_*$ & \,\,\,C$_{\rm r}$ & \,\,\,\, $\nu L_r$ &  $L_{\rm[O~III]}$ & $M_{\rm BH}$ \\
\hline	
\endfirsthead

\multicolumn{3}{c}{{\tablename} \thetable{} -- Continued} \\[0.5ex]
\hline \hline 
& \,\,\,\,\,z & NVSS & [O~III] & \,\,\,\,\,m$_{\rm r}$ & \,\,Dn & \,\,\,$\sigma_*$ & \,\,\,C$_{\rm r}$ & \,\,\,\, $\nu L_r$ &  $L_{\rm[O~III]}$ & $M_{\rm BH}$ \\
\hline
\endhead

\hline
  \multicolumn{10}{c}{{Continued on Next Page}} \\
\endfoot

  \\[-1.8ex] 
\endlastfoot

SDSS~J090100.09+103701.7 & 0.029 &   63.3 &  162.2 & 13.254 &   1.96 & 250 &   3.46 &  39.27 &  39.54 &   8.5 \\
SDSS~J092122.11+545153.9 & 0.045 &   36.6 &   50.5 & 14.217 &   1.95 & 253 &   3.35 &  39.41 &  39.40 &   8.5 \\
SDSS~J092151.48+332406.5 & 0.024 &  117.3 &  109.4 & 13.127 &   1.95 & 227 &   3.34 &  39.34 &  39.17 &   8.4 \\
SDSS~J093957.34+164712.8 & 0.047 &   41.3 &   49.3 & 14.982 &   1.70 & 217 &   2.95 &  39.51 &  39.44 &   8.3 \\
SDSS~J101623.01+601405.6 & 0.031 &   35.0 &   58.1 & 13.244 &   1.95 & 251 &   3.24 &  39.07 &  39.15 &   8.5 \\
SDSS~J104740.48+385553.6 & 0.035 &   55.9 &   49.9 & 13.255 &   1.96 & 304 &   3.36 &  39.39 &  39.19 &   8.9 \\
SDSS~J111125.21+265748.9 & 0.034 &   86.8 &   87.5 & 13.322 &   1.99 & 281 &   2.91 &  39.53 &  39.38 &   8.7 \\
SDSS~J132451.44+362242.7 & 0.017 &  789.4 &  394.5 & 12.662 &   2.01 & 242 &   3.32 &  39.91 &  39.46 &   8.5 \\
SDSS~J133242.54+071938.1 & 0.023 &  152.5 &  169.9 & 12.987 &   1.81 & 237 &   2.86 &  39.45 &  39.35 &   8.4 \\
SDSS~J145222.83+170717.8 & 0.045 &  102.8 &   20.4 & 14.226 &   1.94 & 240 &   3.05 &  39.86 &  39.01 &   8.4 \\
SDSS~J155603.90+242652.9 & 0.043 &  127.0 &   91.9 & 14.064 &   1.86 & 252 &   3.12 &  39.90 &  39.62 &   8.5 \\
SDSS~J155749.61+161836.6 & 0.037 &  113.4 &   70.4 & 13.100 &   2.00 & 328 &   2.99 &  39.73 &  39.38 &   9.0 \\
SDSS~J160332.08+171155.2 & 0.034 &  662.0 &   79.0 & 13.549 &   2.02 & 298 &   3.45 &  40.42 &  39.35 &   8.8 \\
SDSS~J160722.95+135316.4 & 0.034 &   75.1 &   46.3 & 13.554 &   1.99 & 268 &   3.41 &  39.47 &  39.11 &   8.6 \\
\hline
\hline
\end{longtable}
\end{center}
Column description: (1) source name; (2)
redshift; (3) NVSS 1.4 GHz flux density [mJy]; (4) [O~III] flux [in
10$^{-17}$ erg cm$^{-2}$ s$^{-1}$ units]; (5) SDSS DR7 r band AB magnitude; (6)
concentration index $C_r$; (7) Dn(4000) index; (8) stellar velocity dispersion
[\kms]; (9) logarithm of the radio luminosity [erg s$^{-1}$]; (10) logarithm
of the [O~III] line luminosity [erg s$^{-1}$]; (11) logarithm of the black
hole mass [in solar units].
\twocolumn

\appendix
\section{FIRST images of the 219 \FR\ sources.}

\begin{figure*}
\includegraphics[width=6.3cm,height=6.3cm]{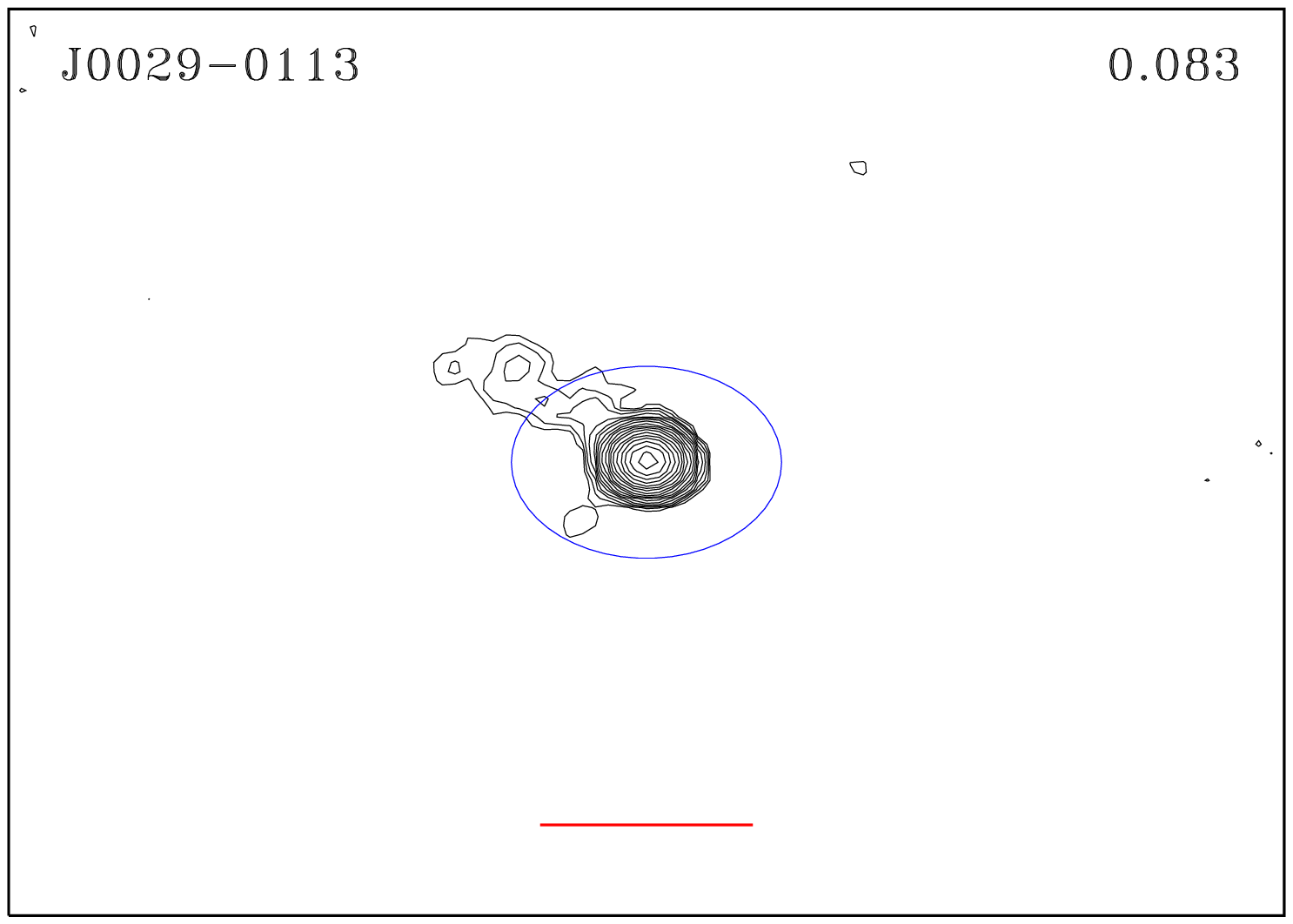} 
\includegraphics[width=6.3cm,height=6.3cm]{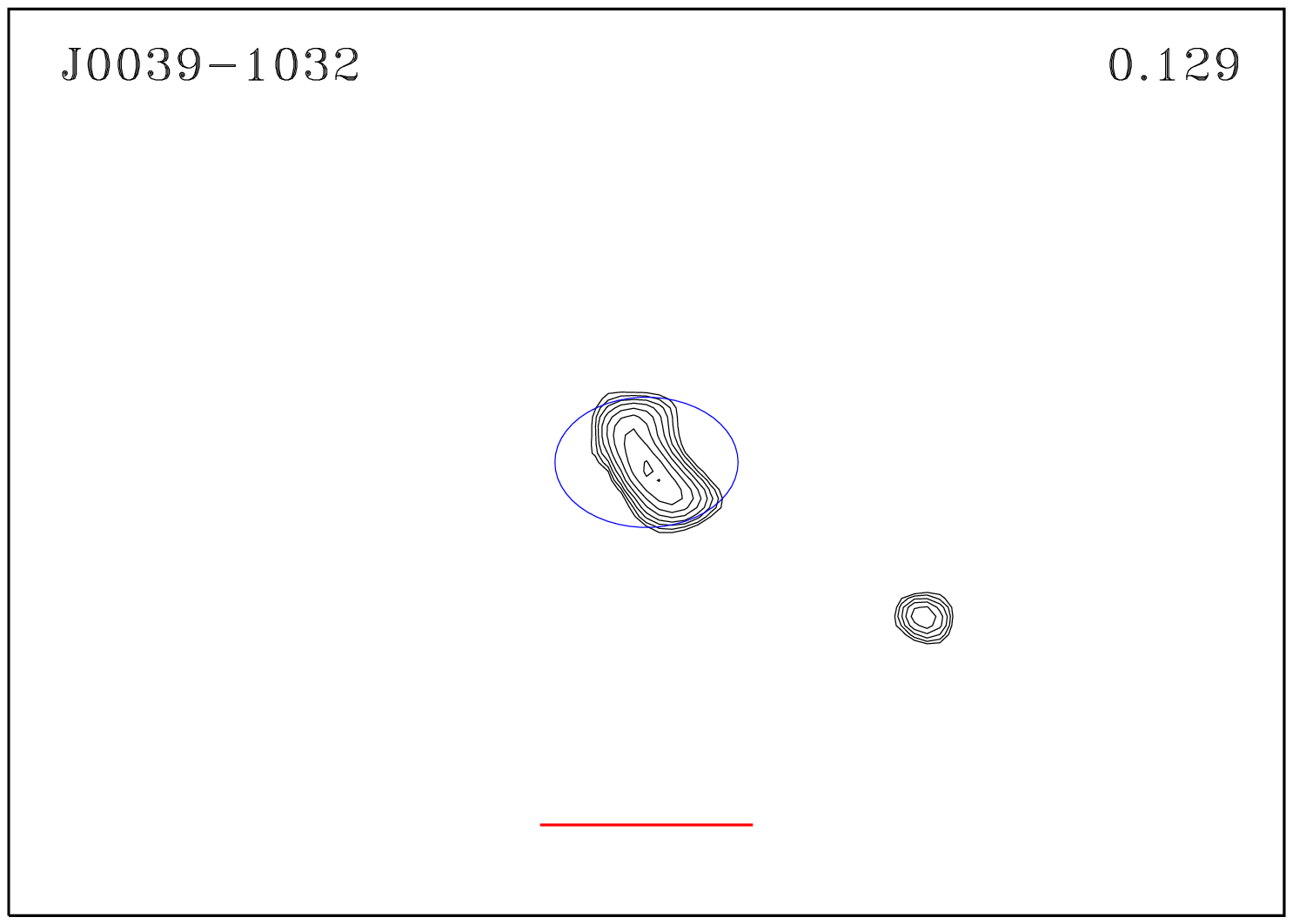} 
\includegraphics[width=6.3cm,height=6.3cm]{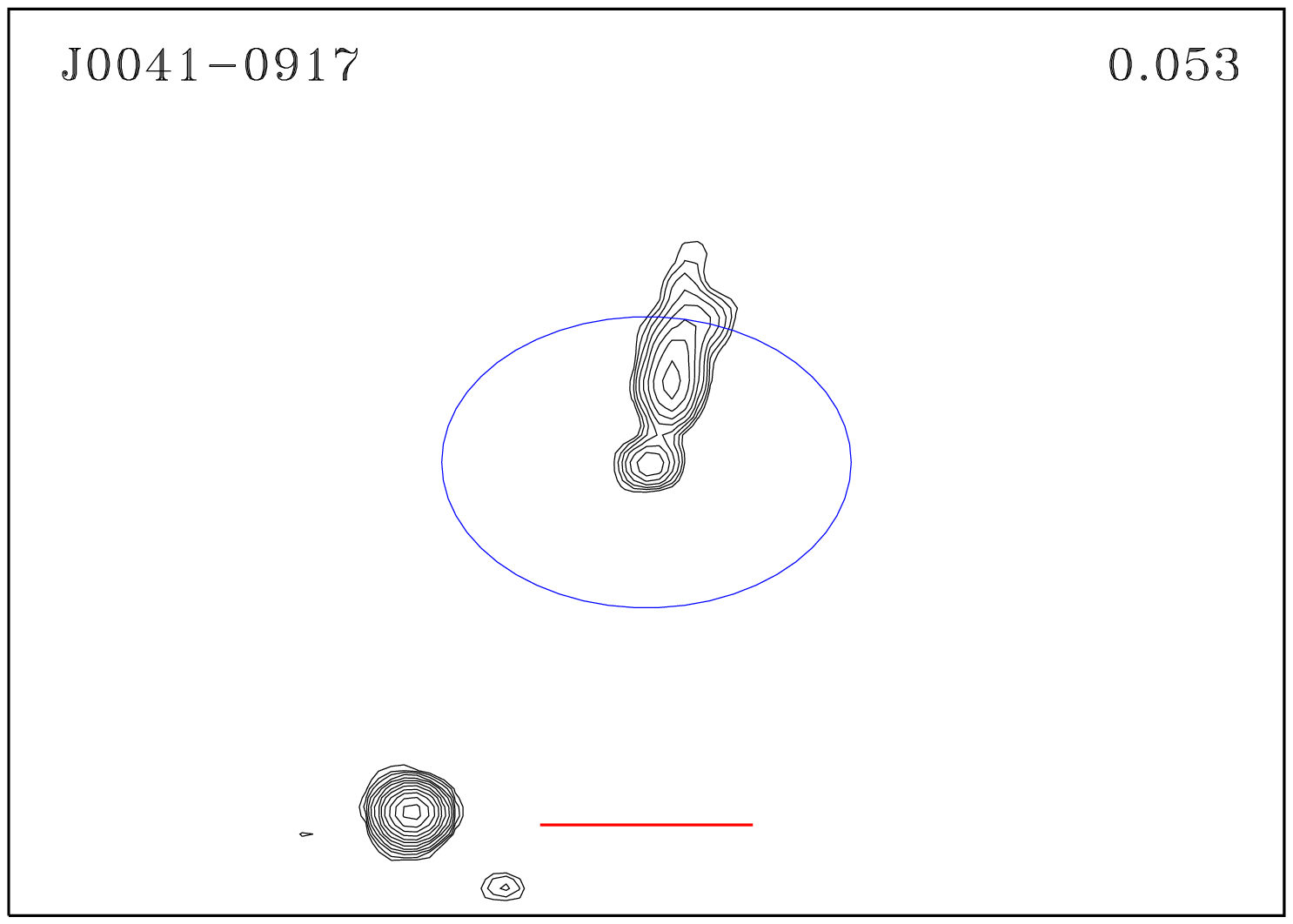} 

\includegraphics[width=6.3cm,height=6.3cm]{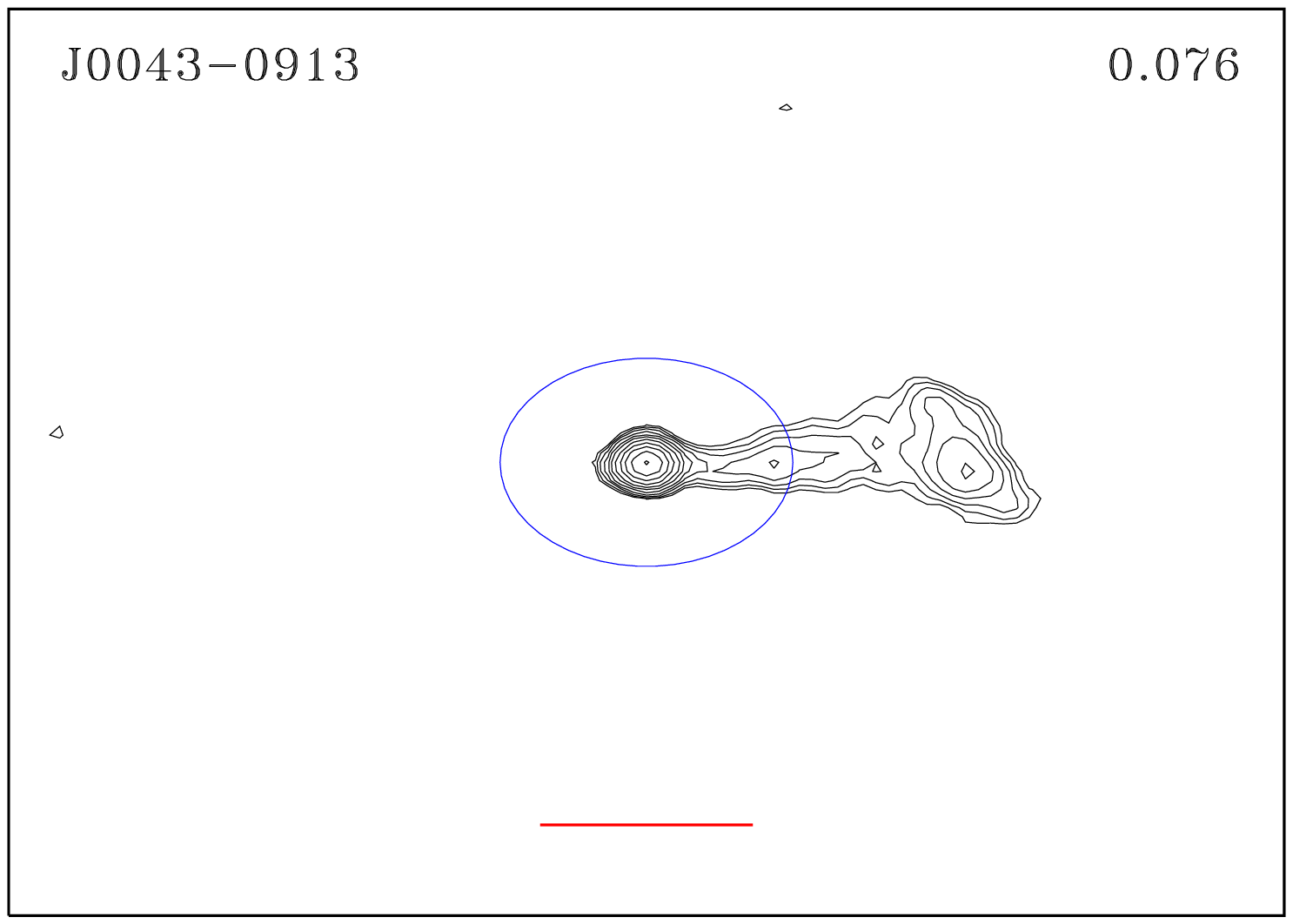} 
\includegraphics[width=6.3cm,height=6.3cm]{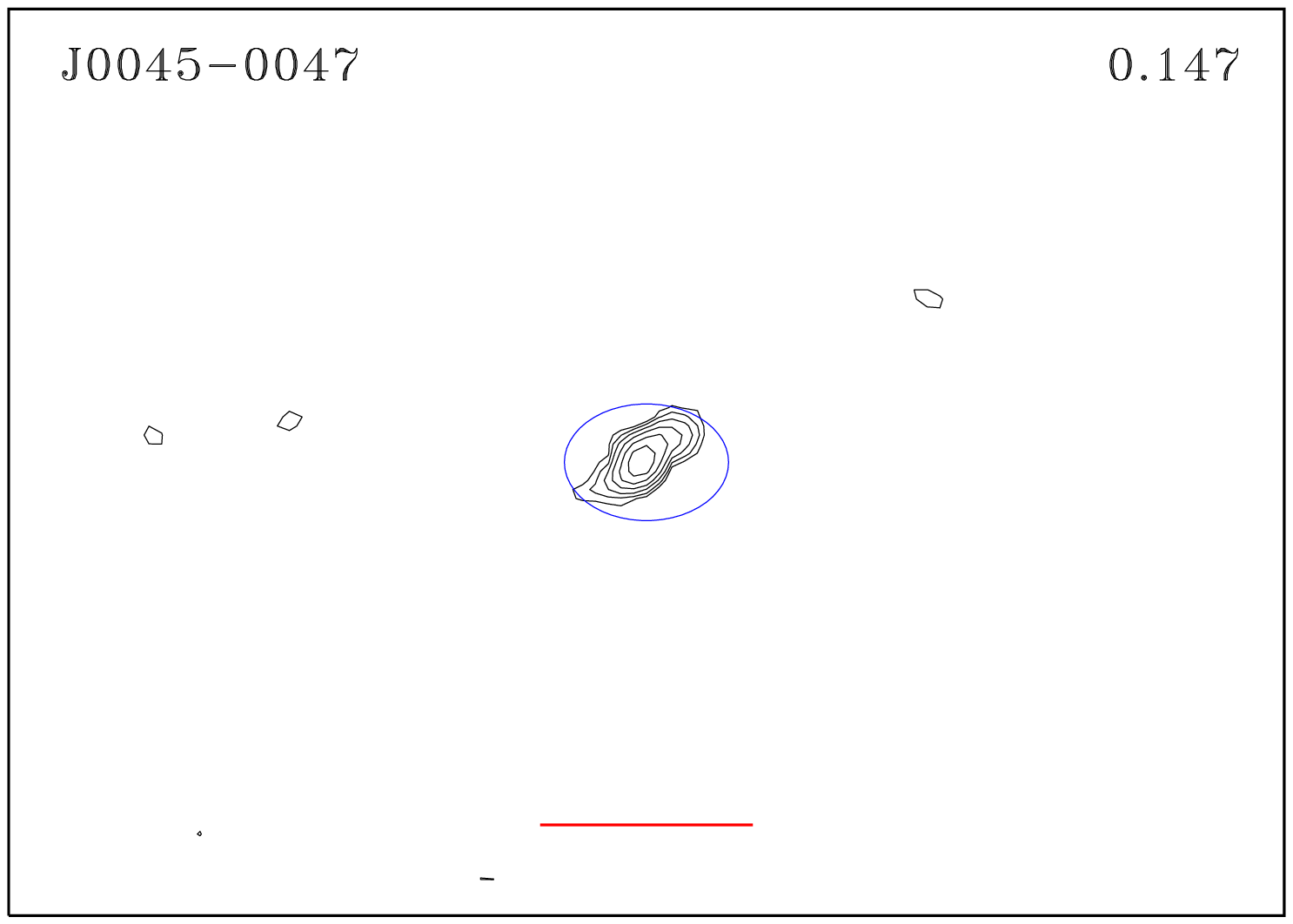} 
\includegraphics[width=6.3cm,height=6.3cm]{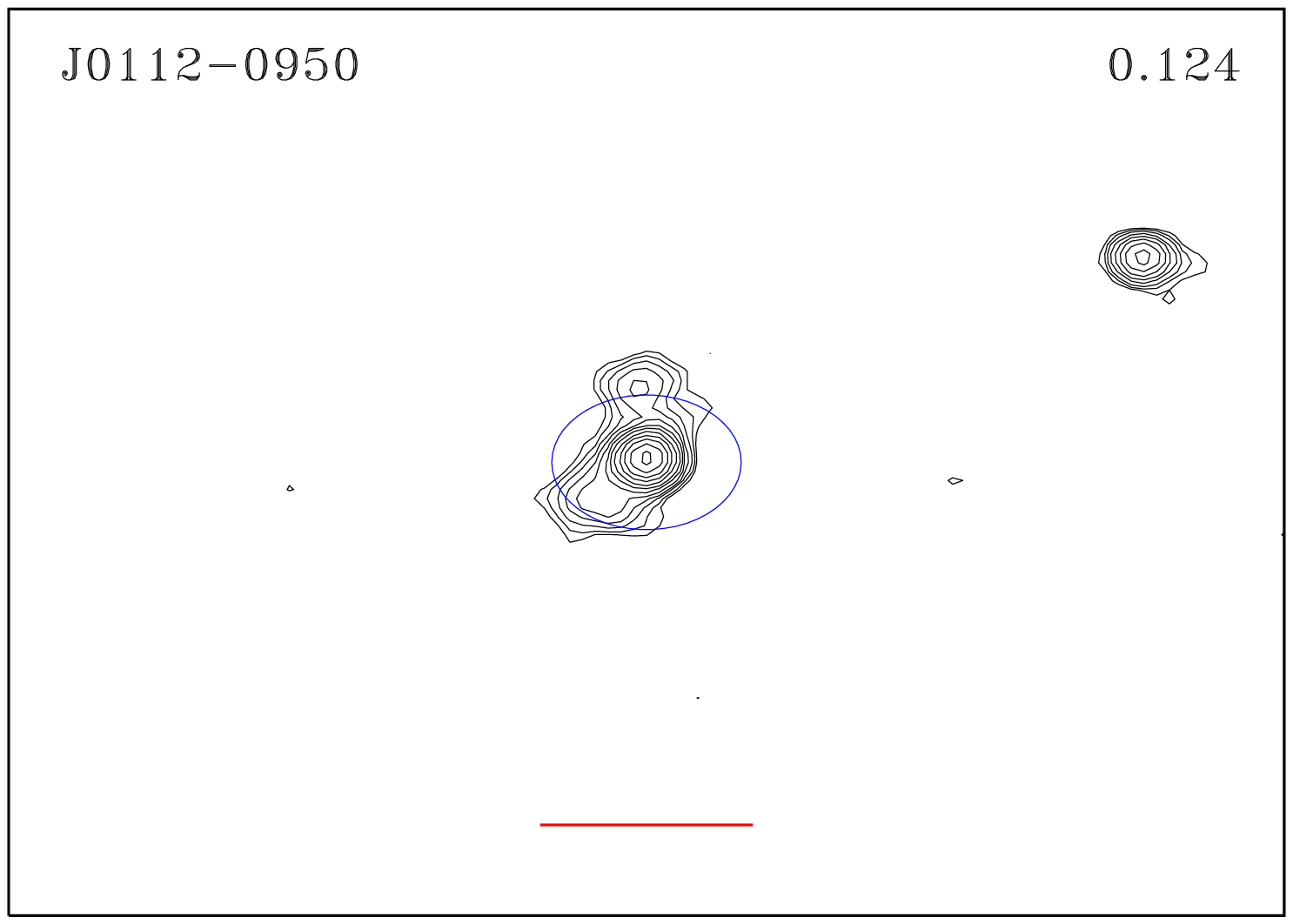} 

\includegraphics[width=6.3cm,height=6.3cm]{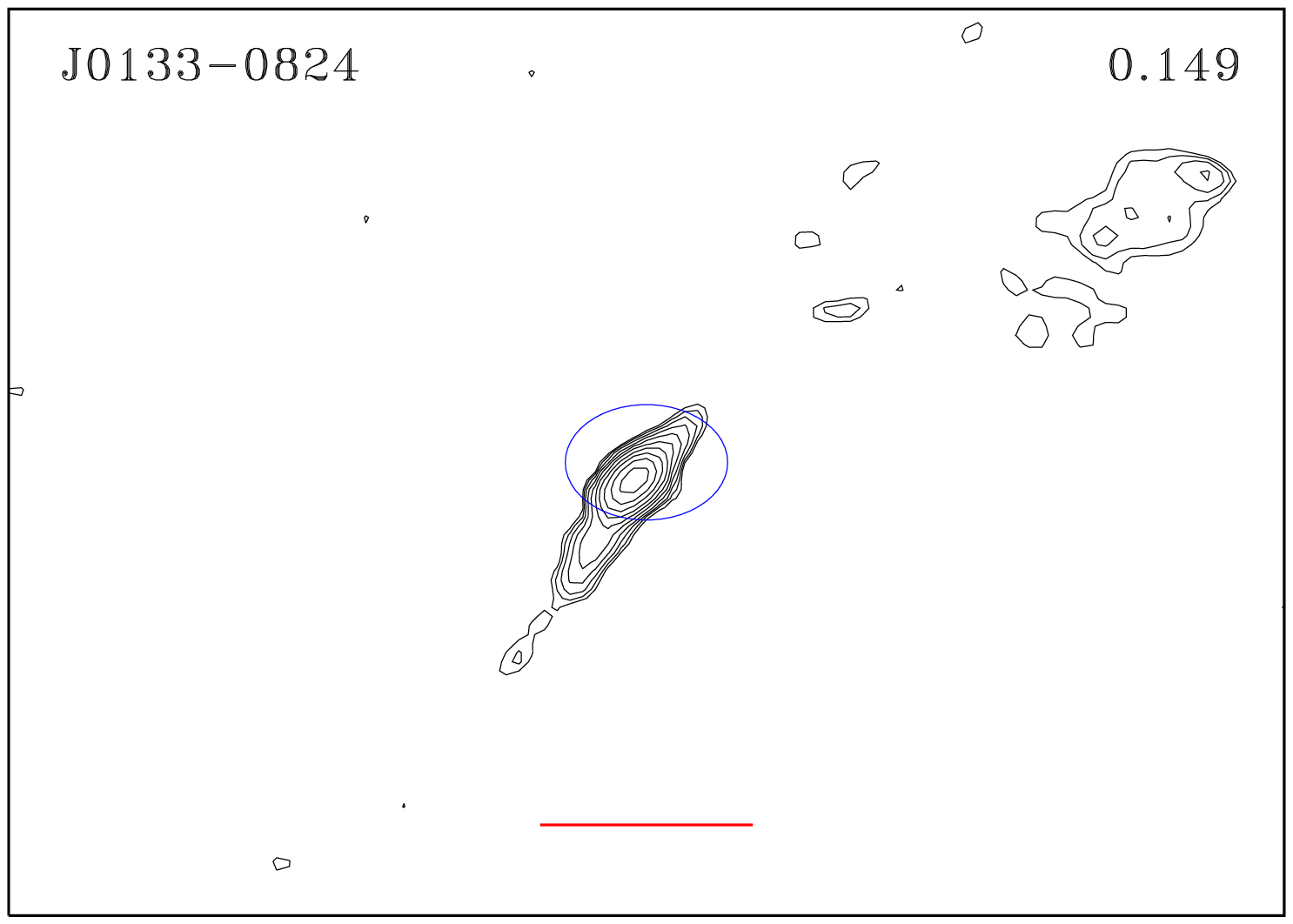} 
\includegraphics[width=6.3cm,height=6.3cm]{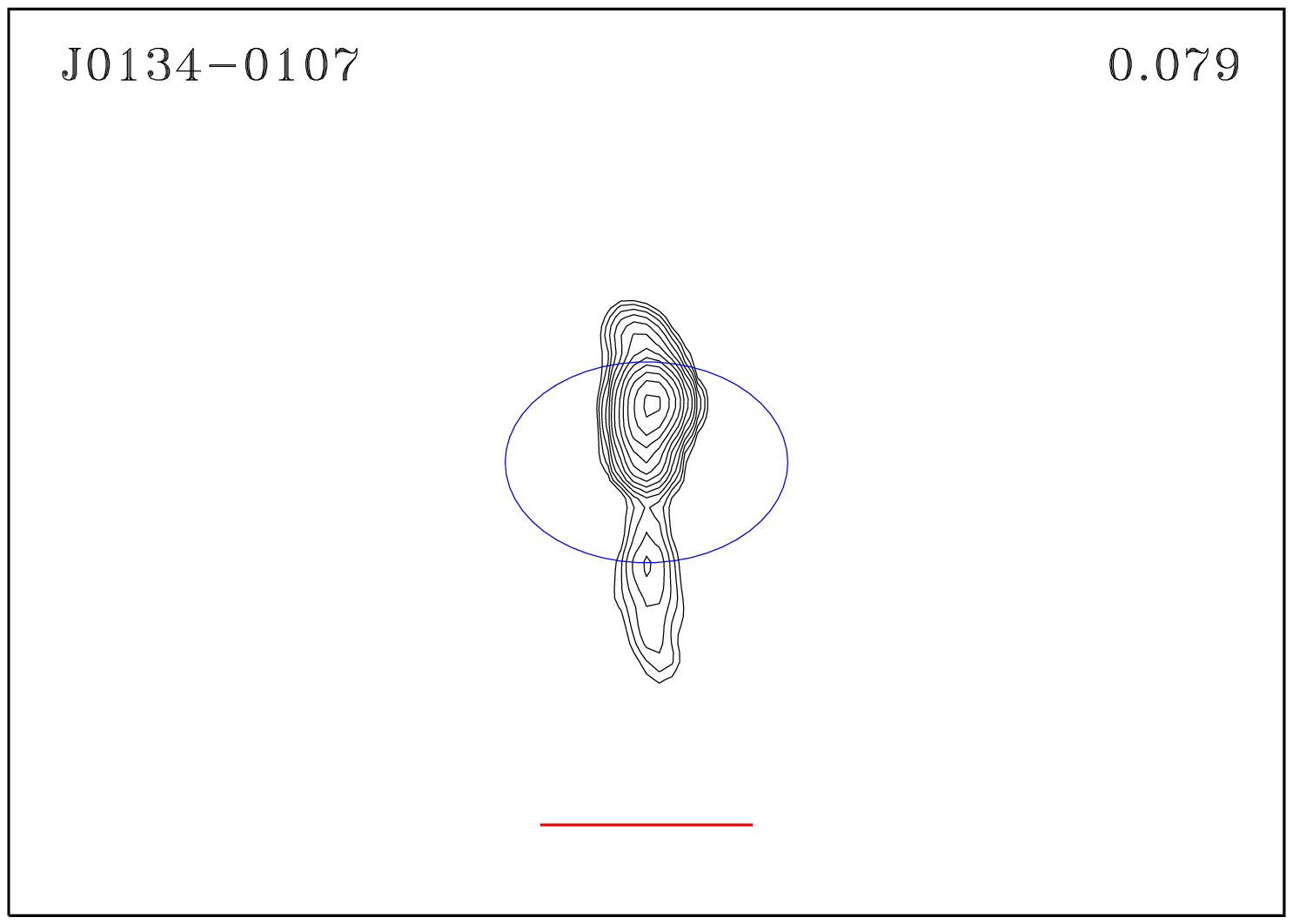} 
\includegraphics[width=6.3cm,height=6.3cm]{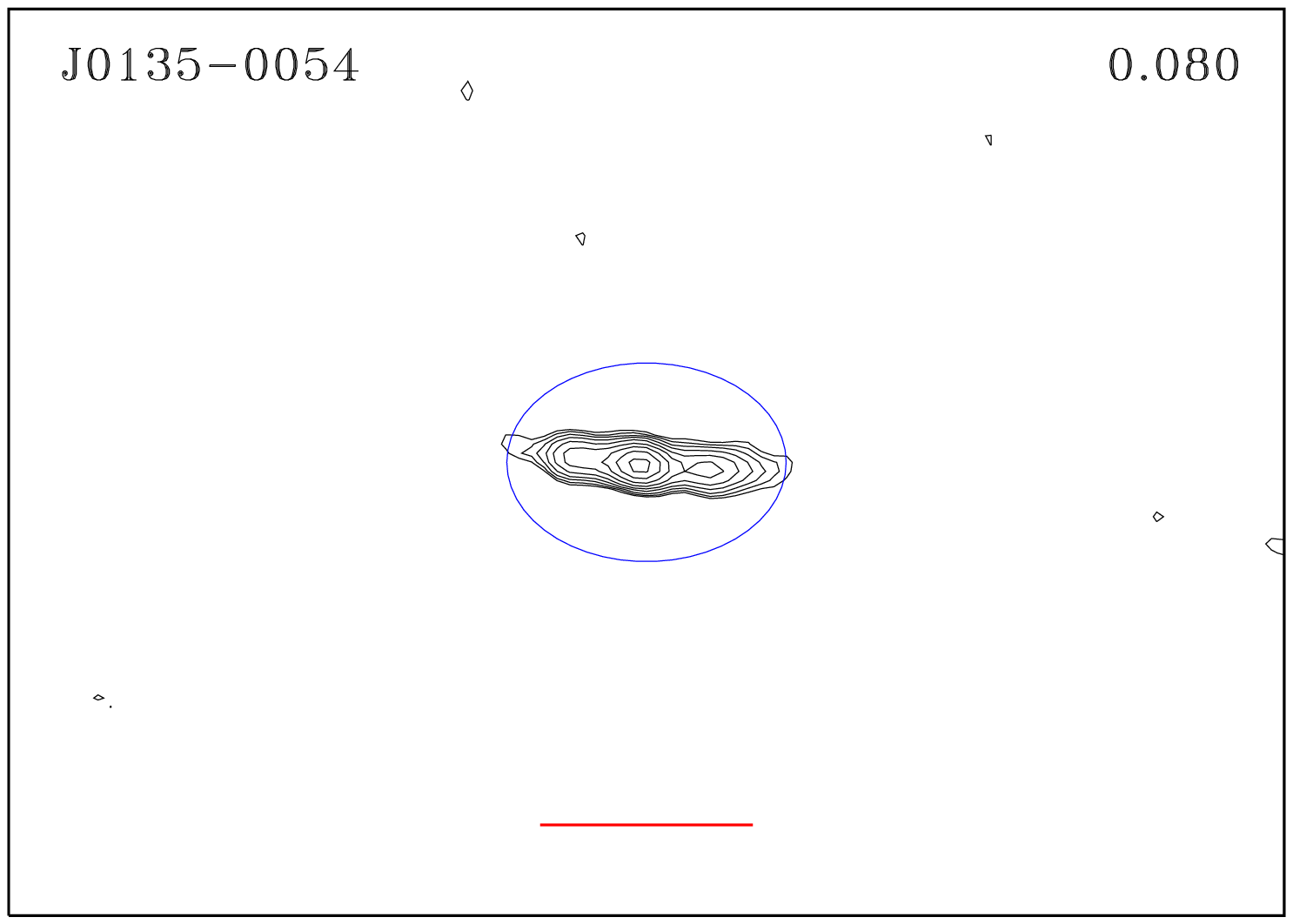} 

\includegraphics[width=6.3cm,height=6.3cm]{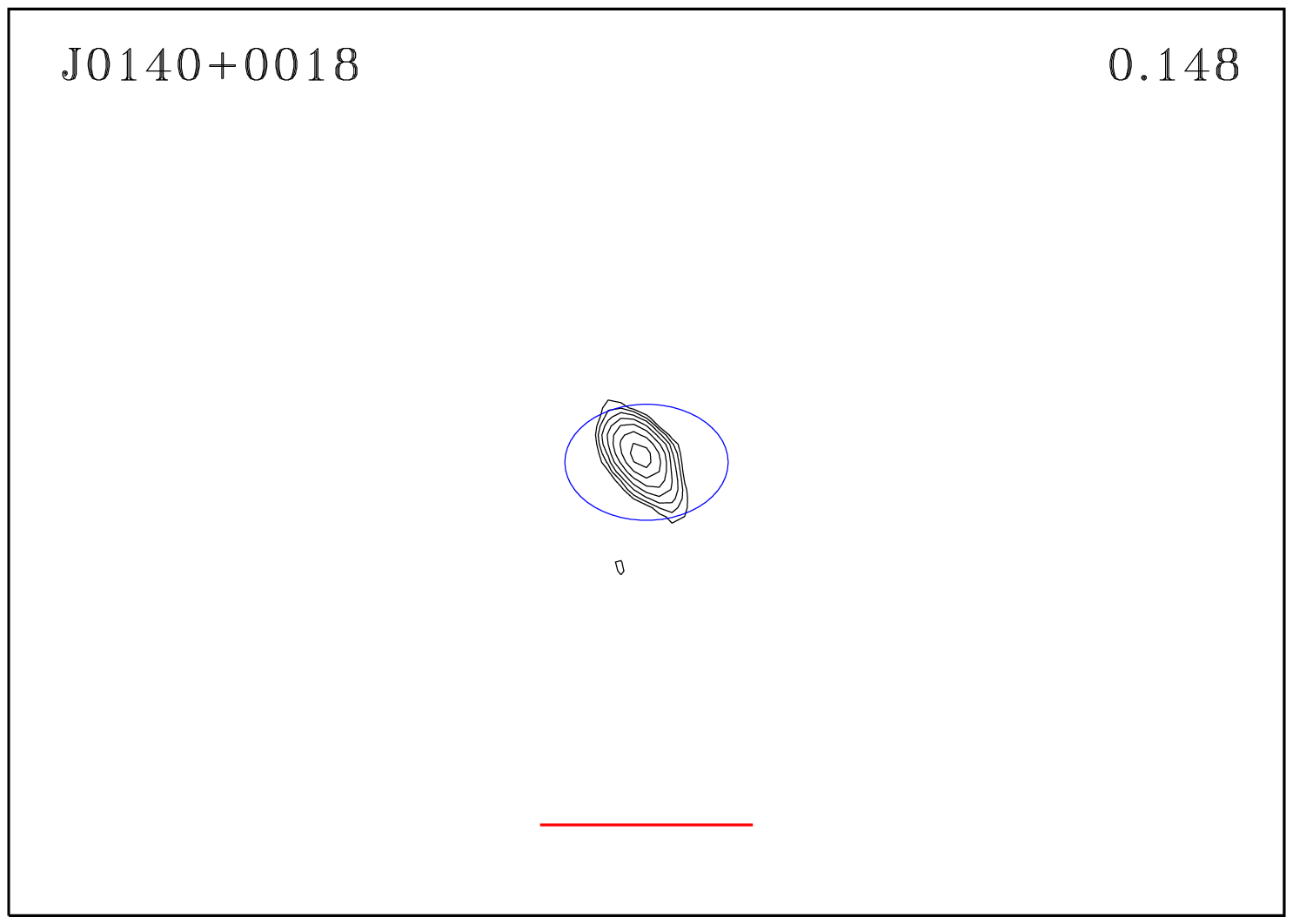} 
\includegraphics[width=6.3cm,height=6.3cm]{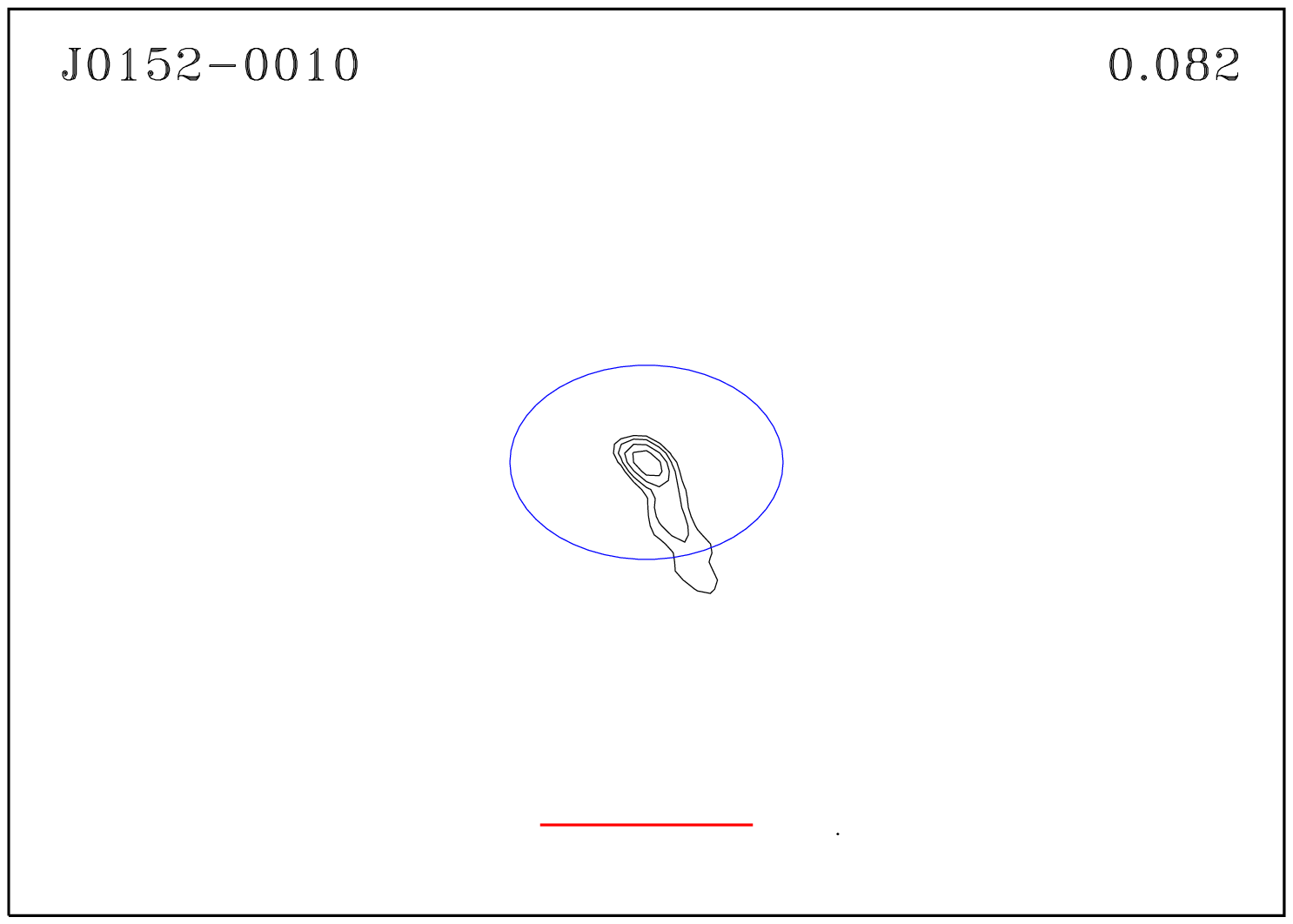} 
\includegraphics[width=6.3cm,height=6.3cm]{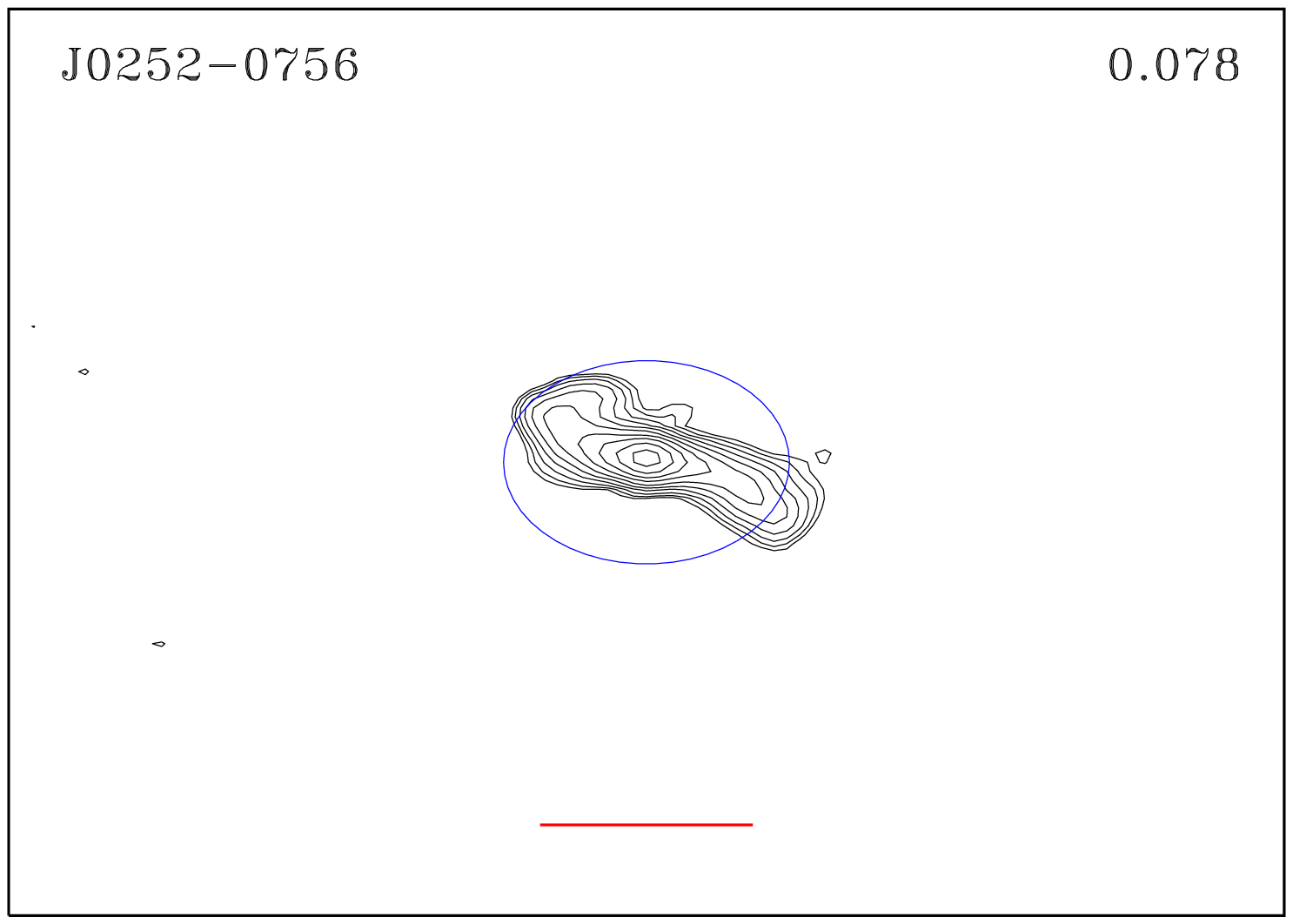} 
\caption{Images of the FR~Is selected. Contours are drawn starting
  from 0.45 mJy/beam and increase with a geometric progression with a common
  ratio of $\sqrt2$. The field of view is 3'$\times$3'; the red tick at the
  bottom is 30$\arcsec$ long. The blue circle is centered on the host galaxy
  and has a radius of 30 kpc. The source ID and redshift are reported in the
  upper corners.}
\label{images3}
\end{figure*}

\addtocounter{figure}{-1}
\begin{figure*}
\includegraphics[width=6.3cm,height=6.3cm]{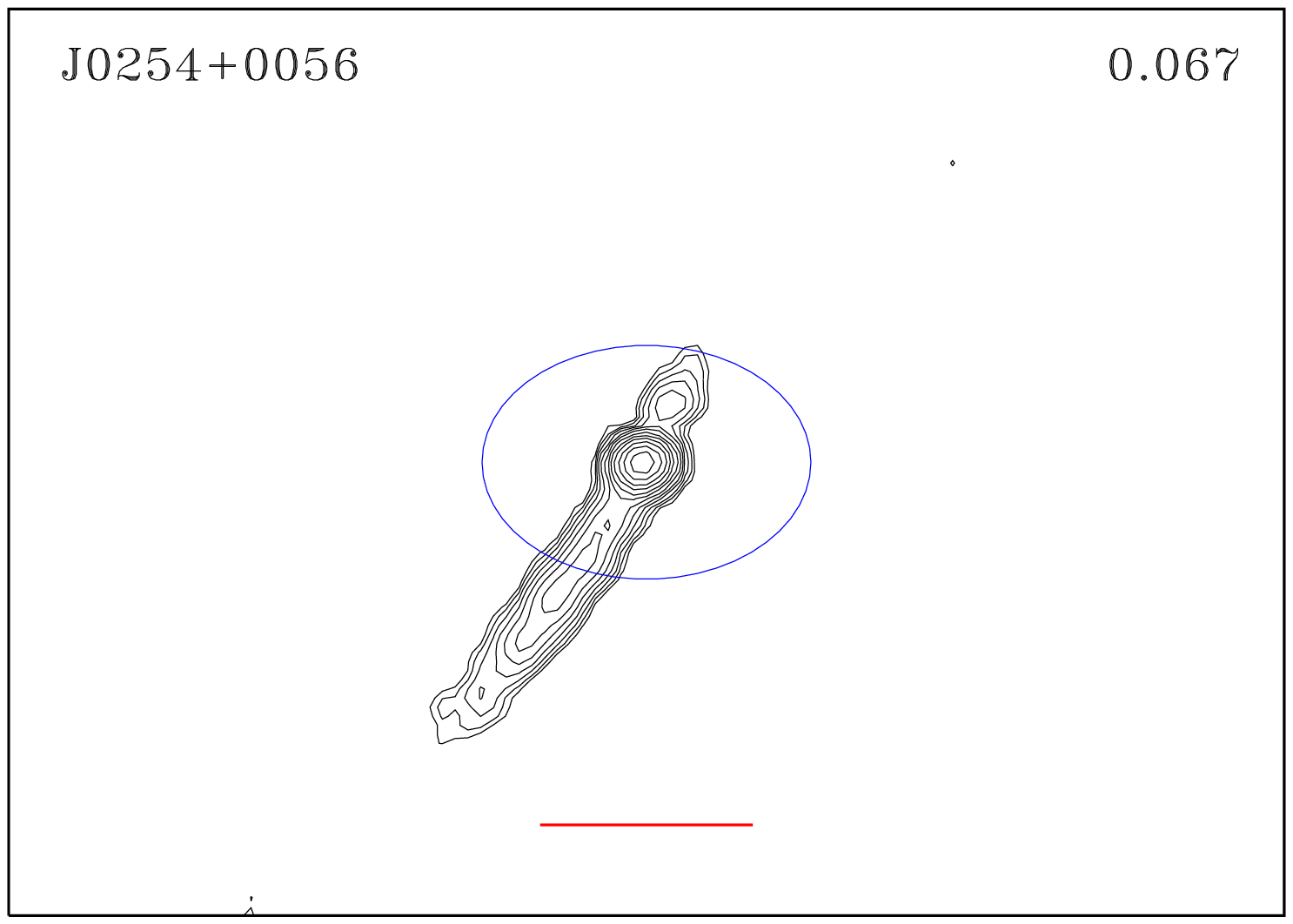} 
\includegraphics[width=6.3cm,height=6.3cm]{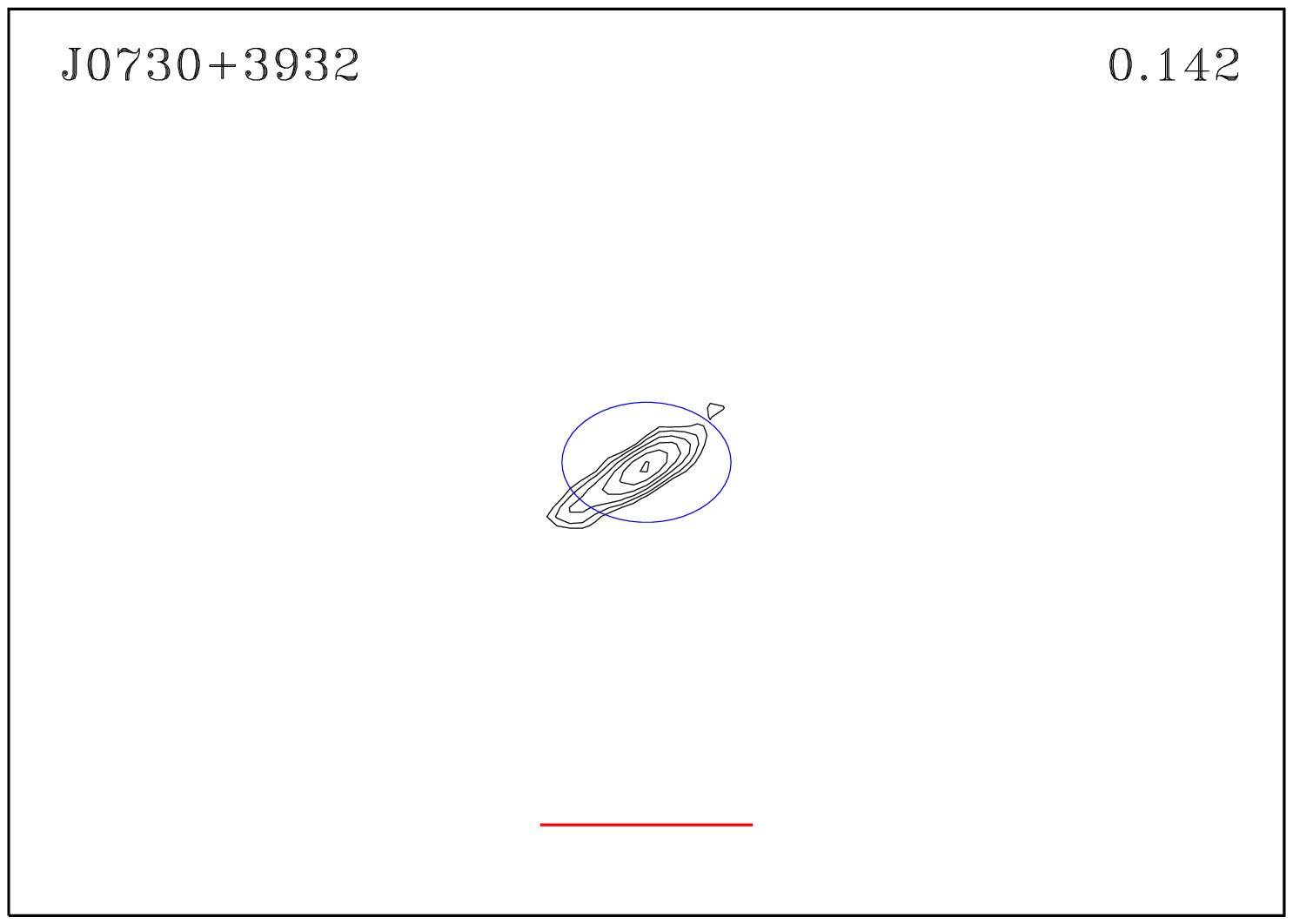} 
\includegraphics[width=6.3cm,height=6.3cm]{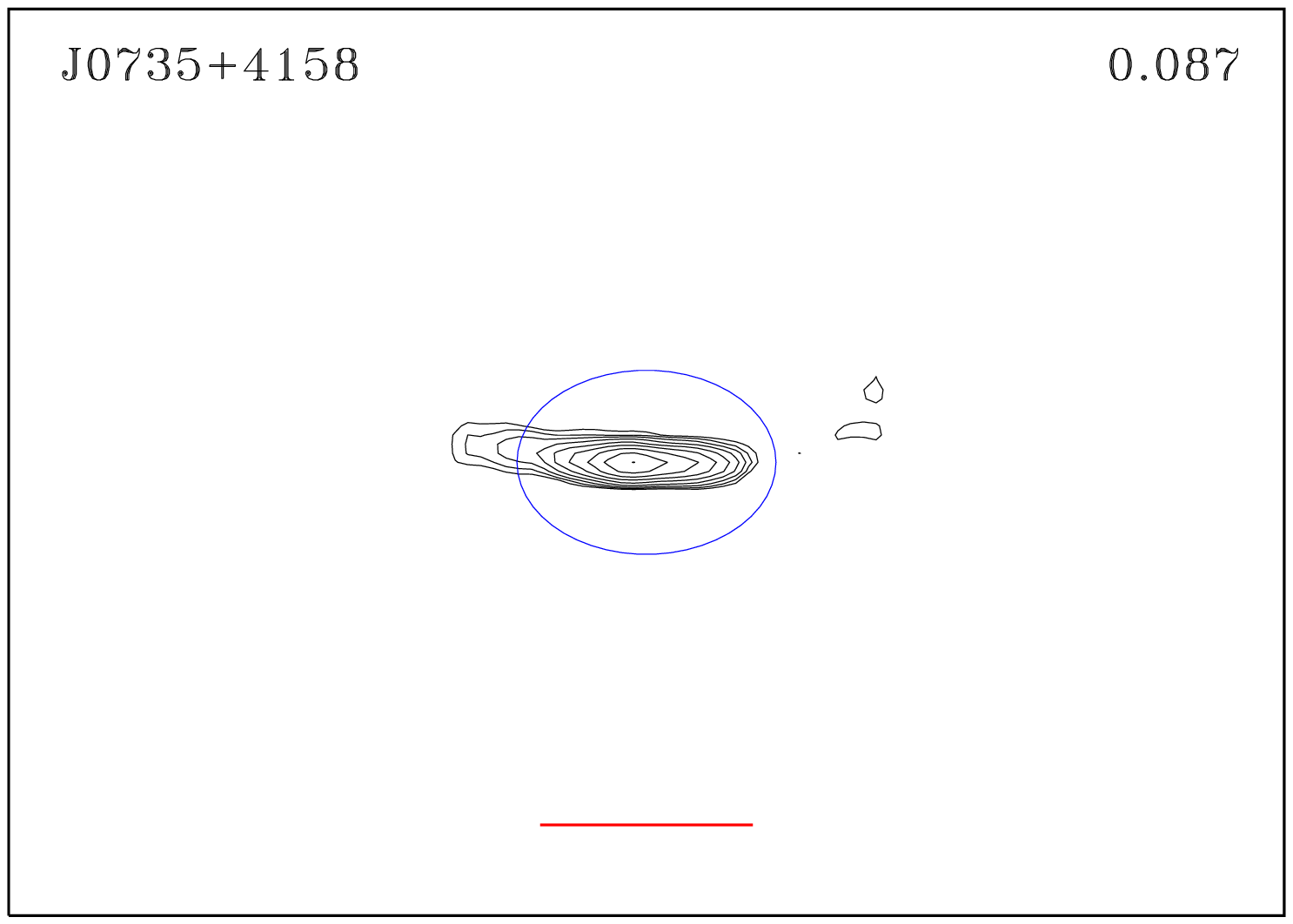} 

\includegraphics[width=6.3cm,height=6.3cm]{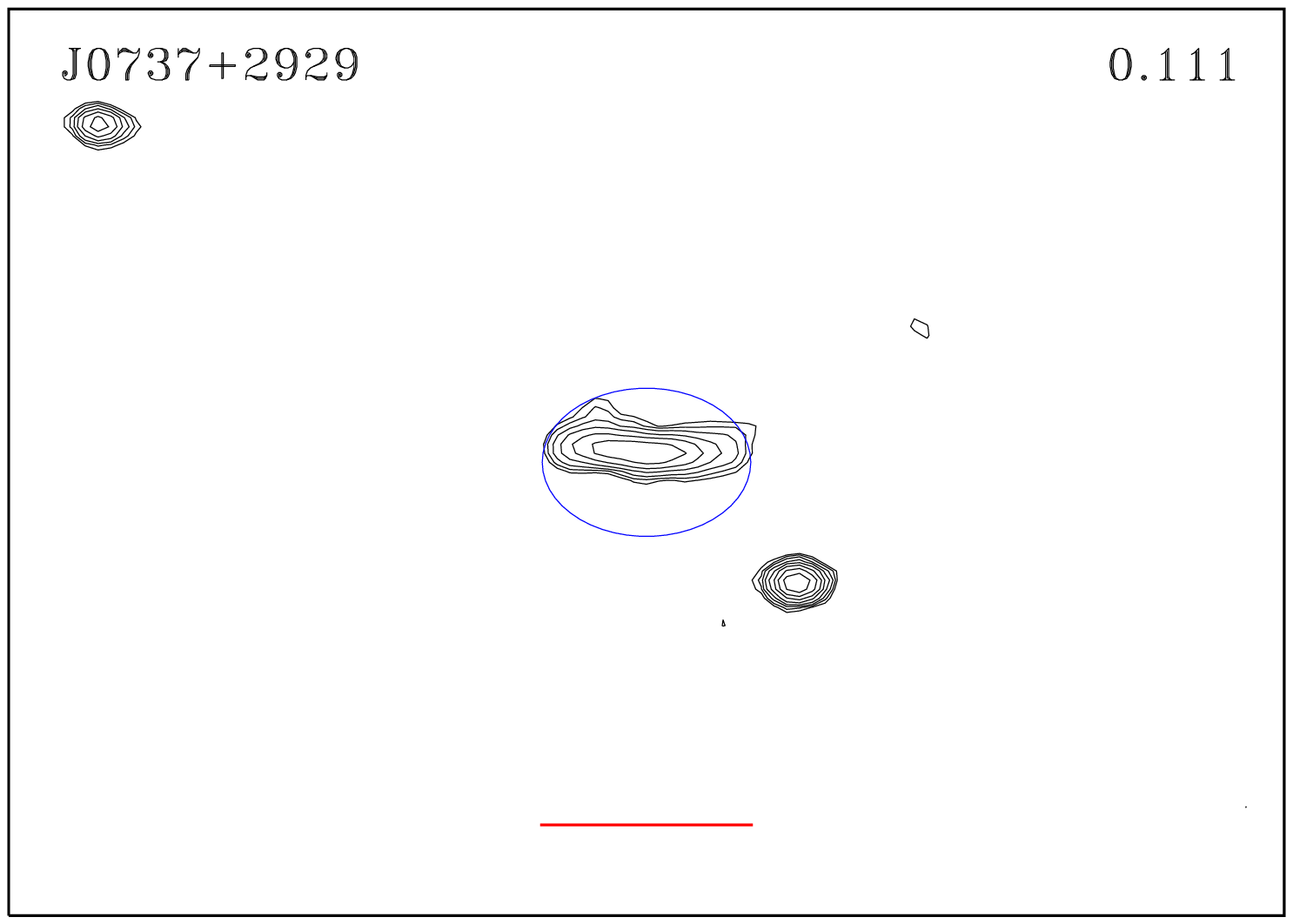} 
\includegraphics[width=6.3cm,height=6.3cm]{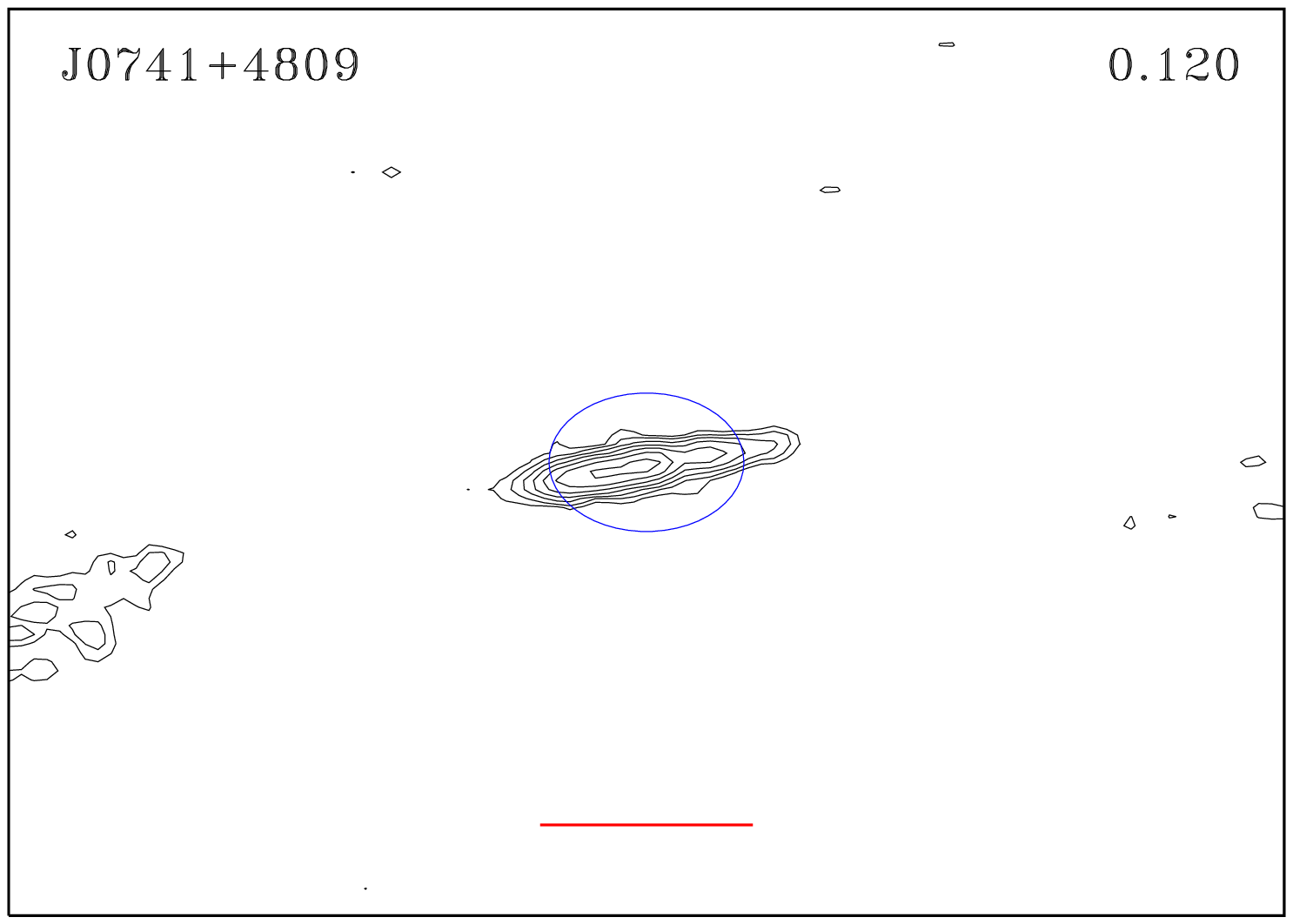} 
\includegraphics[width=6.3cm,height=6.3cm]{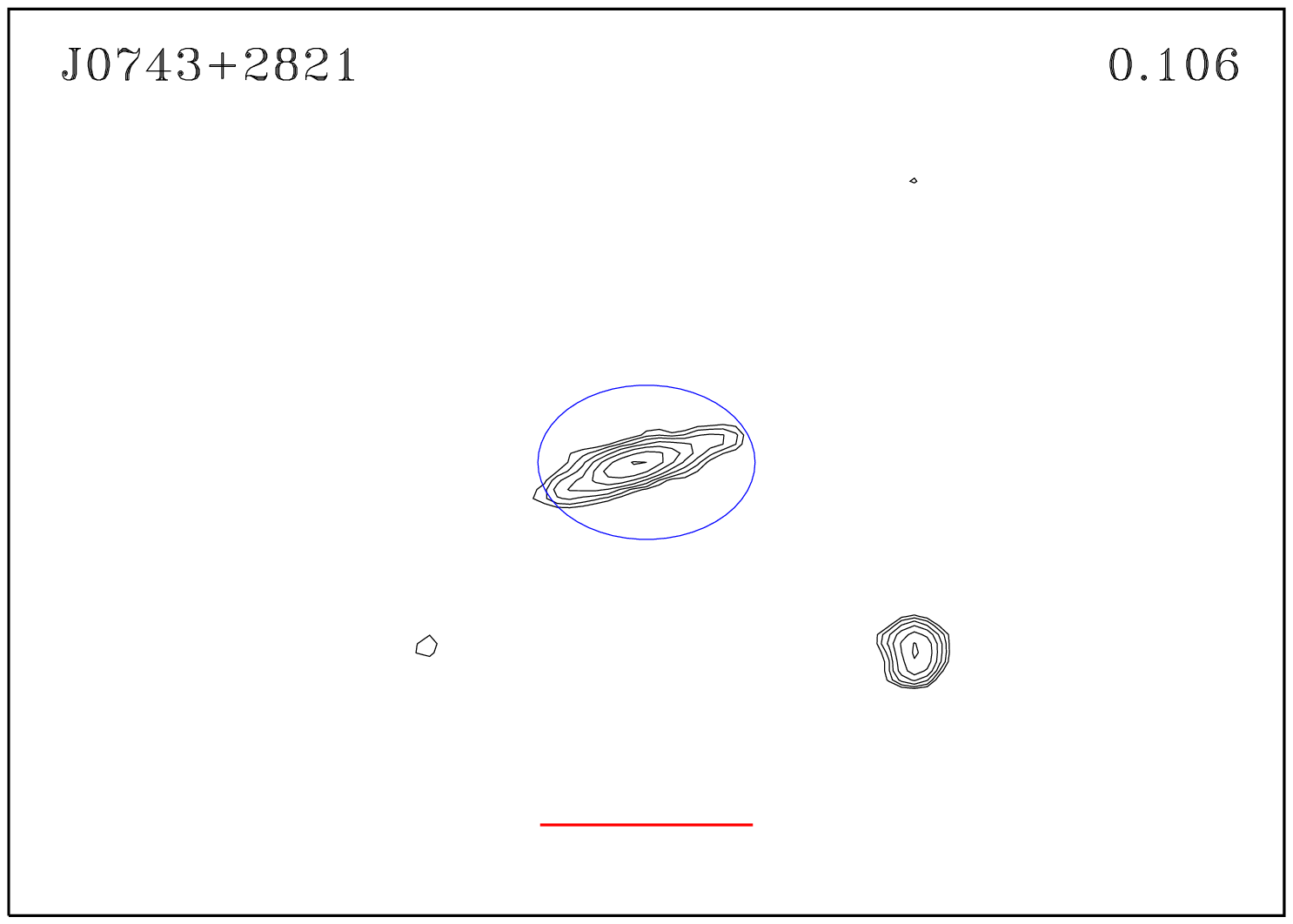} 

\includegraphics[width=6.3cm,height=6.3cm]{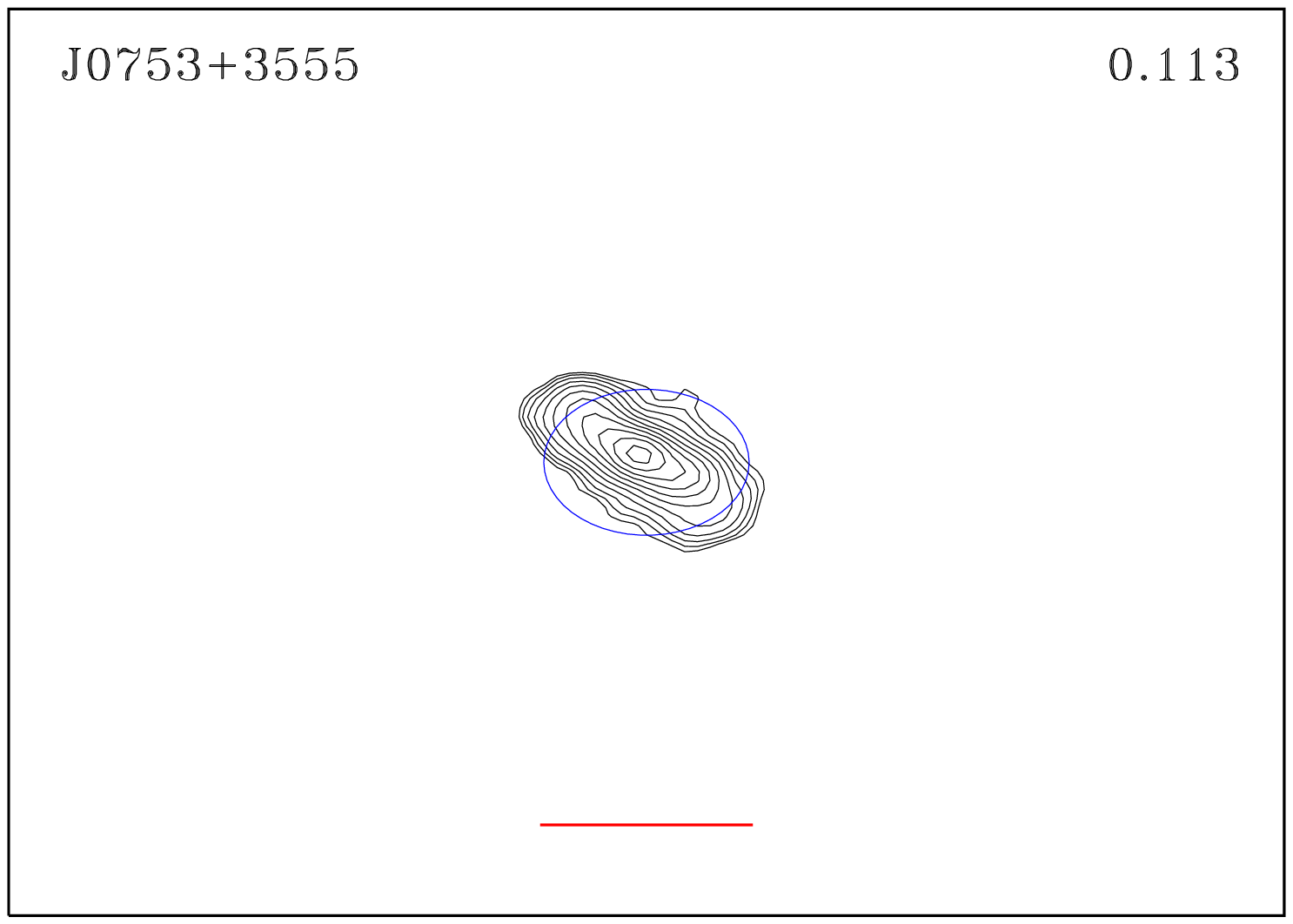} 
\includegraphics[width=6.3cm,height=6.3cm]{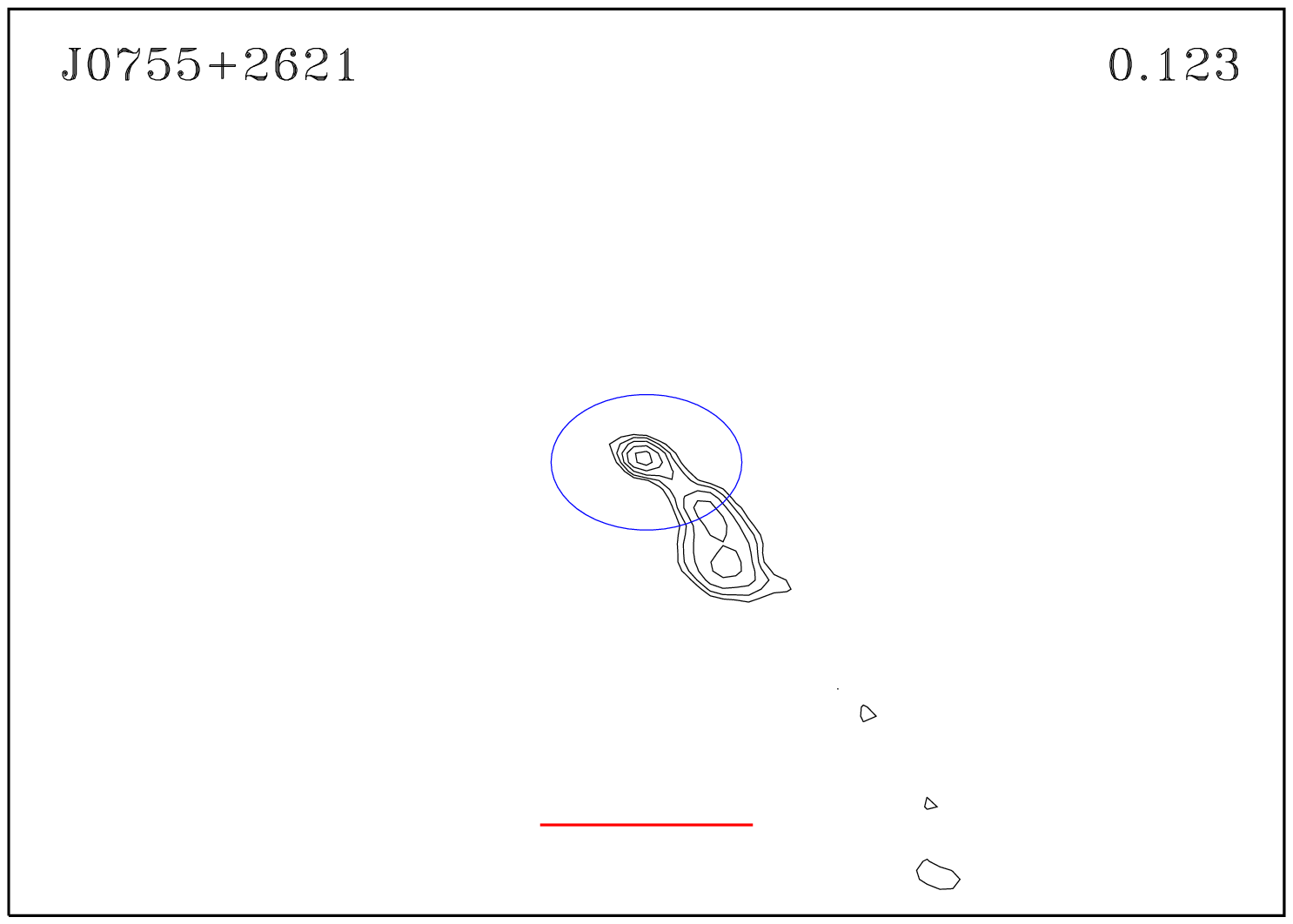} 
\includegraphics[width=6.3cm,height=6.3cm]{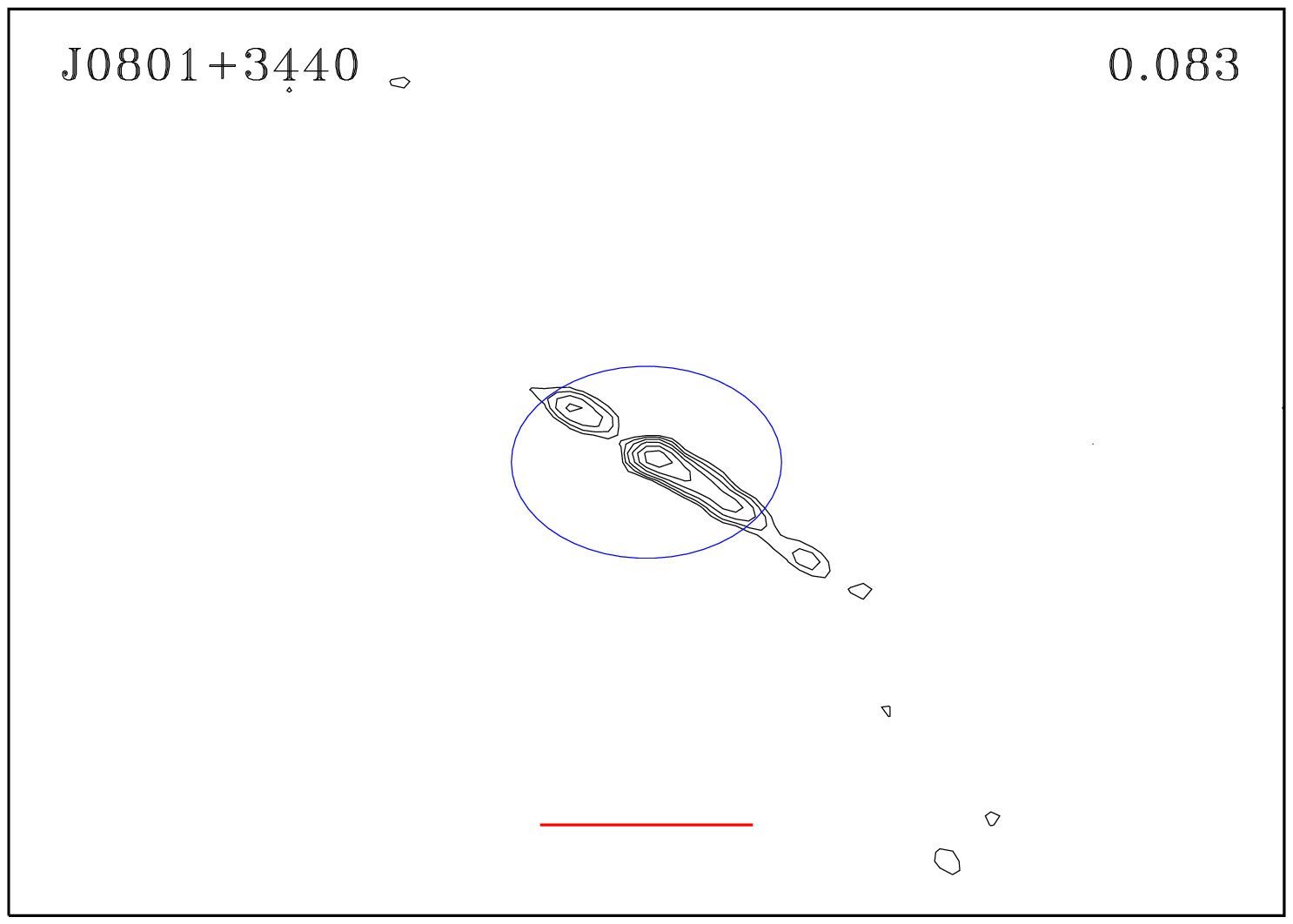} 

\includegraphics[width=6.3cm,height=6.3cm]{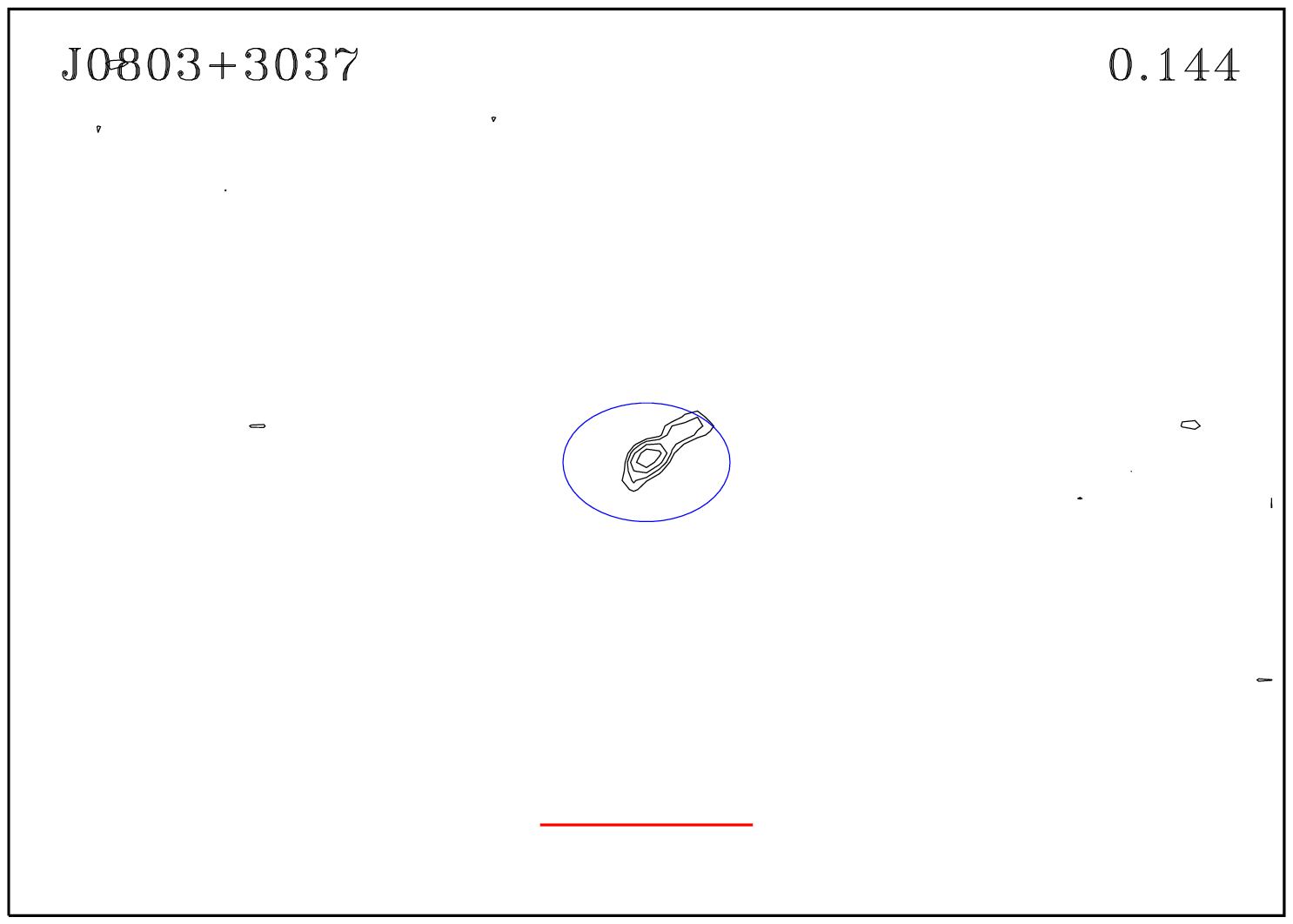} 
\includegraphics[width=6.3cm,height=6.3cm]{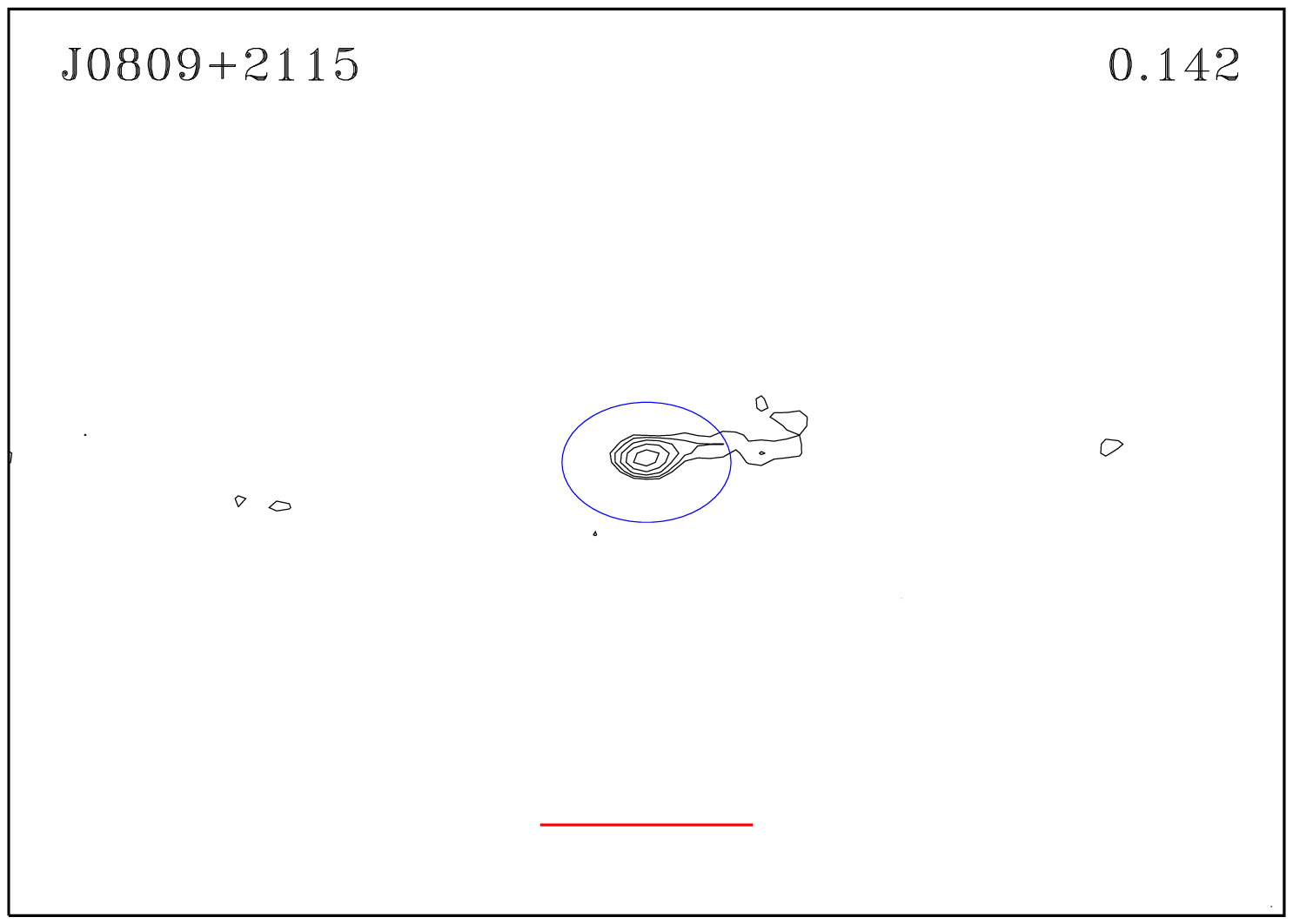} 
\includegraphics[width=6.3cm,height=6.3cm]{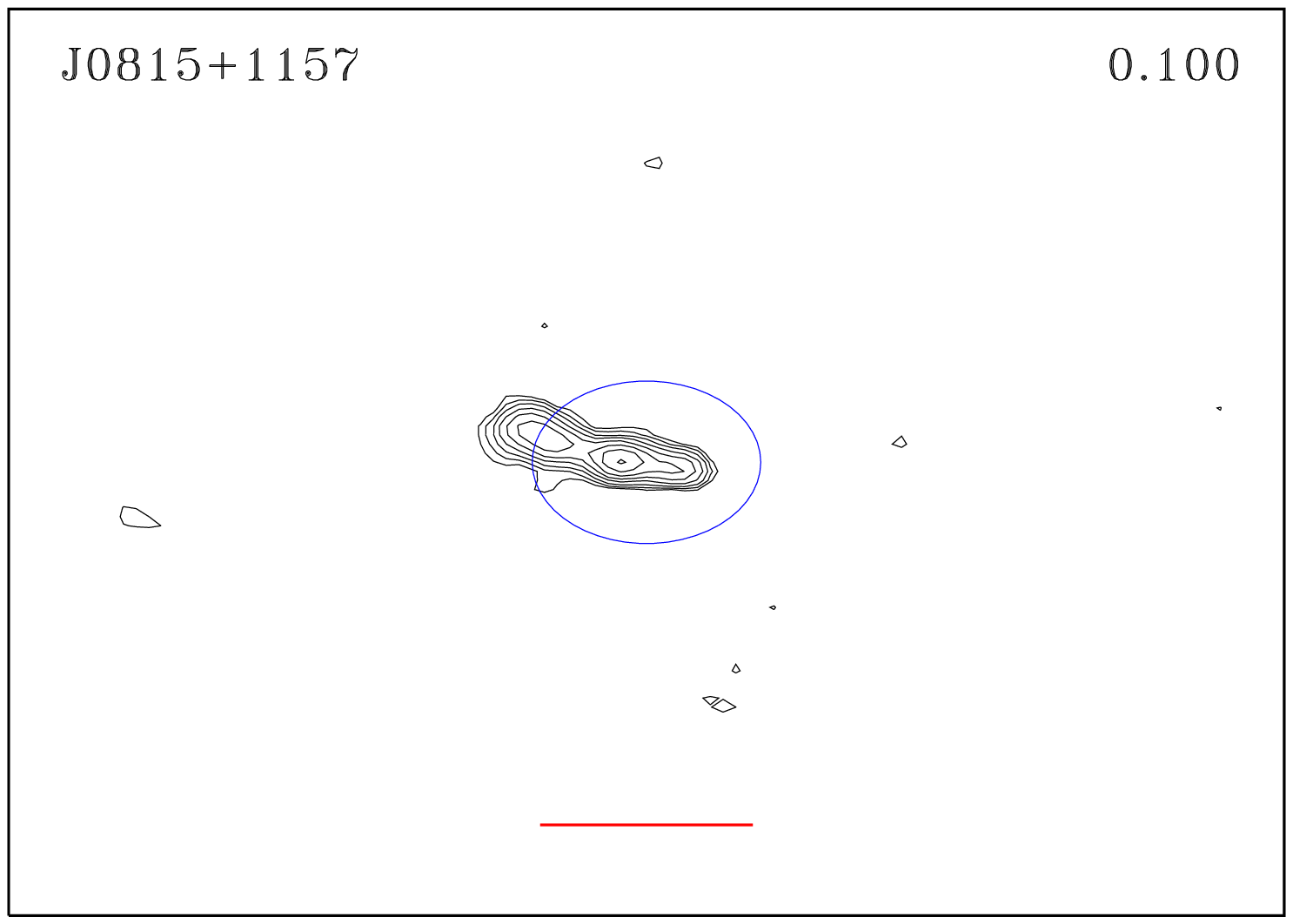} 
\caption{(continued)}
\end{figure*}

\addtocounter{figure}{-1}
\begin{figure*}
\includegraphics[width=6.3cm,height=6.3cm]{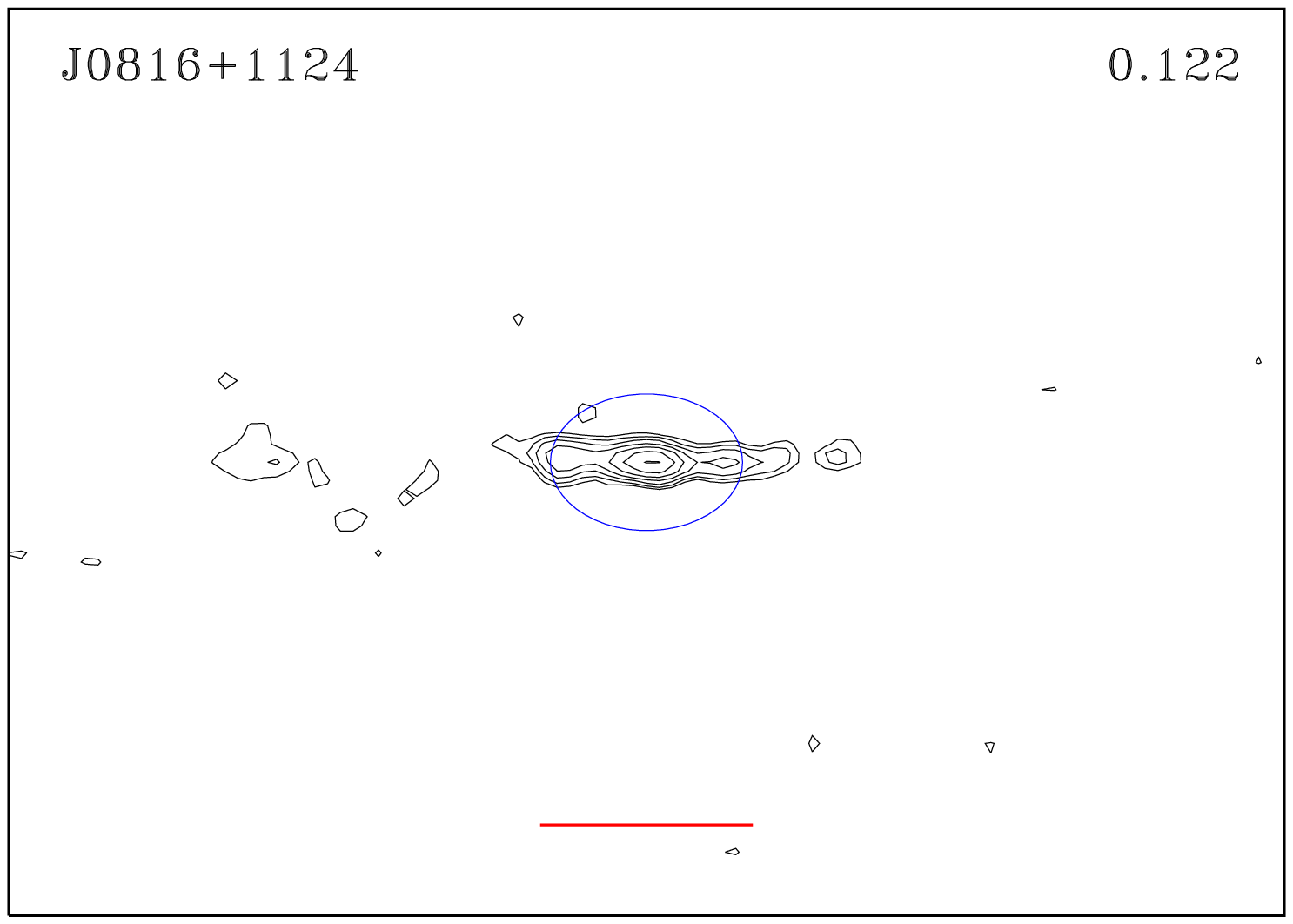} 
\includegraphics[width=6.3cm,height=6.3cm]{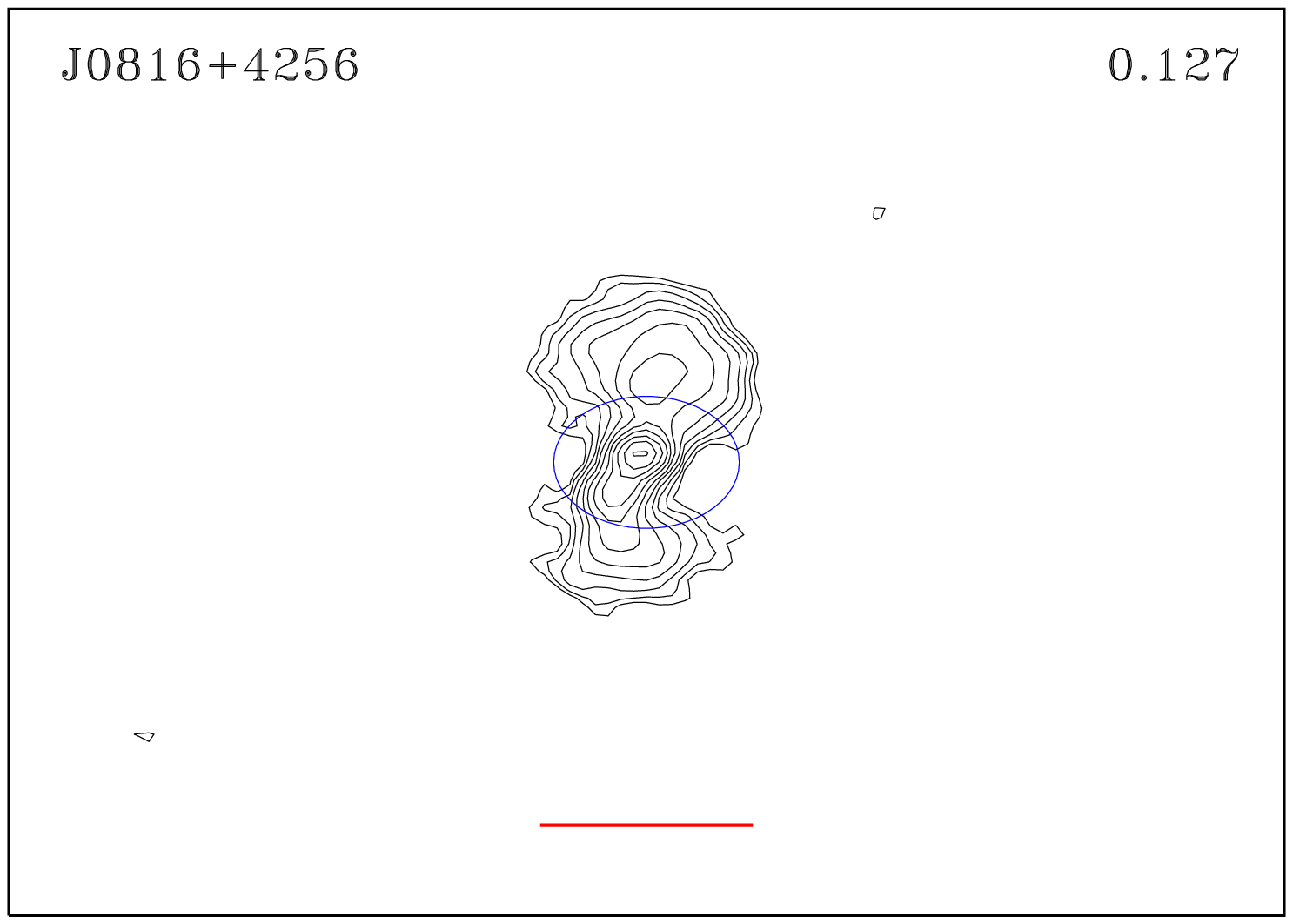} 
\includegraphics[width=6.3cm,height=6.3cm]{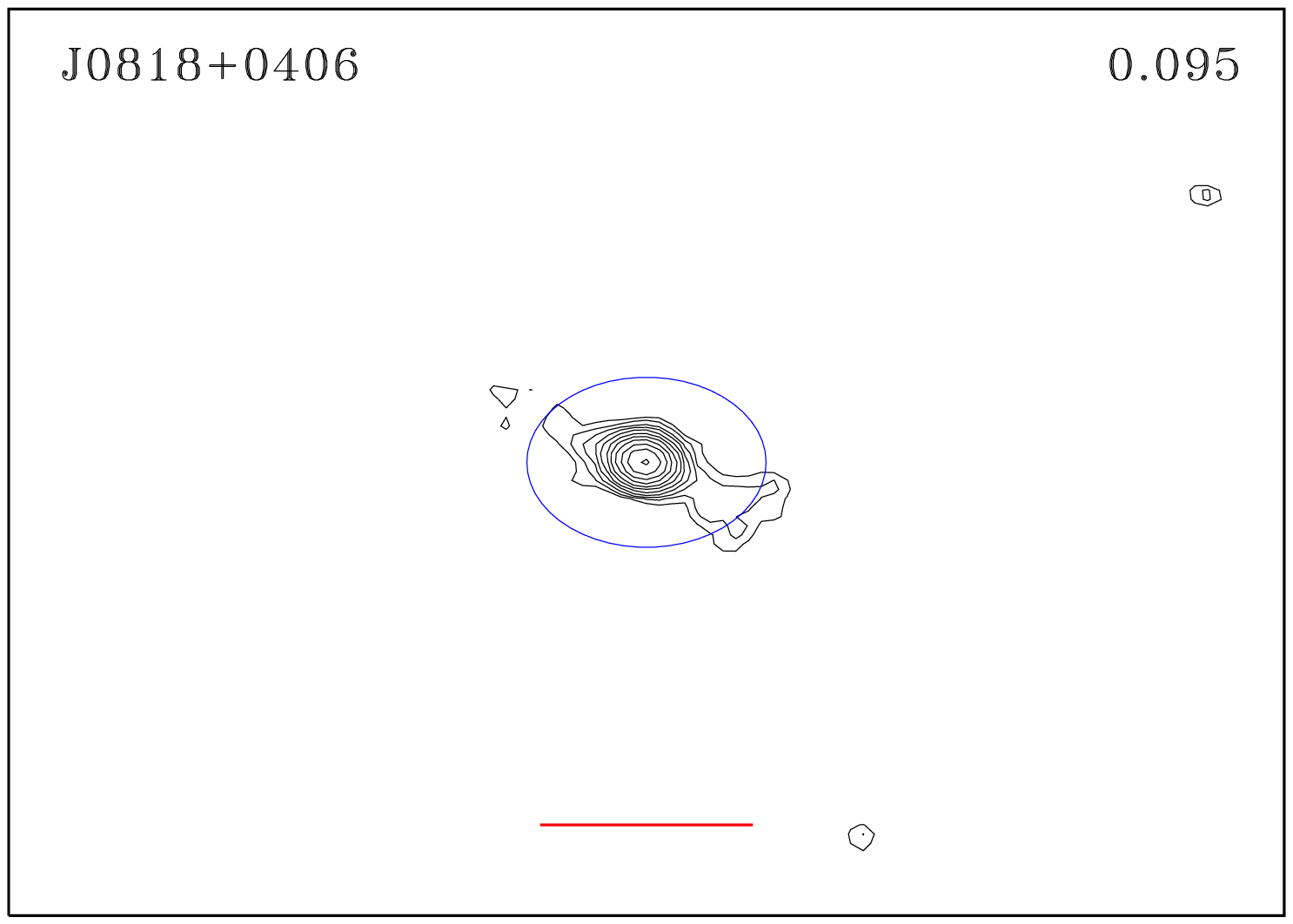} 

\includegraphics[width=6.3cm,height=6.3cm]{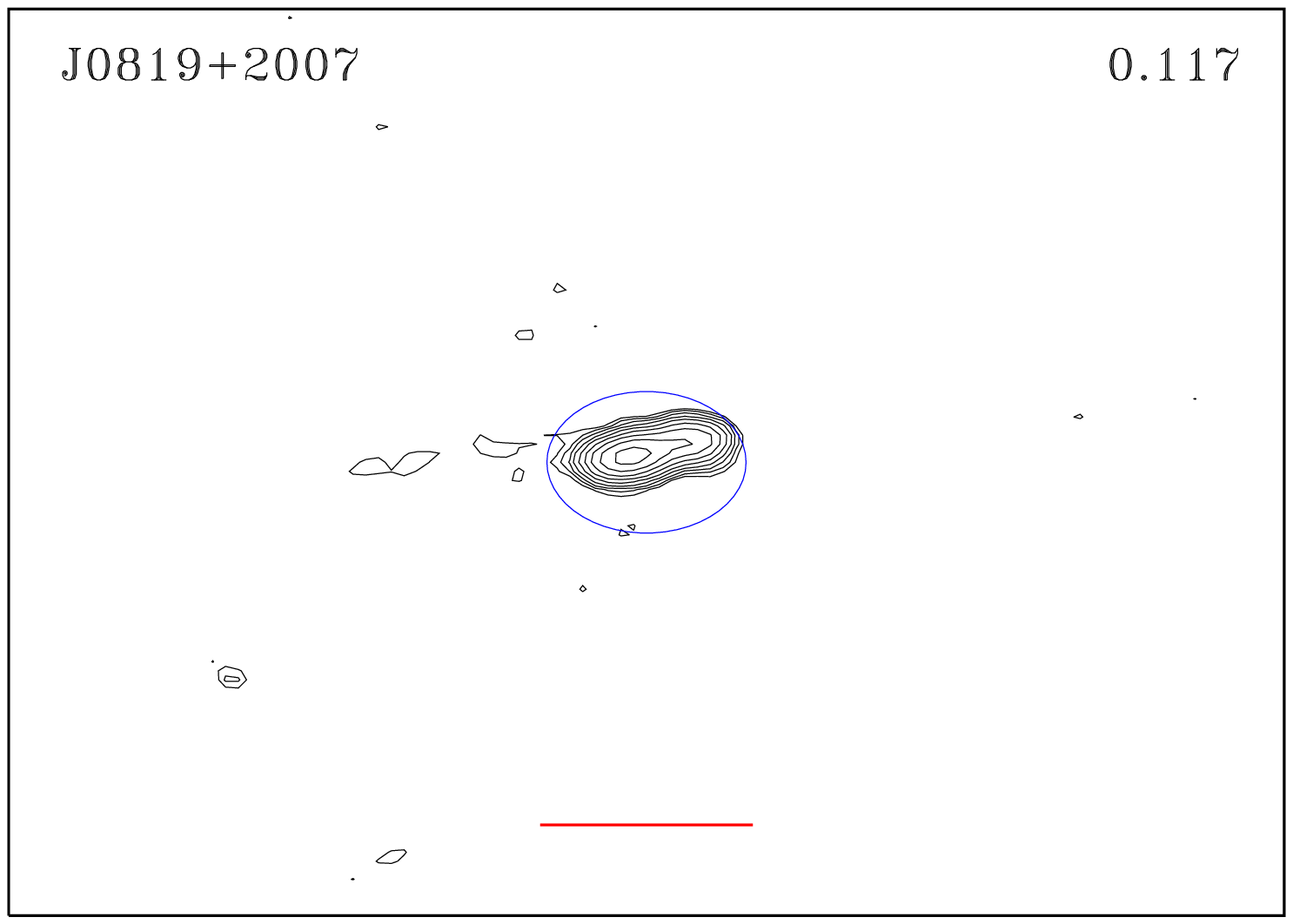} 
\includegraphics[width=6.3cm,height=6.3cm]{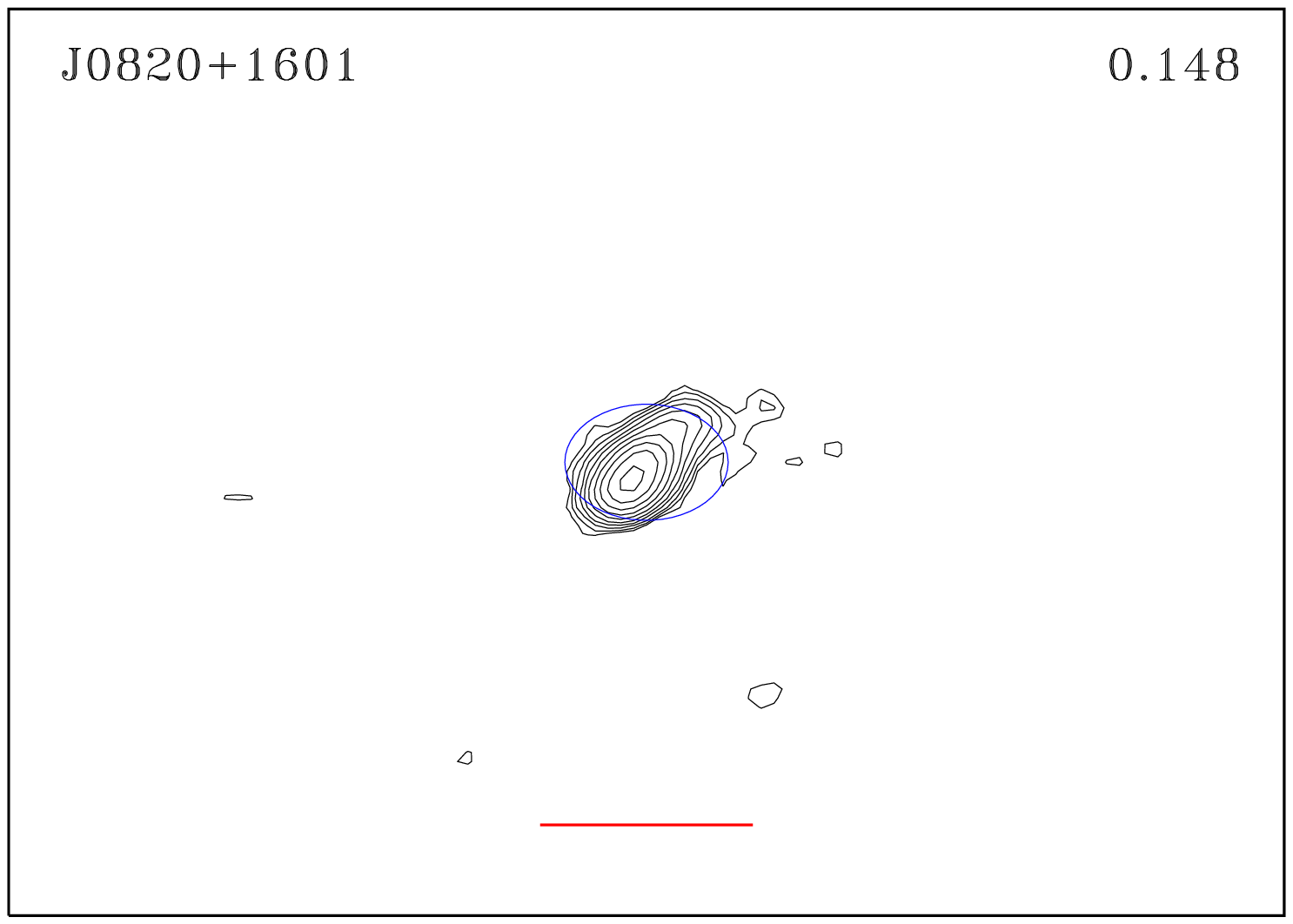} 
\includegraphics[width=6.3cm,height=6.3cm]{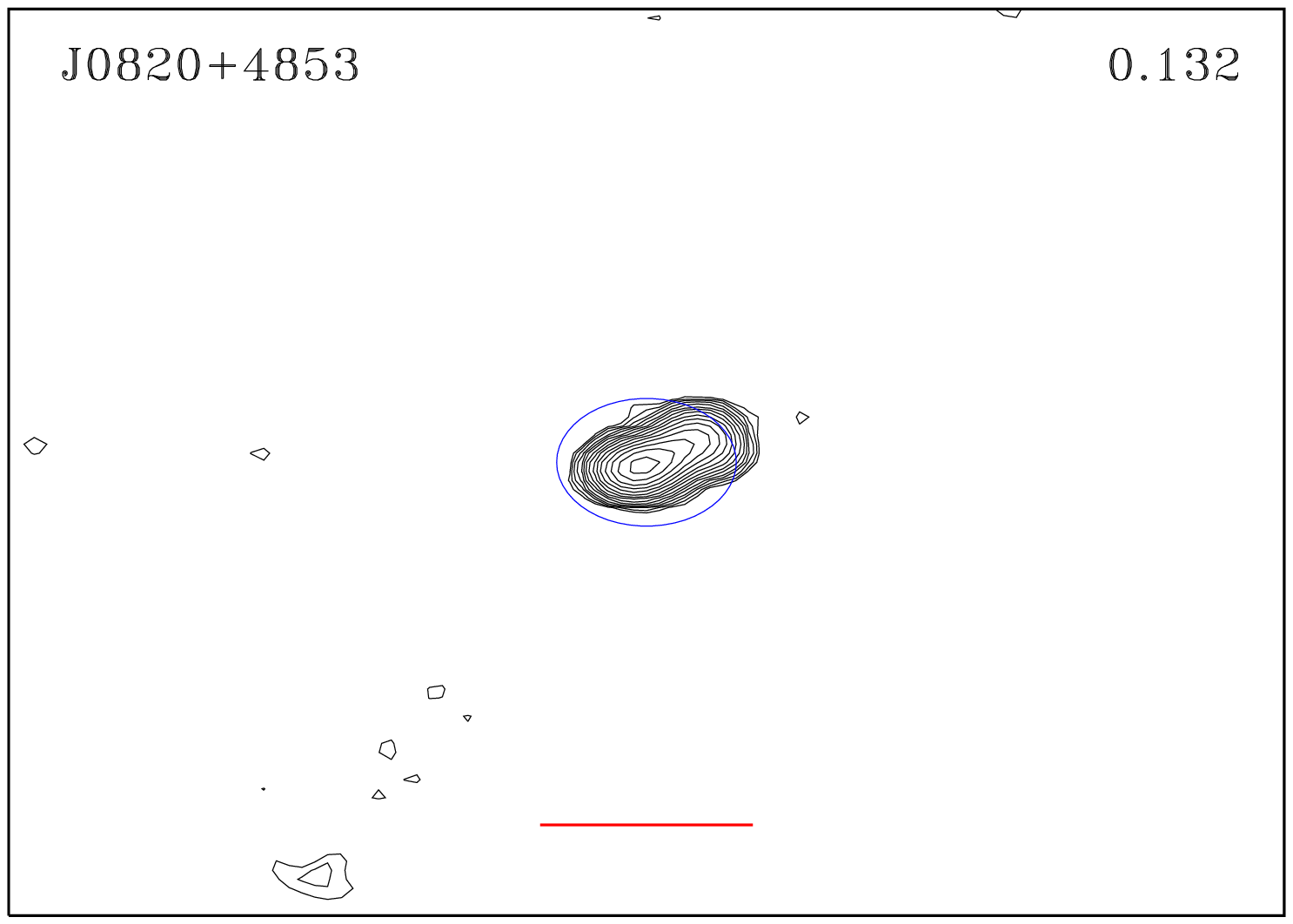} 

\includegraphics[width=6.3cm,height=6.3cm]{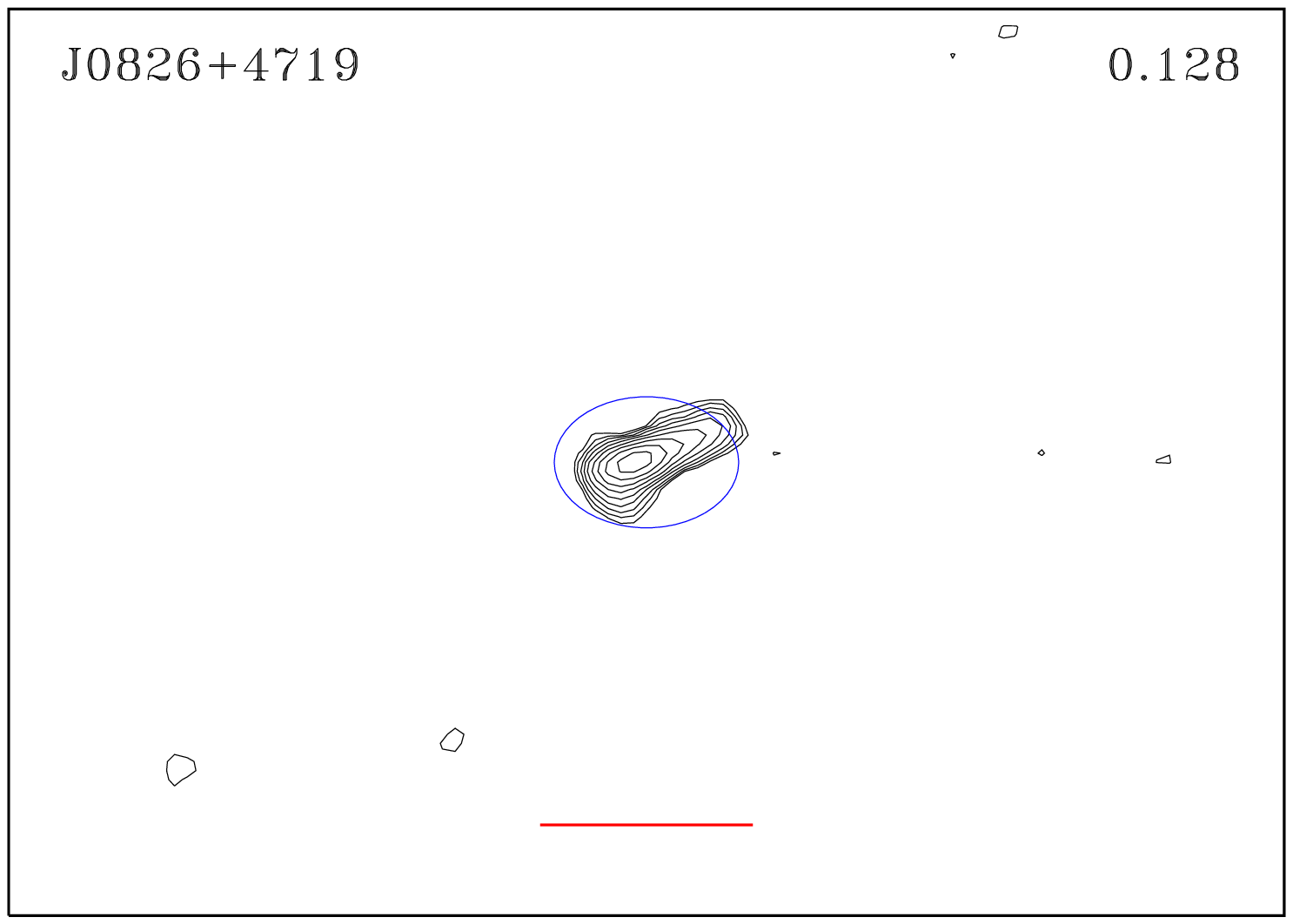} 
\includegraphics[width=6.3cm,height=6.3cm]{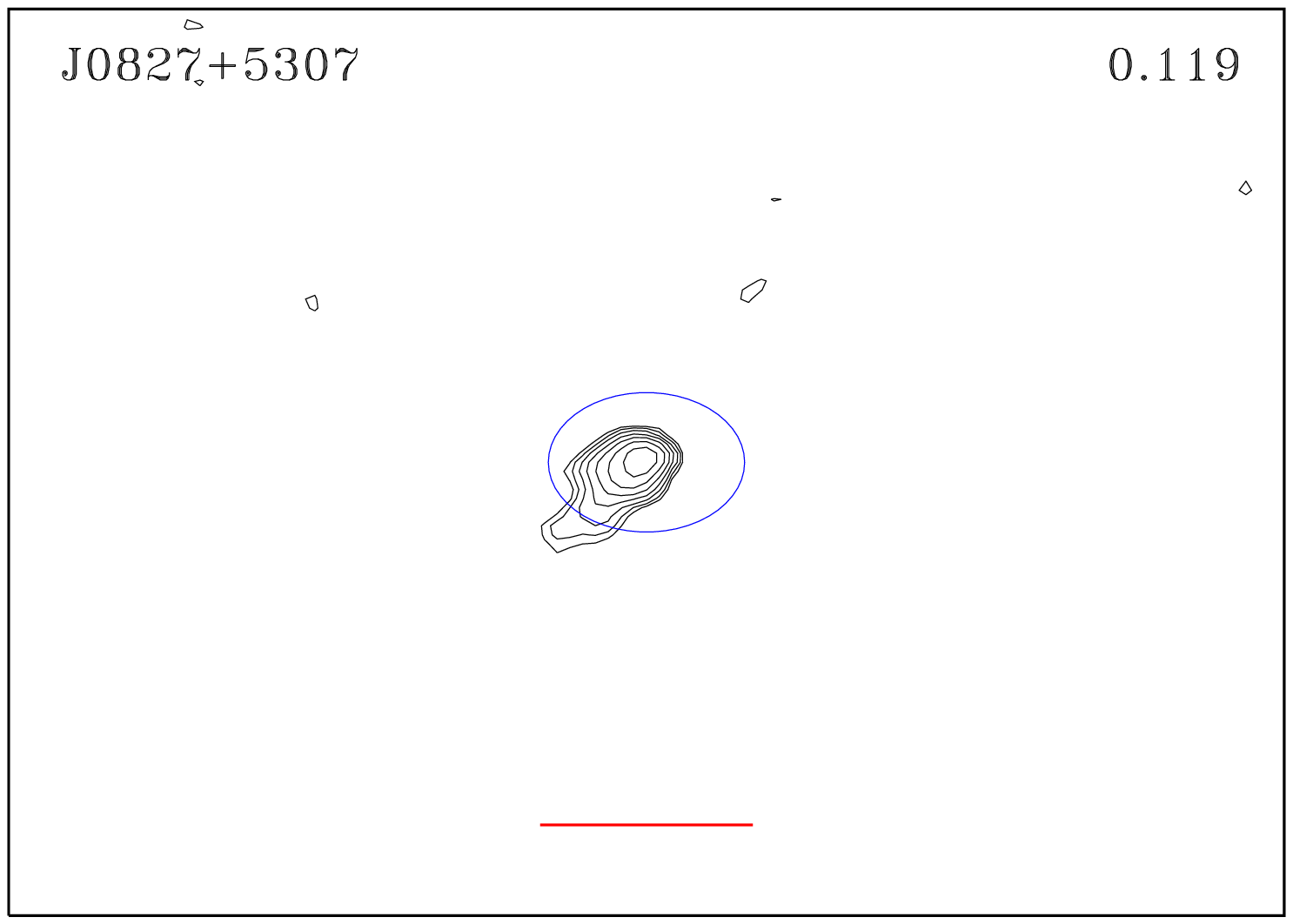} 
\includegraphics[width=6.3cm,height=6.3cm]{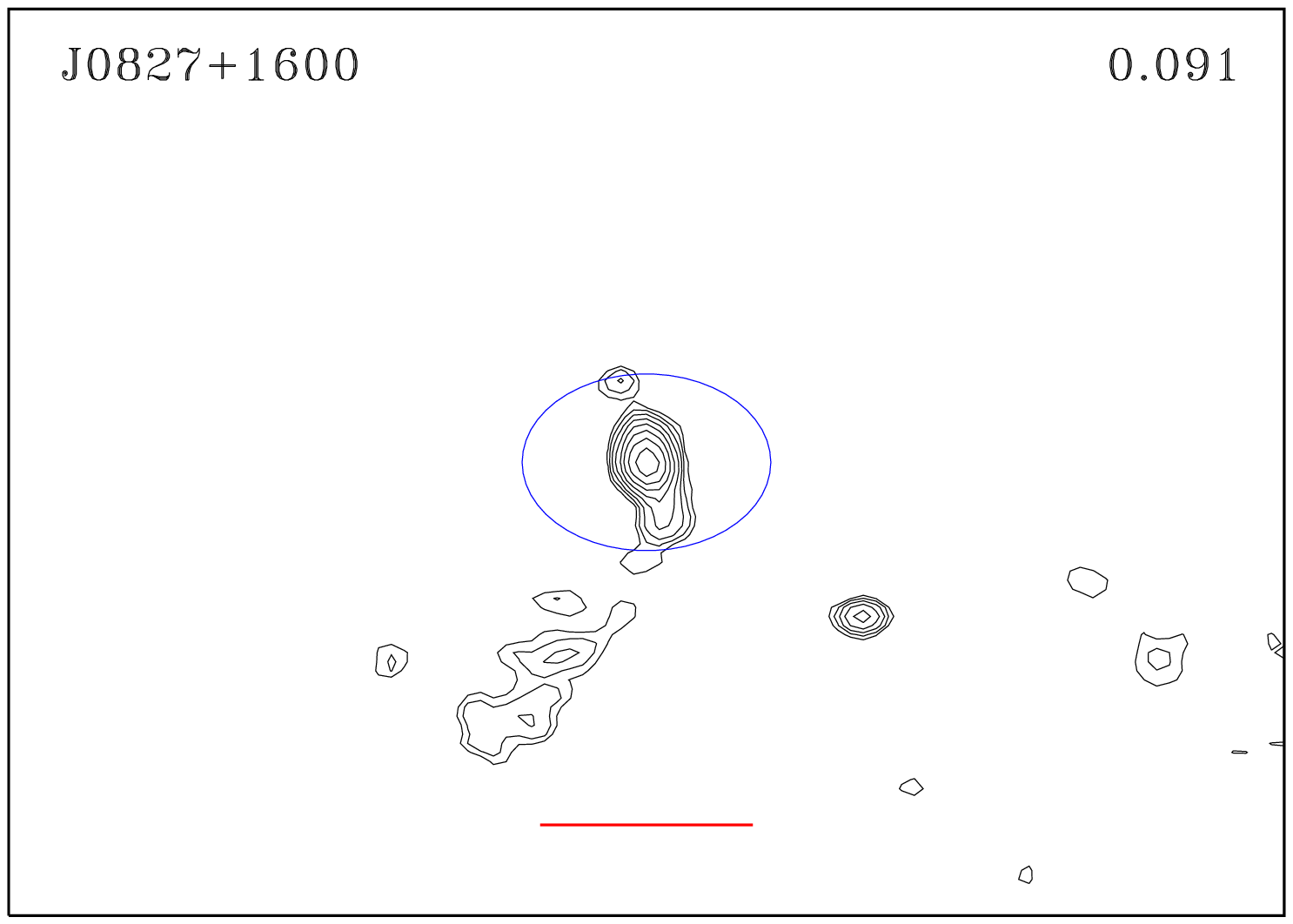} 

\includegraphics[width=6.3cm,height=6.3cm]{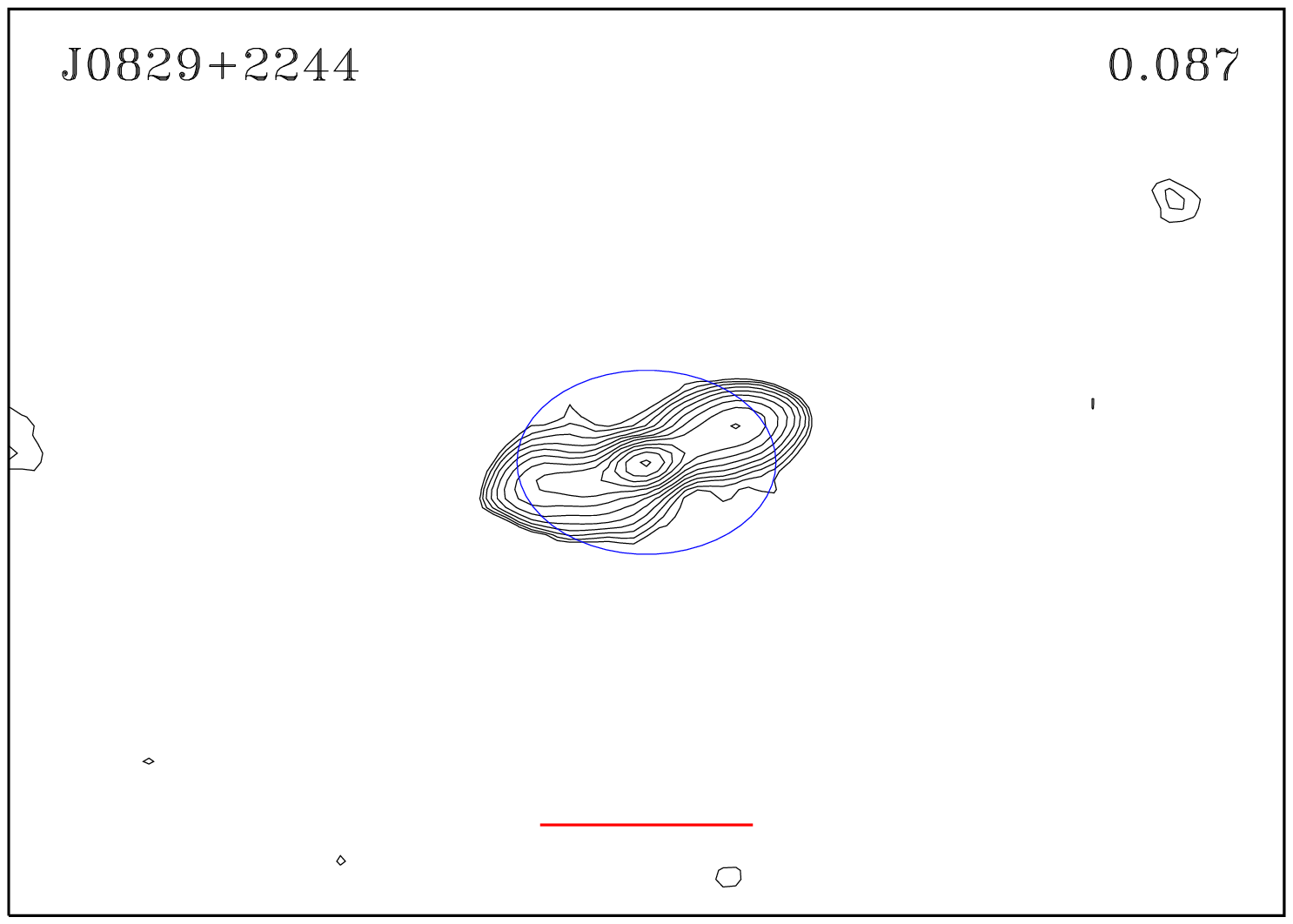} 
\includegraphics[width=6.3cm,height=6.3cm]{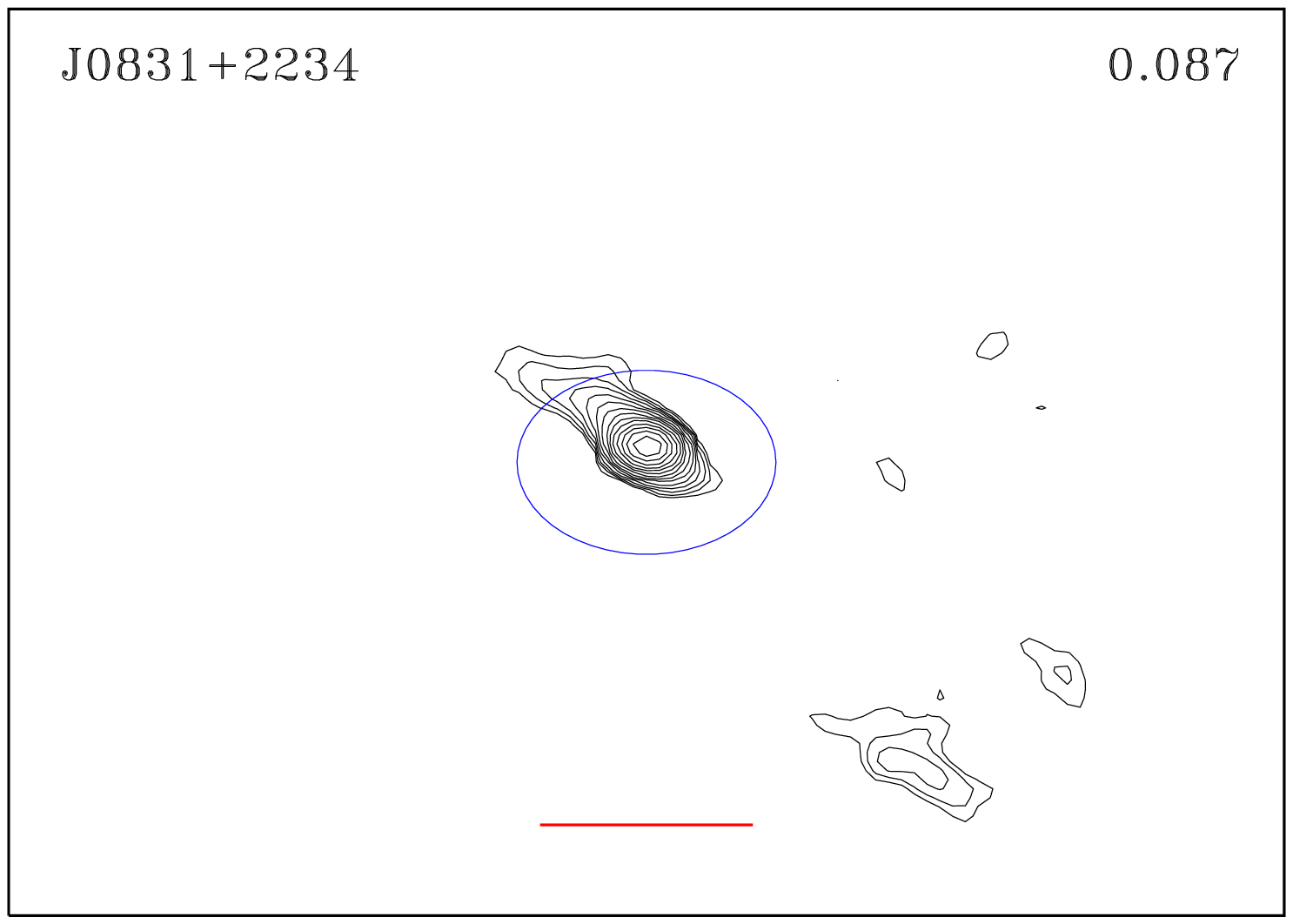} 
\includegraphics[width=6.3cm,height=6.3cm]{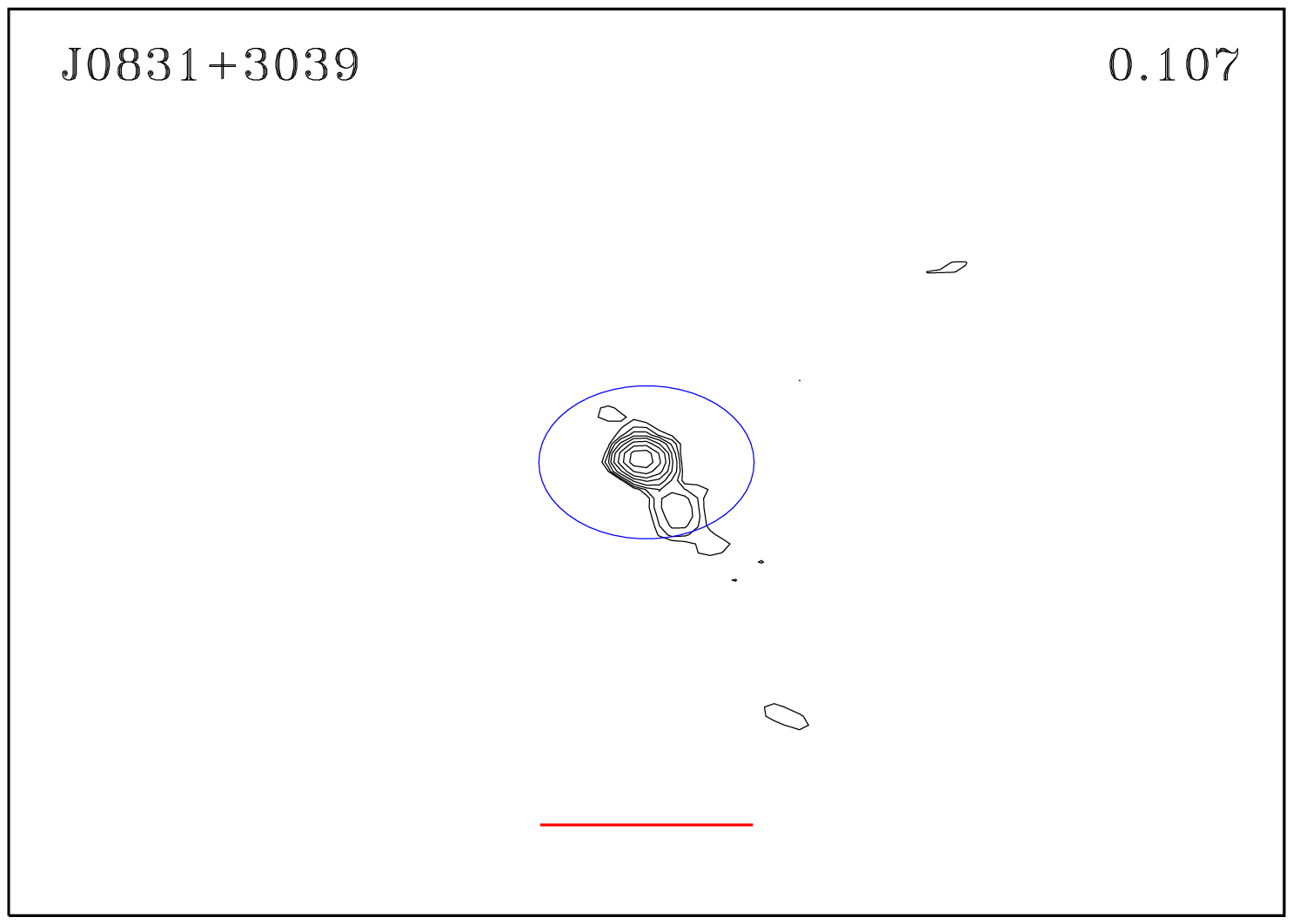} 
\caption{(continued)}
\end{figure*}

\addtocounter{figure}{-1}
\begin{figure*}
\includegraphics[width=6.3cm,height=6.3cm]{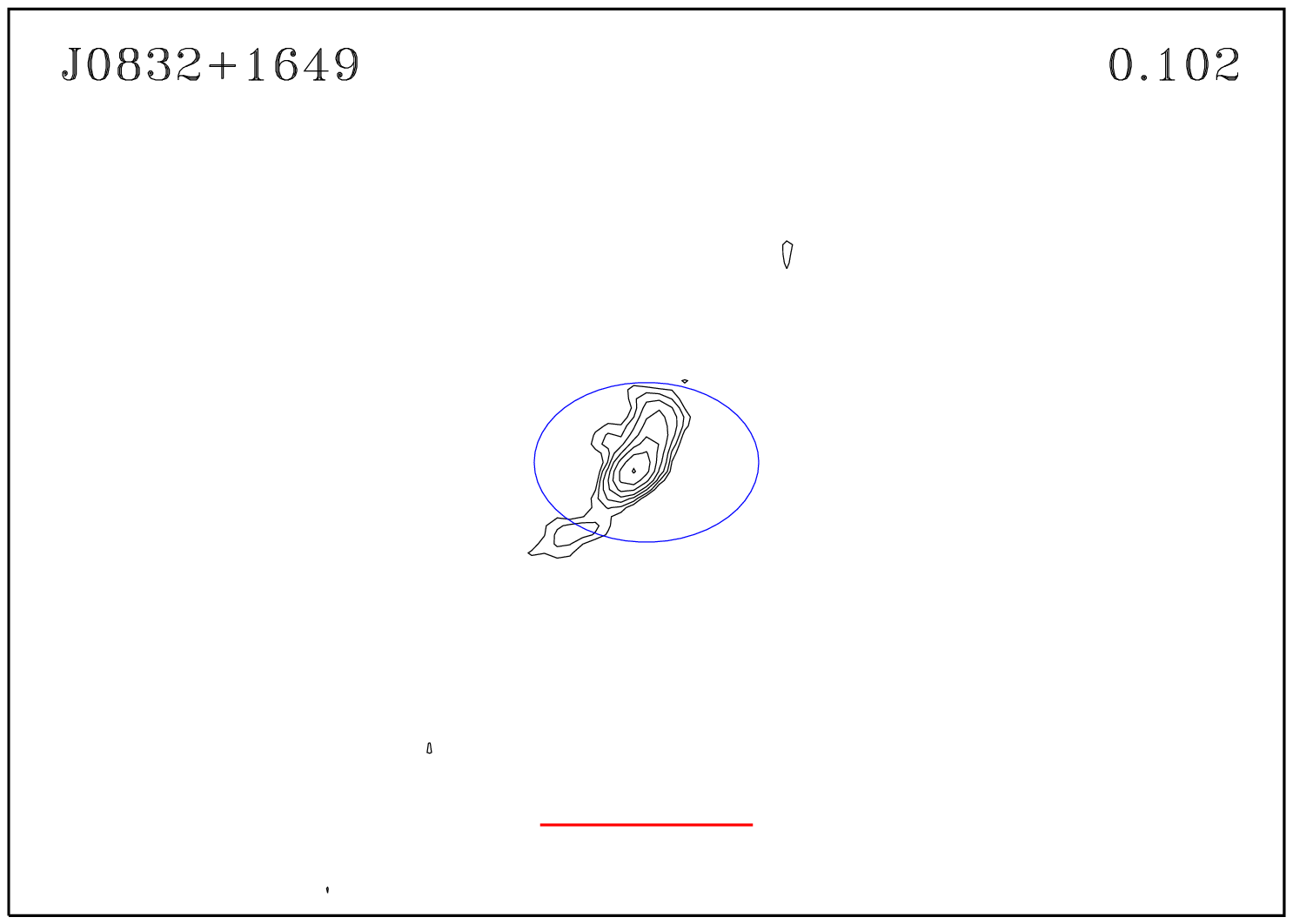} 
\includegraphics[width=6.3cm,height=6.3cm]{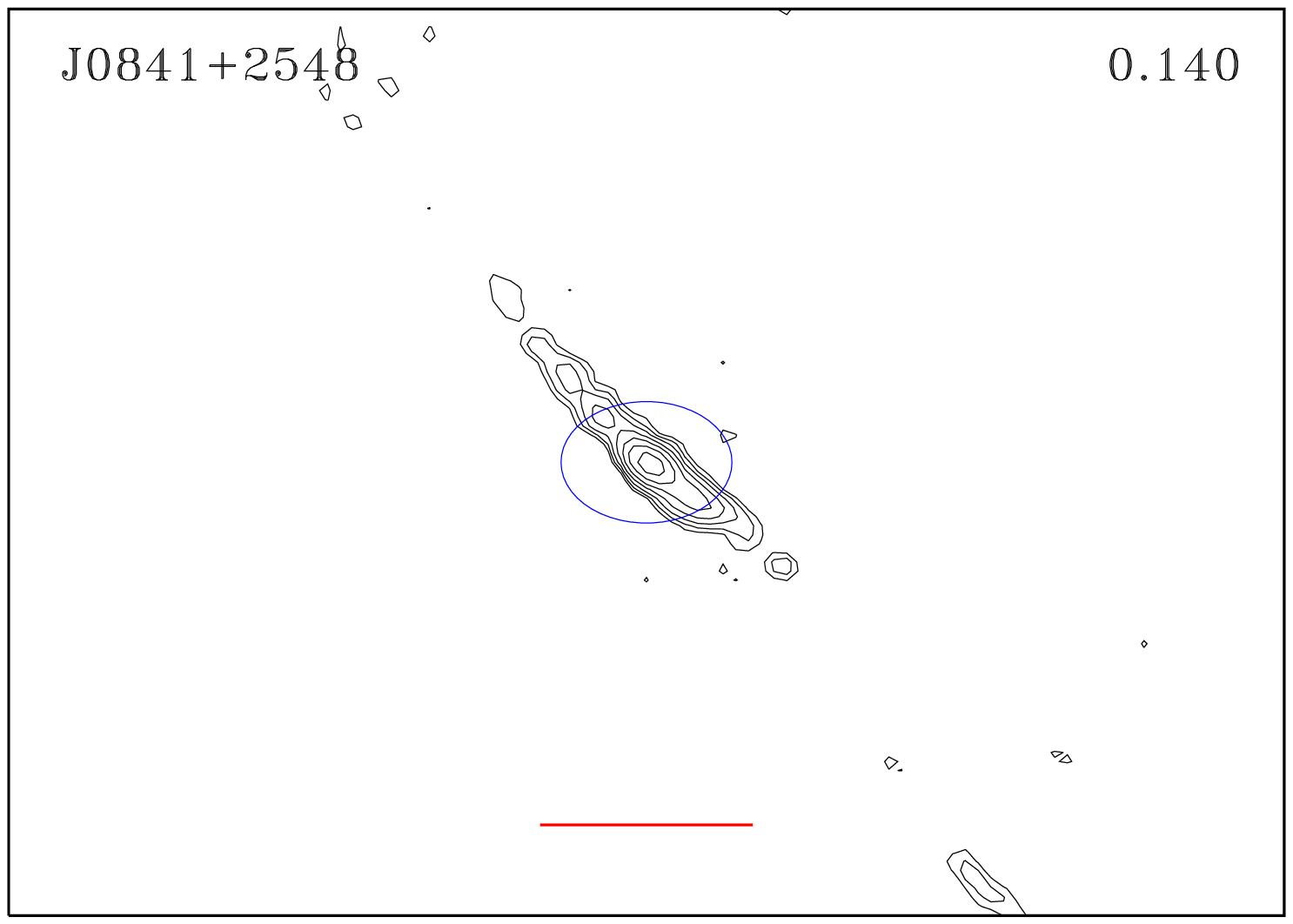} 
\includegraphics[width=6.3cm,height=6.3cm]{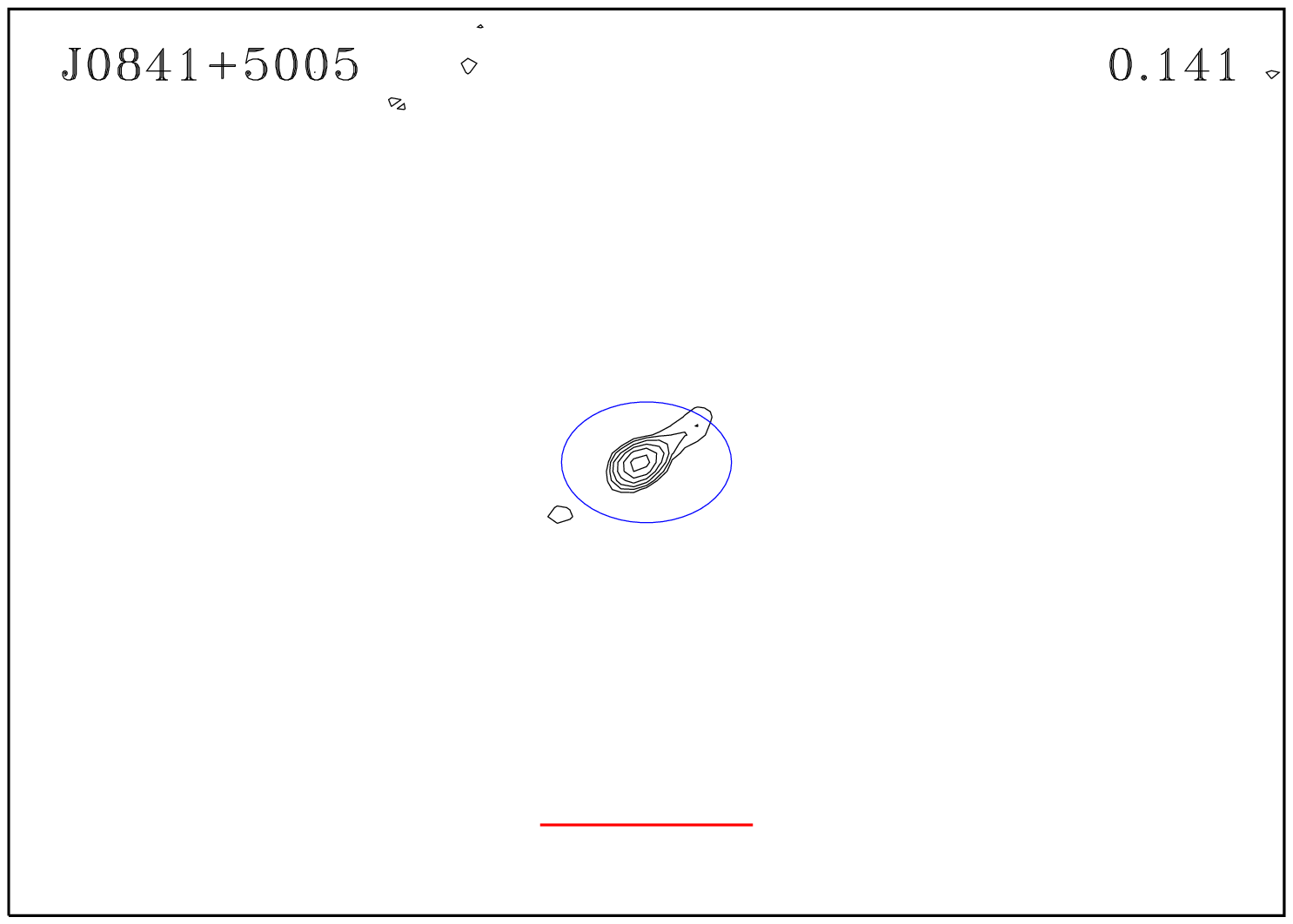} 

\includegraphics[width=6.3cm,height=6.3cm]{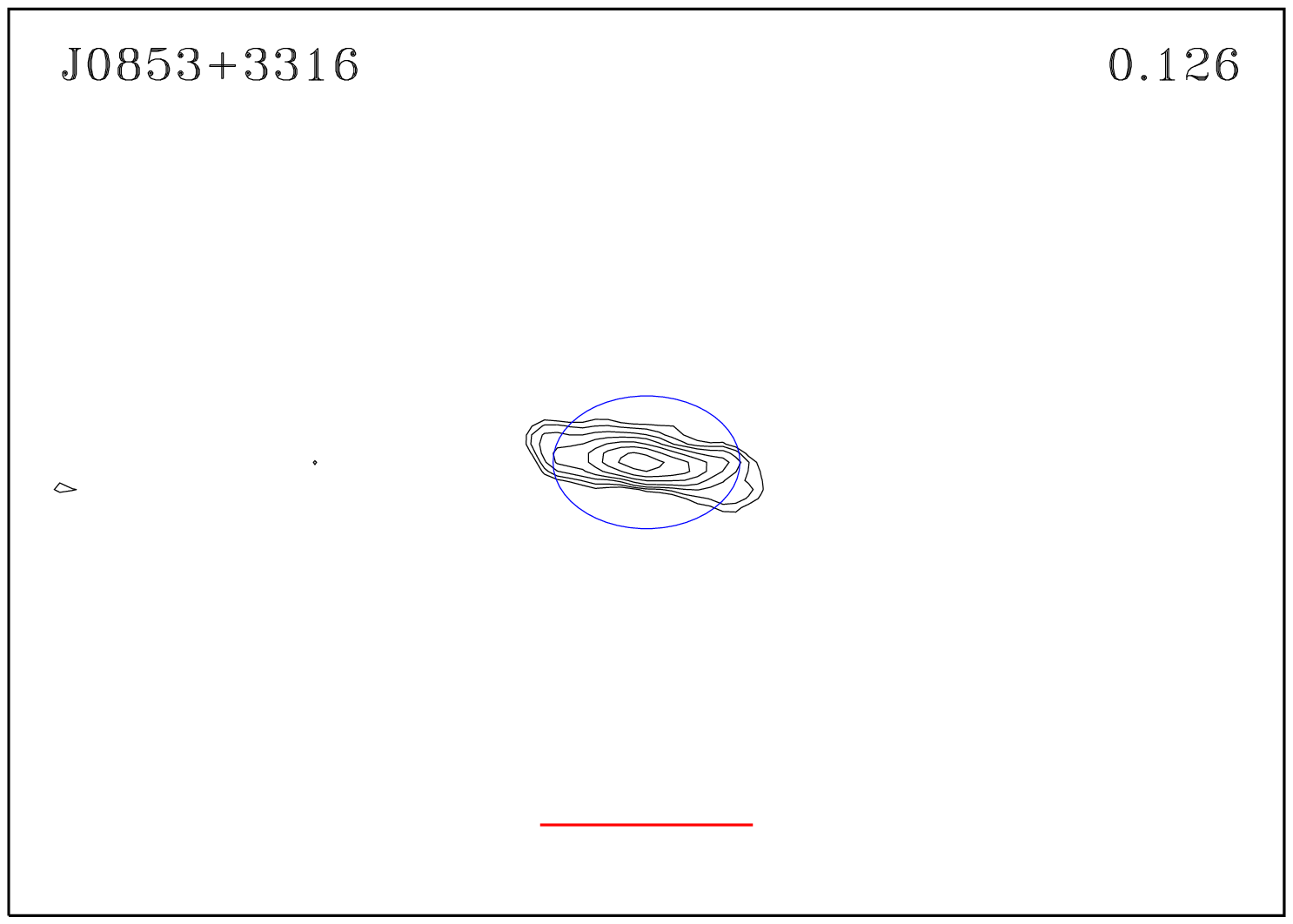} 
\includegraphics[width=6.3cm,height=6.3cm]{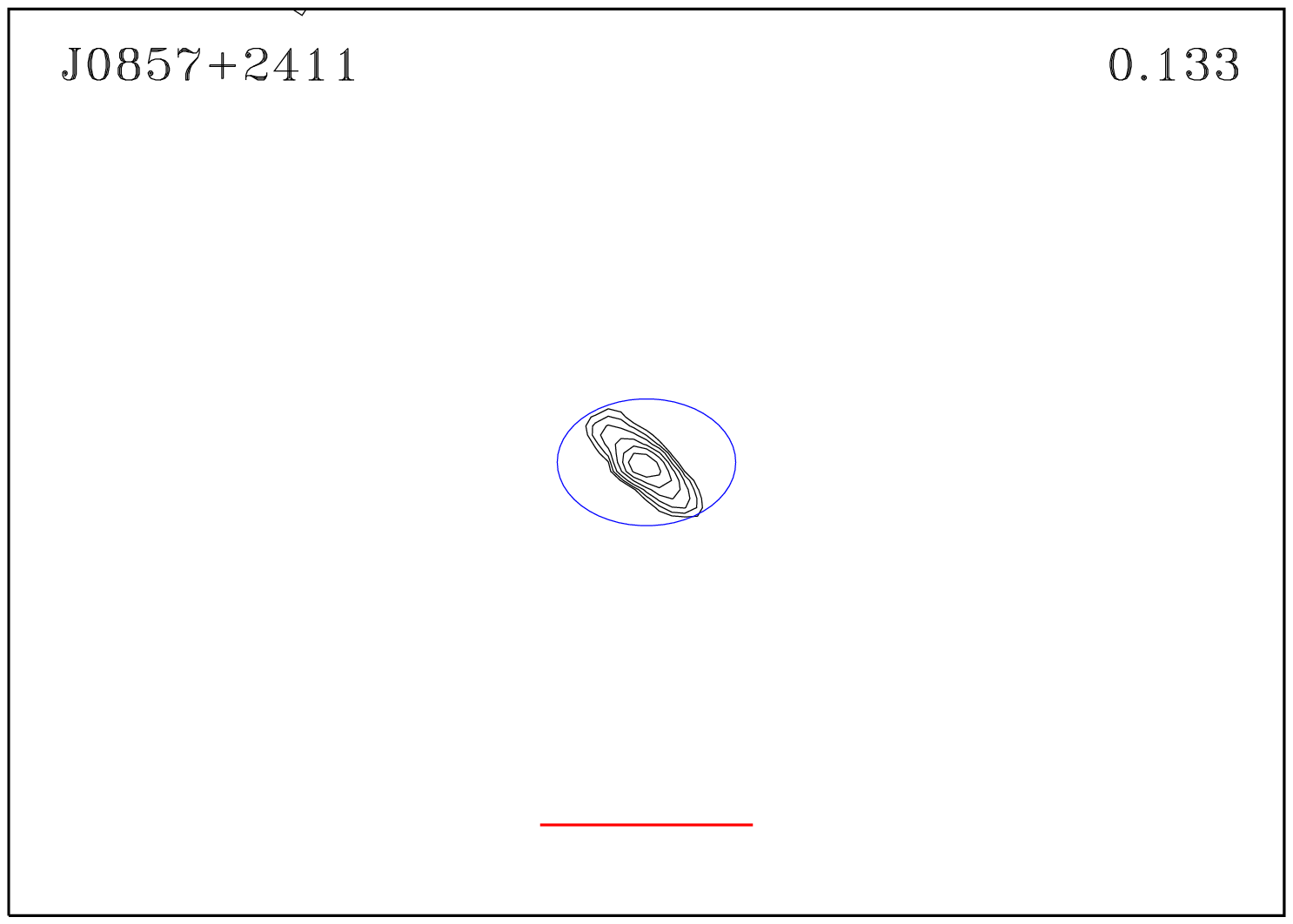} 
\includegraphics[width=6.3cm,height=6.3cm]{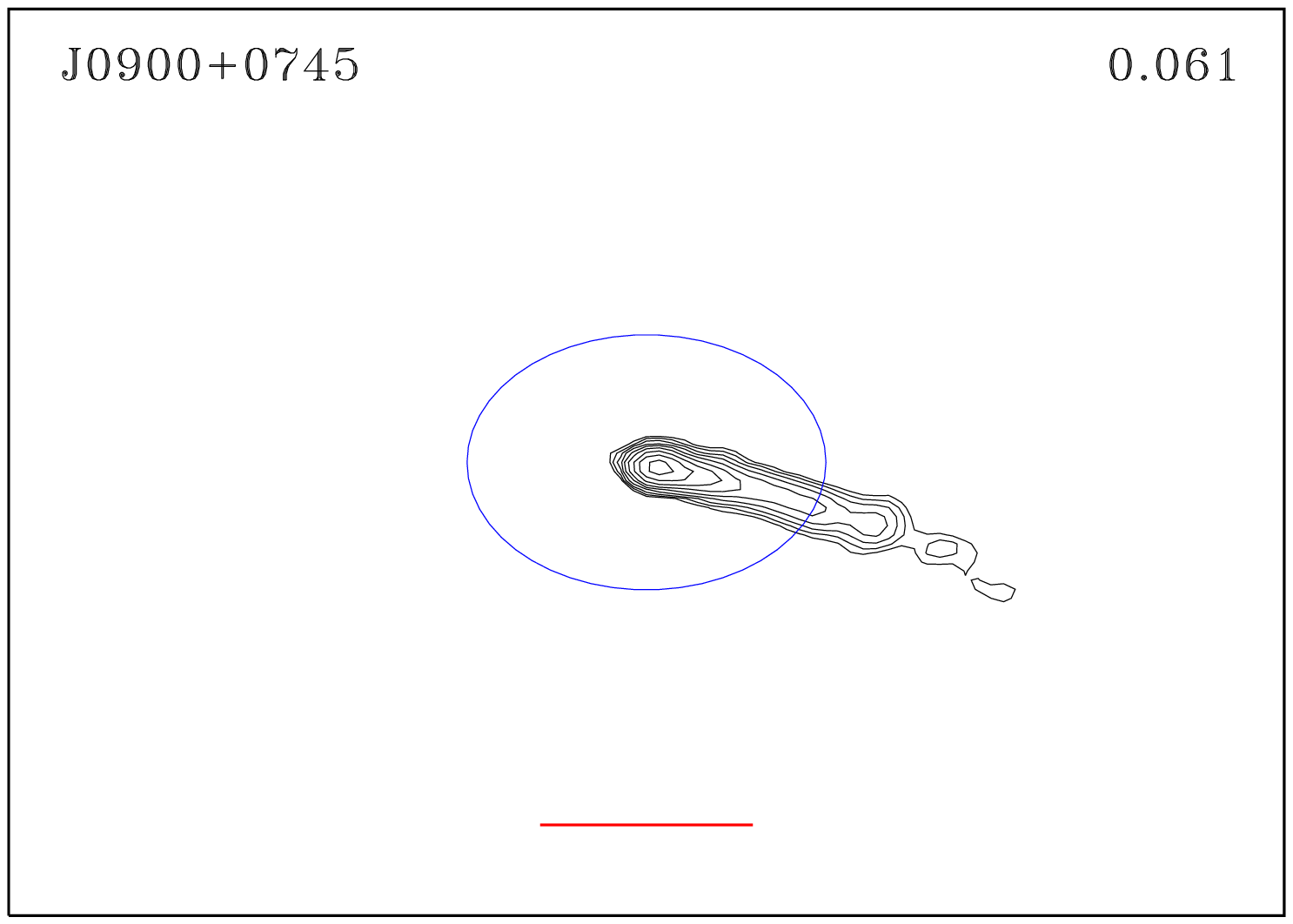} 

\includegraphics[width=6.3cm,height=6.3cm]{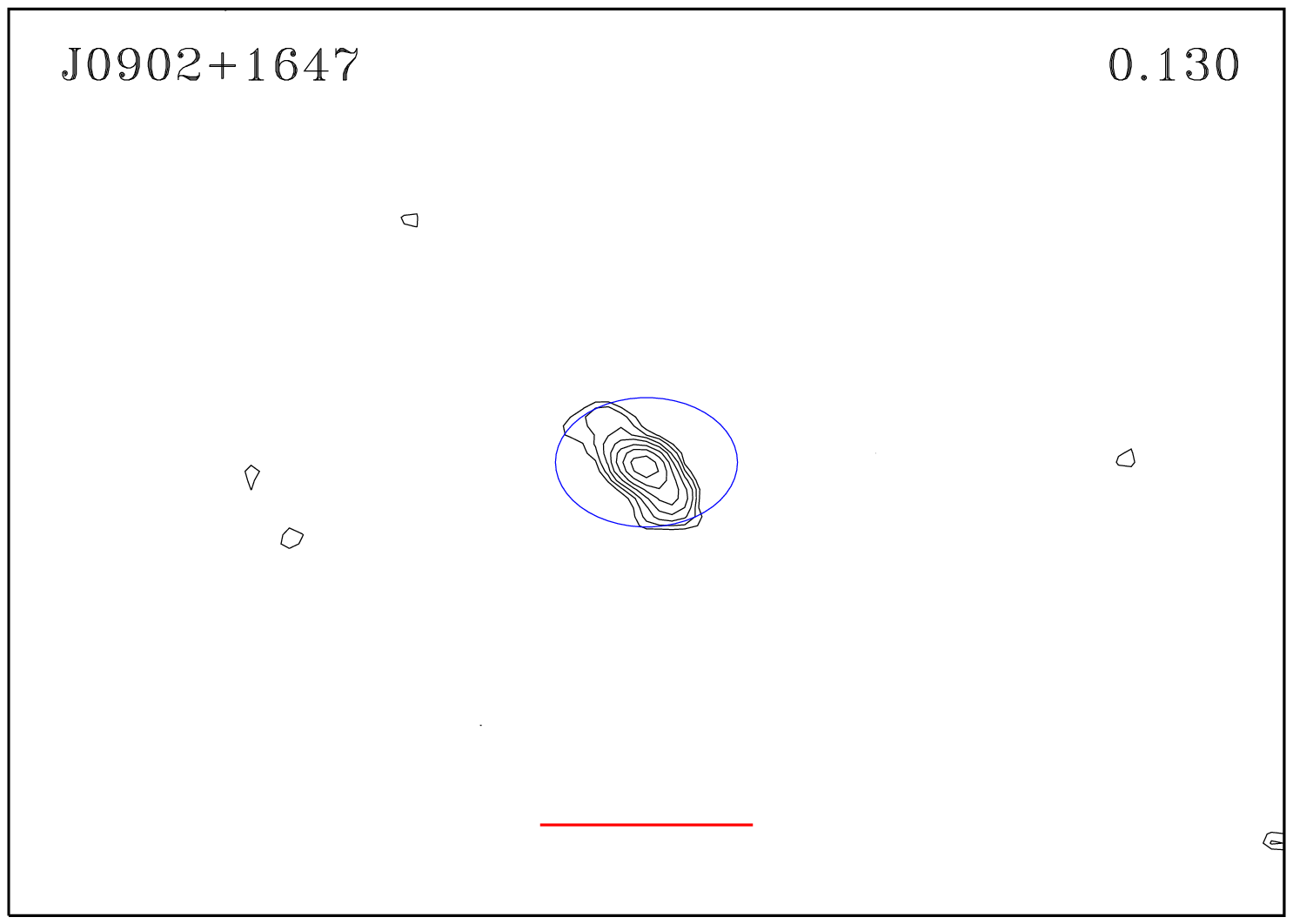} 
\includegraphics[width=6.3cm,height=6.3cm]{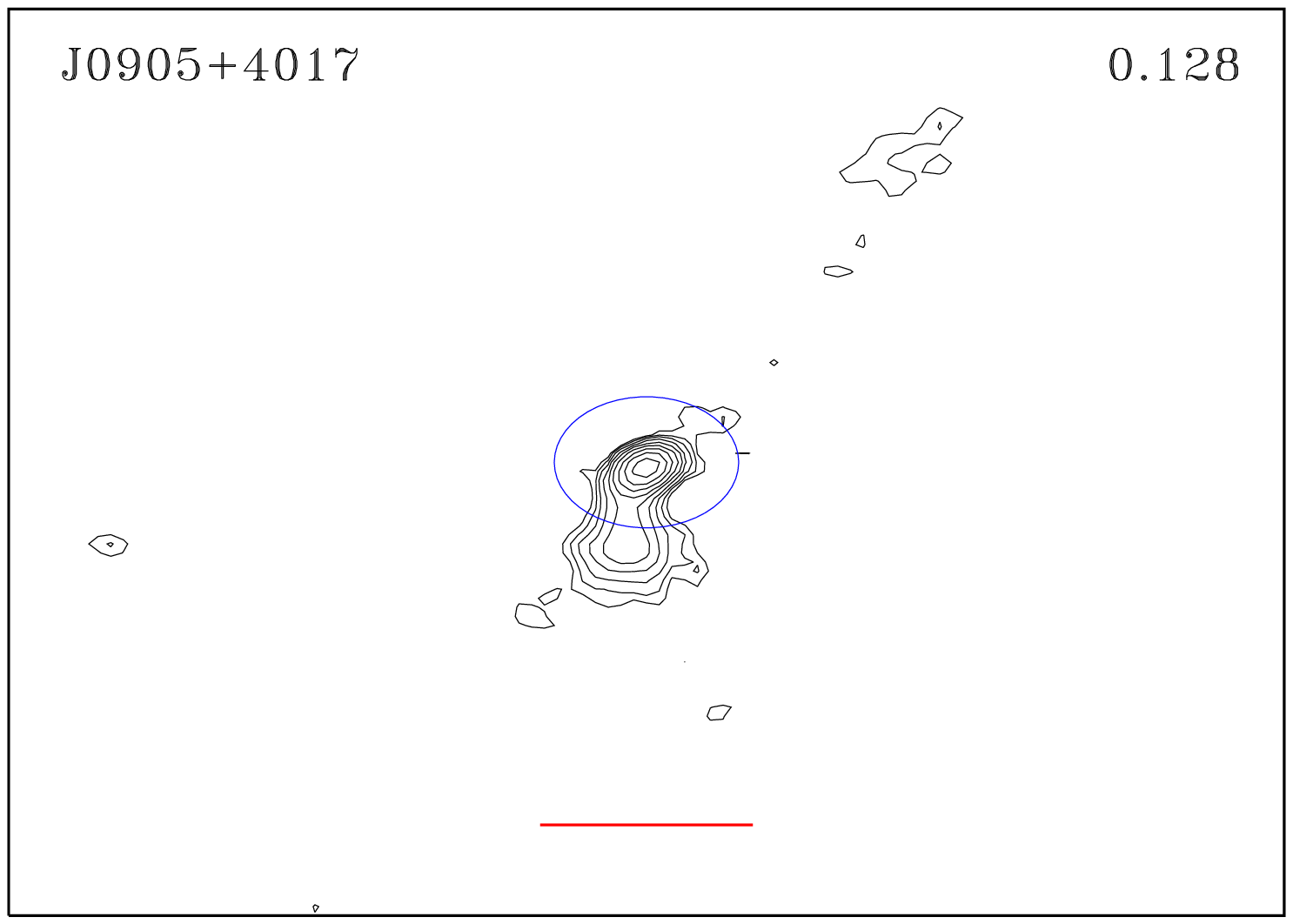} 
\includegraphics[width=6.3cm,height=6.3cm]{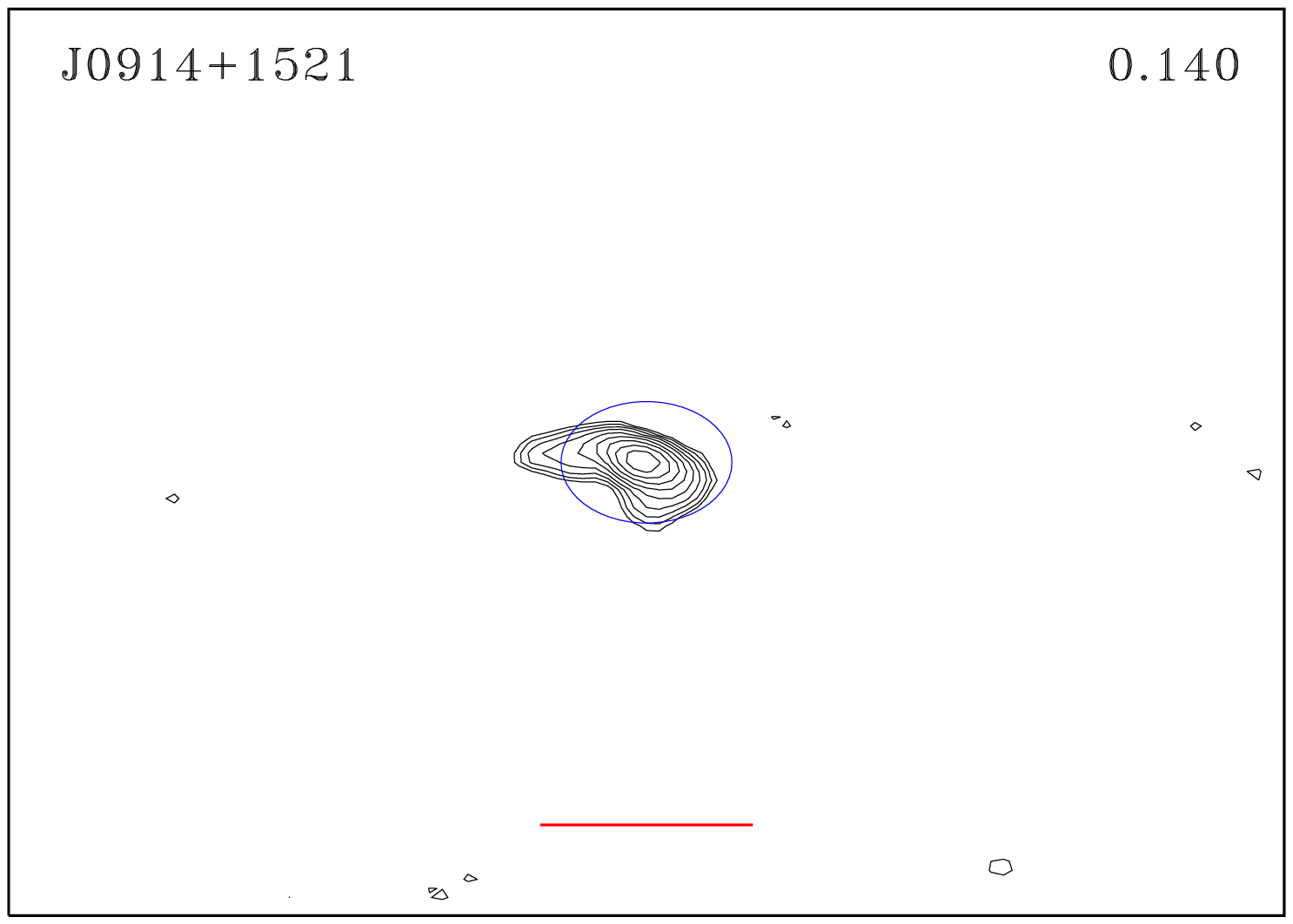} 

\includegraphics[width=6.3cm,height=6.3cm]{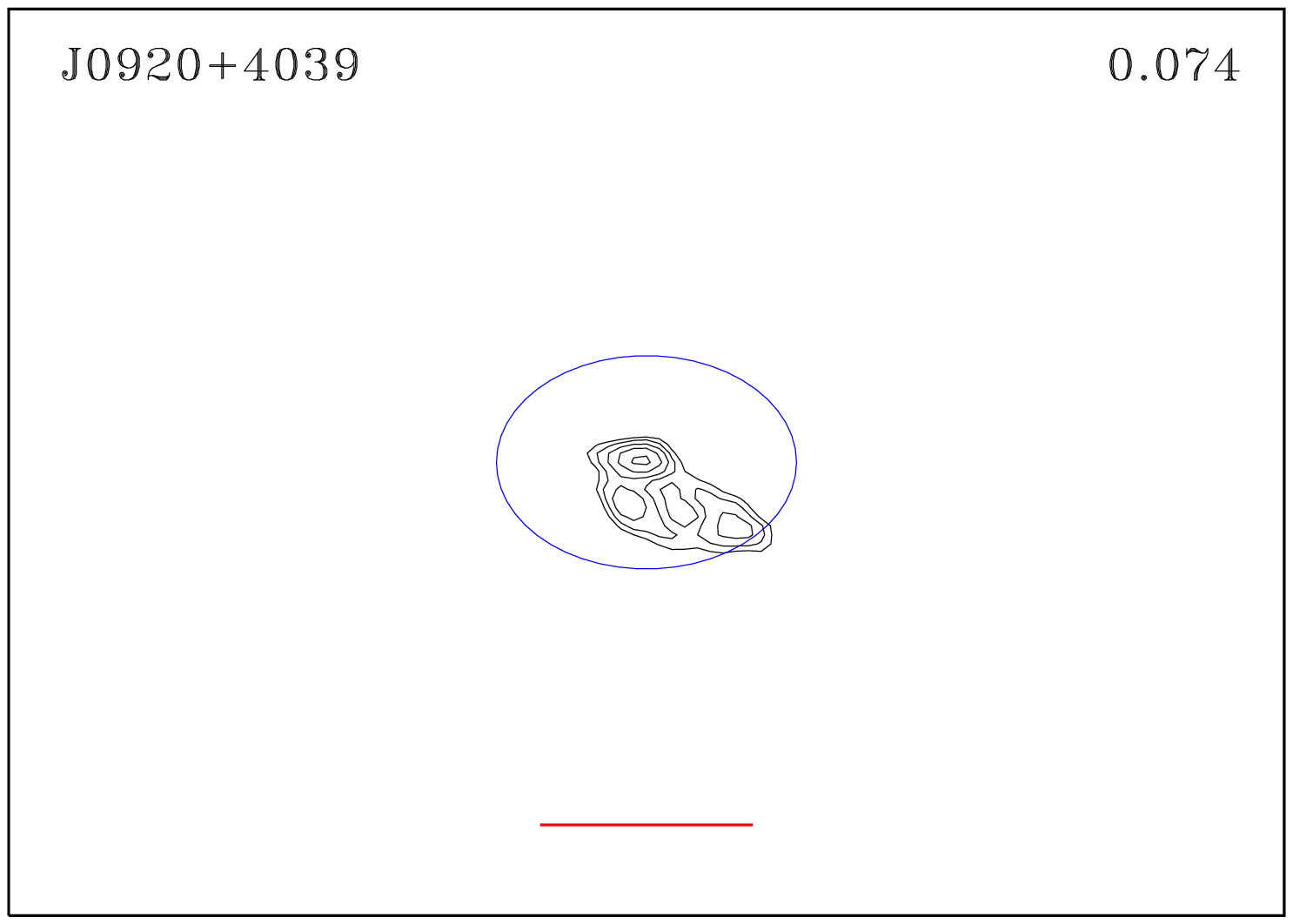} 
\includegraphics[width=6.3cm,height=6.3cm]{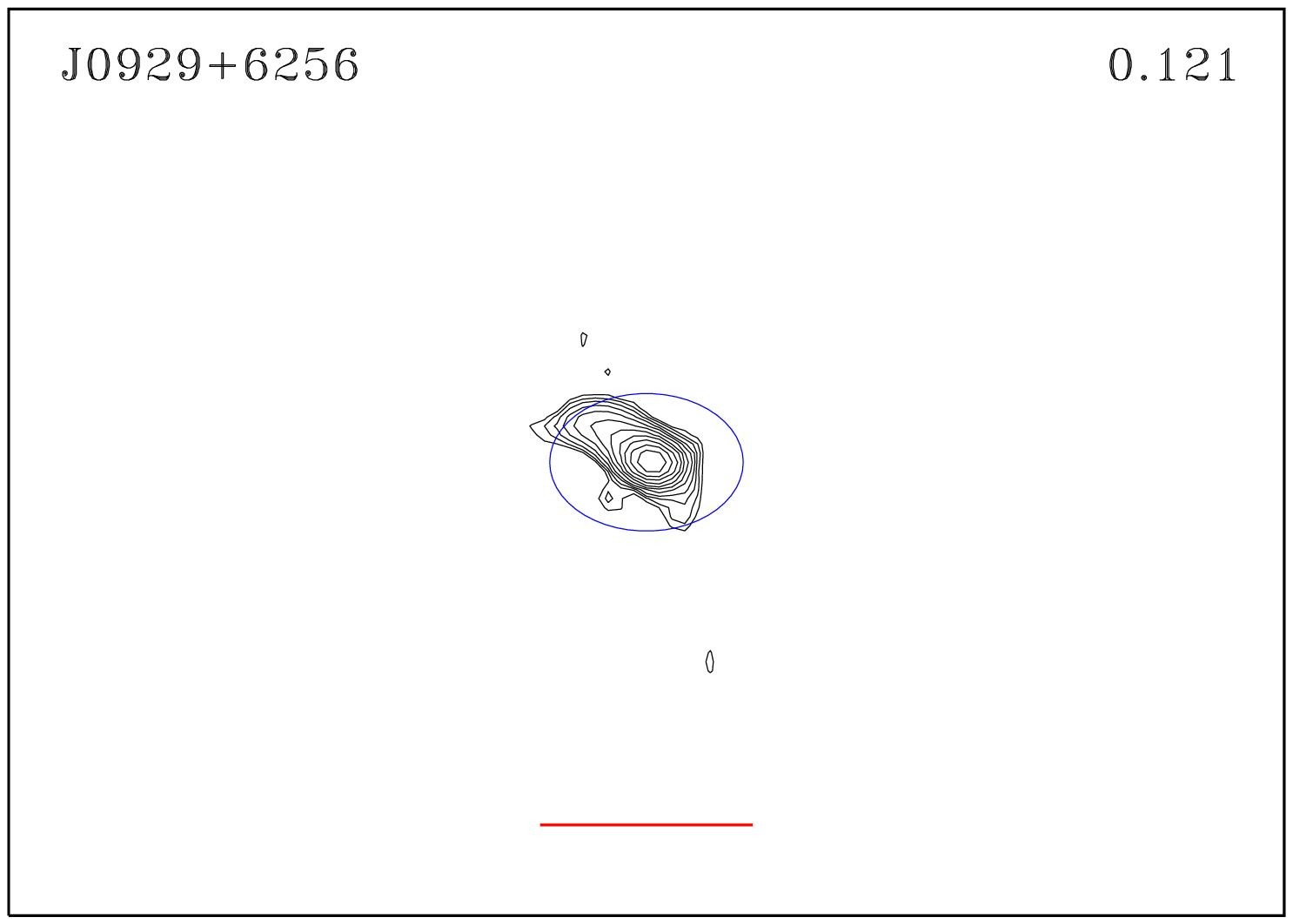} 
\includegraphics[width=6.3cm,height=6.3cm]{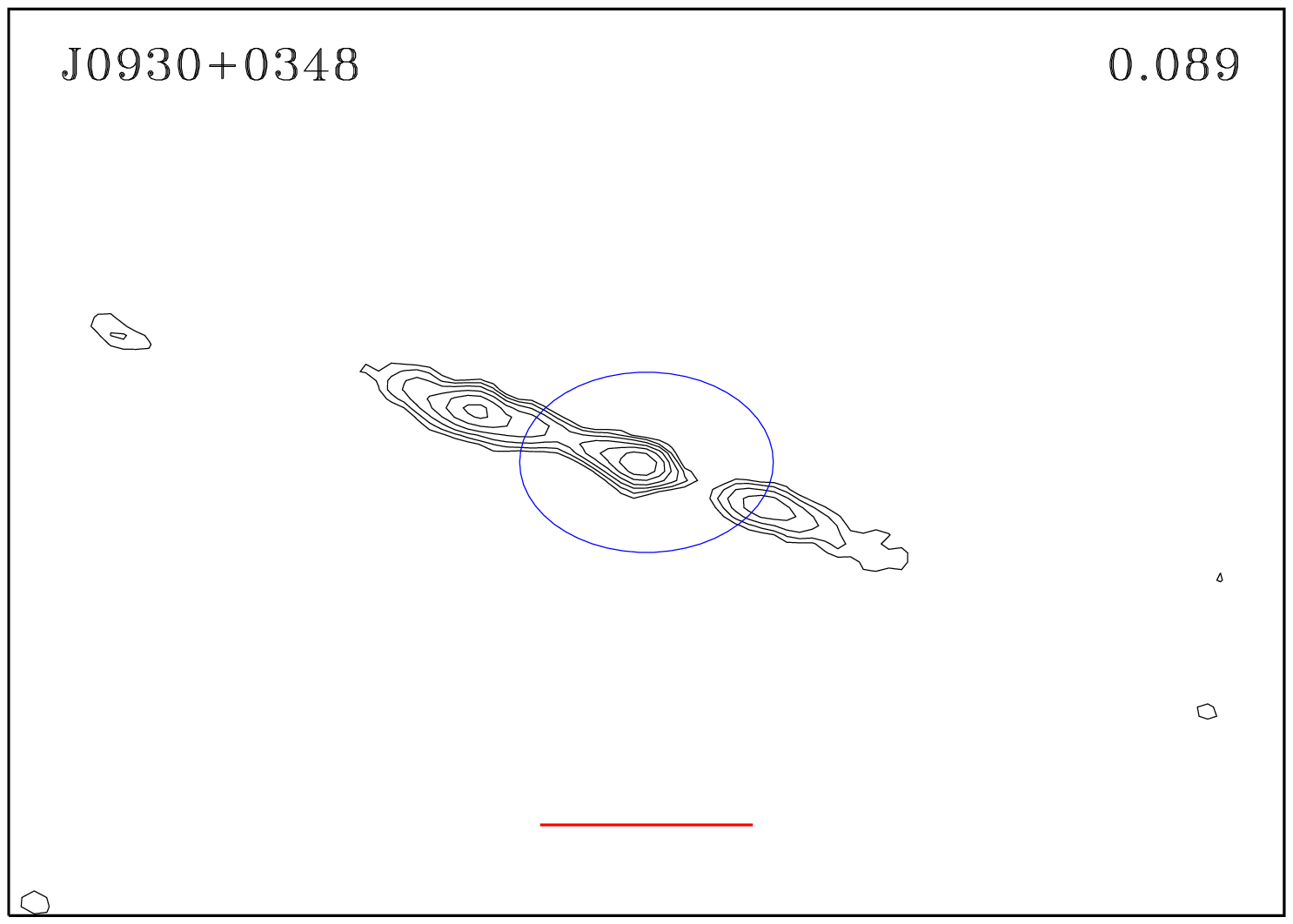} 
\caption{(continued)}
\end{figure*}

\addtocounter{figure}{-1}
\begin{figure*}
\includegraphics[width=6.3cm,height=6.3cm]{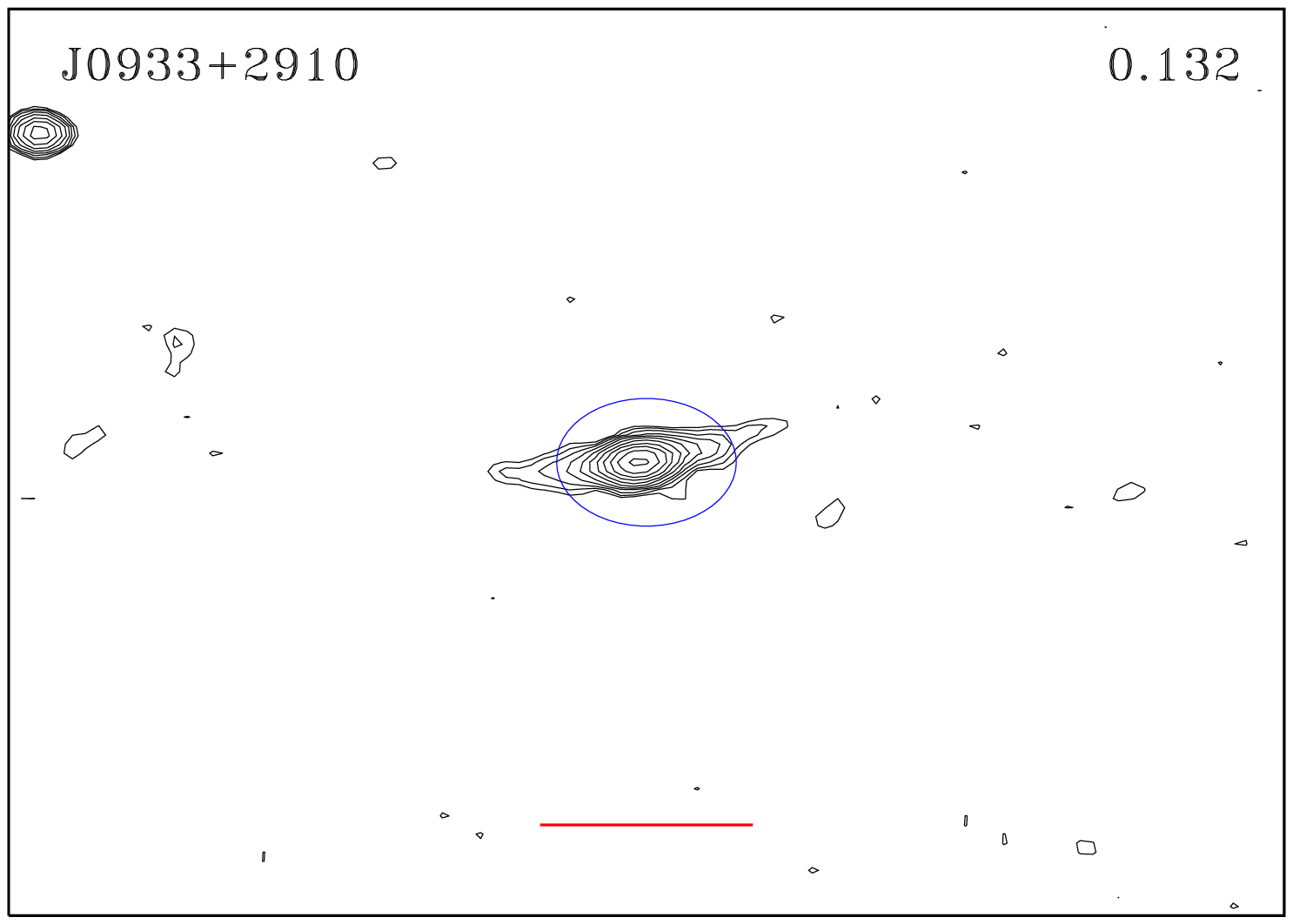} 
\includegraphics[width=6.3cm,height=6.3cm]{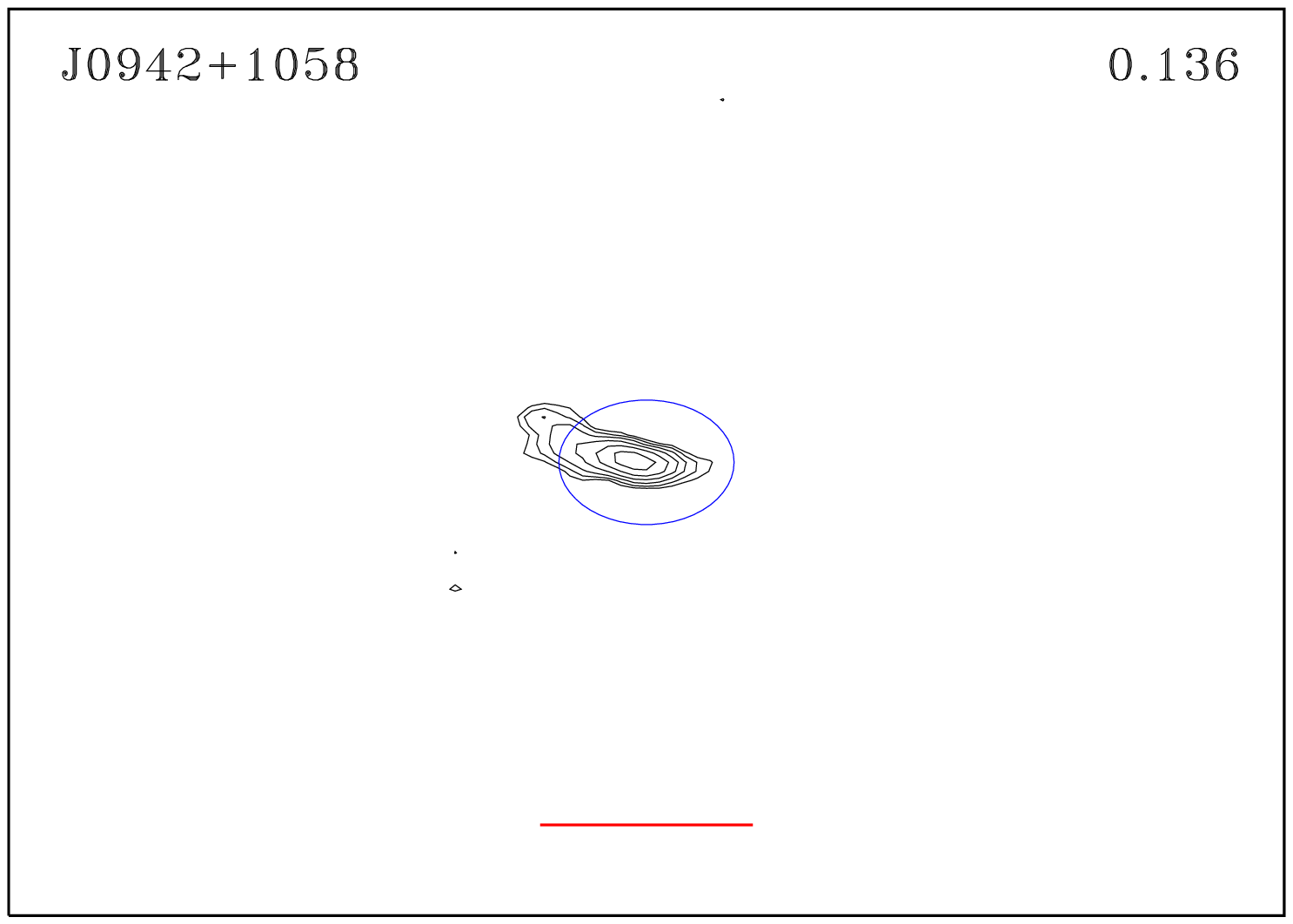} 
\includegraphics[width=6.3cm,height=6.3cm]{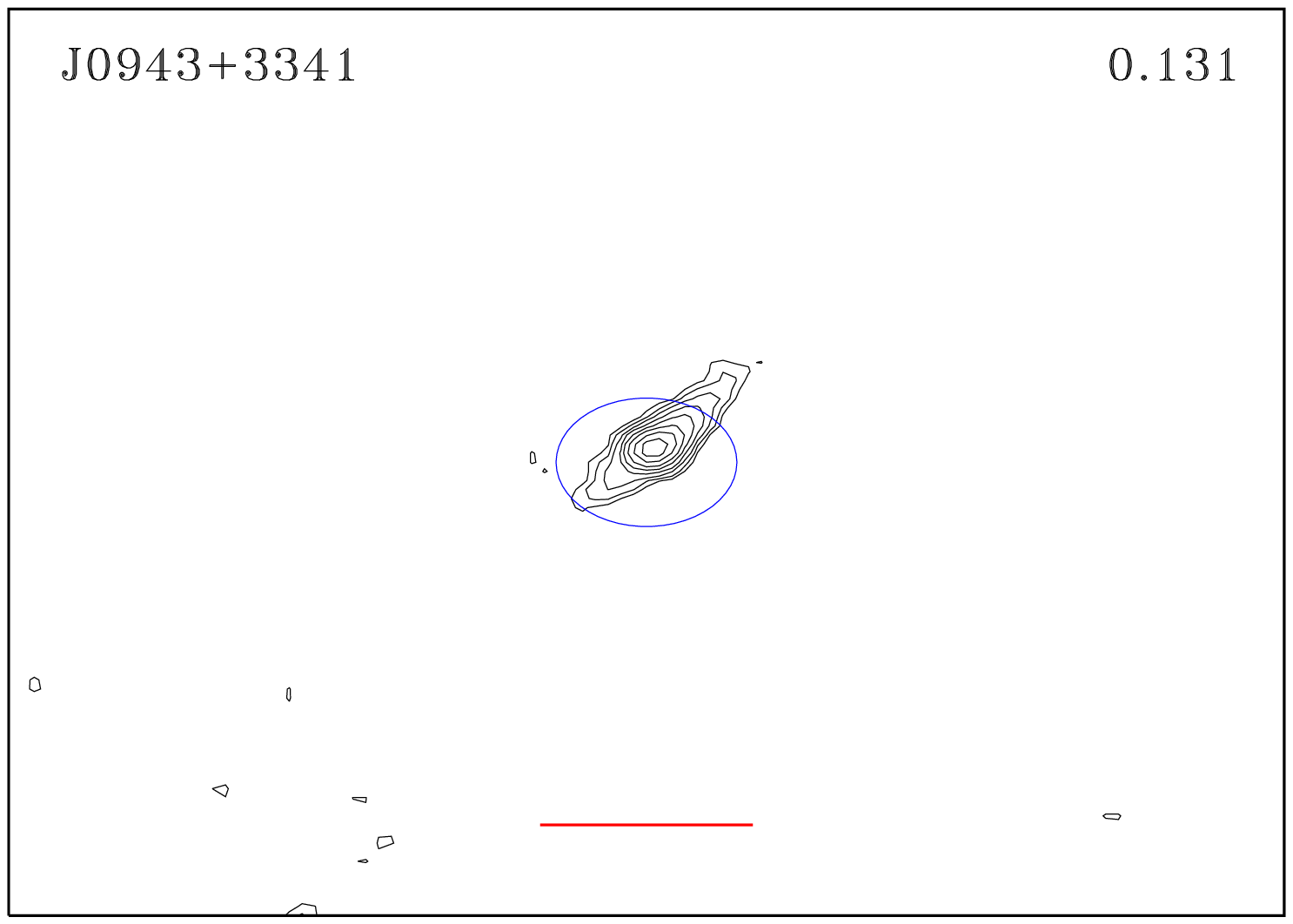} 

\includegraphics[width=6.3cm,height=6.3cm]{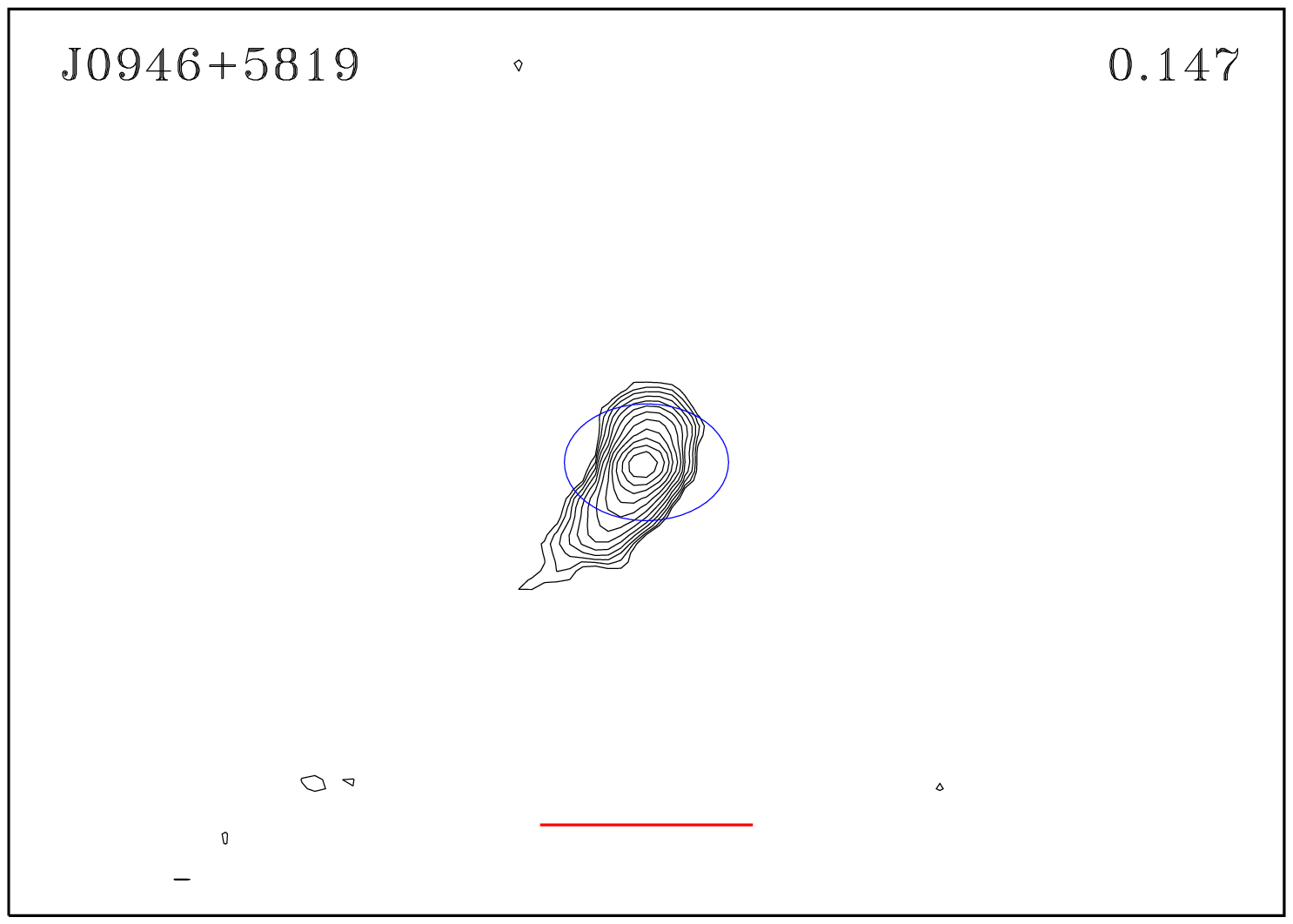} 
\includegraphics[width=6.3cm,height=6.3cm]{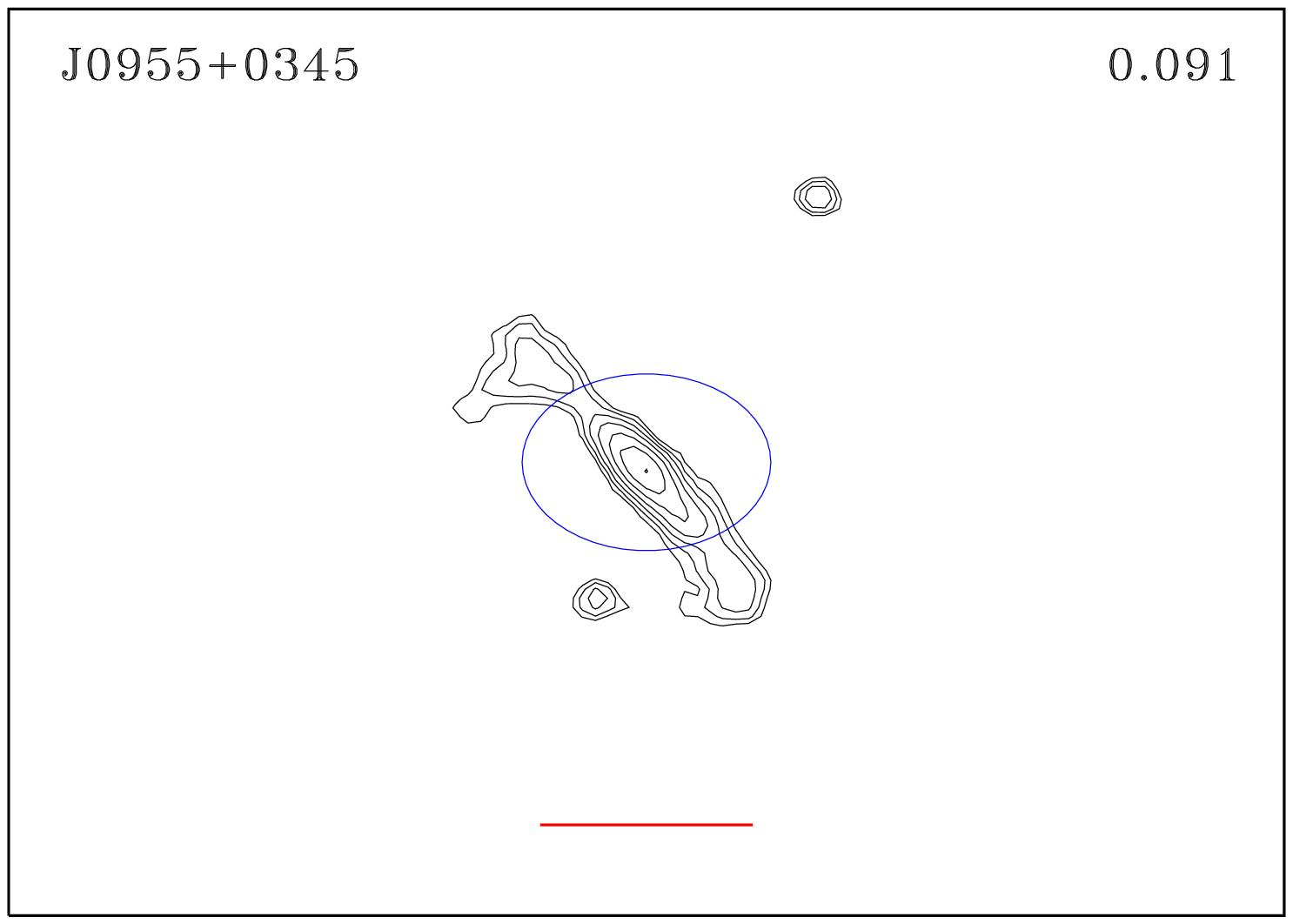} 
\includegraphics[width=6.3cm,height=6.3cm]{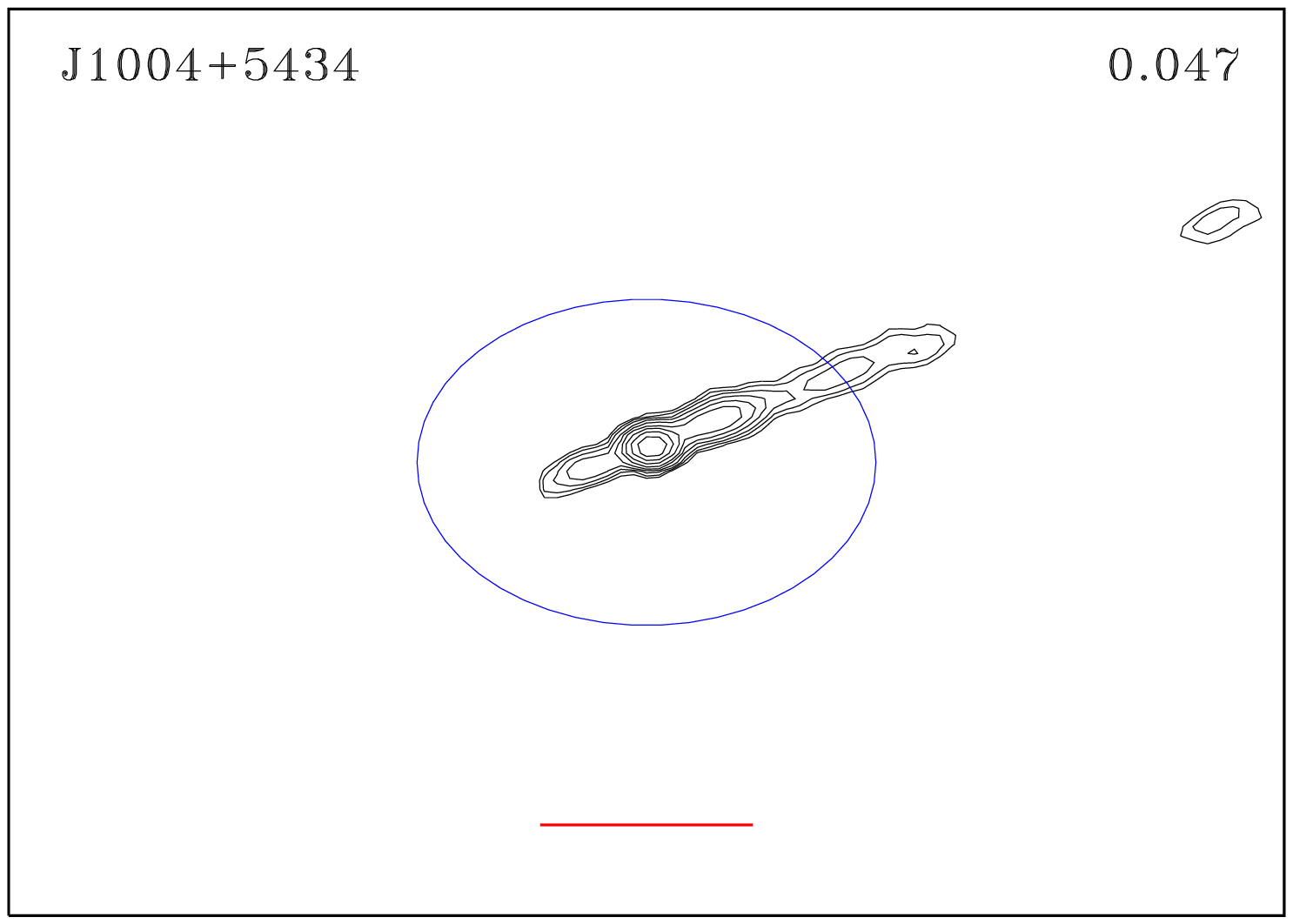} 

\includegraphics[width=6.3cm,height=6.3cm]{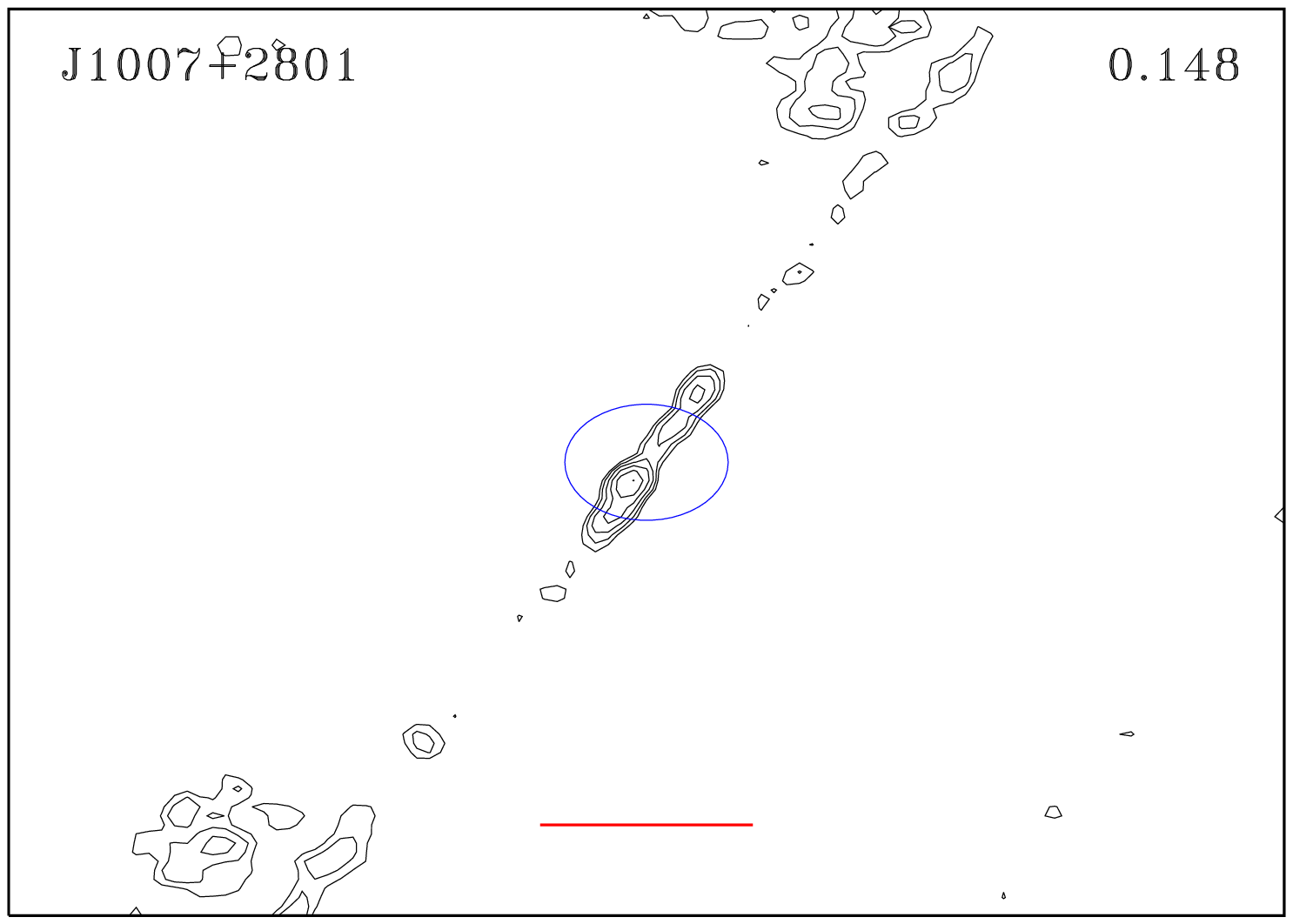} 
\includegraphics[width=6.3cm,height=6.3cm]{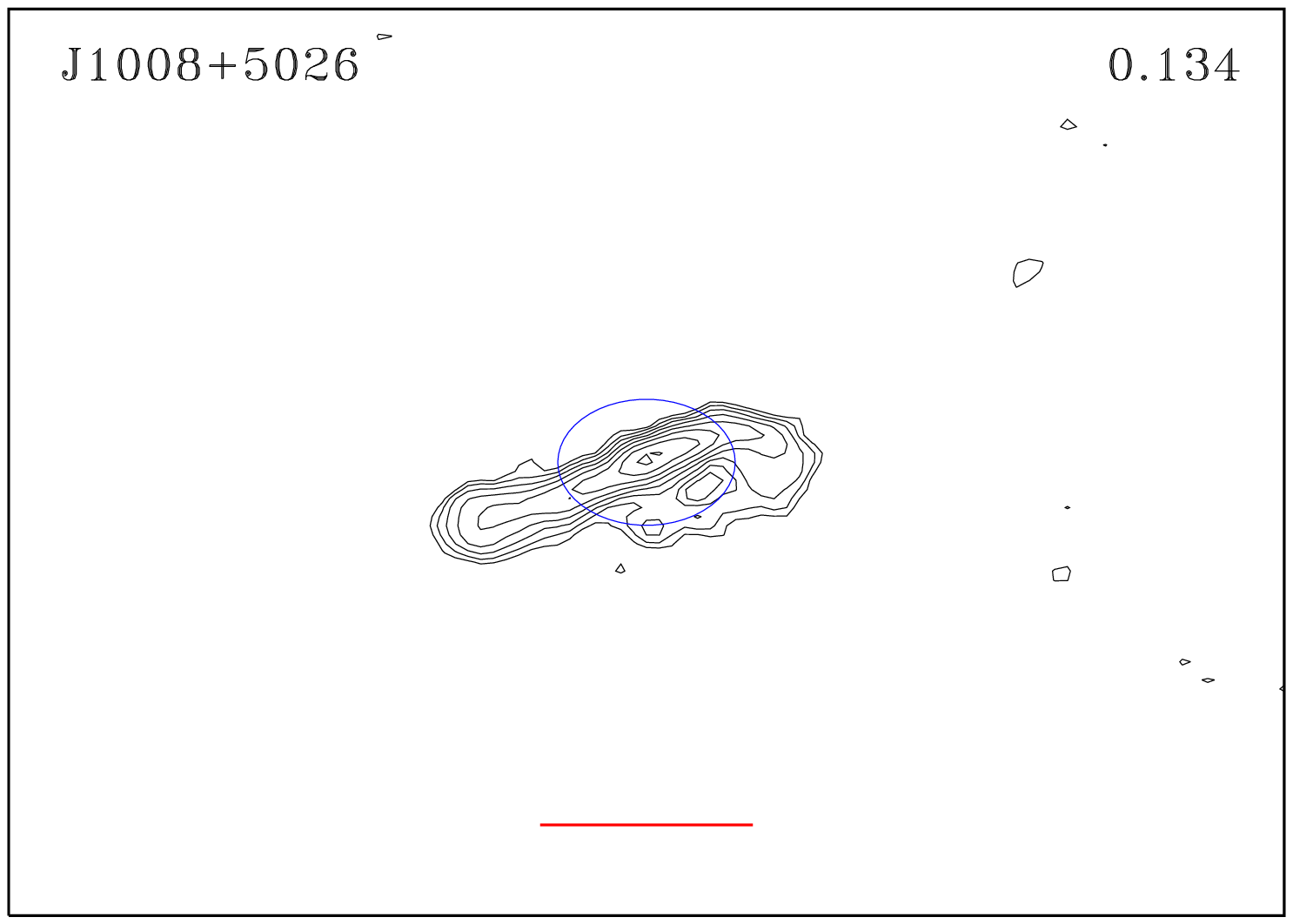} 
\includegraphics[width=6.3cm,height=6.3cm]{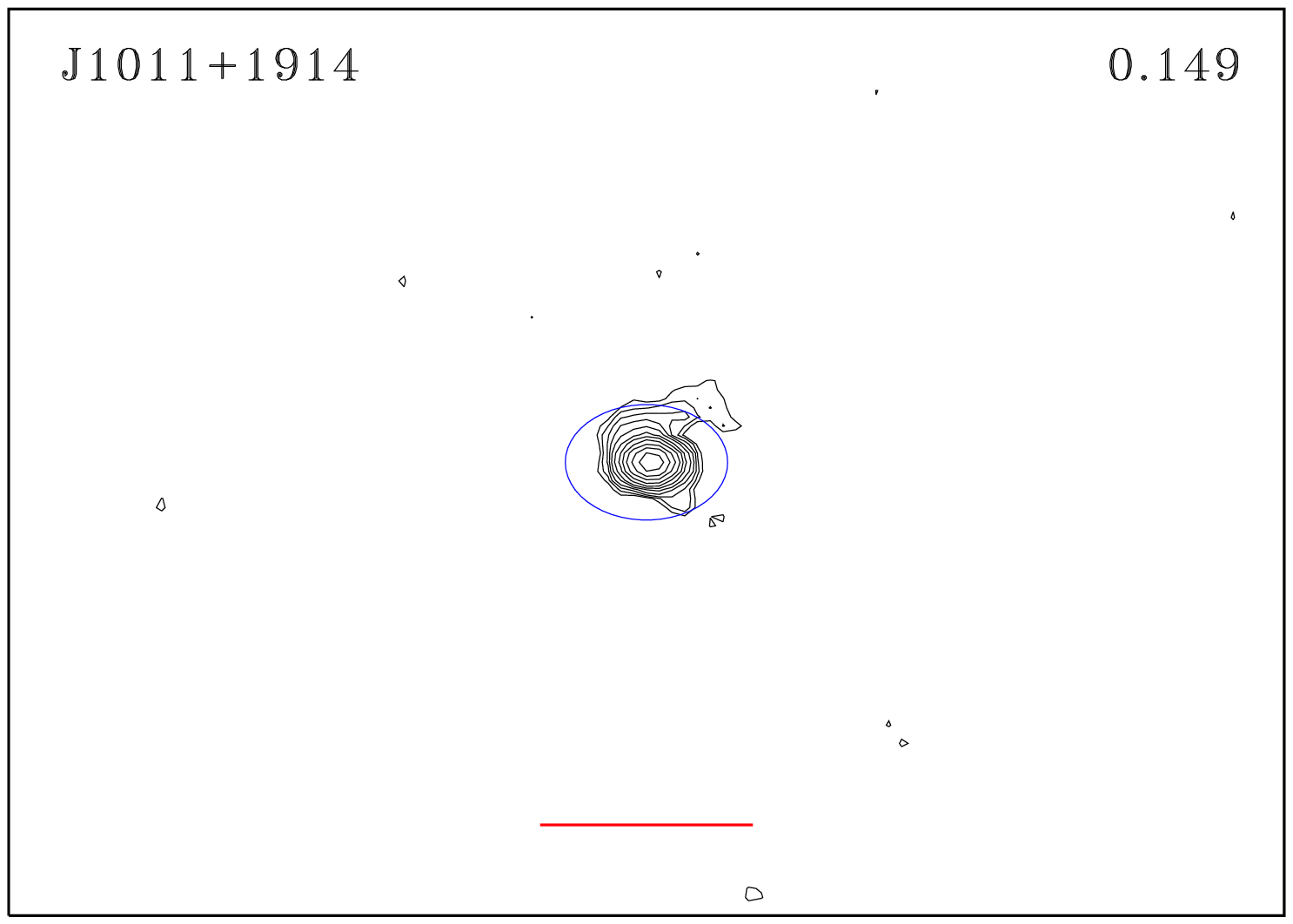} 

\includegraphics[width=6.3cm,height=6.3cm]{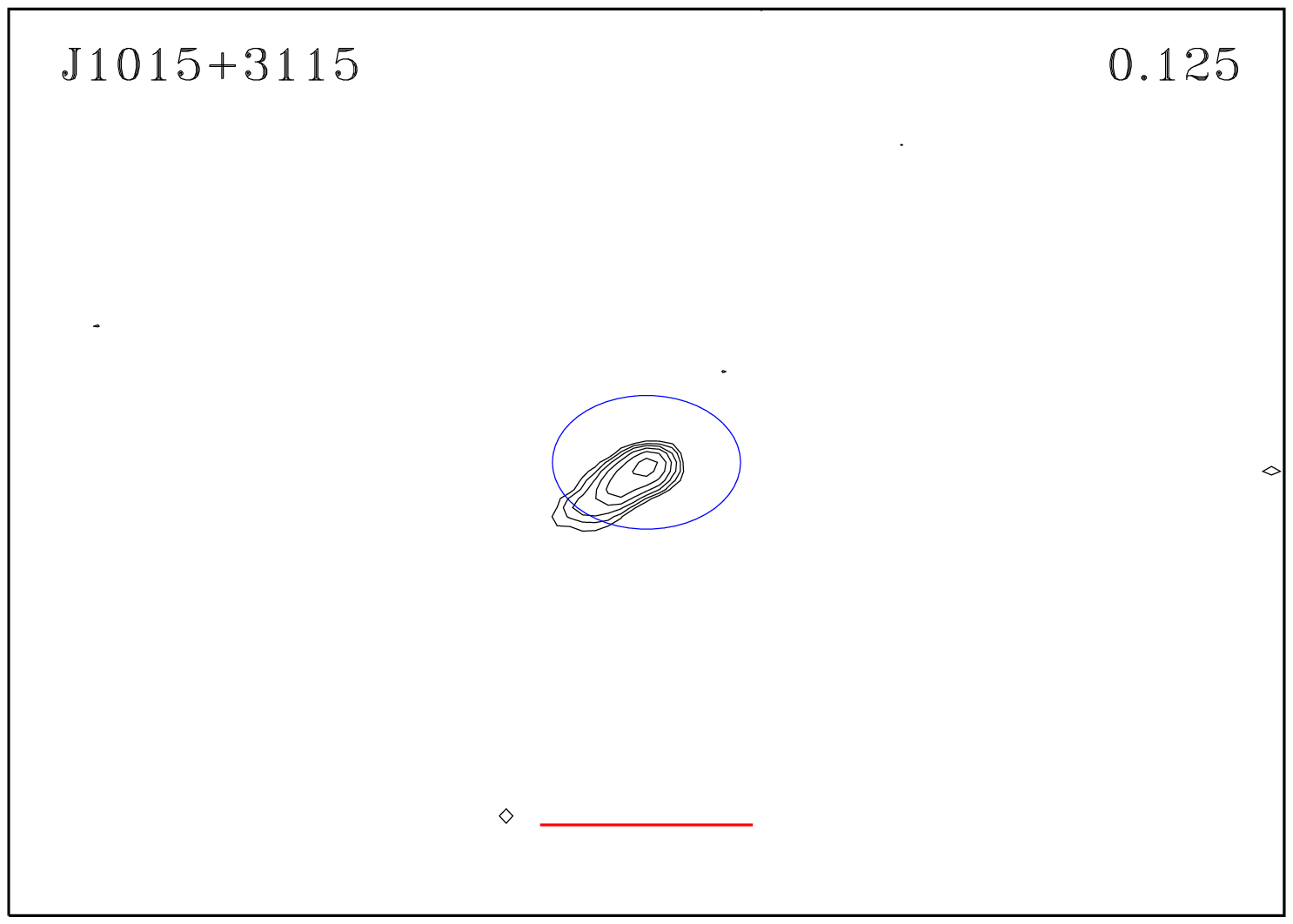} 
\includegraphics[width=6.3cm,height=6.3cm]{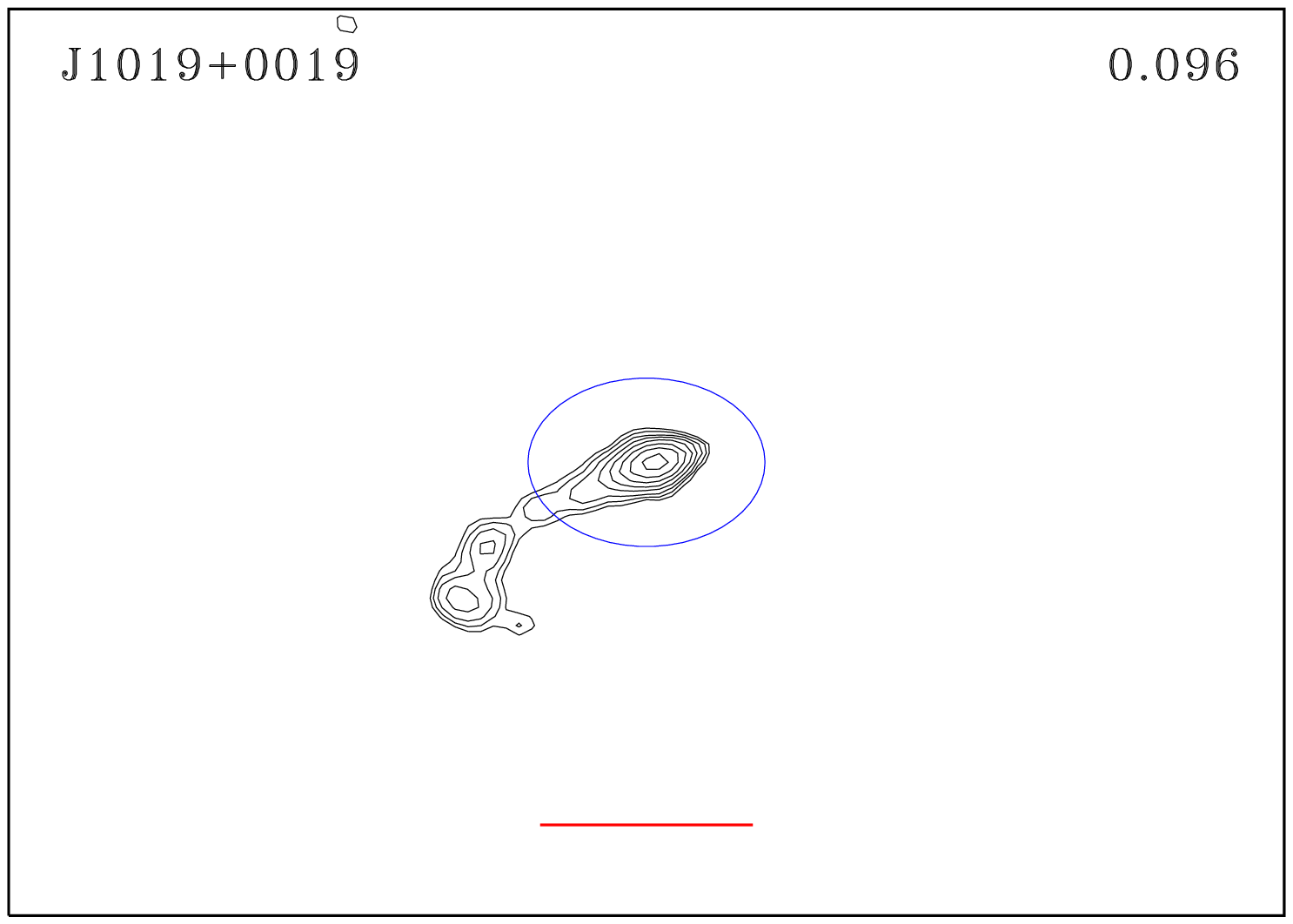} 
\includegraphics[width=6.3cm,height=6.3cm]{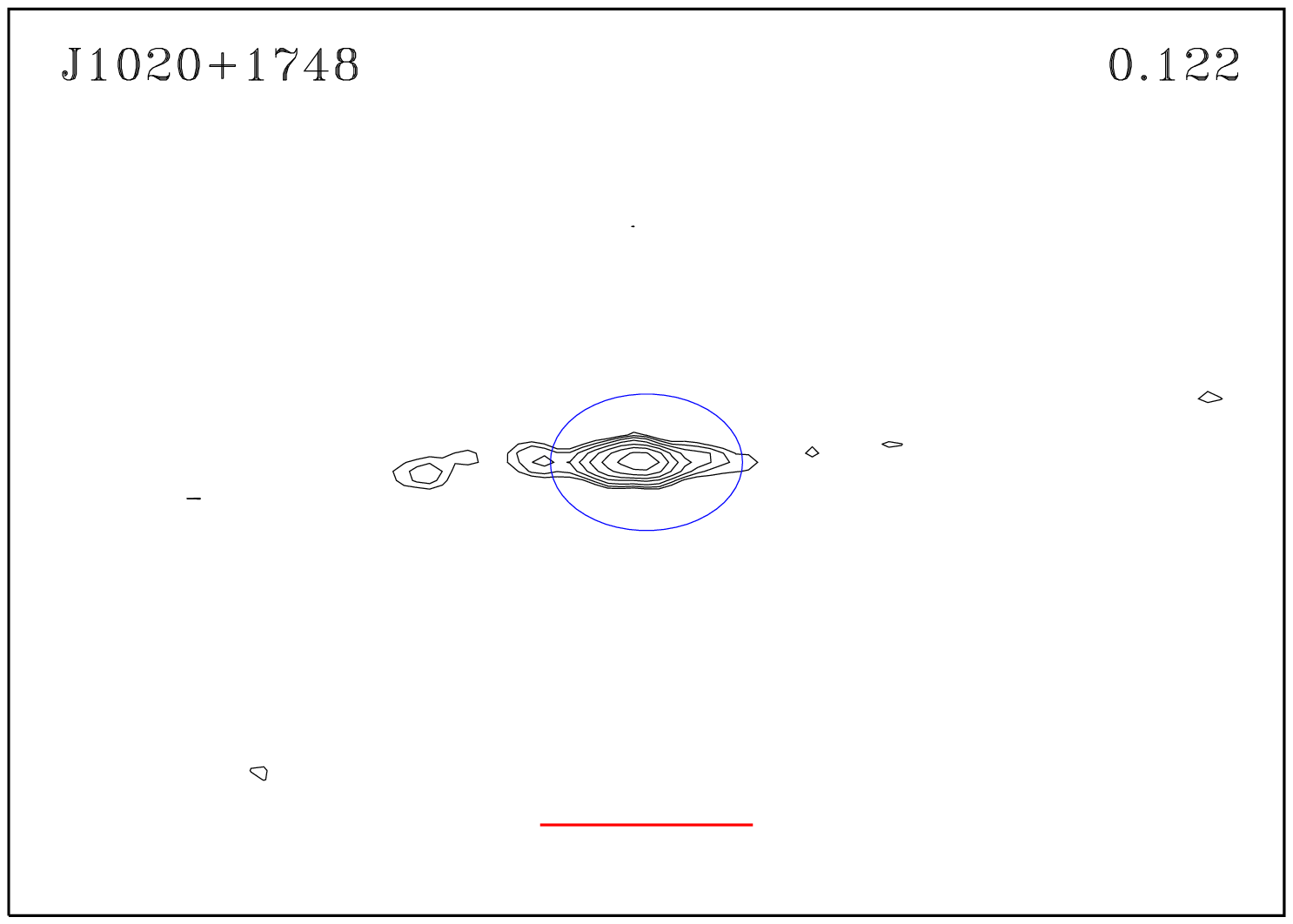} 
\caption{(continued)}
\end{figure*}

\addtocounter{figure}{-1}
\begin{figure*}
\includegraphics[width=6.3cm,height=6.3cm]{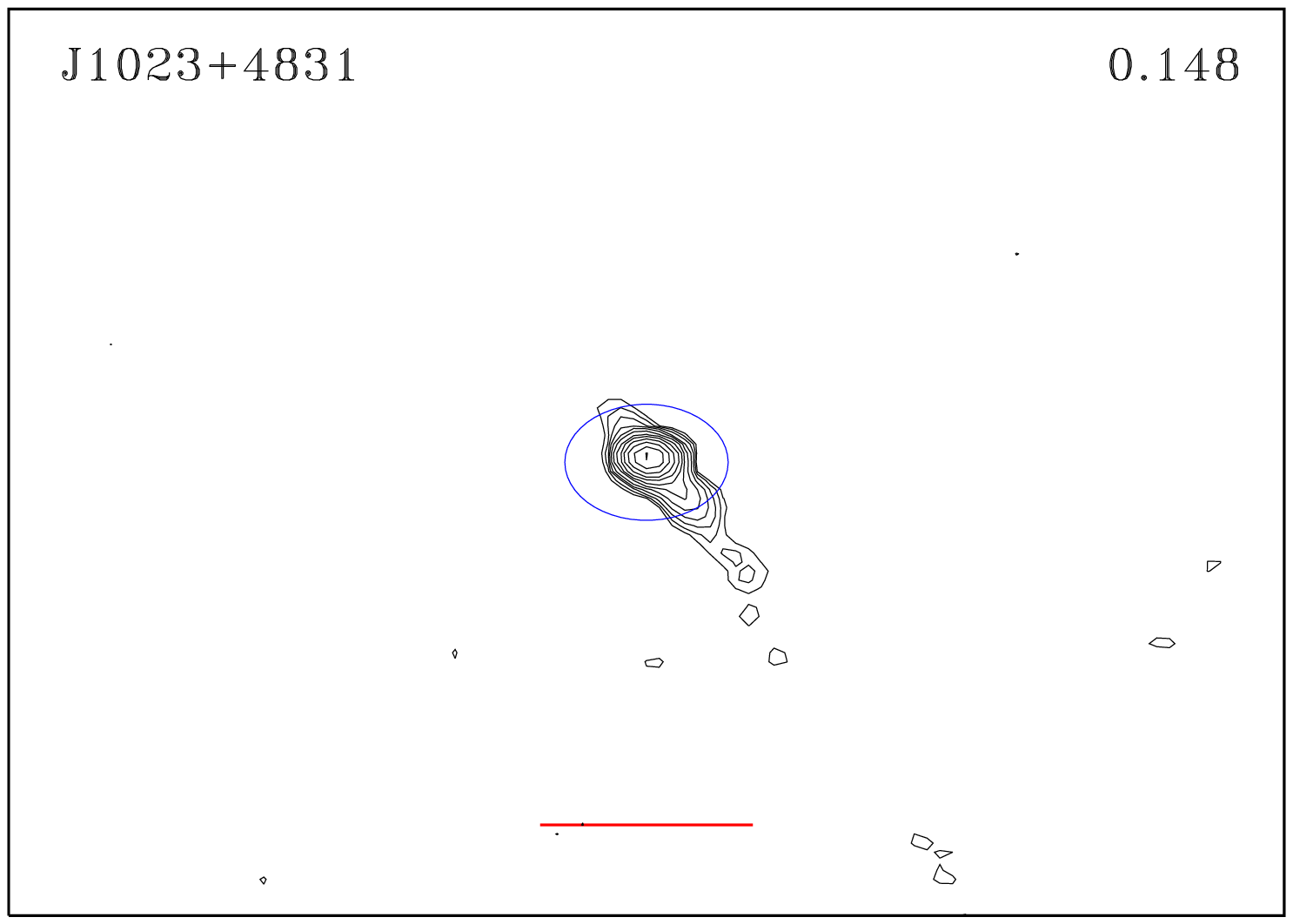} 
\includegraphics[width=6.3cm,height=6.3cm]{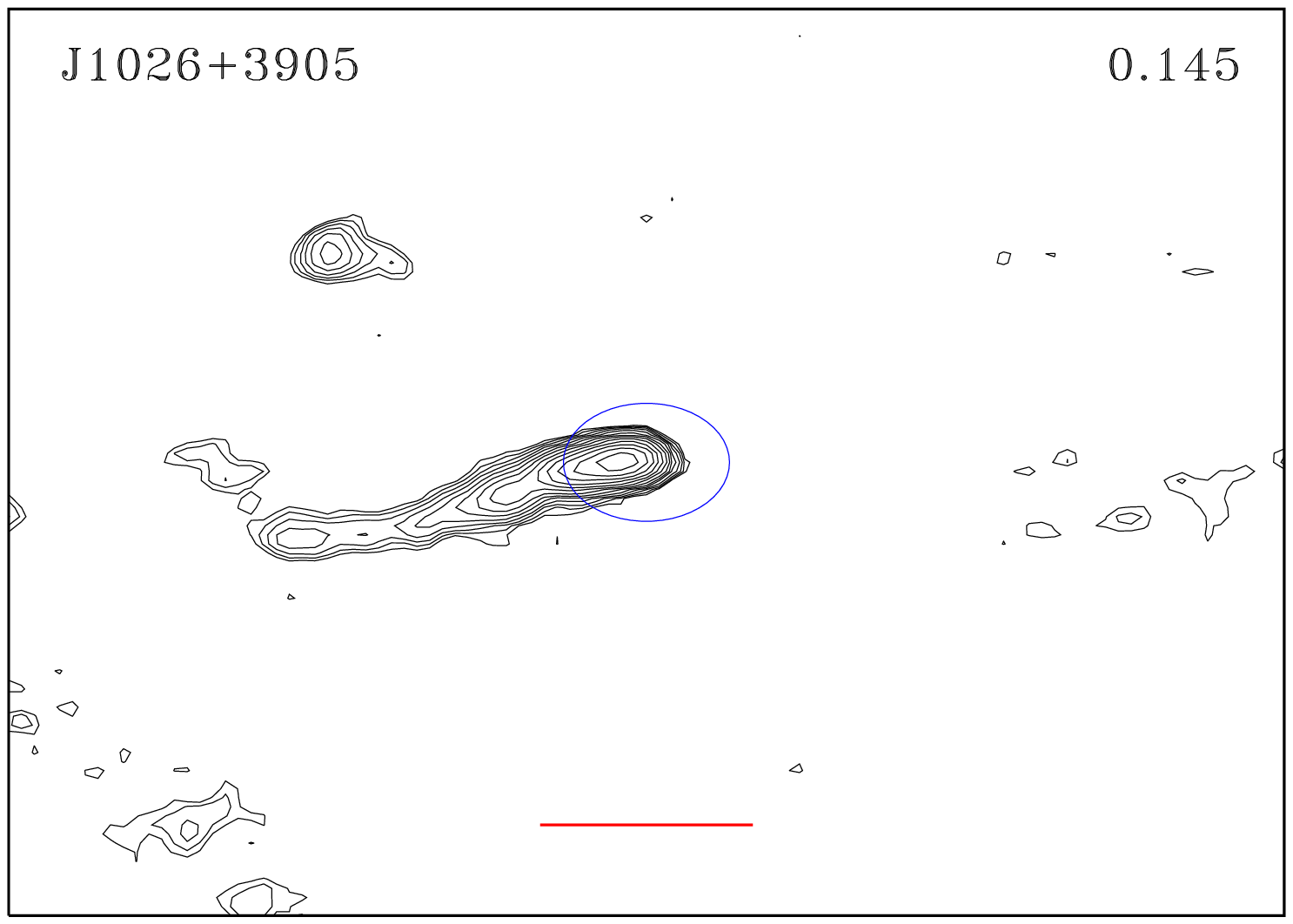} 
\includegraphics[width=6.3cm,height=6.3cm]{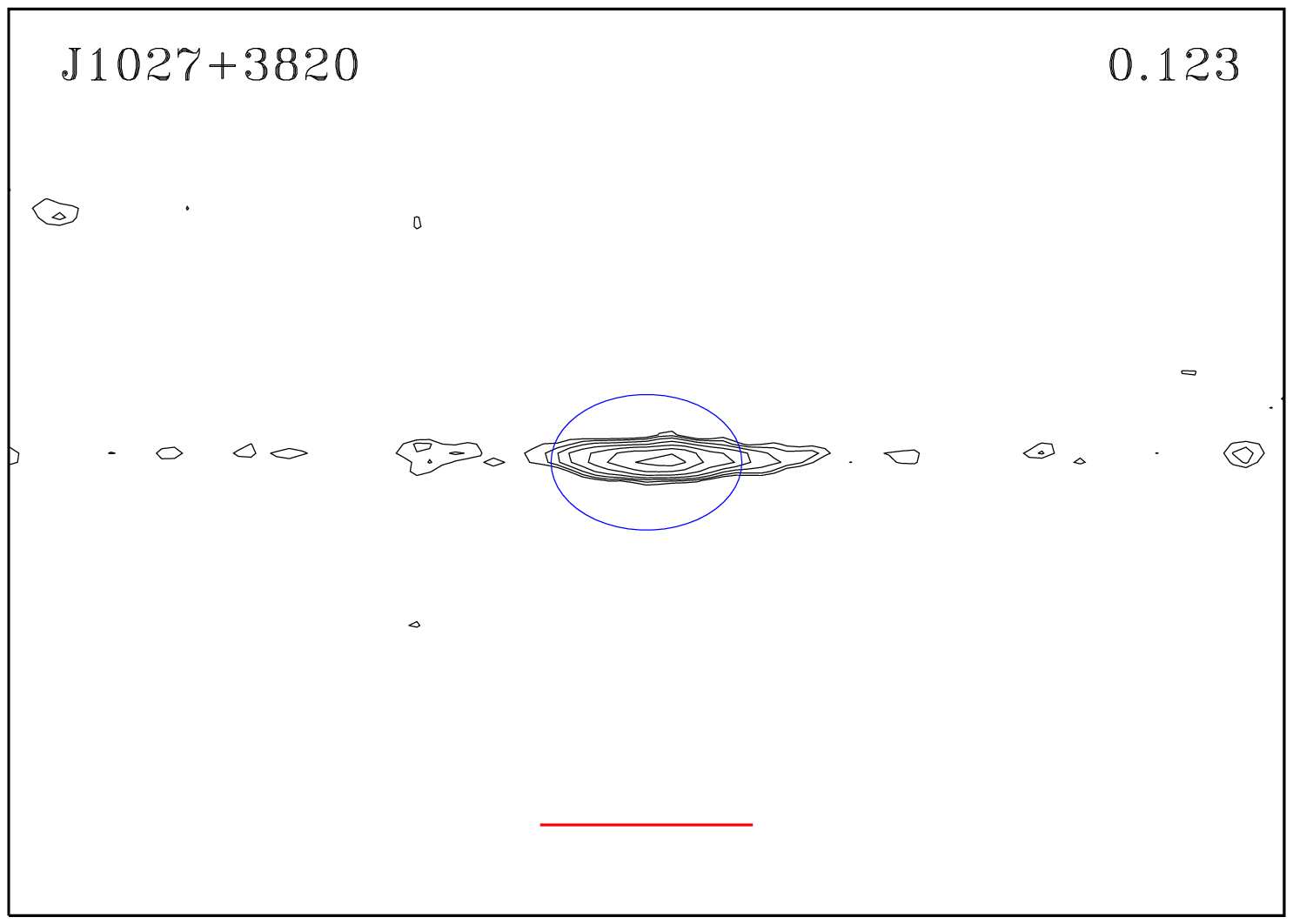} 

\includegraphics[width=6.3cm,height=6.3cm]{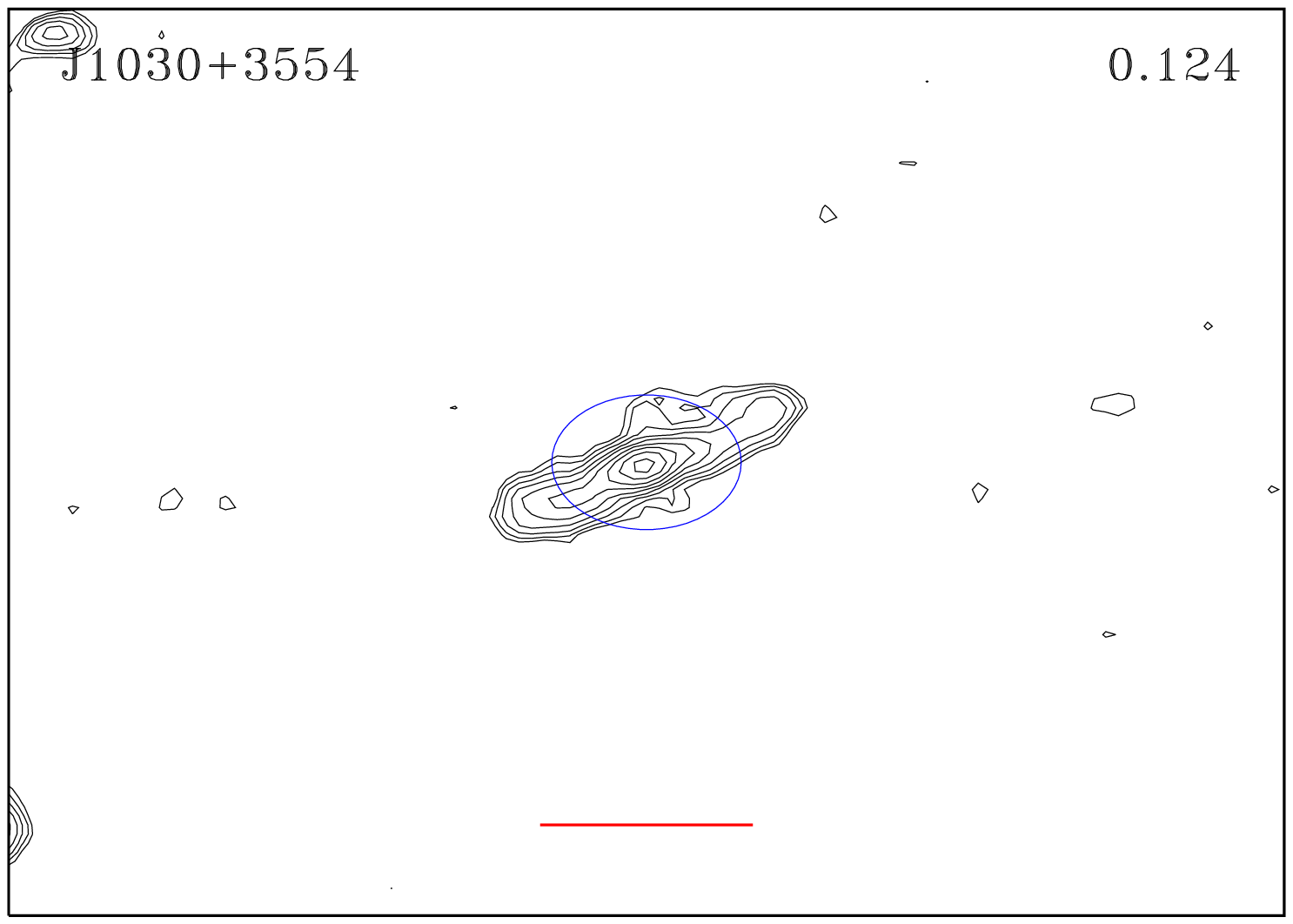} 
\includegraphics[width=6.3cm,height=6.3cm]{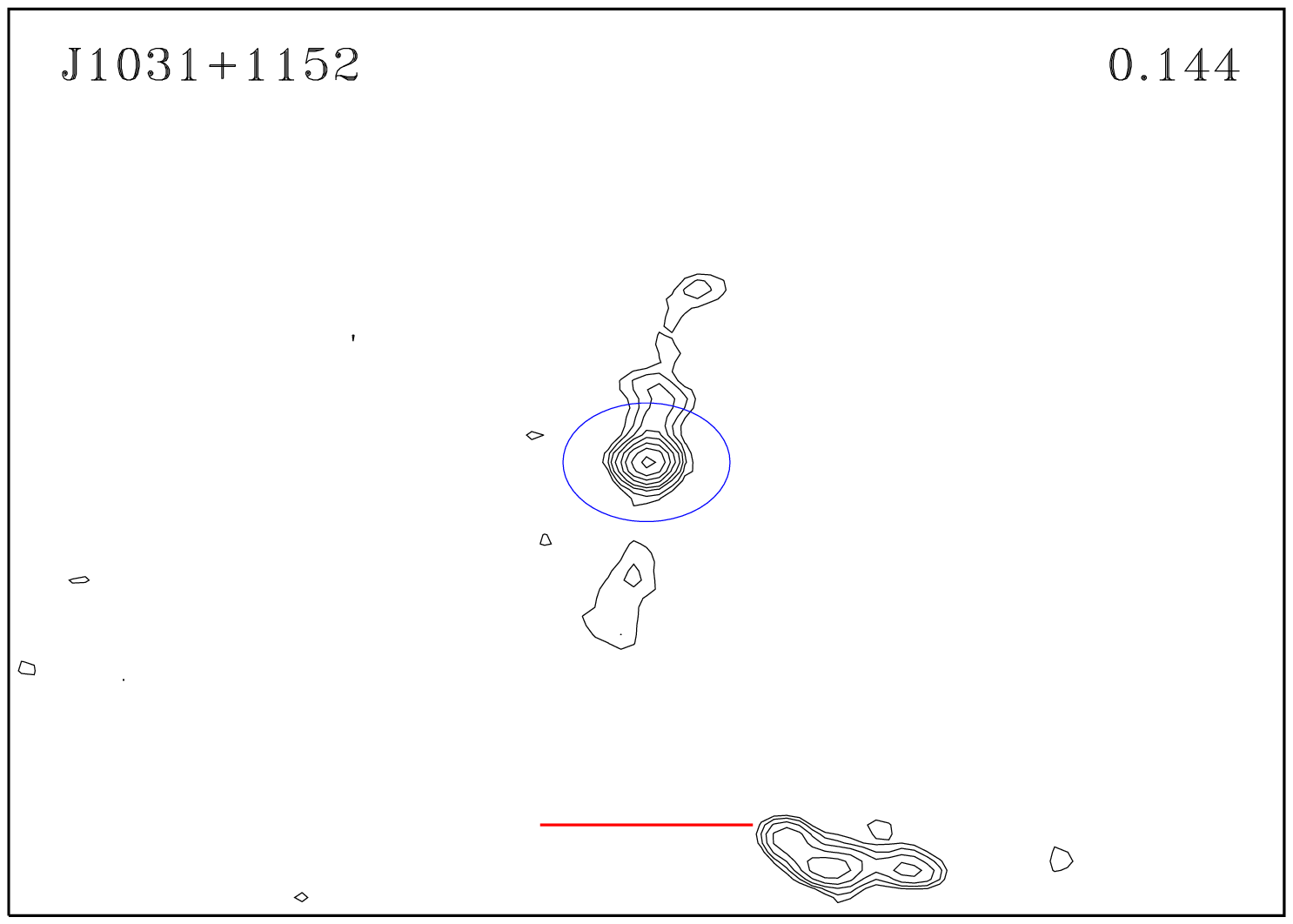} 
\includegraphics[width=6.3cm,height=6.3cm]{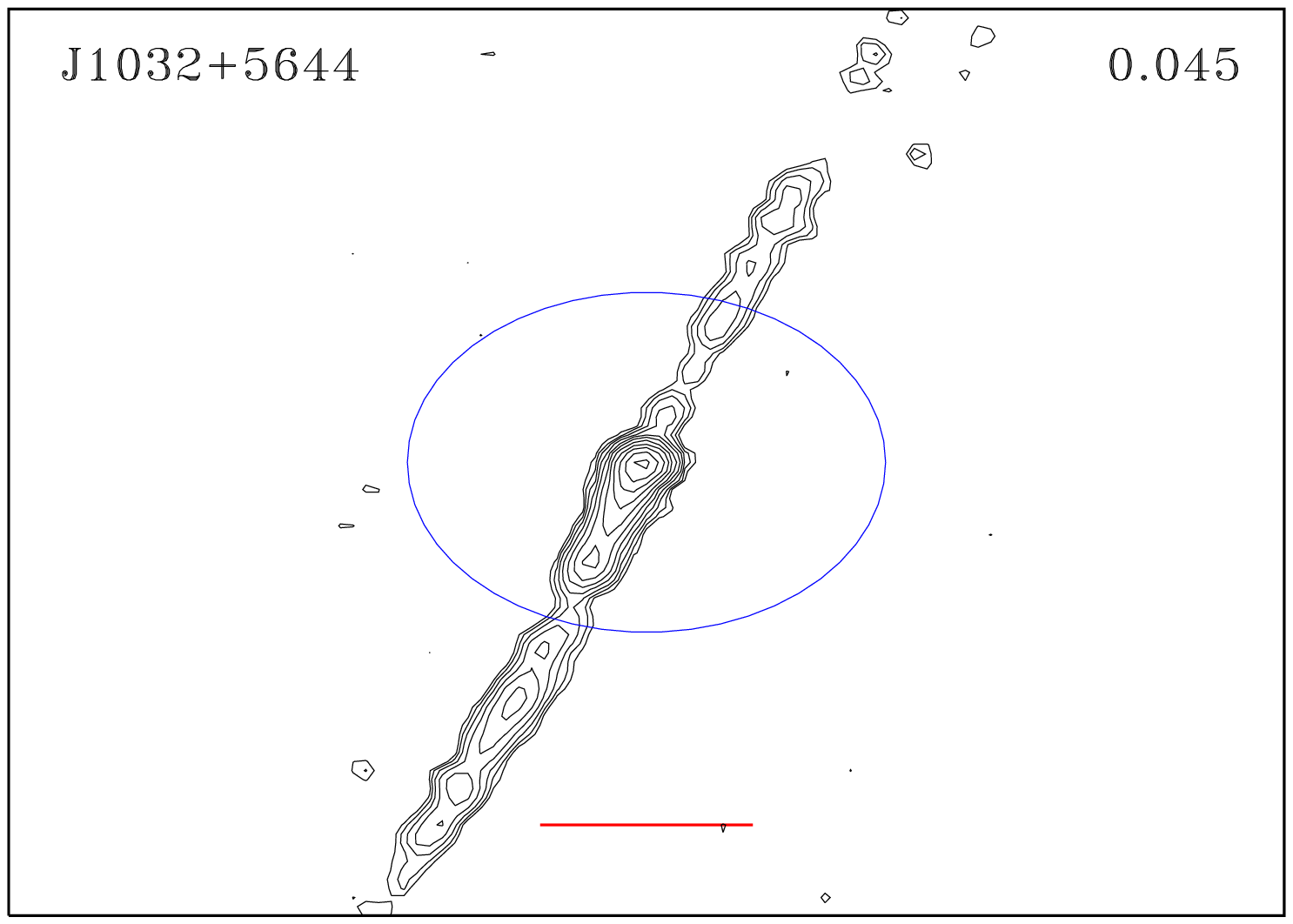} 

\includegraphics[width=6.3cm,height=6.3cm]{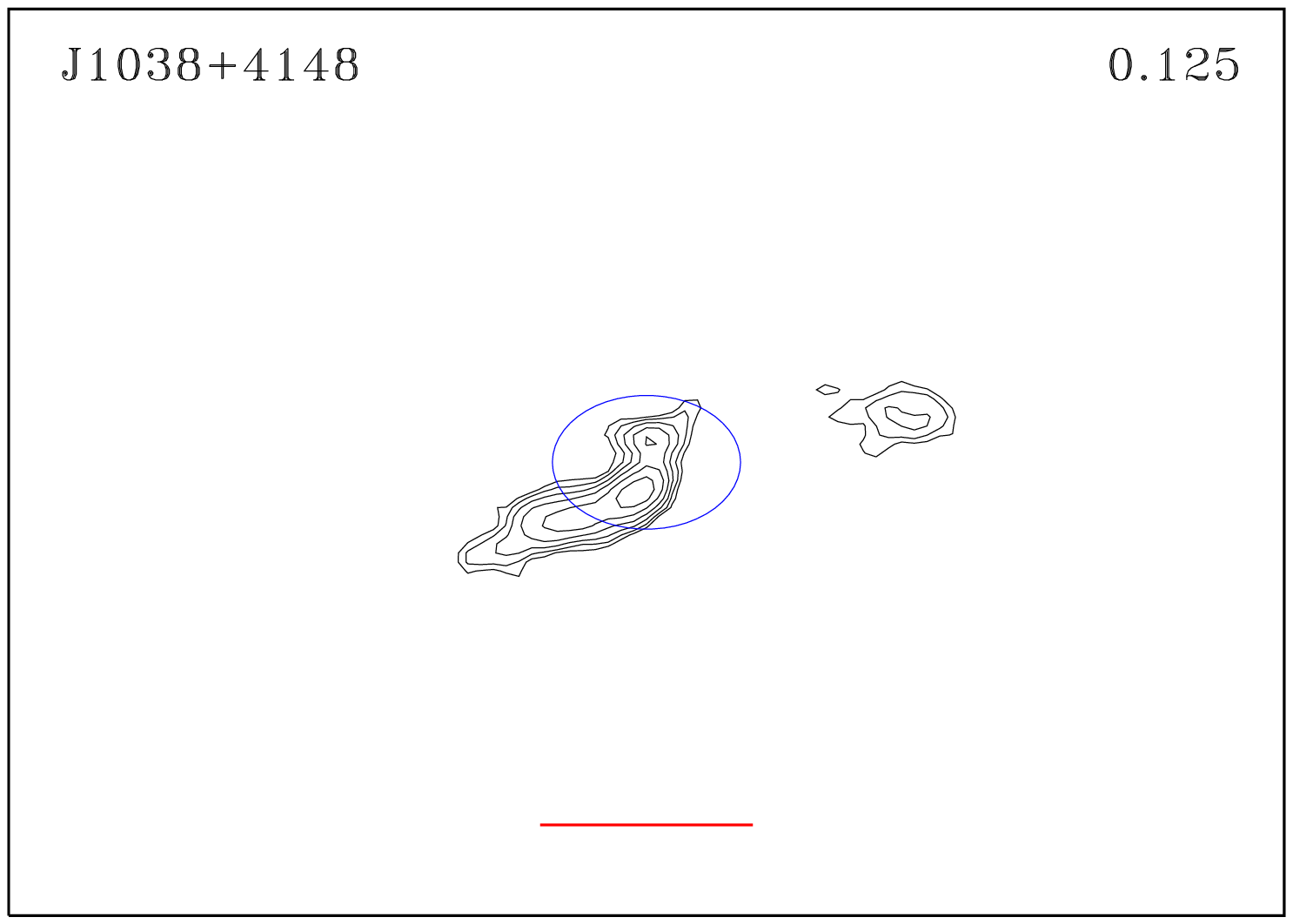} 
\includegraphics[width=6.3cm,height=6.3cm]{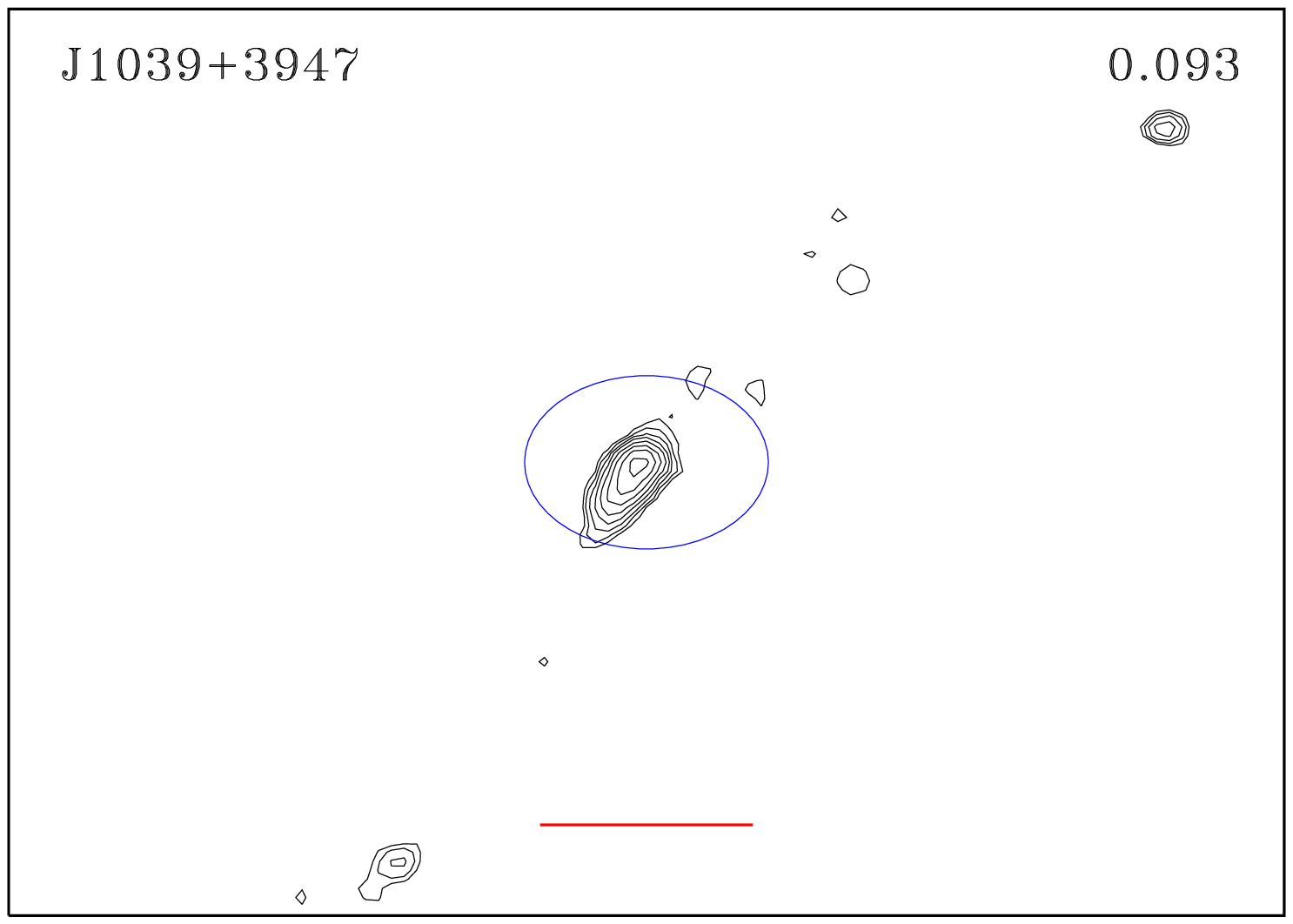} 
\includegraphics[width=6.3cm,height=6.3cm]{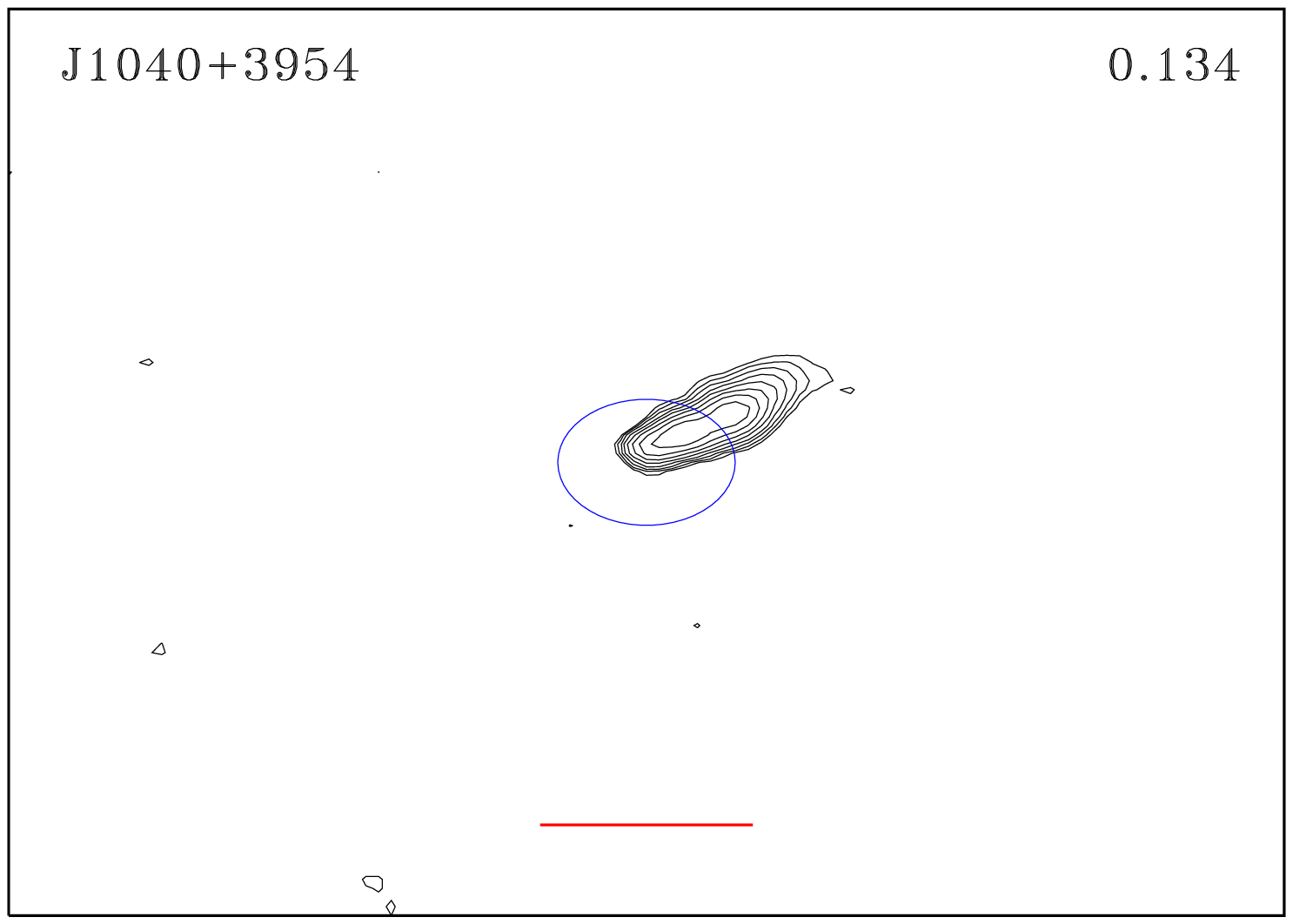} 

\includegraphics[width=6.3cm,height=6.3cm]{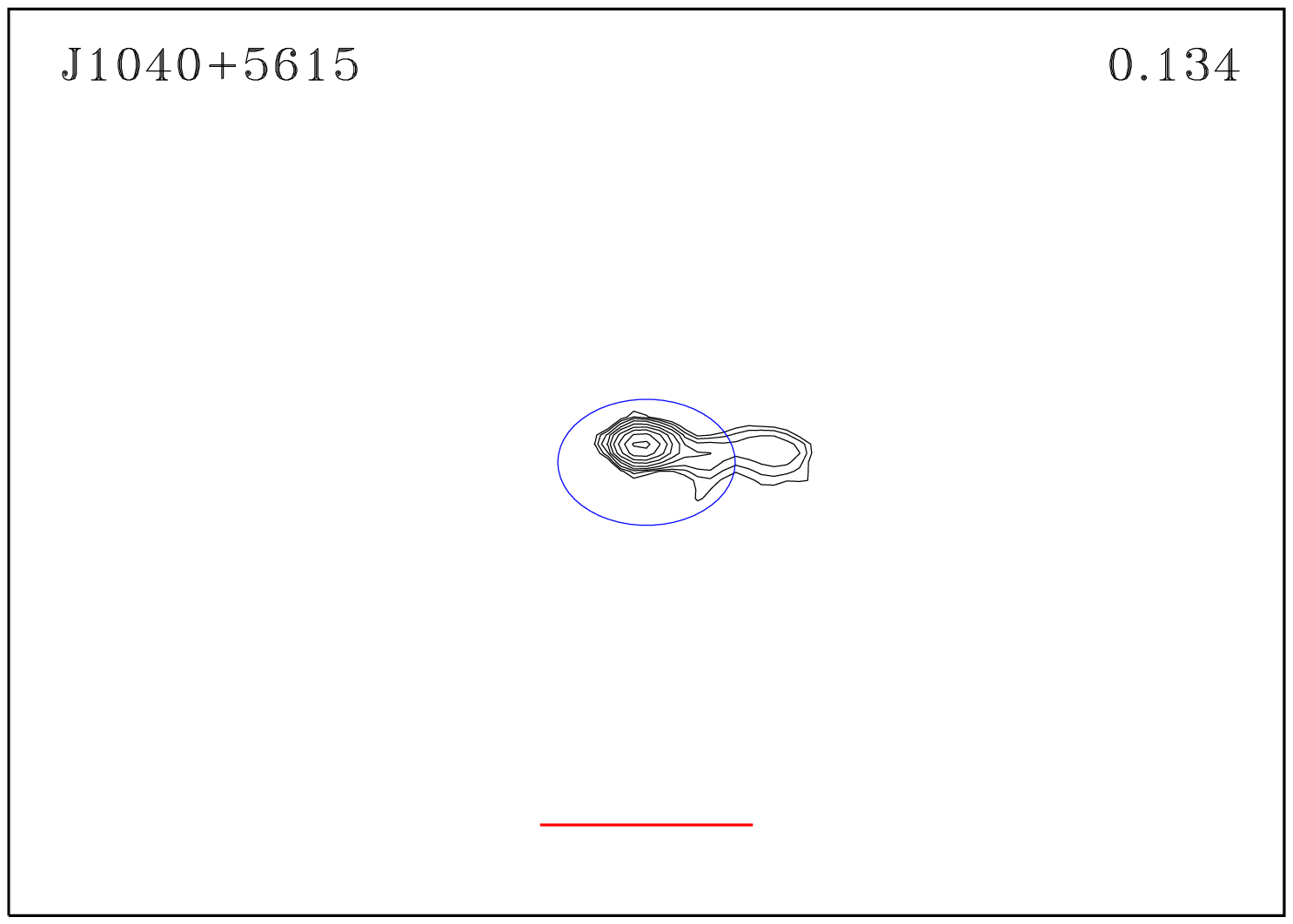} 
\includegraphics[width=6.3cm,height=6.3cm]{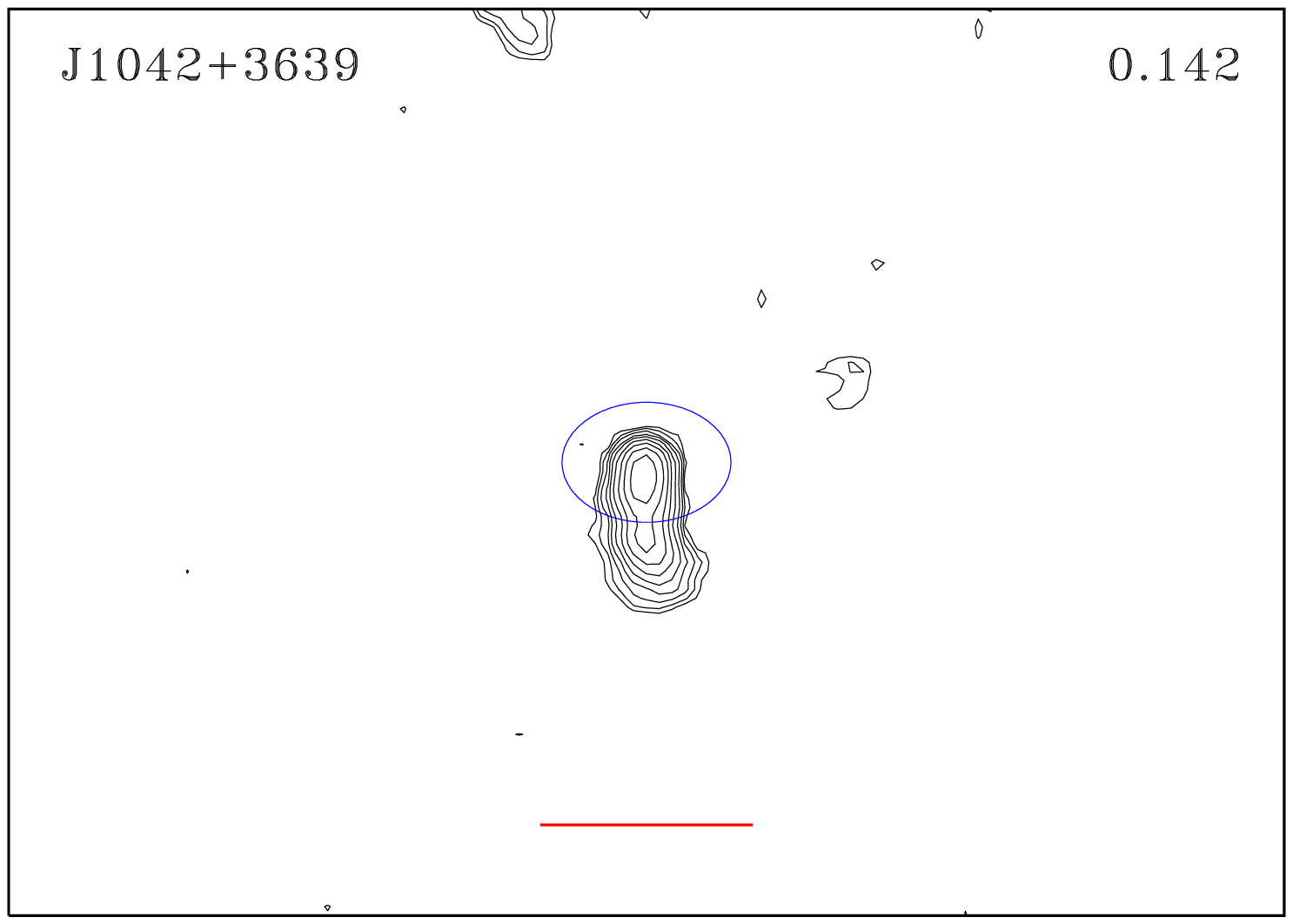} 
\includegraphics[width=6.3cm,height=6.3cm]{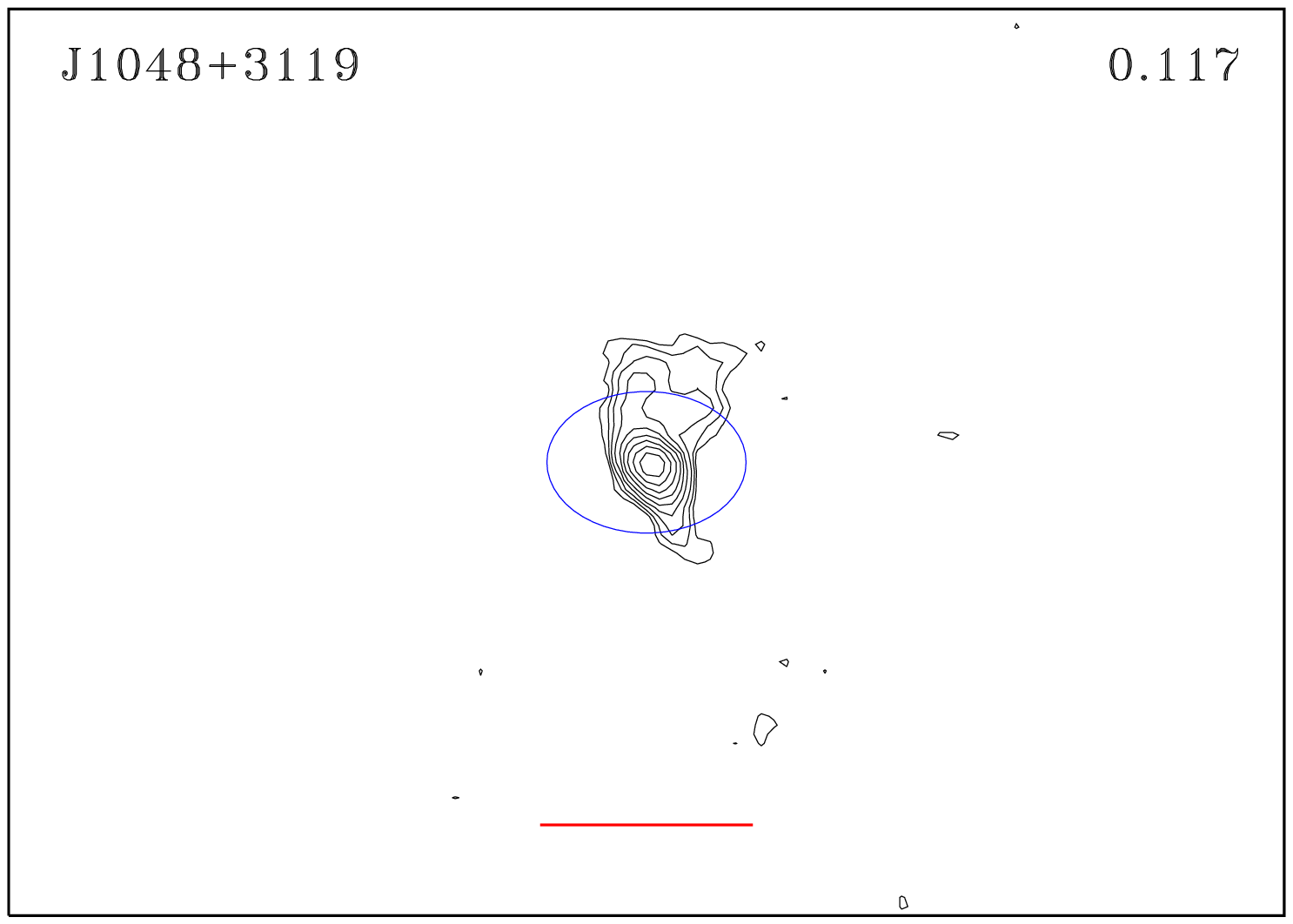} 
\caption{(continued)}
\end{figure*}

\clearpage
\addtocounter{figure}{-1}
\begin{figure*}
\includegraphics[width=6.3cm,height=6.3cm]{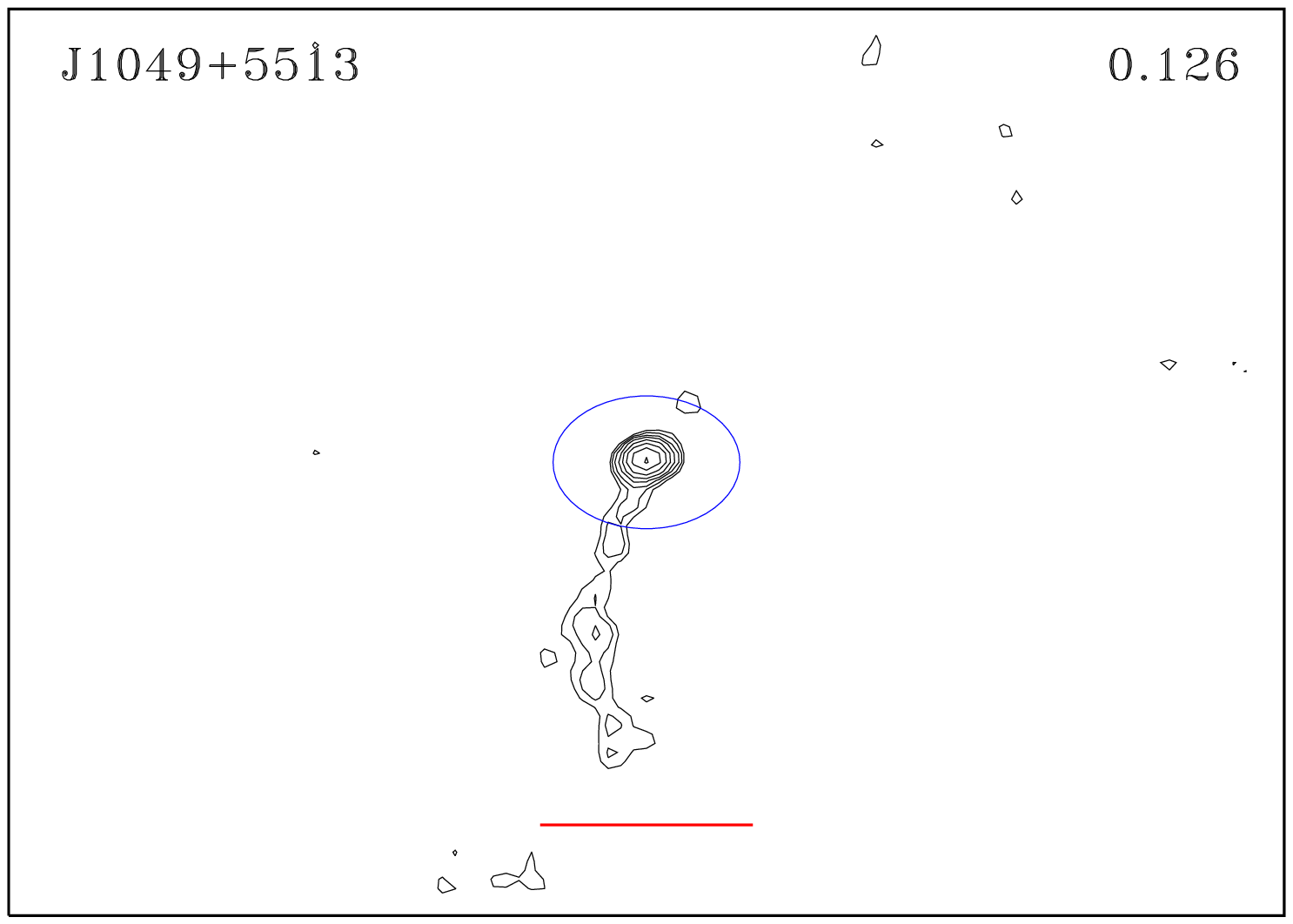} 
\includegraphics[width=6.3cm,height=6.3cm]{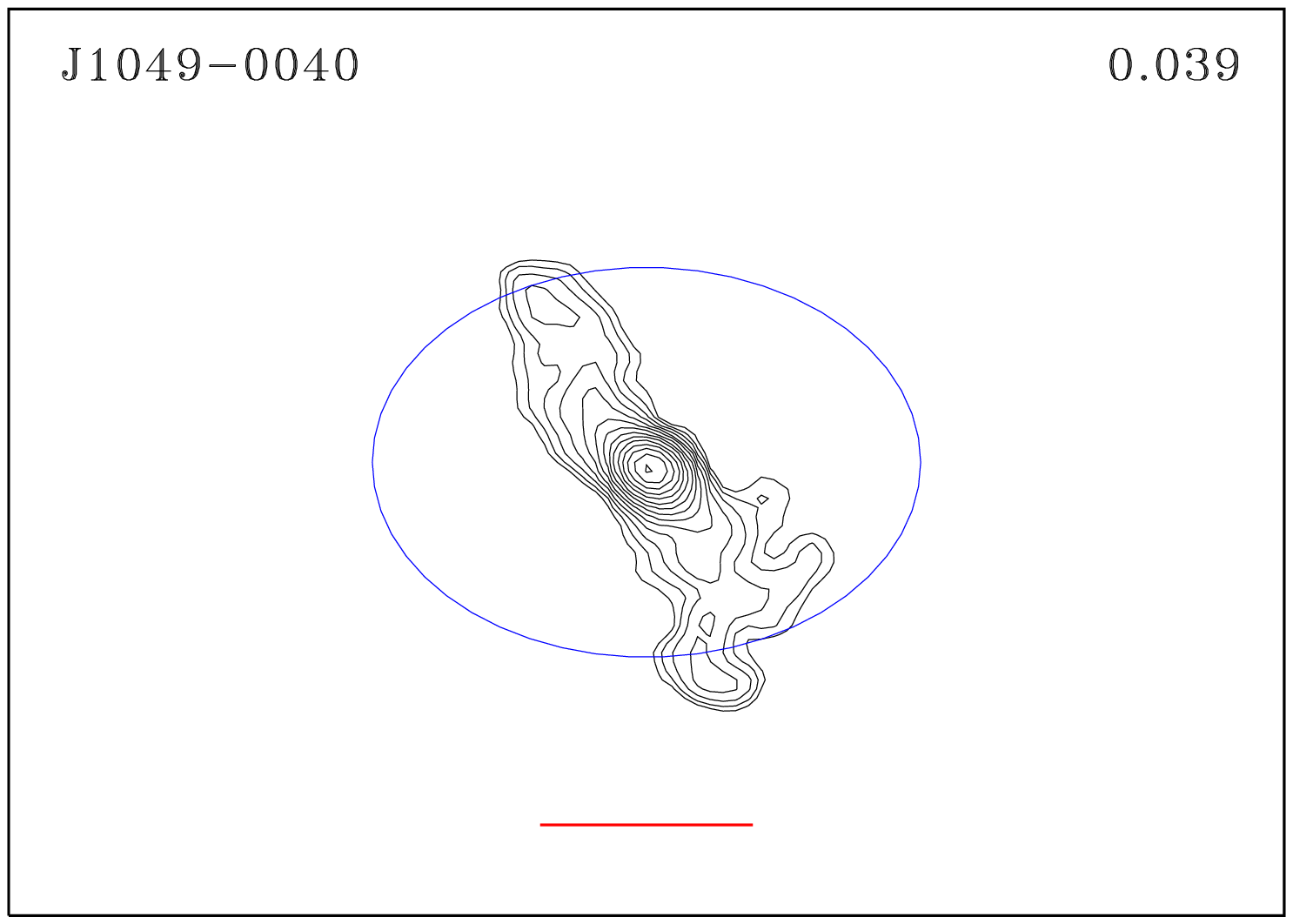} 
\includegraphics[width=6.3cm,height=6.3cm]{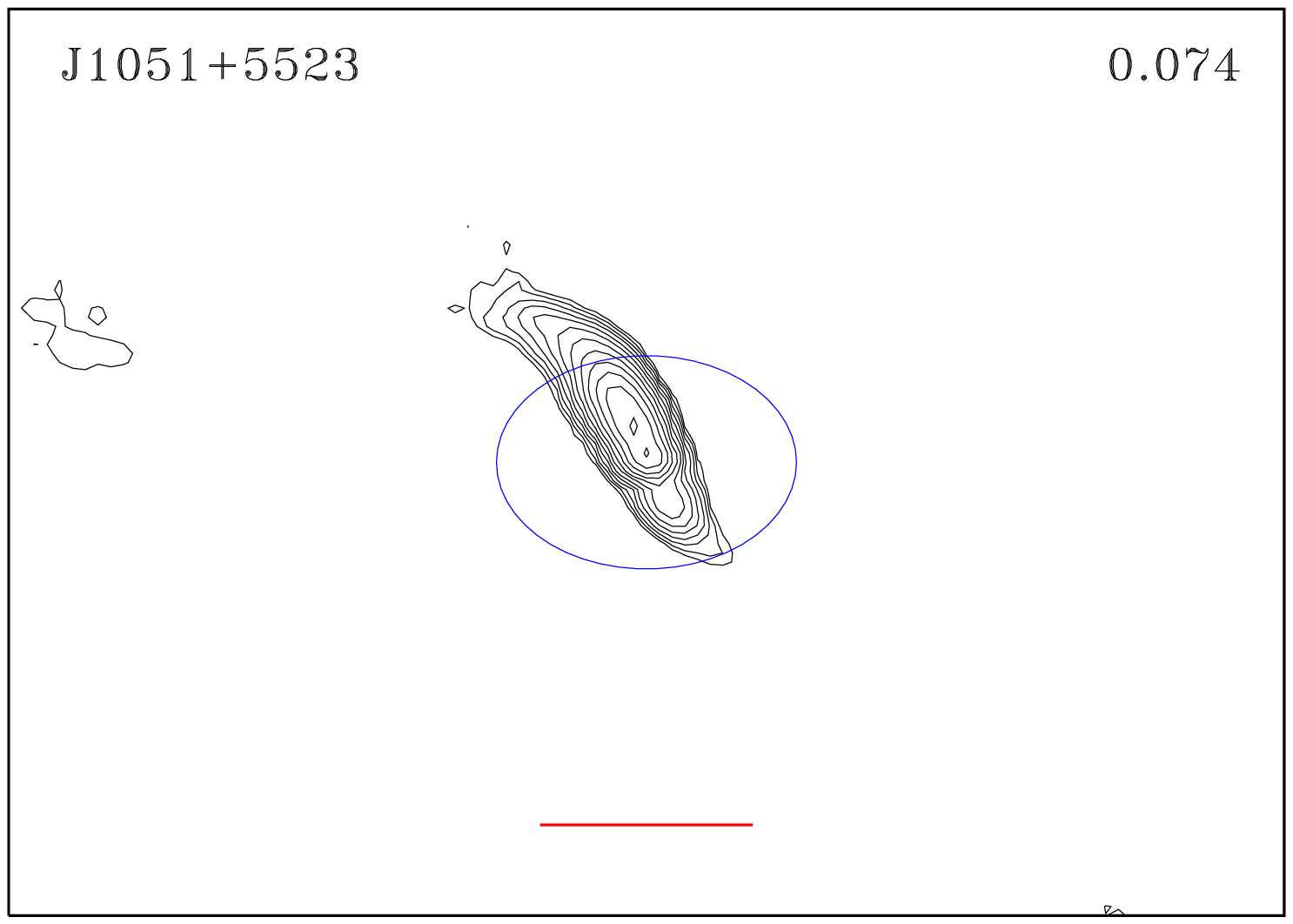} 

\includegraphics[width=6.3cm,height=6.3cm]{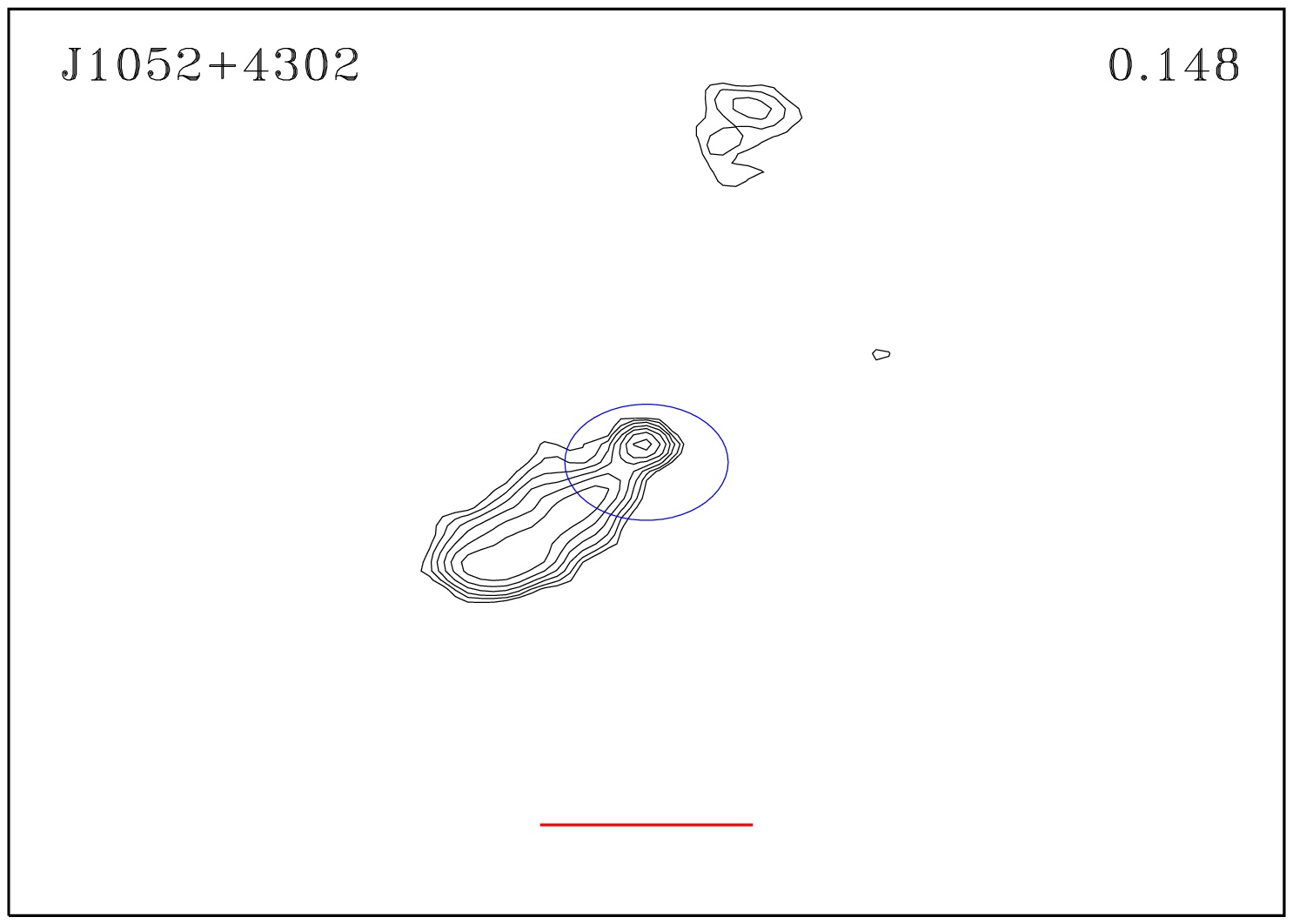} 
\includegraphics[width=6.3cm,height=6.3cm]{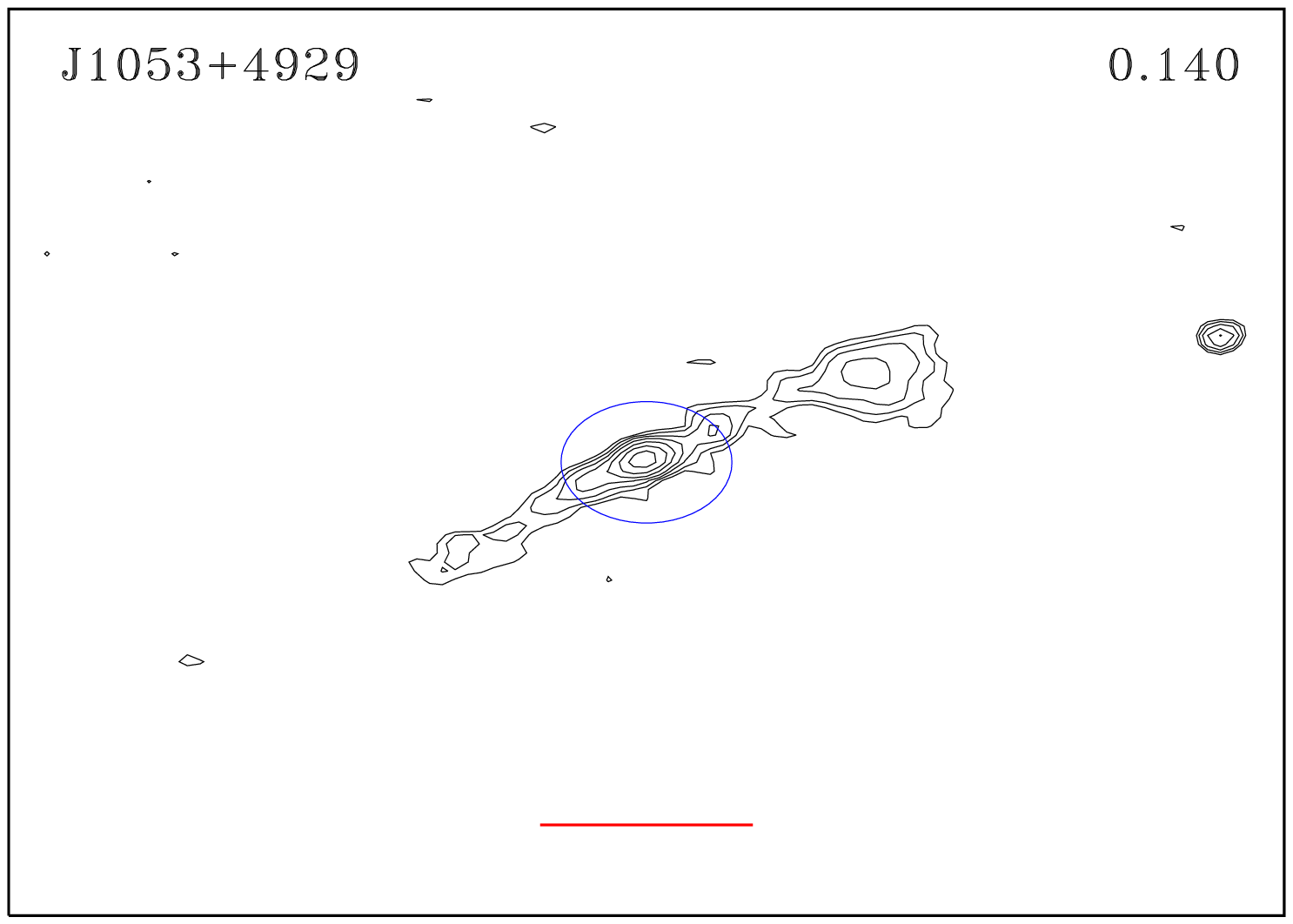} 
\includegraphics[width=6.3cm,height=6.3cm]{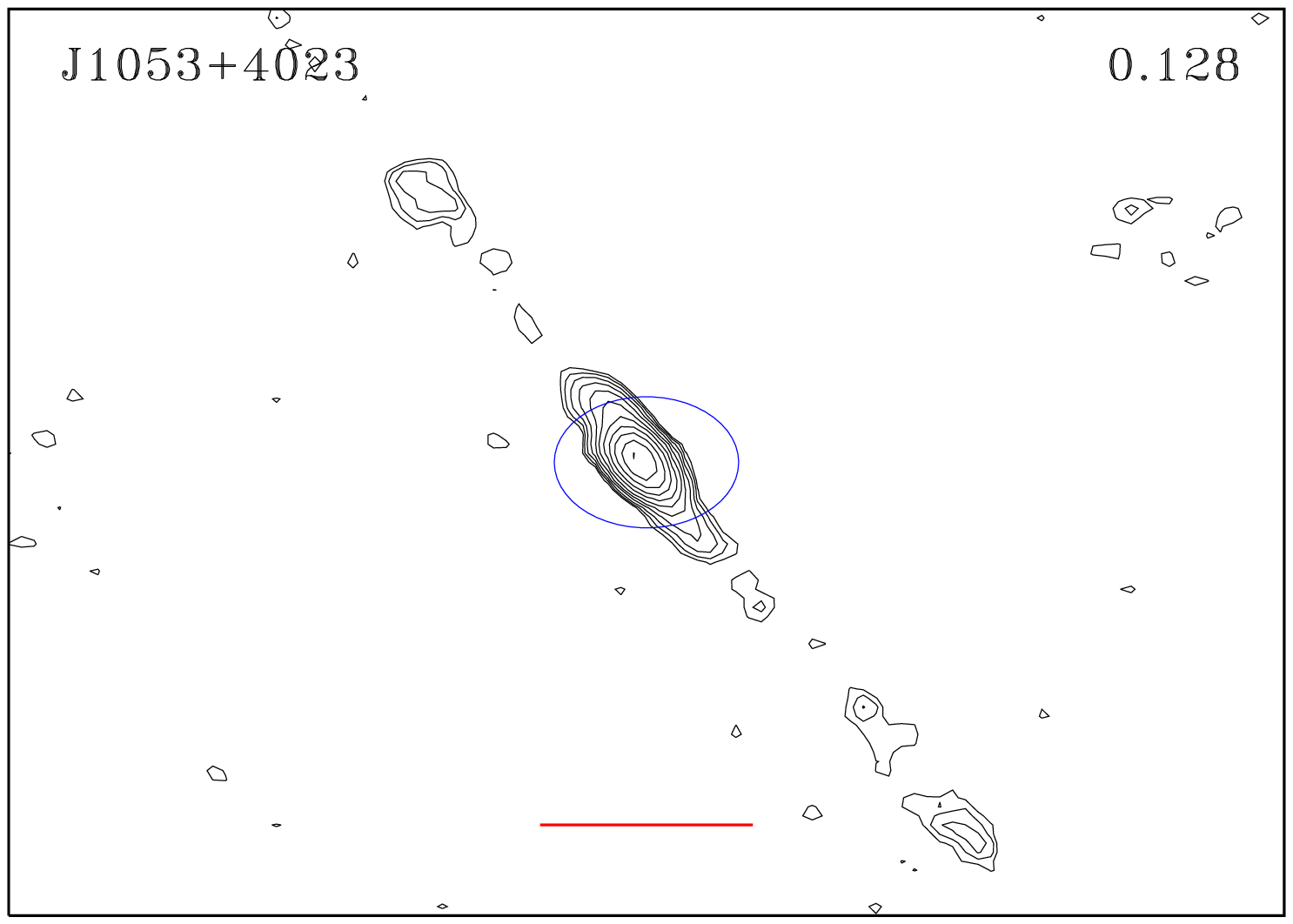} 

\includegraphics[width=6.3cm,height=6.3cm]{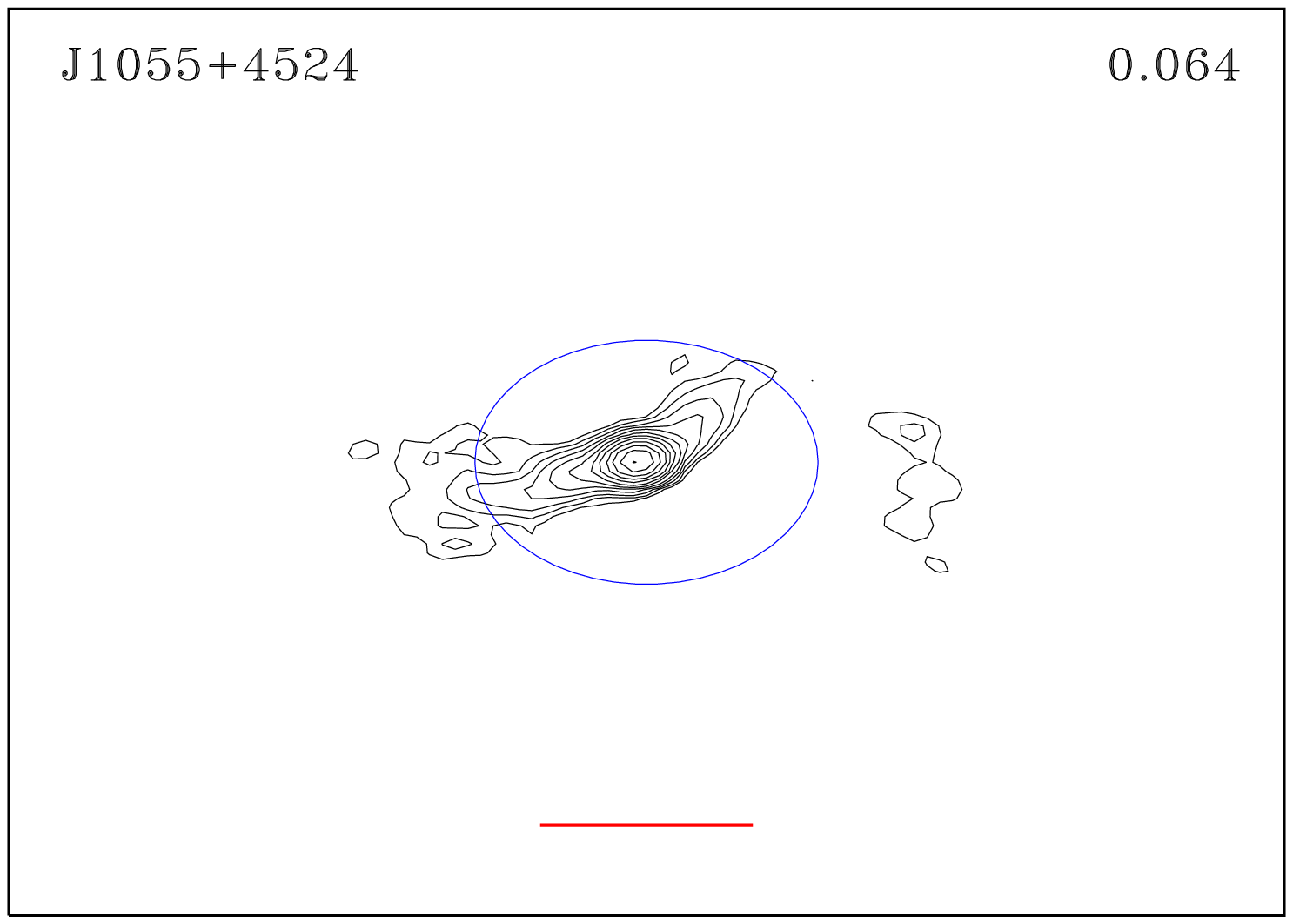} 
\includegraphics[width=6.3cm,height=6.3cm]{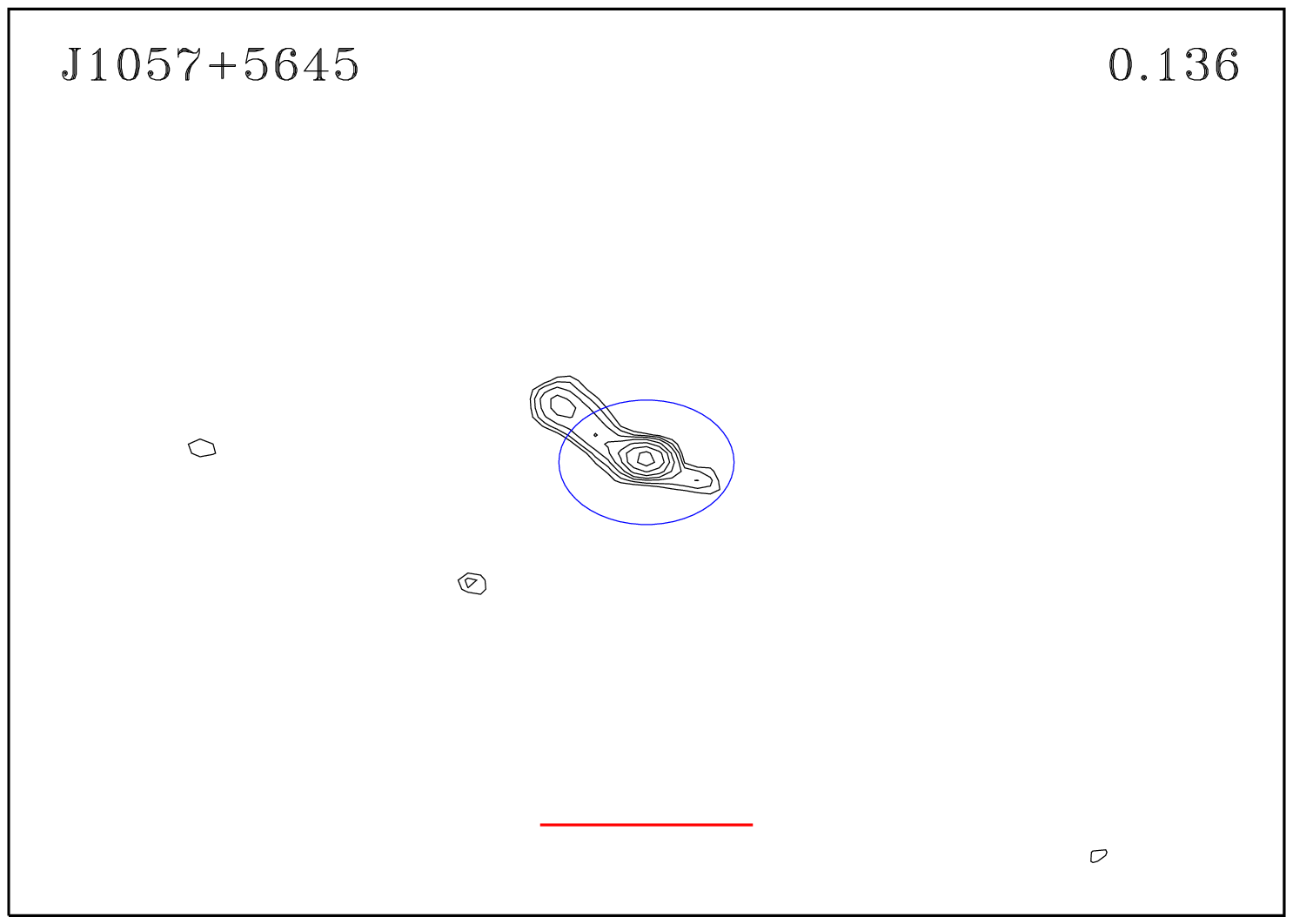} 
\includegraphics[width=6.3cm,height=6.3cm]{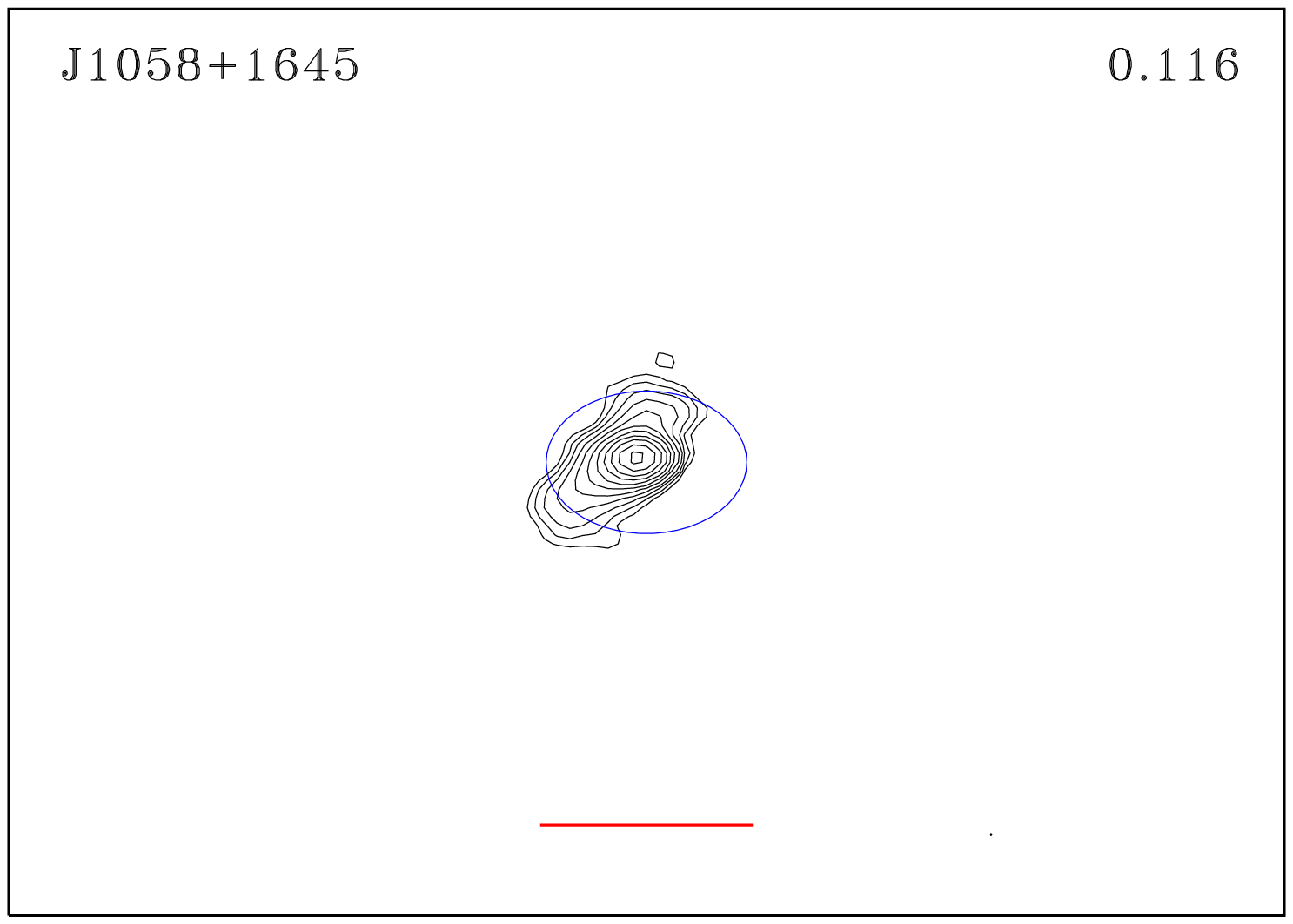} 

\includegraphics[width=6.3cm,height=6.3cm]{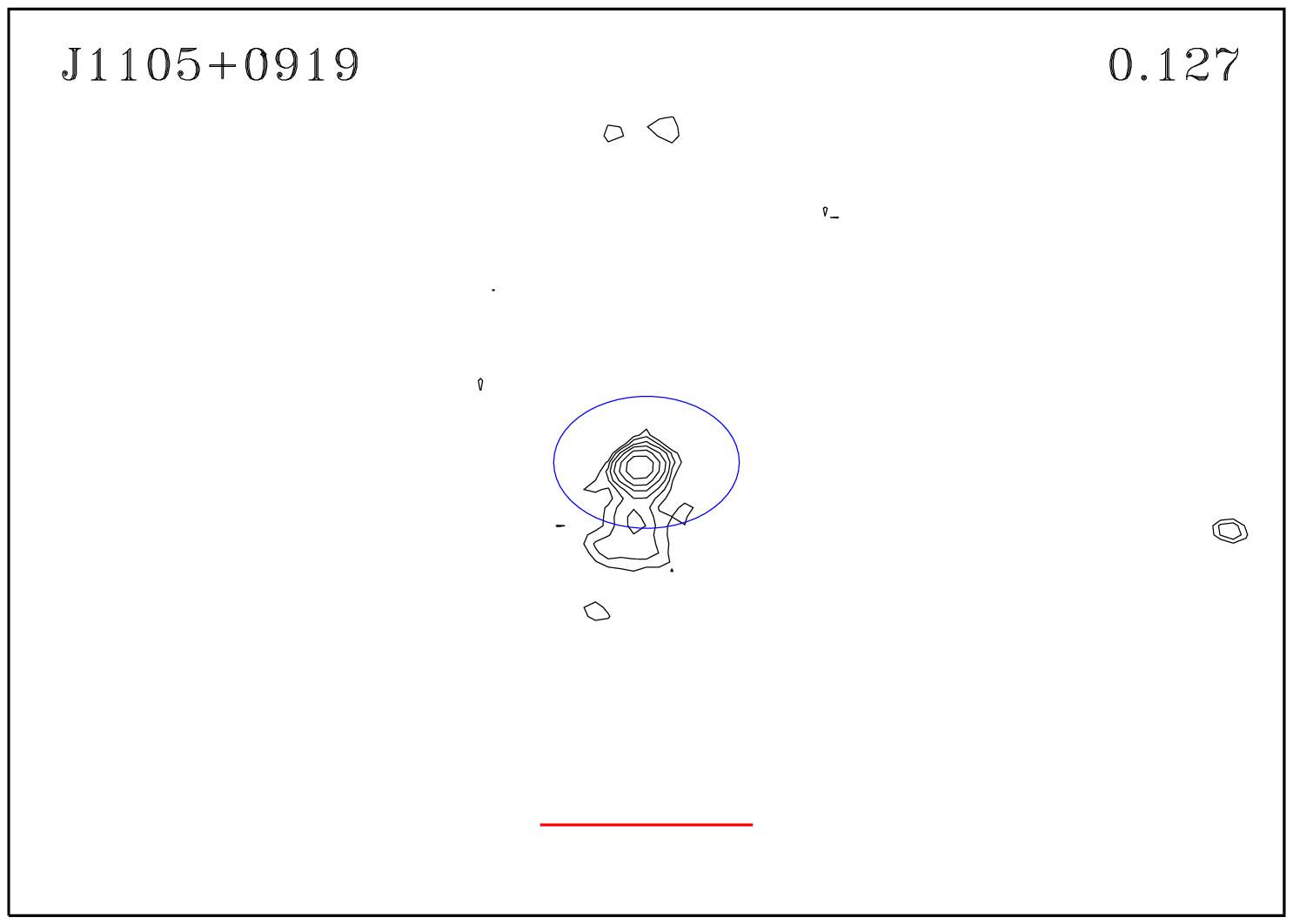} 
\includegraphics[width=6.3cm,height=6.3cm]{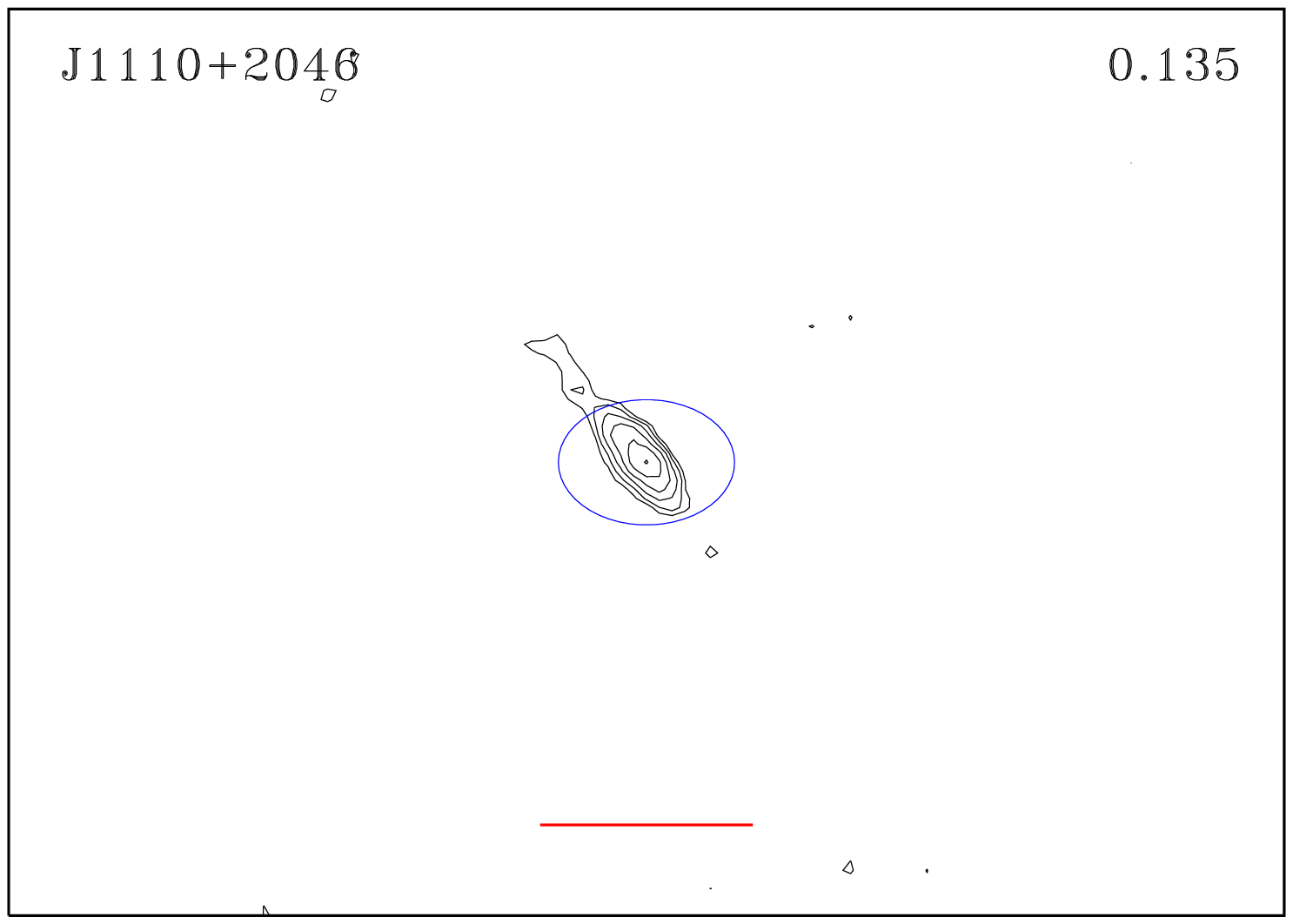} 
\includegraphics[width=6.3cm,height=6.3cm]{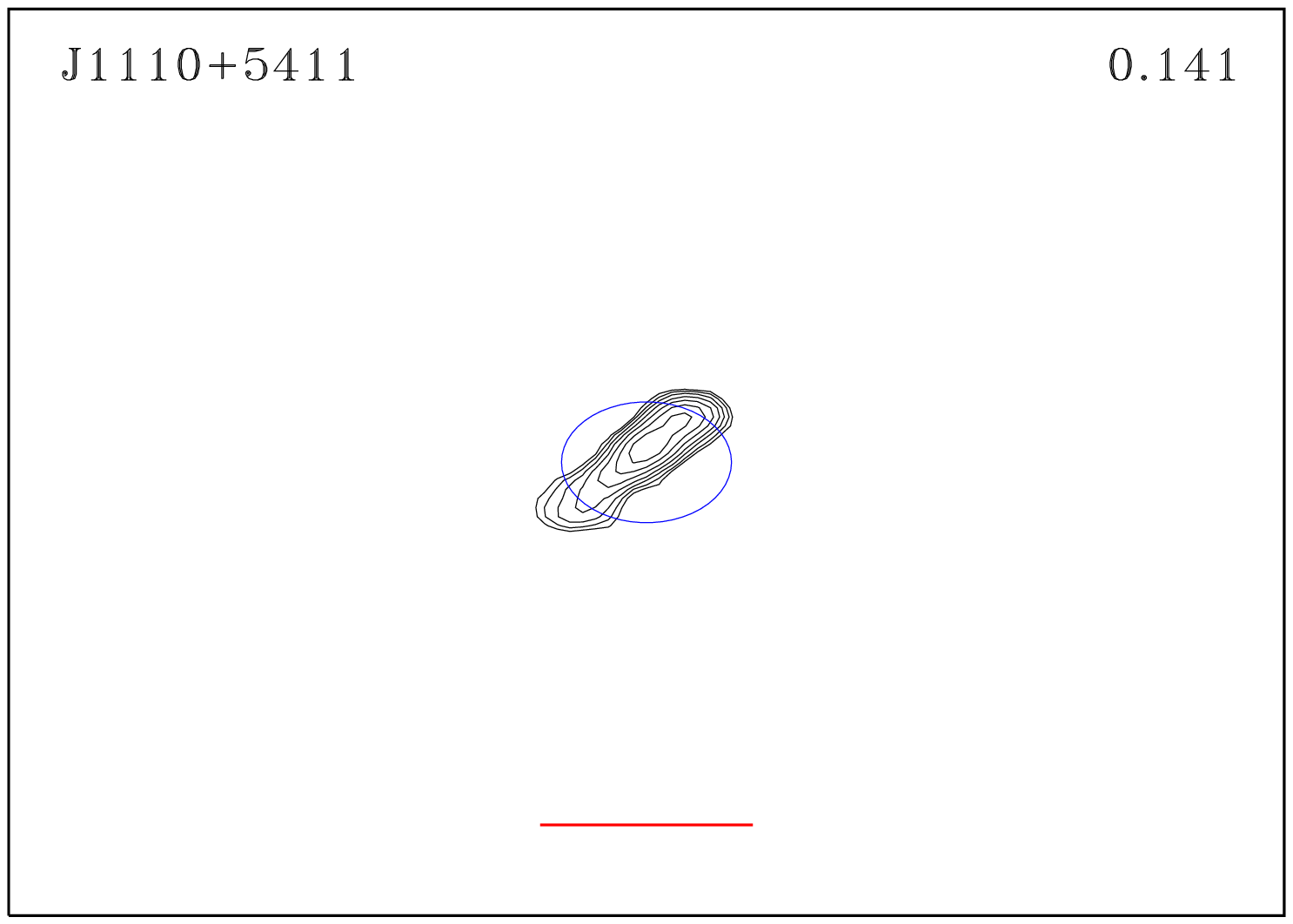} 
\caption{(continued)}
\end{figure*}

\addtocounter{figure}{-1}
\begin{figure*}
\includegraphics[width=6.3cm,height=6.3cm]{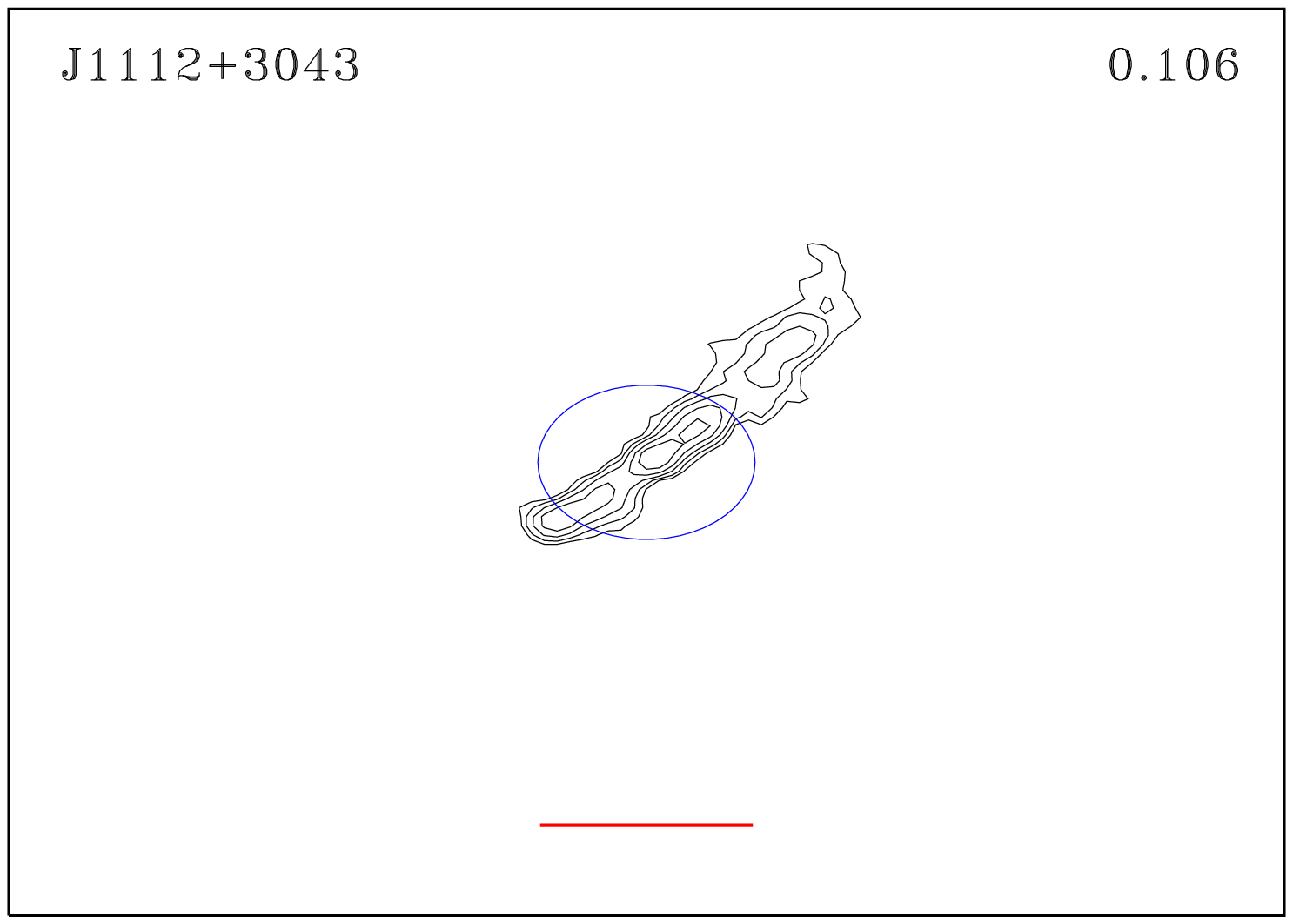} 
\includegraphics[width=6.3cm,height=6.3cm]{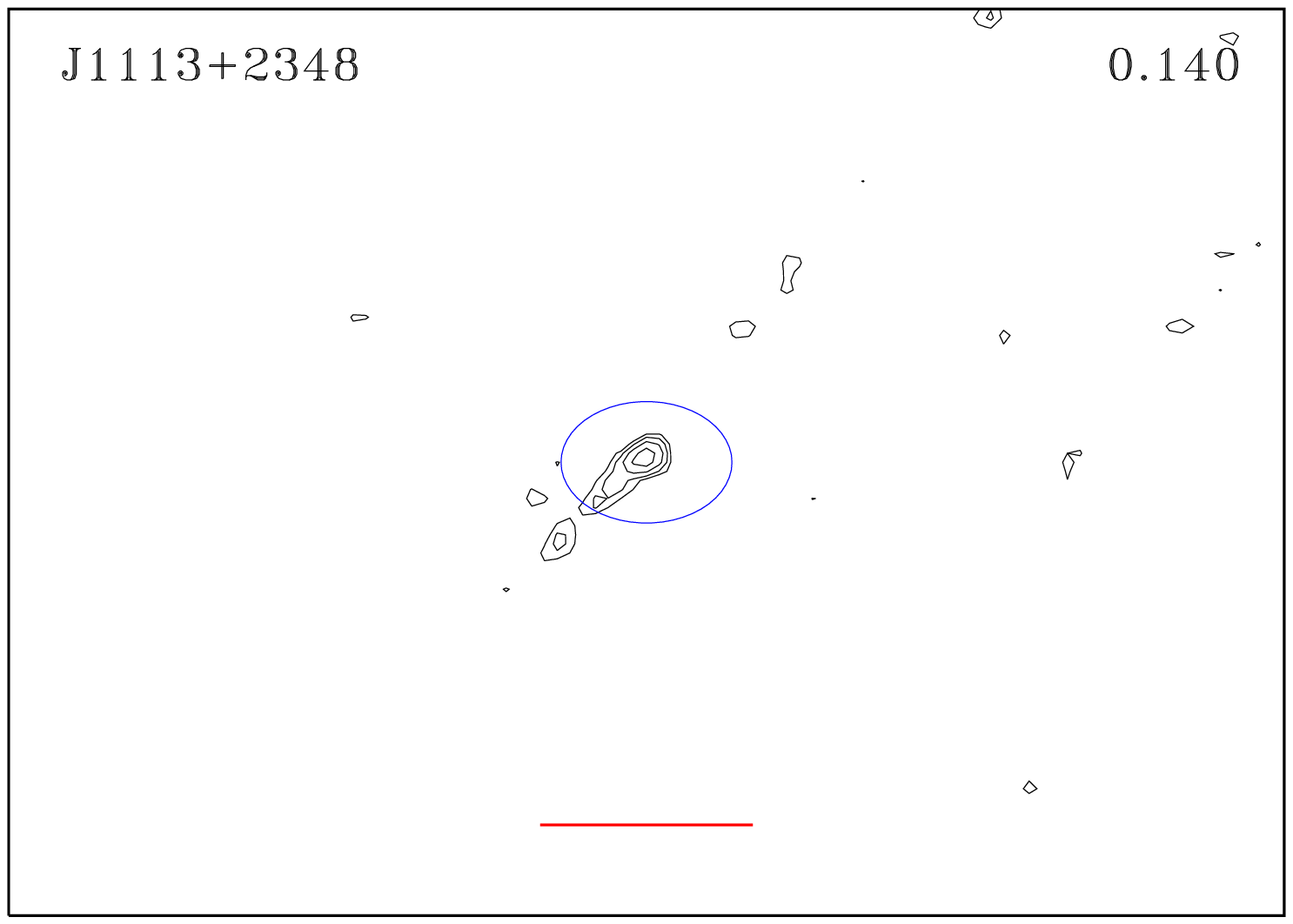} 
\includegraphics[width=6.3cm,height=6.3cm]{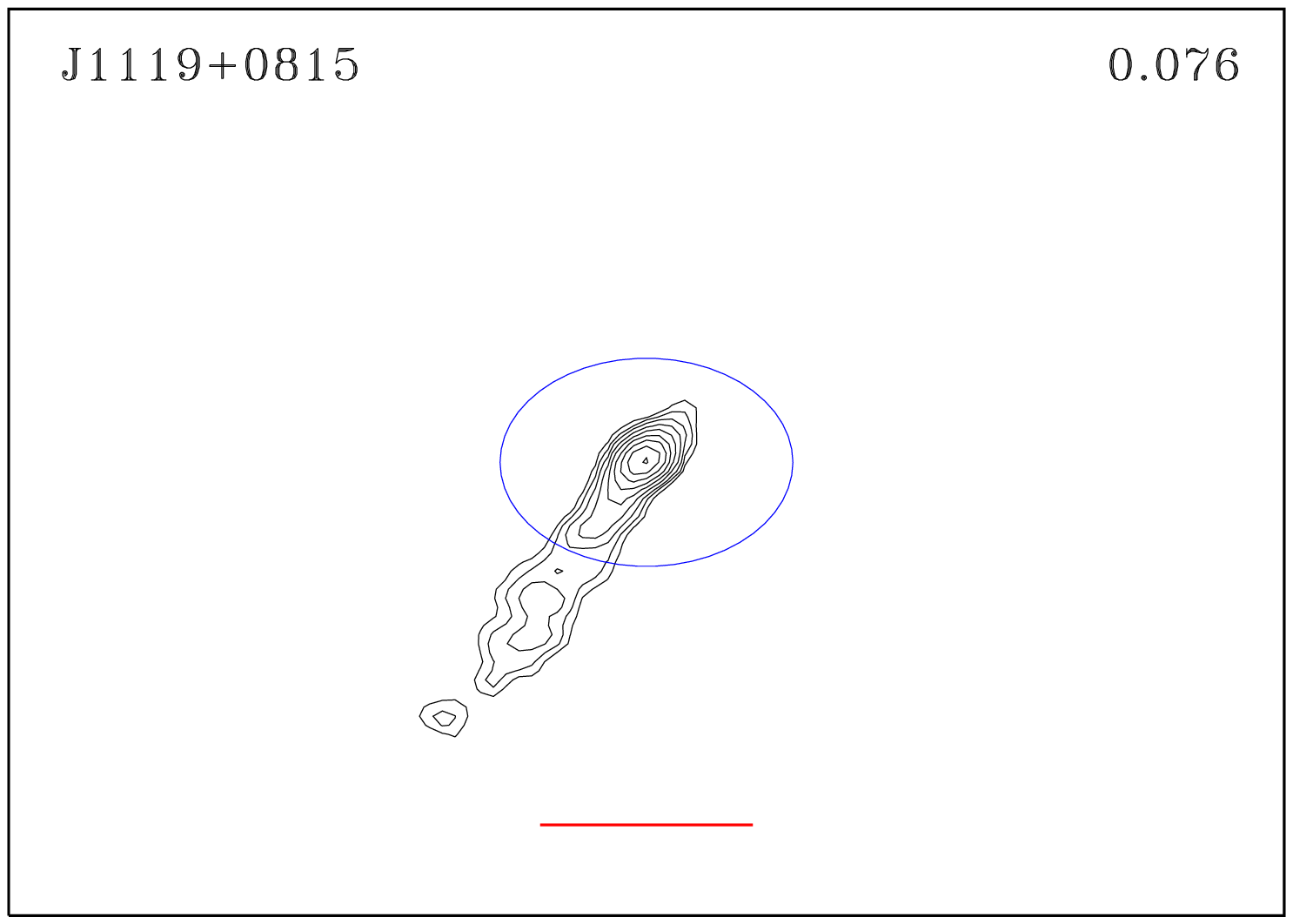} 

\includegraphics[width=6.3cm,height=6.3cm]{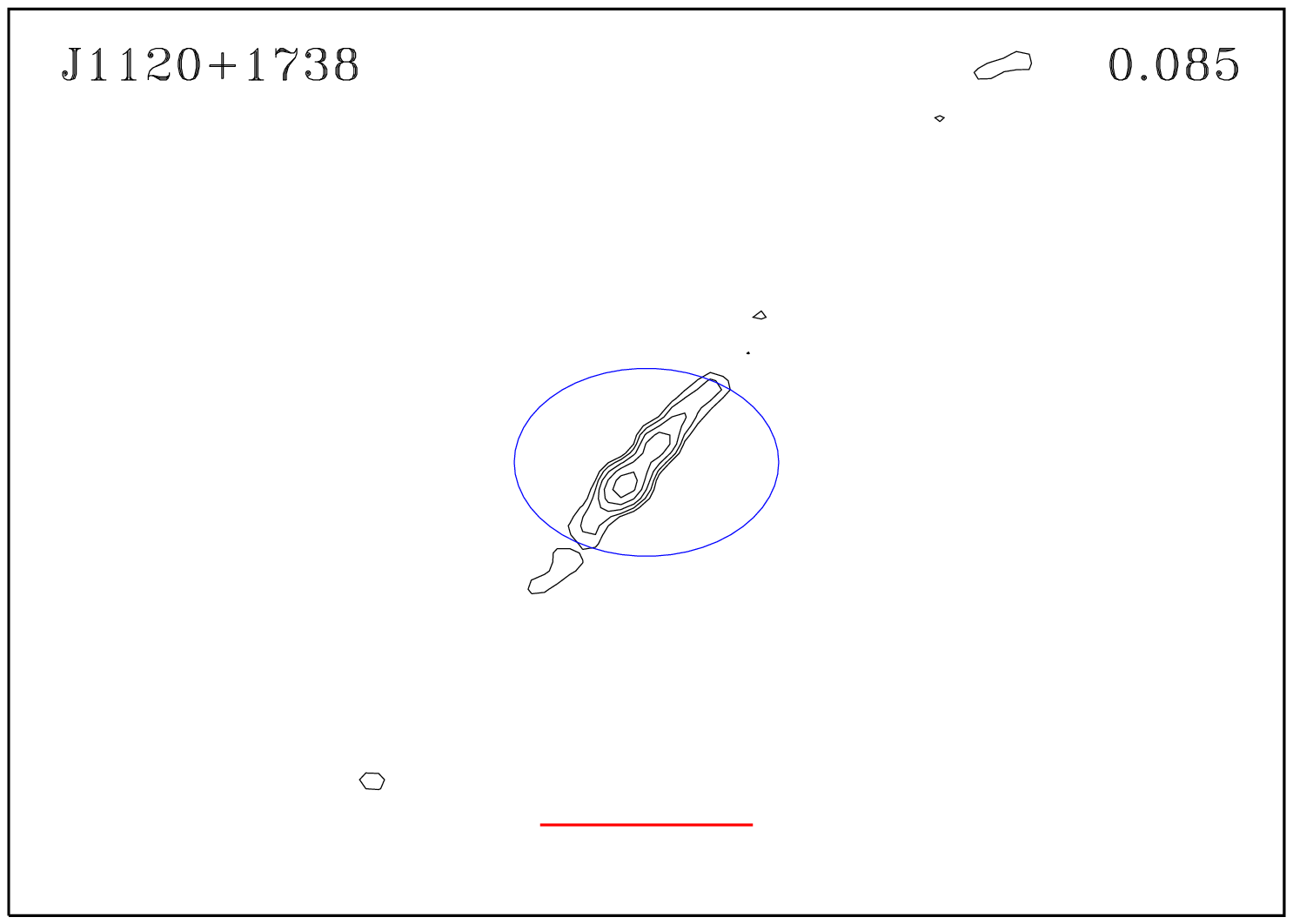} 
\includegraphics[width=6.3cm,height=6.3cm]{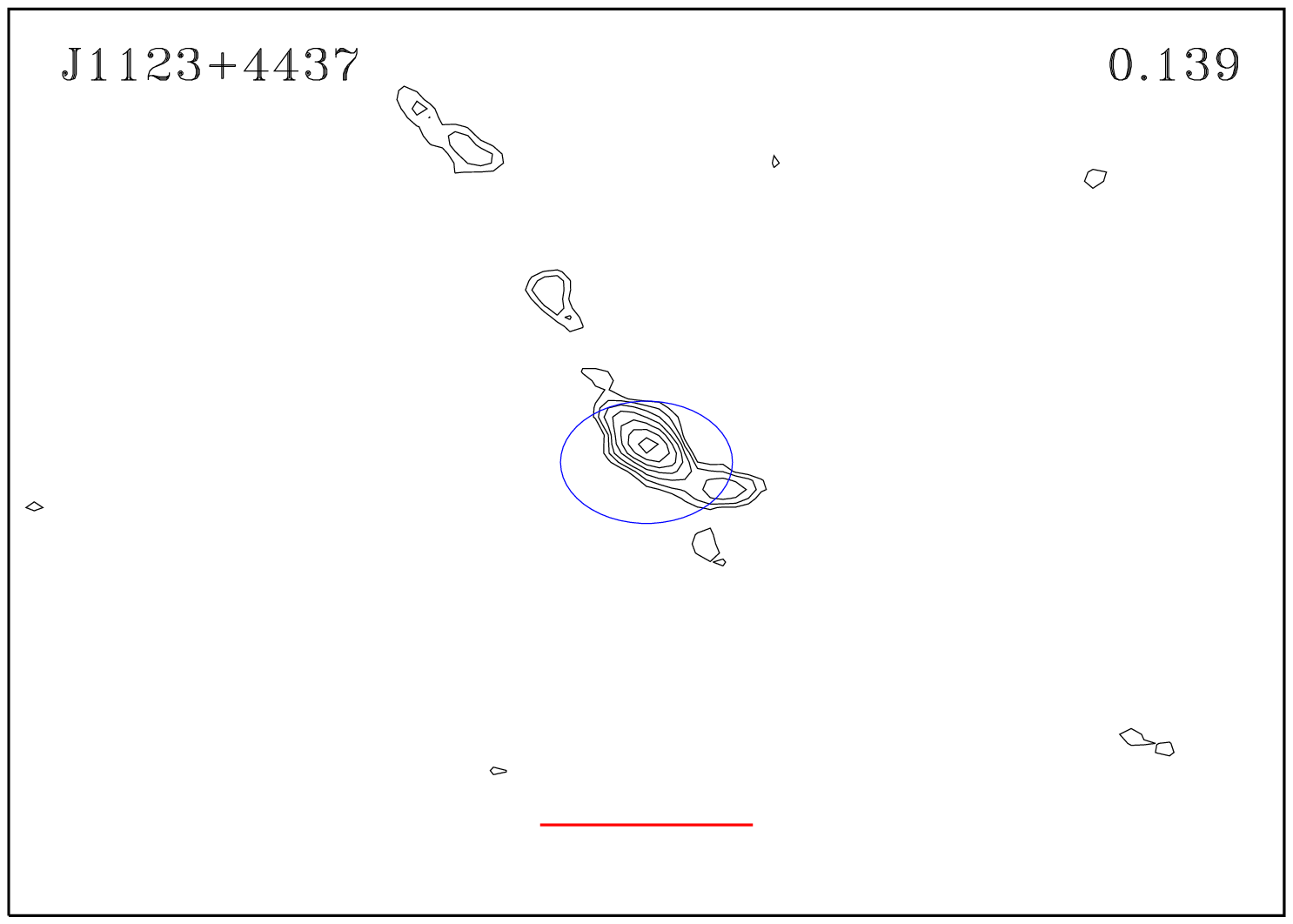} 
\includegraphics[width=6.3cm,height=6.3cm]{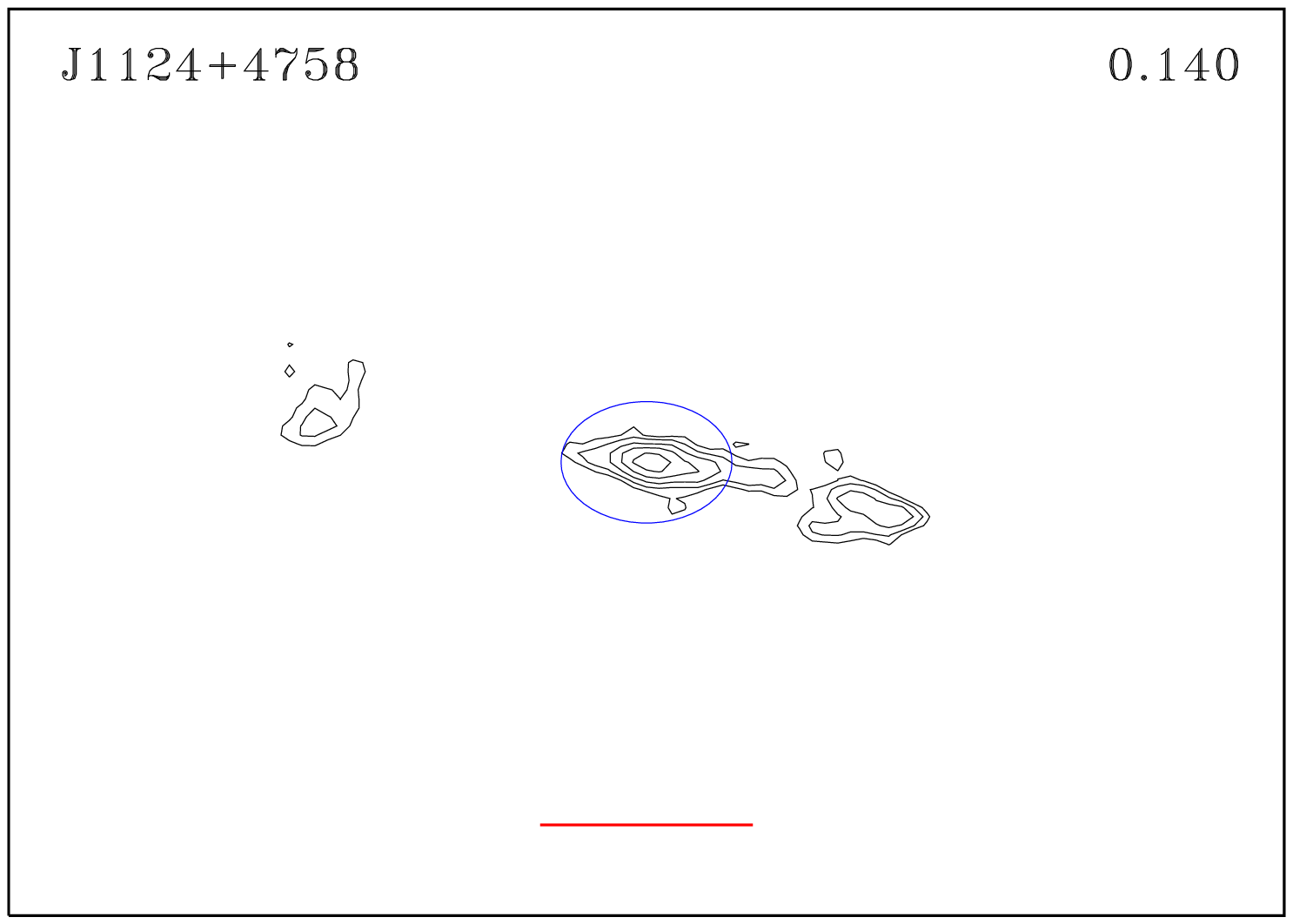} 

\includegraphics[width=6.3cm,height=6.3cm]{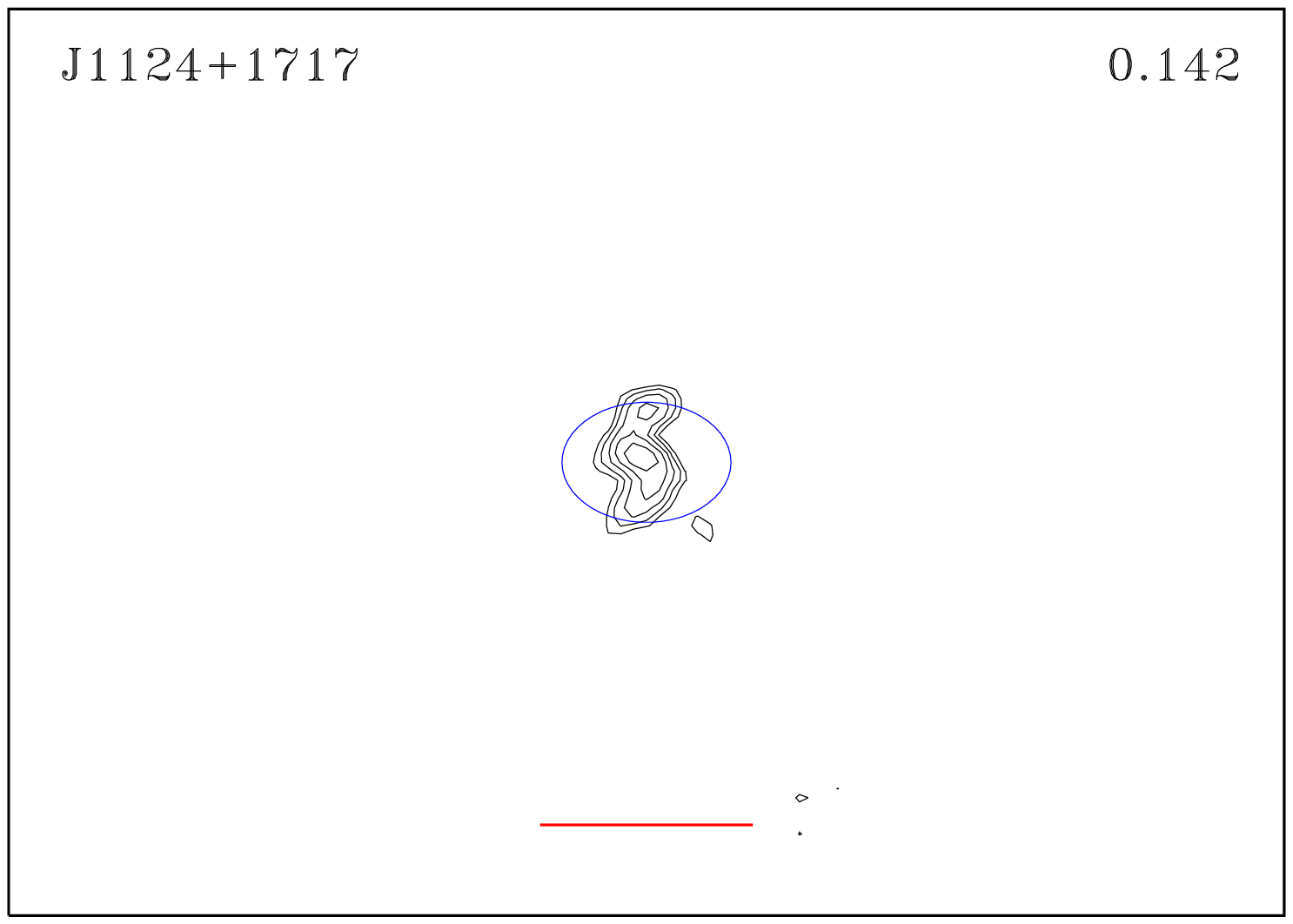} 
\includegraphics[width=6.3cm,height=6.3cm]{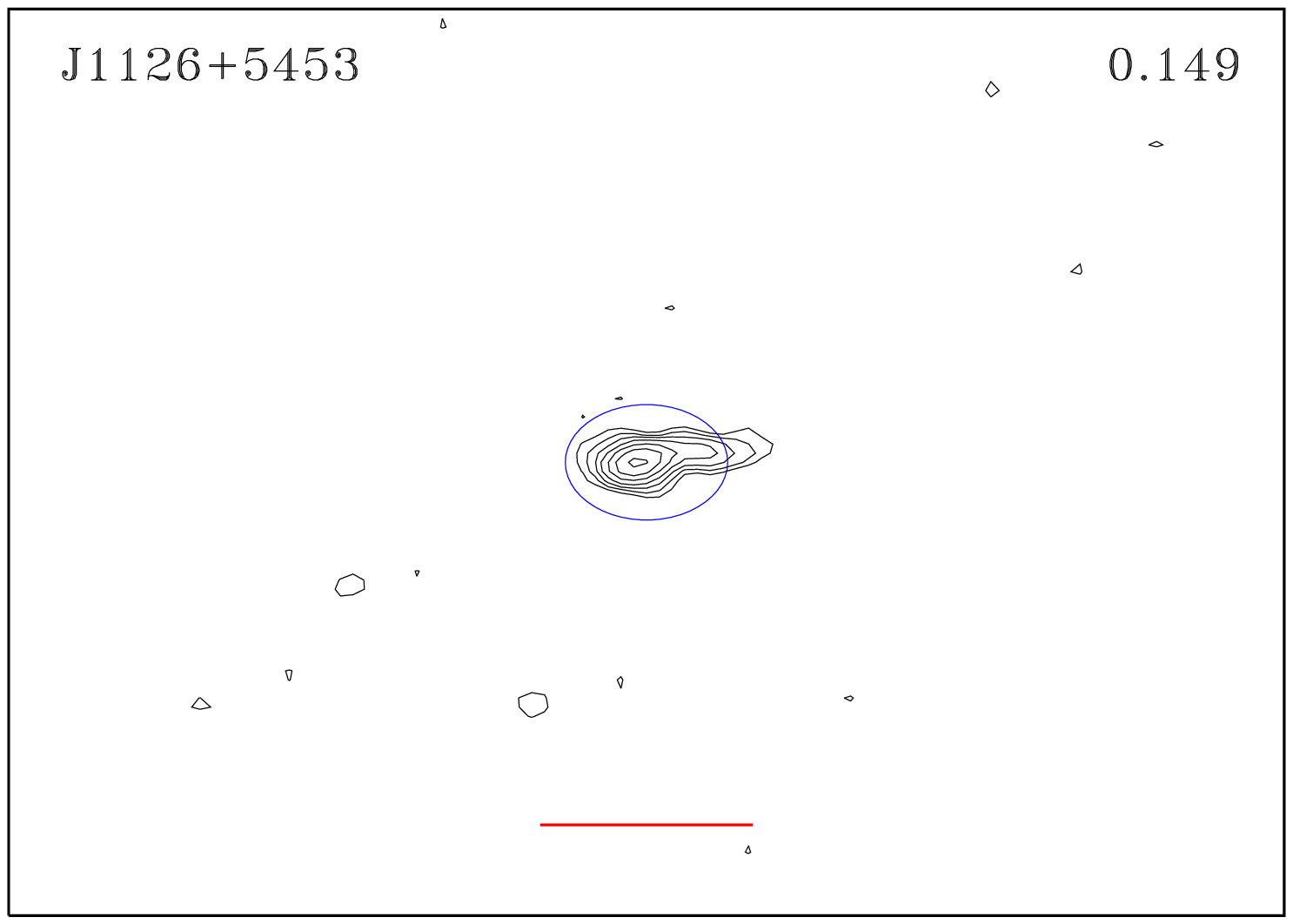} 
\includegraphics[width=6.3cm,height=6.3cm]{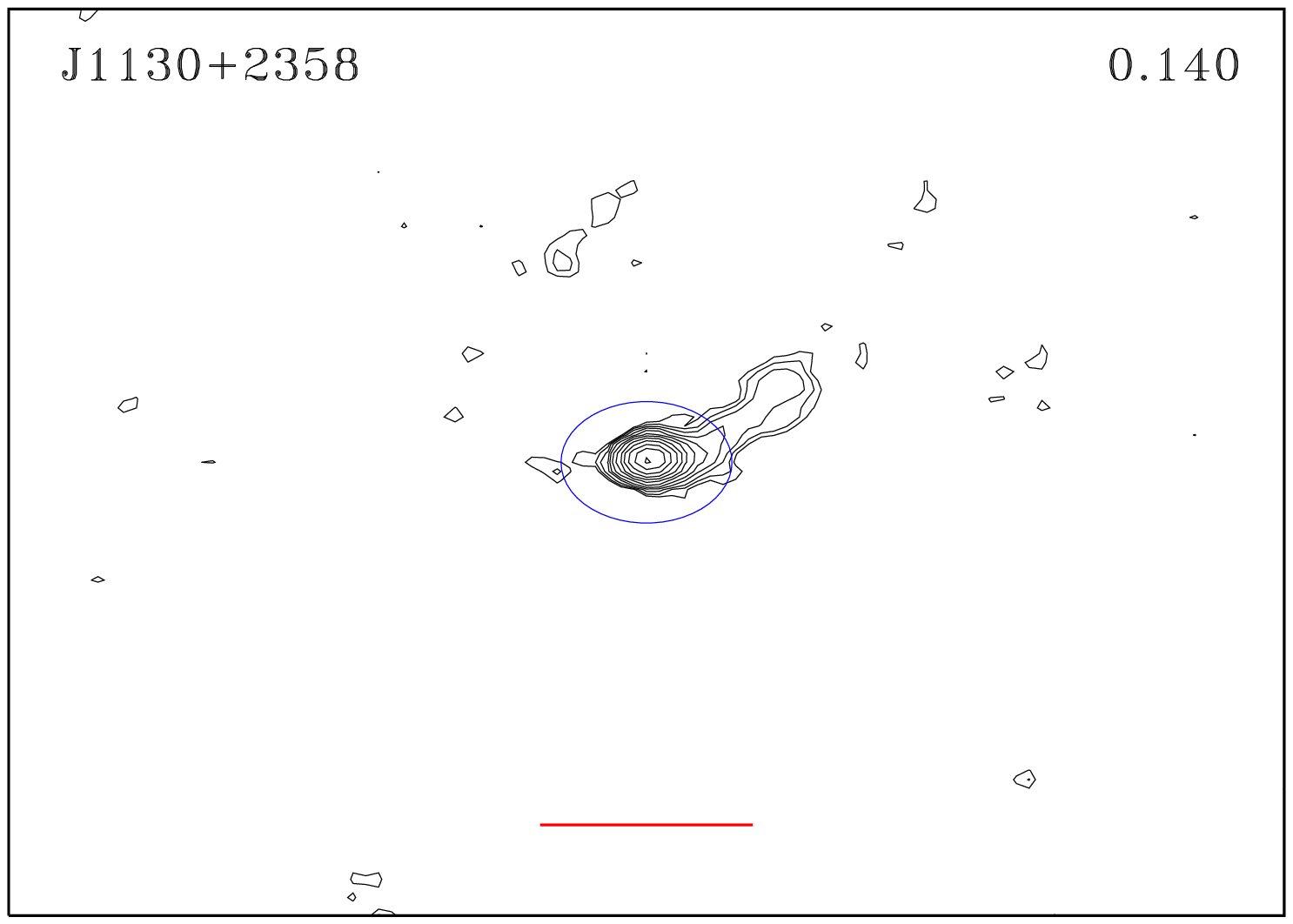} 

\includegraphics[width=6.3cm,height=6.3cm]{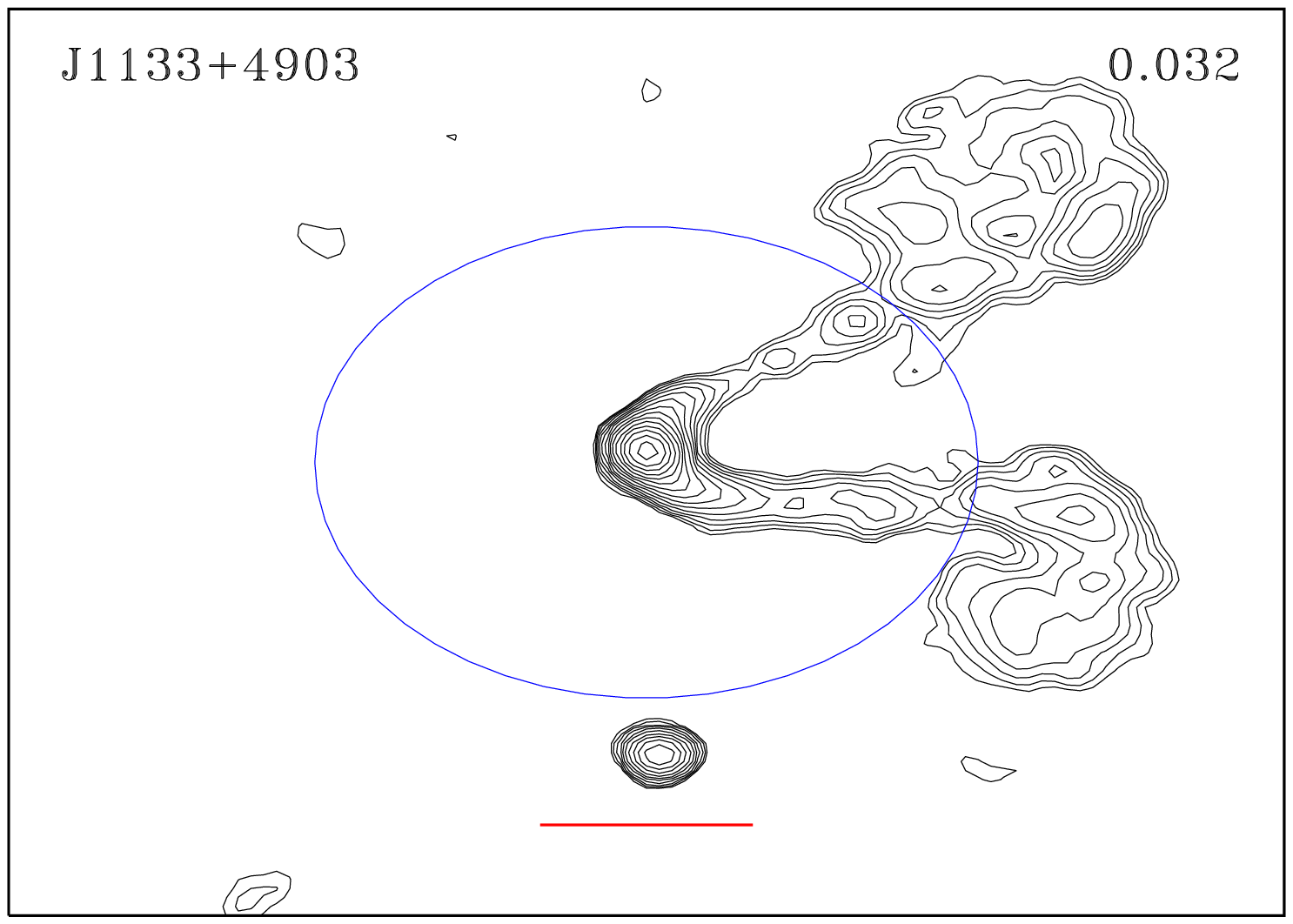} 
\includegraphics[width=6.3cm,height=6.3cm]{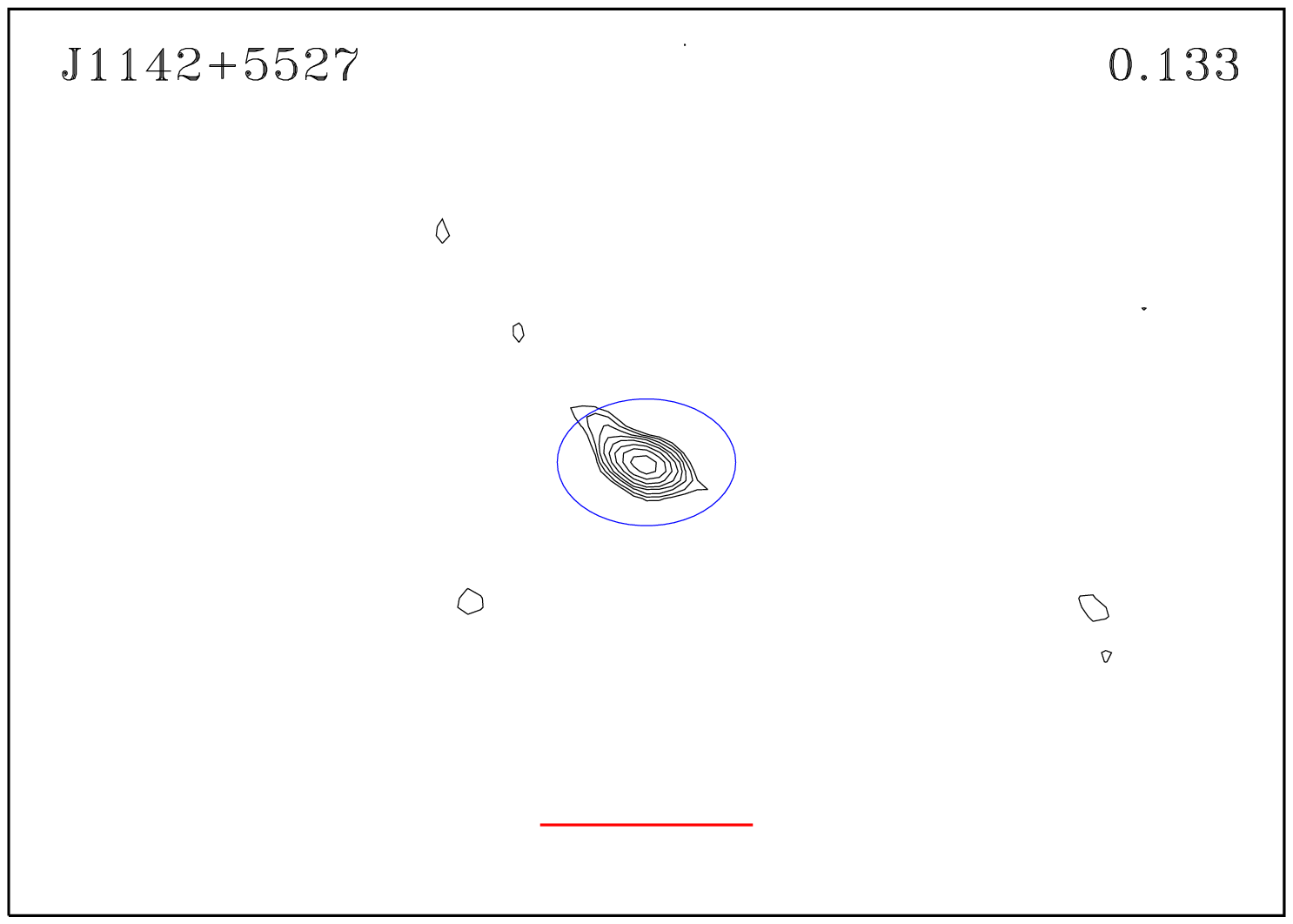} 
\includegraphics[width=6.3cm,height=6.3cm]{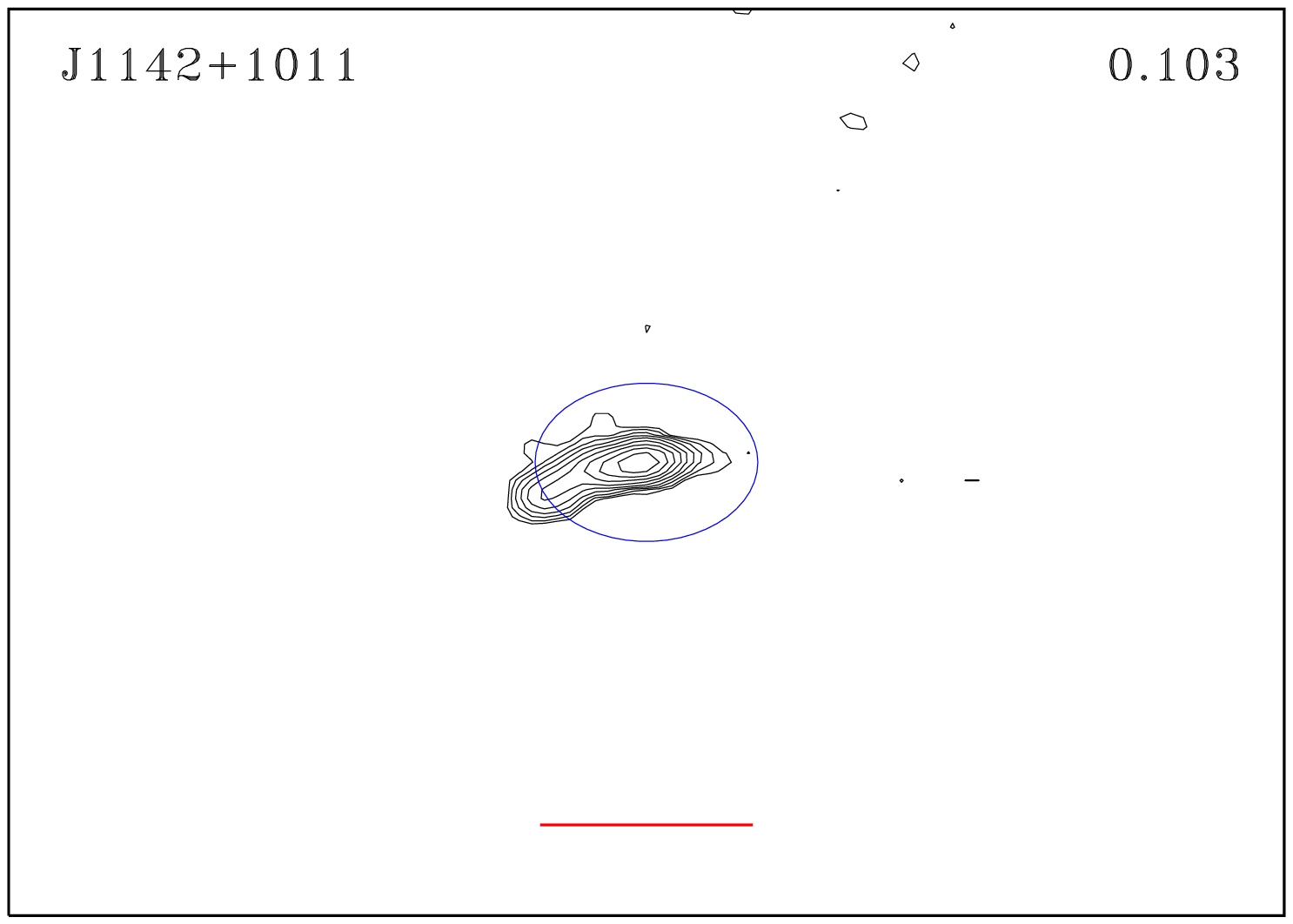} 
\caption{(continued)}
\end{figure*}

\addtocounter{figure}{-1}
\begin{figure*}
\includegraphics[width=6.3cm,height=6.3cm]{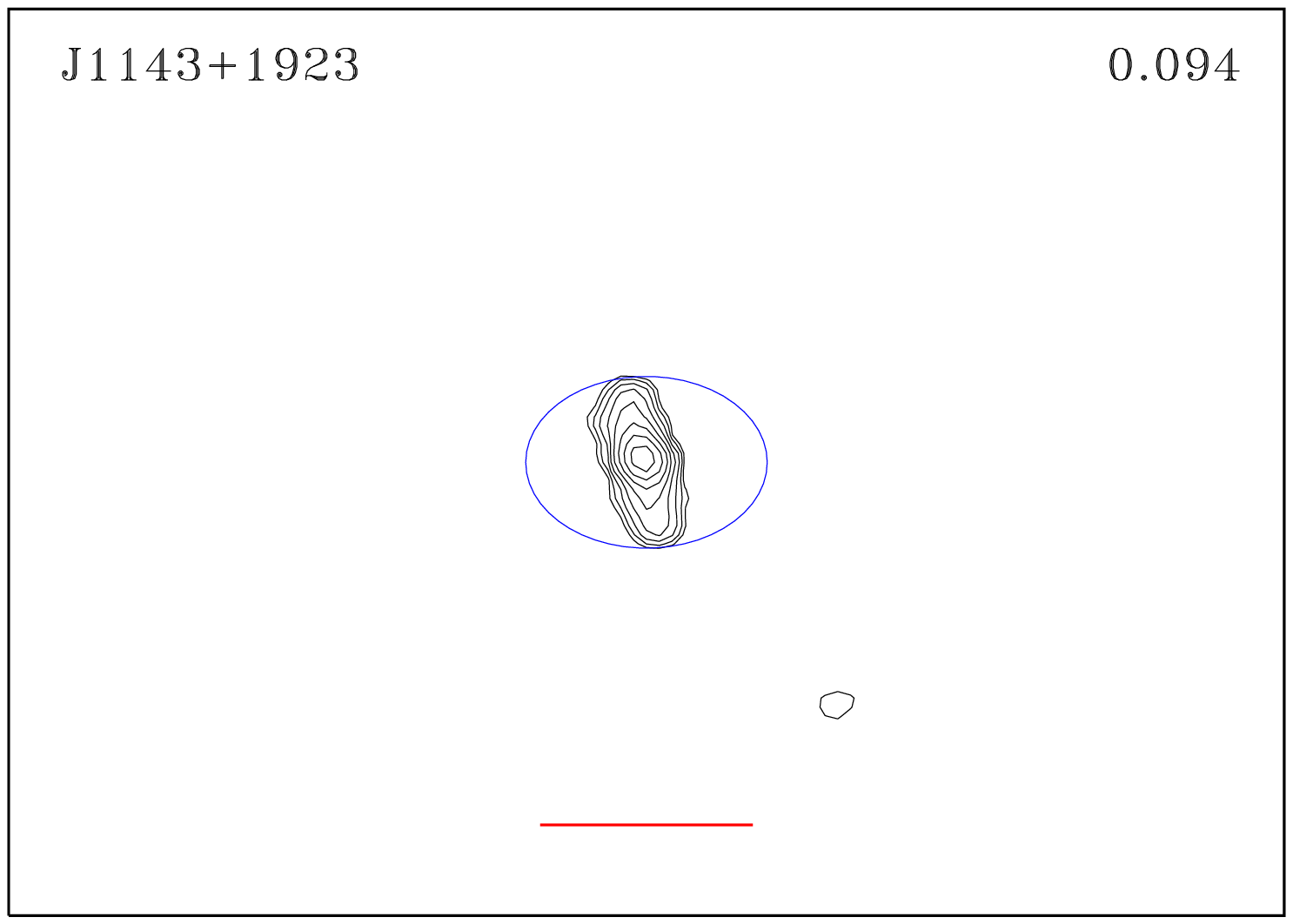} 
\includegraphics[width=6.3cm,height=6.3cm]{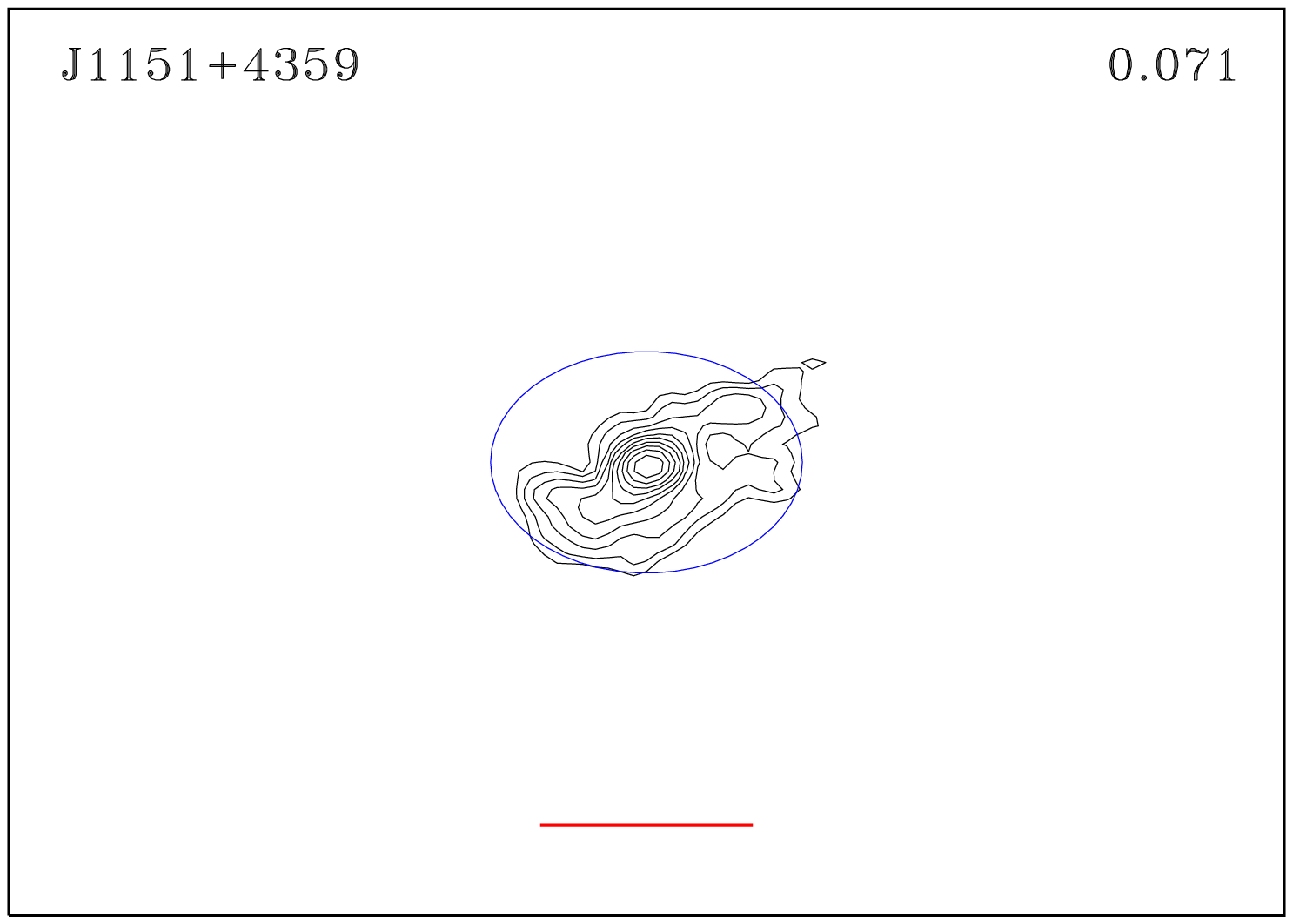} 
\includegraphics[width=6.3cm,height=6.3cm]{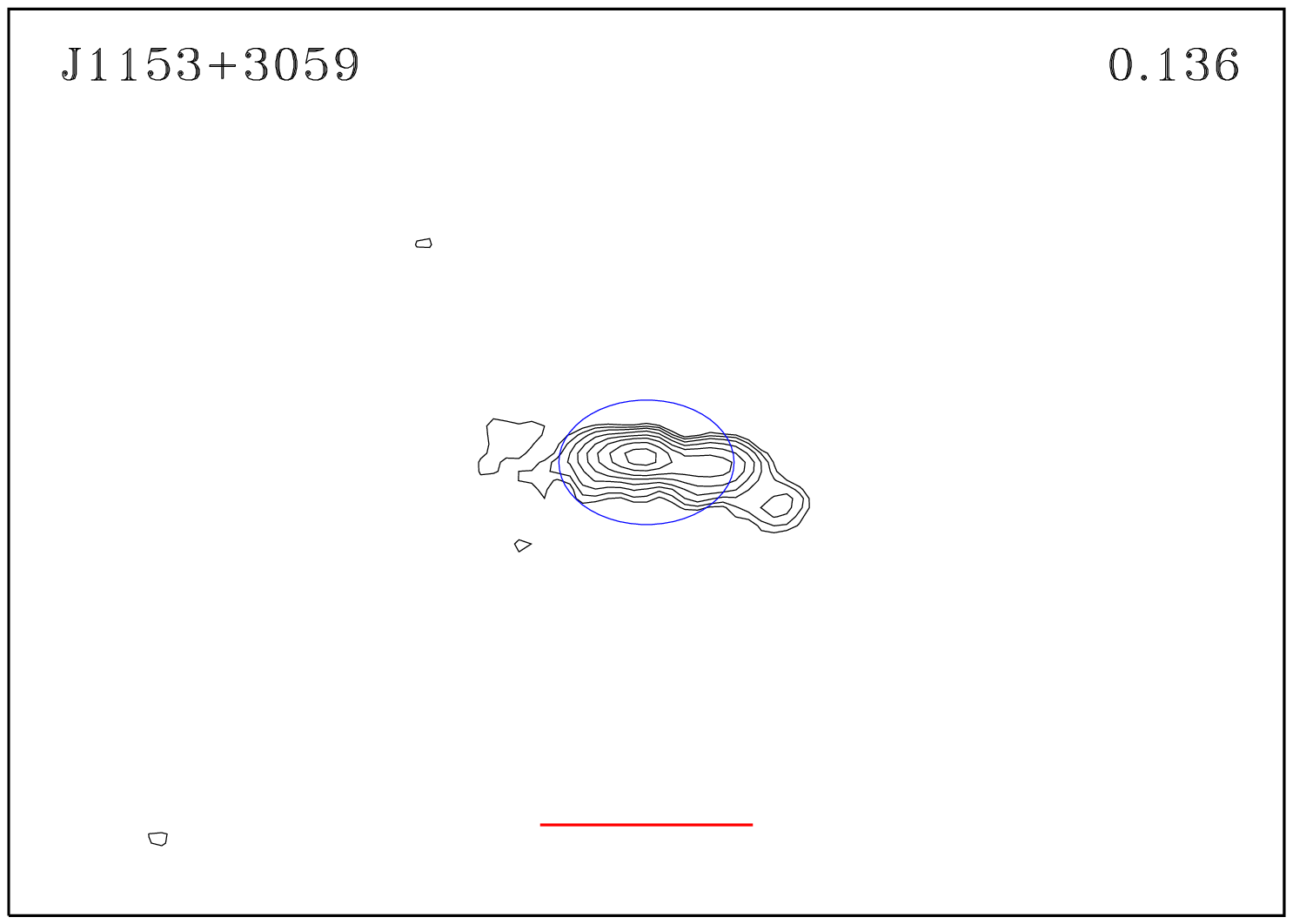} 

\includegraphics[width=6.3cm,height=6.3cm]{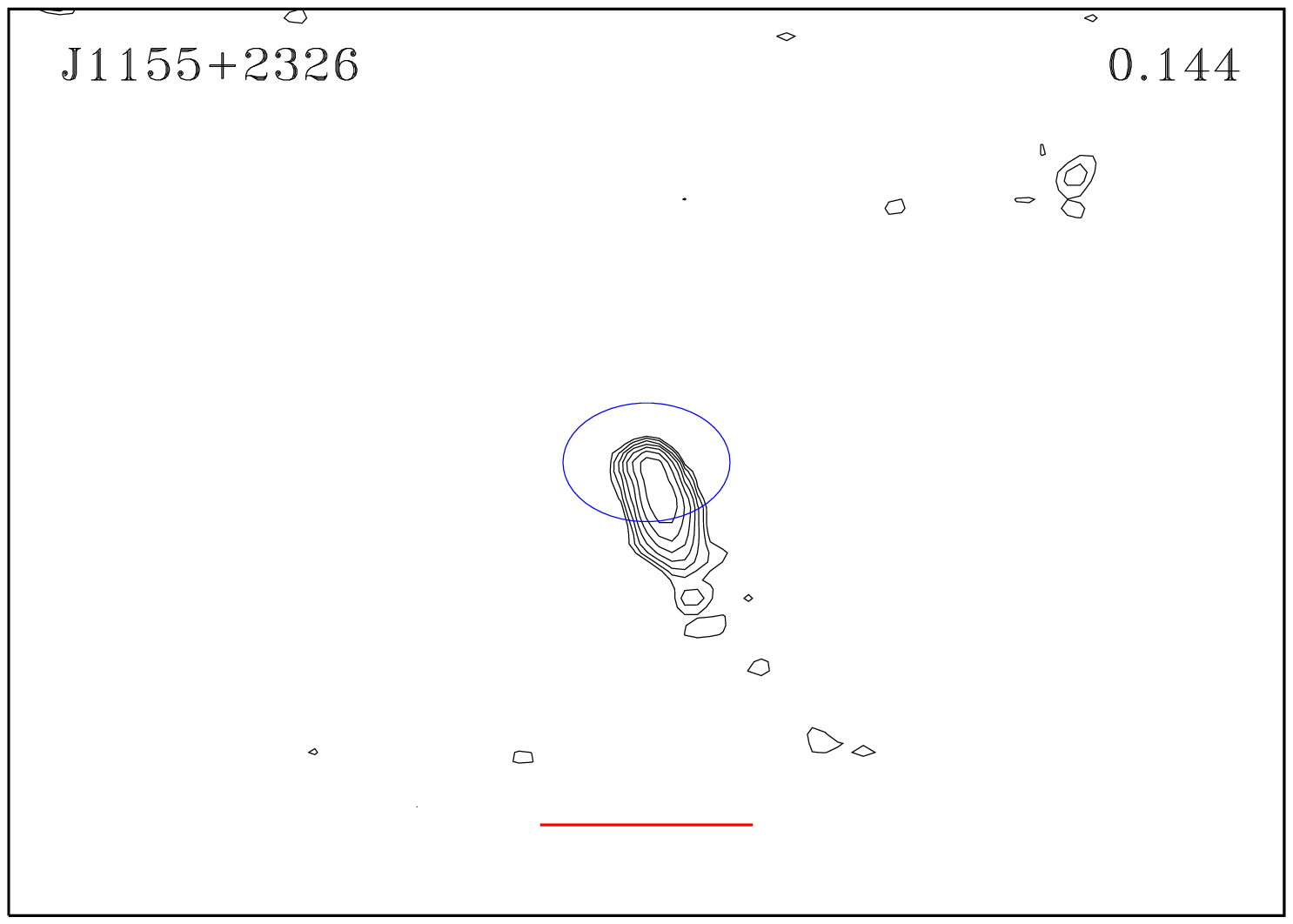} 
\includegraphics[width=6.3cm,height=6.3cm]{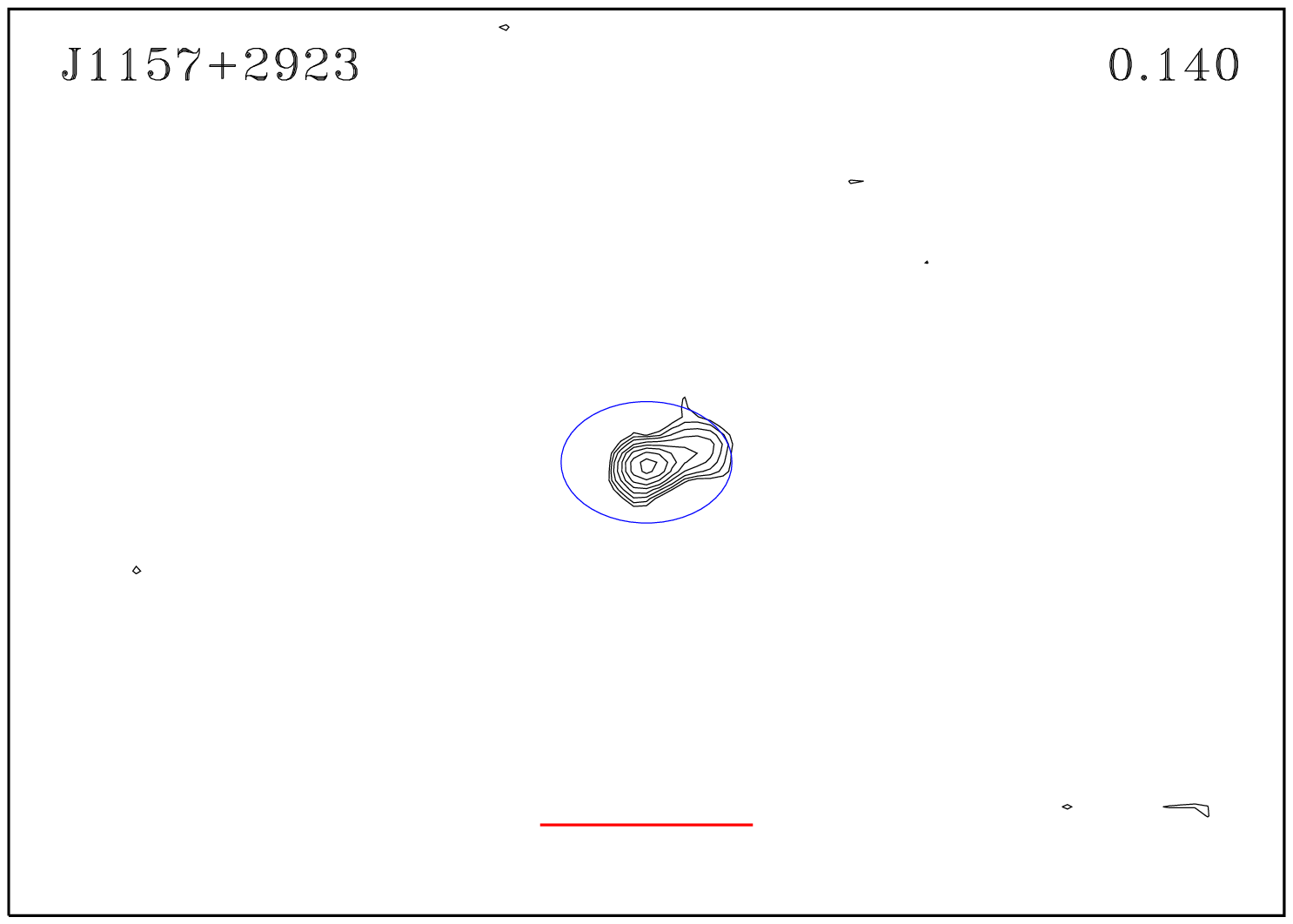} 
\includegraphics[width=6.3cm,height=6.3cm]{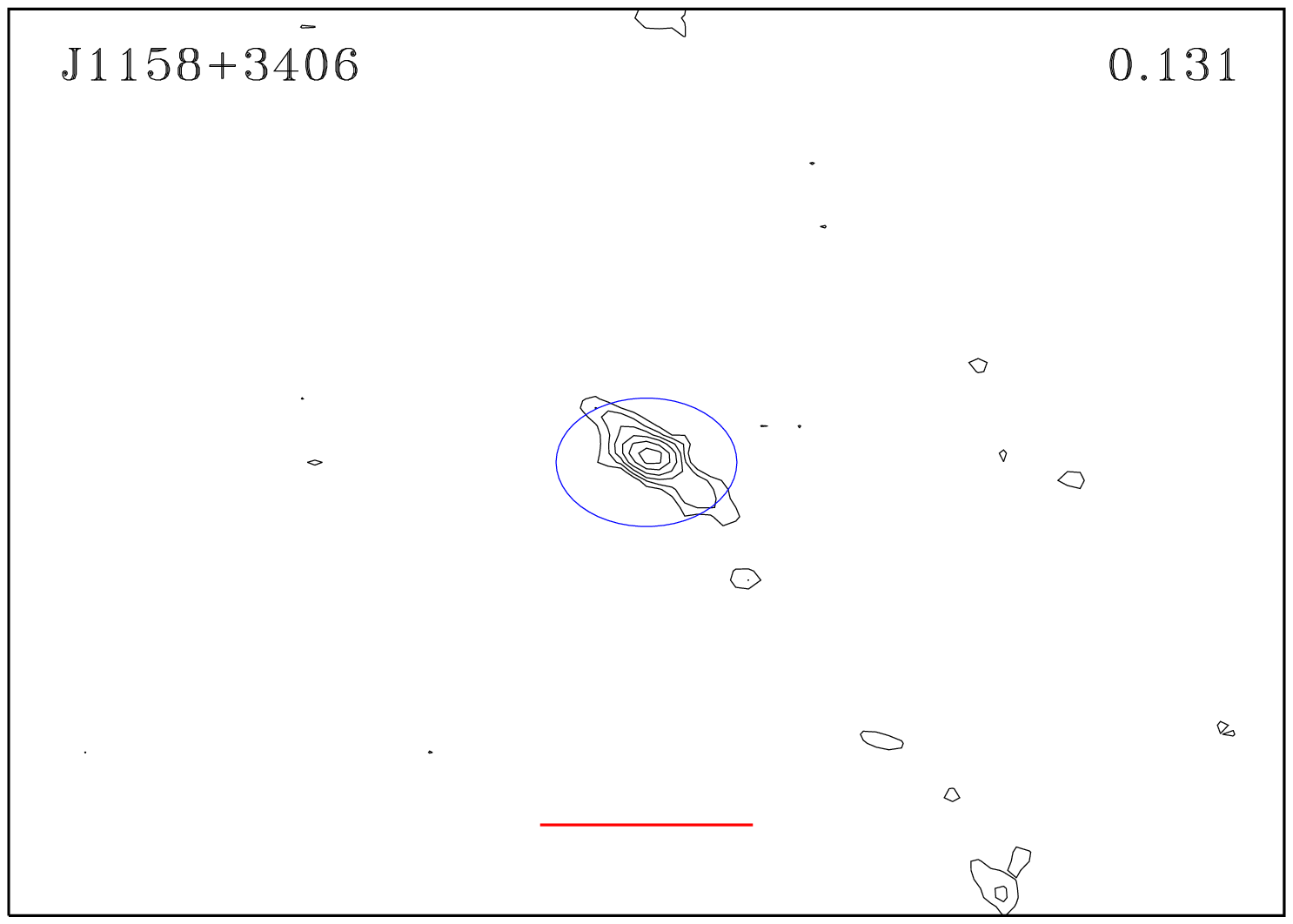} 

\includegraphics[width=6.3cm,height=6.3cm]{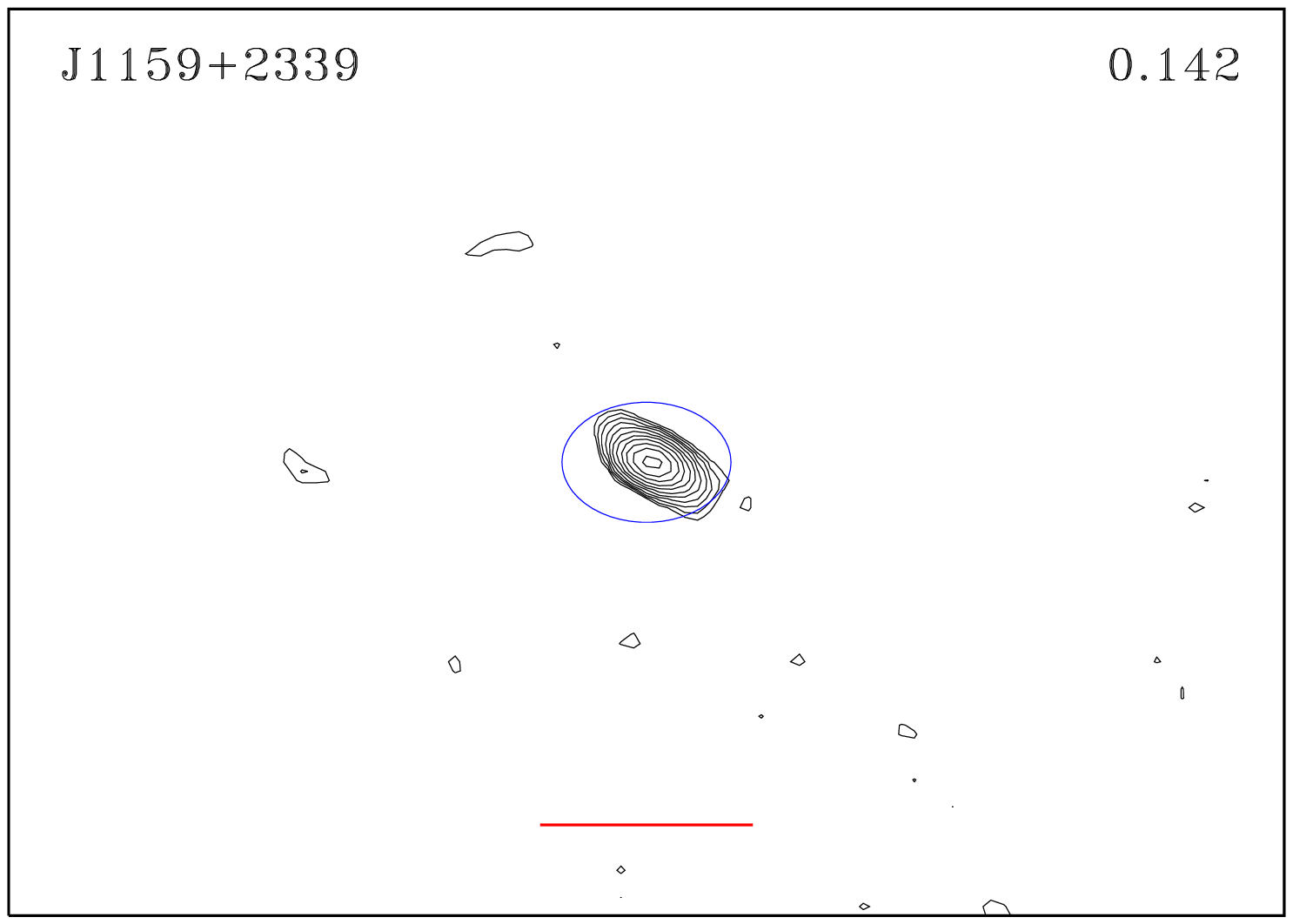} 
\includegraphics[width=6.3cm,height=6.3cm]{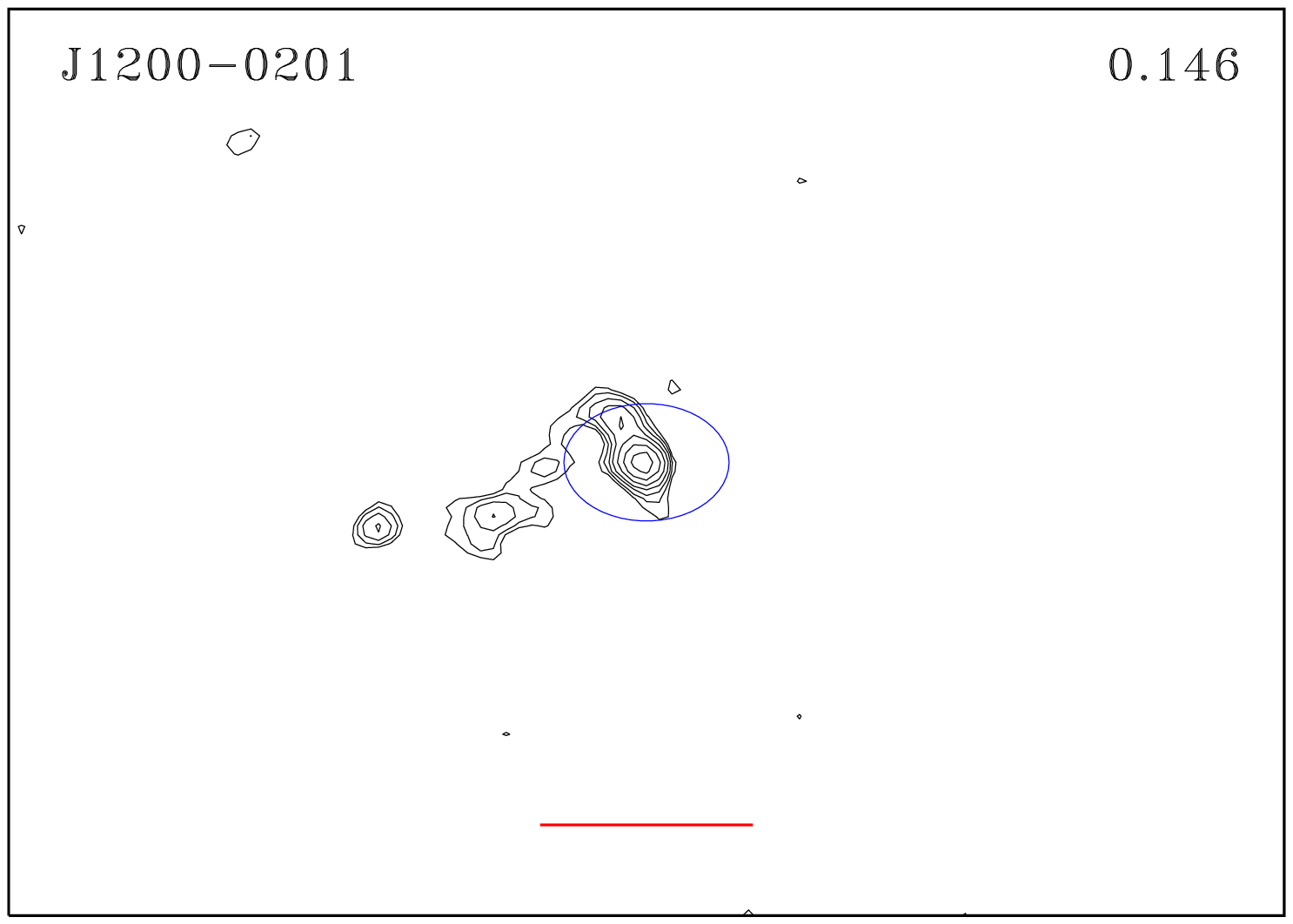} 
\includegraphics[width=6.3cm,height=6.3cm]{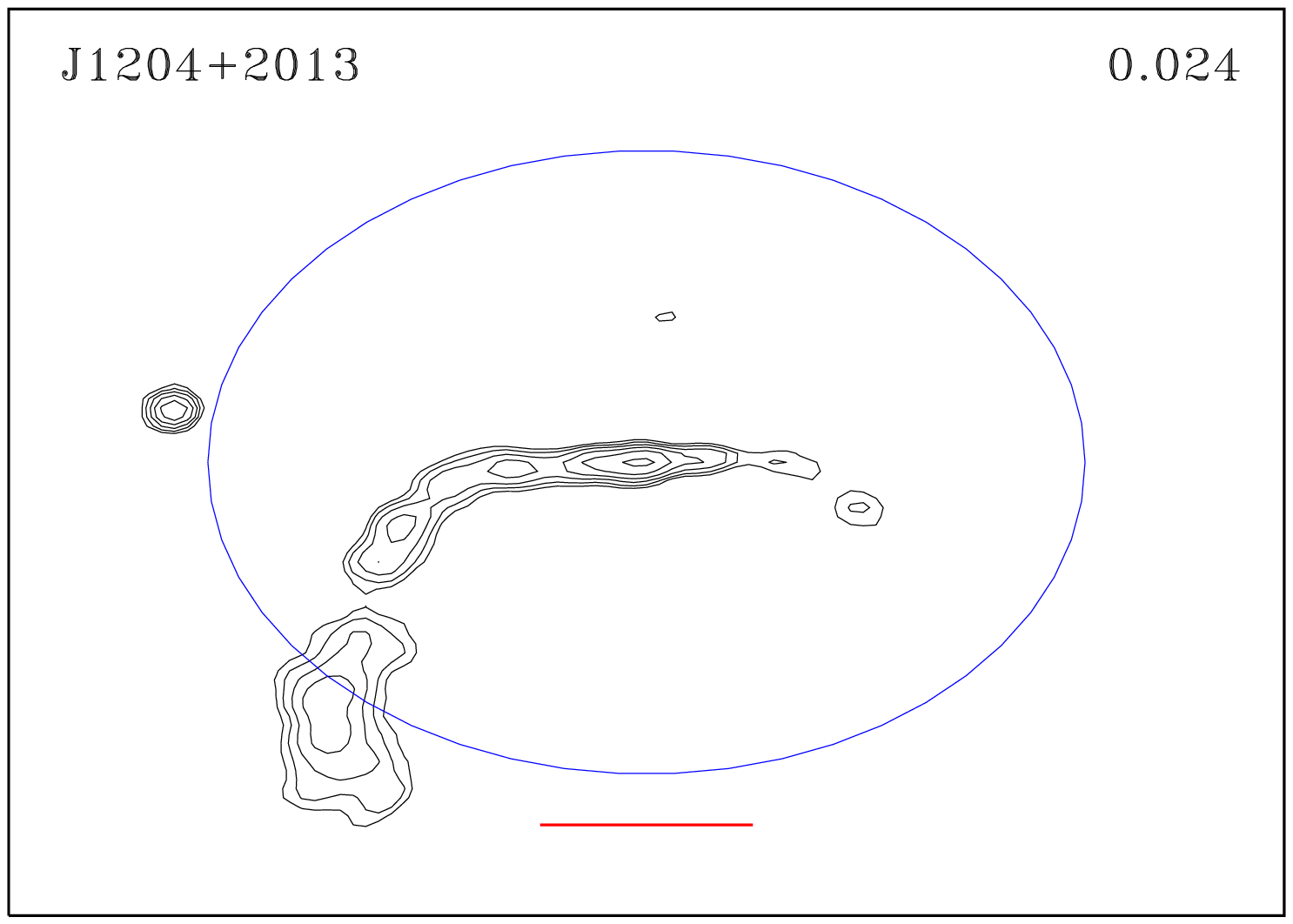} 

\includegraphics[width=6.3cm,height=6.3cm]{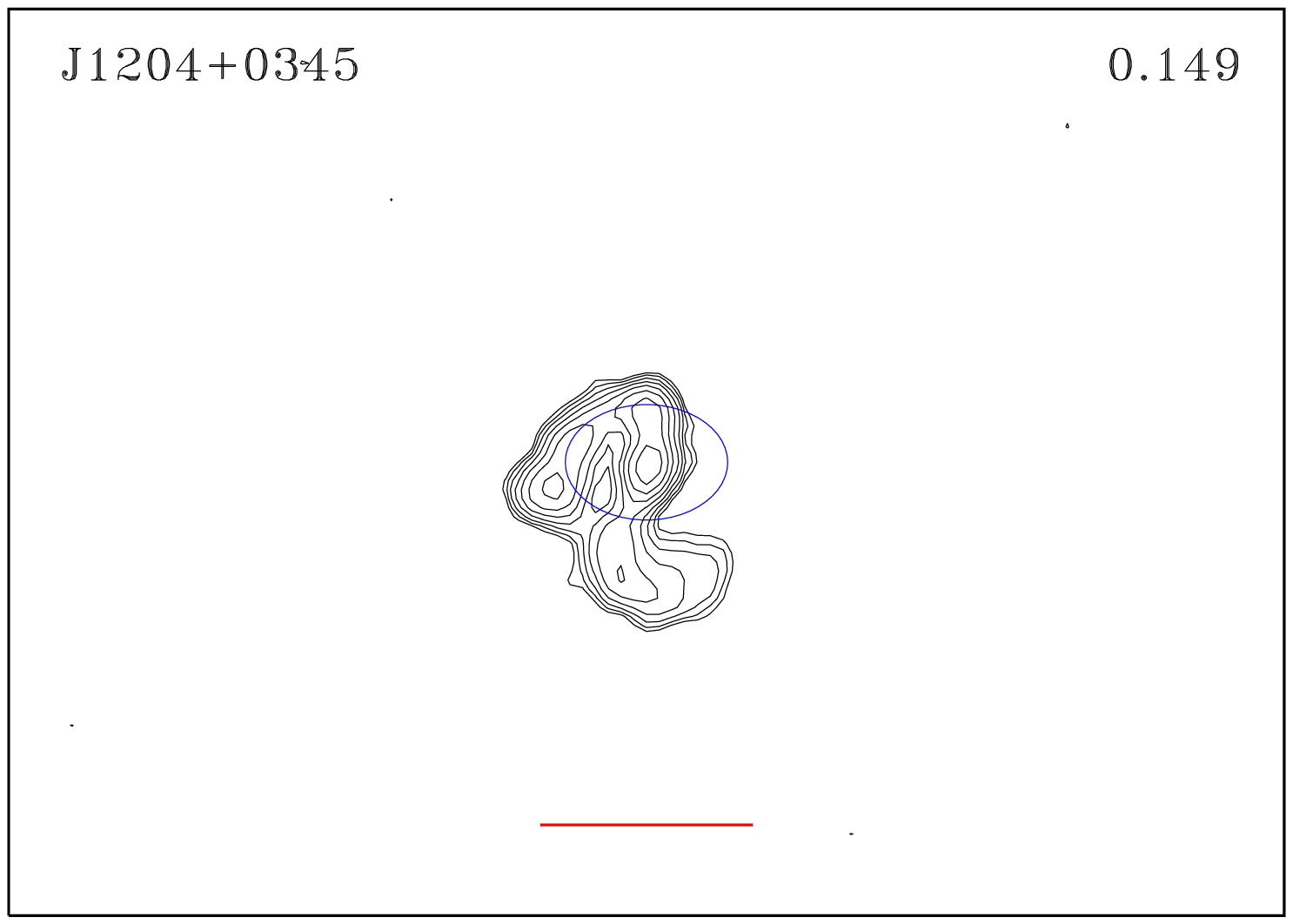} 
\includegraphics[width=6.3cm,height=6.3cm]{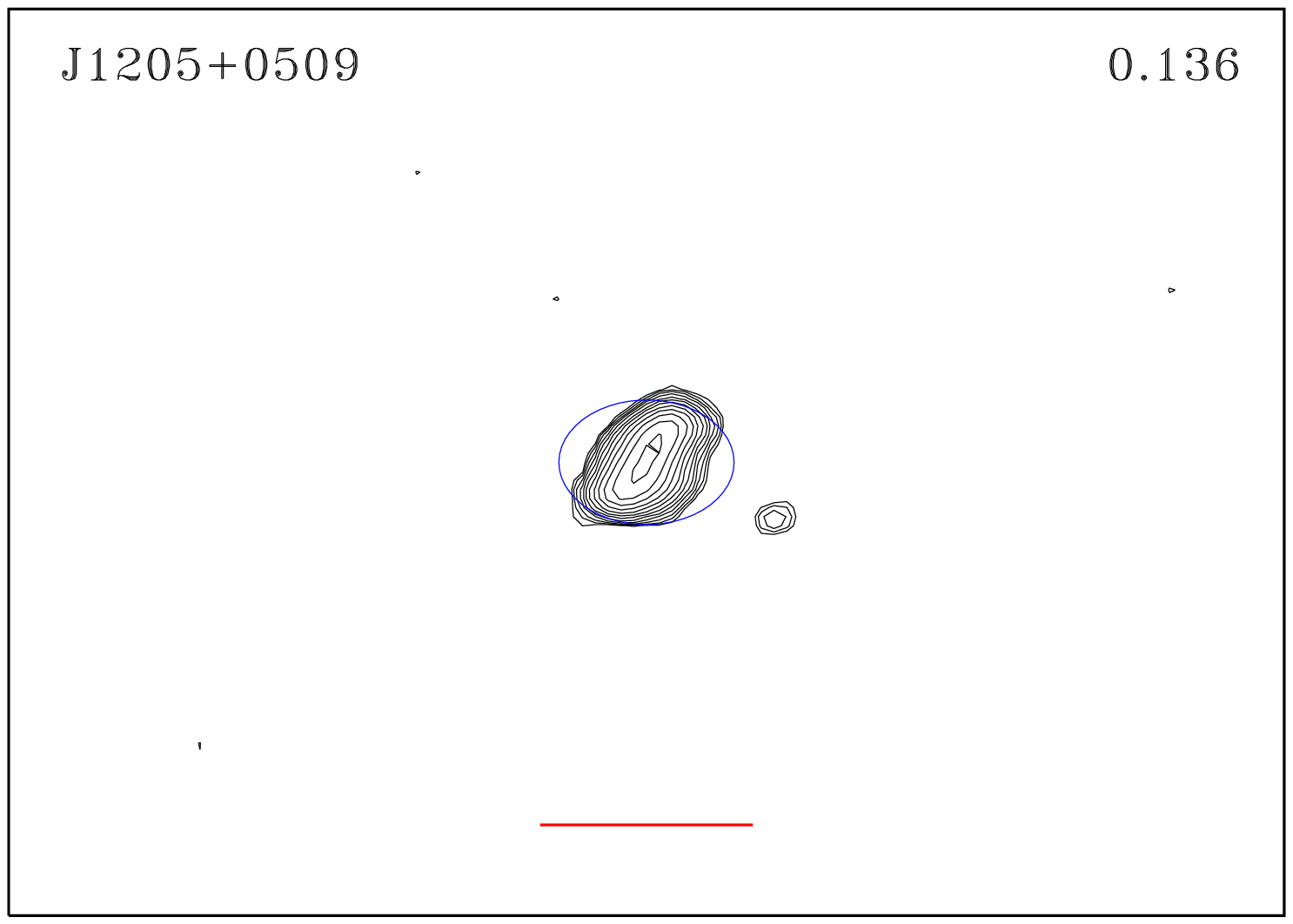} 
\includegraphics[width=6.3cm,height=6.3cm]{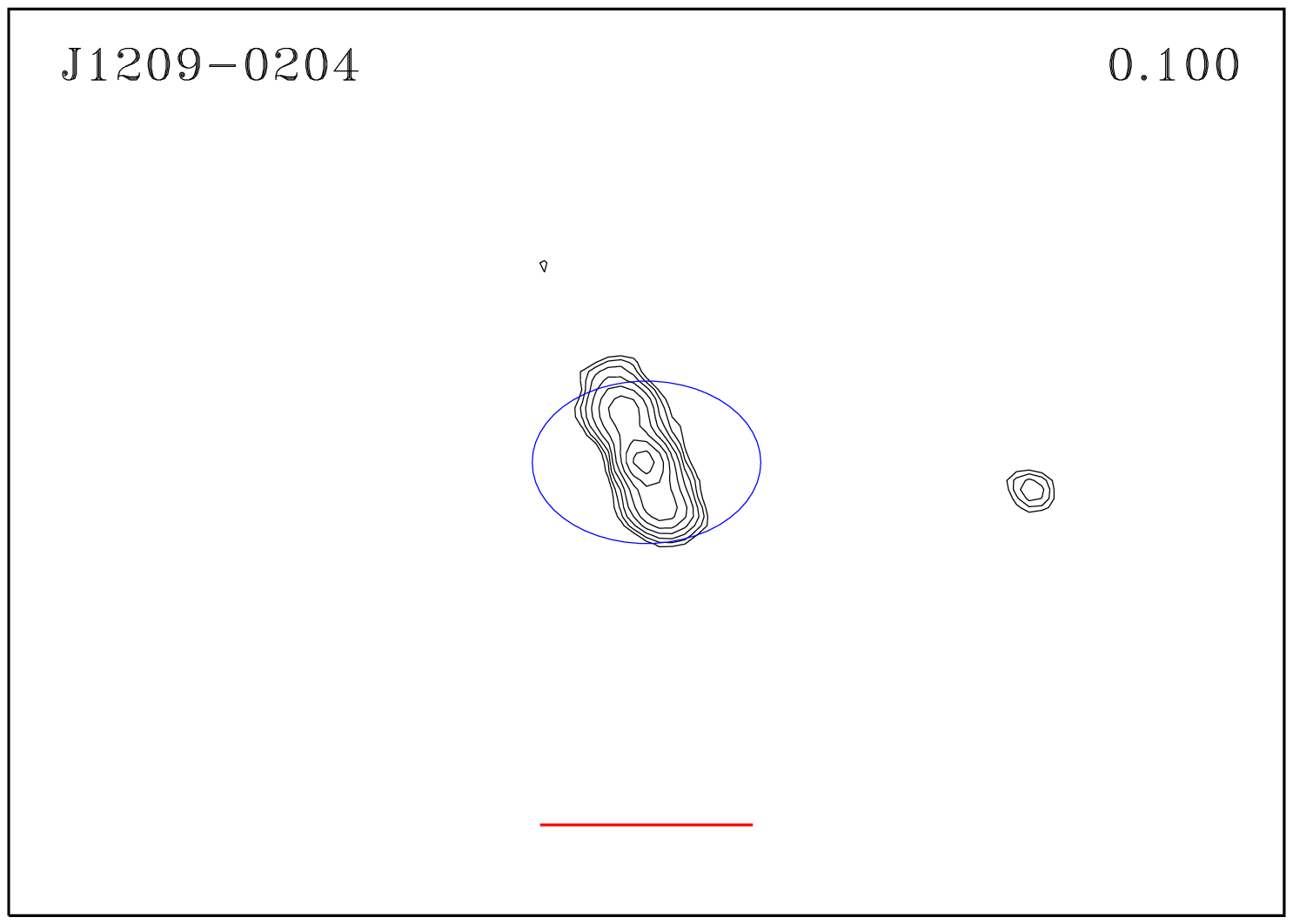} 
\caption{(continued)}
\end{figure*}

\addtocounter{figure}{-1}
\begin{figure*}
\includegraphics[width=6.3cm,height=6.3cm]{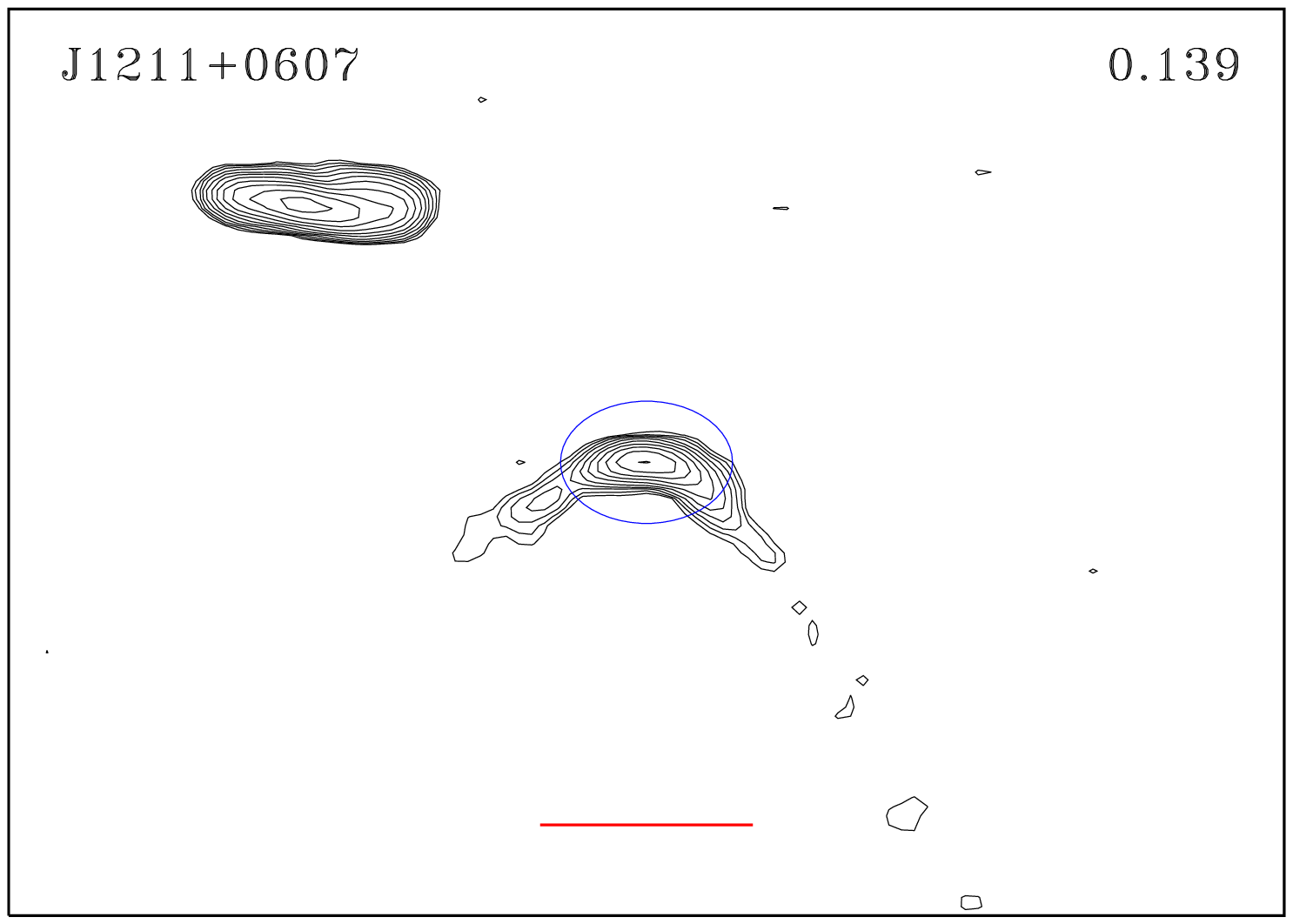} 
\includegraphics[width=6.3cm,height=6.3cm]{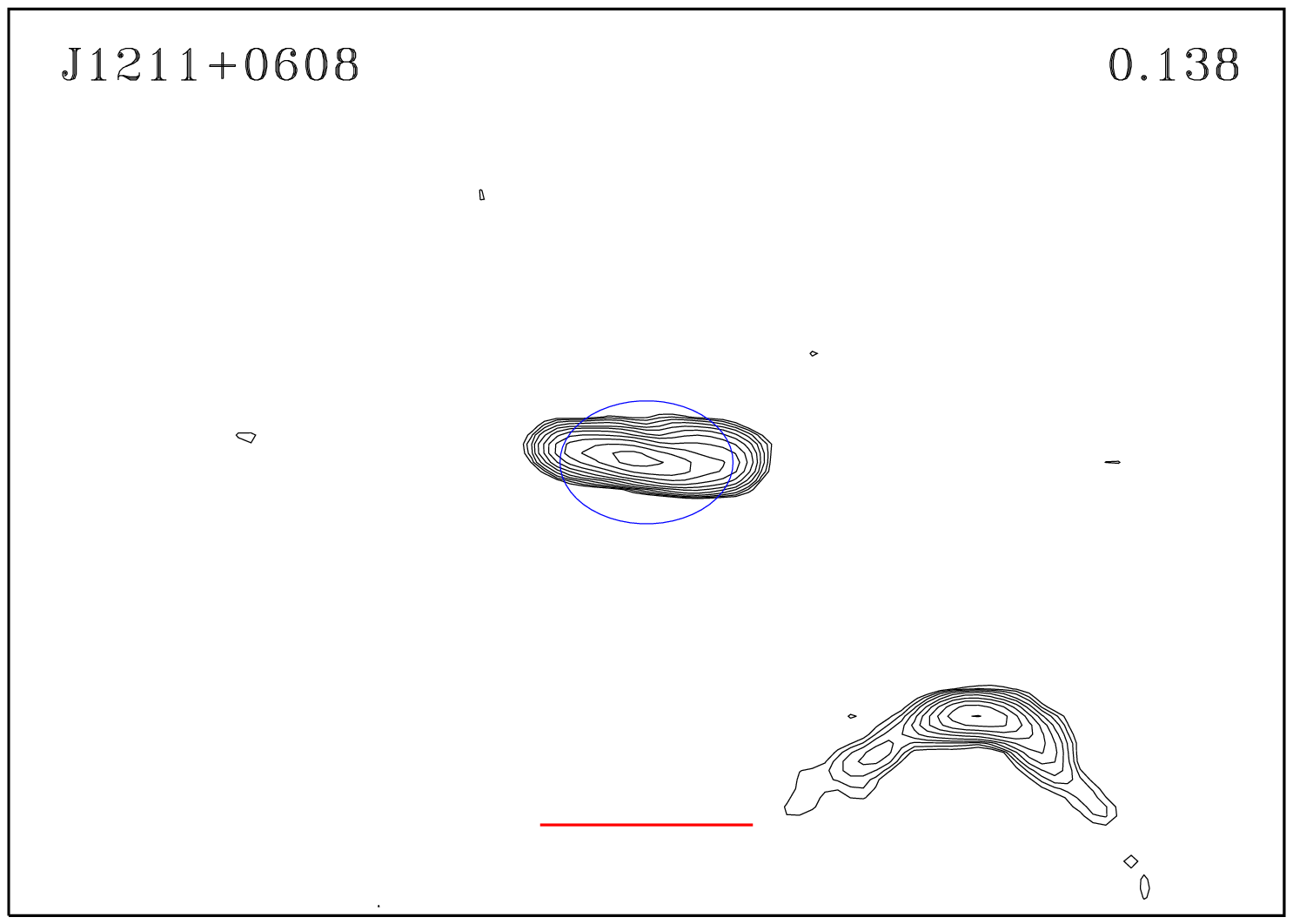} 
\includegraphics[width=6.3cm,height=6.3cm]{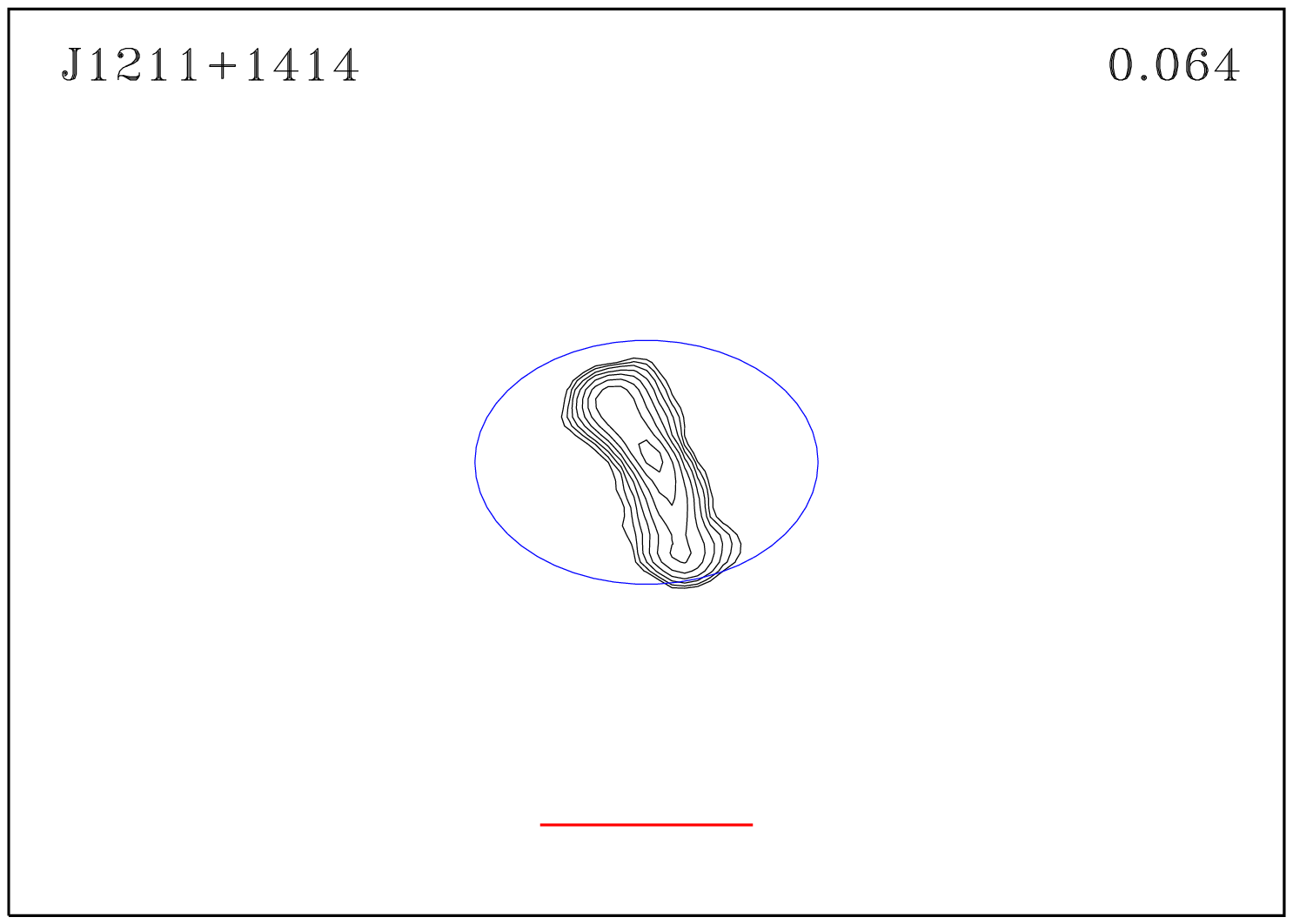} 

\includegraphics[width=6.3cm,height=6.3cm]{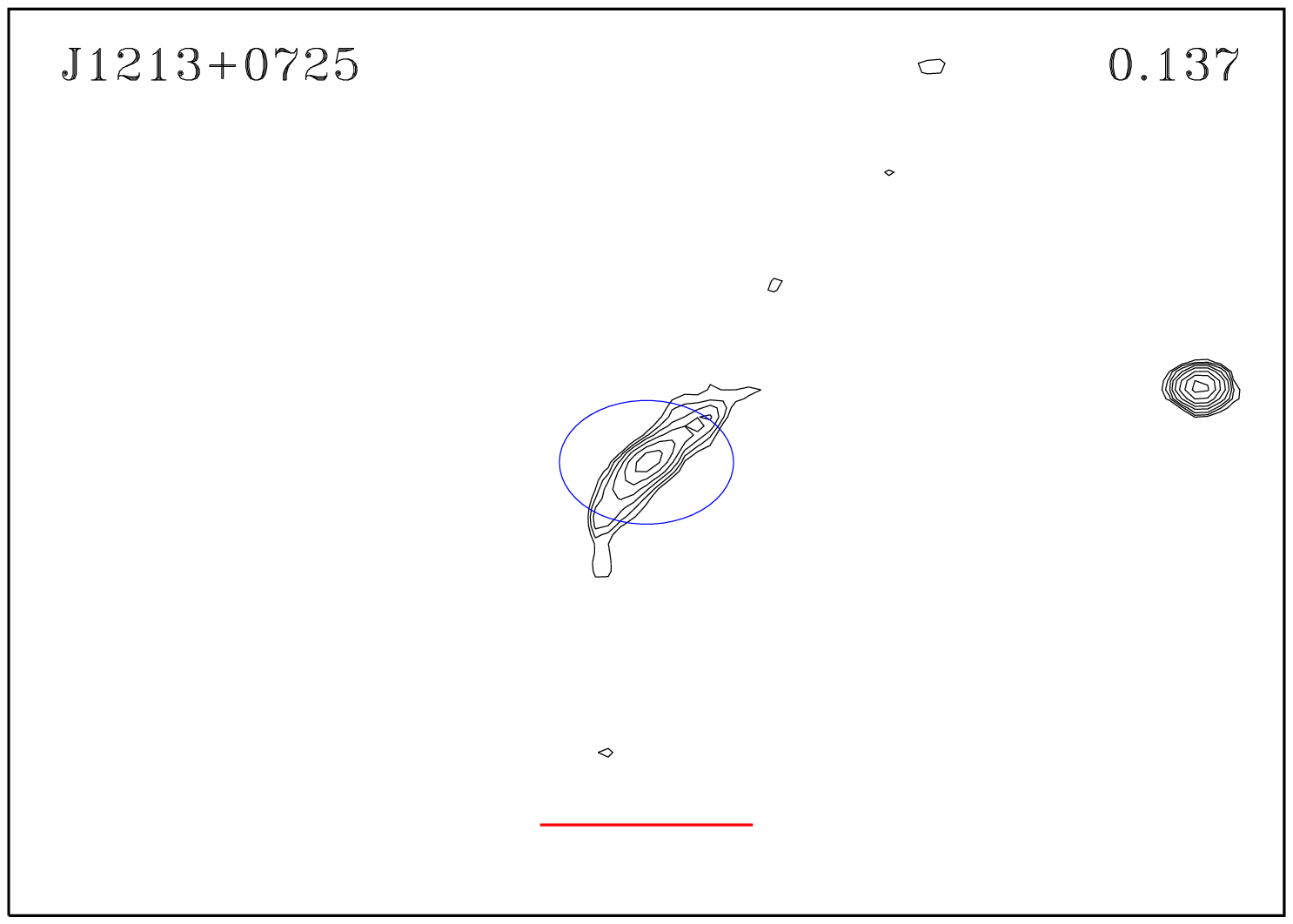} 
\includegraphics[width=6.3cm,height=6.3cm]{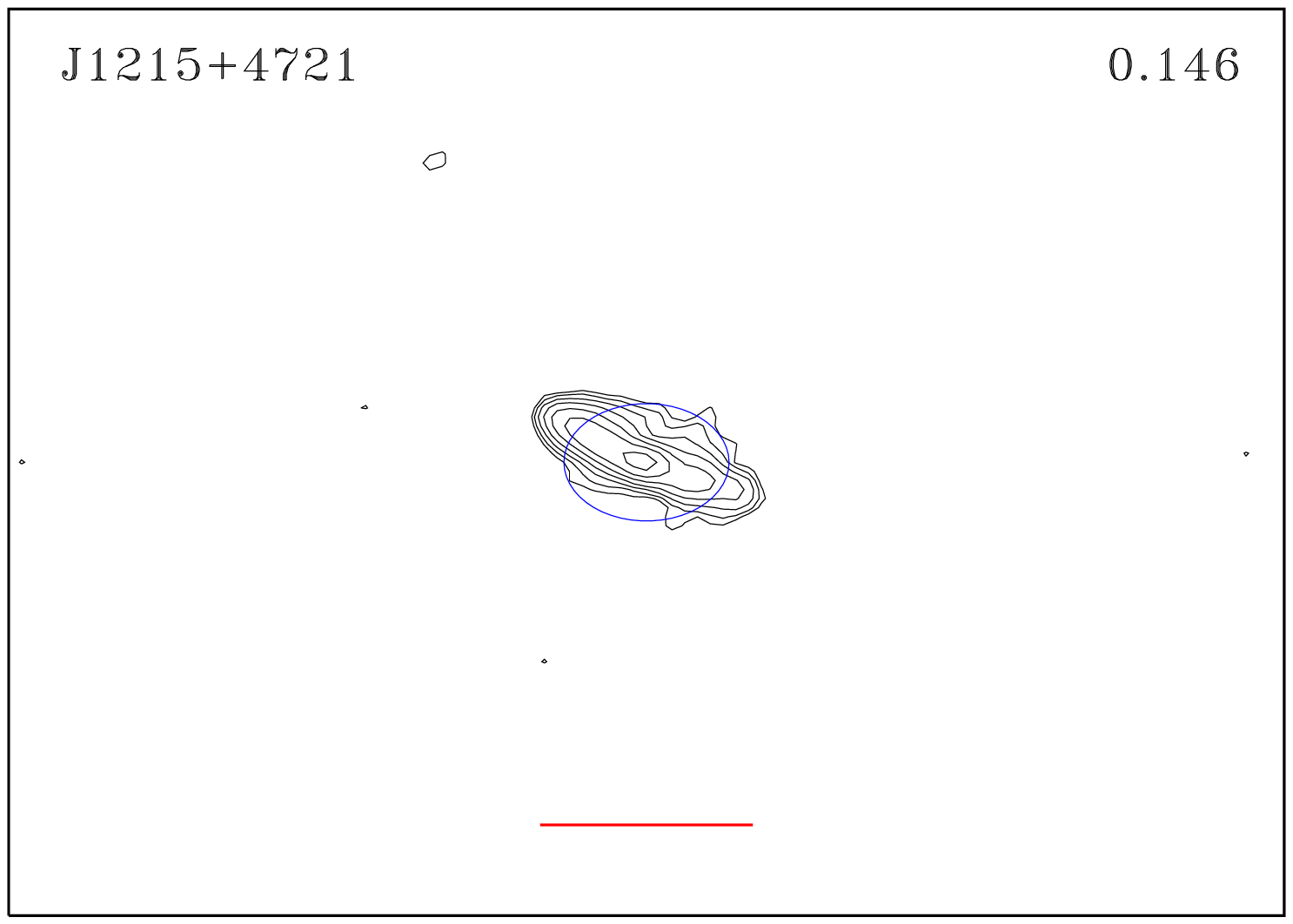} 
\includegraphics[width=6.3cm,height=6.3cm]{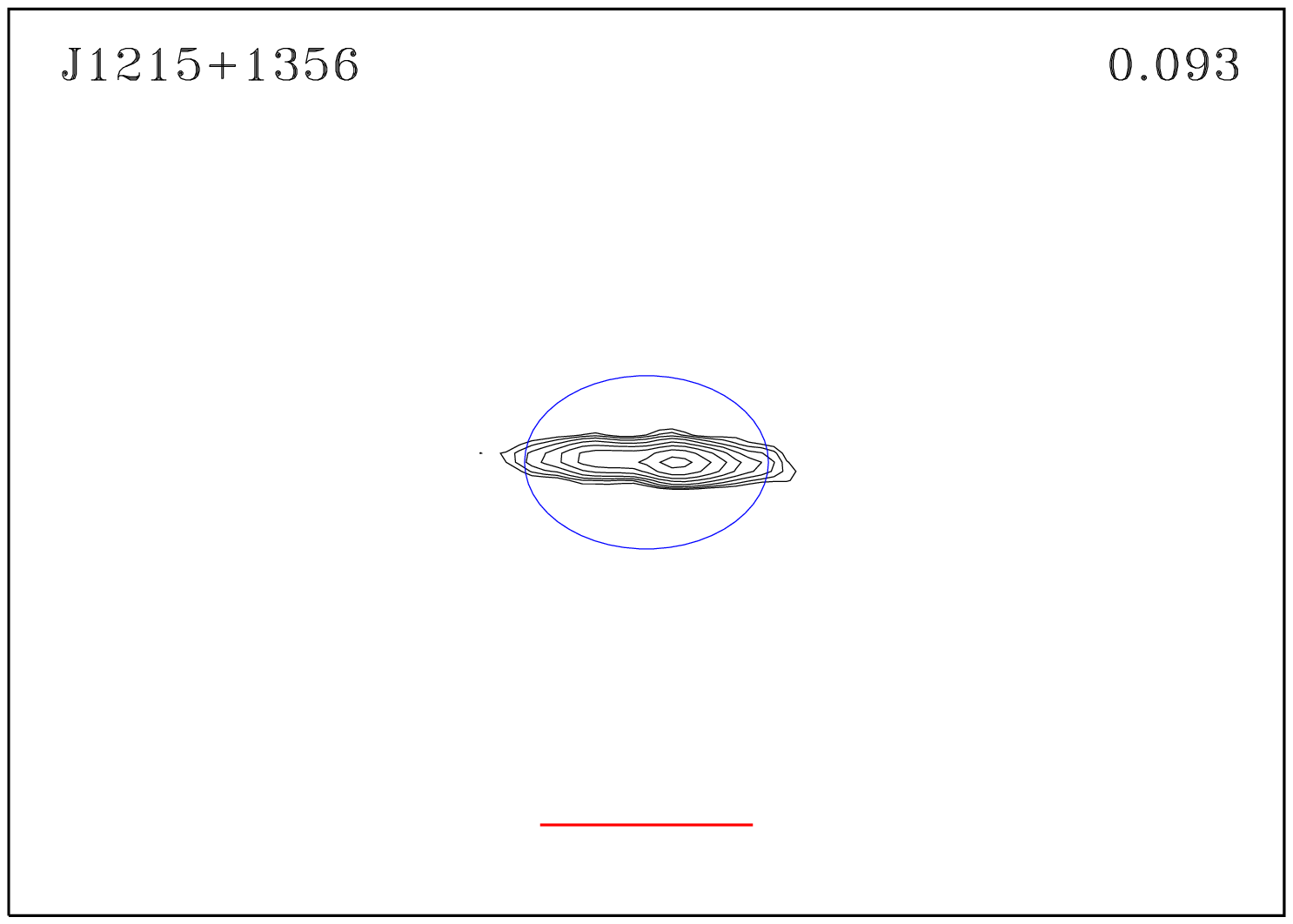} 

\includegraphics[width=6.3cm,height=6.3cm]{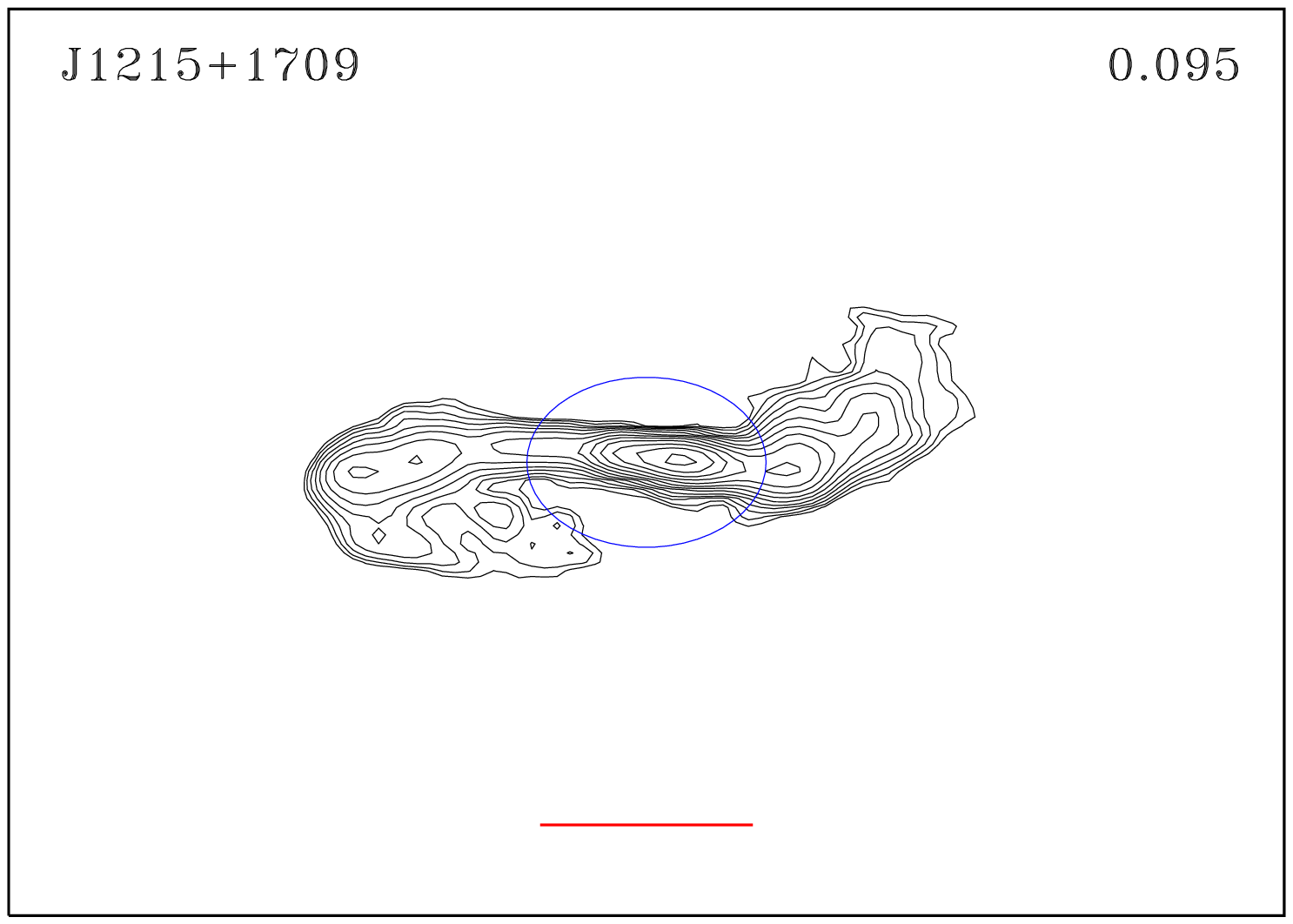} 
\includegraphics[width=6.3cm,height=6.3cm]{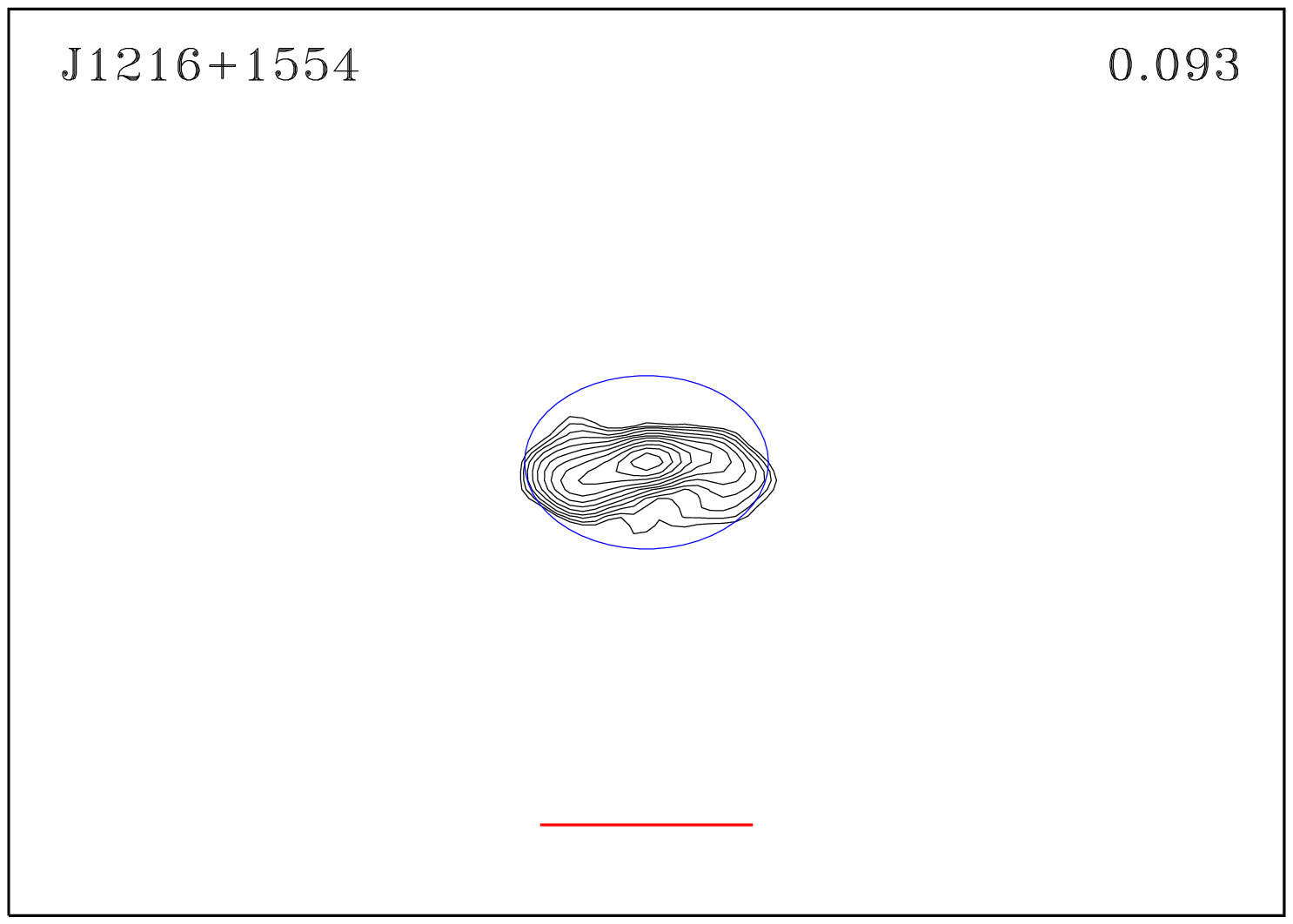} 
\includegraphics[width=6.3cm,height=6.3cm]{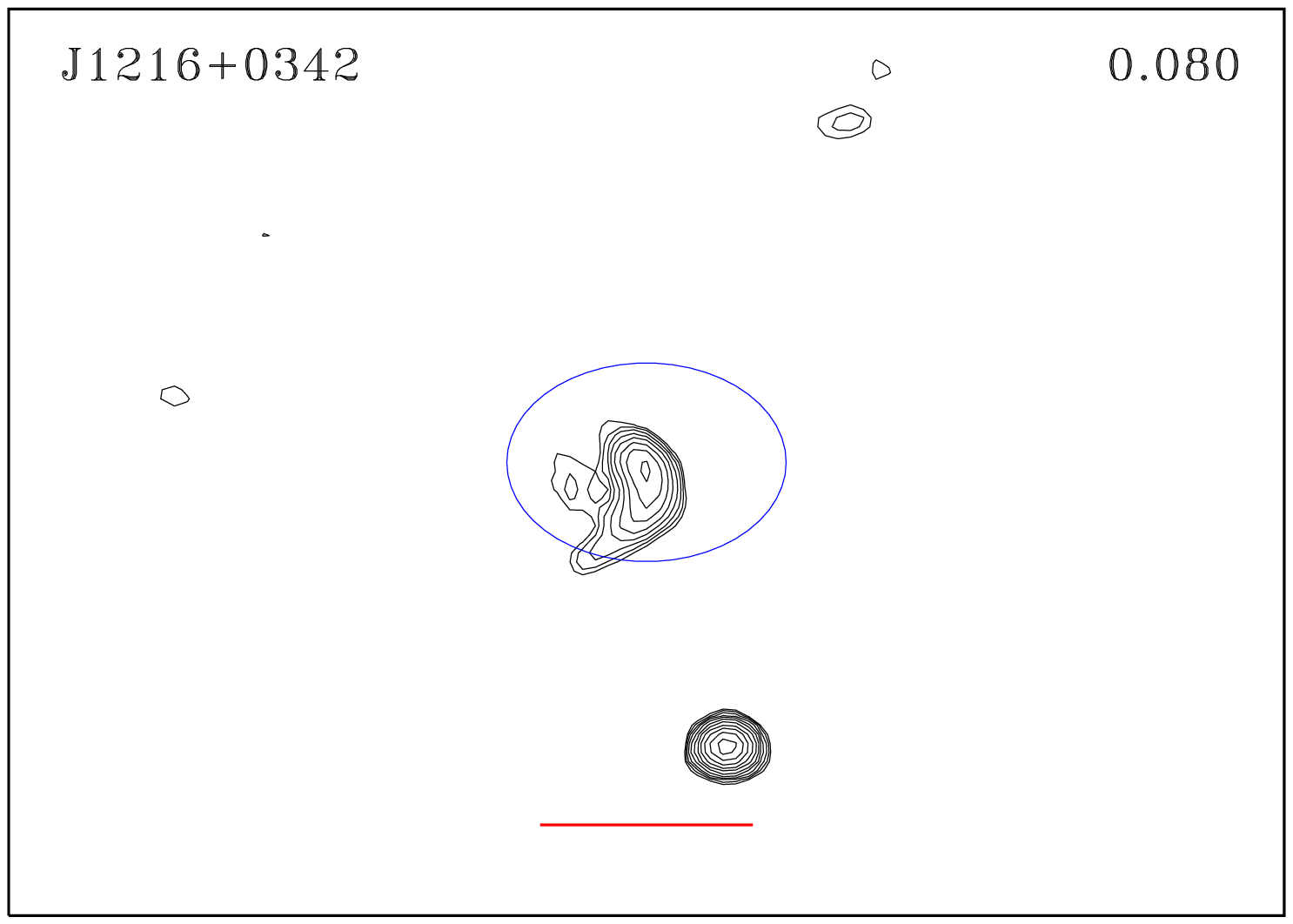} 

\includegraphics[width=6.3cm,height=6.3cm]{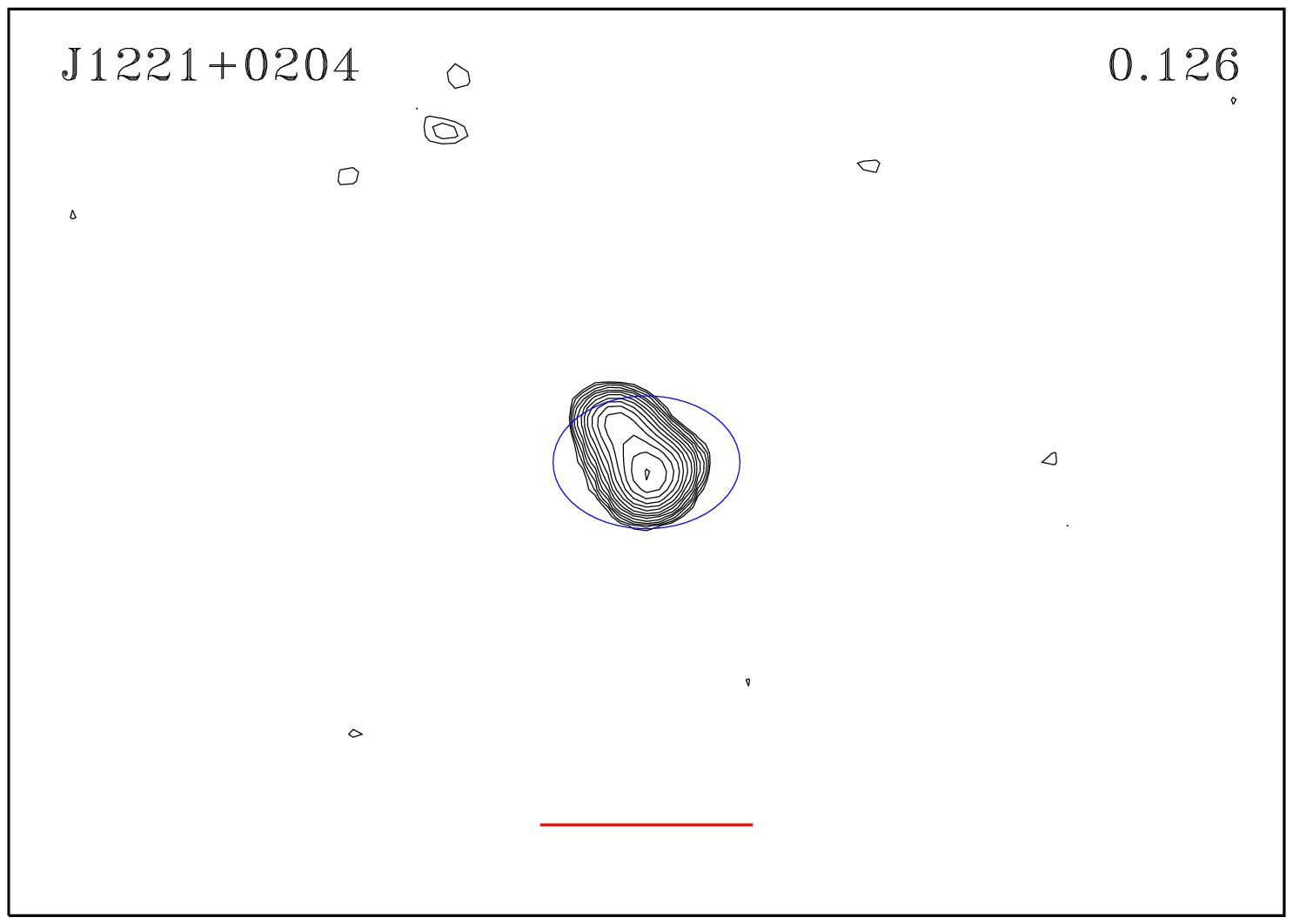} 
\includegraphics[width=6.3cm,height=6.3cm]{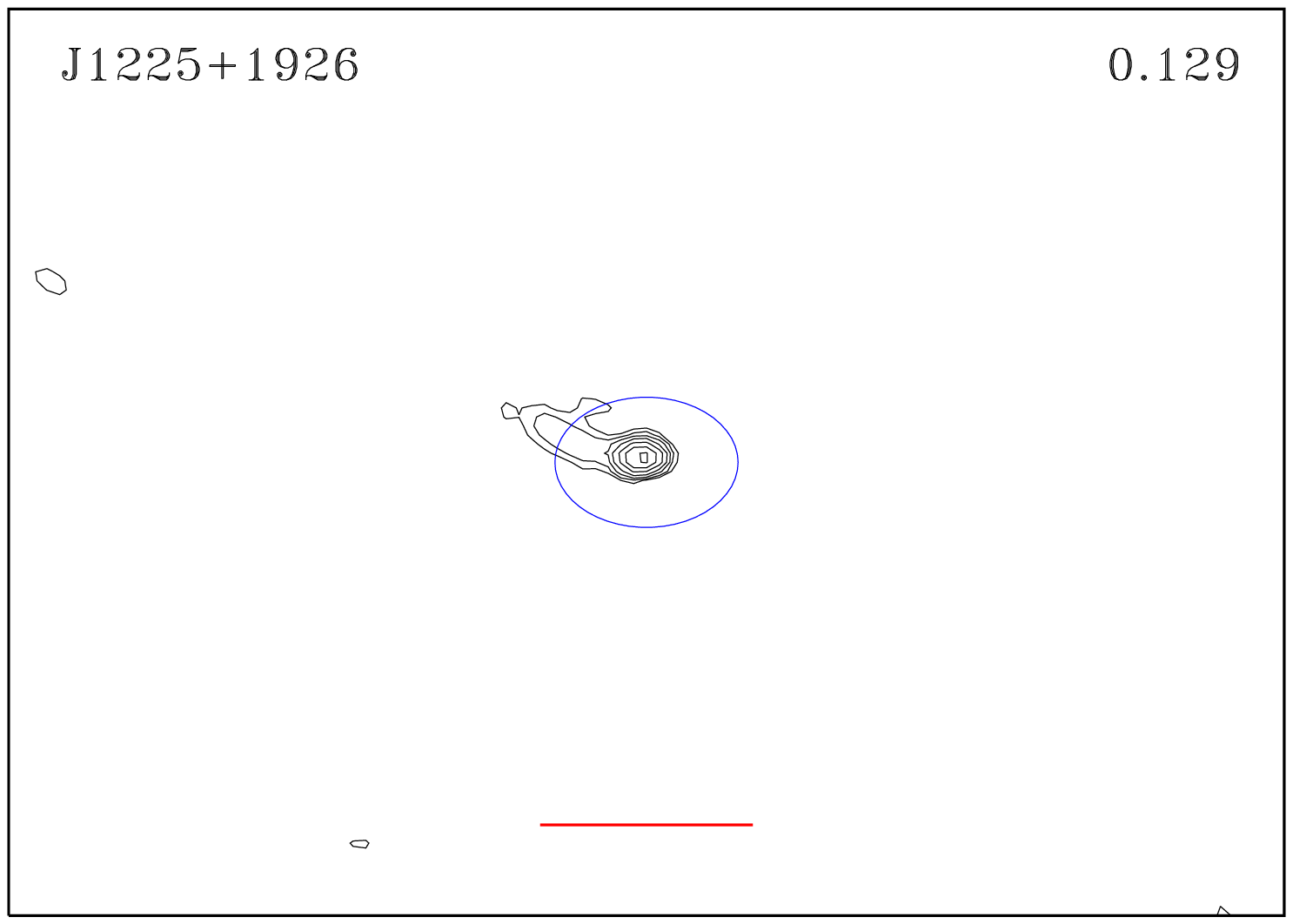} 
\includegraphics[width=6.3cm,height=6.3cm]{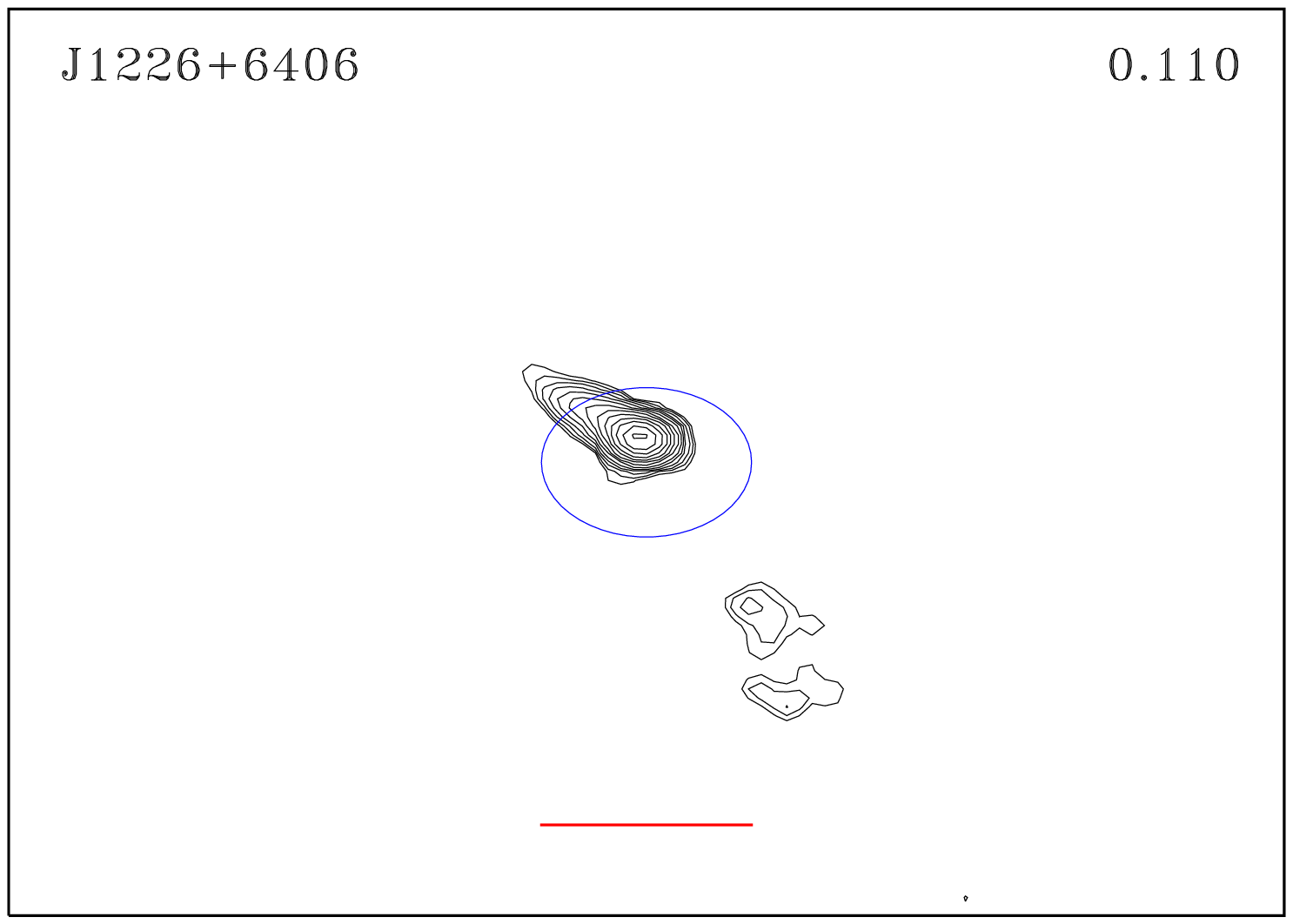} 
\caption{(continued)}
\end{figure*}

\addtocounter{figure}{-1}
\begin{figure*}

\includegraphics[width=6.3cm,height=6.3cm]{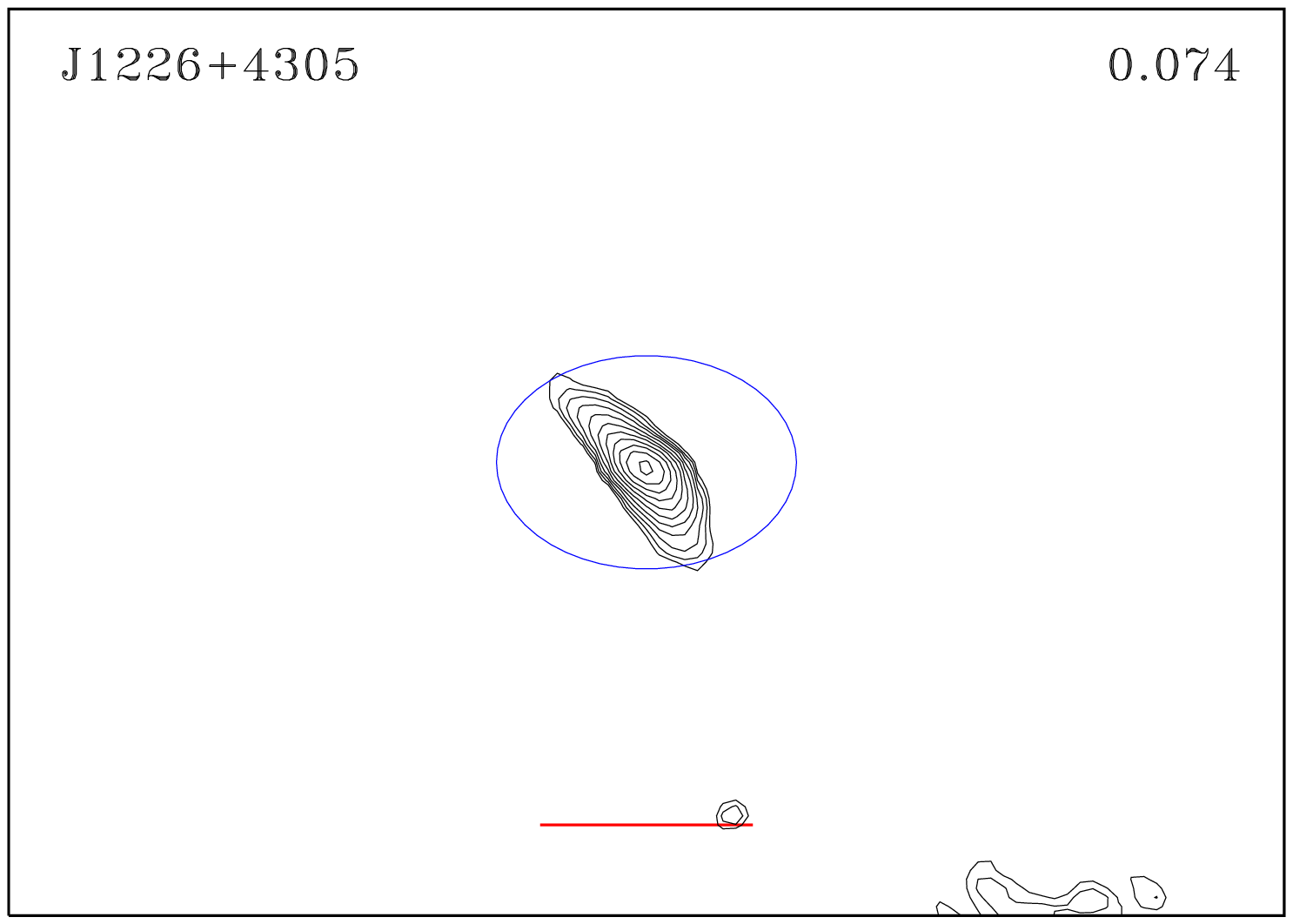} 
\includegraphics[width=6.3cm,height=6.3cm]{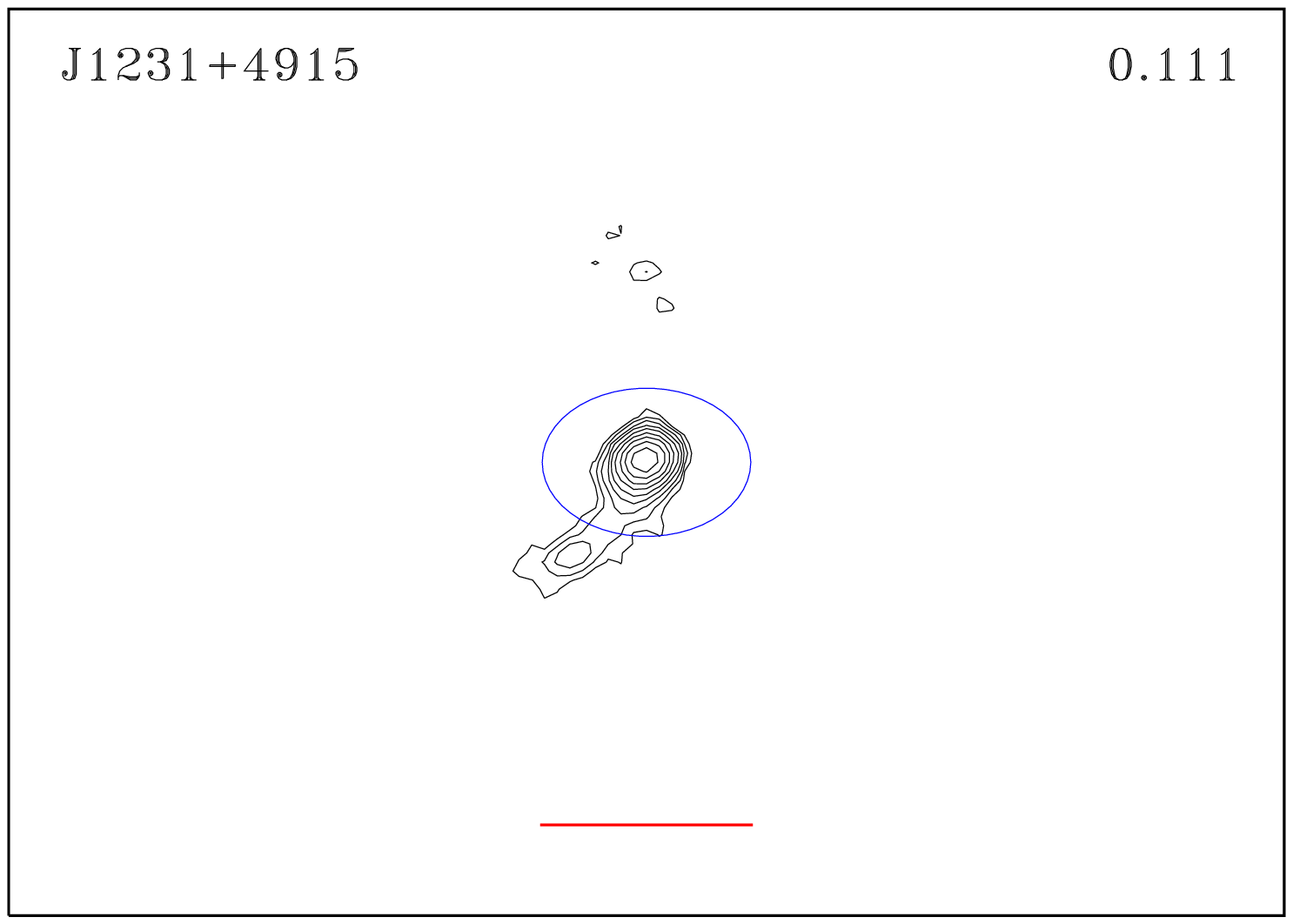} 
\includegraphics[width=6.3cm,height=6.3cm]{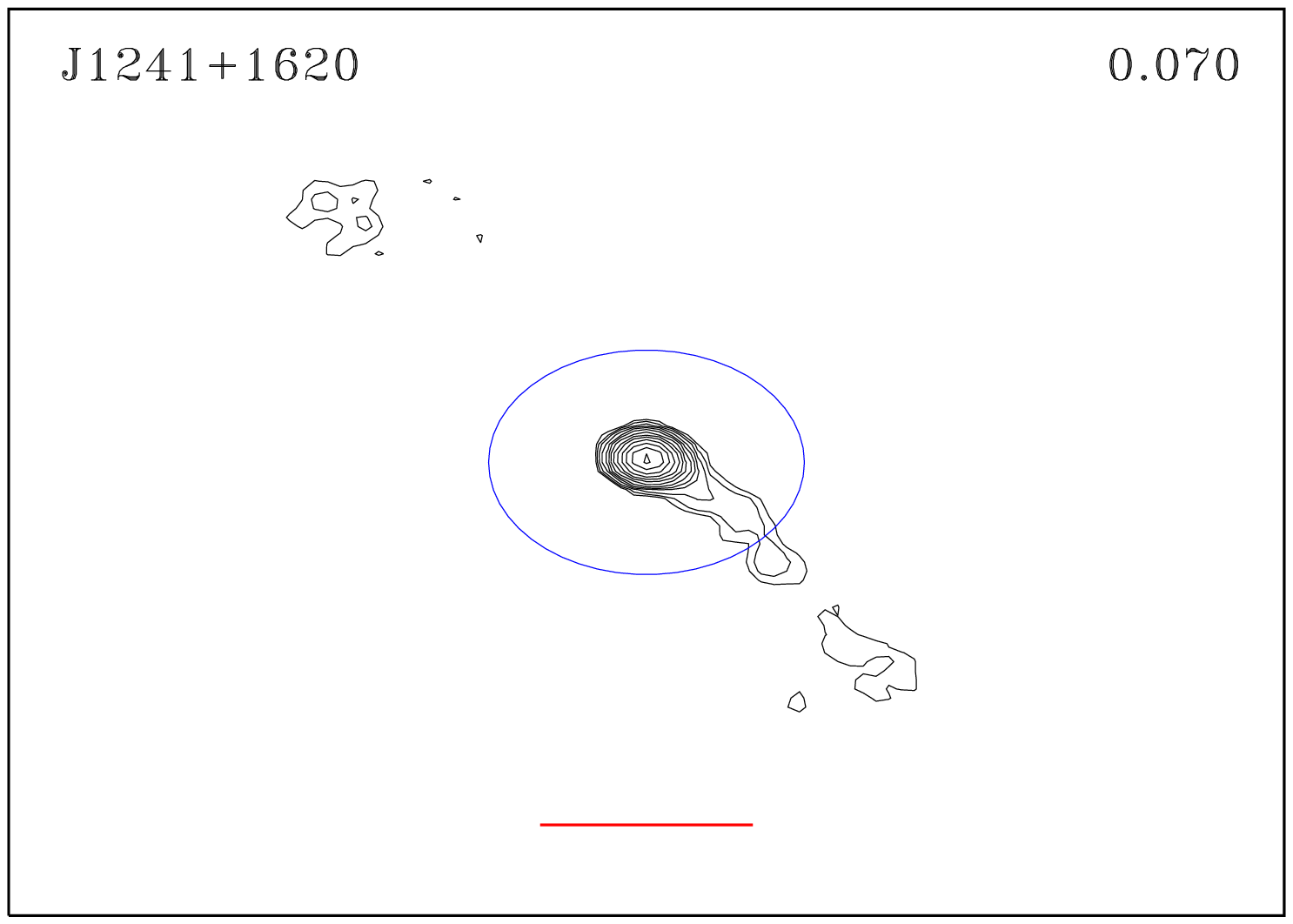} 

\includegraphics[width=6.3cm,height=6.3cm]{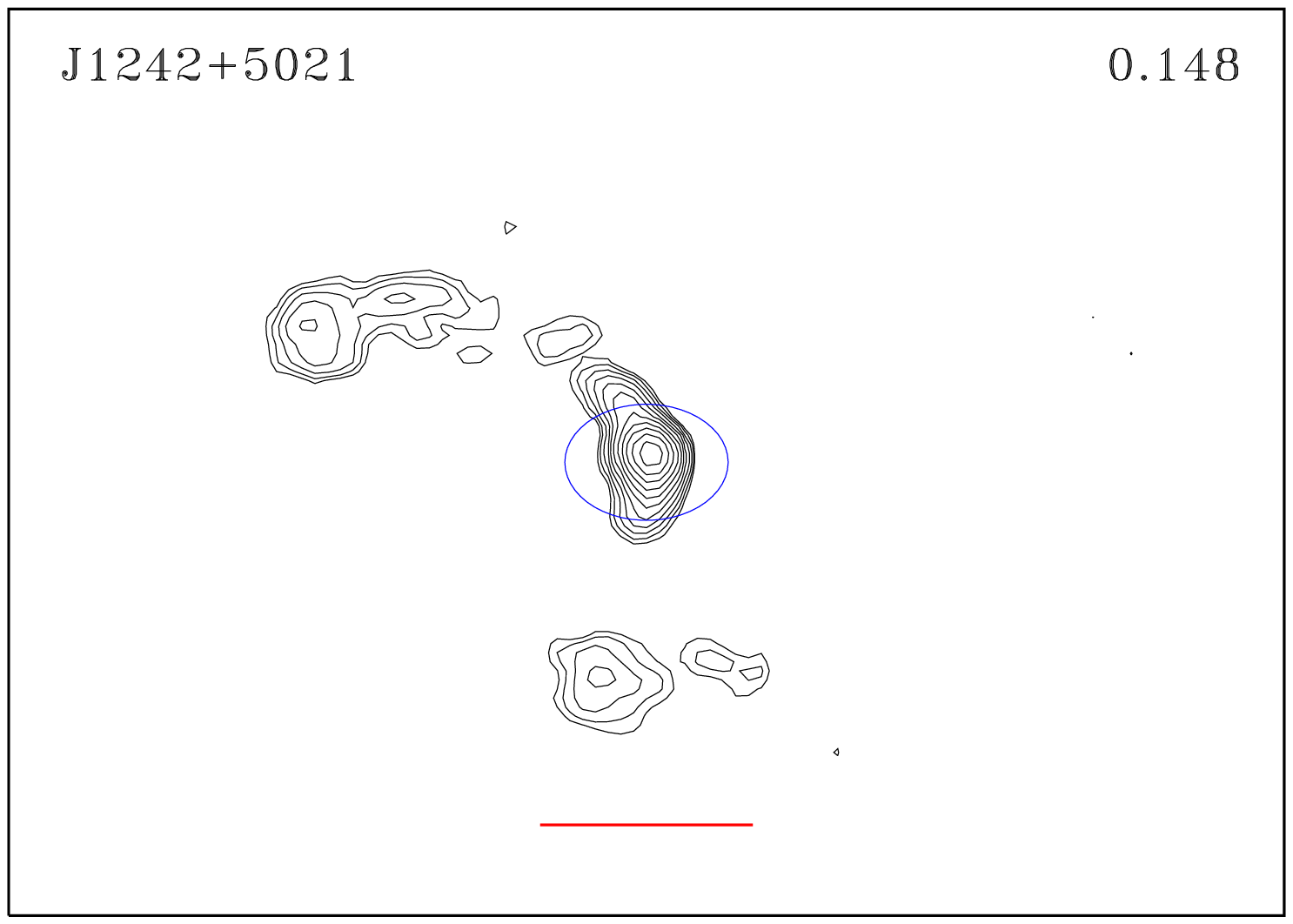} 
\includegraphics[width=6.3cm,height=6.3cm]{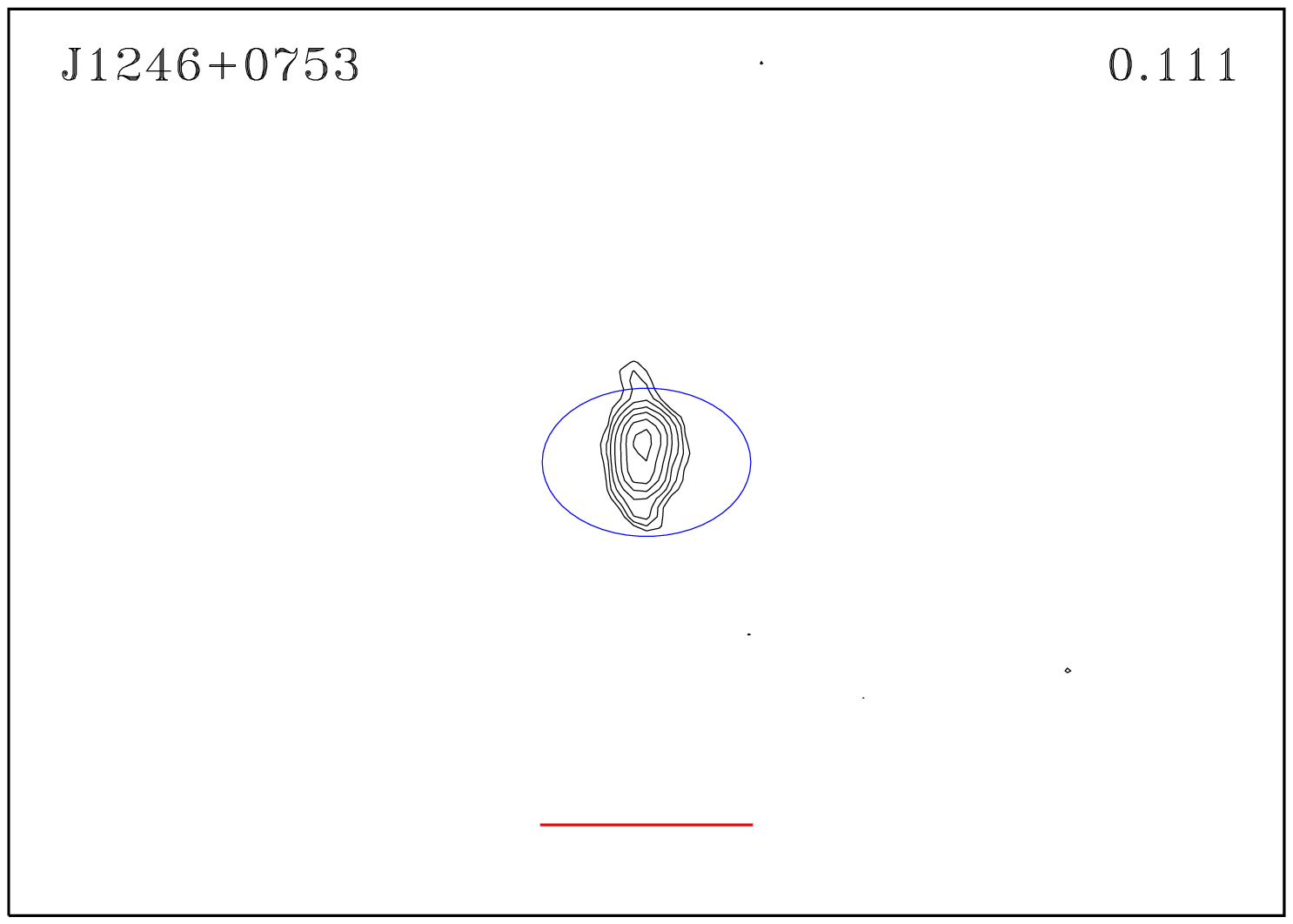} 
\includegraphics[width=6.3cm,height=6.3cm]{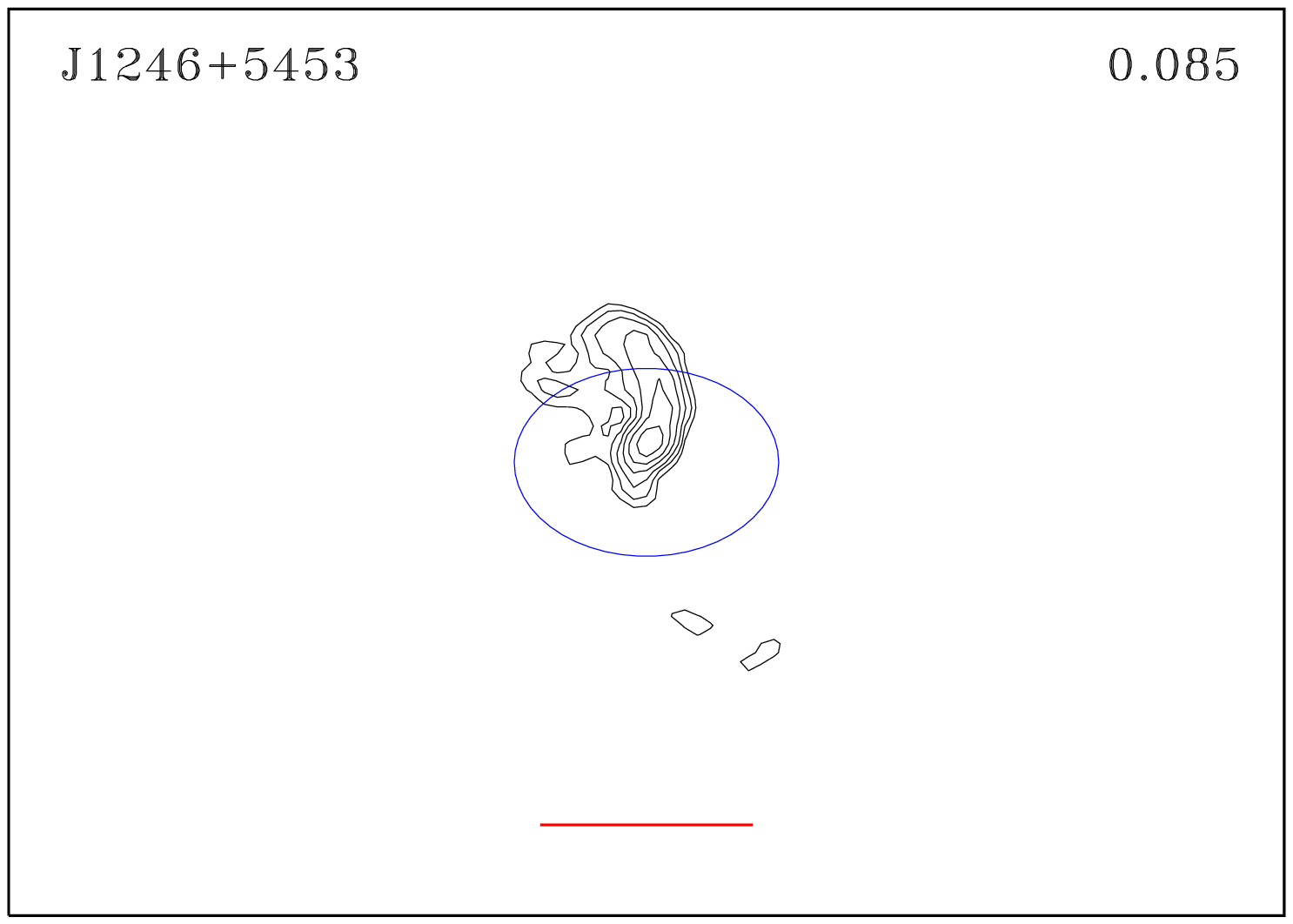} 

\includegraphics[width=6.3cm,height=6.3cm]{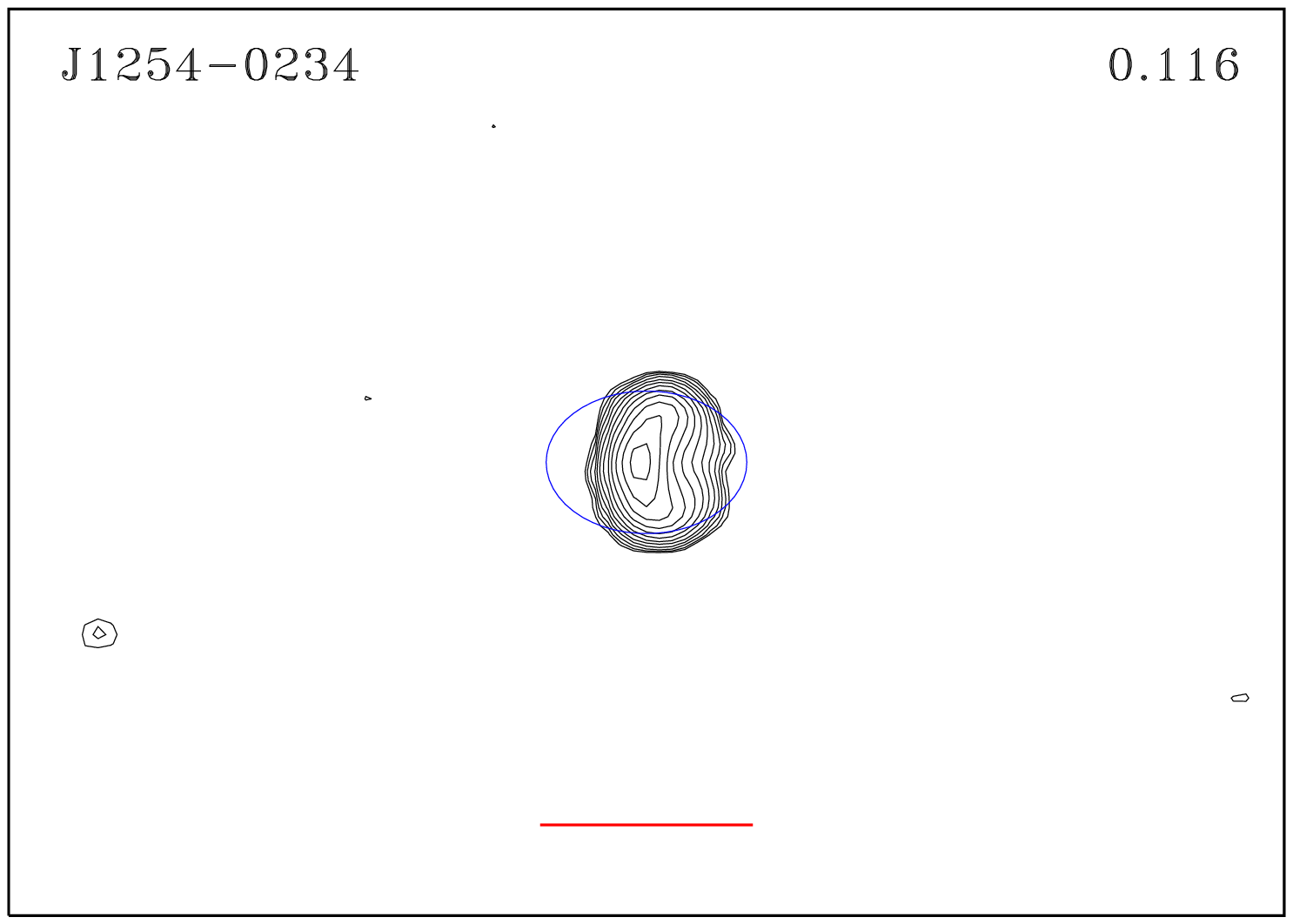} 
\includegraphics[width=6.3cm,height=6.3cm]{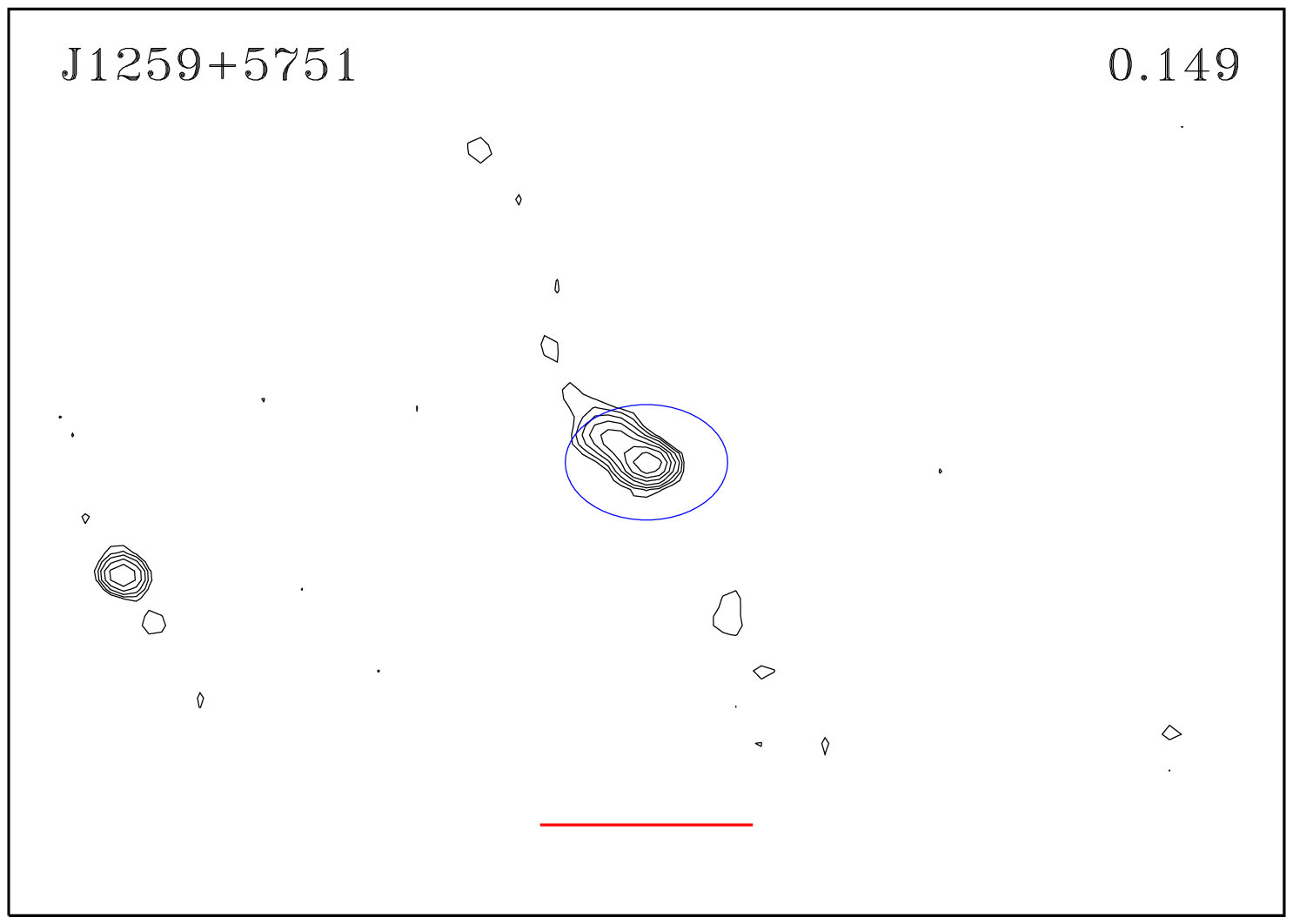} 
\includegraphics[width=6.3cm,height=6.3cm]{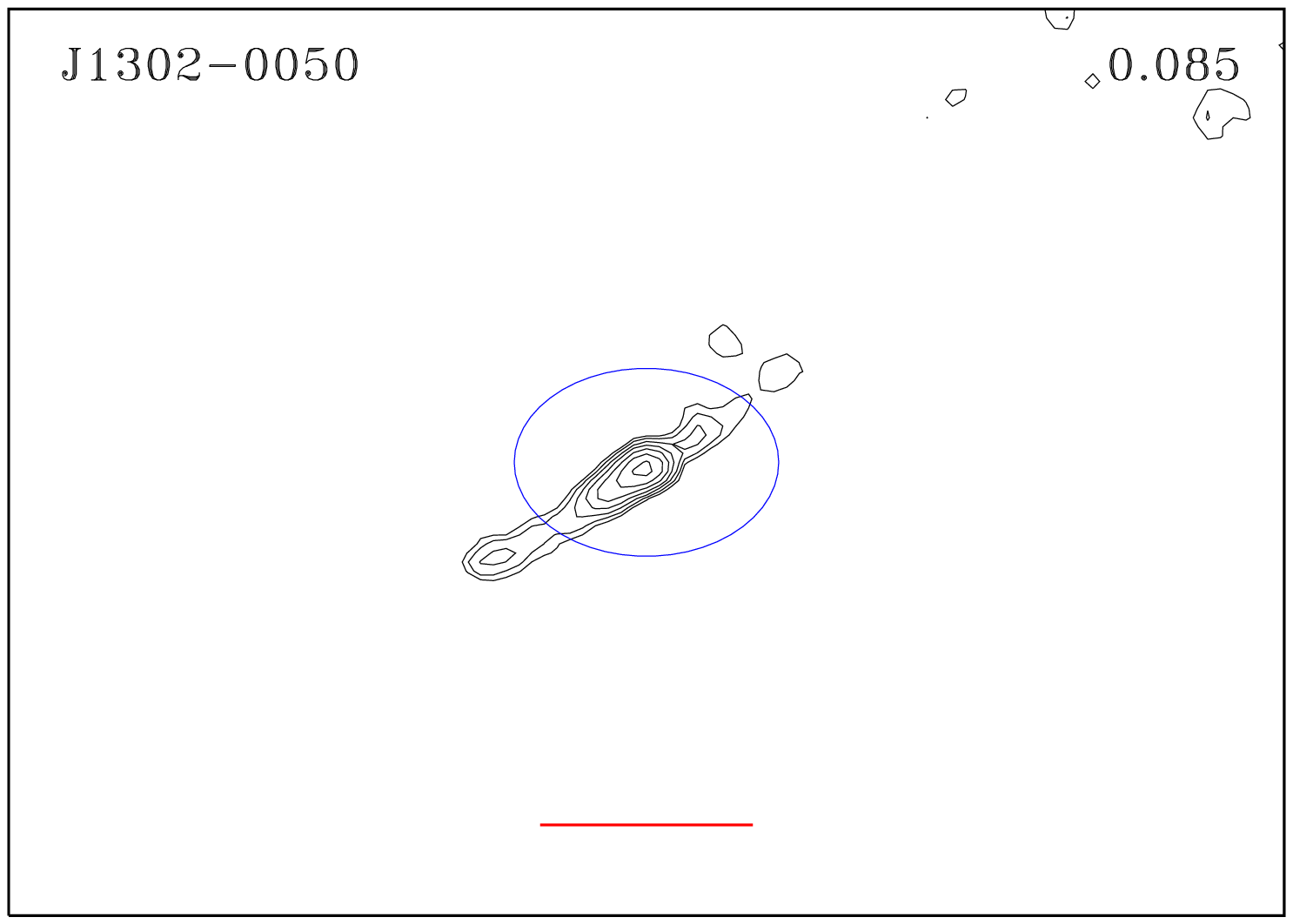} 

\includegraphics[width=6.3cm,height=6.3cm]{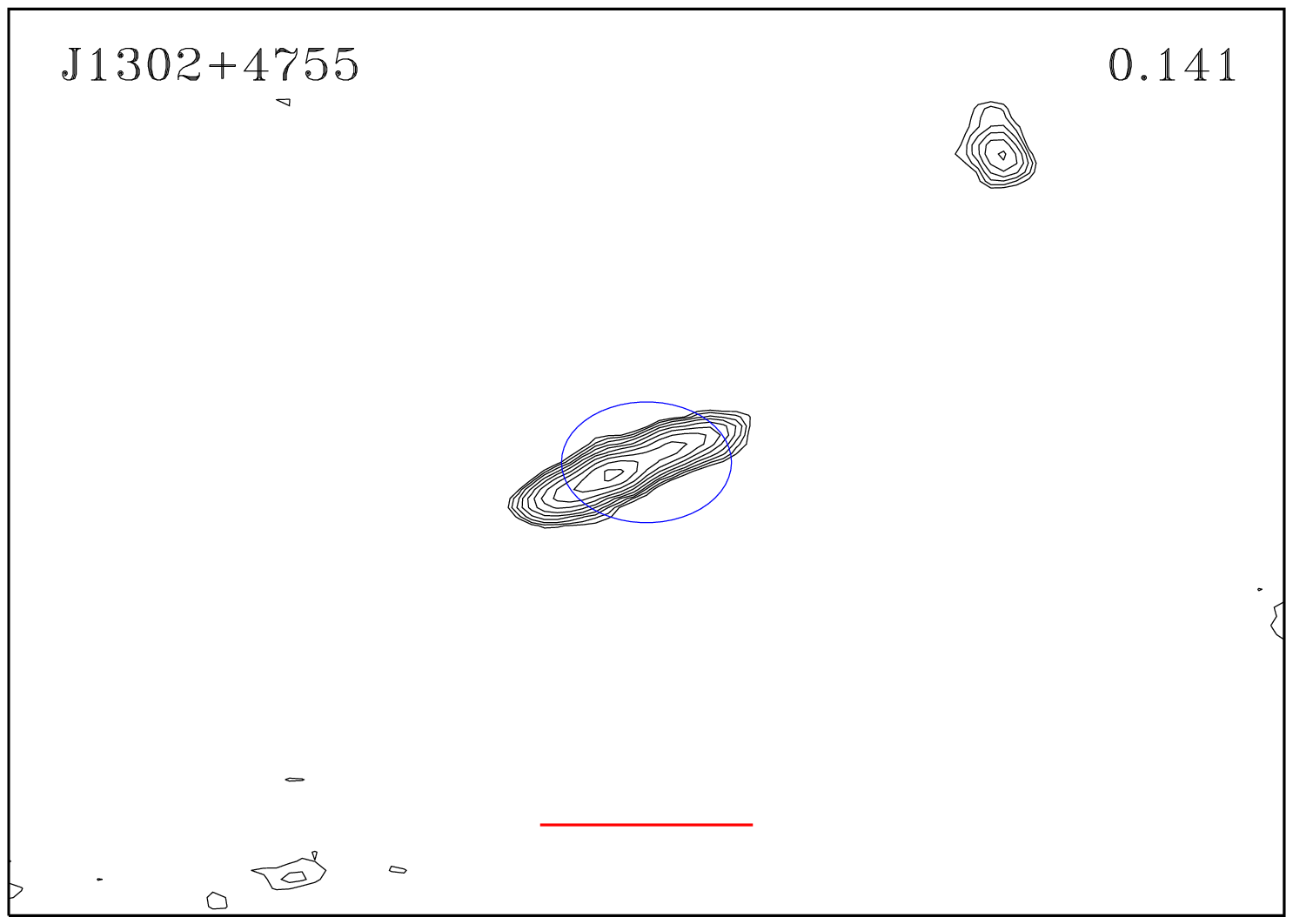} 
\includegraphics[width=6.3cm,height=6.3cm]{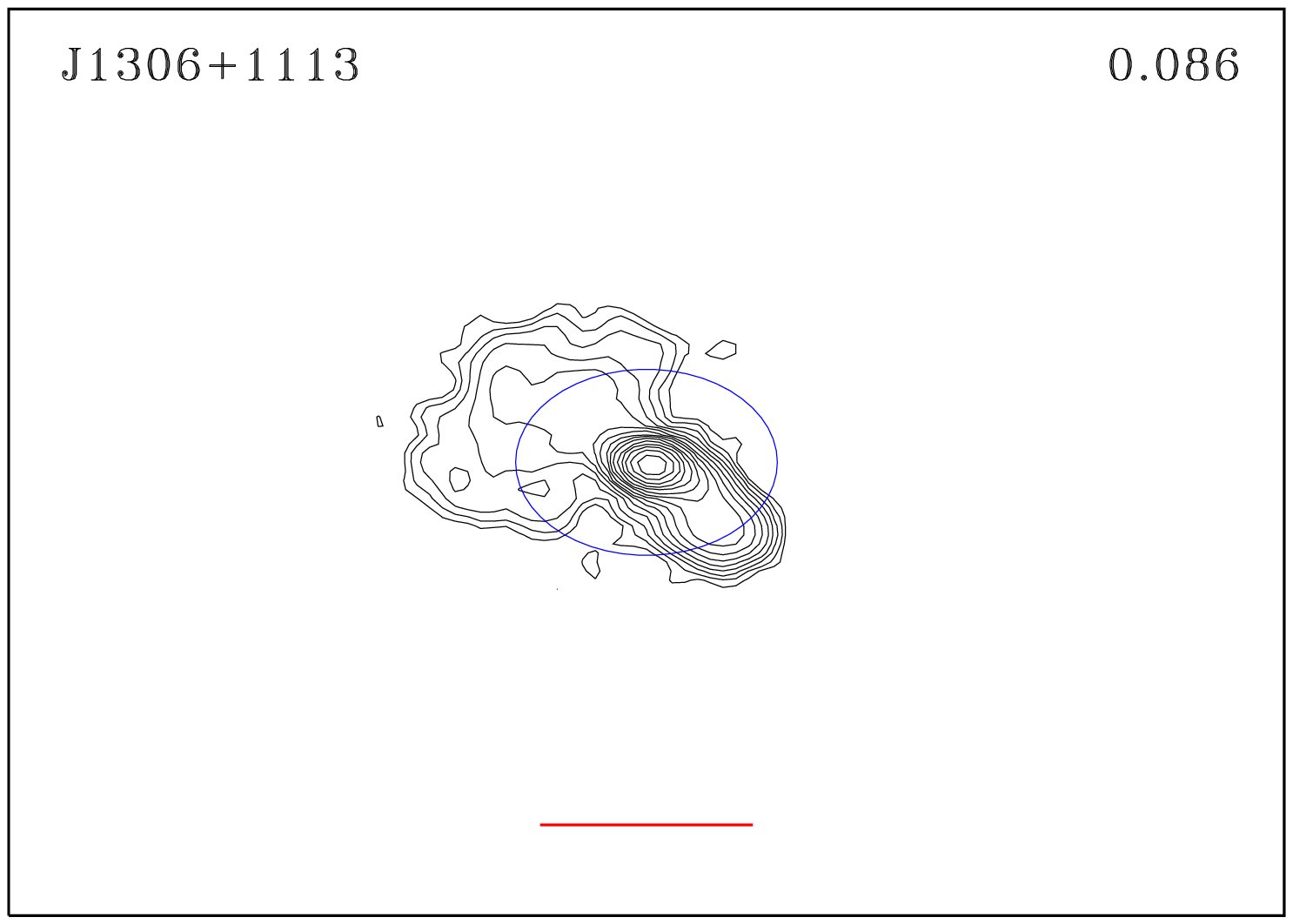} 
\includegraphics[width=6.3cm,height=6.3cm]{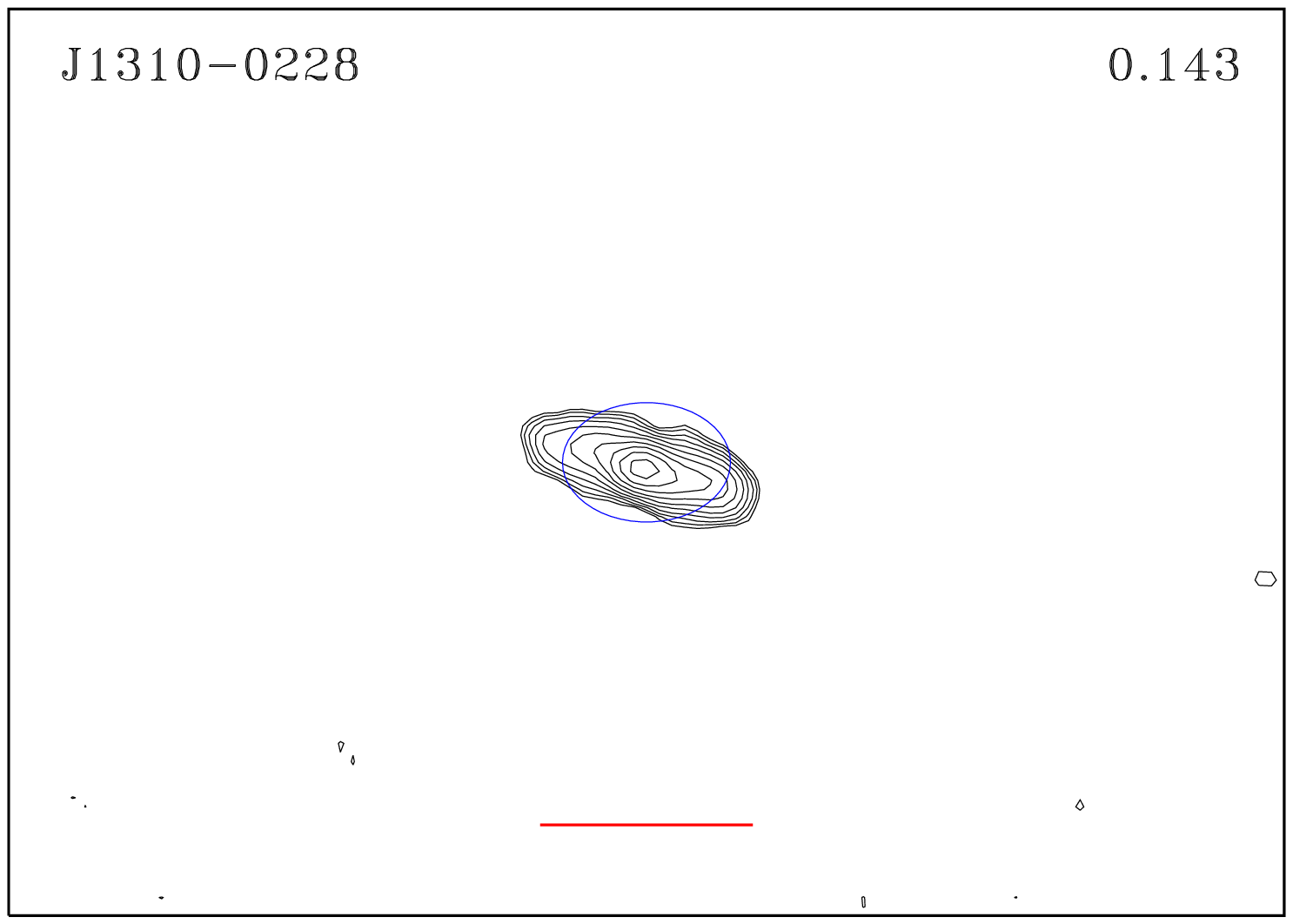} 
\caption{(continued)}
\end{figure*}

\addtocounter{figure}{-1}
\begin{figure*}
\includegraphics[width=6.3cm,height=6.3cm]{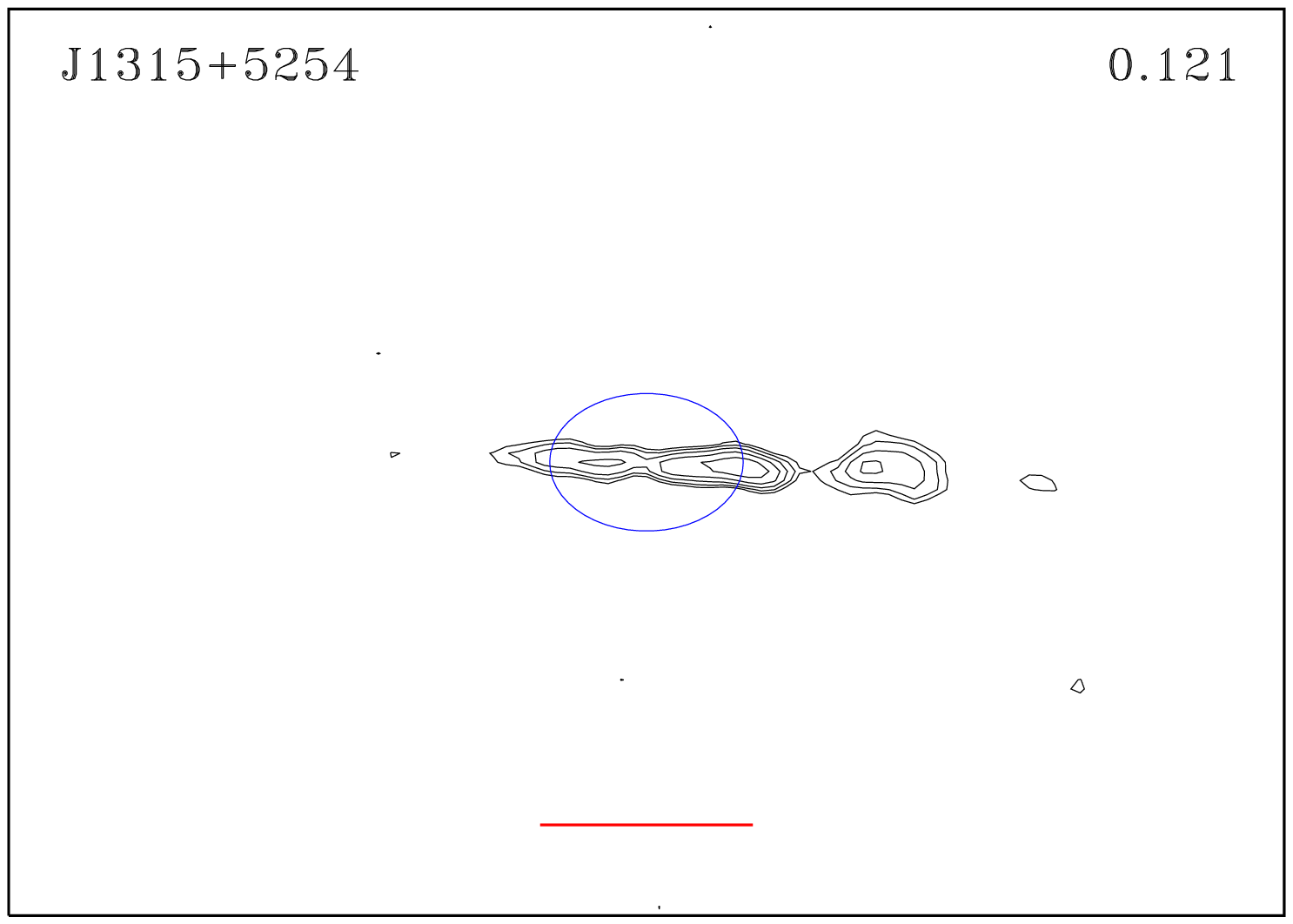} 
\includegraphics[width=6.3cm,height=6.3cm]{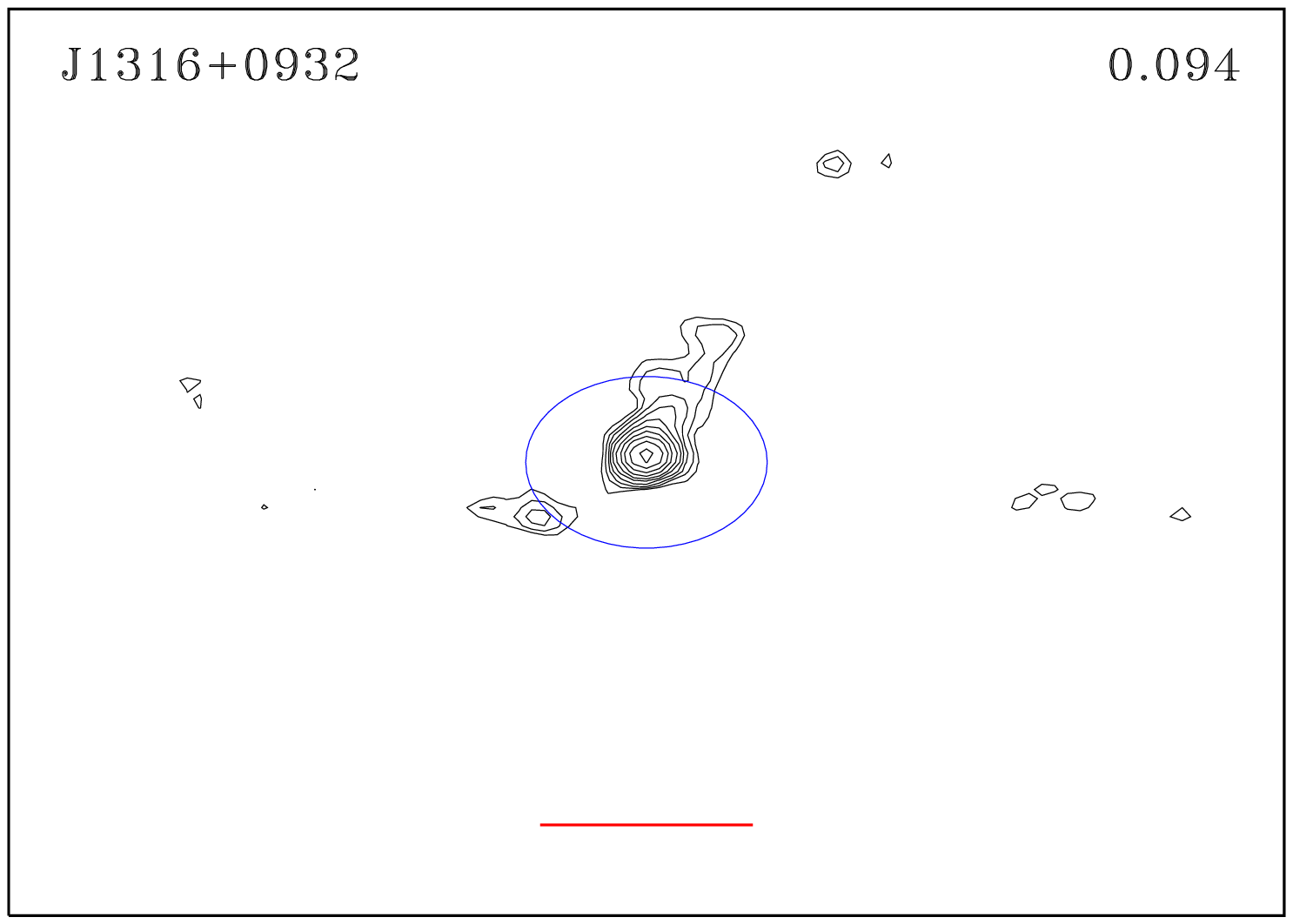} 
\includegraphics[width=6.3cm,height=6.3cm]{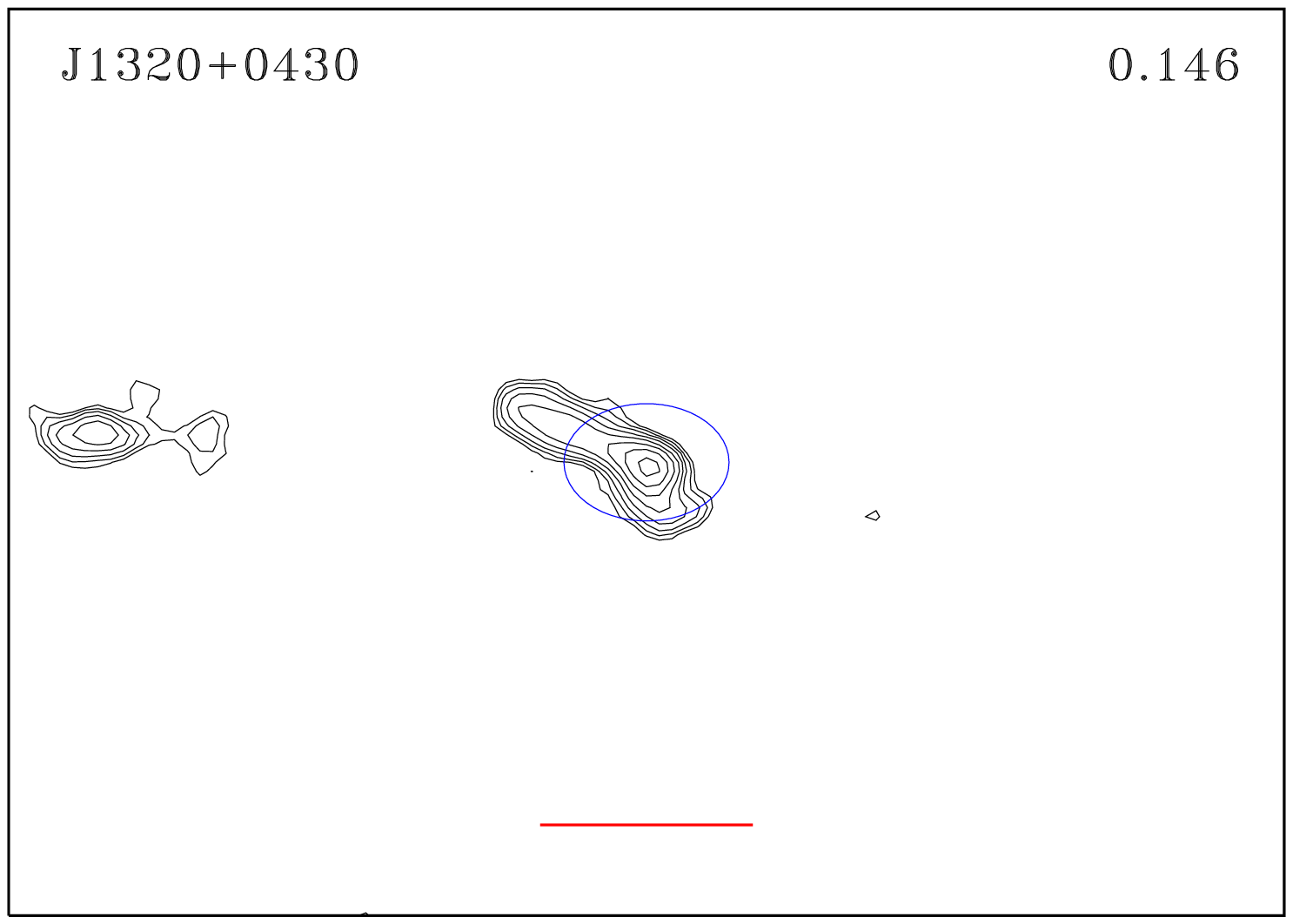} 

\includegraphics[width=6.3cm,height=6.3cm]{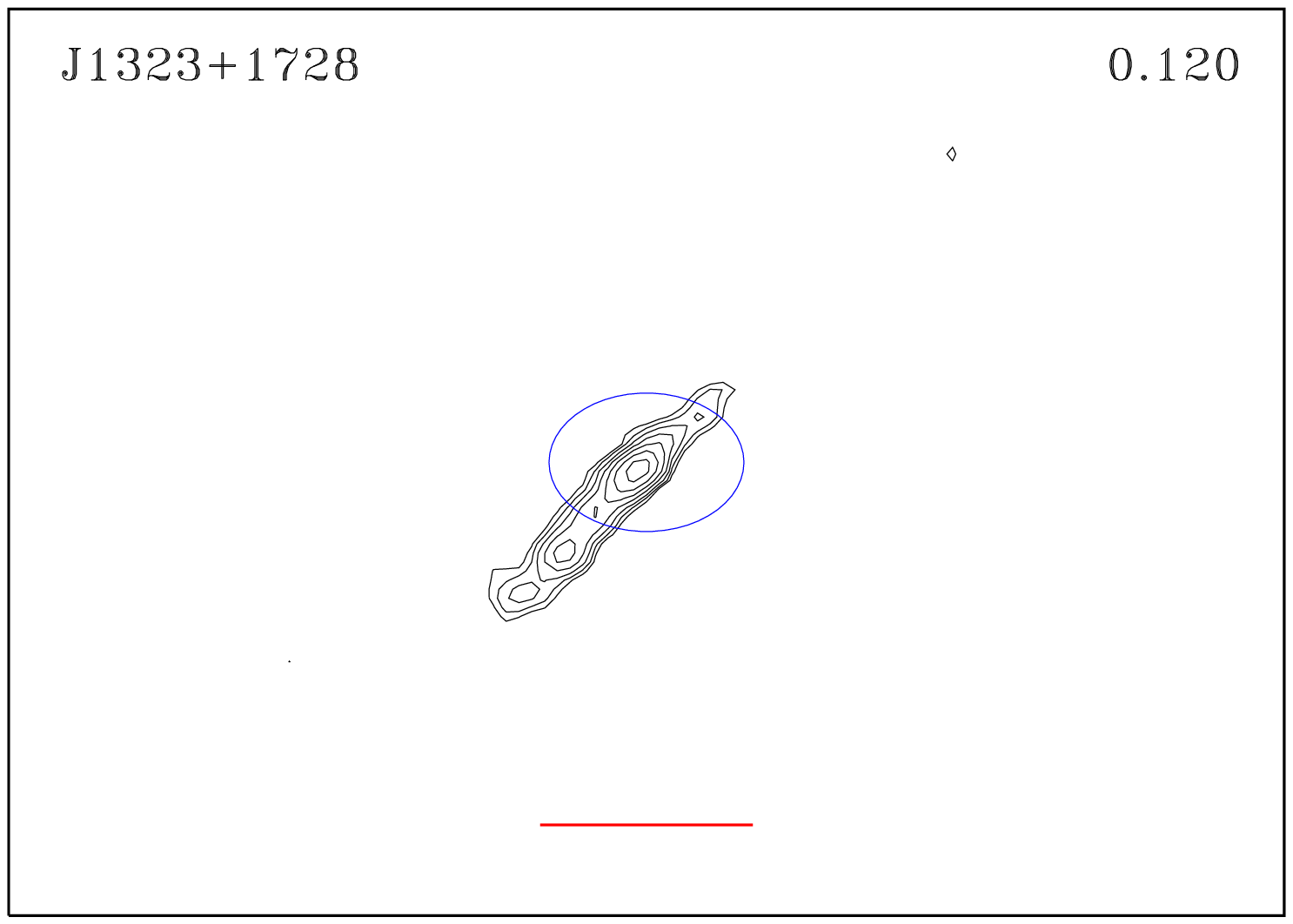} 
\includegraphics[width=6.3cm,height=6.3cm]{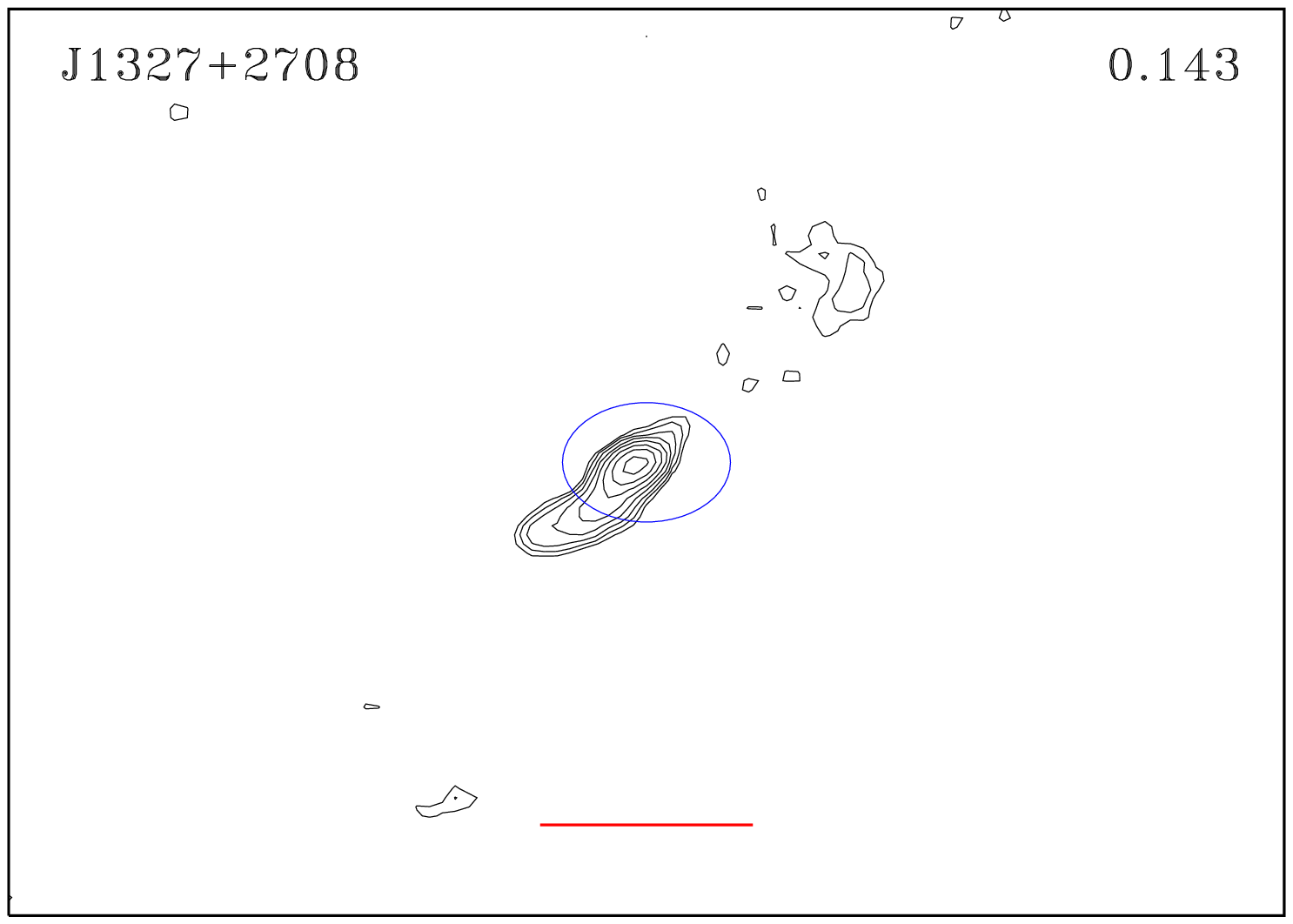} 
\includegraphics[width=6.3cm,height=6.3cm]{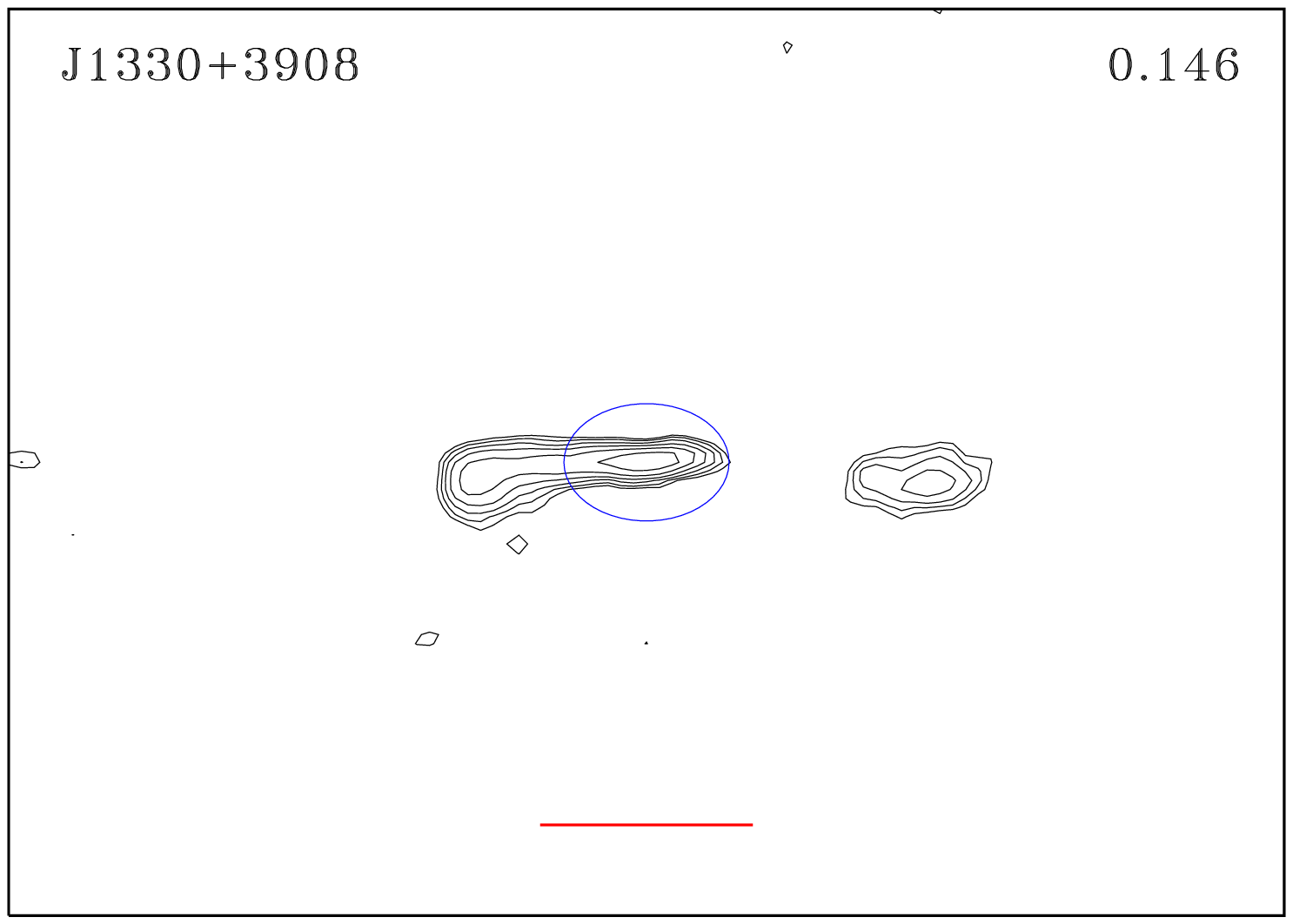} 

\includegraphics[width=6.3cm,height=6.3cm]{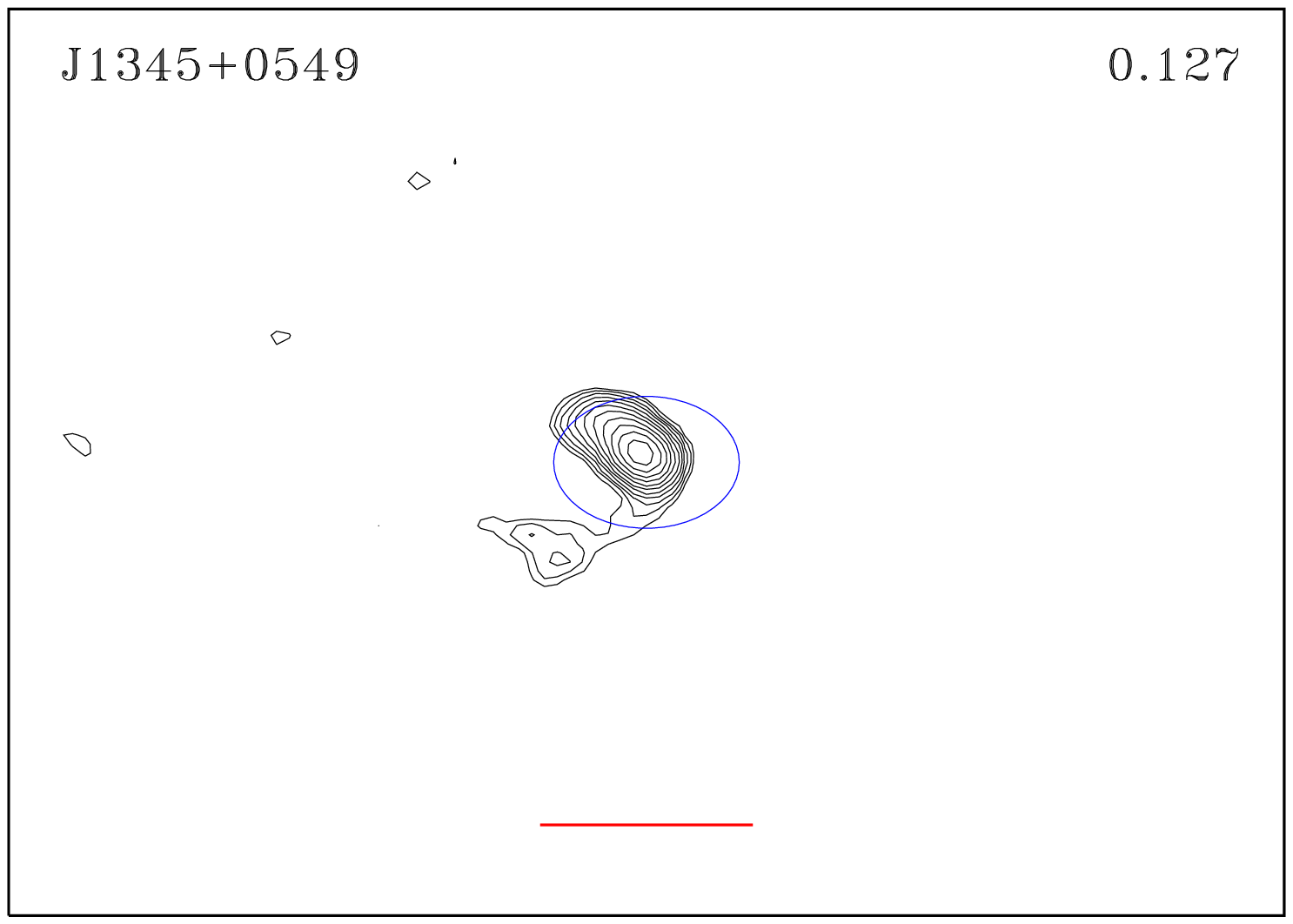} 
\includegraphics[width=6.3cm,height=6.3cm]{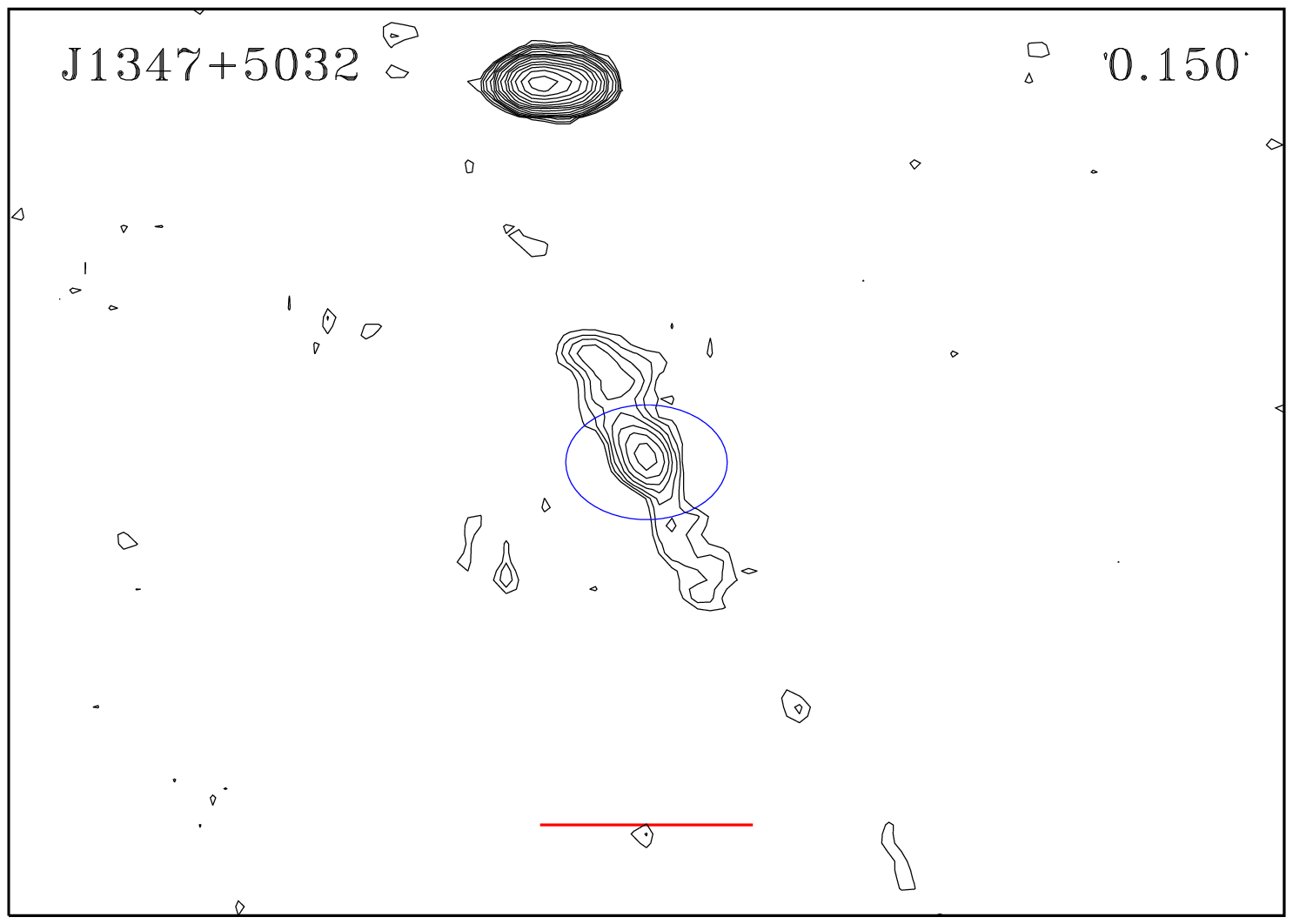} 
\includegraphics[width=6.3cm,height=6.3cm]{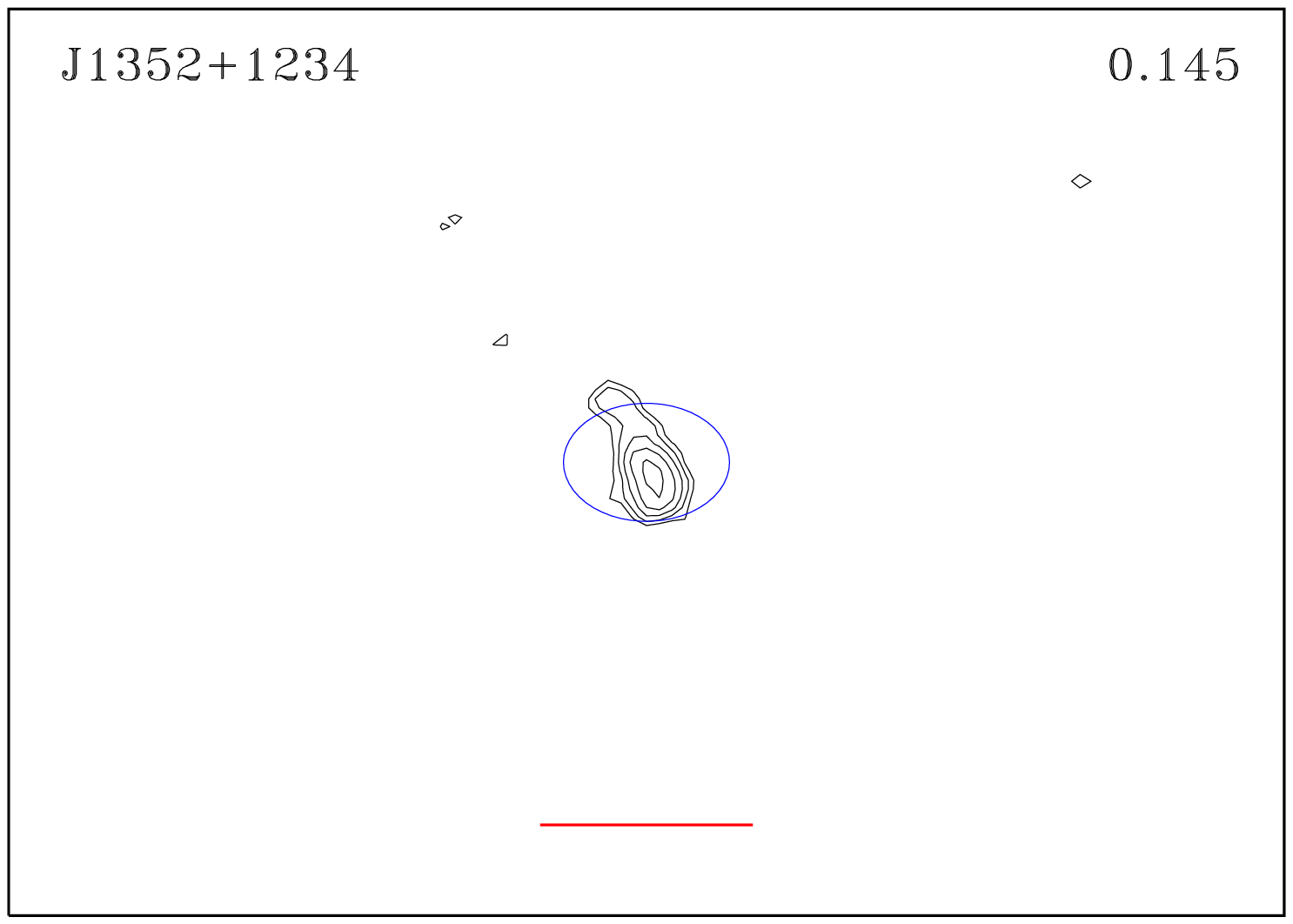} 

\includegraphics[width=6.3cm,height=6.3cm]{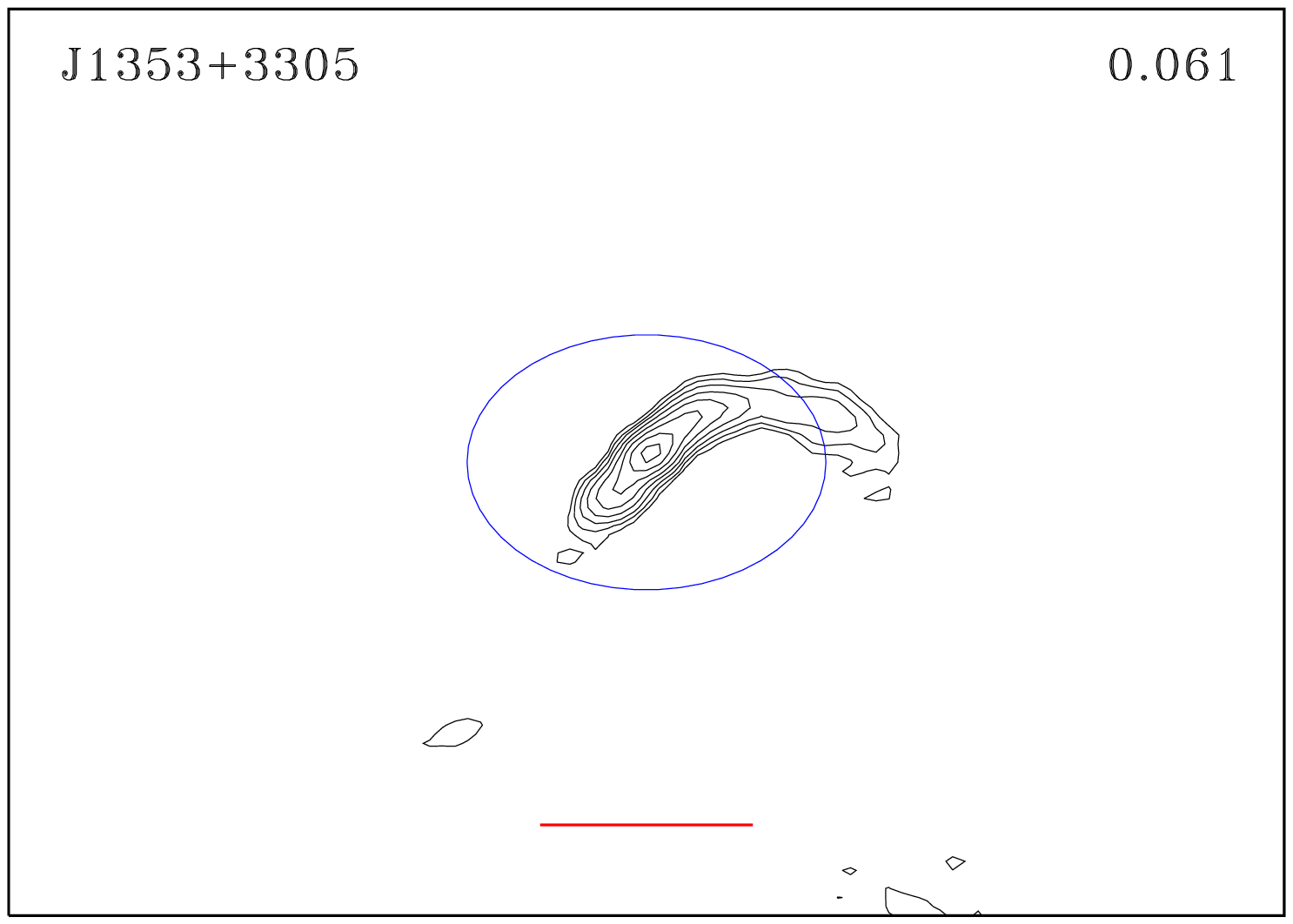} 
\includegraphics[width=6.3cm,height=6.3cm]{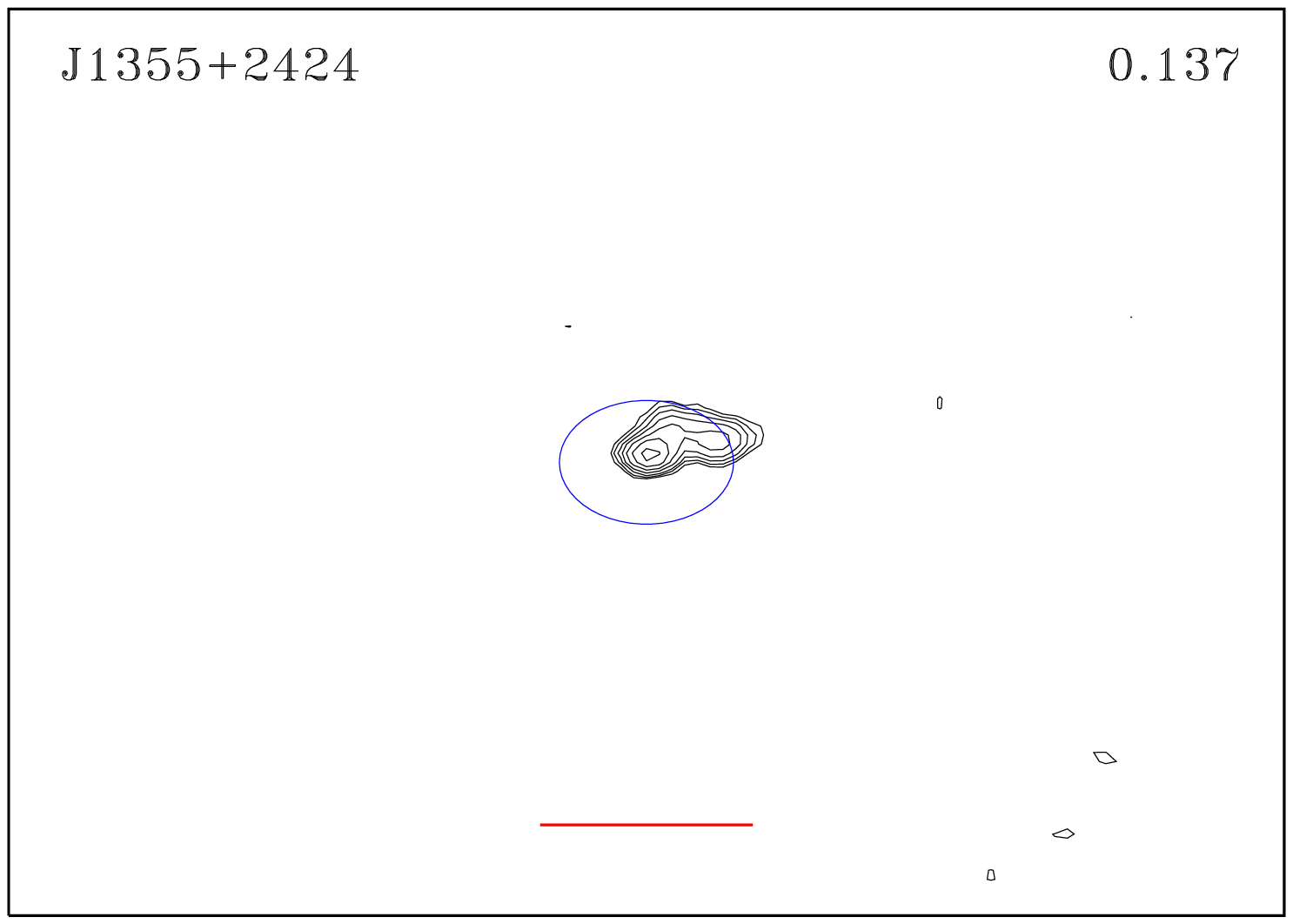} 
\includegraphics[width=6.3cm,height=6.3cm]{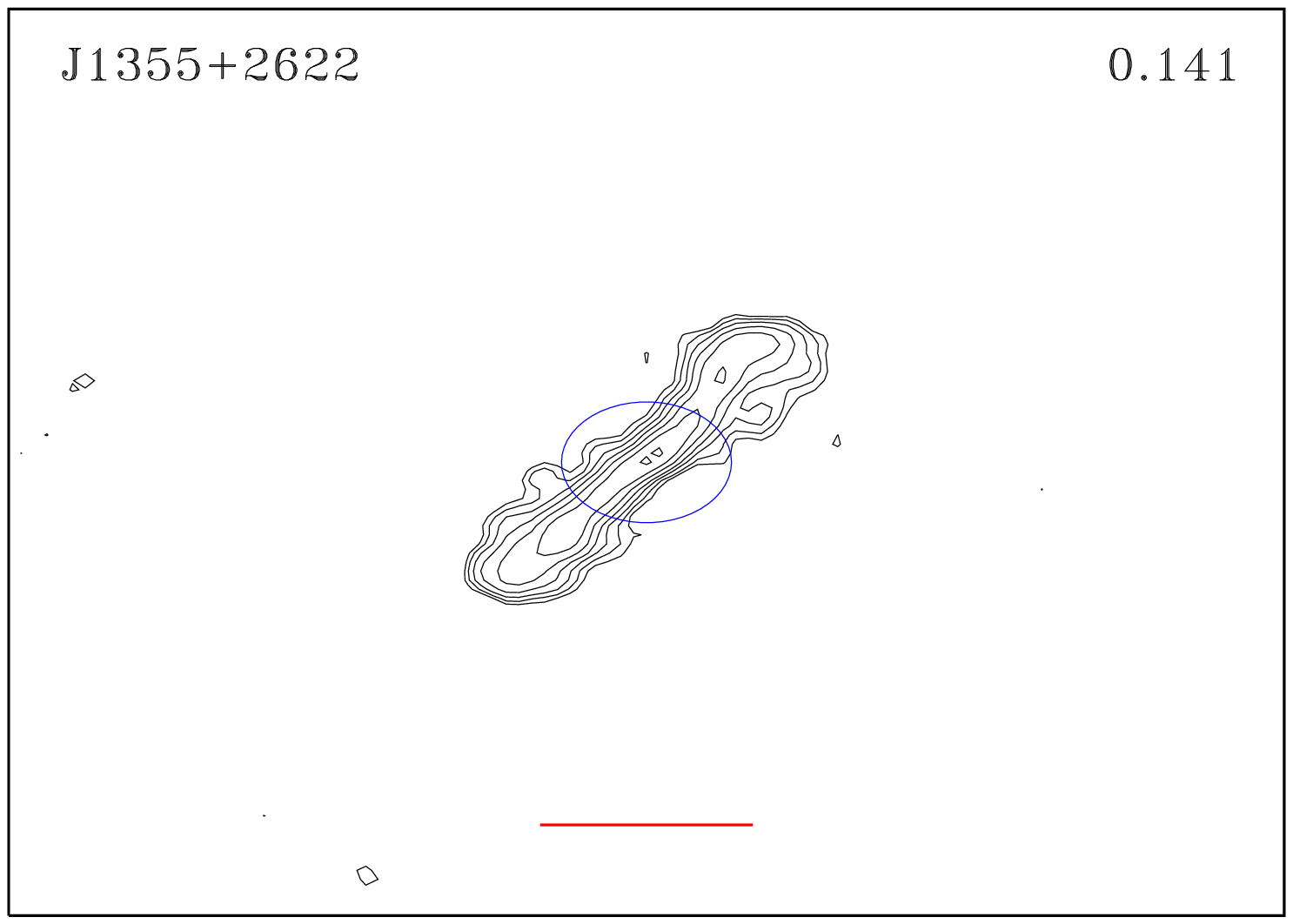} 
\caption{(continued)}
\end{figure*}

\addtocounter{figure}{-1}
\begin{figure*}
\includegraphics[width=6.3cm,height=6.3cm]{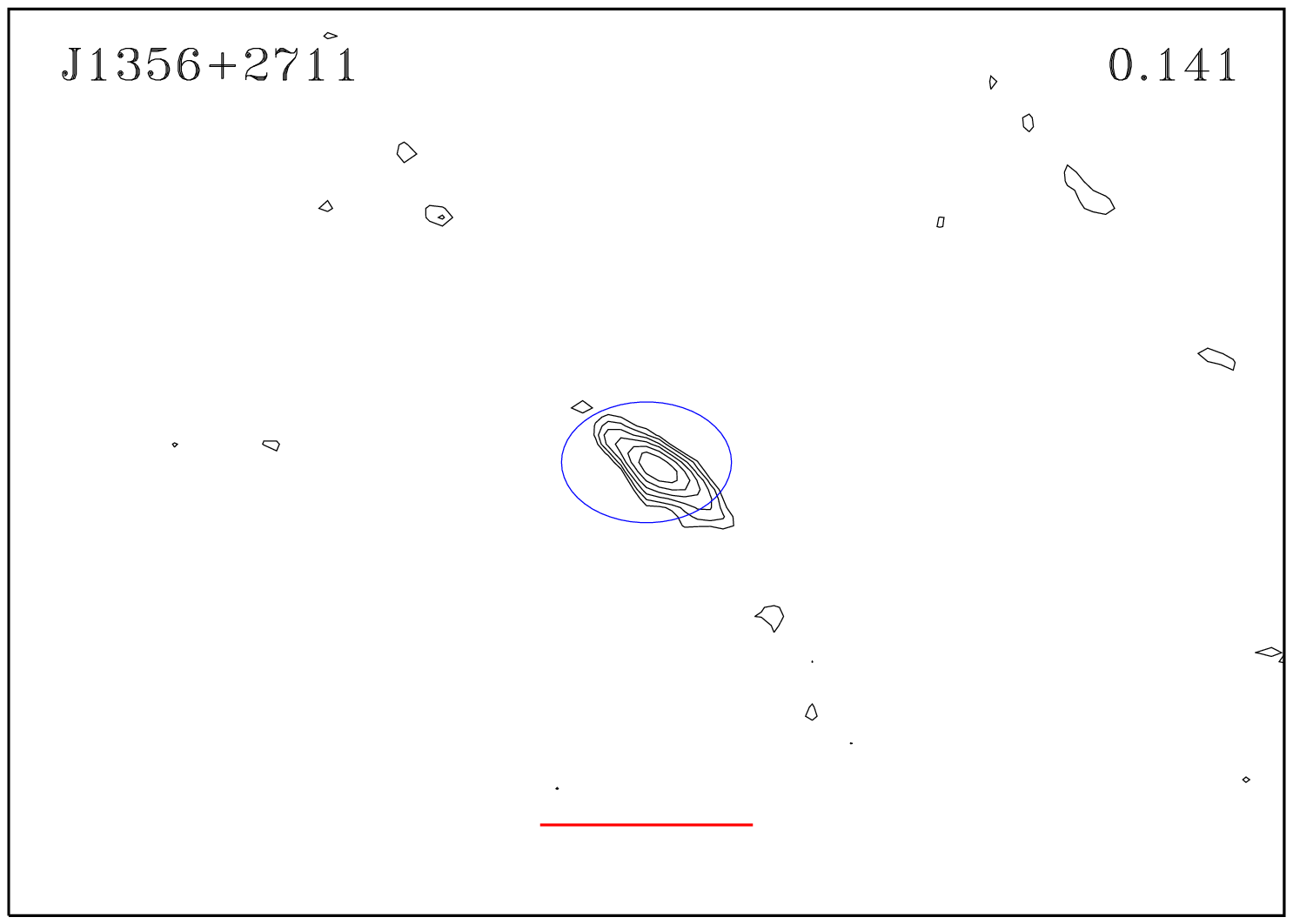} 
\includegraphics[width=6.3cm,height=6.3cm]{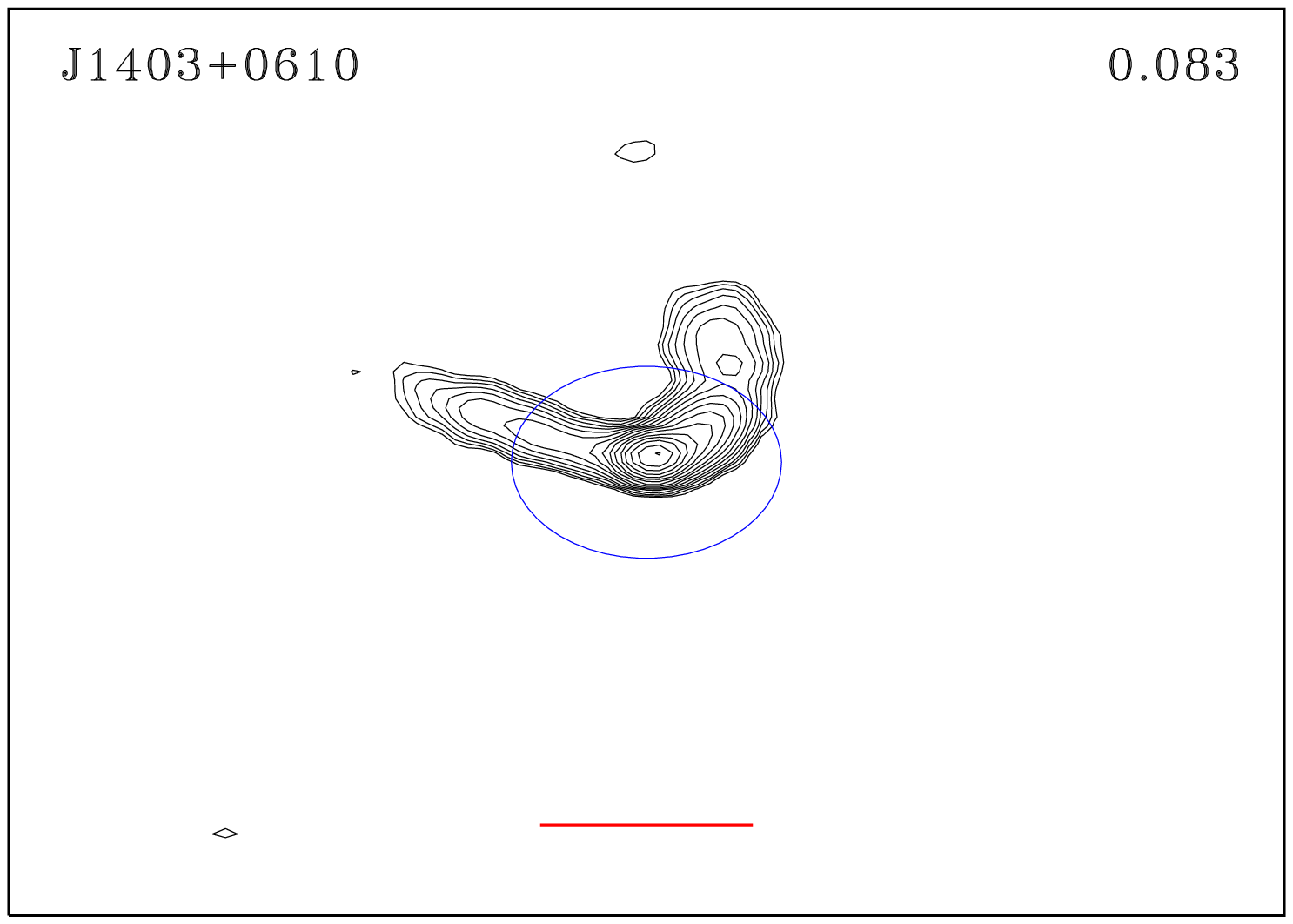} 
\includegraphics[width=6.3cm,height=6.3cm]{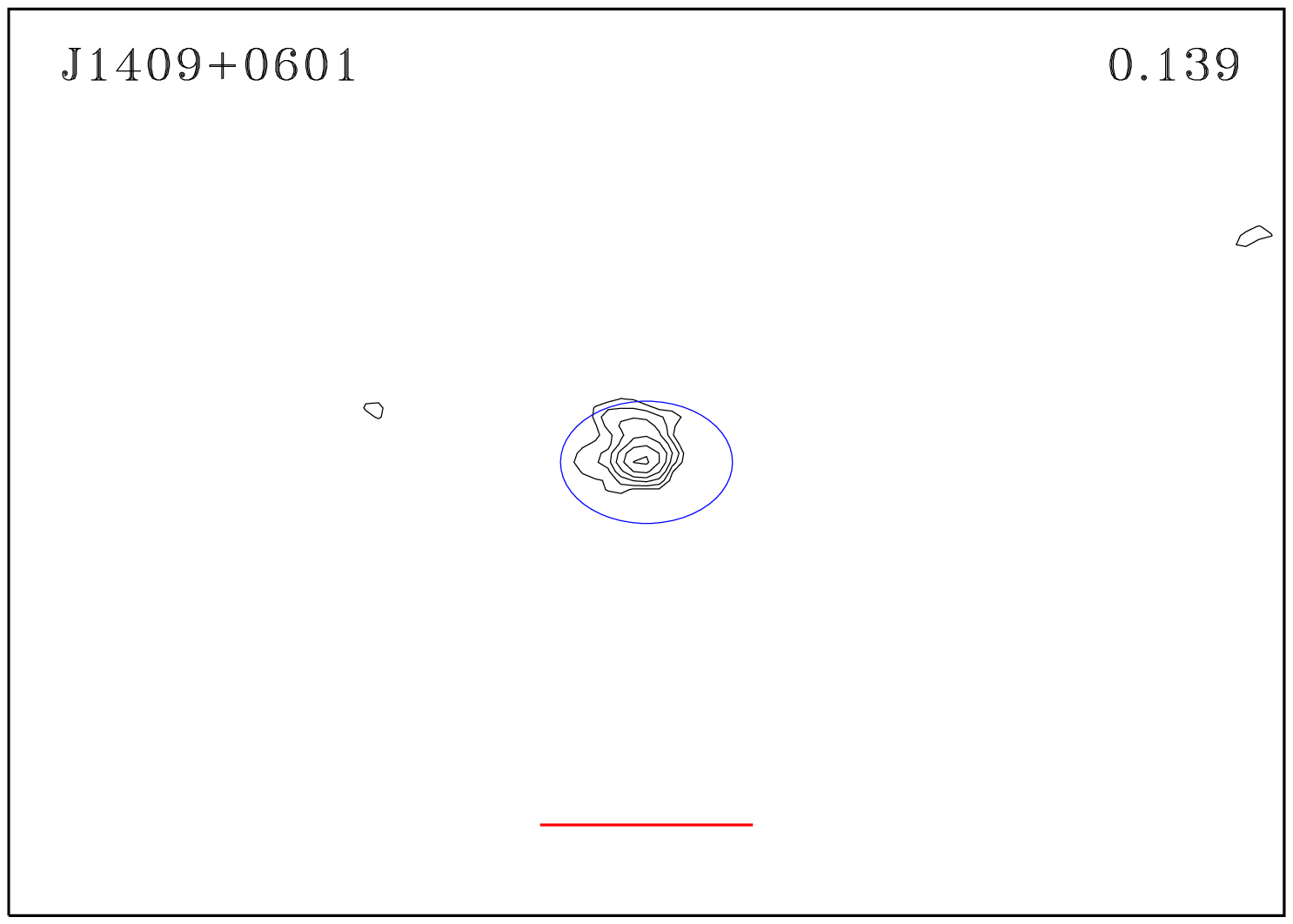} 

\includegraphics[width=6.3cm,height=6.3cm]{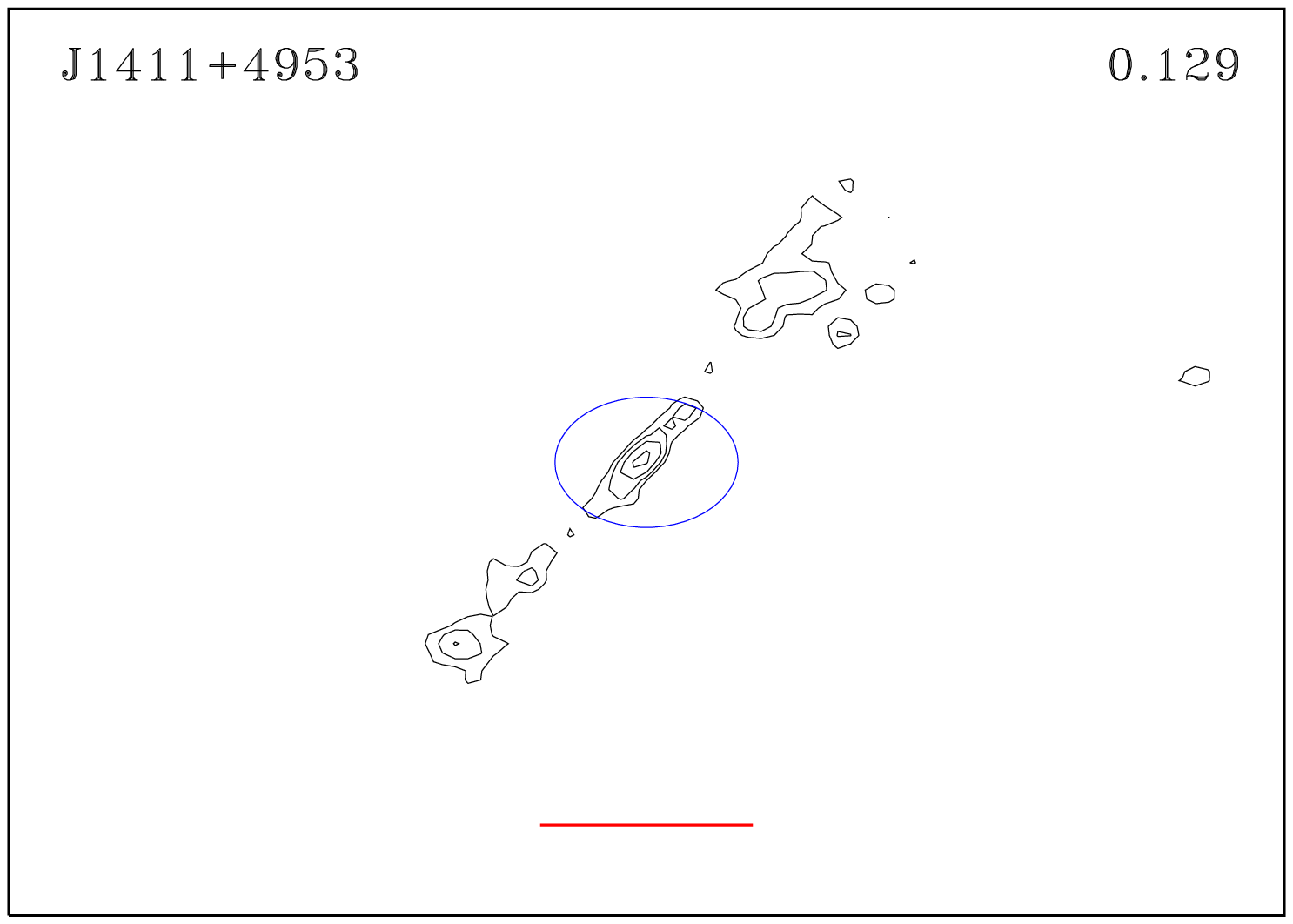} 
\includegraphics[width=6.3cm,height=6.3cm]{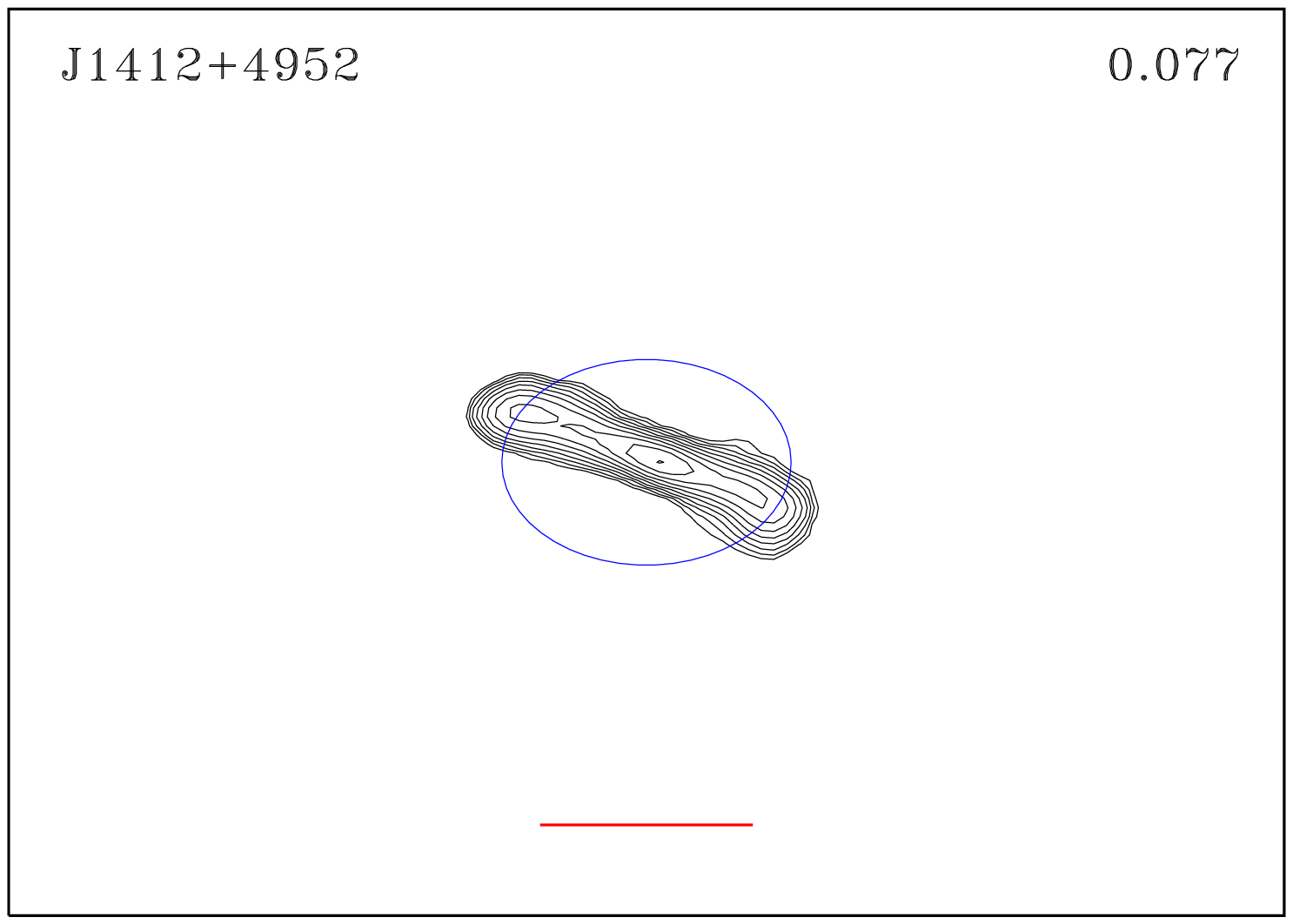} 
\includegraphics[width=6.3cm,height=6.3cm]{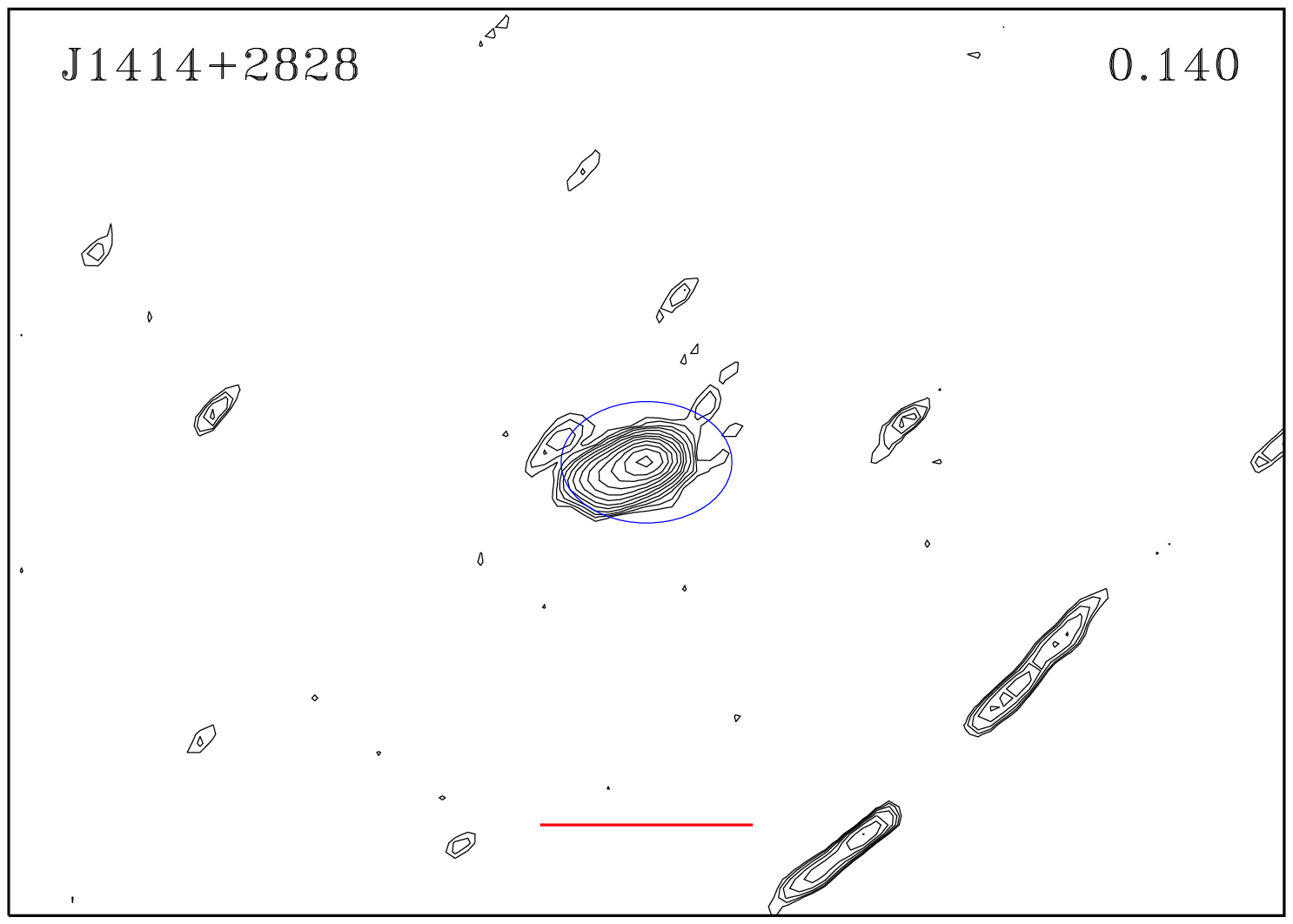} 

\includegraphics[width=6.3cm,height=6.3cm]{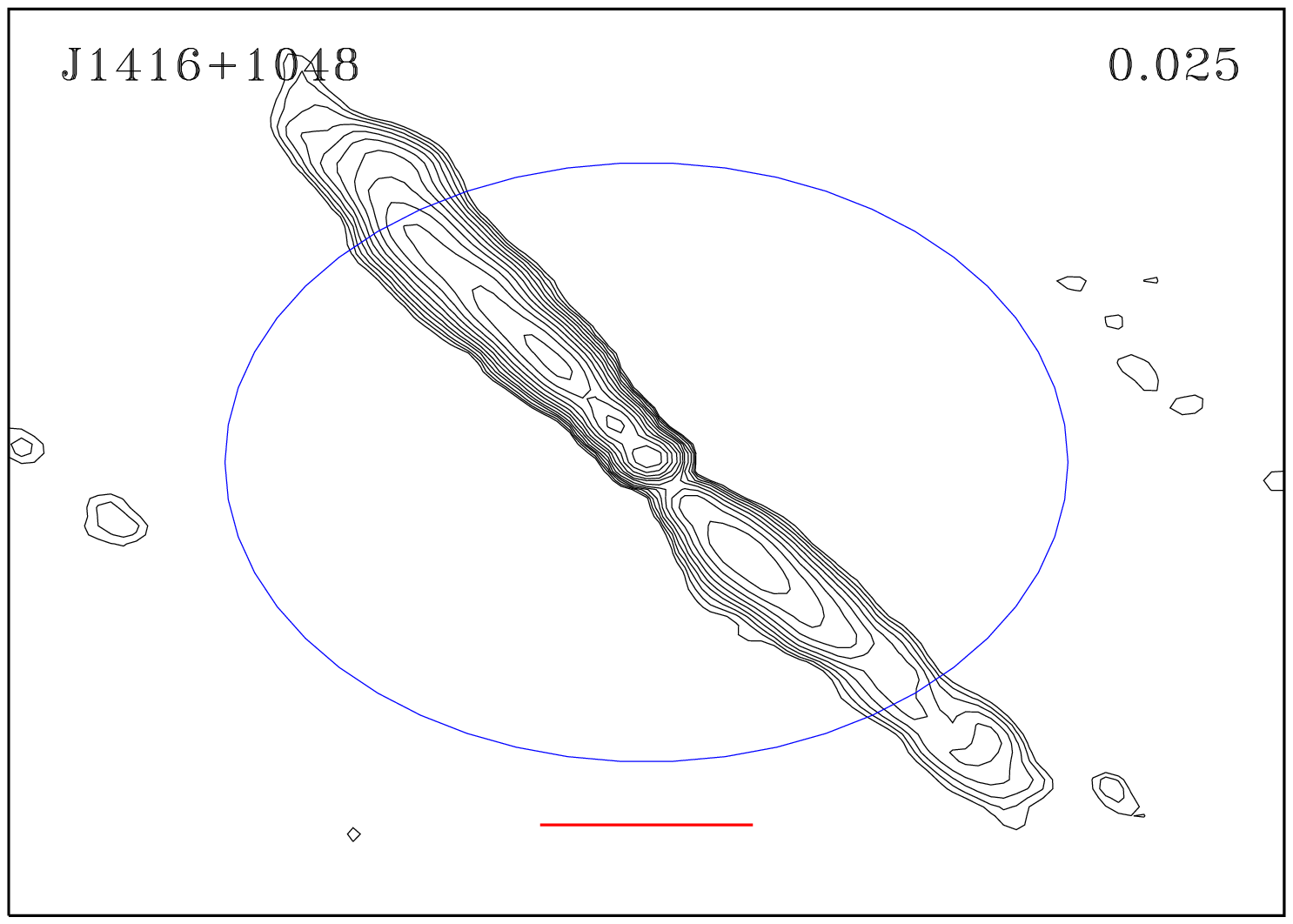} 
\includegraphics[width=6.3cm,height=6.3cm]{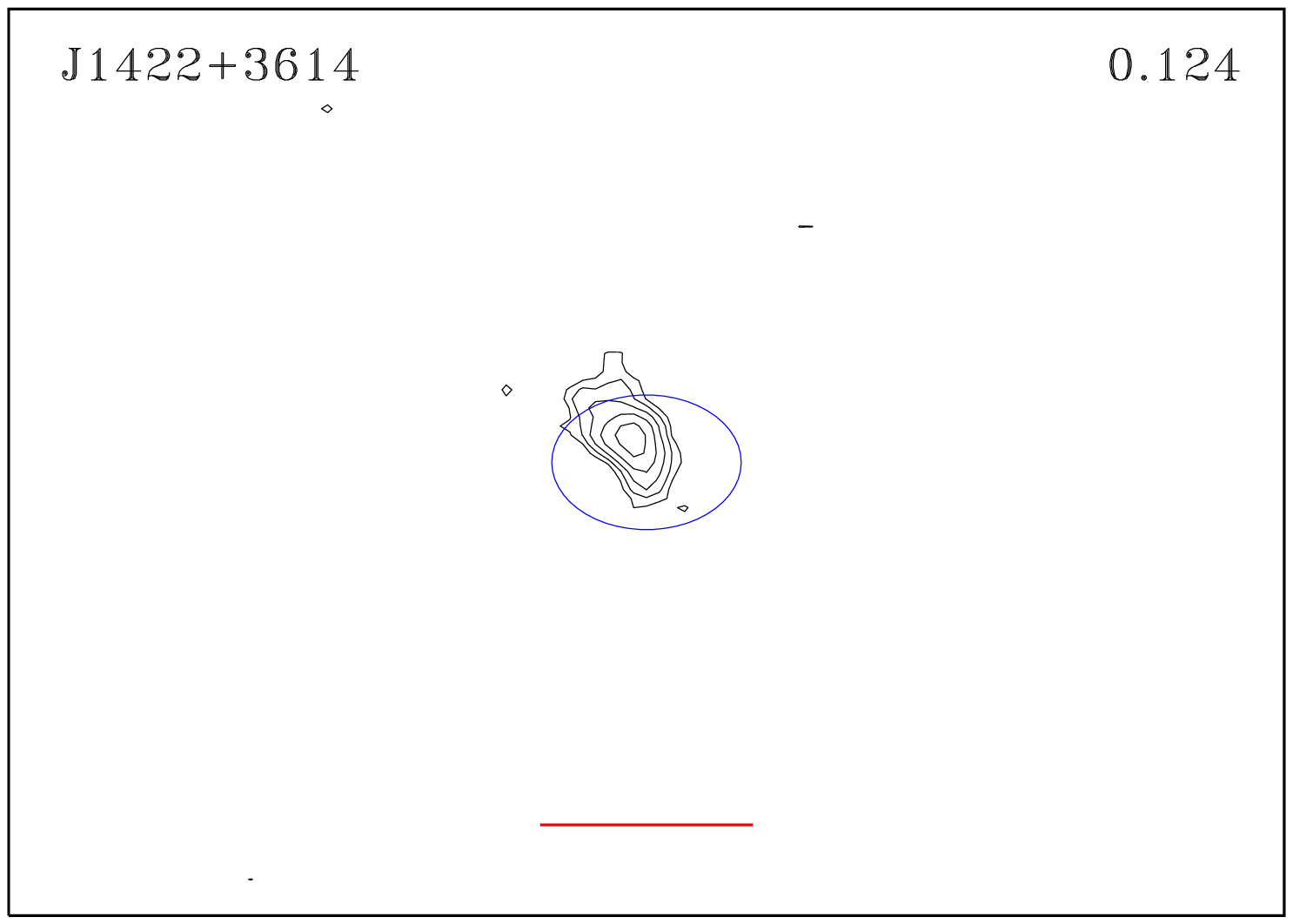} 
\includegraphics[width=6.3cm,height=6.3cm]{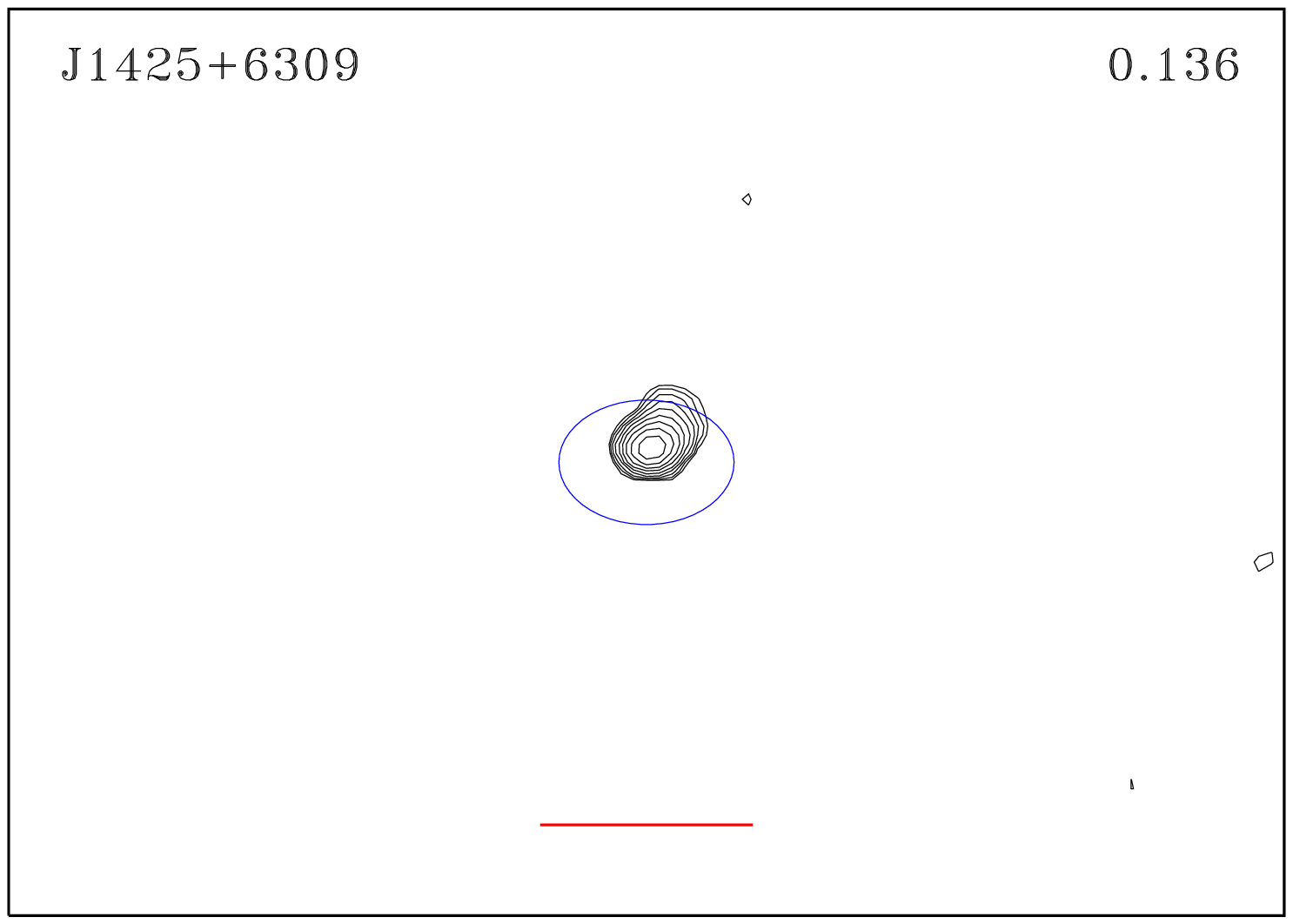} 

\includegraphics[width=6.3cm,height=6.3cm]{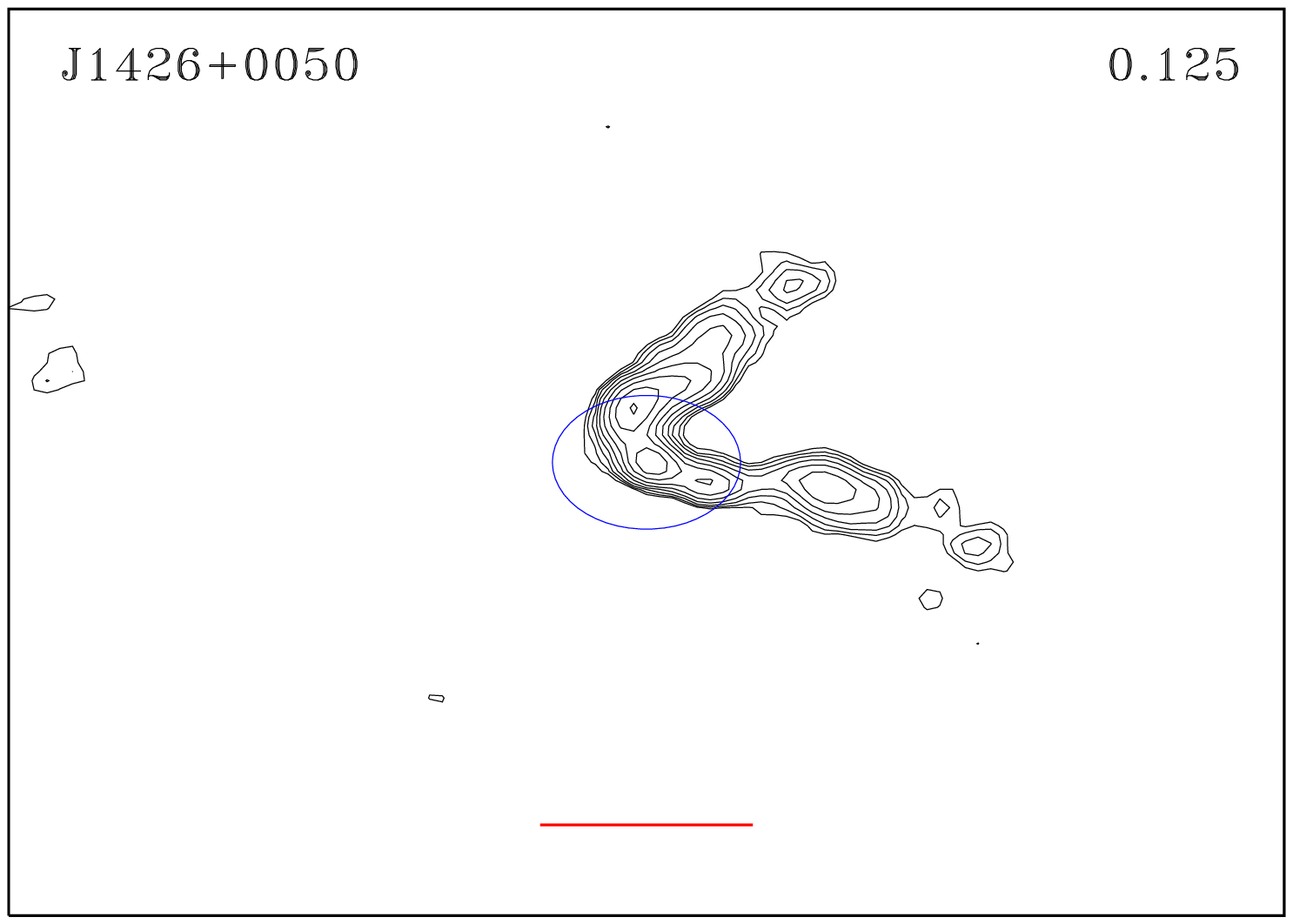} 
\includegraphics[width=6.3cm,height=6.3cm]{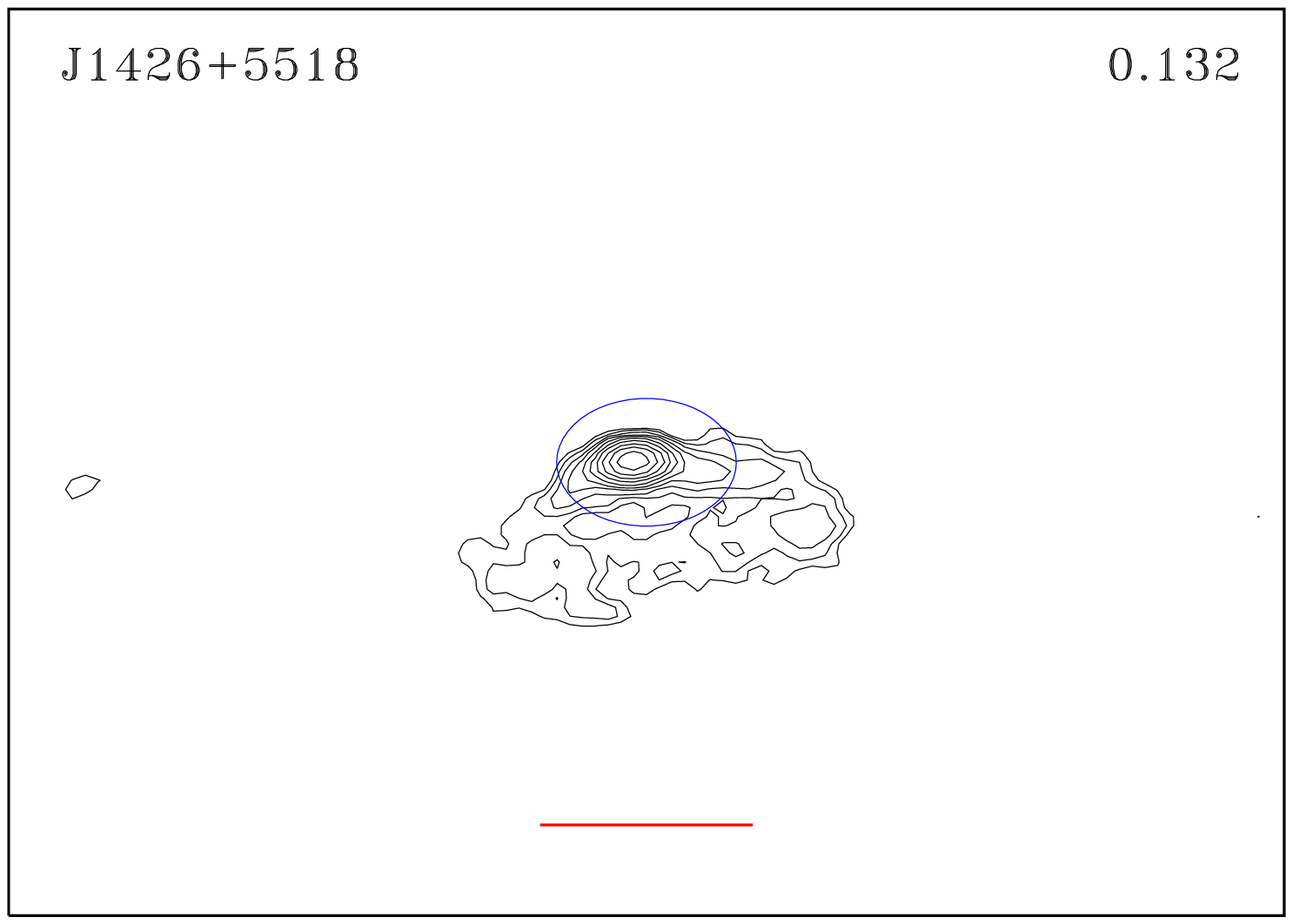} 
\includegraphics[width=6.3cm,height=6.3cm]{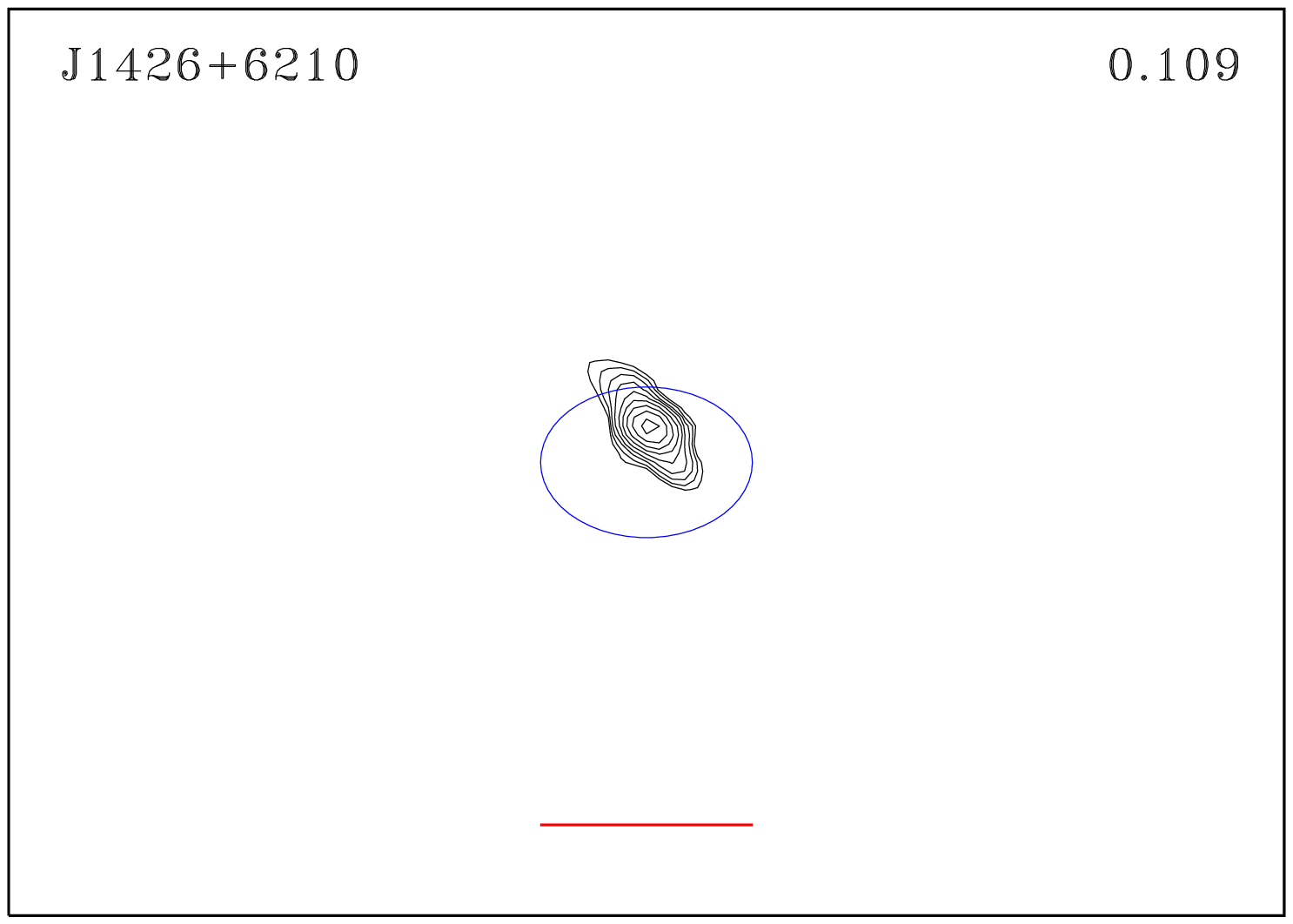} 
\caption{(continued)}
\end{figure*}

\addtocounter{figure}{-1}
\begin{figure*}
\includegraphics[width=6.3cm,height=6.3cm]{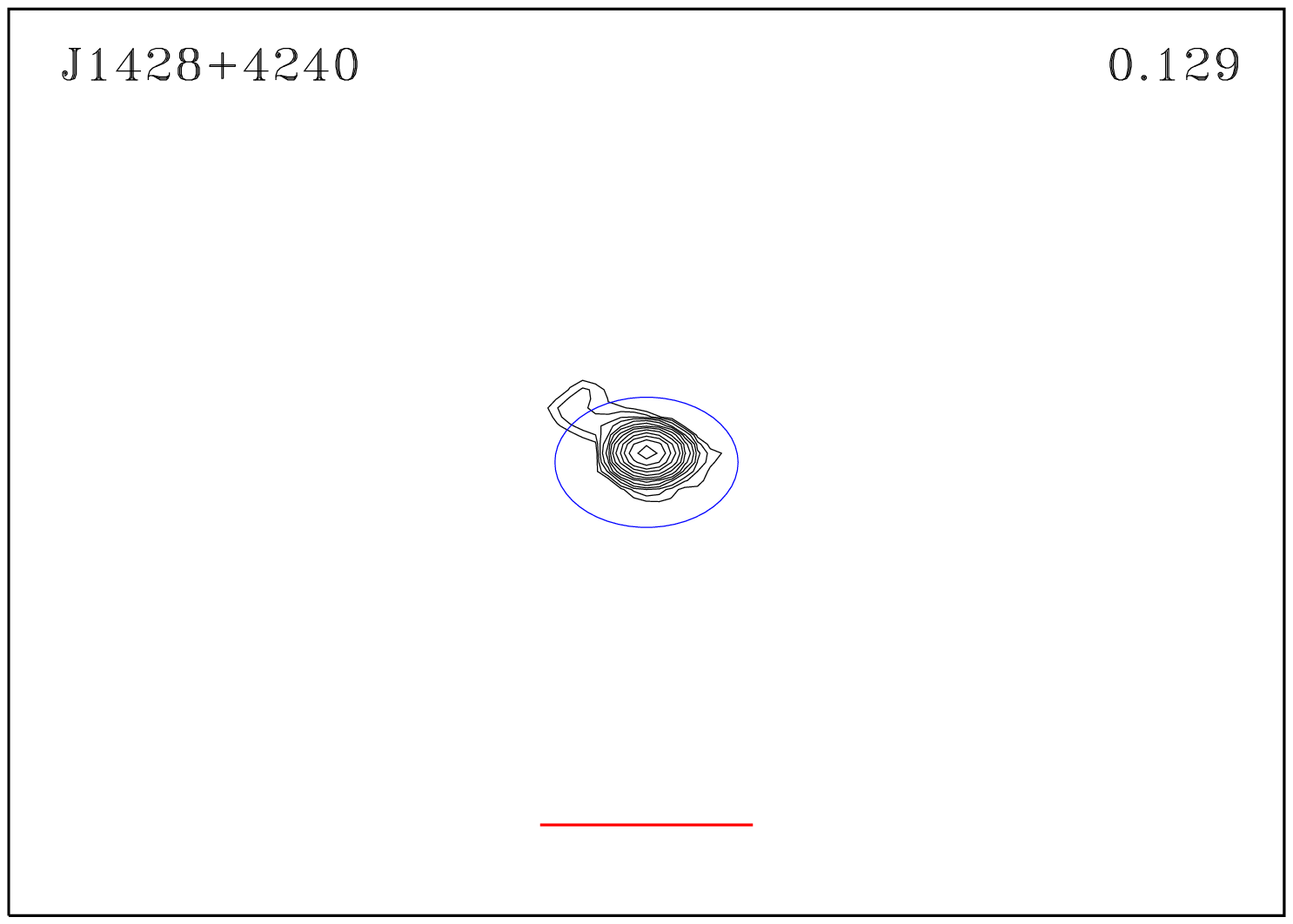} 
\includegraphics[width=6.3cm,height=6.3cm]{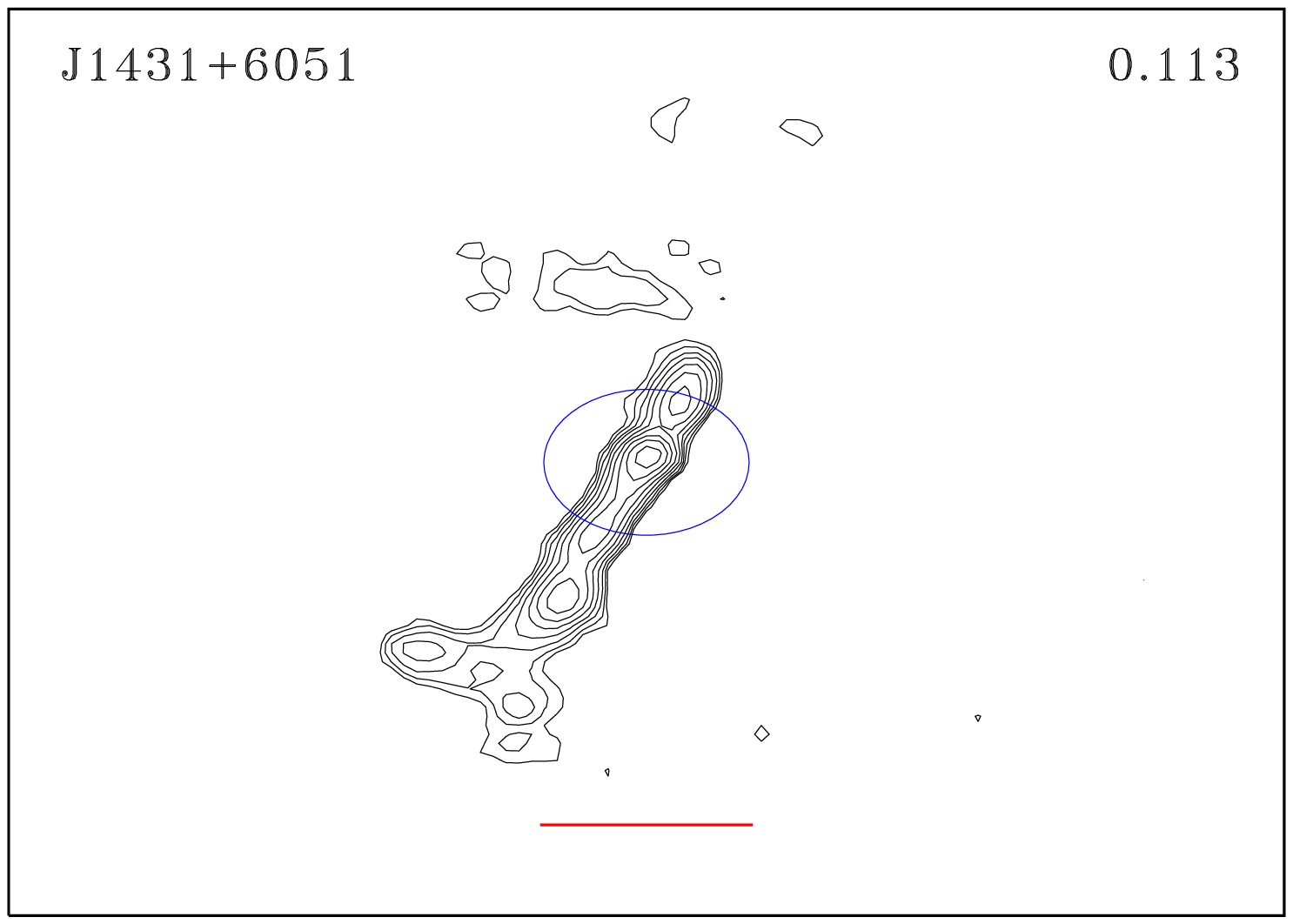} 
\includegraphics[width=6.3cm,height=6.3cm]{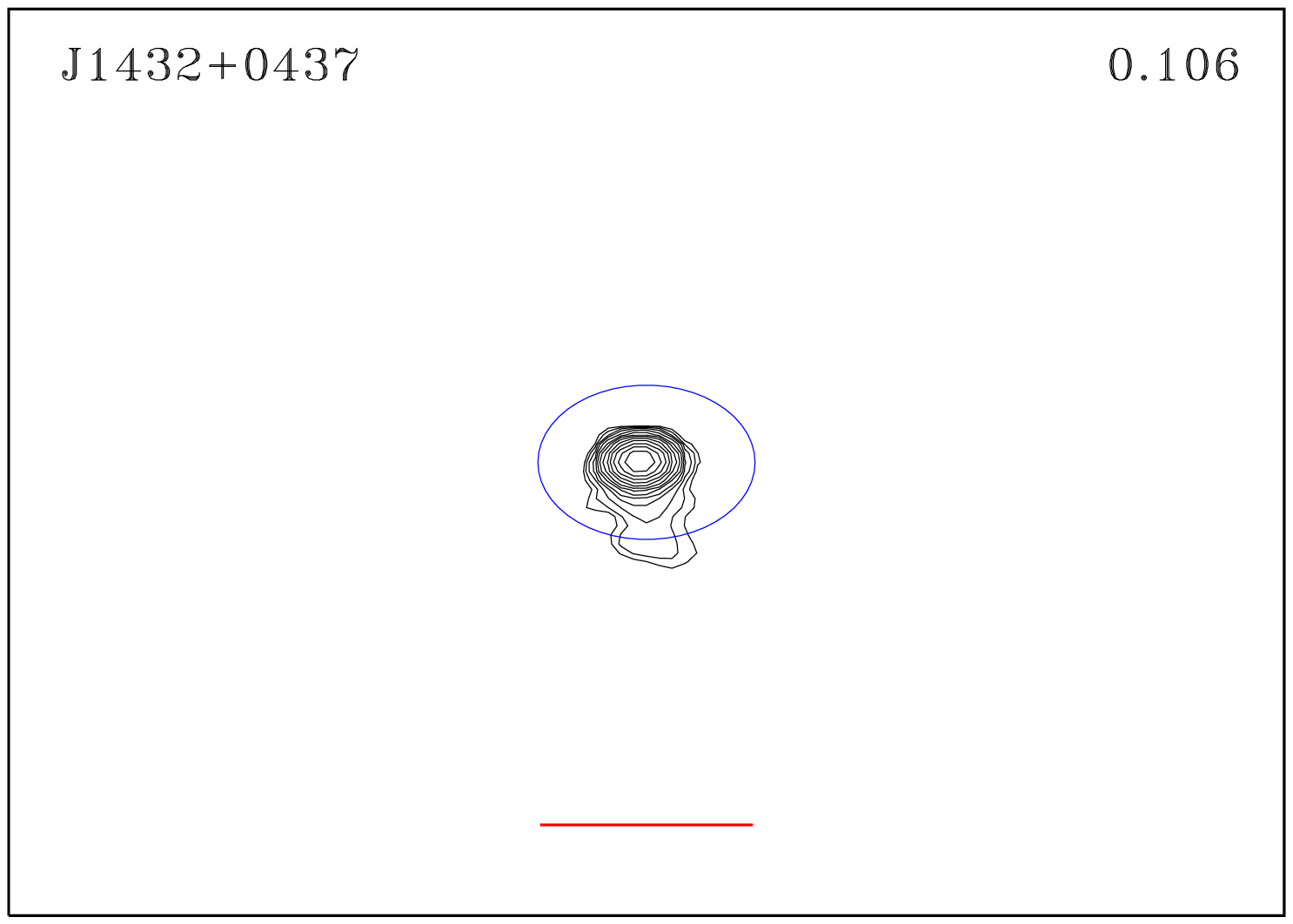} 

\includegraphics[width=6.3cm,height=6.3cm]{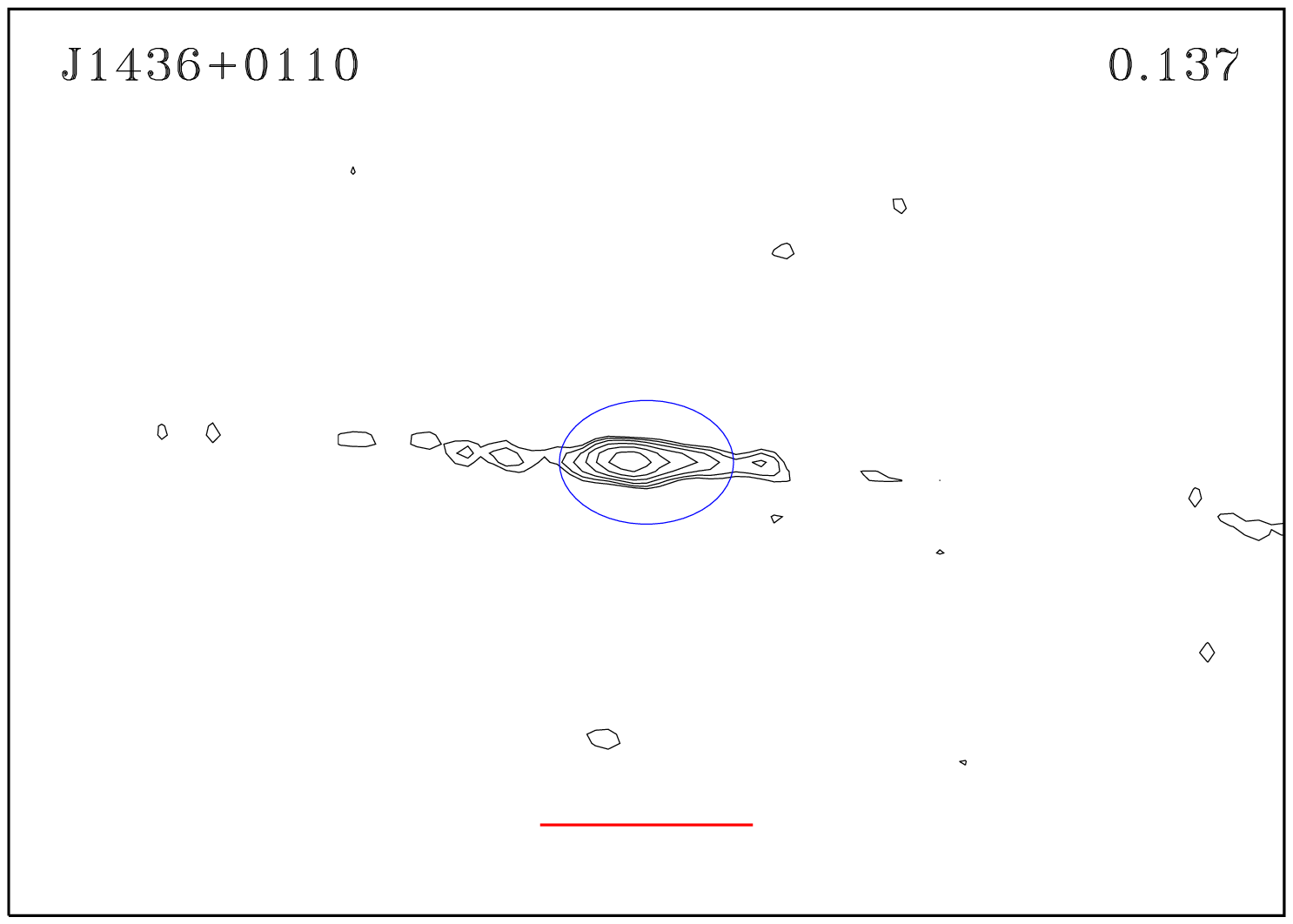} 
\includegraphics[width=6.3cm,height=6.3cm]{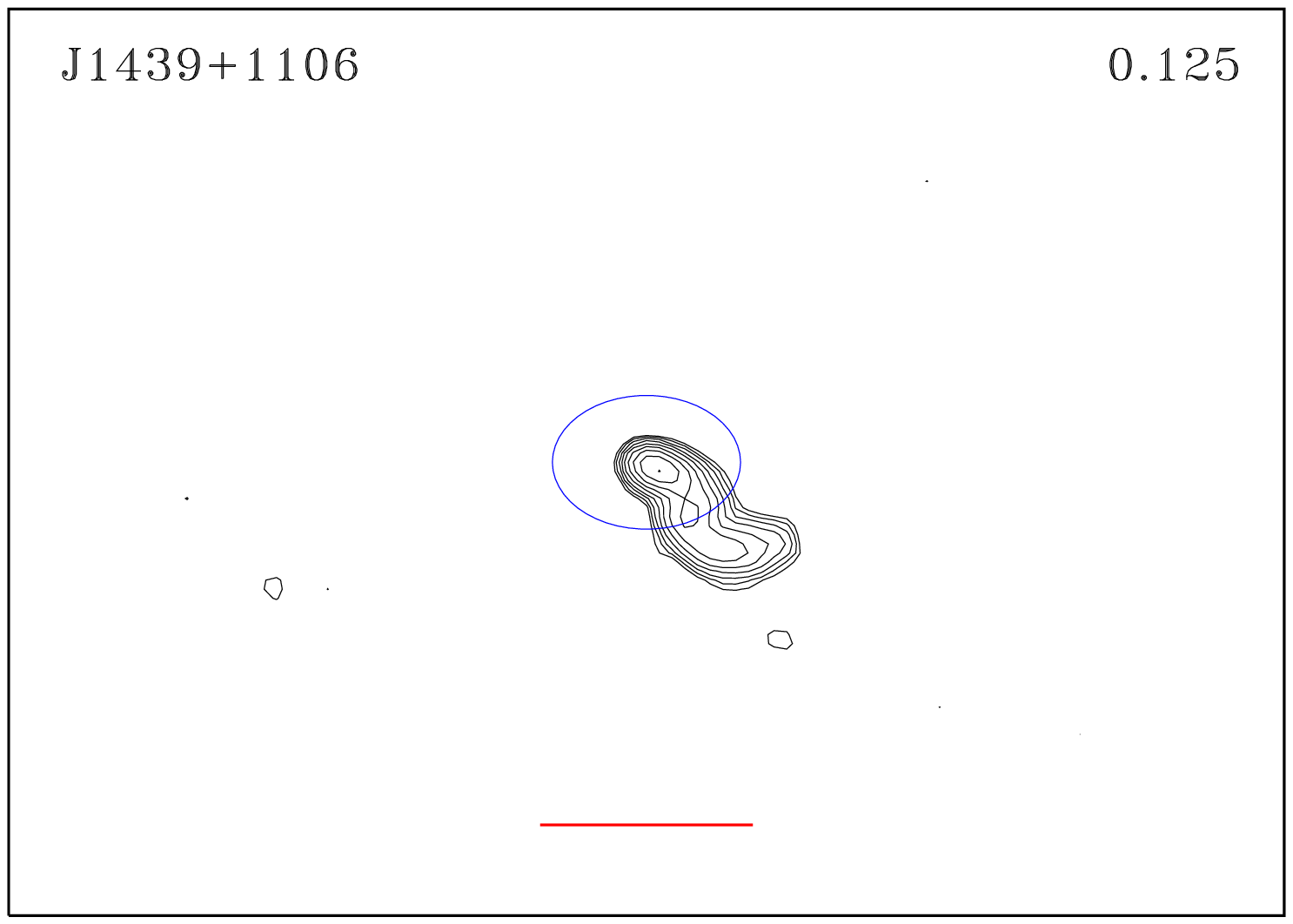} 
\includegraphics[width=6.3cm,height=6.3cm]{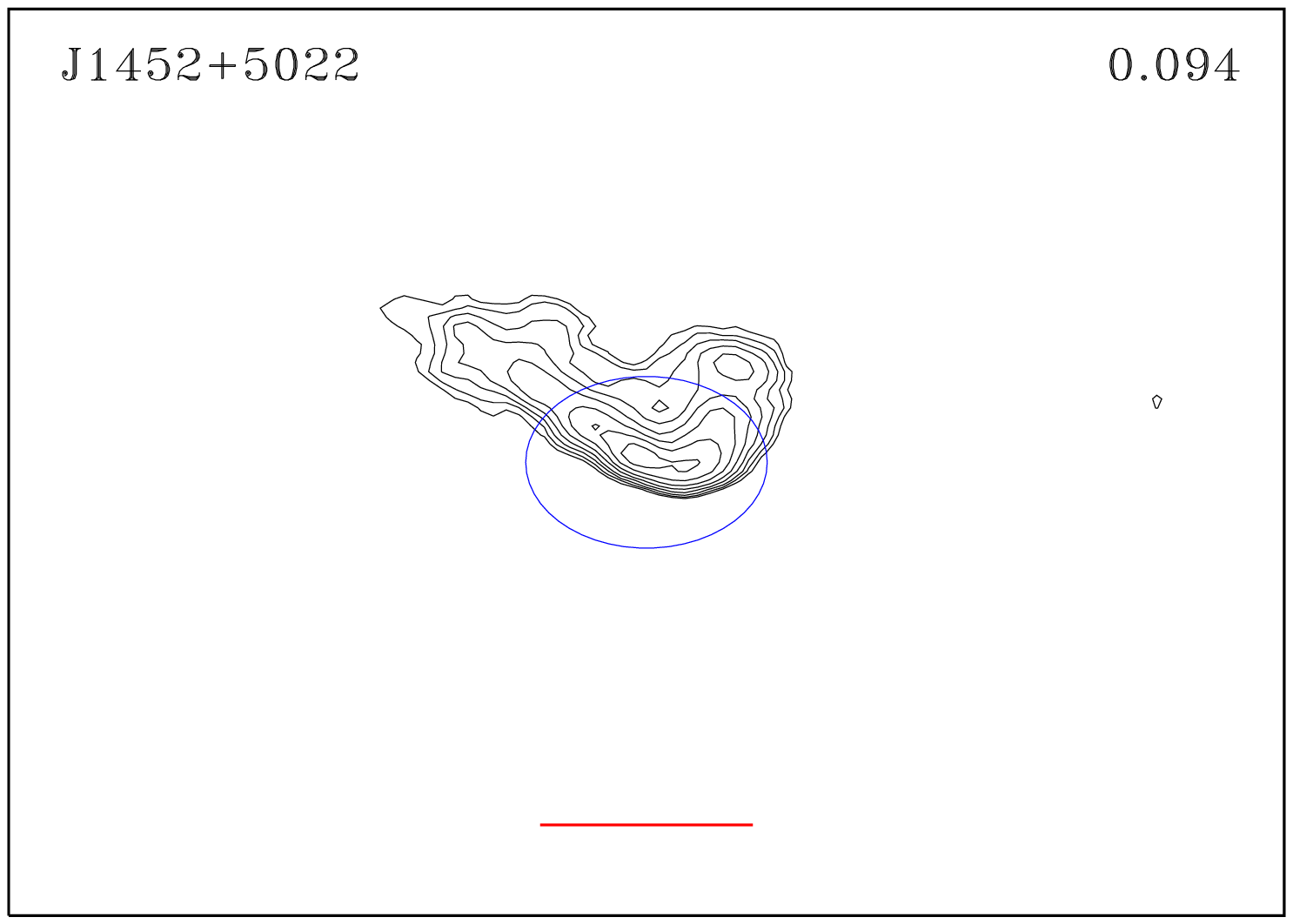} 

\includegraphics[width=6.3cm,height=6.3cm]{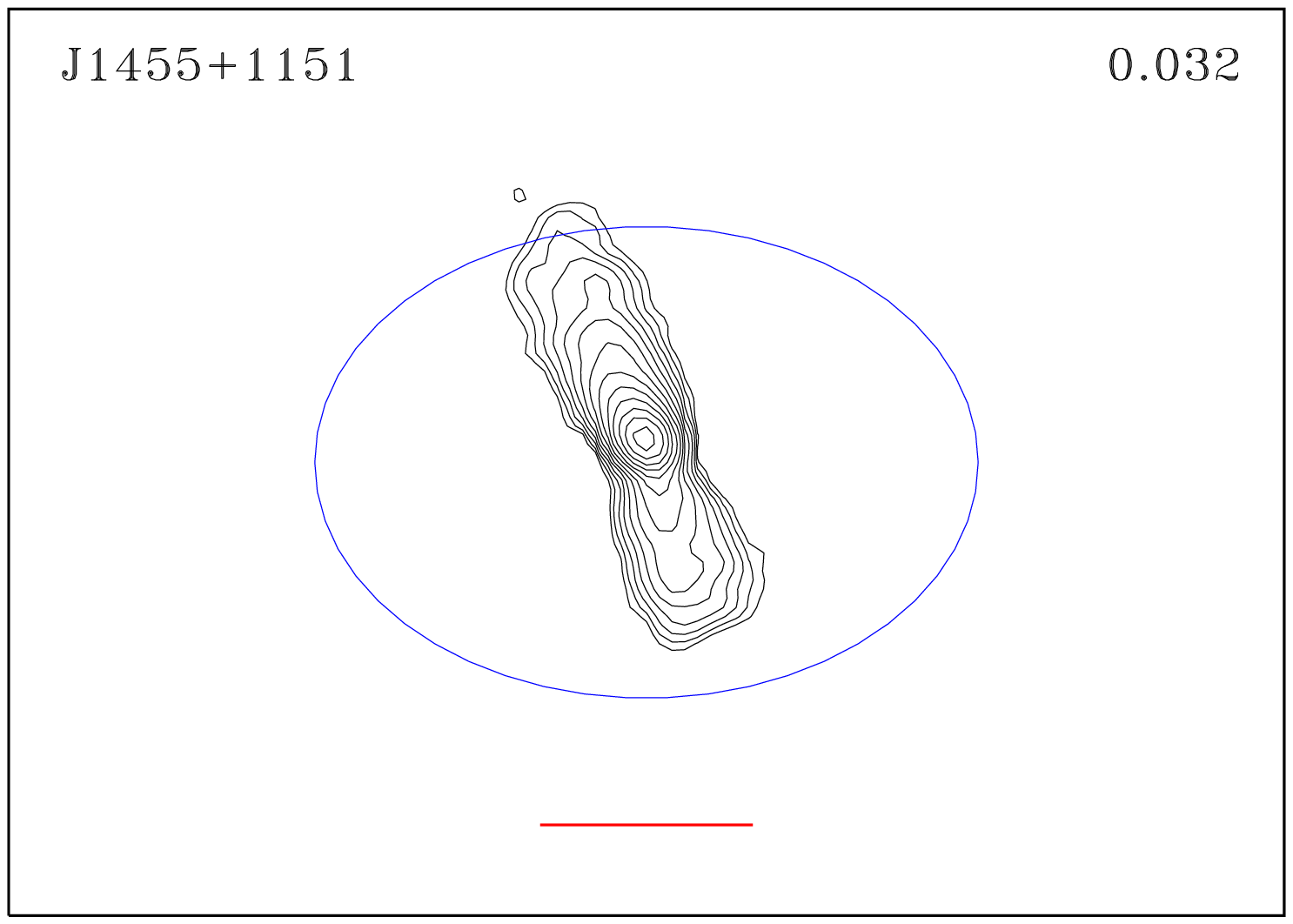} 
\includegraphics[width=6.3cm,height=6.3cm]{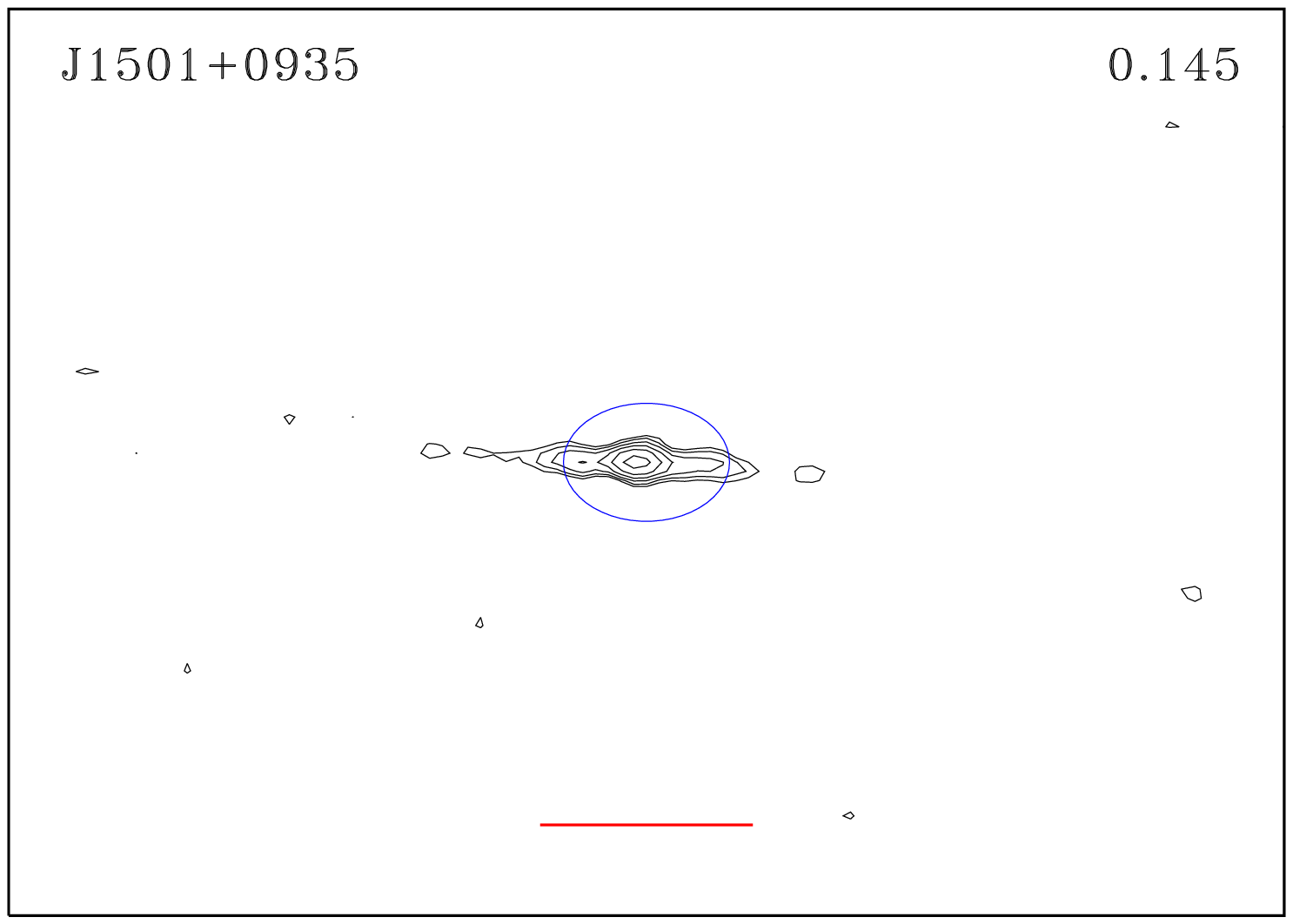} 
\includegraphics[width=6.3cm,height=6.3cm]{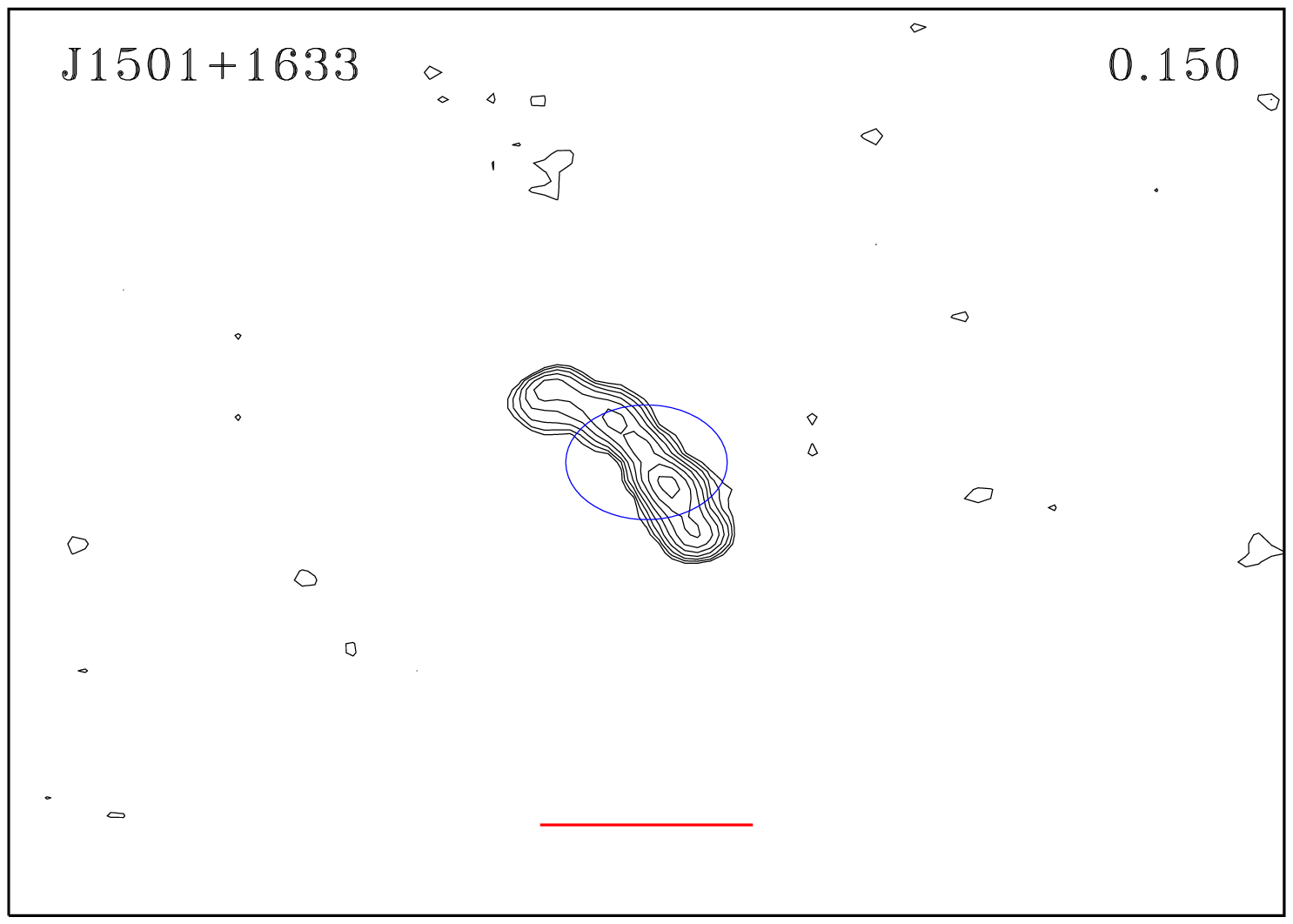} 

\includegraphics[width=6.3cm,height=6.3cm]{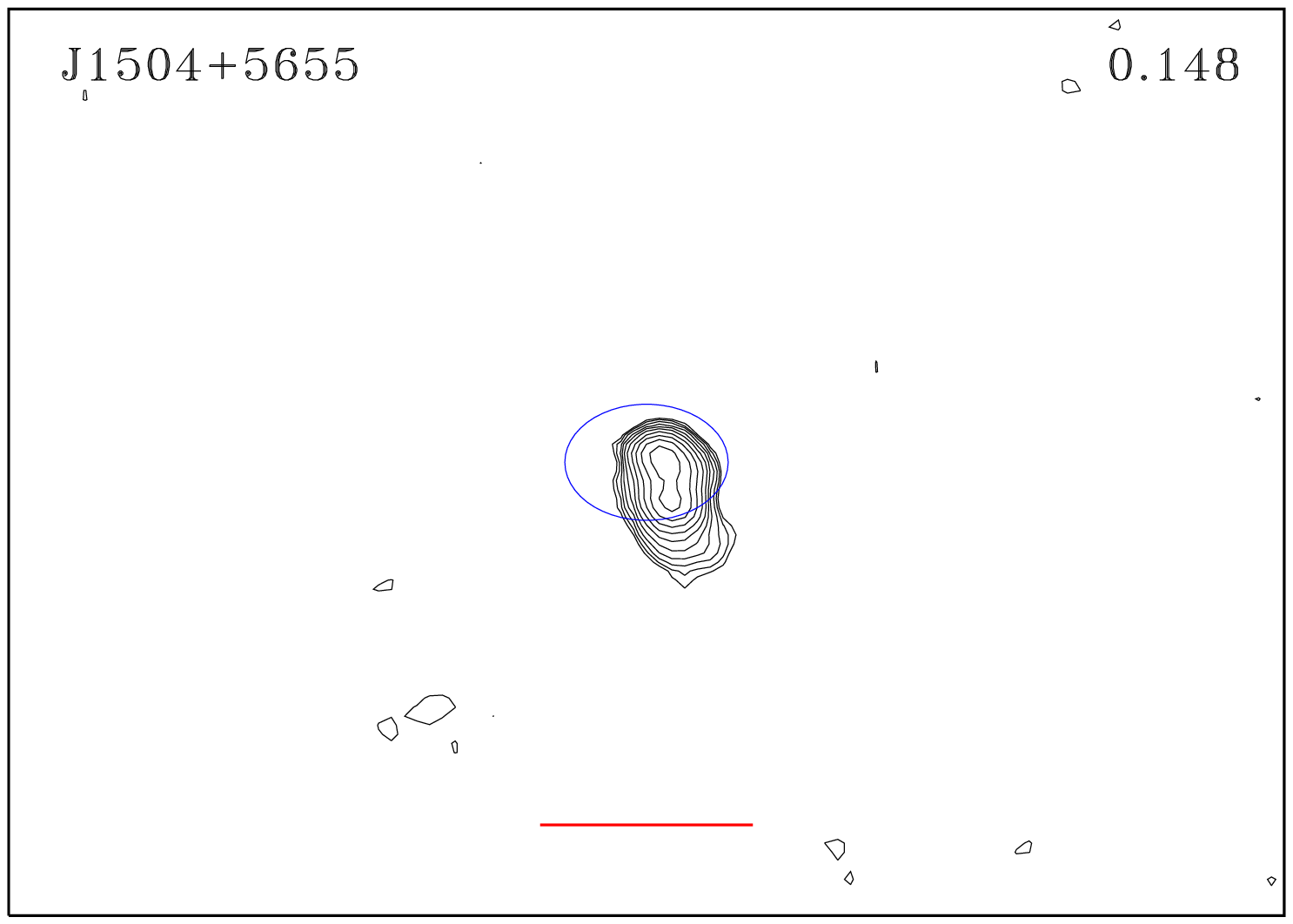} 
\includegraphics[width=6.3cm,height=6.3cm]{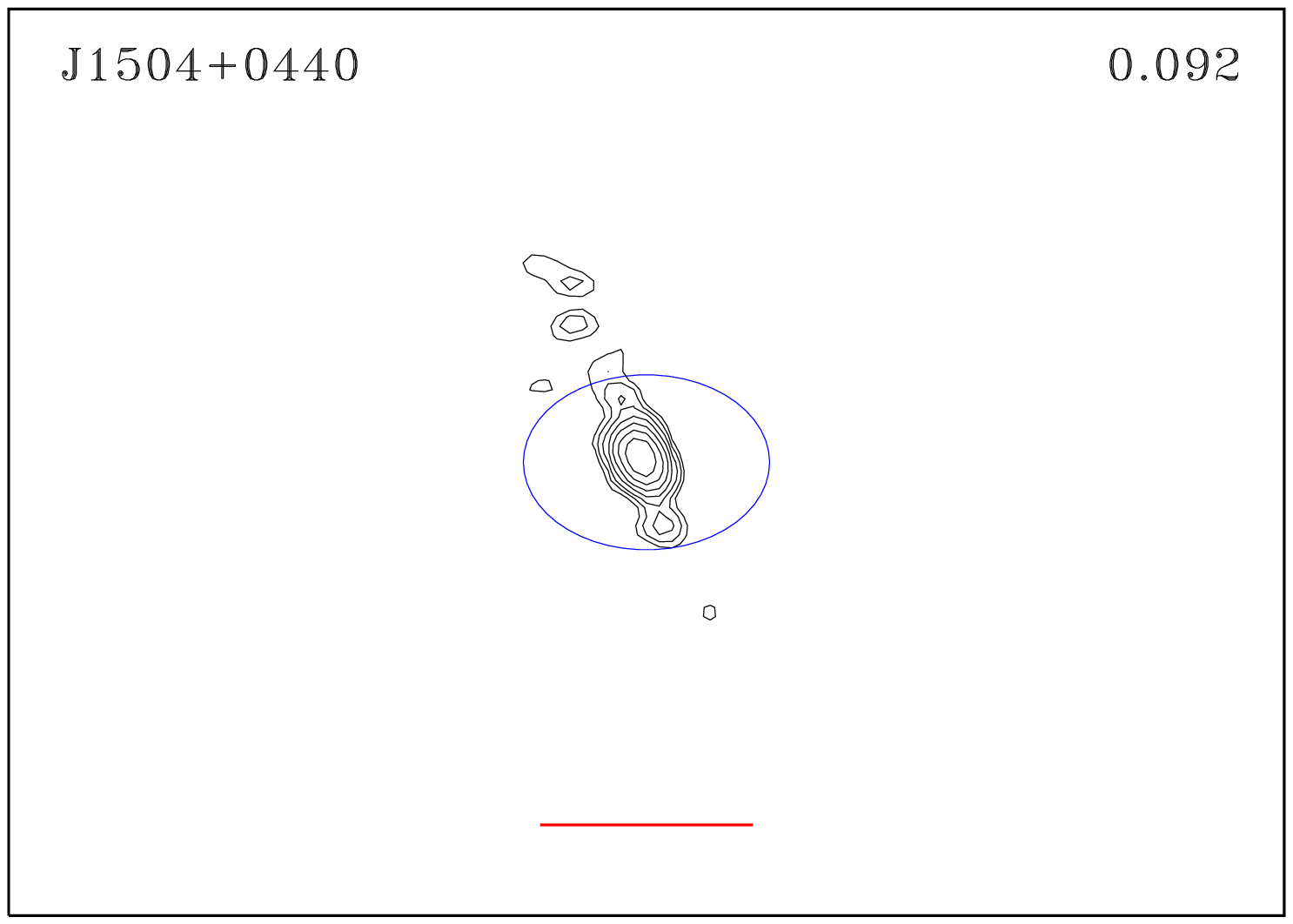} 
\includegraphics[width=6.3cm,height=6.3cm]{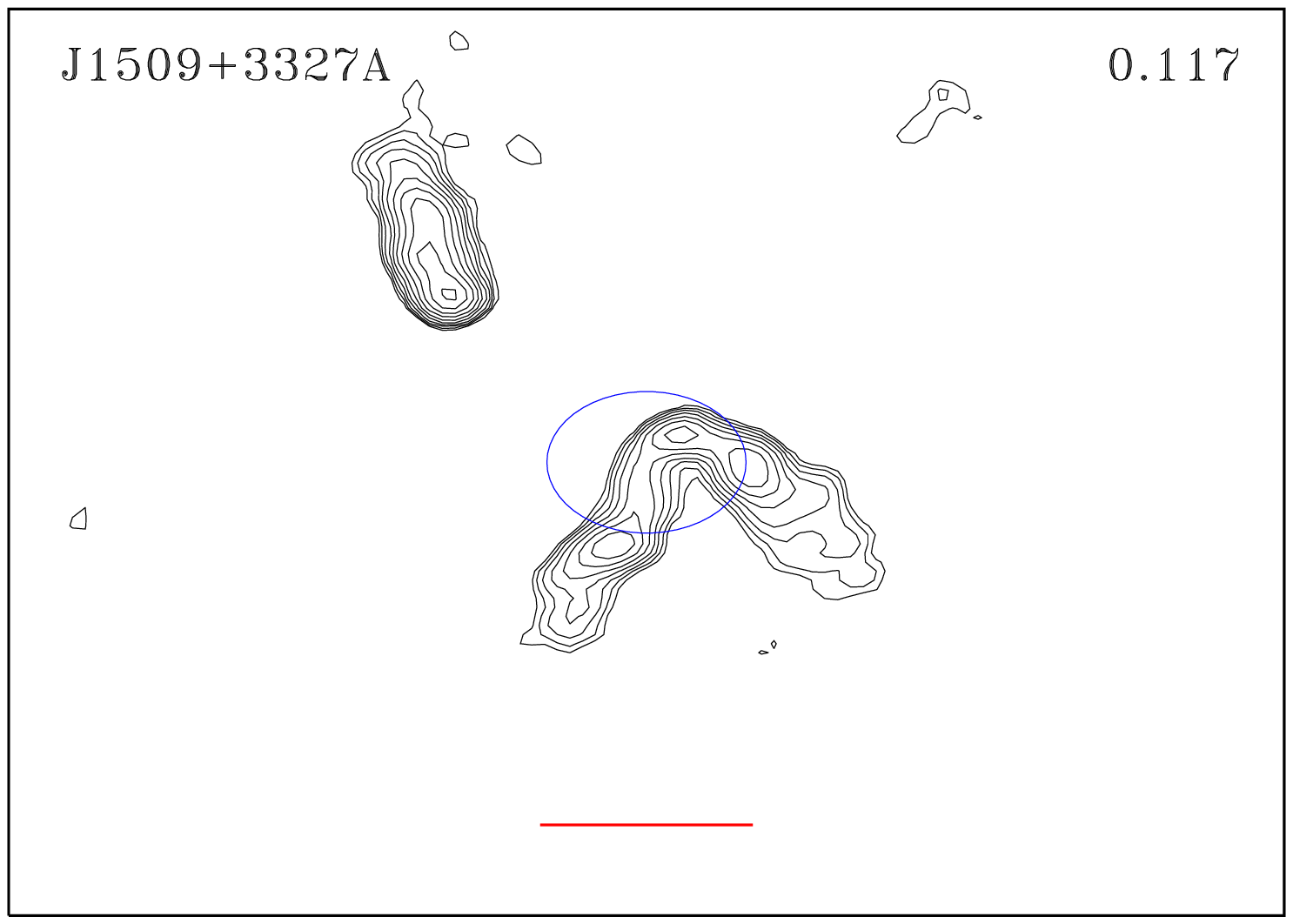} 
\caption{(continued)}
\end{figure*}

\addtocounter{figure}{-1}
\begin{figure*}
\includegraphics[width=6.3cm,height=6.3cm]{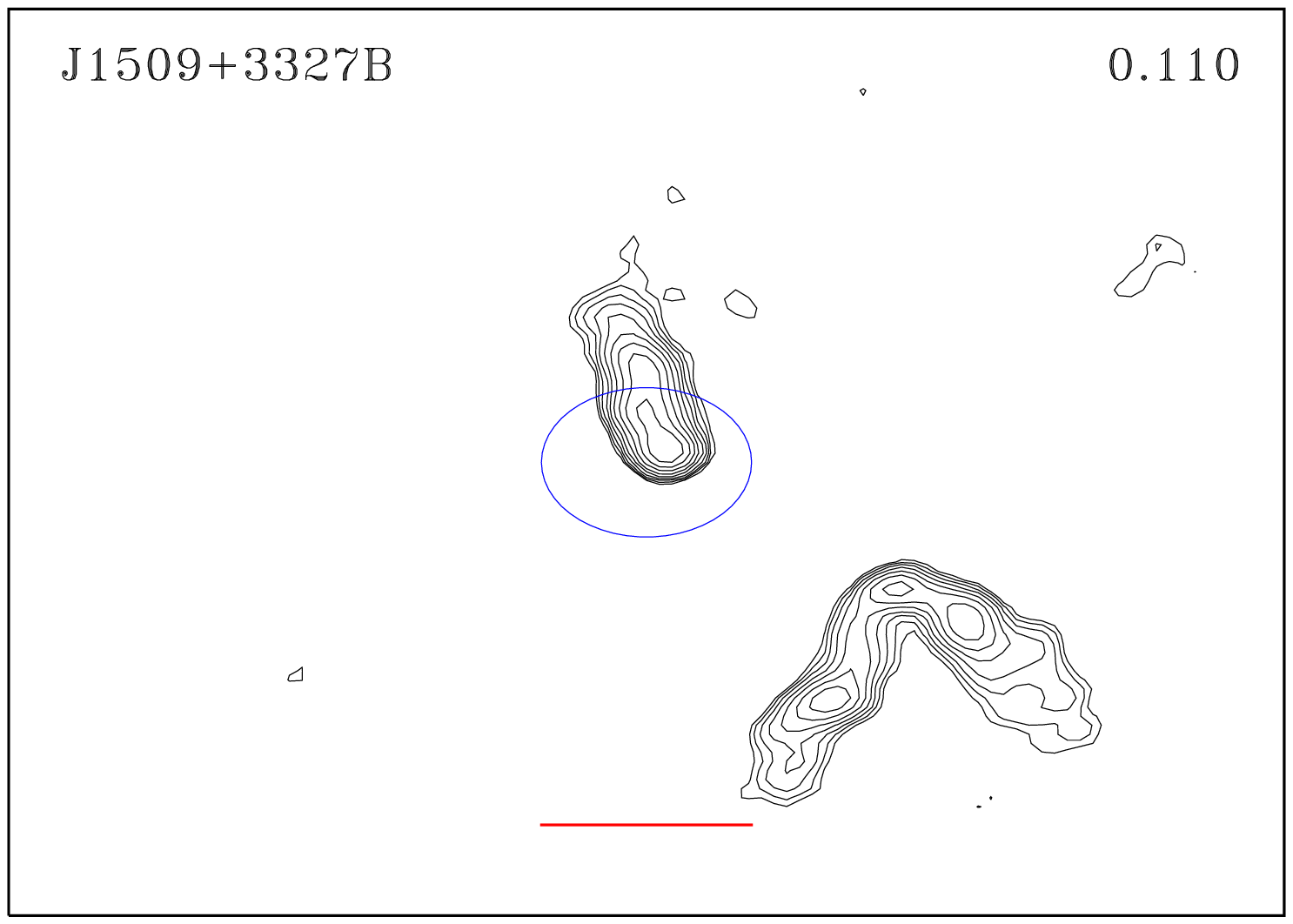} 
\includegraphics[width=6.3cm,height=6.3cm]{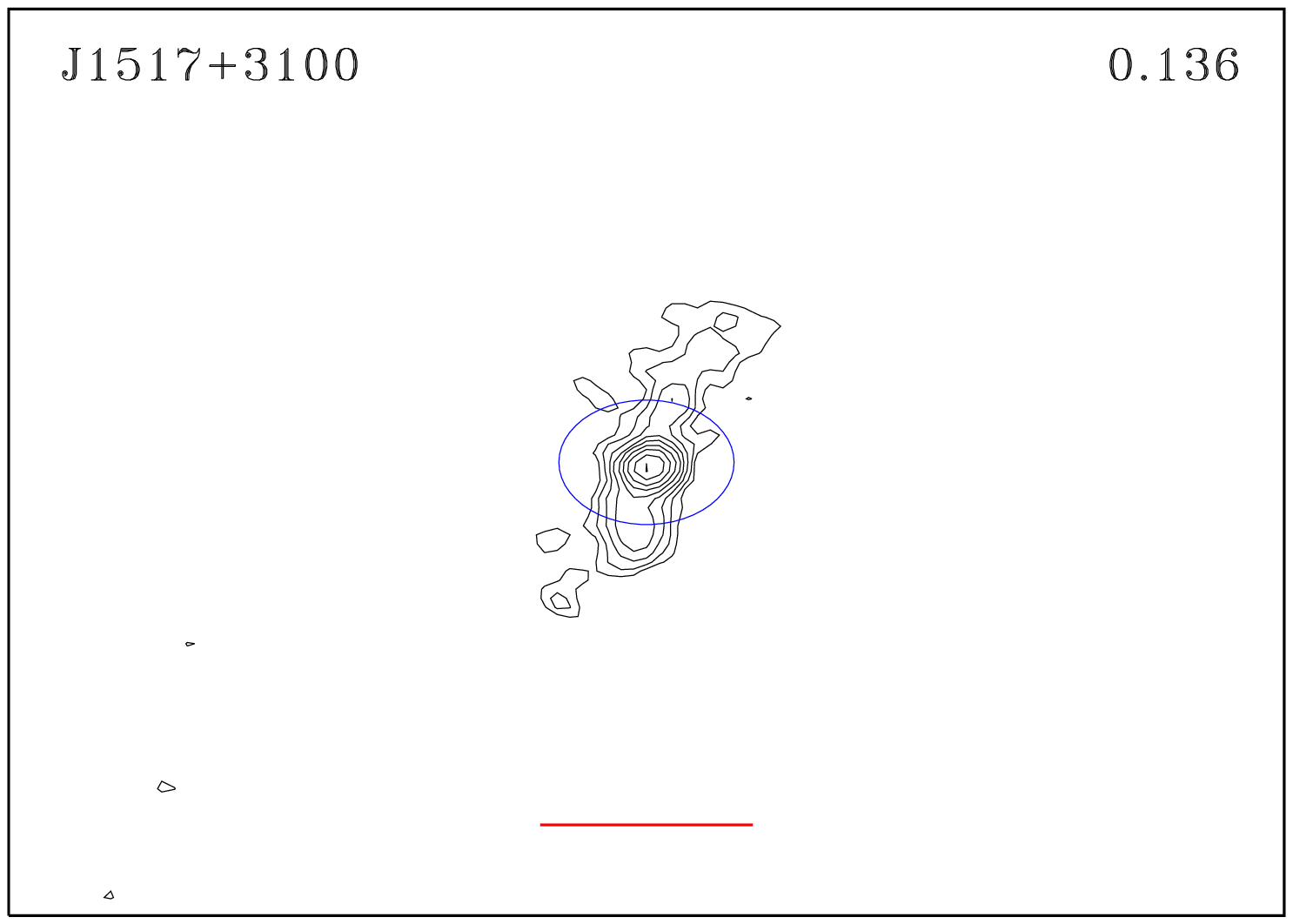} 
\includegraphics[width=6.3cm,height=6.3cm]{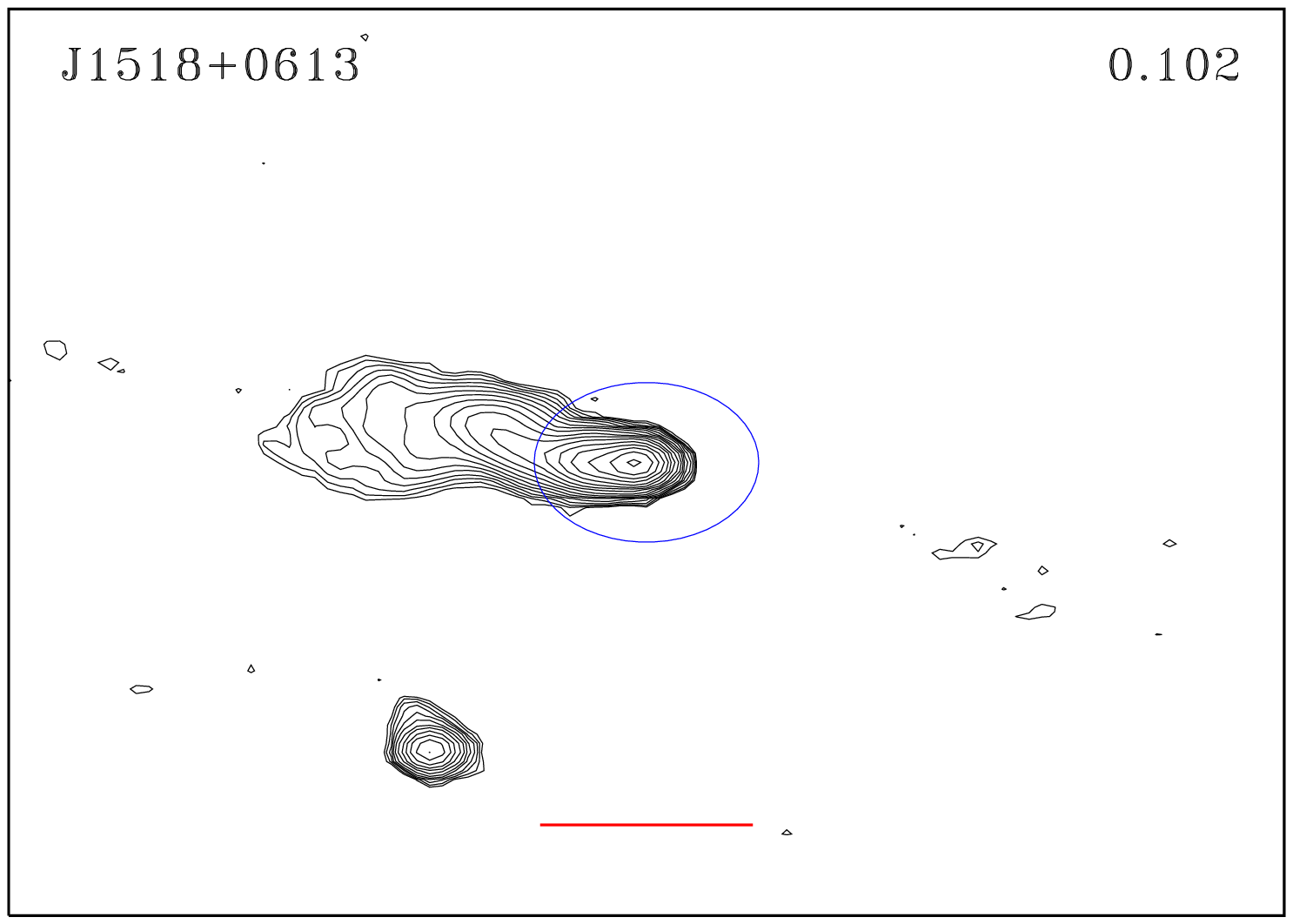} 

\includegraphics[width=6.3cm,height=6.3cm]{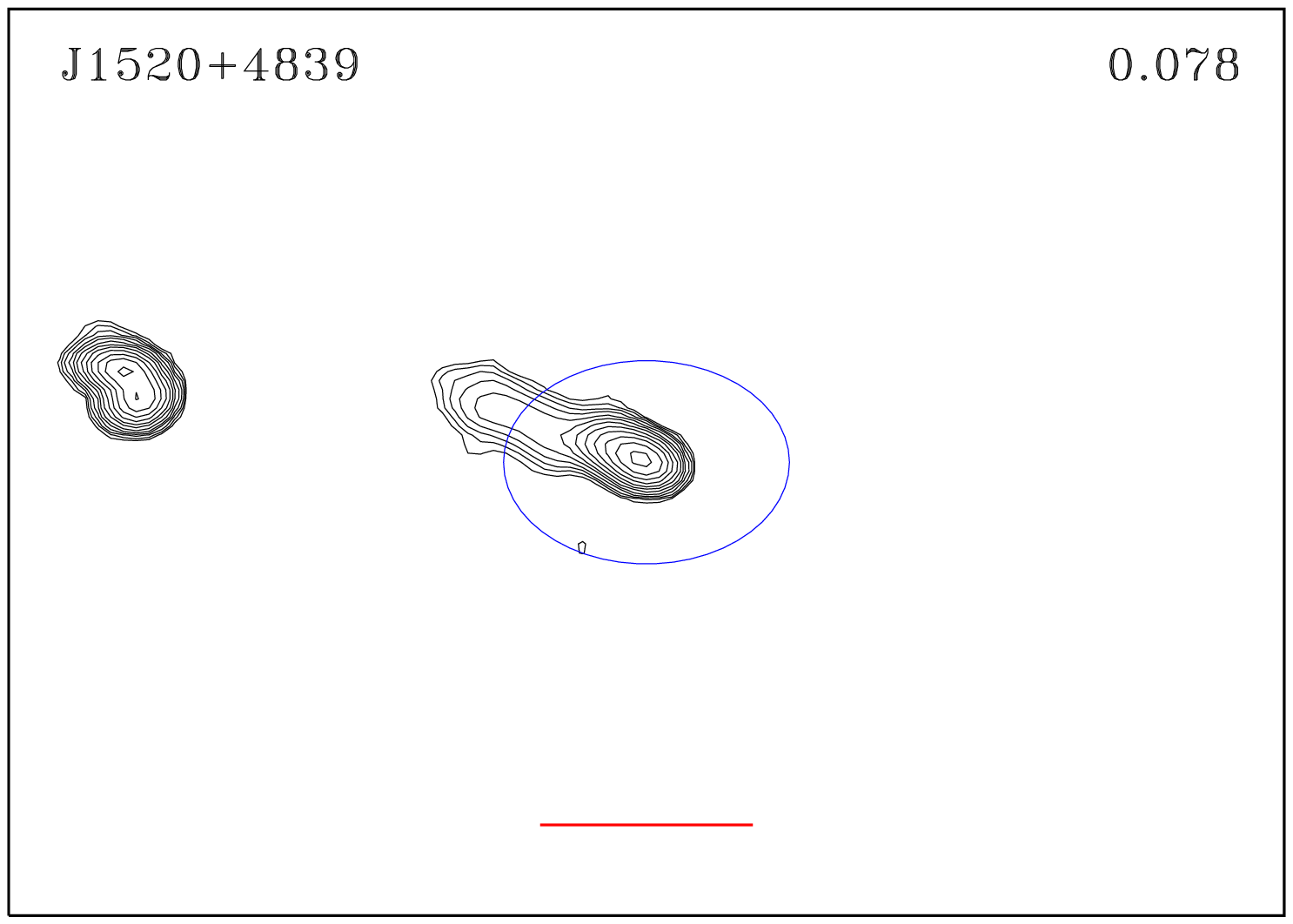} 
\includegraphics[width=6.3cm,height=6.3cm]{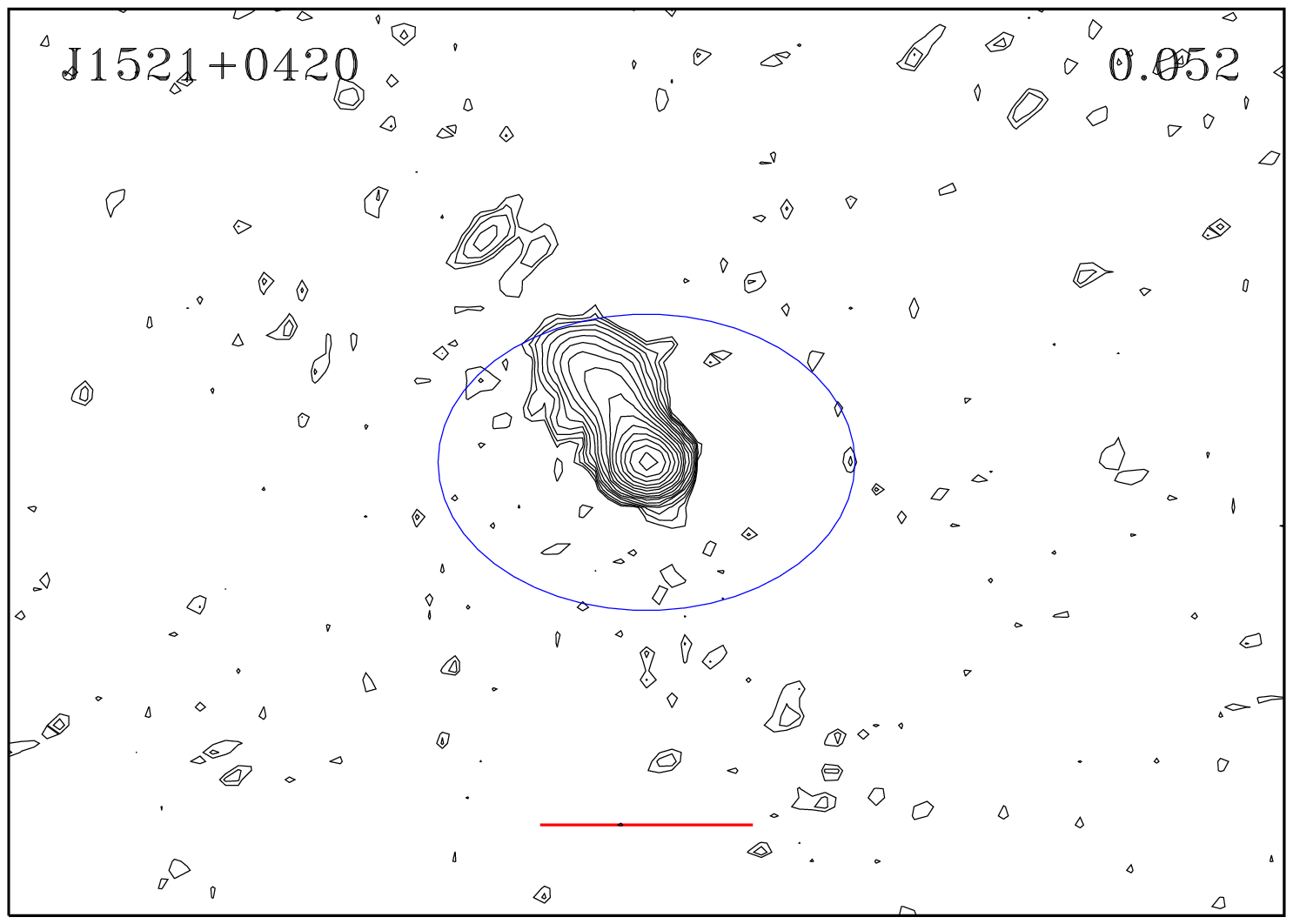} 
\includegraphics[width=6.3cm,height=6.3cm]{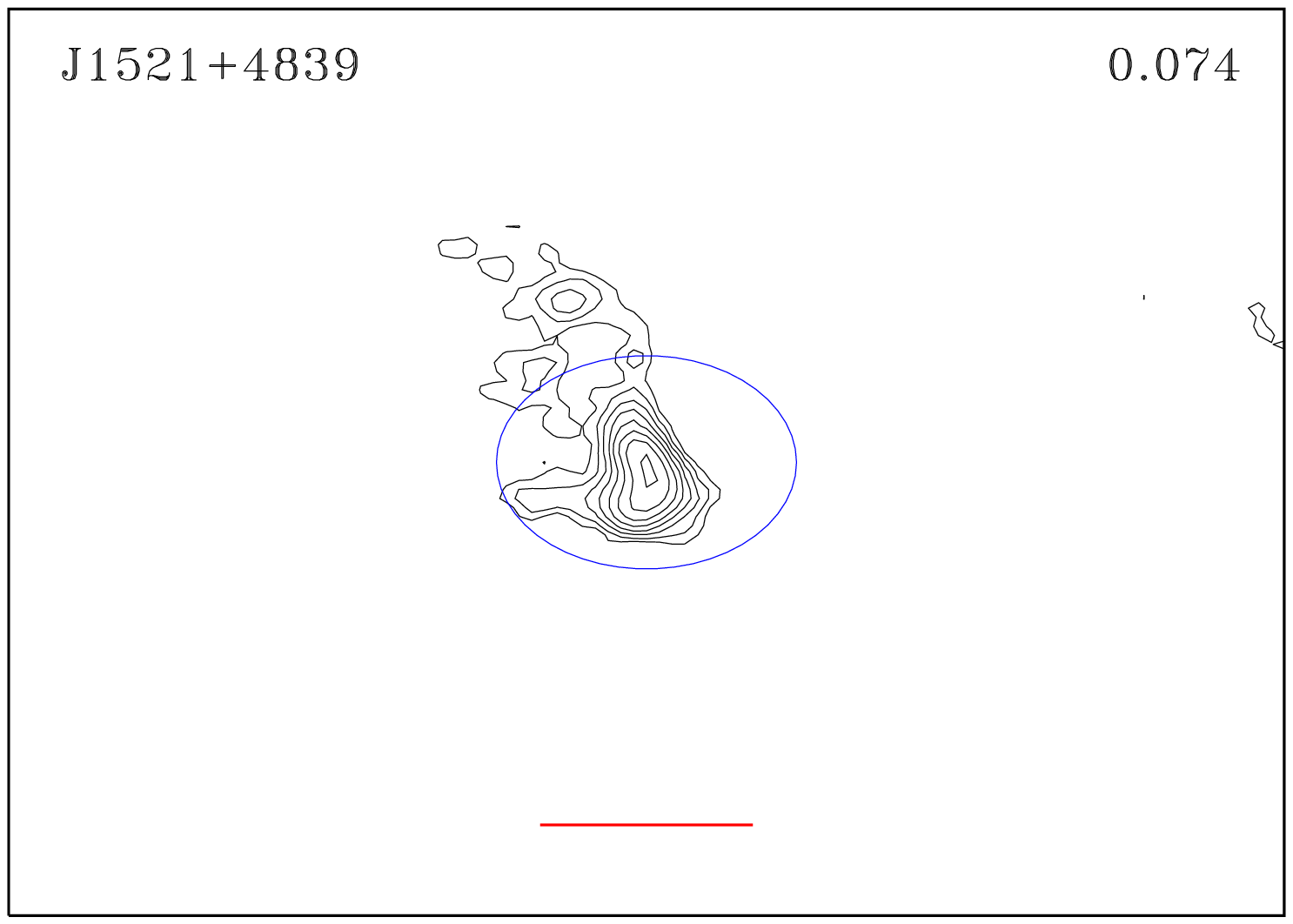} 

\includegraphics[width=6.3cm,height=6.3cm]{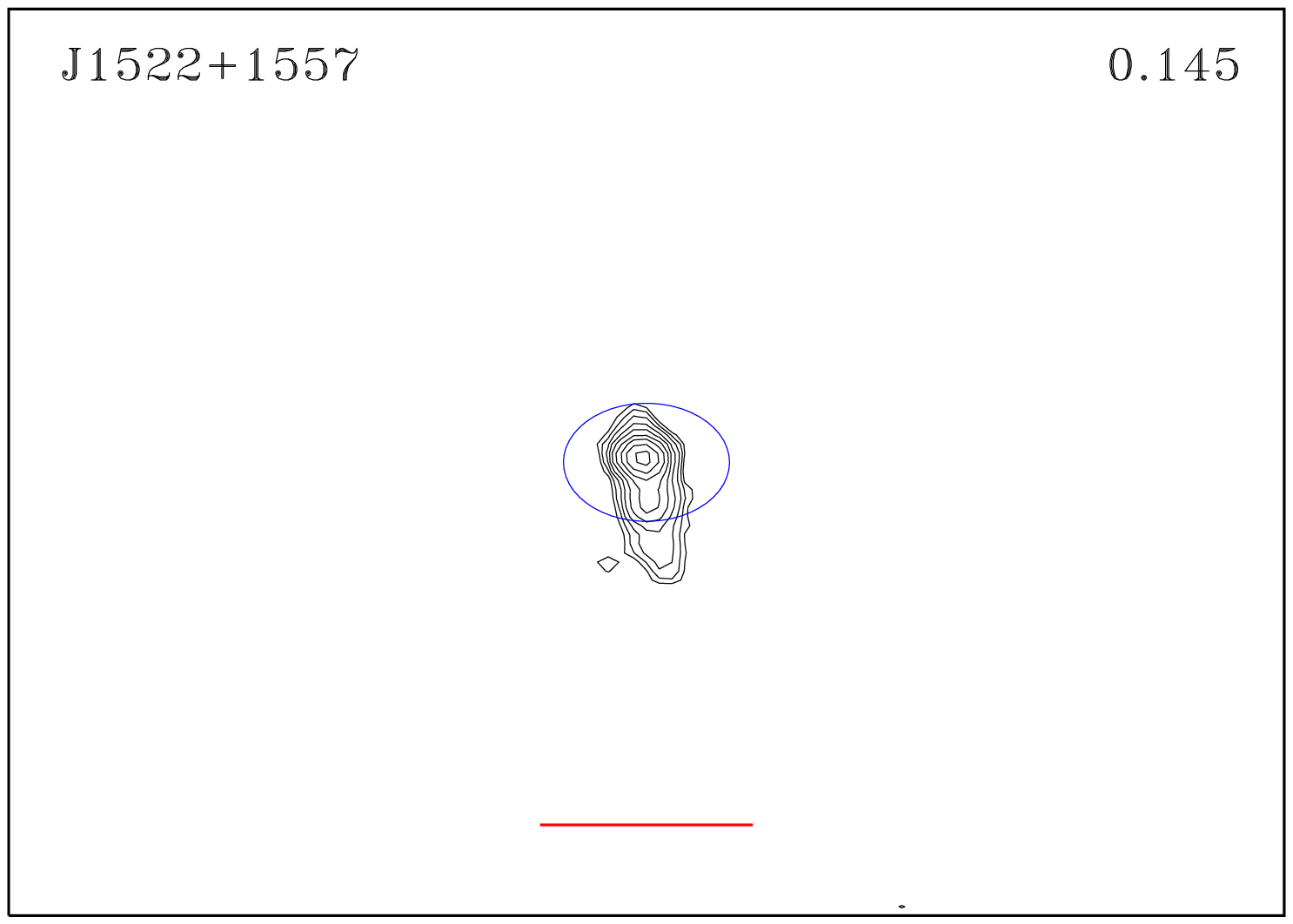} 
\includegraphics[width=6.3cm,height=6.3cm]{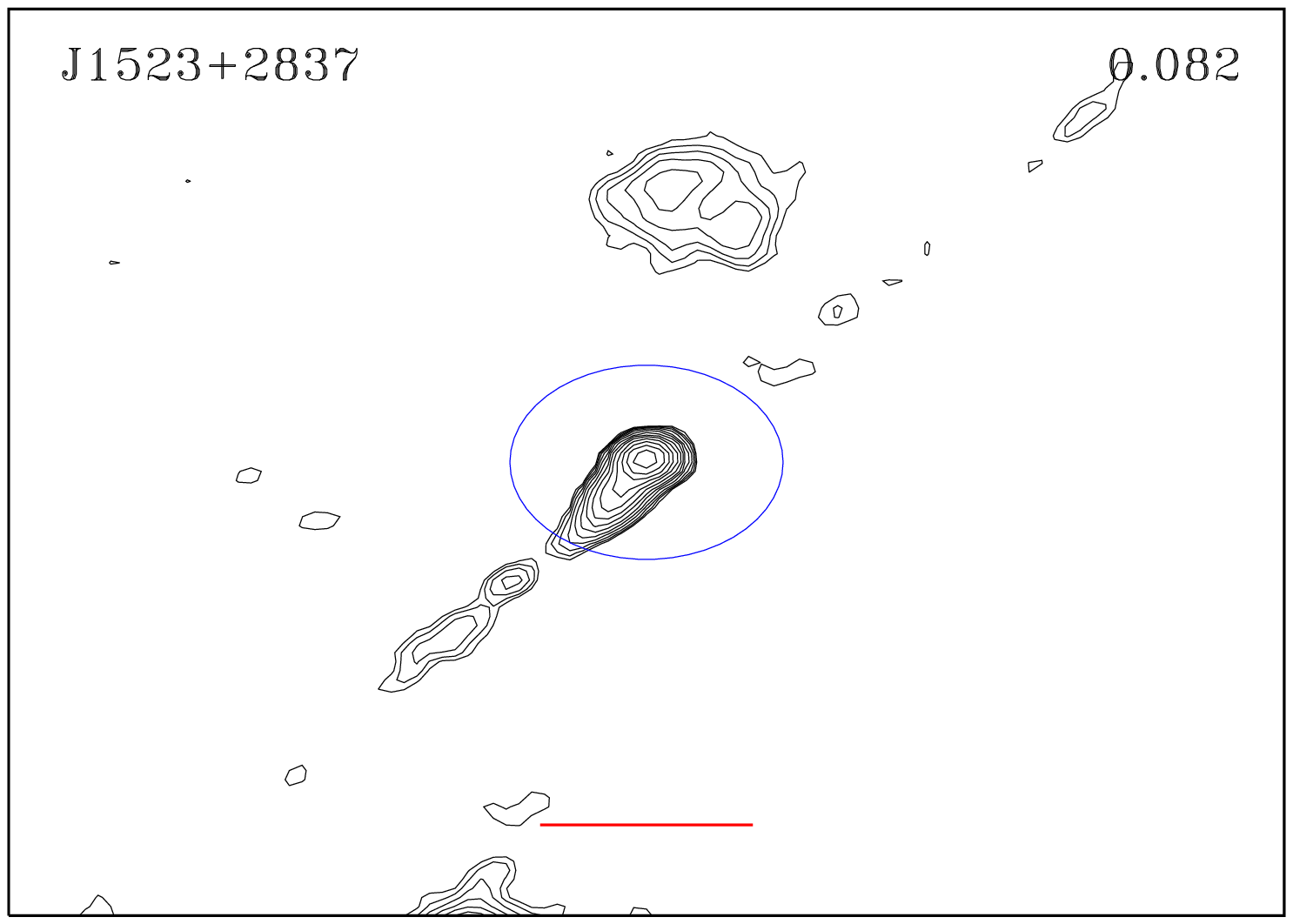} 
\includegraphics[width=6.3cm,height=6.3cm]{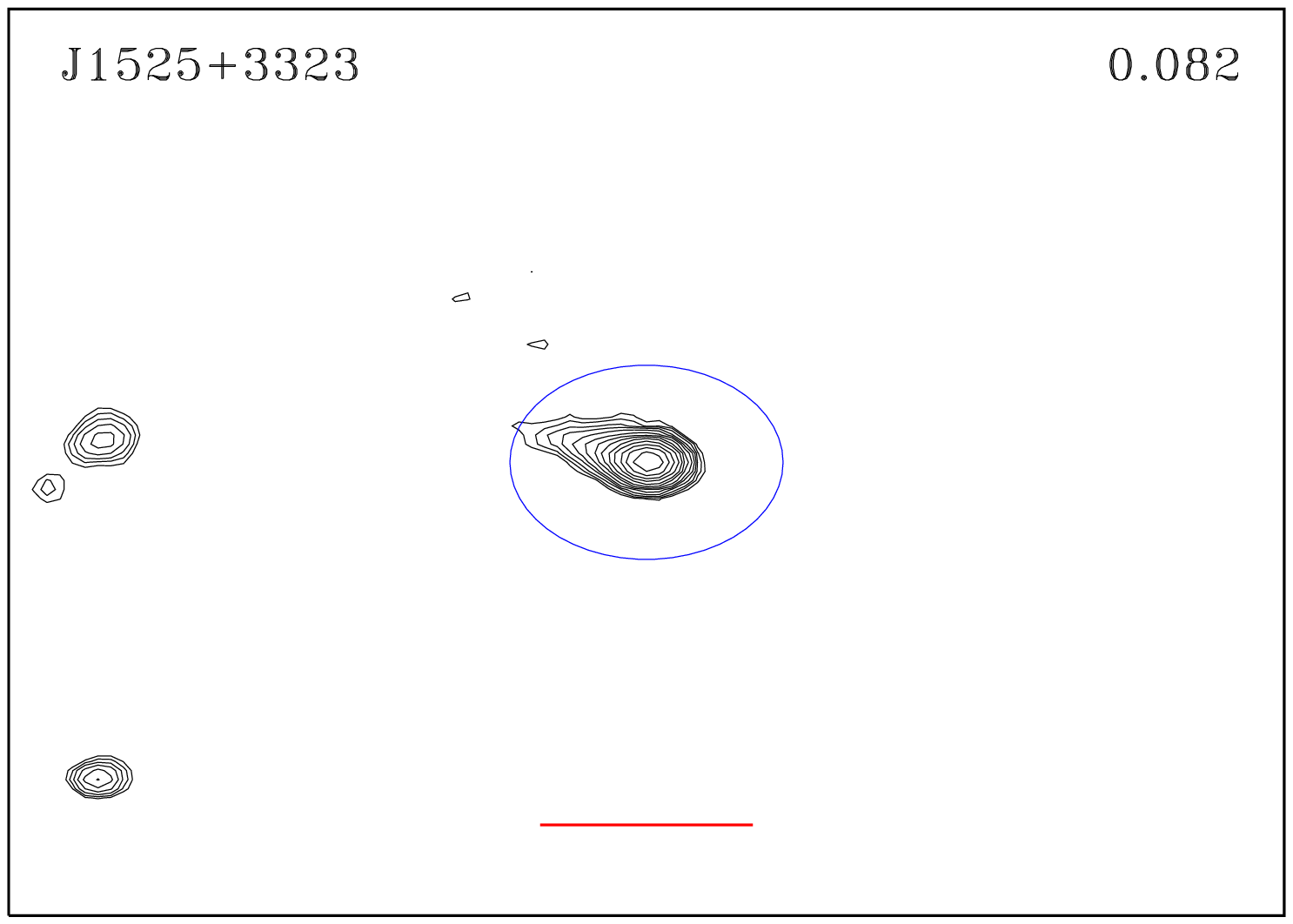} 

\includegraphics[width=6.3cm,height=6.3cm]{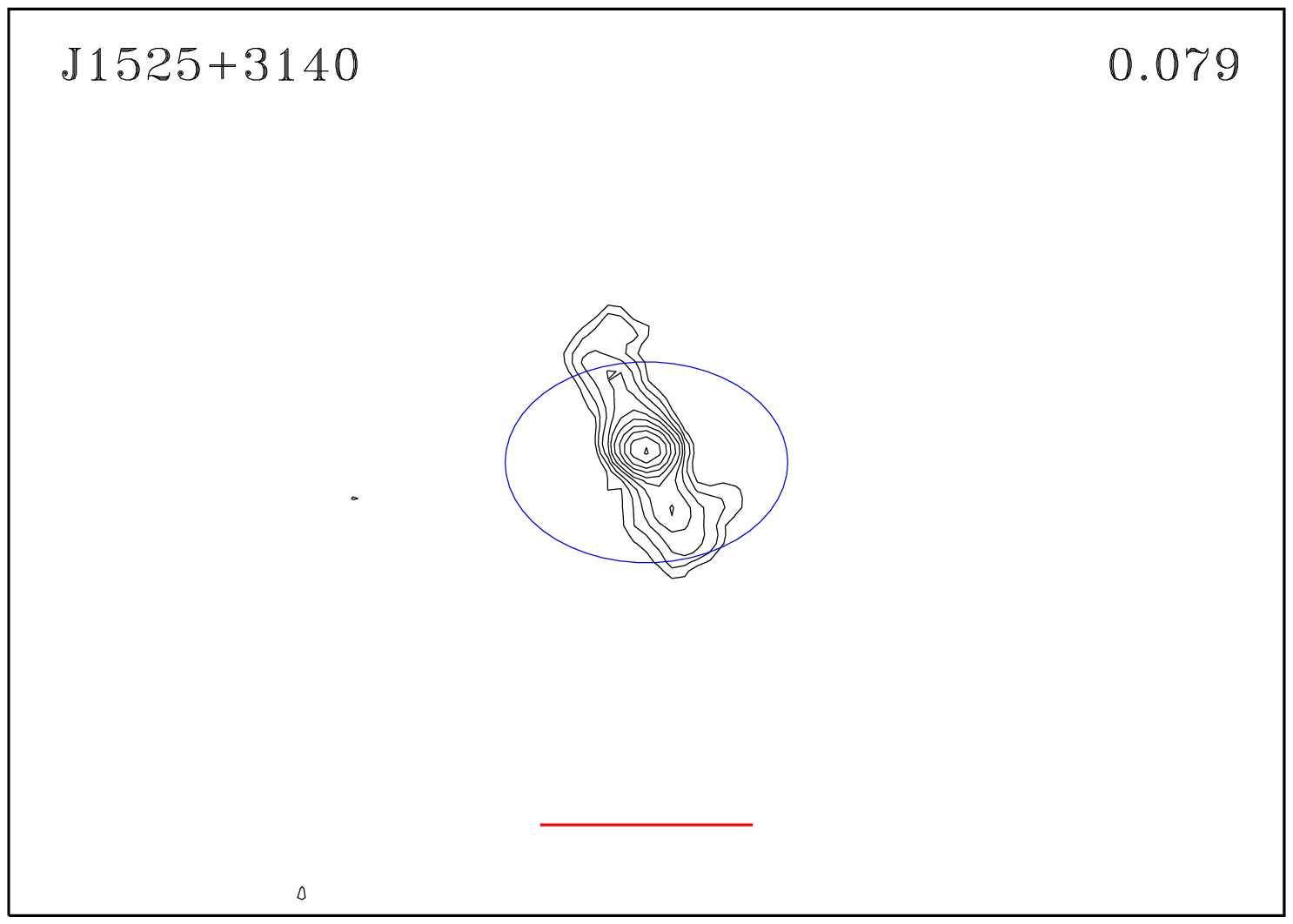} 
\includegraphics[width=6.3cm,height=6.3cm]{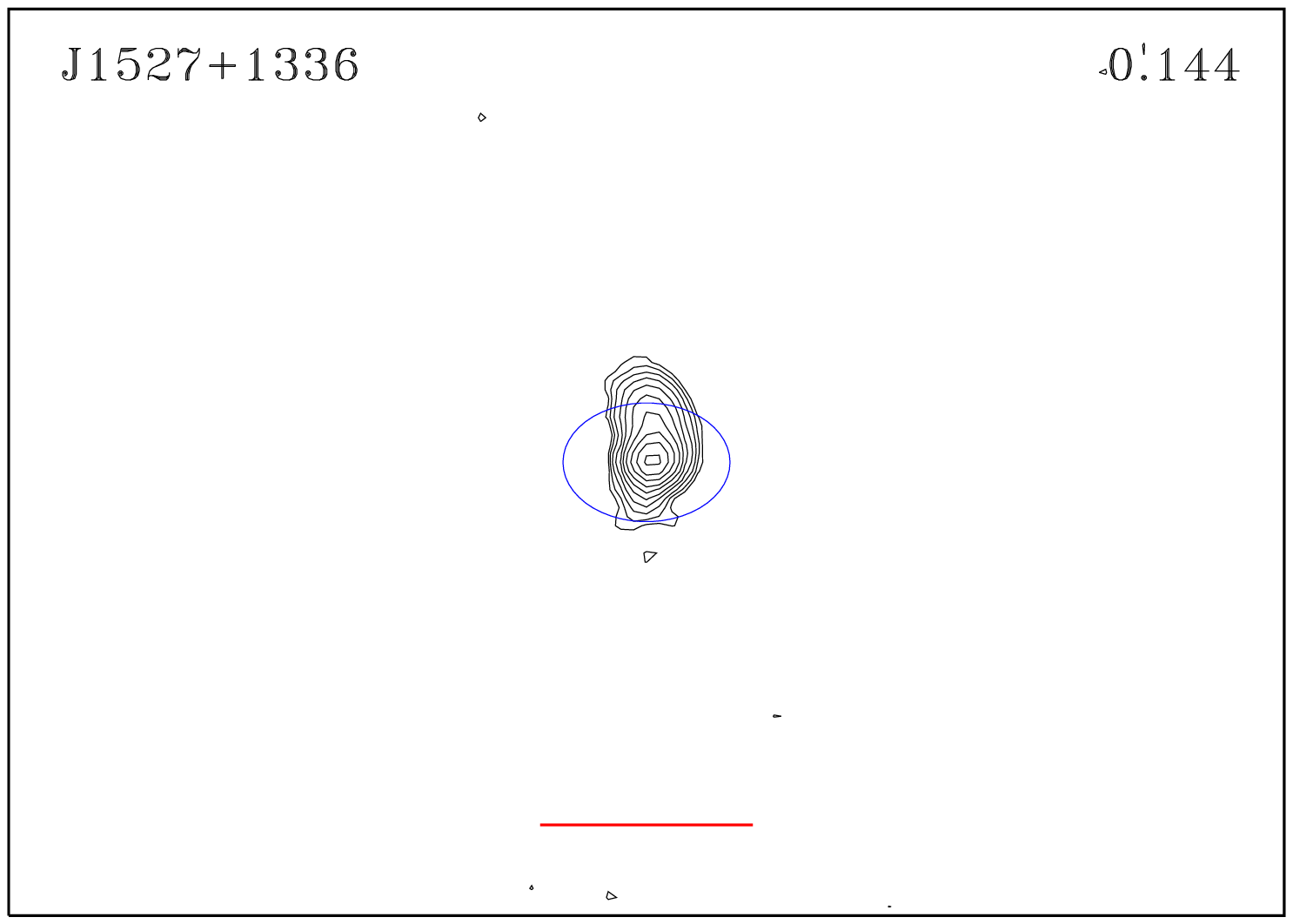} 
\includegraphics[width=6.3cm,height=6.3cm]{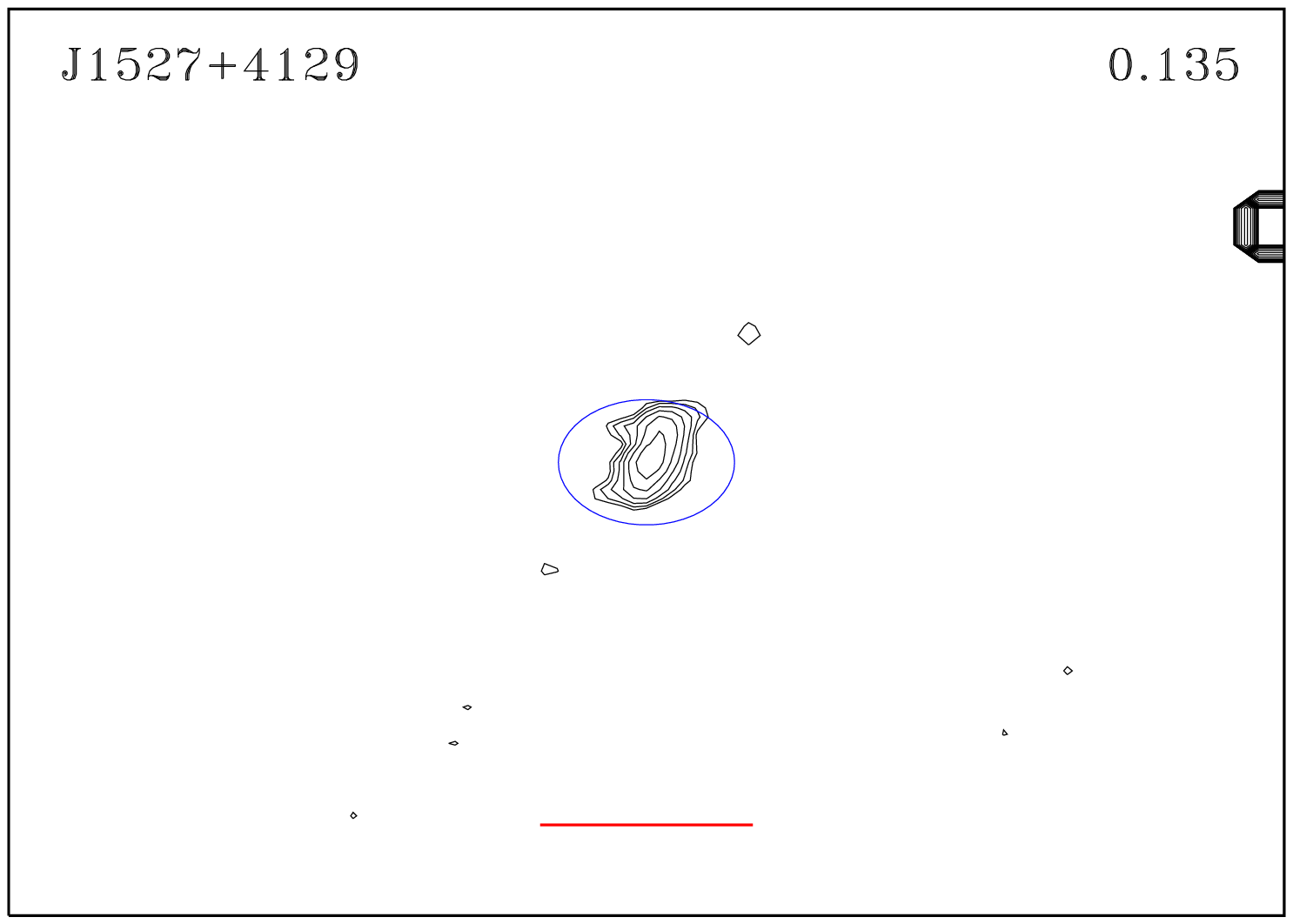} 
\caption{(continued)}
\end{figure*}

\addtocounter{figure}{-1}
\begin{figure*}
\includegraphics[width=6.3cm,height=6.3cm]{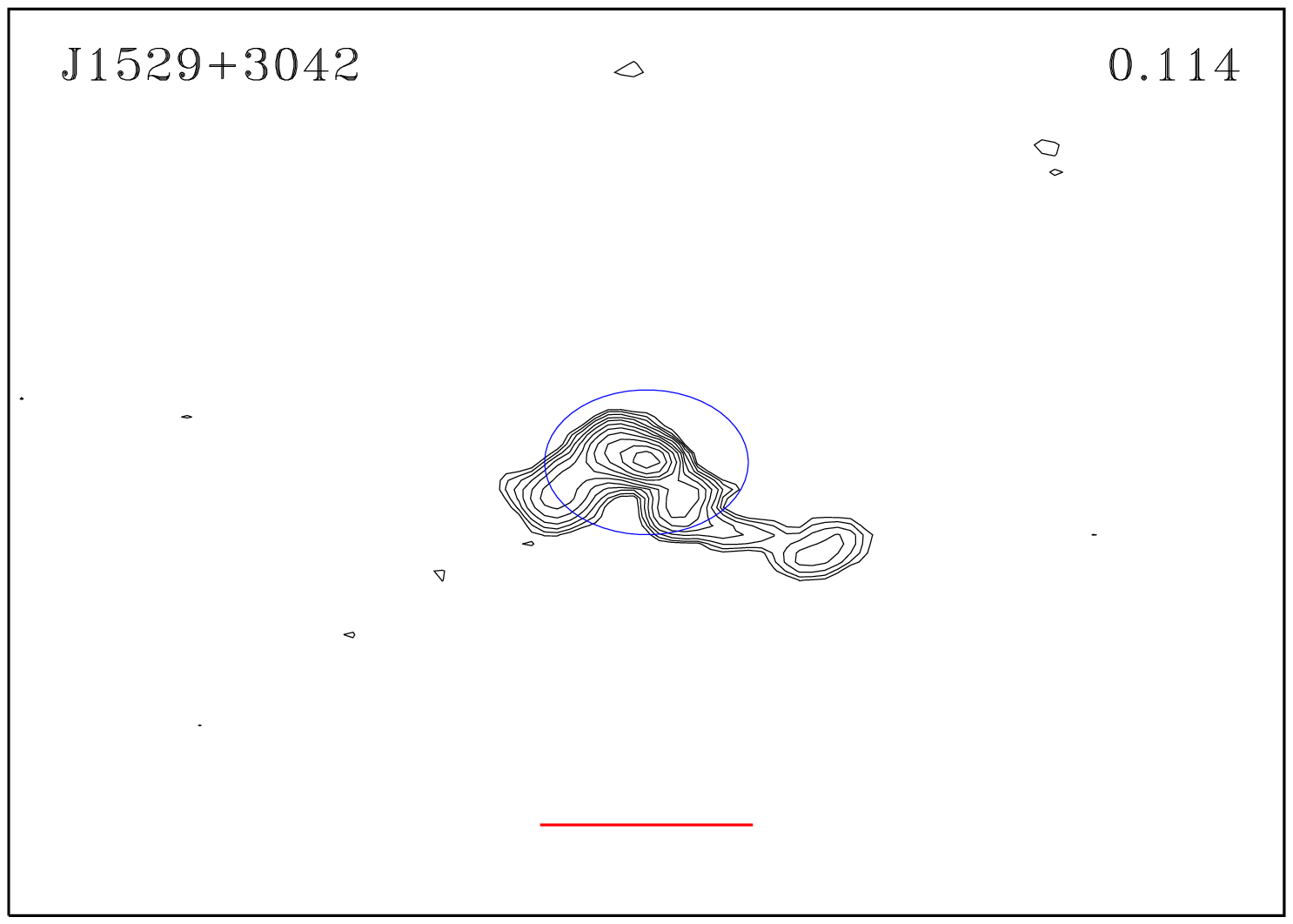} 
\includegraphics[width=6.3cm,height=6.3cm]{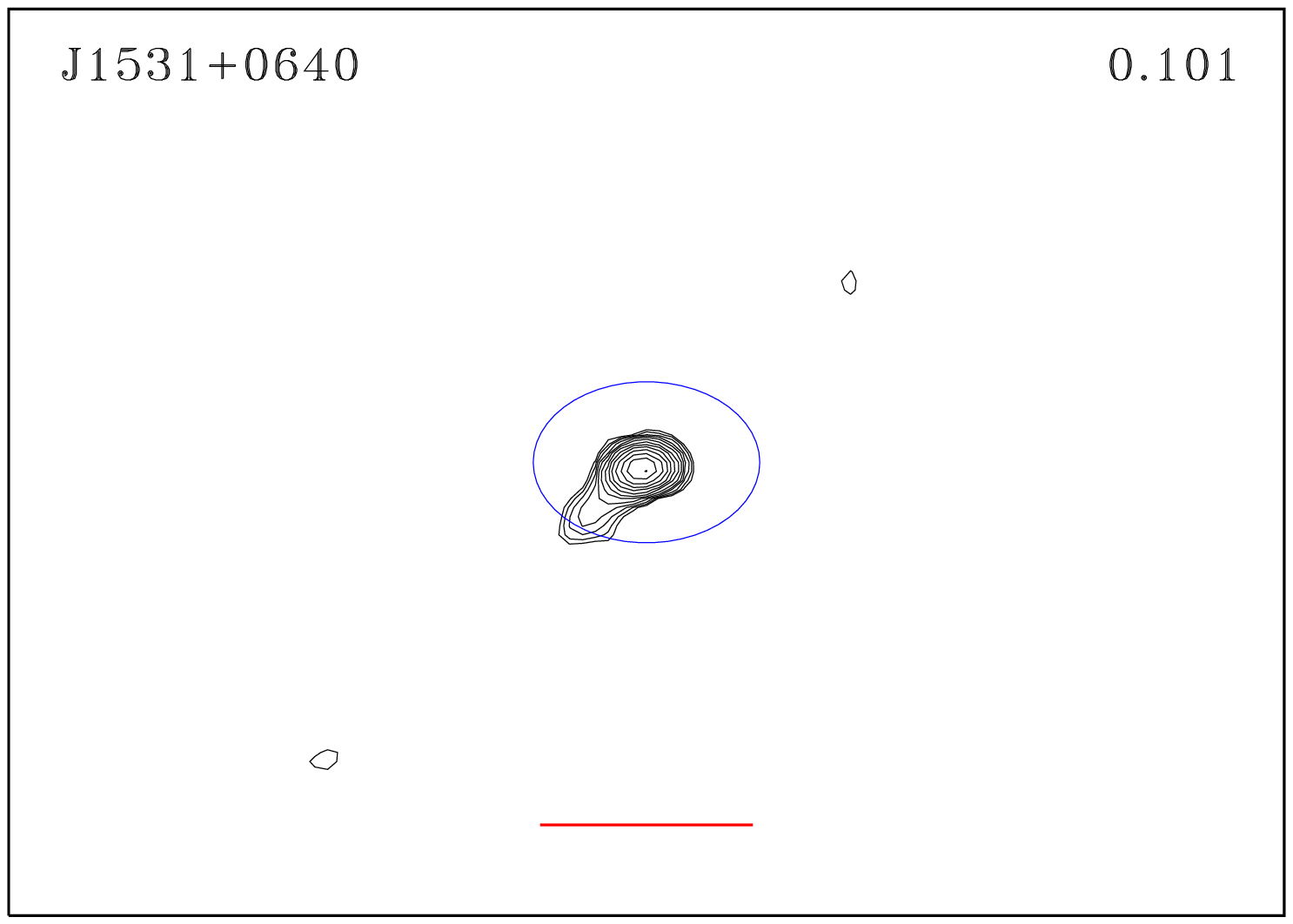} 
\includegraphics[width=6.3cm,height=6.3cm]{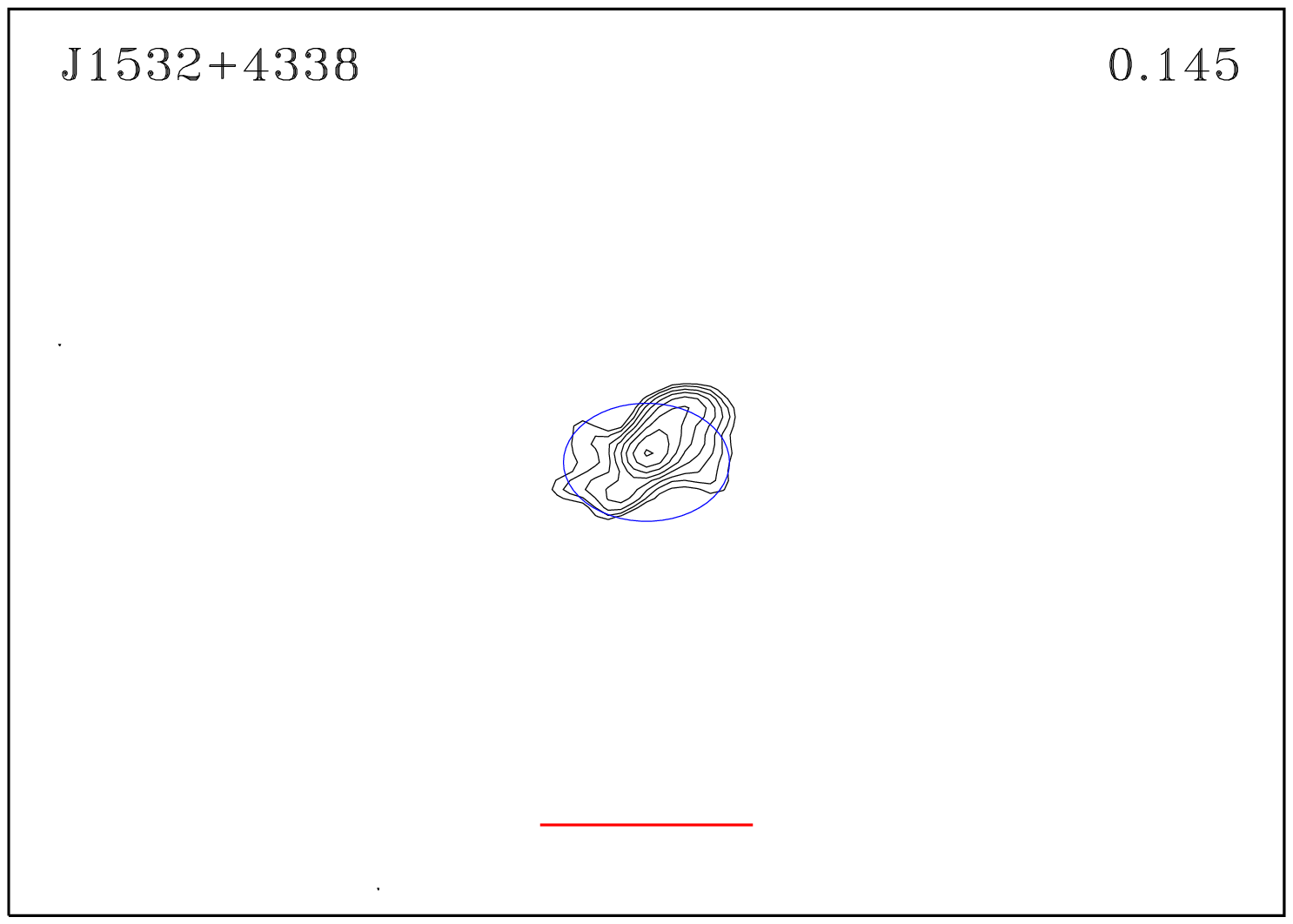} 

\includegraphics[width=6.3cm,height=6.3cm]{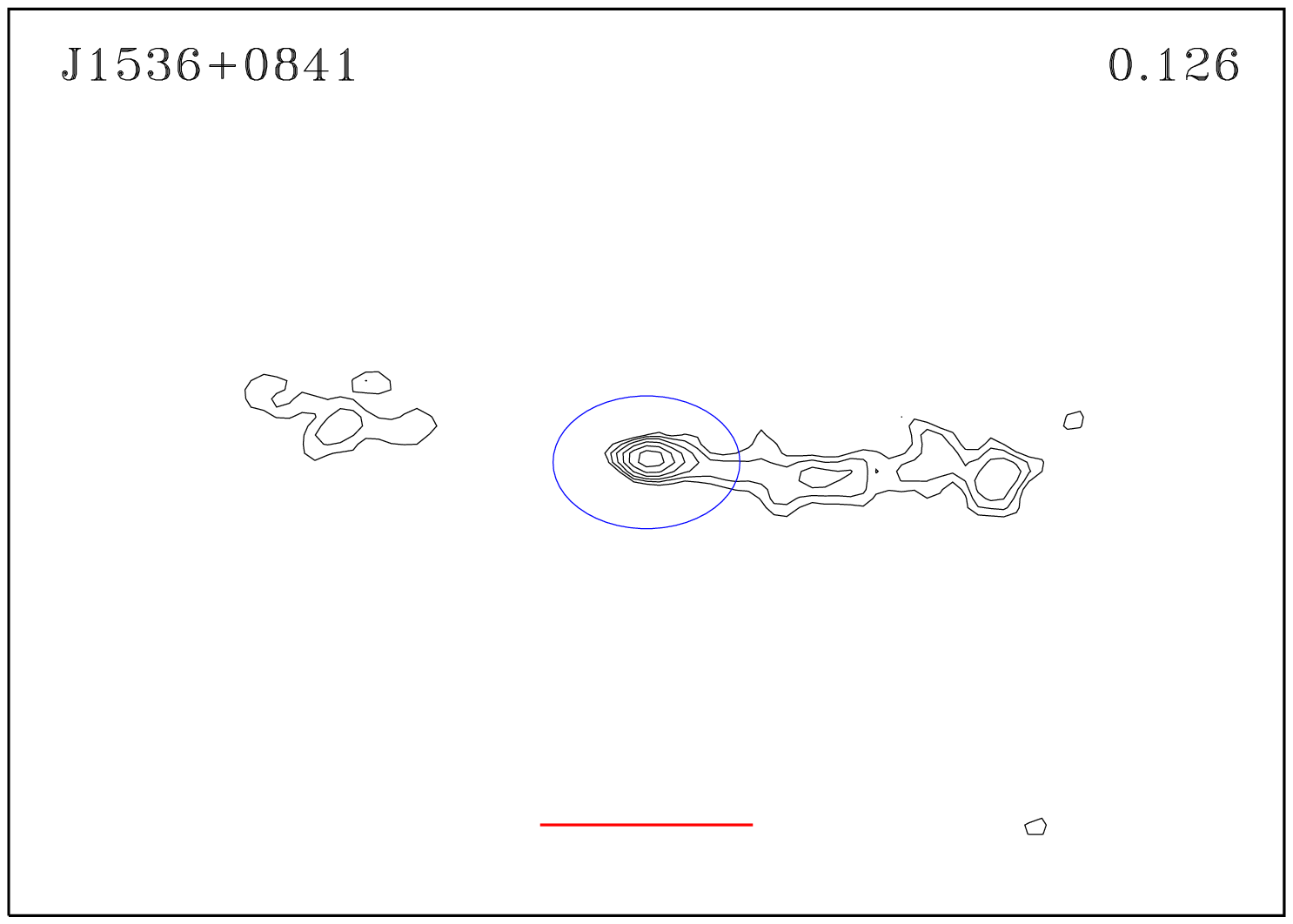} 
\includegraphics[width=6.3cm,height=6.3cm]{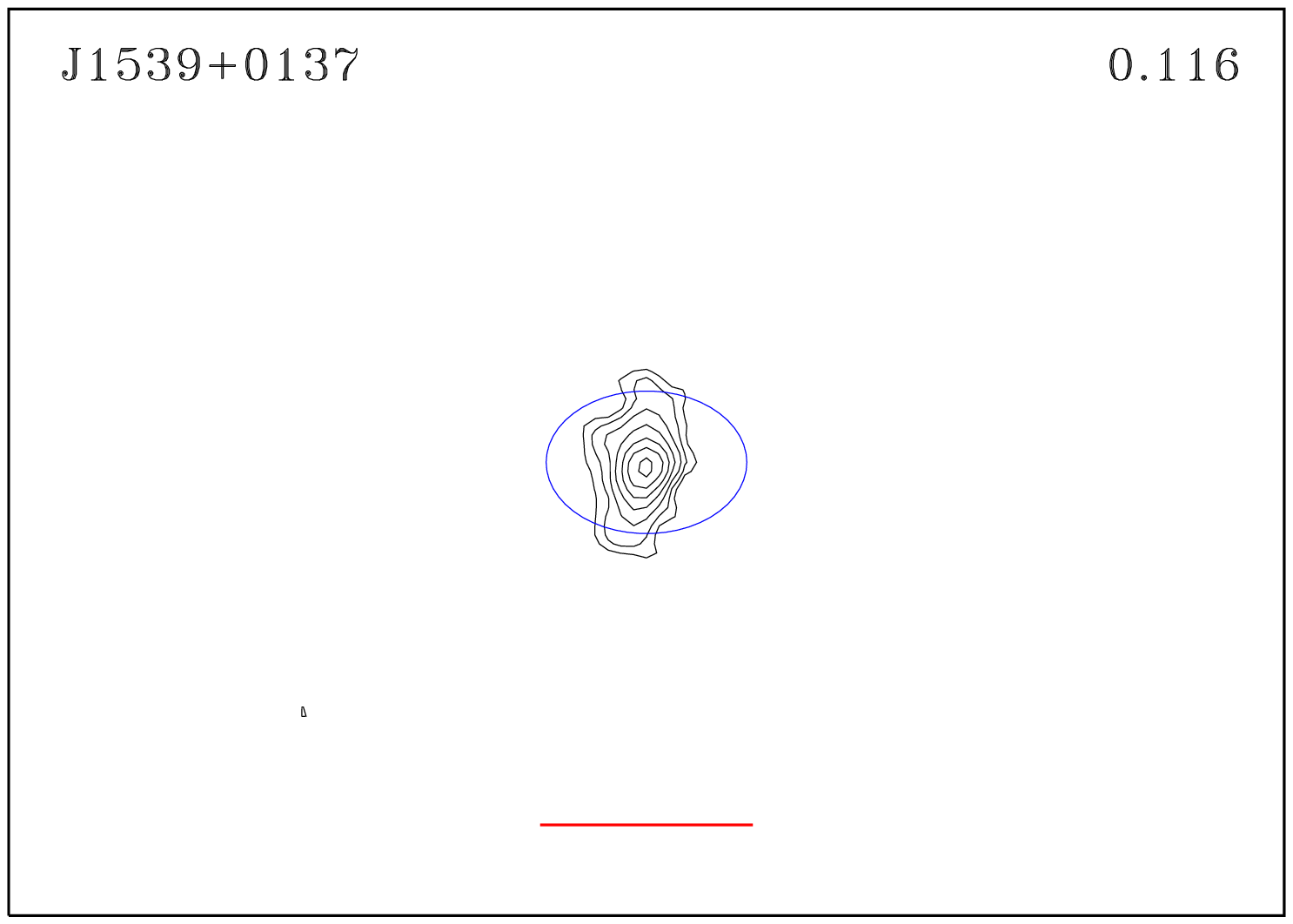} 
\includegraphics[width=6.3cm,height=6.3cm]{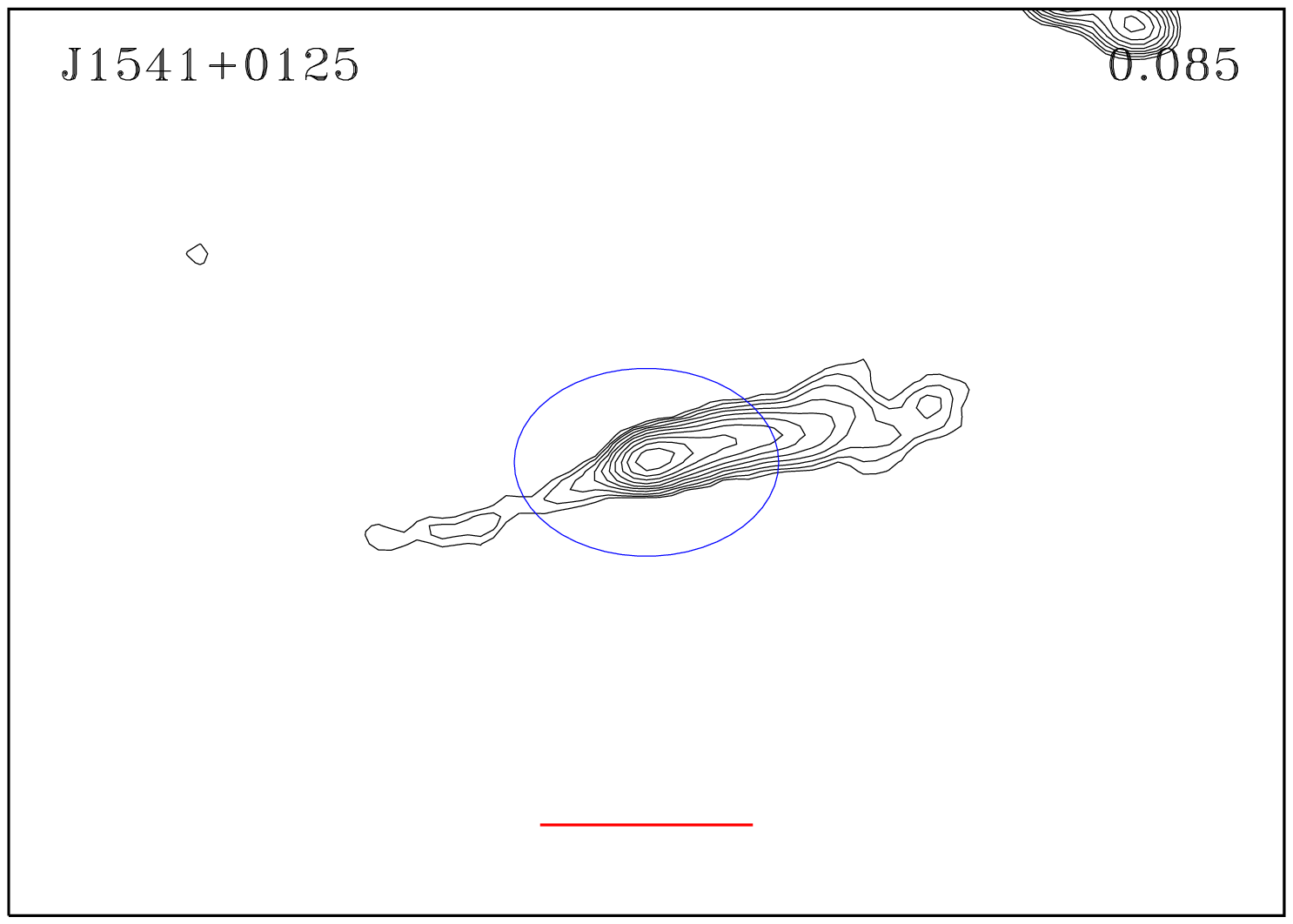} 

\includegraphics[width=6.3cm,height=6.3cm]{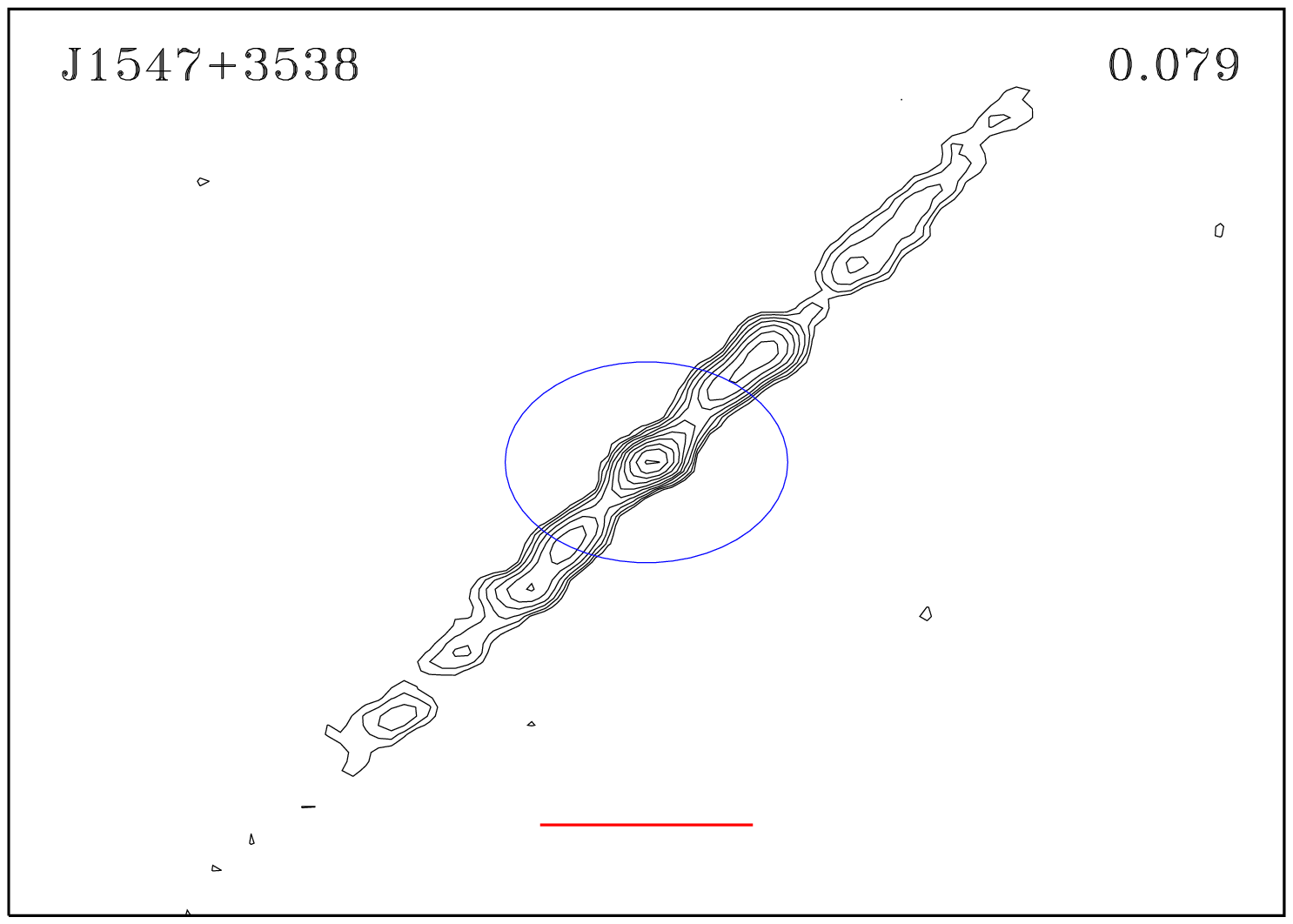} 
\includegraphics[width=6.3cm,height=6.3cm]{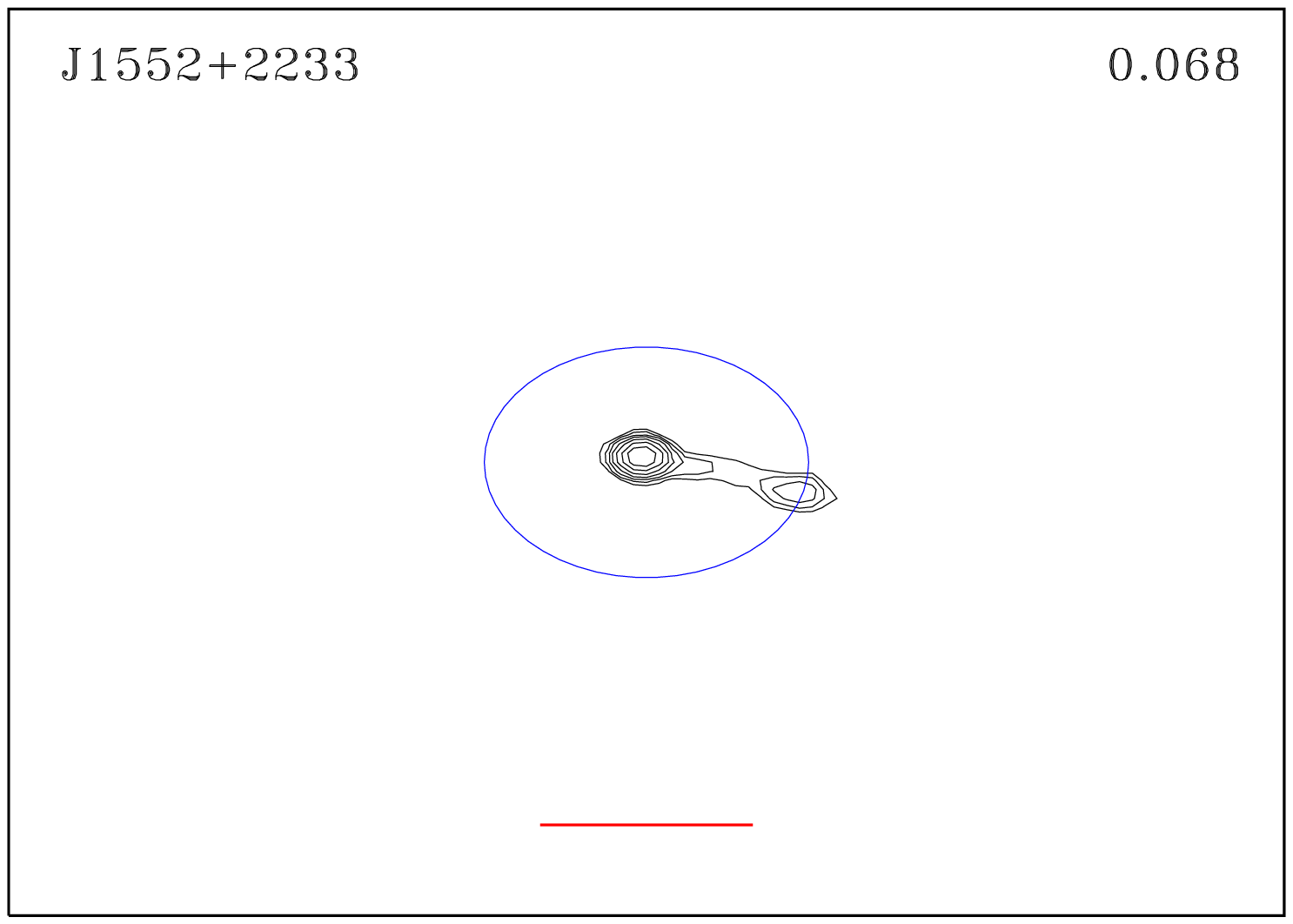} 
\includegraphics[width=6.3cm,height=6.3cm]{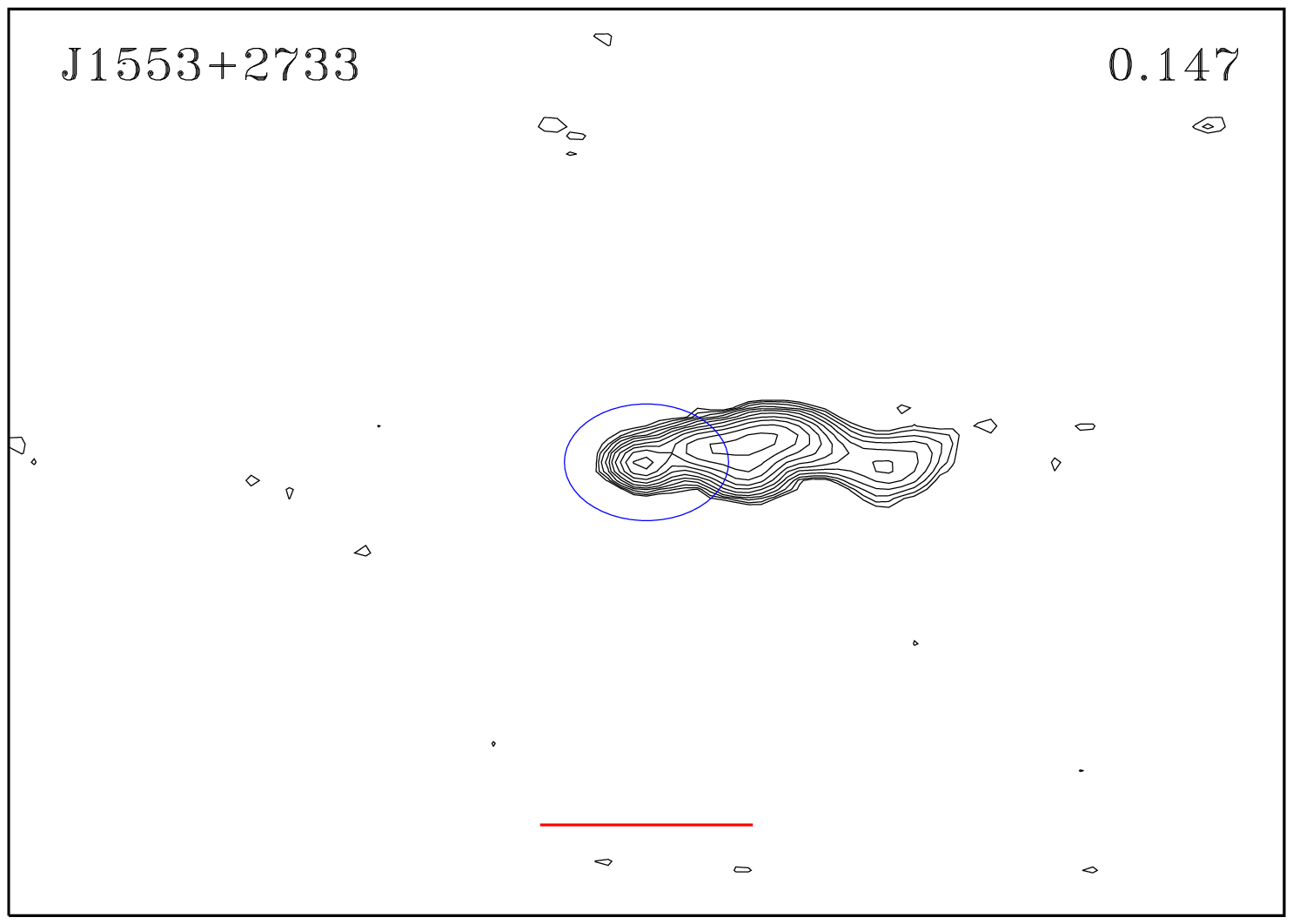} 

\includegraphics[width=6.3cm,height=6.3cm]{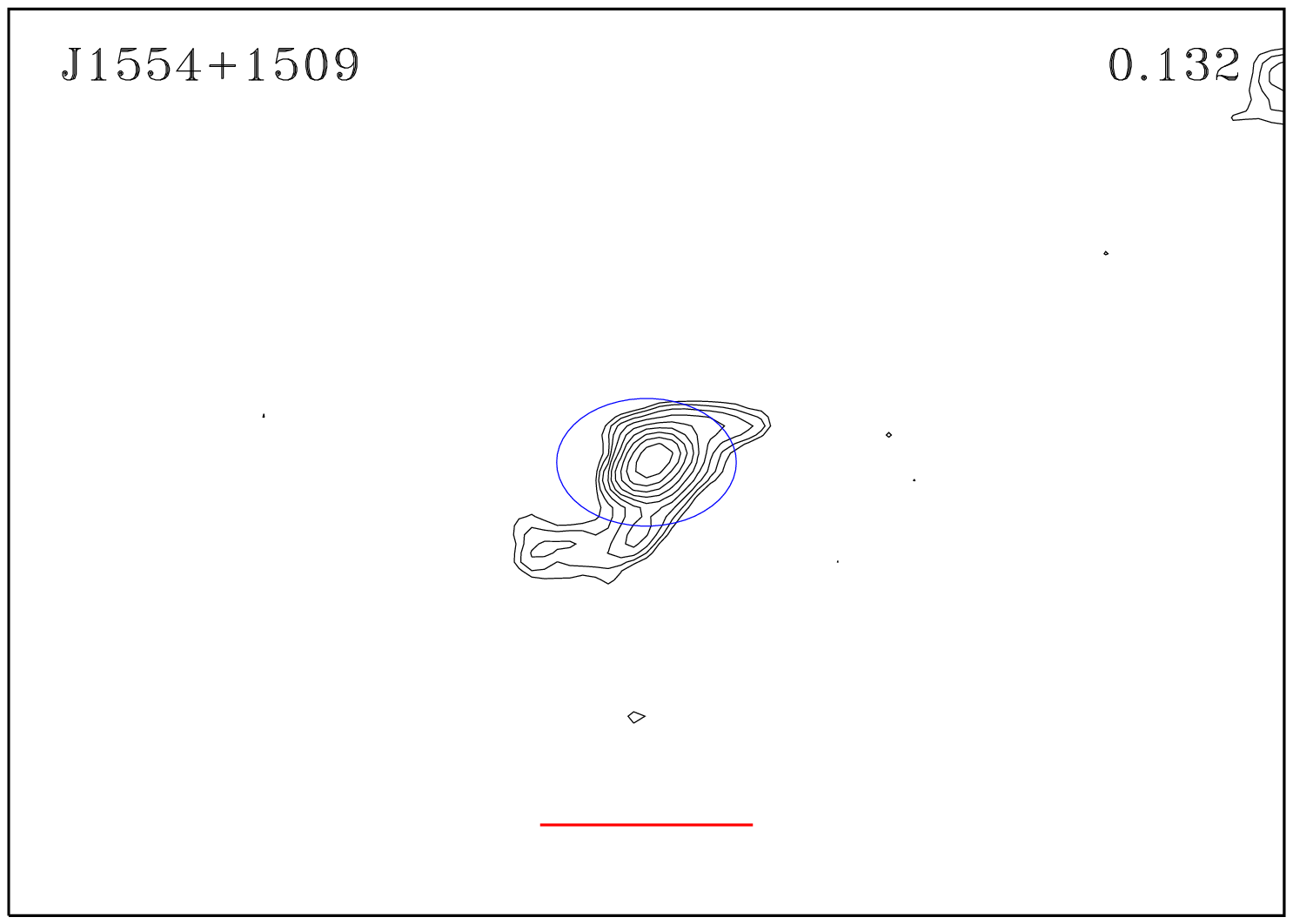} 
\includegraphics[width=6.3cm,height=6.3cm]{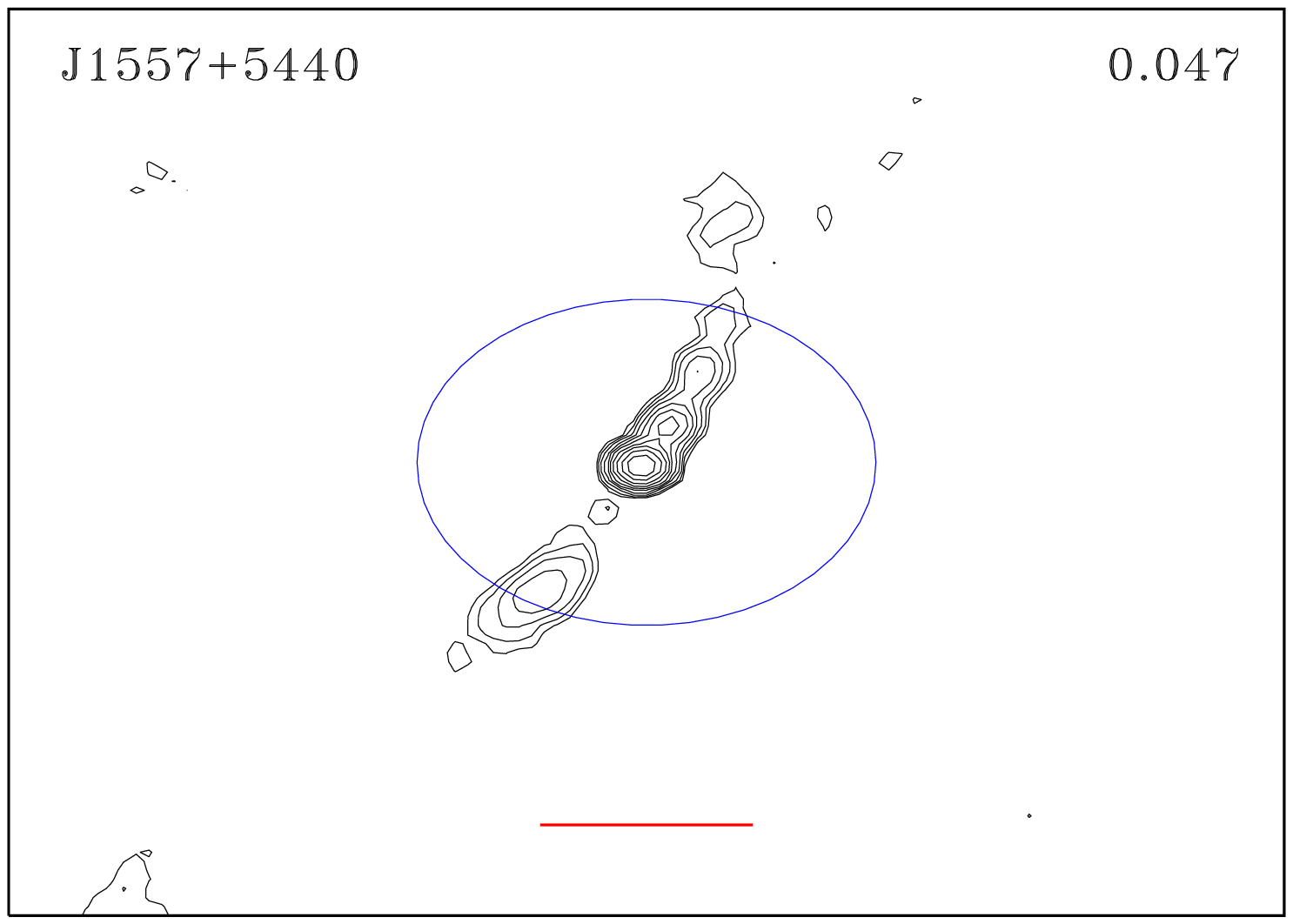} 
\includegraphics[width=6.3cm,height=6.3cm]{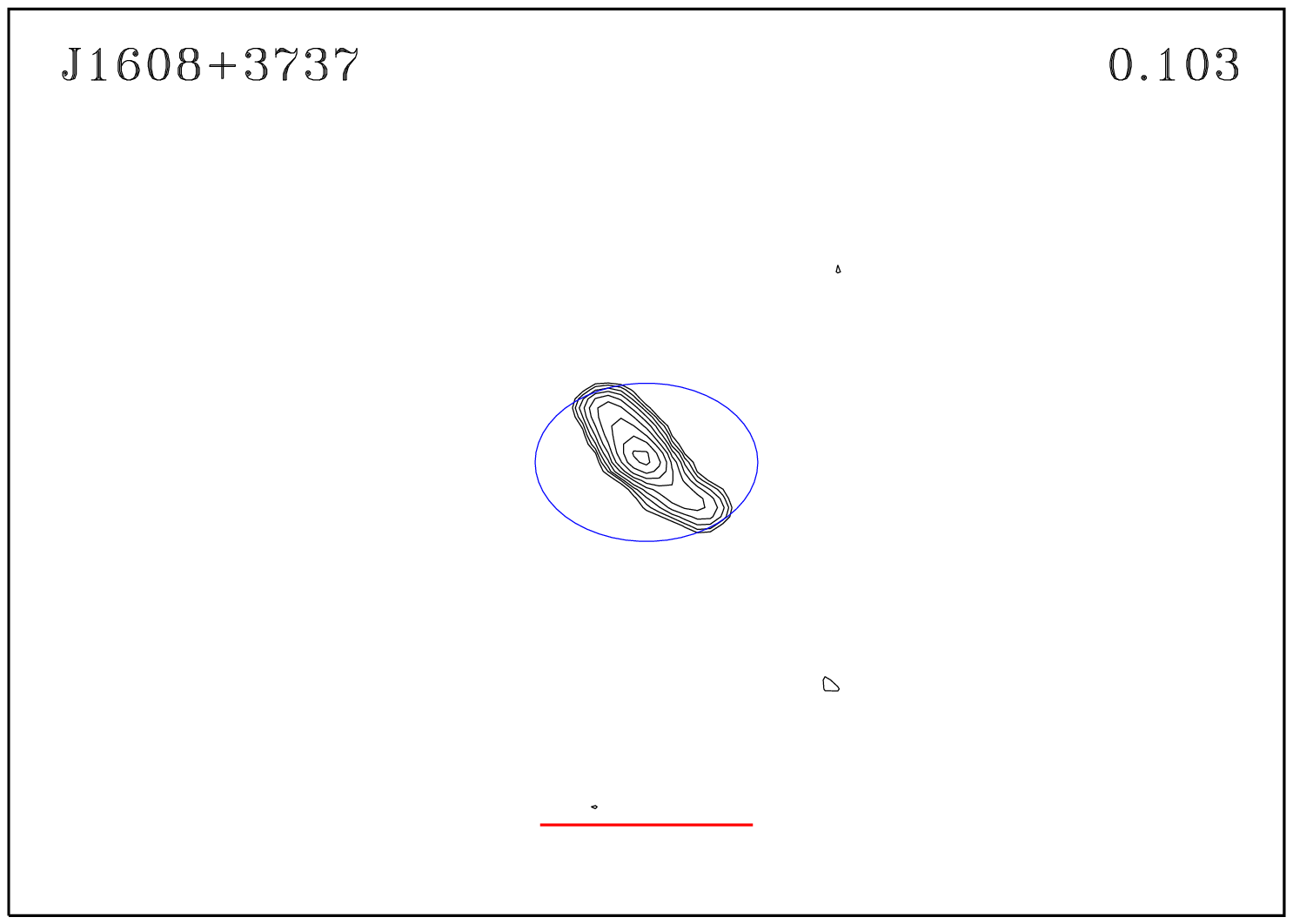} 
\caption{(continued)}
\end{figure*}

\addtocounter{figure}{-1}
\begin{figure*}
\includegraphics[width=6.3cm,height=6.3cm]{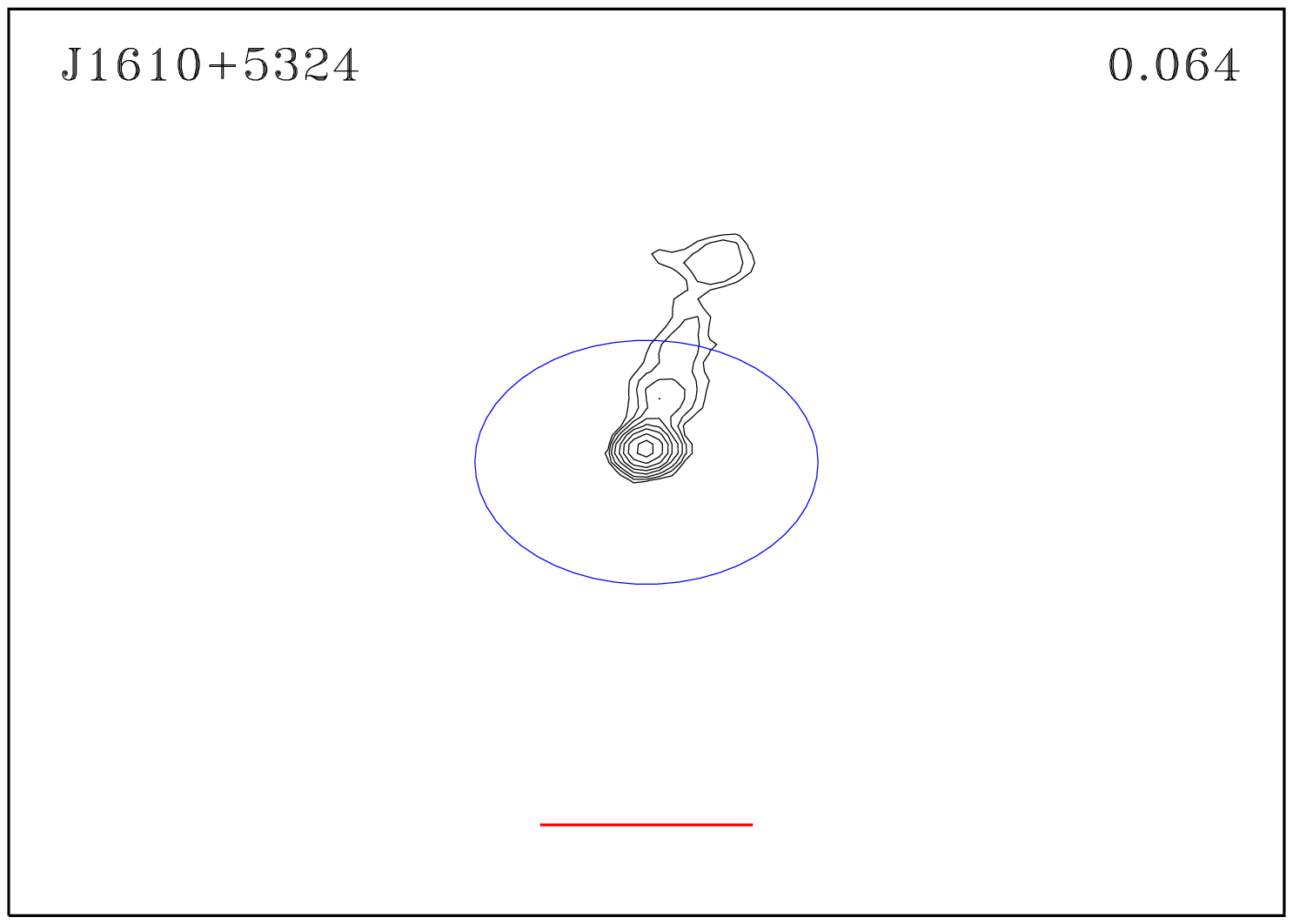} 
\includegraphics[width=6.3cm,height=6.3cm]{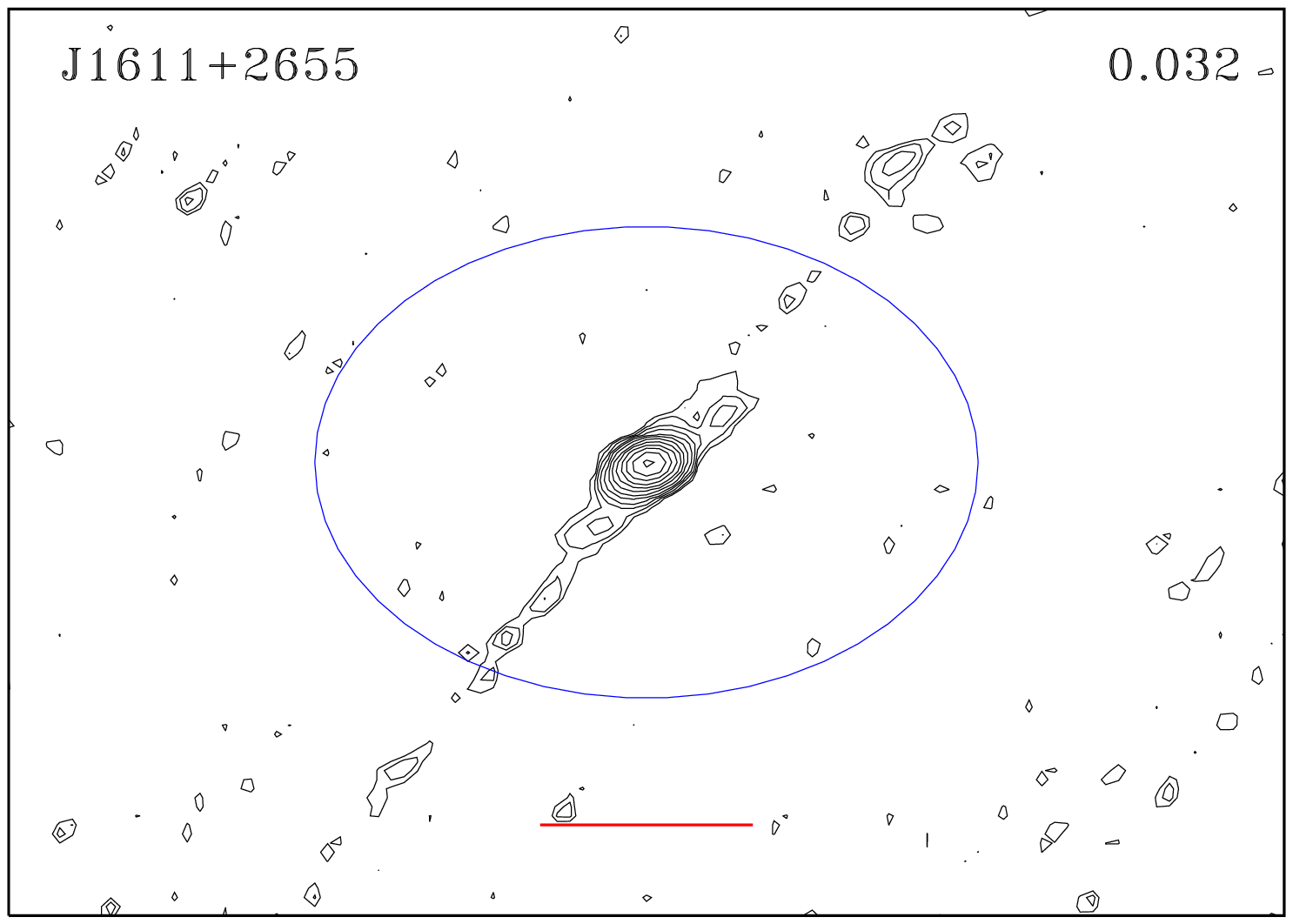} 
\includegraphics[width=6.3cm,height=6.3cm]{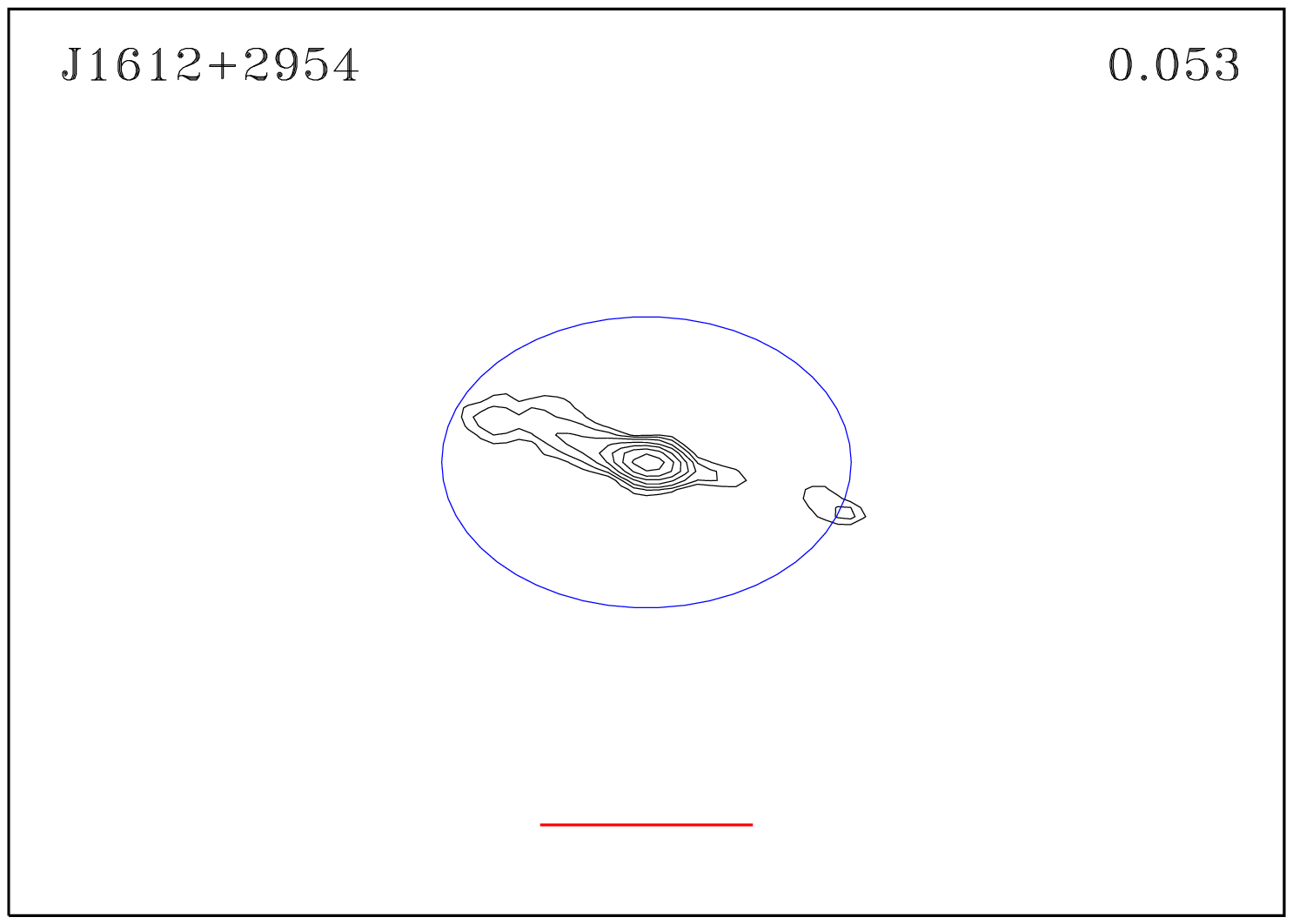} 

\includegraphics[width=6.3cm,height=6.3cm]{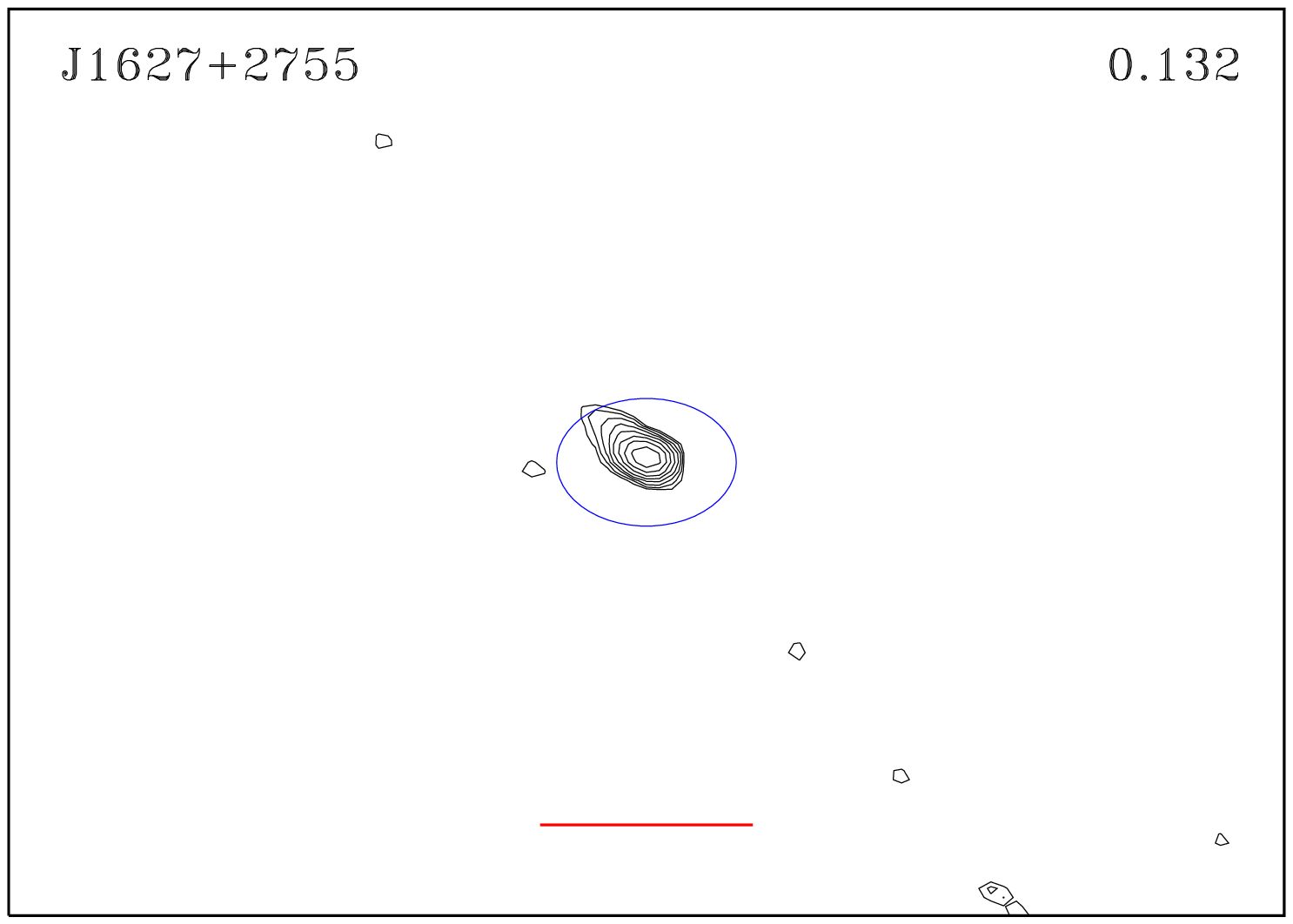} 
\includegraphics[width=6.3cm,height=6.3cm]{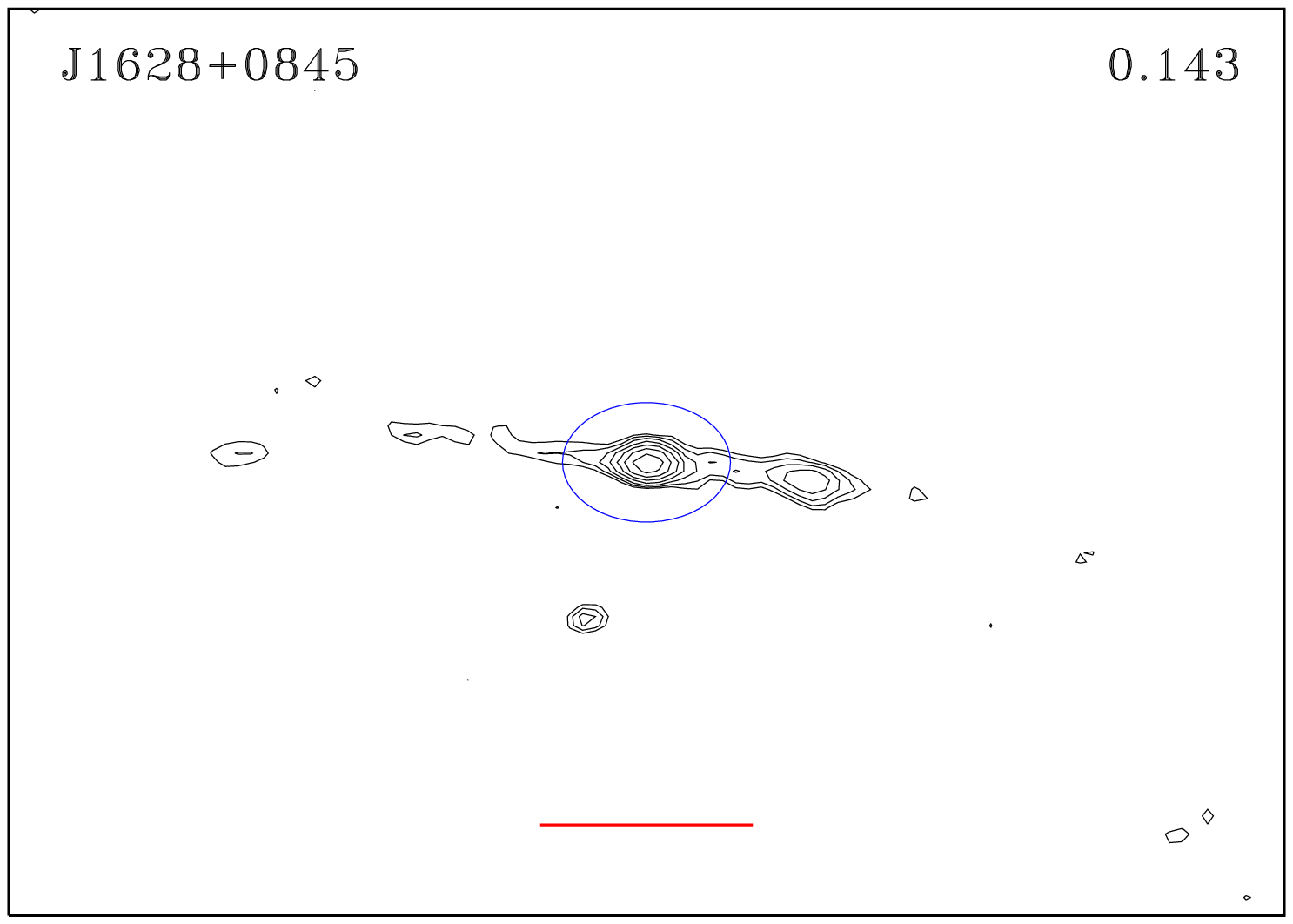} 
\includegraphics[width=6.3cm,height=6.3cm]{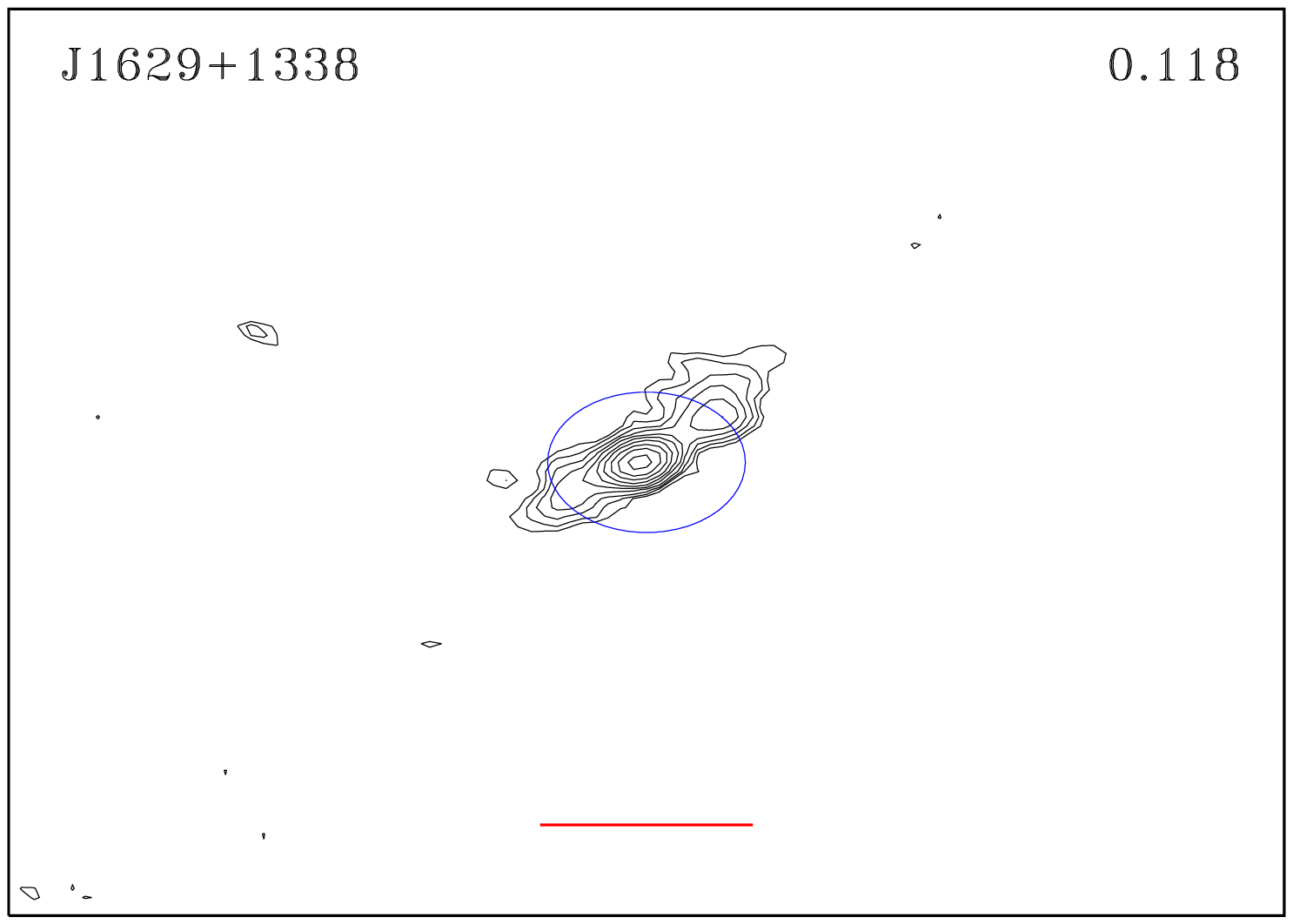} 

\includegraphics[width=6.3cm,height=6.3cm]{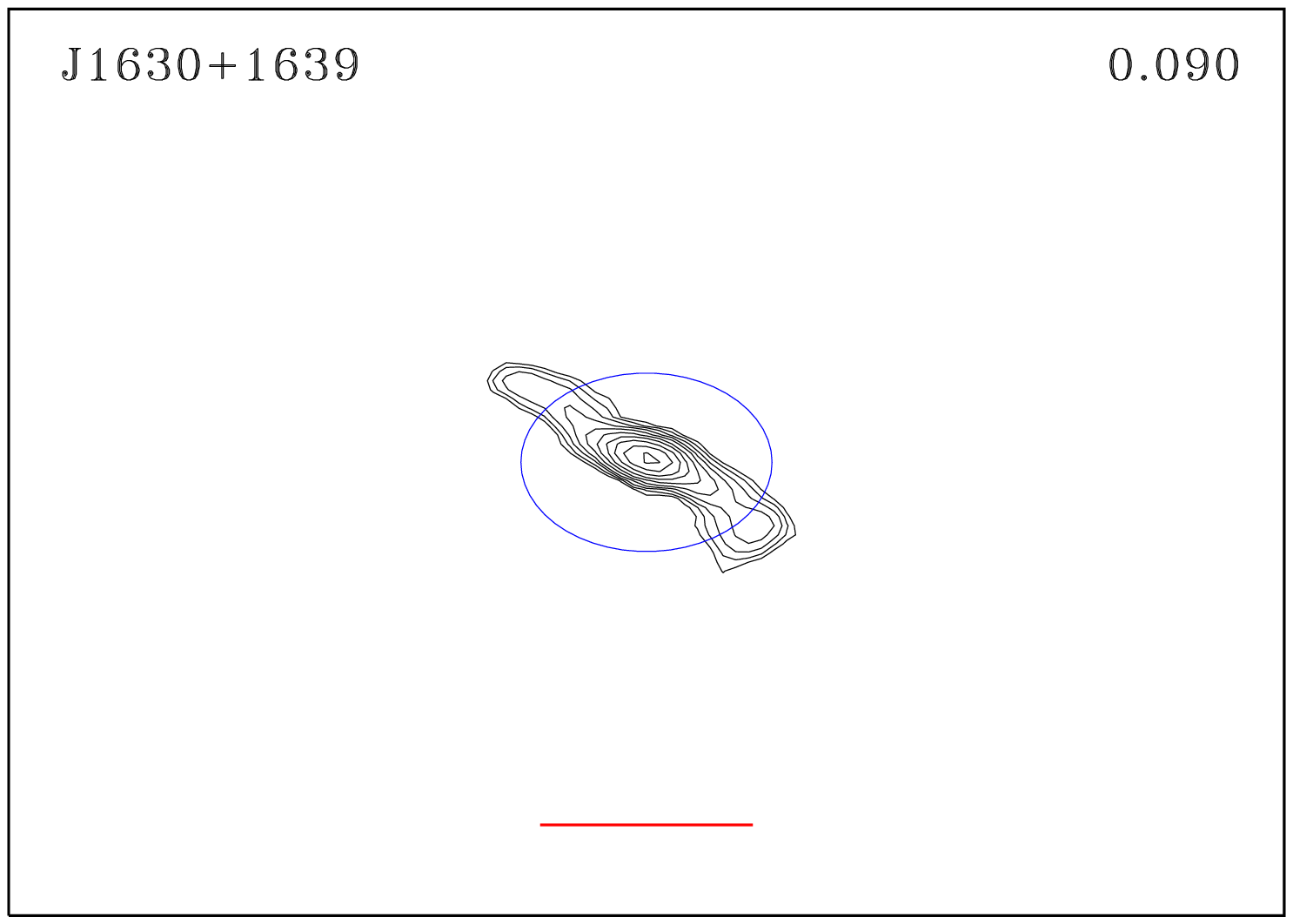} 
\includegraphics[width=6.3cm,height=6.3cm]{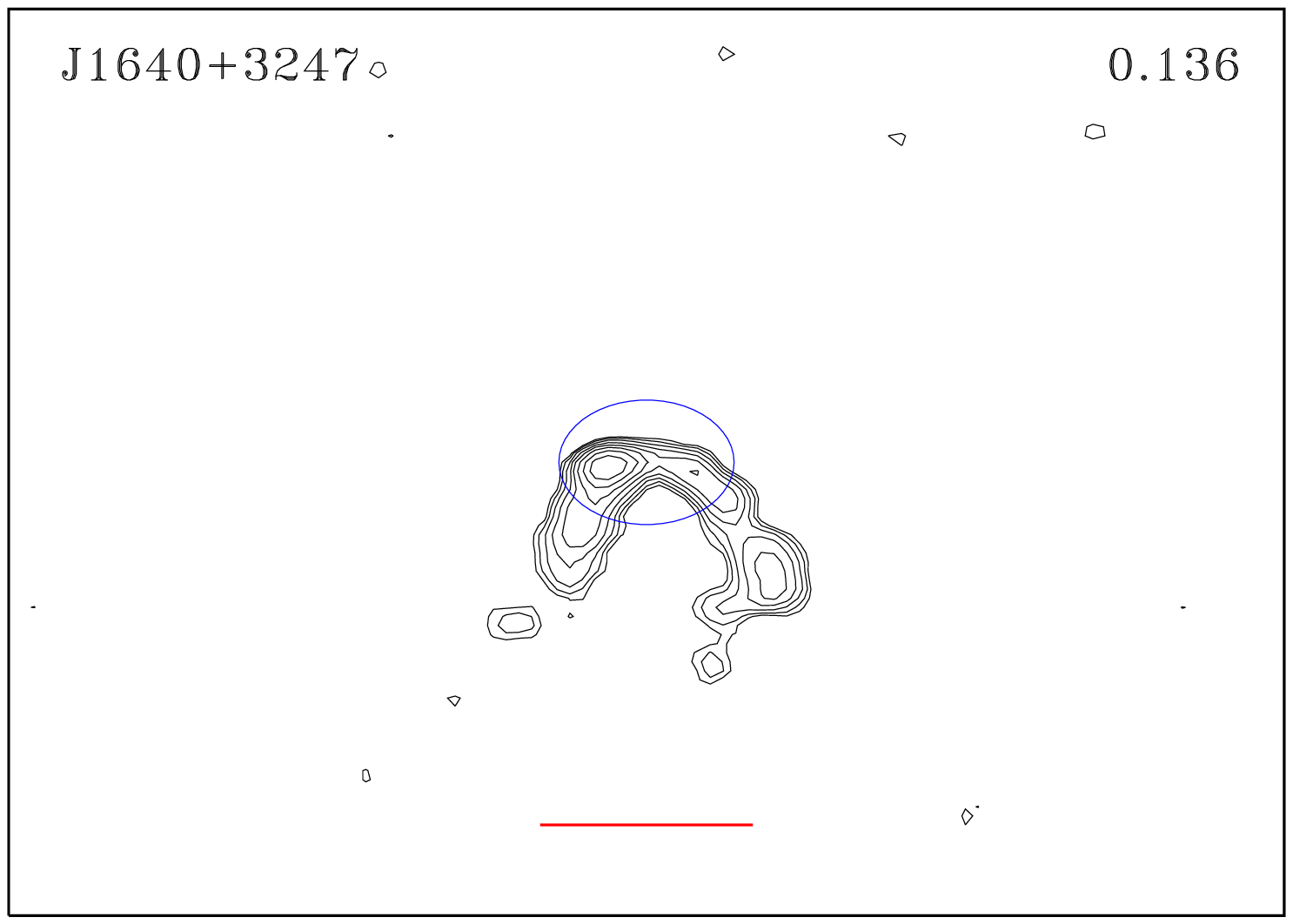} 
\includegraphics[width=6.3cm,height=6.3cm]{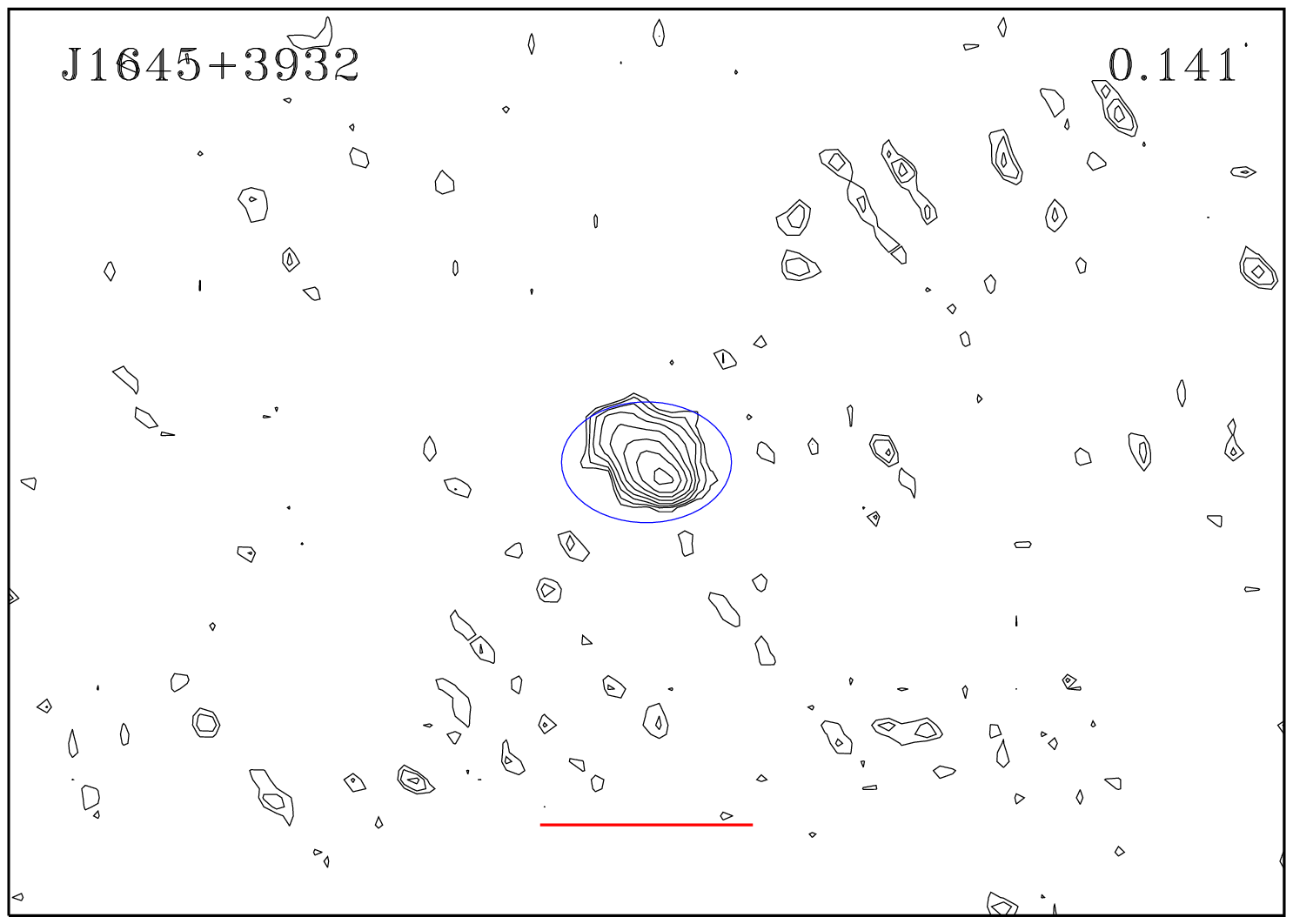} 

\includegraphics[width=6.3cm,height=6.3cm]{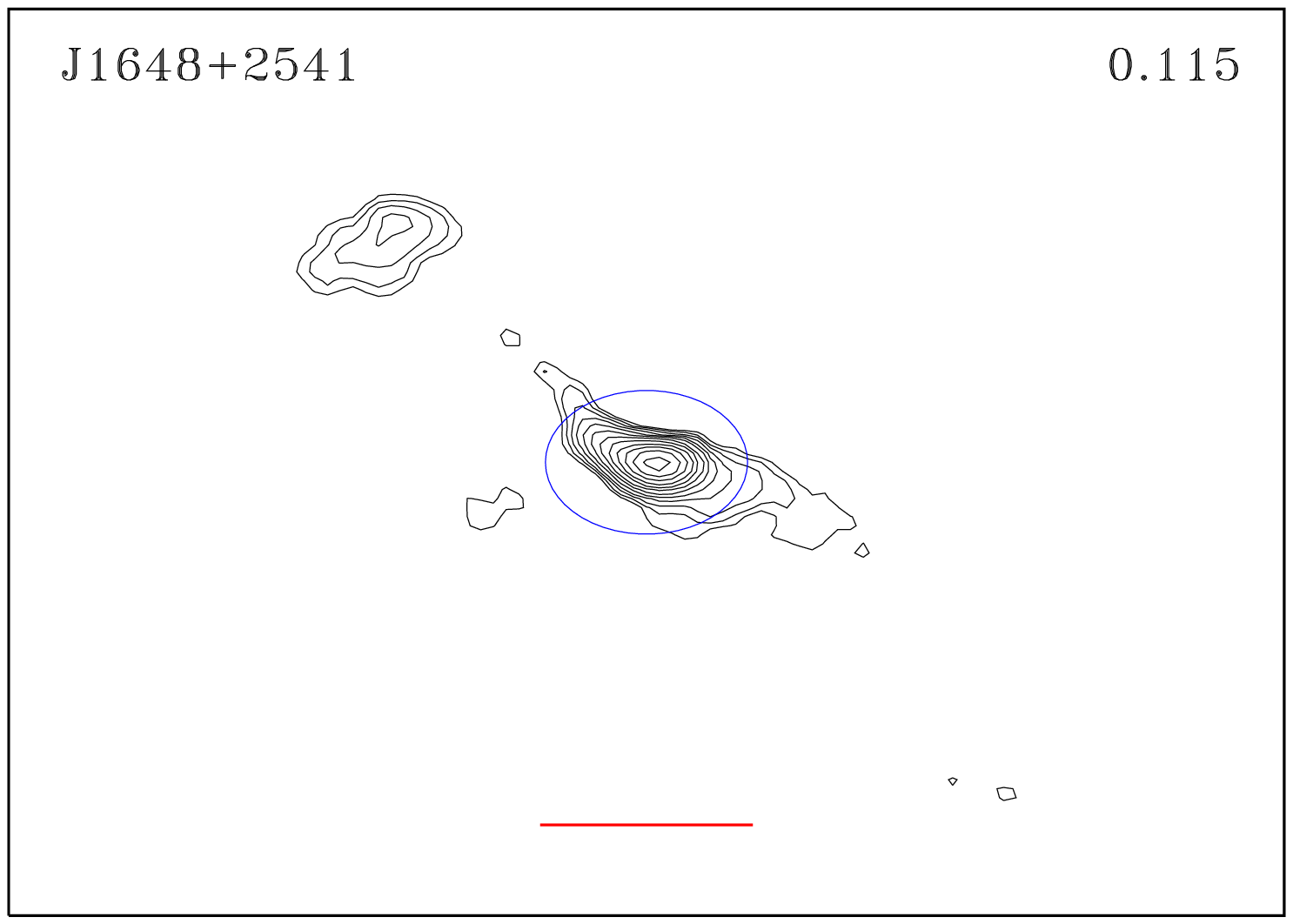} 
\includegraphics[width=6.3cm,height=6.3cm]{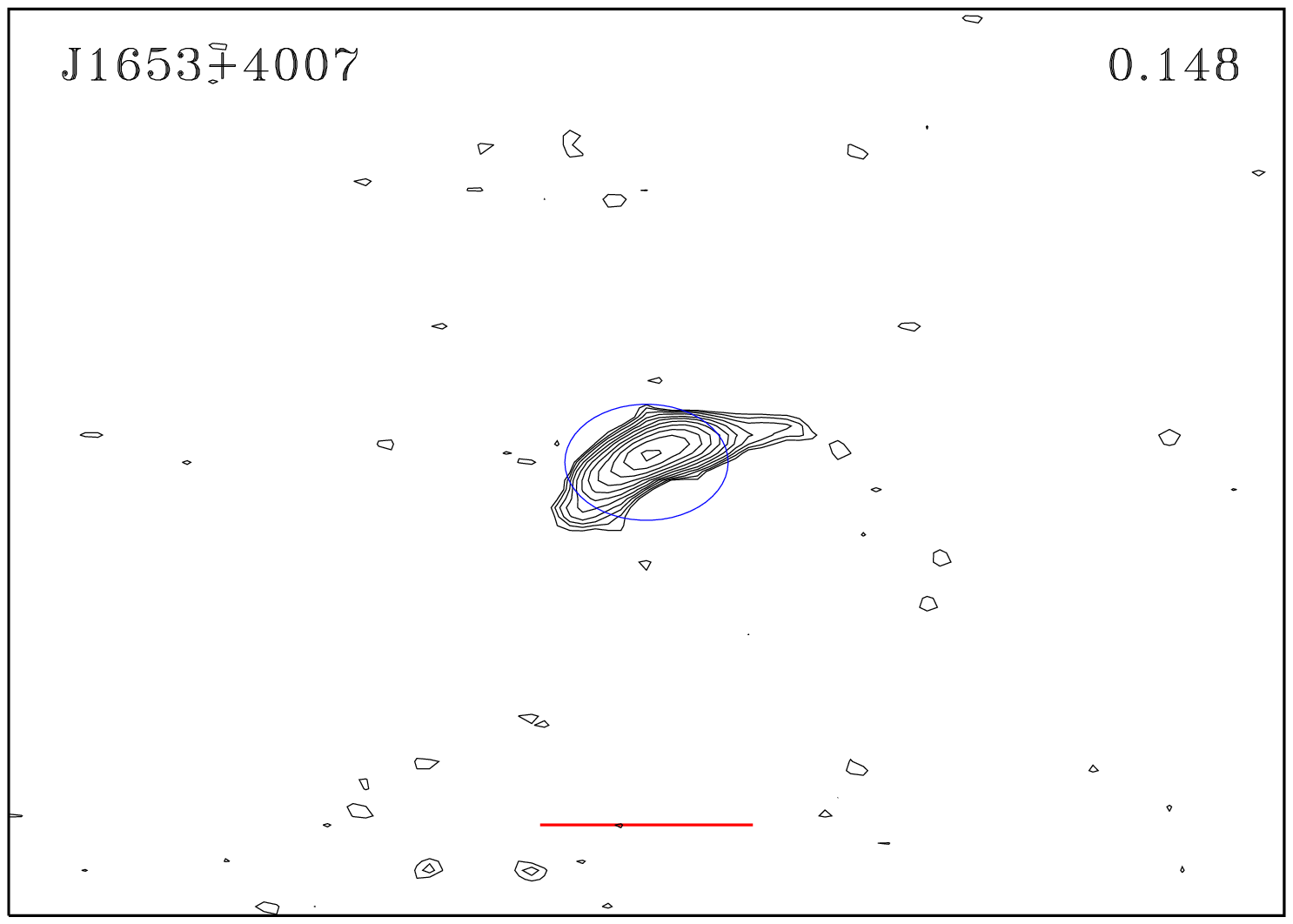} 
\includegraphics[width=6.3cm,height=6.3cm]{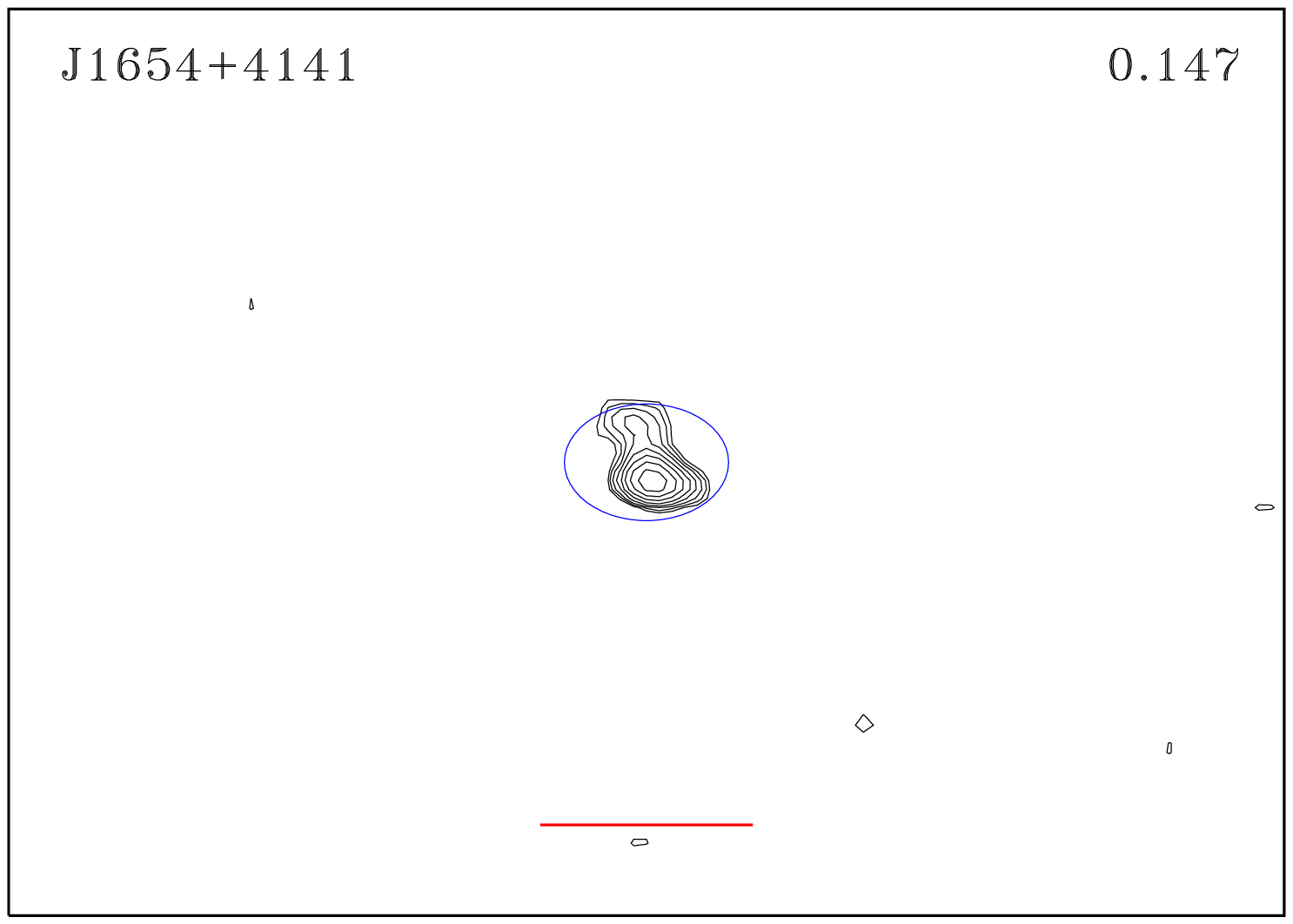} 
\caption{(continued)}
\end{figure*}

\addtocounter{figure}{-1}
\begin{figure*}
\includegraphics[width=6.3cm,height=6.3cm]{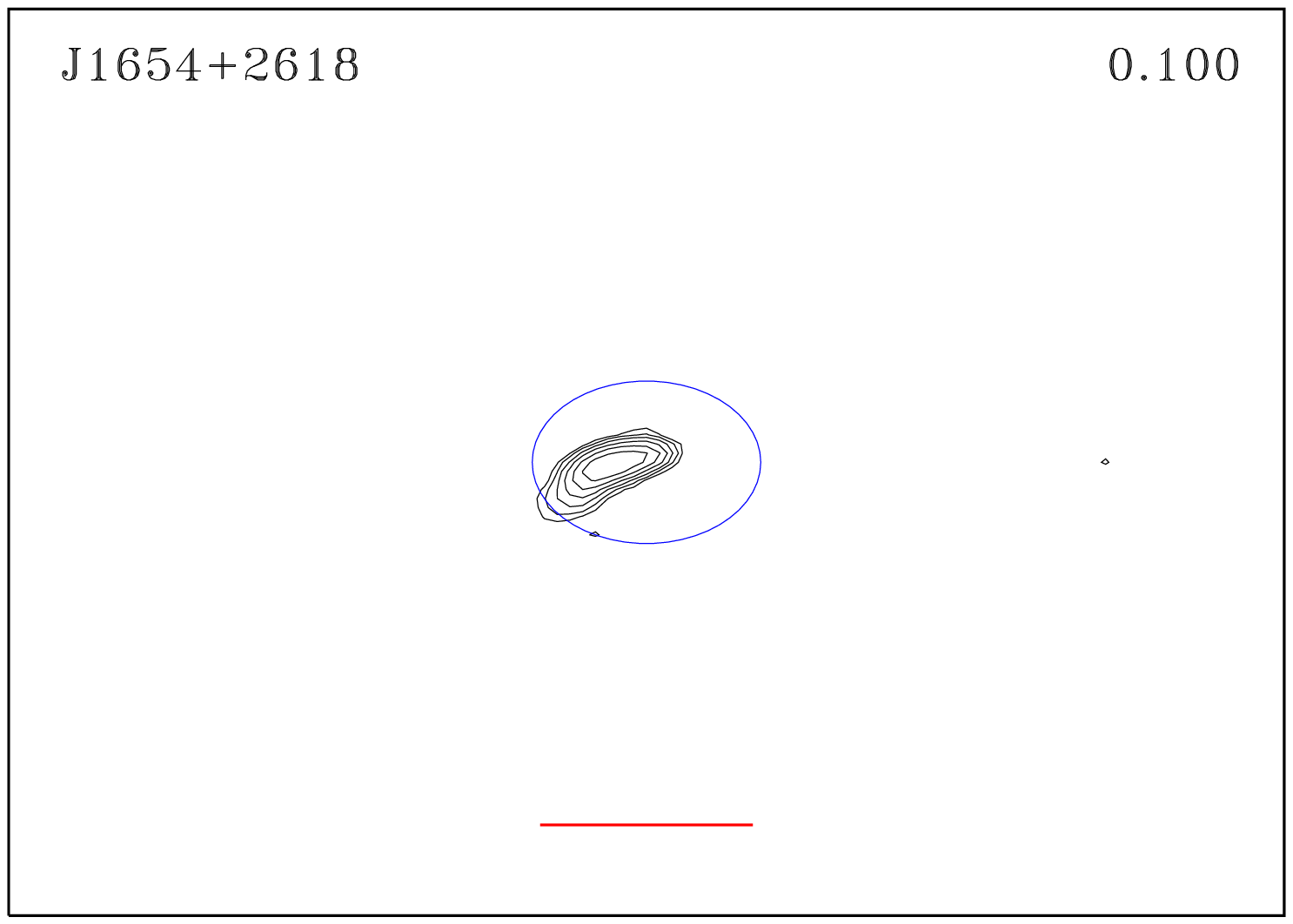} 
\includegraphics[width=6.3cm,height=6.3cm]{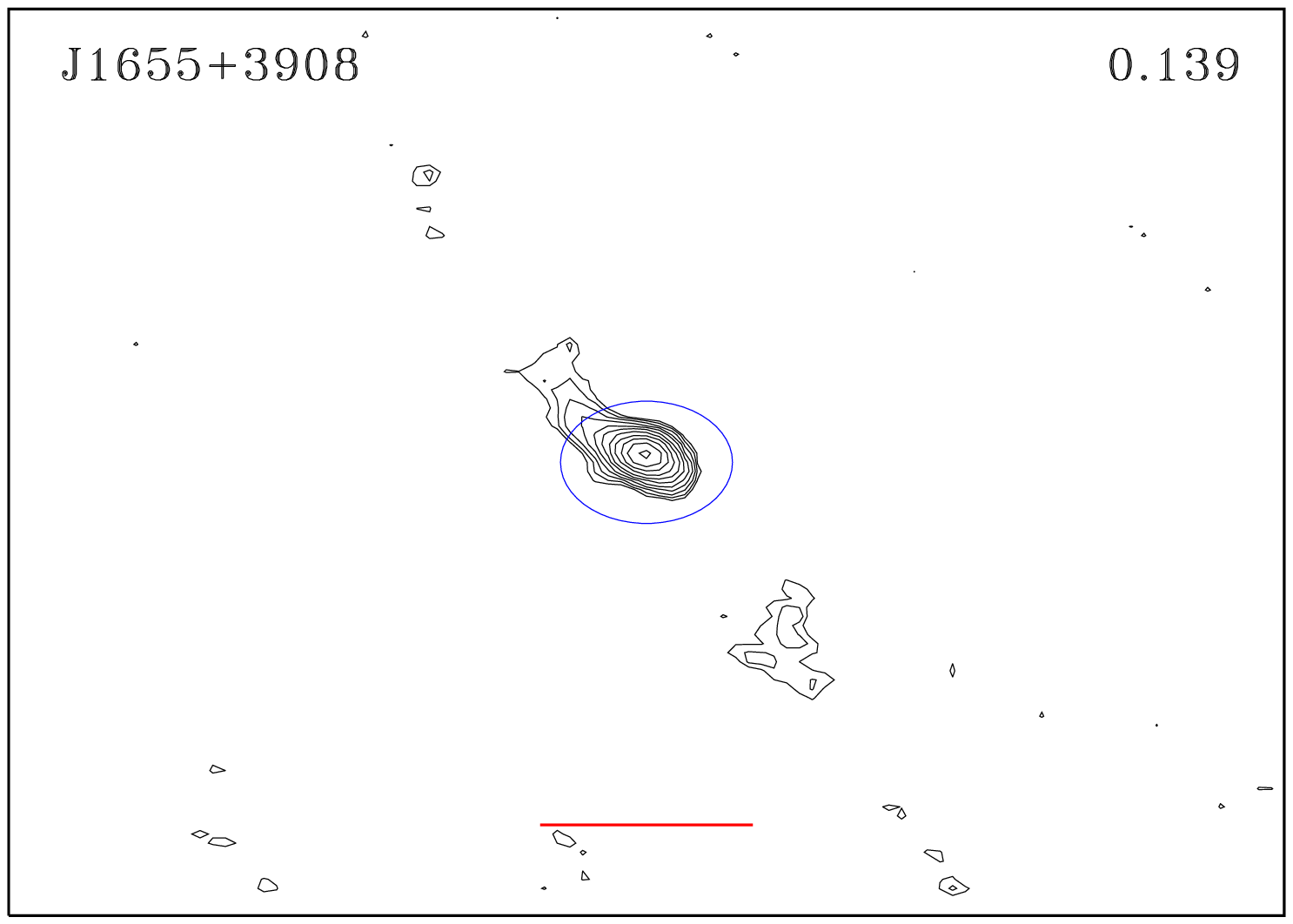} 
\includegraphics[width=6.3cm,height=6.3cm]{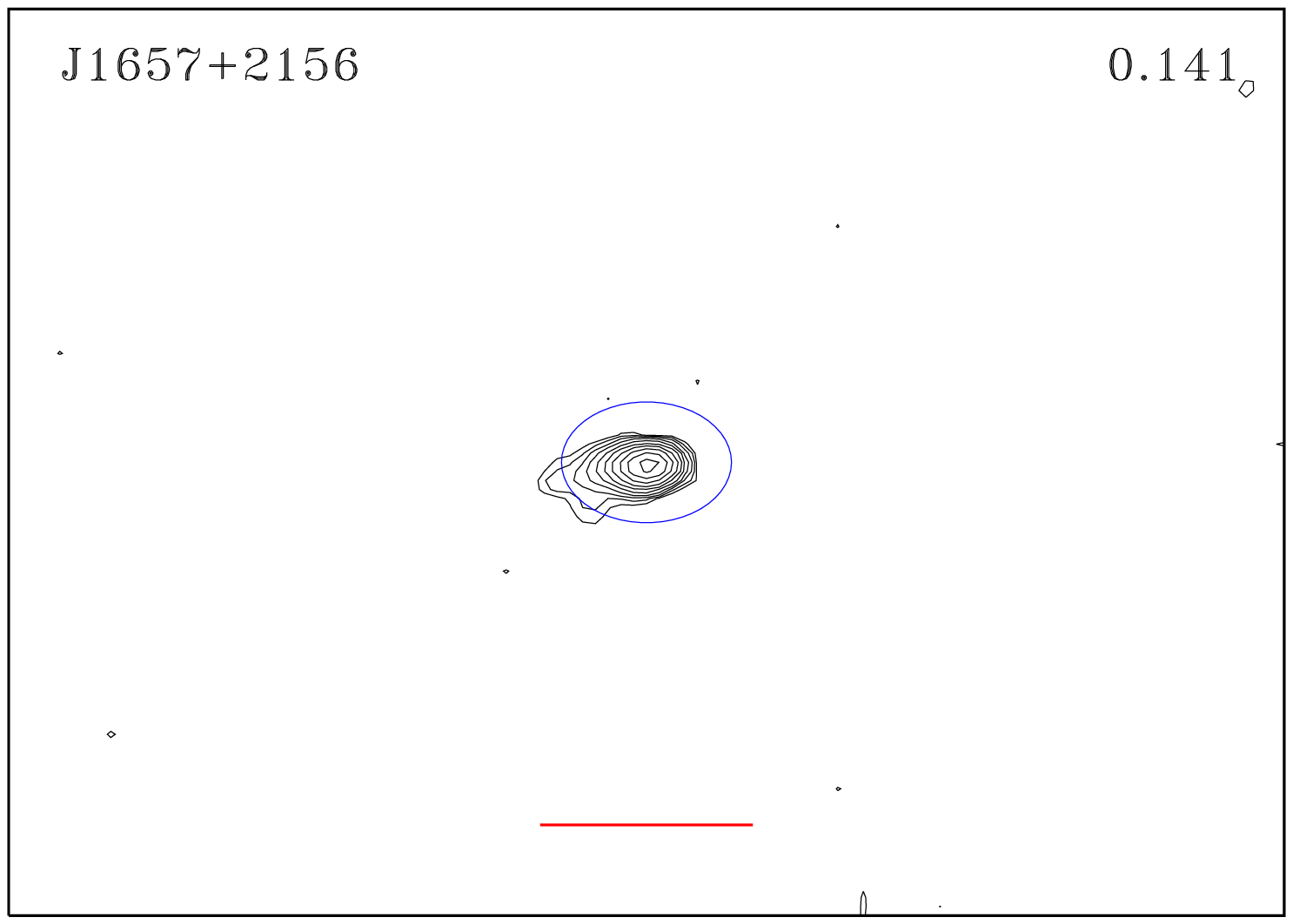} 

\includegraphics[width=6.3cm,height=6.3cm]{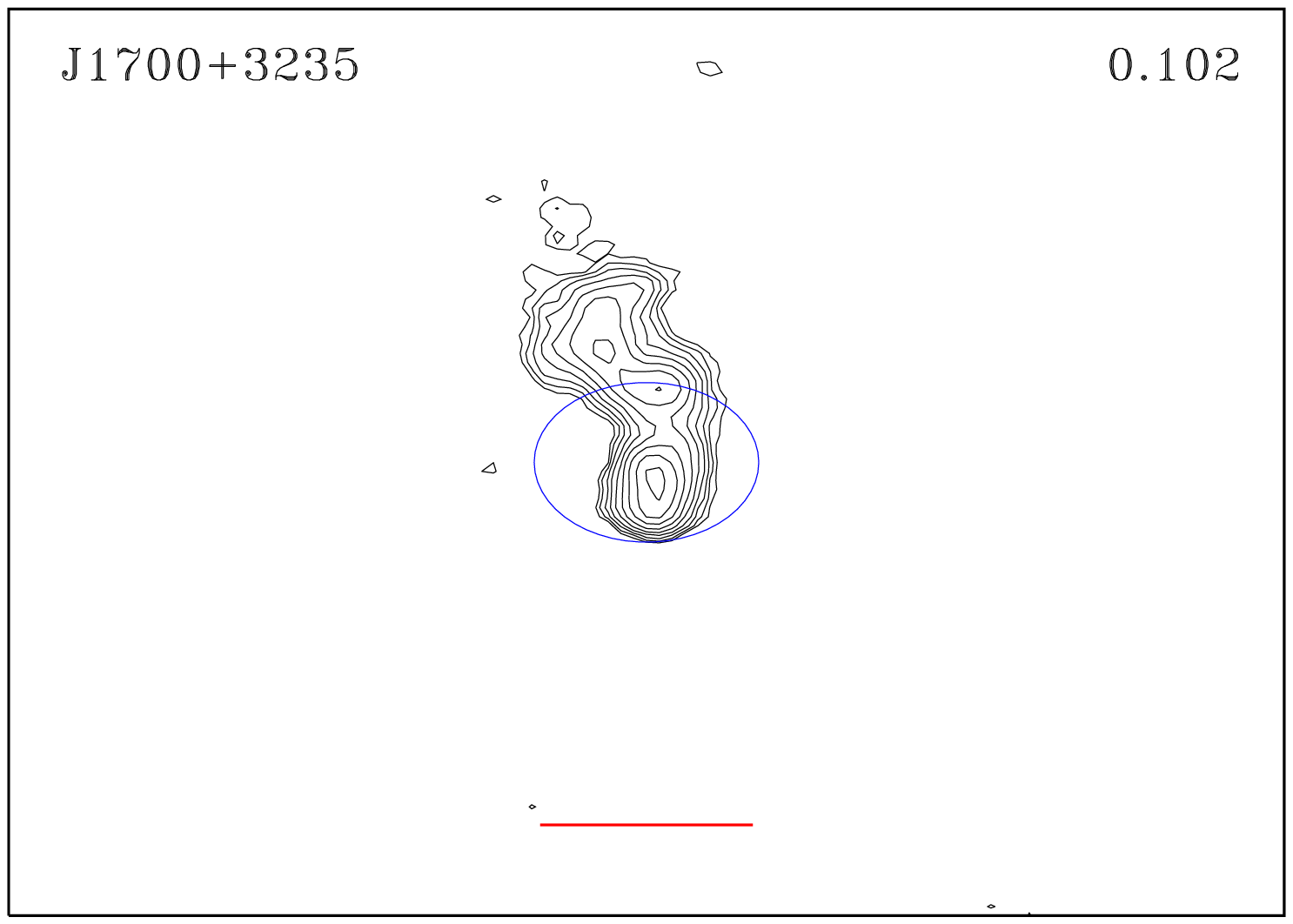} 
\includegraphics[width=6.3cm,height=6.3cm]{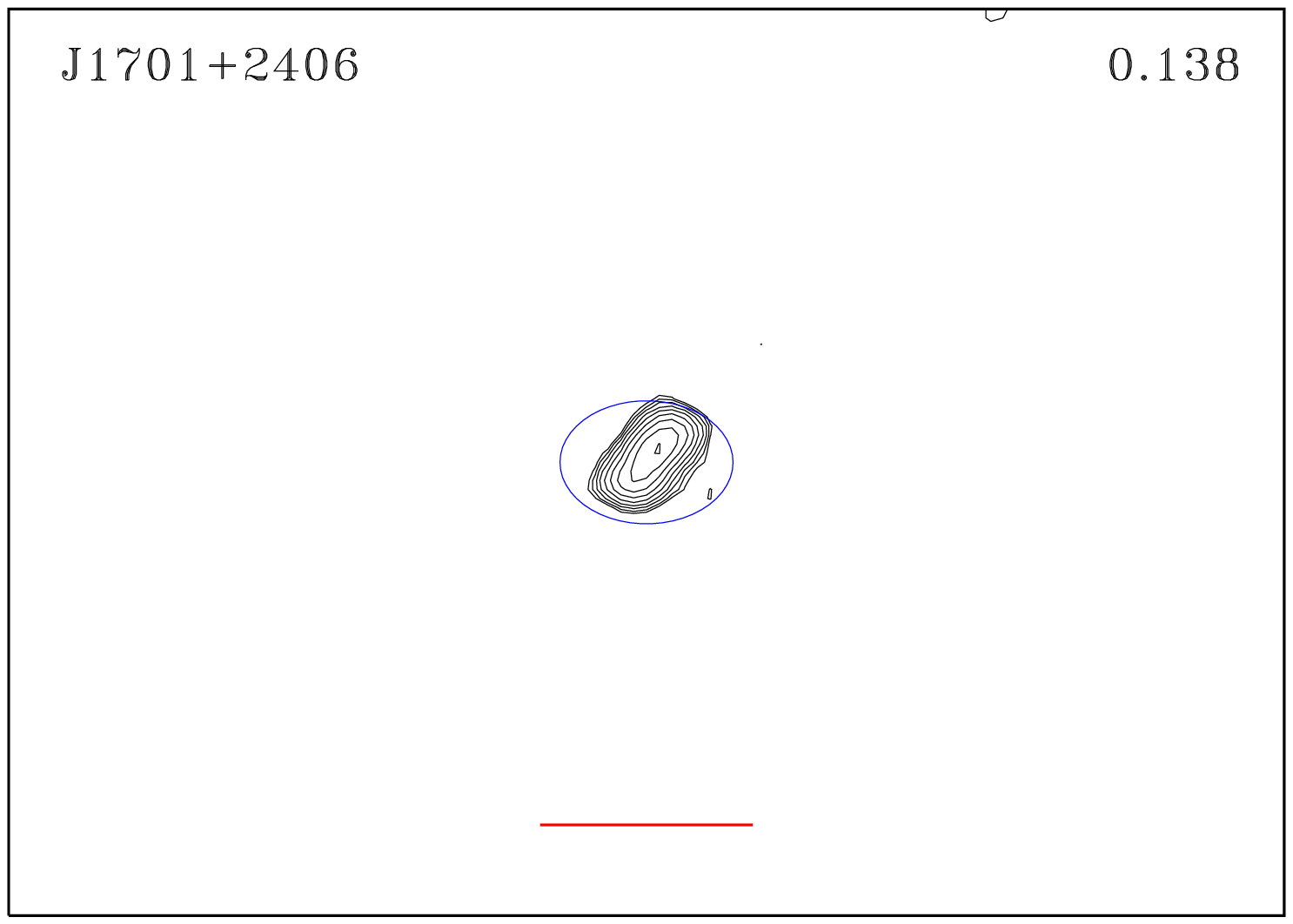} 
\includegraphics[width=6.3cm,height=6.3cm]{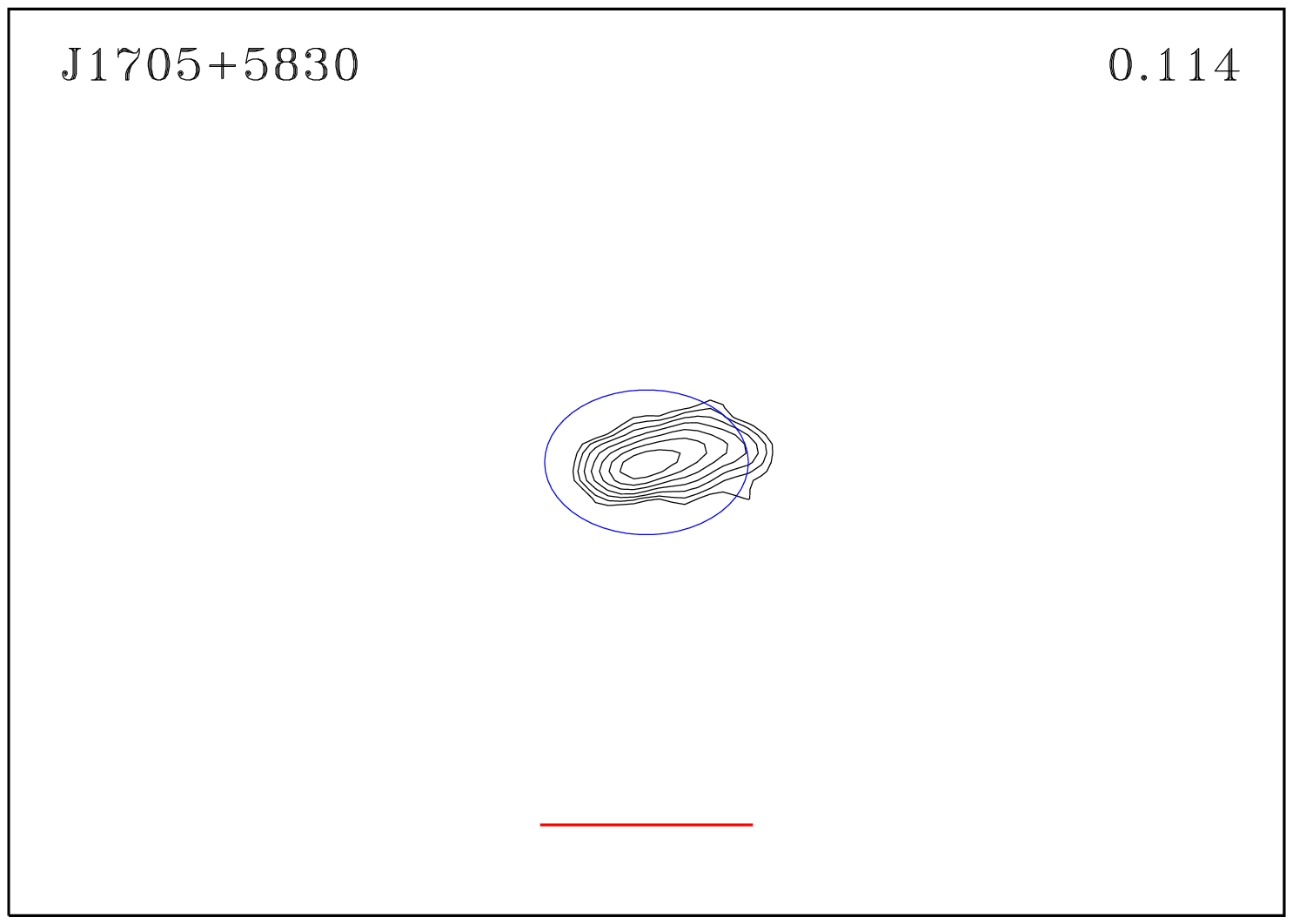} 

\includegraphics[width=6.3cm,height=6.3cm]{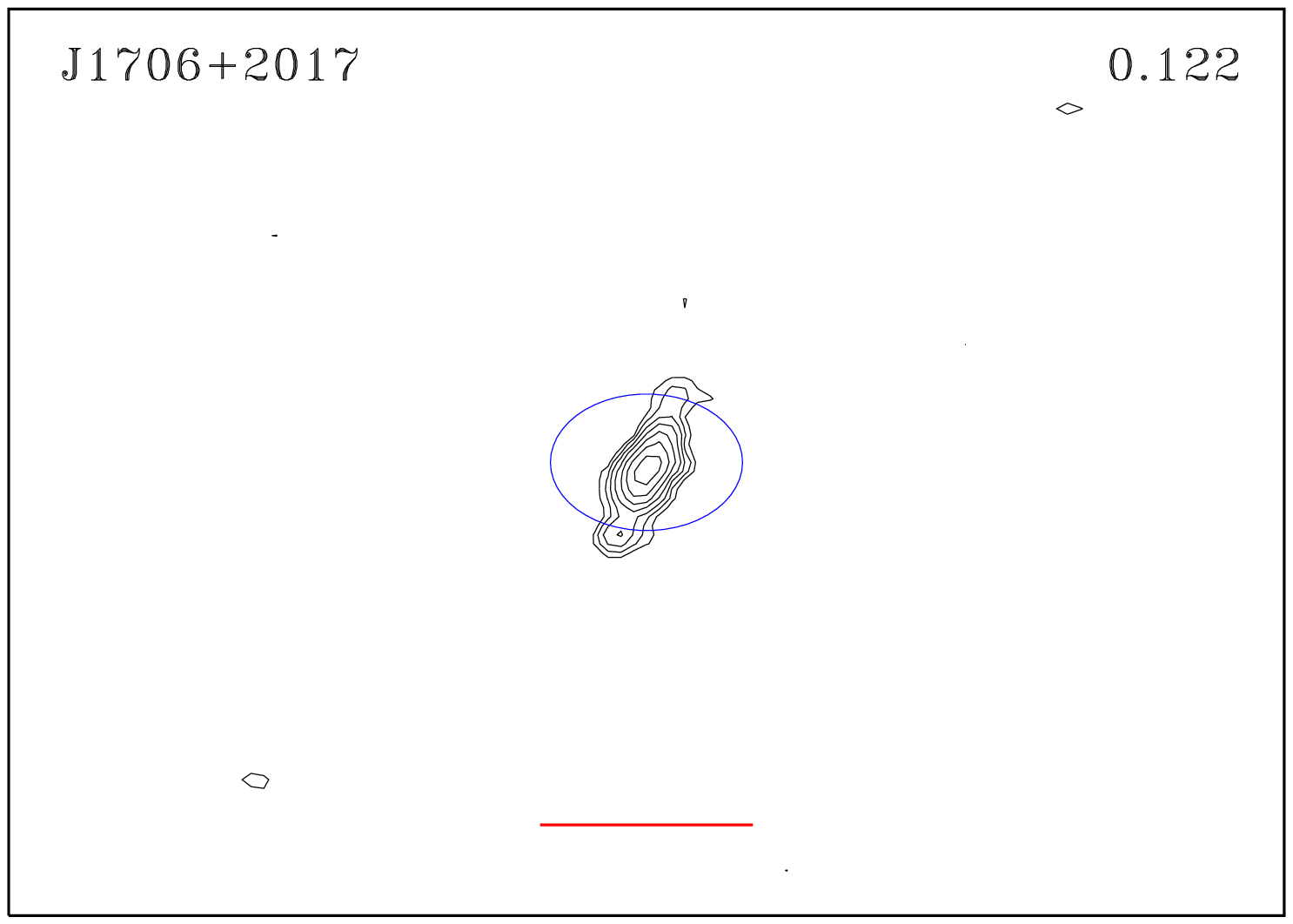} 
\includegraphics[width=6.3cm,height=6.3cm]{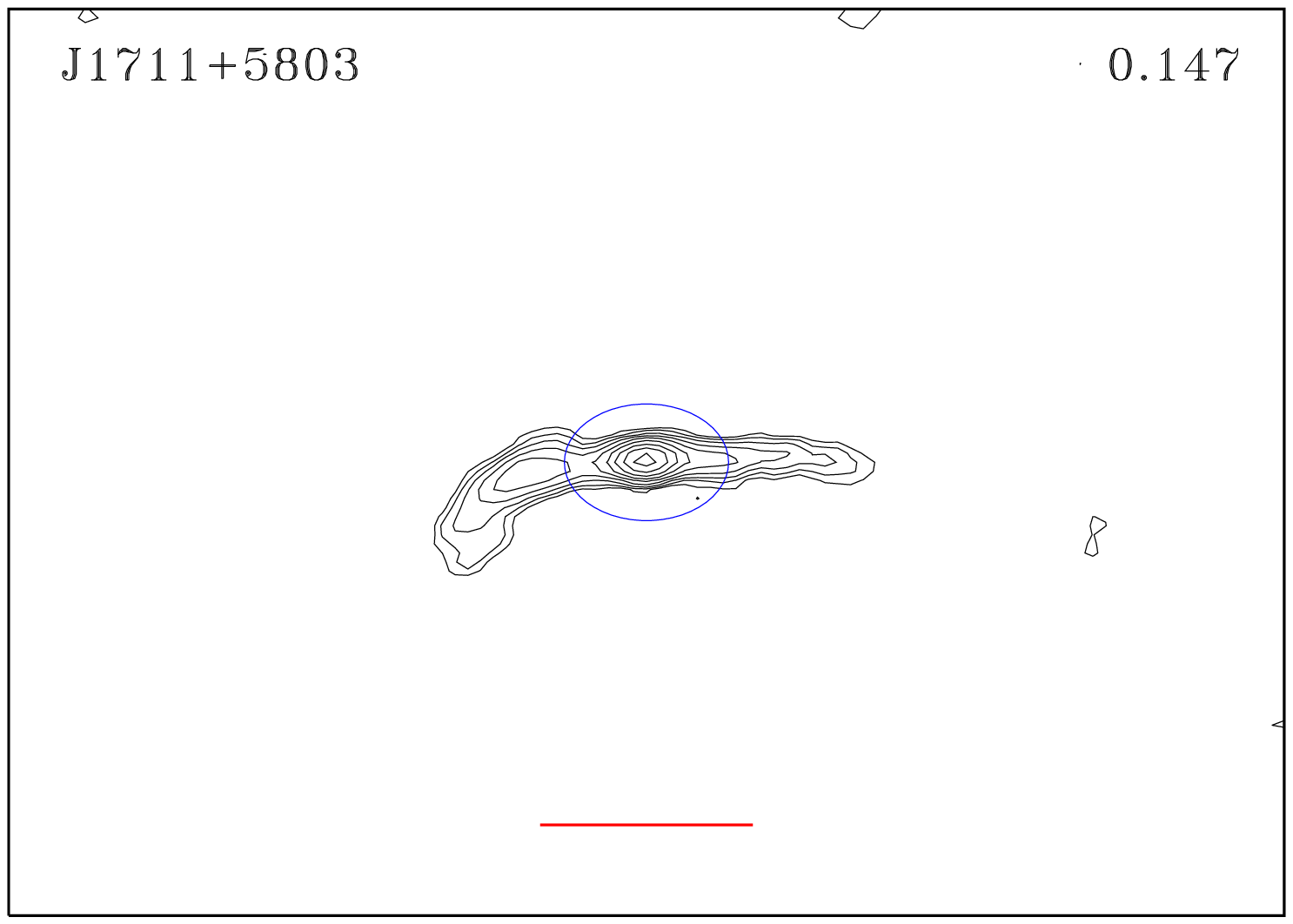} 
\includegraphics[width=6.3cm,height=6.3cm]{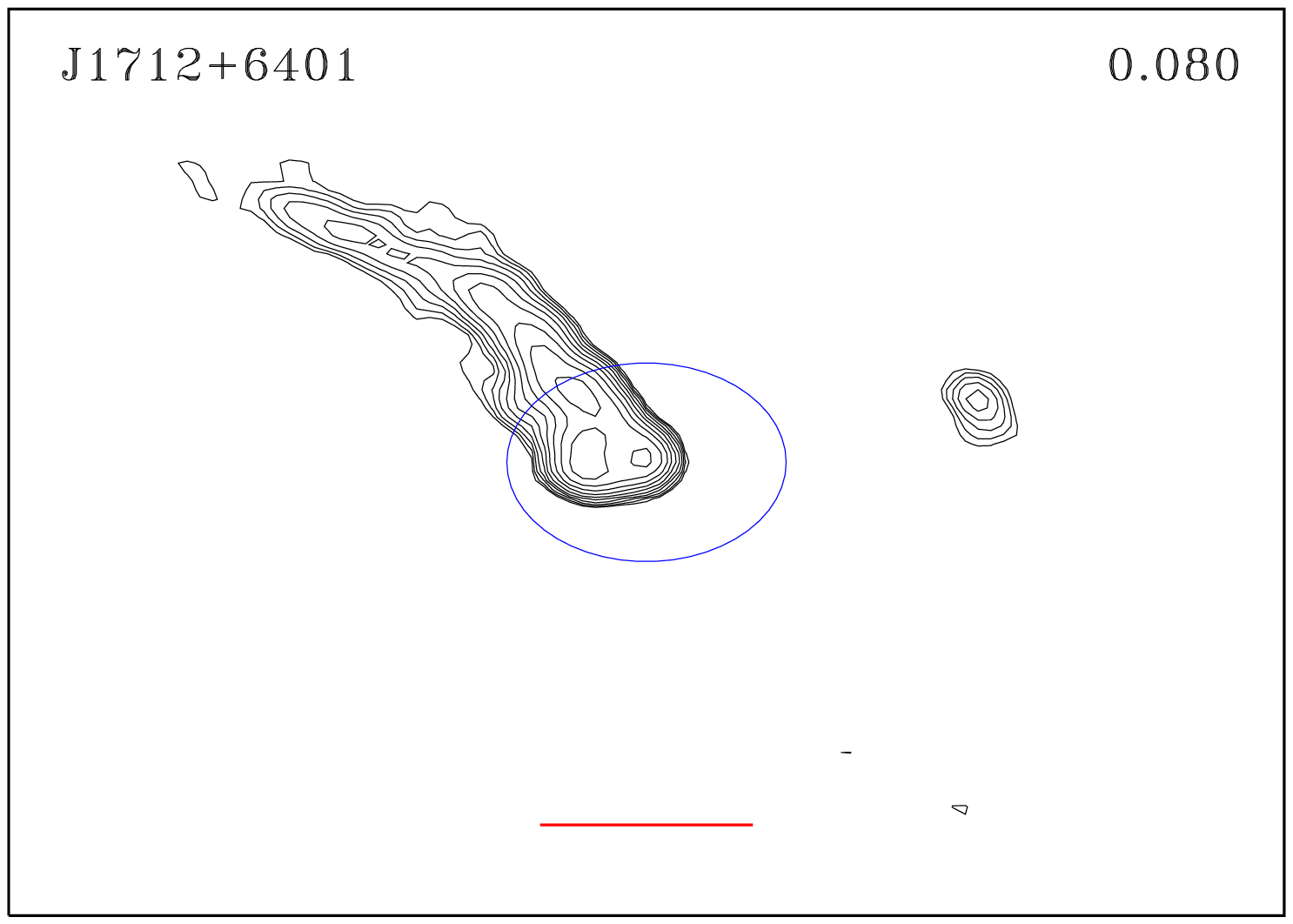} 

\includegraphics[width=6.3cm,height=6.3cm]{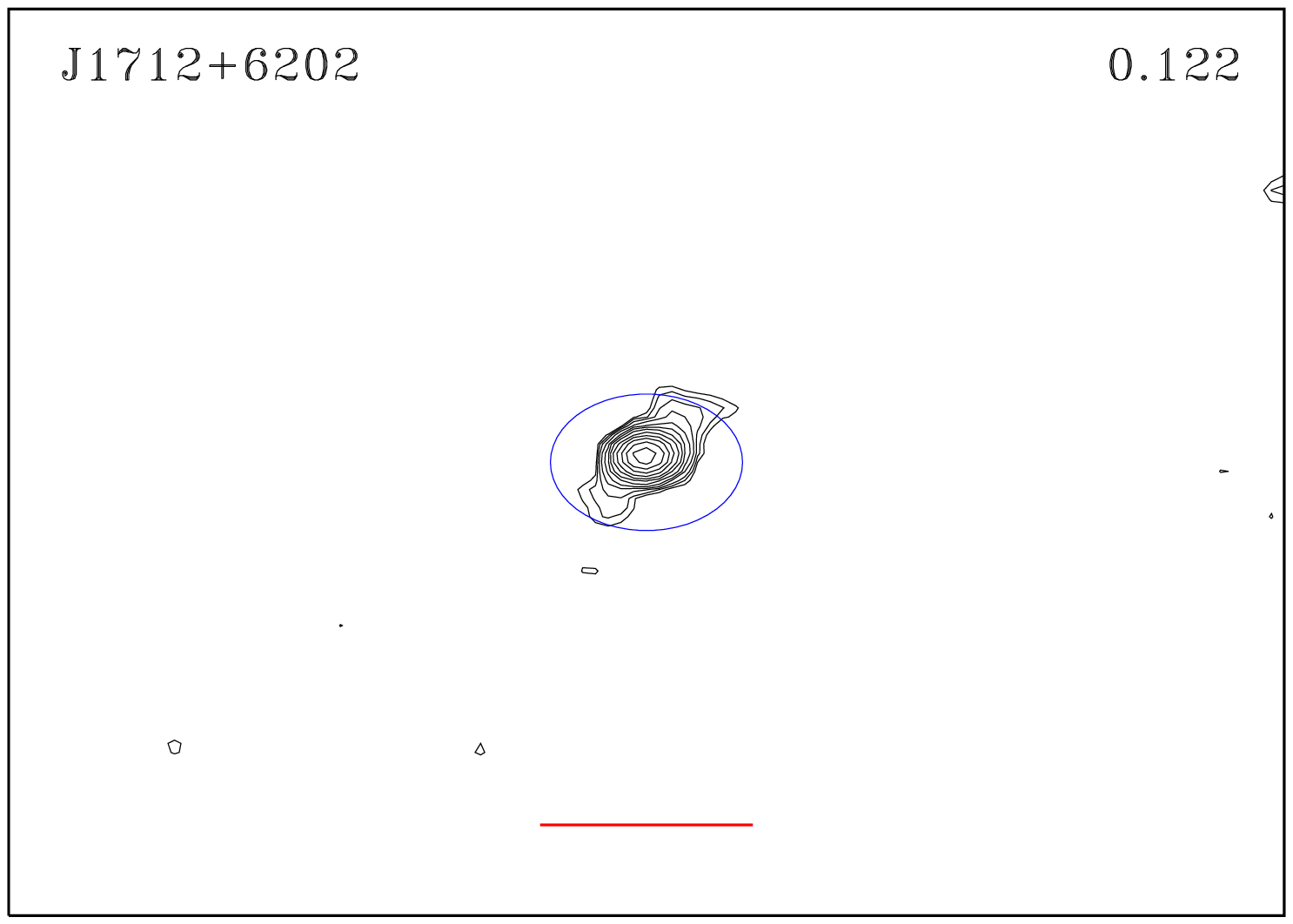} 
\includegraphics[width=6.3cm,height=6.3cm]{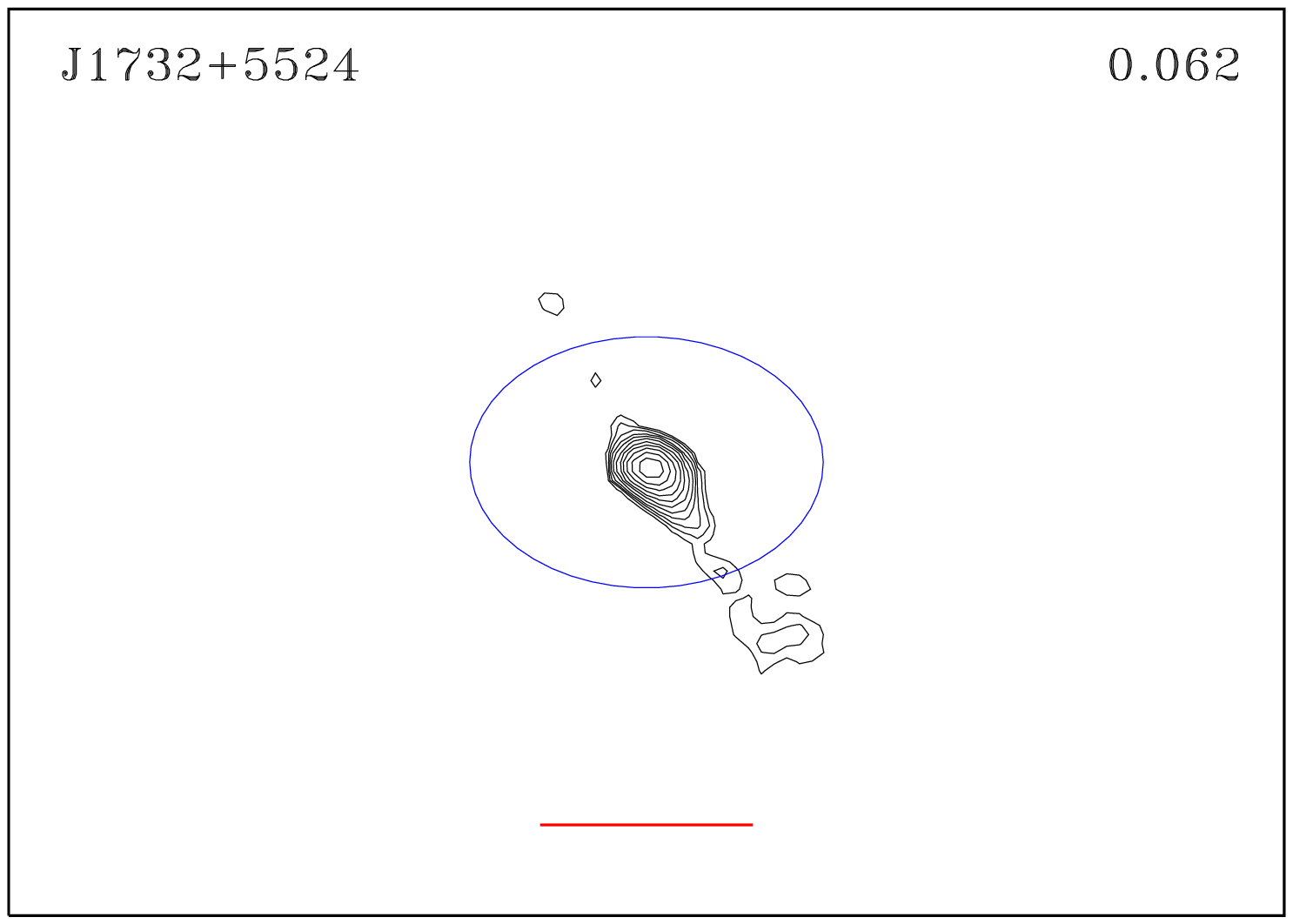} 
\includegraphics[width=6.3cm,height=6.3cm]{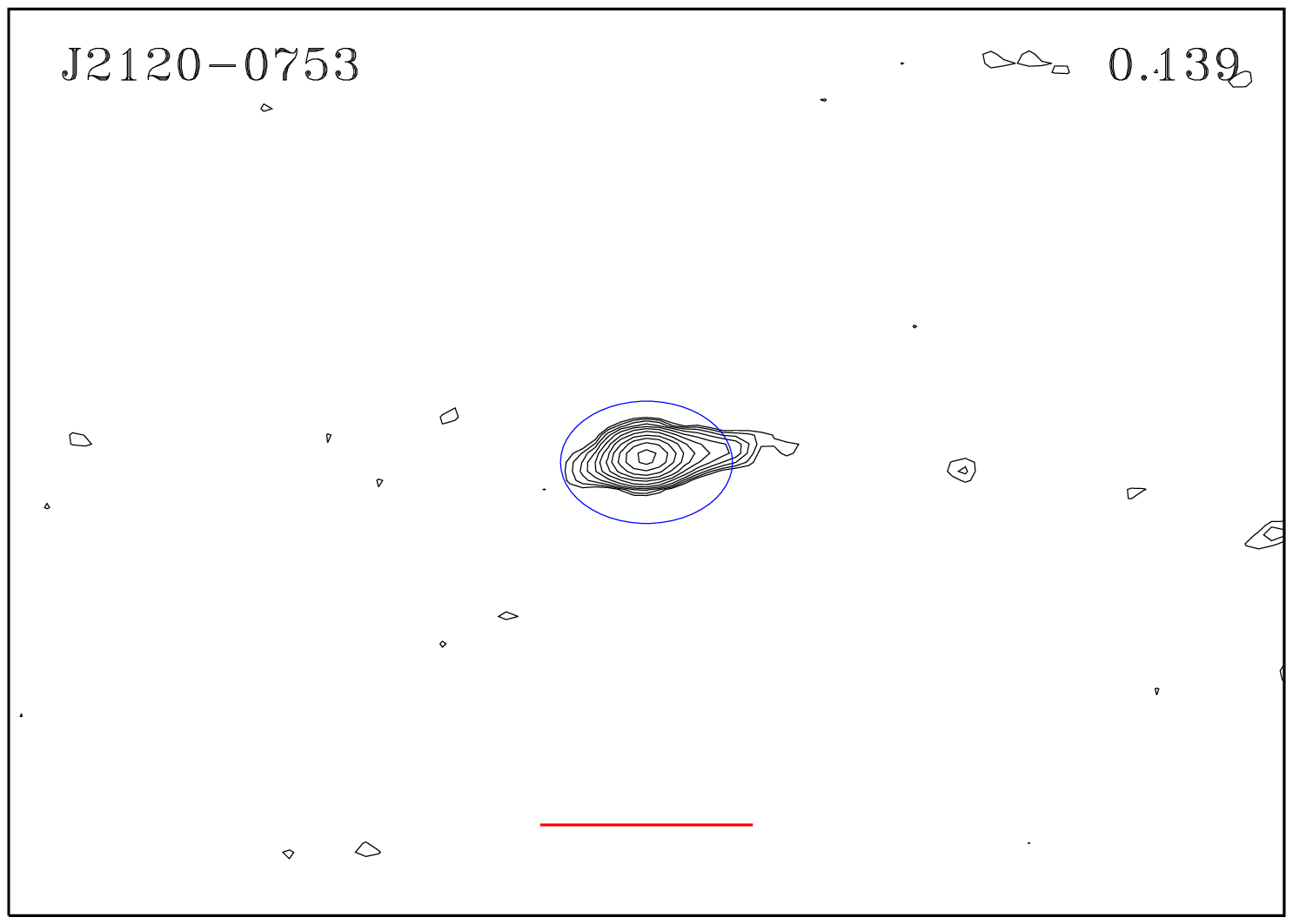} 
\caption{(continued)}
\end{figure*}

\addtocounter{figure}{-1}
\begin{figure*}
\includegraphics[width=6.3cm,height=6.3cm]{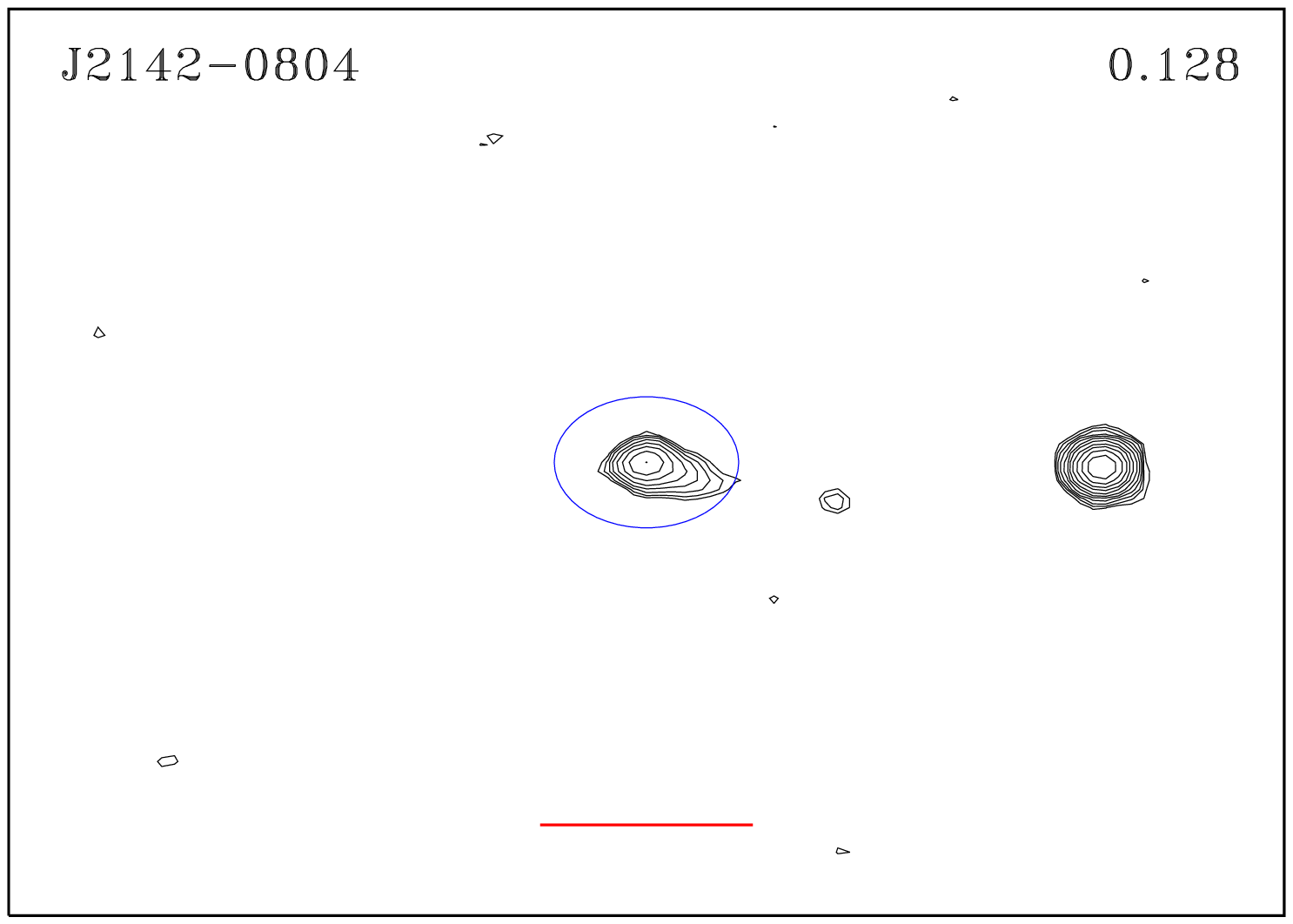} 
\includegraphics[width=6.3cm,height=6.3cm]{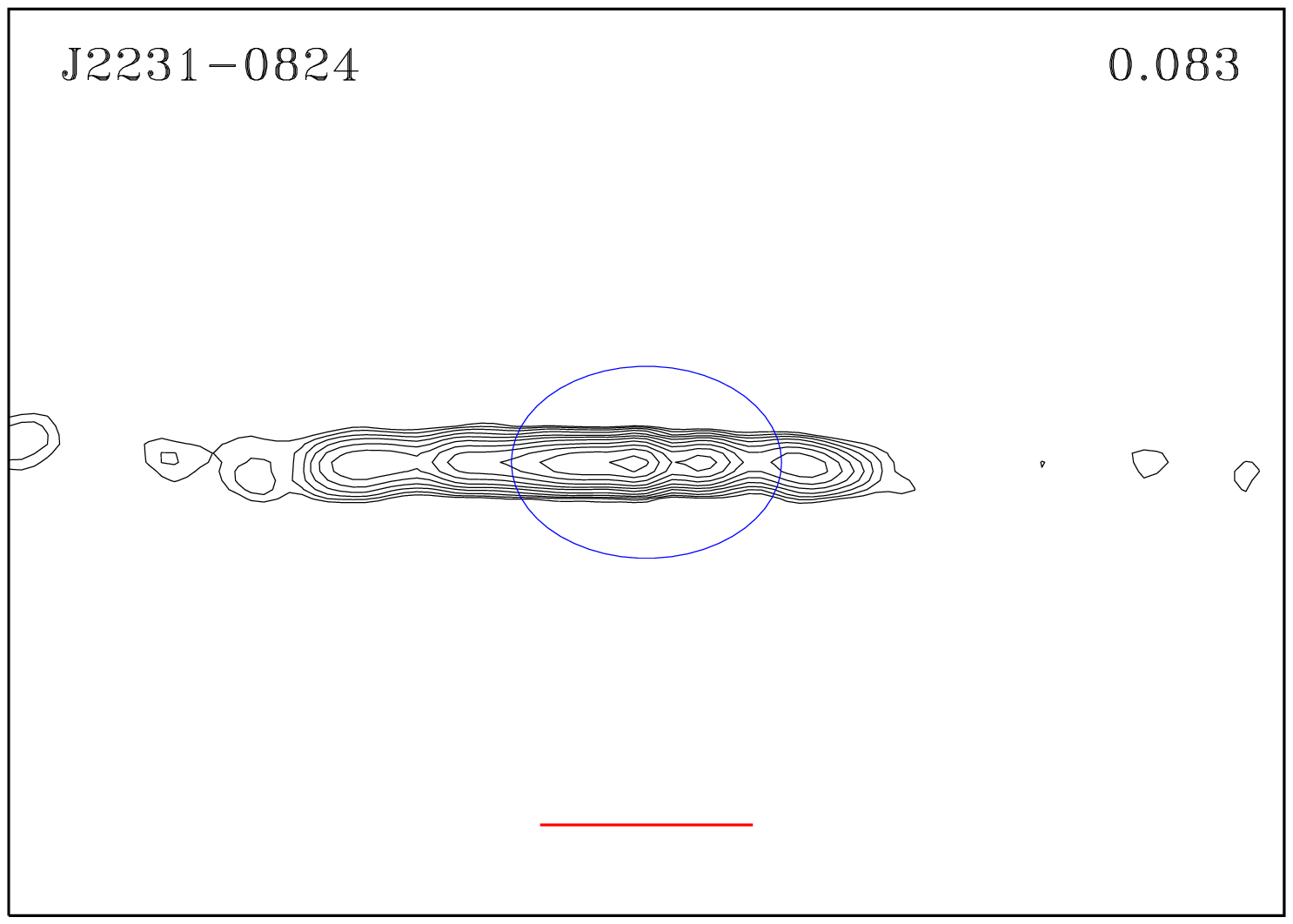} 
\includegraphics[width=6.3cm,height=6.3cm]{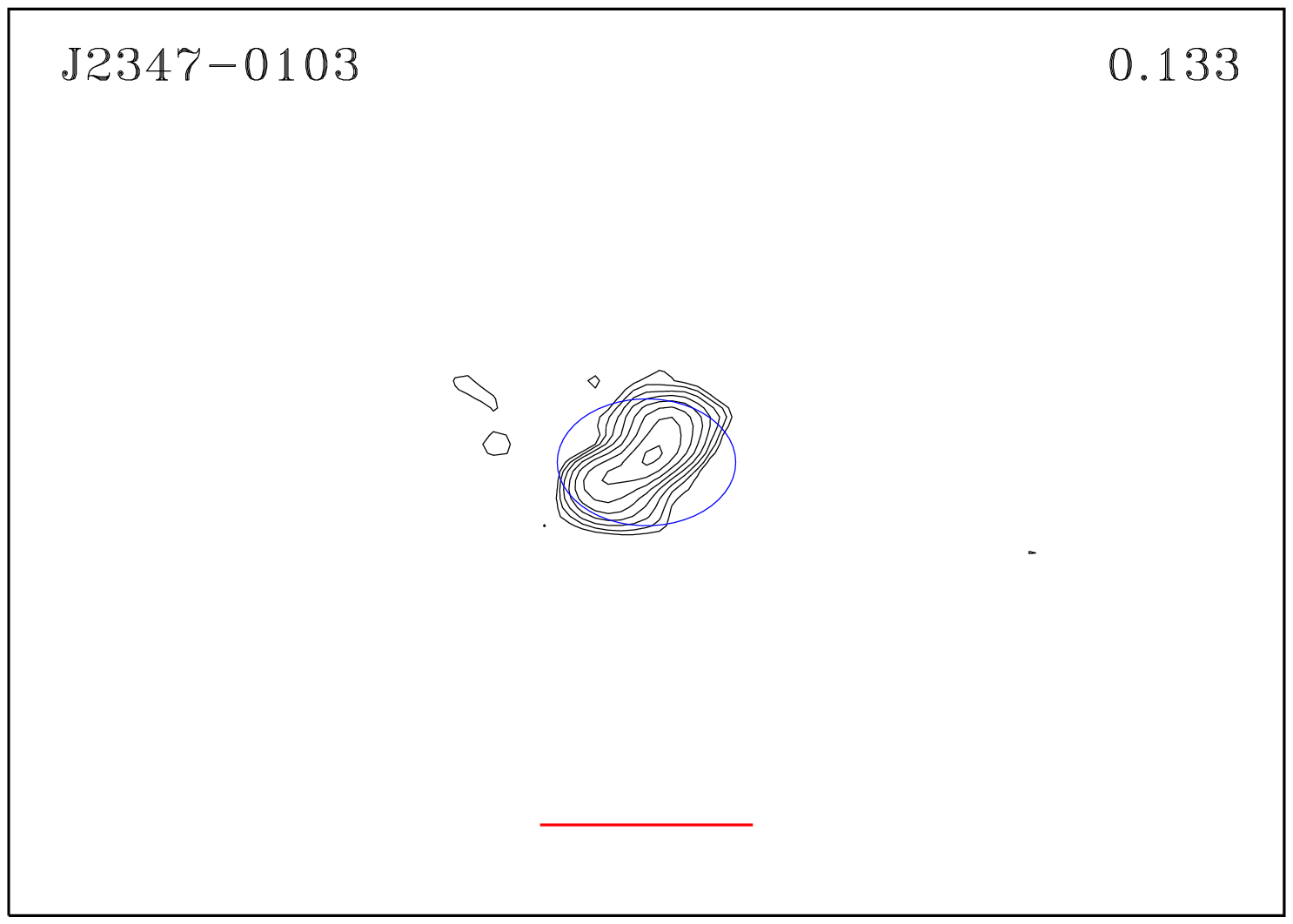} 
\caption{(continued)}
\end{figure*}

\end{document}